\documentclass[11pt,a4paper]{report}

\usepackage[margin=2.5cm]{geometry}
\usepackage{eso-pic}  

\usepackage[utf8]{inputenc}
\usepackage[T1]{fontenc}
\usepackage{lmodern}
\usepackage{microtype}

\usepackage{amsmath,amssymb,amsfonts,amsthm}
\usepackage{graphicx}
\graphicspath{%
  {./}%
  {./fig/}%
  {./fig/chapter11/}%
}
\usepackage{bm}

\usepackage{tikz}
\usetikzlibrary{shapes,arrows.meta,positioning,calc,fit,backgrounds,patterns,decorations.pathreplacing}
\usepackage{pgfplots}
\pgfplotsset{compat=1.18}
\usepgfplotslibrary{fillbetween}

\usepackage{booktabs}
\usepackage{array}
\usepackage{enumitem}
\usepackage{multicol}
\usepackage{ragged2e}

\usepackage{listings}
\definecolor{codegreen}{rgb}{0,0.5,0}
\definecolor{codegray}{rgb}{0.5,0.5,0.5}
\definecolor{codepurple}{rgb}{0.58,0,0.82}
\definecolor{backcolor}{rgb}{0.97,0.97,0.97}

\lstdefinestyle{pythonstyle}{
    language=Python,
    backgroundcolor=\color{backcolor},
    commentstyle=\color{codegreen},
    keywordstyle=\color{blue!80!black},
    numberstyle=\tiny\color{codegray},
    stringstyle=\color{codepurple},
    basicstyle=\small\ttfamily,
    breakatwhitespace=false,
    breaklines=true,
    captionpos=b,
    keepspaces=true,
    numbers=left,
    numbersep=5pt,
    showspaces=false,
    showstringspaces=false,
    showtabs=false,
    tabsize=4,
    frame=single,
    framerule=0.4pt,
    rulecolor=\color{codegray},
    aboveskip=0.8em,
    belowskip=0.8em,
}
\usepackage[most]{tcolorbox}

\usepackage{xcolor}
\definecolor{uzhblue}{rgb}{0,0,0.5}
\definecolor{uzhgreylight}{RGB}{236,238,241}
\definecolor{uzhgreydark}{RGB}{81,86,94}
\definecolor{harvardcrimson}{RGB}{153,0,0}
\definecolor{darkgreen}{RGB}{0,100,0}
\definecolor{darkred}{RGB}{139,0,0}
\definecolor{softblue}{RGB}{70,130,180}
\definecolor{softorange}{RGB}{230,159,0}
\definecolor{softgreen}{RGB}{0,158,115}

\usepackage{hyperref}
\hypersetup{
    colorlinks=true,
    linkcolor=uzhblue,
    citecolor=uzhblue,
    urlcolor=harvardcrimson,
    hypertexnames=false,
    breaklinks=true       
}
\PassOptionsToPackage{hyphens}{url}
\usepackage{natbib}
\usepackage{algorithm}
\usepackage{algorithmic}

\usepackage{float}

\usepackage{pifont}

\newcommand{\x}{\bm{x}}
\newcommand{\h}{\bm{h}}
\newcommand{\y}{\bm{y}}
\newcommand{\w}{\bm{w}}
\newcommand{\bb}{\bm{b}}
\newcommand{\z}{\bm{z}}
\newcommand{\W}{\bm{W}}
\newcommand{\Wh}{\bm{W}_{hh}}
\newcommand{\Wx}{\bm{W}_{xh}}

\newcommand{\X}{\bm{X}}
\renewcommand{\a}{\bm{a}}
\newcommand{\R}{\mathbb{R}}
\newcommand{\E}[1]{\mathbb{E}\!\left[#1\right]}
\DeclareMathOperator*{\argmin}{arg\,min}
\DeclareMathOperator*{\argmax}{arg\,max}
\newcommand{\emphc}[1]{\textcolor{harvardcrimson}{\textbf{#1}}}
\DeclareUrlCommand\tpath{\urlstyle{tt}}   
\newcolumntype{L}[1]{>{\RaggedRight\arraybackslash}p{#1}}


\newtcolorbox{definitionbox}[1][]{
    colback=uzhgreylight,
    colframe=uzhblue,
    fonttitle=\bfseries,
    title={#1},
    rounded corners,
    boxrule=0.8pt,
}

\newtcolorbox{remarkbox}[1][]{
    colback=softgreen!8,
    colframe=darkgreen,
    fonttitle=\bfseries,
    title={#1},
    rounded corners,
    boxrule=0.8pt,
    before upper=\sloppy,  
}

\newtcolorbox{keyinsightbox}[1][]{
    colback=harvardcrimson!5,
    colframe=harvardcrimson,
    fonttitle=\bfseries,
    title={#1},
    rounded corners,
    boxrule=0.8pt,
}

\usepackage{titlesec}
\titleformat{\chapter}[display]
  {\normalfont\huge\bfseries\color{uzhblue}}
  {\chaptertitlename\ \thechapter}{20pt}{\Huge}
\titleformat{\section}
  {\normalfont\Large\bfseries\color{uzhblue}}
  {\thesection}{1em}{}
\titleformat{\subsection}
  {\normalfont\large\bfseries\color{uzhblue!80}}
  {\thesubsection}{1em}{}

\newcommand{\manuscriptmonthyear}{%
  \ifcase\month\or January\or February\or March\or April\or May\or June\or
  July\or August\or September\or October\or November\or December\fi
  ~\number\year}

\begin{document}
\setcounter{tocdepth}{2}
\hypersetup{pageanchor=false}

\begin{titlepage}
  \thispagestyle{empty}
  \AddToShipoutPictureBG*{%
    \AtPageUpperLeft{%
      \raisebox{-\paperheight}{%
        \makebox[\paperwidth][c]{%
          \includegraphics[width=\paperwidth,height=\paperheight]%
                          {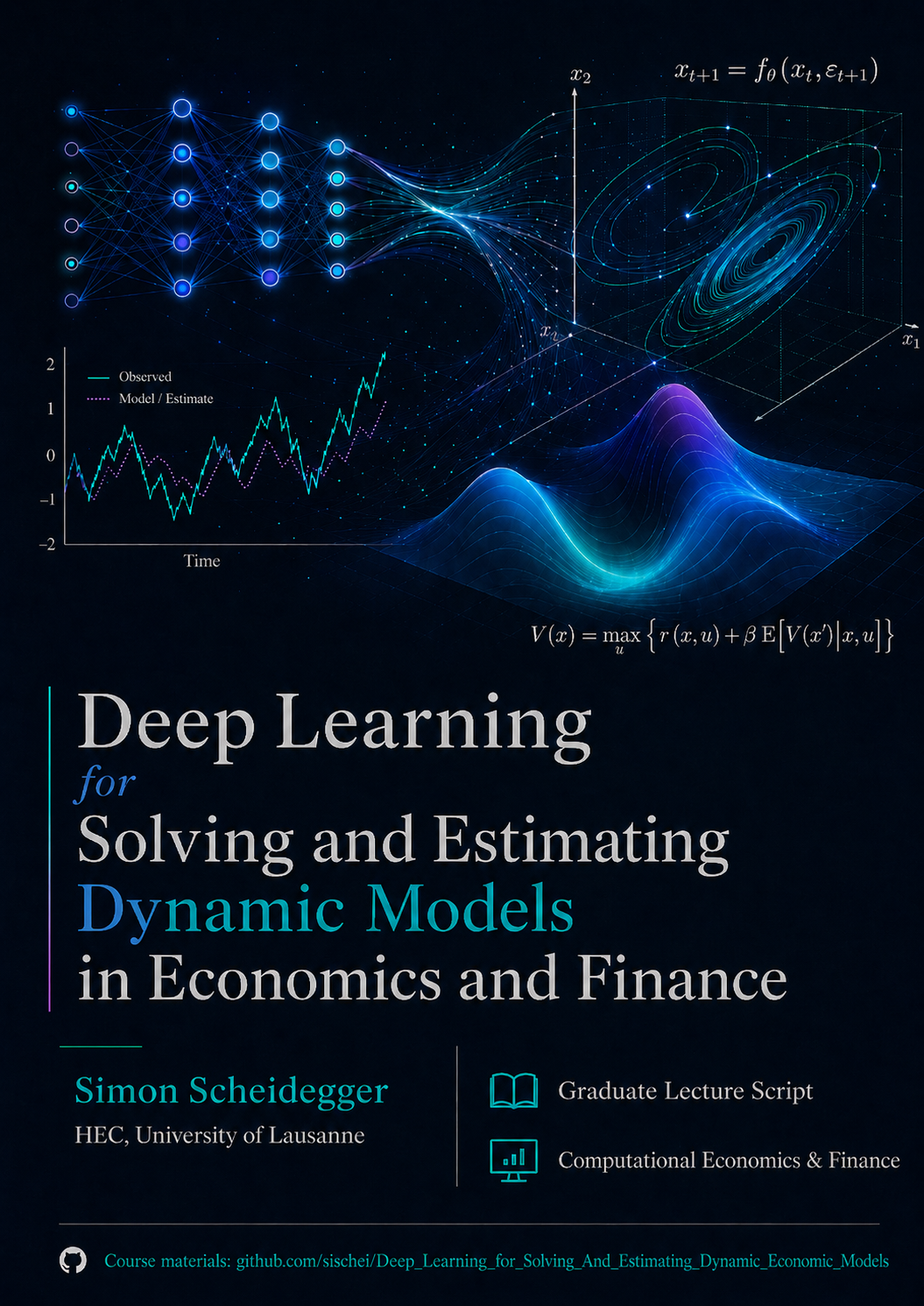}%
        }%
      }%
    }%
  }
  \null
\end{titlepage}

\begin{titlepage}
\centering
\vspace*{2cm}
{\Huge\bfseries\color{uzhblue} Deep Learning for Solving and\\[0.3em] Estimating Dynamic Models in\\[0.3em] Economics and Finance\par}
\vspace{0.9cm}
{\Large\bfseries\color{uzhgreydark} \manuscriptmonthyear\par}
\vspace{1.3cm}
{\large Simon Scheidegger\\[0.5em]
\normalsize HEC, University of Lausanne\par}
\vfill
{\small The most recent version of this manuscript, together with the companion code, is maintained at:\\[3pt]
\url{https://github.com/sischei/Deep_Learning_for_Solving_And_Estimating_Dynamic_Economic_Models}\par}
\vspace{0.6cm}
{\footnotesize This version: \manuscriptmonthyear.\\[3pt]
\copyright{} \number\year{} Simon Scheidegger.  Code under MIT, text/figures under CC0 1.0 Universal:\\
\url{https://creativecommons.org/publicdomain/zero/1.0/}\par}
\end{titlepage}

\thispagestyle{empty}
\vspace*{2cm}
\begin{center}
{\Large\bfseries\color{uzhblue} Abstract\par}
\end{center}
\vspace{1.0cm}
\begin{quote}
This script offers an implementation-oriented introduction to deep learning methods for solving and estimating high-dimensional dynamic stochastic models in economics and finance.  Its starting point is the curse of dimensionality: heterogeneous-agent economies, overlapping-generations models with aggregate risk, continuous-time models with occasionally binding constraints, climate-economy models, and macro-finance environments with many assets and frictions all generate state and parameter spaces that strain classical tensor-product grid methods.

The exposition is organized around four complementary methodologies.  Deep Equilibrium Nets embed discrete-time equilibrium conditions directly into neural-network loss functions.  Physics-Informed Neural Networks approximate continuous-time Hamilton--Jacobi--Bellman, Kolmogorov forward, and related partial differential equations.  Deep surrogate models provide fast, differentiable approximations to expensive structural models, while Gaussian processes add a probabilistic layer that quantifies approximation uncertainty; together they support estimation, sensitivity analysis, and constrained policy design.  Gaussian-process-based dynamic programming, combined with active learning and dimension reduction, extends value-function iteration to settings with very large continuous state spaces.

Applications range from representative-agent and international real business cycle models to overlapping-generations and heterogeneous-agent economies, continuous-time macro-finance, structural estimation by simulated method of moments, and climate economics under uncertainty.  The emphasis throughout is practical: economic restrictions are translated into trainable residuals, algorithms are paired with diagnostics, and companion notebooks in TensorFlow and PyTorch invite students to experiment, modify, and learn by doing.

These notes are not intended as an exhaustive textbook.  They are a deliberately subjective and inevitably incomplete snapshot of a rapidly evolving field.  The selection of architectures, references, and case studies reflects one researcher's judgment about useful entry points at the time of writing.  Their purpose is to equip PhD students and researchers with enough conceptual and computational machinery to engage with this frontier hands-on.
\end{quote}
\clearpage

\thispagestyle{empty}
\vspace*{\fill}
\begin{quote}
\centering
``The machine, the frozen form of a living intelligence, is the power that expands the potential of your life by raising the productivity of your time.''

\medskip
--- Ayn Rand, \emph{Atlas Shrugged} (1957), Galt's Speech (Part Three, Chapter VII)
\end{quote}
\vspace*{\fill}
\clearpage

\hypersetup{pageanchor=true}
\pagenumbering{arabic}

\tableofcontents

\chapter*{Preface}
\addcontentsline{toc}{chapter}{Preface}

Quantitative economics and finance increasingly rely on models that are richer, more realistic, and more computationally demanding than anything attempted a generation ago.  In macroeconomics, heterogeneous-agent economies, overlapping-generations models with aggregate risk, continuous-time models with occasionally binding constraints, and integrated assessment models coupling climate and economic dynamics all share the challenge of high-dimensional state spaces where traditional grid-based numerical methods are computationally infeasible.  Macro-finance and asset pricing raise the same challenge from a different direction: production economies with many risky assets and leverage or collateral constraints, sovereign-default and international macro-finance models with occasionally binding frictions, dynamic portfolio choice with transaction costs and multiple illiquid assets, and the large cross-section of test assets that modern empirical asset pricing must confront all push state and parameter spaces well past the reach of tensor-product grids.  Deep learning provides a promising new set of tools for addressing these challenges.

This script grew out of courses taught at multiple universities and central bank seminars over the past several years.  It offers an implementation-oriented introduction to deep learning methods for solving and estimating dynamic stochastic economic models, drawing heavily on the author's own research but also on the rapidly growing body of work by many colleagues in this field.  Although the companion course is delivered as 18 lectures, the manuscript is written to be self-contained and suitable for independent study.  Throughout the text, the exposition follows chapters and sections rather than the calendar of the live course.  The list of references is necessarily incomplete; the literature is expanding faster than any single manuscript can track, and the selection of topics reflects one researcher's judgment about which methods give economists and finance researchers some of the most useful entry points today.  For comprehensive surveys of deep learning for dynamic economic models, the reader is referred to \citet{fernandezvillaverde2024taming}; for machine learning in empirical asset pricing and finance, to \citet{gukellyxiu2020}, \citet{nagel2021mlap}, and \citet{kellyxiu2023fml}; and to the references therein.  The present script aims to complement such surveys by providing concrete algorithms, working code, and step-by-step examples that readers can adapt to their own research problems.

\paragraph{What these notes cover.}
The exposition is organized around four complementary methodologies:
\begin{enumerate}[itemsep=2pt]
\item \textbf{Deep Equilibrium Nets} (DEQNs), which embed the equilibrium conditions of a dynamic stochastic model directly into the loss function of a neural network, replacing the traditional numerical solution step with gradient-based optimization \citep{azinovicDEEPEQUILIBRIUMNETS2022}.  We apply this framework to representative-agent models, international business cycles, overlapping generations, and heterogeneous-agent economies, including the histogram-based treatment of distributions building on \citet{young2010} and the later sequence-space extension of \citet{azinovicyangzemlicka2025sequencespace}.
\item \textbf{Physics-Informed Neural Networks} (PINNs), which approximate Hamilton--Jacobi--Bellman equations and Kolmogorov forward equations in continuous time without computing conditional expectations \citep{sirignano2018dgm, raissi2019physics}.  We develop this machinery for consumption-savings problems and option pricing, and compare it with finite-difference HJB--KFE methods for continuous-time heterogeneous-agent models in the tradition of \citet{achdou2022income} and with the Economic Model Informed Neural Networks (EMINNs) of \citet{gu2024masterequations}.
\item \textbf{Deep surrogate models and Gaussian processes}, which construct fast, differentiable approximations of expensive structural models, enabling rapid estimation, uncertainty quantification, and global sensitivity analysis \citep{SCHEIDEGGER201968, chen2026Deep}.  Once trained, the same surrogates can also be used to \emph{design constrained optimal policies}, including in dynamic stochastic heterogeneous-agent models: a long search over policy parameters that would require thousands of full re-solves of the structural model collapses into a small optimization on the surrogate.  The climate-economics chapter uses this construction to derive Pareto-improving carbon-tax rules in an OLG--IAM with deep uncertainty \citep{kubler2025using}.
\item \textbf{GP-based dynamic programming}, which embeds Gaussian process regression into a value function iteration algorithm \citep{engel2005reinforcement, deisenroth2009gaussian, rennerscheidegger_2018} and, when combined with active subspaces \citep{constantine2015active}, scales to economies with up to 500 continuous state variables \citep{SCHEIDEGGER201968}.
\end{enumerate}
The unifying thesis is that economic structure drives the learning problem rather than being separated from it.  In DEQNs, PINNs, and EMINNs the structure is embedded directly into a neural-network loss as residual terms (equilibrium conditions, PDEs, Bellman equations), making the training objective itself unsupervised; in the surrogate and GP-based estimation chapters the structure instead appears upstream, in the model that generates the (input, output) training pairs the surrogate then learns in a standard supervised regression.  Either way, the four methodologies above are different surfaces for the same idea.

\paragraph{How these notes are organized.}
At a high level, Chapters~1--6 build the discrete-time toolkit (foundations, DEQNs, IRBC, architecture search and loss normalization, OLG, heterogeneous agents). Chapters~7--8 turn to continuous-time methods (PINNs, HJB--KFE for heterogeneous agents). Chapter~9 develops the deep-surrogate and Gaussian-process toolkit and deploys it on two distinct oracles: surrogates of the structural model itself, used downstream for estimation, uncertainty quantification, and policy search, and surrogates of the Bellman operator, used inside GP value-function iteration to solve high-dimensional dynamic programs.  Chapter~10 then uses these surrogates for structural estimation via SMM. Chapter~11 applies the toolbox to climate economics under deep uncertainty, and Chapter~12 closes by comparing methods and highlighting open problems. The next chapter, \emph{How to Read This Script}, expands this overview into a full reading guide: cover-to-cover and selective paths, notation and cross-references, the role of the companion notebooks, and the visual conventions used throughout.

\paragraph{Audience and prerequisites.}
These notes are aimed at PhD students in economics, finance, and computational social science, as well as researchers at central banks and policy institutions who wish to apply these methods to their own models.  We assume familiarity with basic econometrics, Python programming, and undergraduate-level calculus and probability.  For readers who need a refresher on these foundations, we recommend \citet{goodfellow2016deep} for deep learning, \citet{judd1998numerical} for numerical methods in economics, the open-source QuantEcon lectures (\url{https://quantecon.org}) for Python programming and quantitative economics, and the Econ-ARK toolkit (\url{https://econ-ark.org}) for heterogeneous-agent modeling.

\paragraph{Relation to the literature.}
These notes complement several existing references.  The comprehensive survey by \citet{fernandezvillaverde2024taming} provides a broad overview of deep learning for economics, while \citet{gukellyxiu2020}, \citet{nagel2021mlap}, and \citet{kellyxiu2023fml} survey the rapidly growing machine-learning literature in empirical asset pricing and finance; \citet{judd1998numerical} remains the standard reference for classical numerical methods; the textbooks by \citet{stokeylucas1989} and \citet{ljungqvist2018recursive} cover the economic theory that underpins the models we solve; the open first-year Ph.D.\ textbook by \citet{azzimontikrusellmckaymukoyama2026macro} provides a general introduction to state-of-the-art macroeconomics; and the open dynamic-programming volumes by \citet{sargentstachurski2026dp} provide a modern computational treatment of recursive methods and operator-based analysis.  The present script is distinguished by its focus on \emph{implementation}: every method is accompanied by working code, and the exposition is driven by concrete economic applications rather than abstract theory.

\paragraph{Software and reproducibility.}
All code examples are available as executable Jupyter notebooks in TensorFlow and PyTorch, hosted on the companion GitHub repository.\footnote{\url{https://github.com/sischei/Deep_Learning_for_Solving_And_Estimating_Dynamic_Economic_Models}}  The notebooks are designed to run on standard hardware; GPU acceleration is beneficial but not required for the classroom-scale examples.  Each chapter indicates which notebooks accompany it.

\paragraph{A snapshot of a fast-moving field.}
Deep learning for quantitative economics and finance is evolving at an extraordinary pace, with new architectures, training algorithms, and applications appearing month after month.  Both this script and the accompanying lecture suite are therefore best read as a snapshot of a highly dynamically evolving field rather than as a definitive treatment, and they are necessarily incomplete: the algorithms, models, and references collected here are those that, at the time of writing, strike the author as the most useful entry points for an economist or finance researcher beginning to work with deep learning, but the frontier is moving quickly enough that some choices will inevitably look dated within a year or two, and many worthy contributions have had to be left out.  The companion GitHub repository will track revisions as the field matures, and readers are encouraged to consult it for the most recent versions of the code and references.

\paragraph{Feedback and errata.}
This script is a living document.  Corrections, suggestions, and pull requests are welcome on the companion GitHub repository's issue tracker.\footnote{\url{https://github.com/sischei/Deep_Learning_for_Solving_And_Estimating_Dynamic_Economic_Models/issues}}  Direct correspondence may be addressed to \href{mailto:simon.scheidegger@unil.ch}{simon.scheidegger@unil.ch}.

\paragraph{Acknowledgments.}
I am grateful to the many seminar and course participants whose questions and comments have shaped this material.  I also thank Marlon Azinovic, Johannes Brumm, Christopher Carroll, Hui Chen, Antoine Didisheim, Richard Evans, Jes\'us Fern\'andez-Villaverde, Aleksandra Friedl, Luca Gaegauf, Kenneth Gillingham, Pavel Ievlev, Mitsuru Igami, Felix K\"ubler, Gianni Lombardo, Maria Pia Lombardo, Marko Mlikota, Galo Nu\~no, Jonathan Payne, Philipp Renner, Andreas Schaab, Frank Schorfheide, Anna Smirnova, Oliver Surbek, Fabio Trojani, Yucheng Yang, and Jan {\v Z}emli{\v c}ka for the research and discussions that underpin this work.  Part of this manuscript was written while visiting the Bank for International Settlements (BIS); I thank the BIS for its hospitality.  This script also benefitted substantially from Anthropic's Claude as a writing and coding assistant, in particular for polishing earlier drafts of the manuscript and for modernising the companion code examples that had aged across several previous iterations of the course.  Most importantly, I want to thank my wife, Michaela, and my daughter, Kira, for making this work possible through their unwavering support.  The responsibility for any remaining errors of course rests with the author.

\vspace{1em}
\noindent\textit{Simon Scheidegger}\\*
\textit{HEC, University of Lausanne, 2026}

\chapter*{How to Read This Script}
\addcontentsline{toc}{chapter}{How to Read This Script}

The four methodologies sketched in the Preface, namely DEQNs, PINNs, deep surrogates with Gaussian processes, and GP-based dynamic programming, share a common philosophy but differ in their prerequisites, dependencies, and natural reading order.  This short orientation translates that conceptual map into a practical reading plan: how the chapters, companion notebooks, notation tables, and visual conventions fit together, and how the manuscript can be used either as a sequential textbook for a reader encountering these tools for the first time or as a reference one can drop into at the level of a single method.

Although the live course groups the material into eight classroom sessions, the manuscript is organized by conceptual dependency rather than by calendar time.  Each main chapter follows the same rhythm: the economic problem first, then the computational obstacle that classical methods leave open, then the neural-network or GP construction that closes it, then the algorithmic implementation, and finally exercises and notebooks that turn the method into running code.  The notebooks are not an optional appendix; they are part of the exposition.  They make explicit which residuals are minimized, which variables enter as states or pseudo-states, how expectations are approximated, and how accuracy is checked in practice, and the chapters are written on the assumption that the reader will at least skim the corresponding code.

\paragraph{Cover to cover.}
A reader new to both deep learning and computational economics should read the script in order.  Chapter~1 supplies the machine-learning vocabulary used everywhere else: losses, gradients, backpropagation, architectures, optimization, and generalization.  Chapters~2--6 then develop the discrete-time equilibrium toolkit, moving from representative-agent DEQNs to international business cycles, loss balancing, overlapping generations, and heterogeneous-agent distributions.  Chapters~7--8 shift from conditional expectations to continuous-time PDE residuals, first through PINNs and then through HJB--KFE systems.  Chapters~9--10 introduce surrogate models, Gaussian processes, active subspaces, and structural estimation.  Chapter~11 uses these pieces in a climate-economics application, and Chapter~12 closes by comparing methods and highlighting open problems.

\paragraph{Selective reading paths.}
Readers who already know part of the material can take a shorter route, but the dependencies still matter:
\begin{itemize}[itemsep=2pt]
\item \textbf{Discrete-time equilibrium methods:} read Chapters~1--6.  This path covers the DEQN template, expectation approximation, high-dimensional DSGE examples, loss normalization, OLG models, and histogram-based heterogeneous-agent distributions.
\item \textbf{Continuous-time methods:} read Chapter~1 for neural-network basics, Chapter~7 for PINNs and PDE residuals, and Chapter~8 for the continuous-time heterogeneous-agent HJB--KFE system.  Chapter~6 is useful background if the distribution $\mu_t$ is unfamiliar.
\item \textbf{Surrogates and estimation:} read Chapter~1 for optimization and approximation, Chapter~9 for deep surrogates, GPs, active subspaces, and GP-based dynamic programming, and Chapter~10 for the SMM pipeline.  Chapter~2 is the minimum DEQN background for the Brock--Mirman examples used there.
\item \textbf{Climate economics:} read Chapter~11 together with Chapter~2 for DEQN solution logic and Chapter~9 for surrogate-based uncertainty quantification.  The climate chapter is written as an application chapter, not as a replacement for those methodological chapters.
\end{itemize}

These reading paths are minimum self-sufficient sets: chapters not listed are not strictly required for the path, but several later sections cross-reference earlier chapters more deeply, and the chapter introductions flag those dependencies explicitly when they arise.

\paragraph{Code and reproducibility.}
Each chapter is accompanied by executable Jupyter notebooks that implement the methods described in the text.  The notebooks are listed at the end of each chapter and are available on the companion GitHub repository.\footnote{\url{https://github.com/sischei/Deep_Learning_for_Solving_And_Estimating_Dynamic_Economic_Models}}  The classroom-scale examples are intentionally smaller than research-scale calibrations: their role is to make the residuals, training loops, diagnostics, and validation logic transparent.  Production-scale settings, when relevant, are noted in comments within the notebooks.  A useful workflow is to read the chapter once for the economic and mathematical structure, run or inspect the notebook to see the implementation, and then return to the chapter's exercises to check whether the approximation and validation choices are understood.

\paragraph{Notation and cross-references.}
The notation tables following this orientation are meant to reduce ambiguity across chapters.  Some symbols are necessarily reused because the script covers several literatures: for example, $\rho$ can denote neural-network parameters, an optimizer coefficient, or a continuous-time discount rate depending on context.  When a symbol changes role, the notation table records the chapter-specific convention.  Cross-references should be used actively: many later chapters refer back to the same core objects, such as residual losses, ergodic sampling, market-clearing errors, HJB residuals, or GP posterior variances.  The goal is not to memorize every symbol on first reading, but to use the notation tables as a stable reference while moving between chapters and notebooks.

\paragraph{Visual conventions.}
The script uses three colored callout boxes plus an algorithm box.  Each color is consistent across the entire manuscript and signals a particular kind of content (the first instance of each box appears in Chapter~\ref{ch:intro}):

\begin{itemize}[itemsep=3pt]
\item \textbf{Blue boxes (Definitions and Algorithms).}  Used for formal definitions, key constructions, and step-by-step pseudocode.  When a section introduces a new method, the algorithmic core is typically displayed in a blue box.
\item \textbf{Green boxes (Remarks and Worked Examples).}  Used for practical guidance, worked numerical examples, and short discussions that contextualize a result.  Green boxes can be skipped on a first read without losing the main thread.
\item \textbf{Crimson boxes (Key Insights).}  Used sparingly to flag the highest-level take-aways of a section, ideas the reader should remember even after the implementation details fade.
\item \textbf{Numbered algorithms.}  Inside blue boxes labeled ``Algorithm: \ldots,'' the pseudocode follows the conventions of the \texttt{algorithmic} package: \textsc{Input} / \textsc{Output} statements, indented \textsc{For} / \textsc{If} blocks, and arrow-style assignments.
\end{itemize}

A few additional typographic conventions: file paths and code identifiers are set in \texttt{monospace}; emphasized terms appear in \emphc{bold crimson}; cross-references to other chapters use the form (Ch.~$N$) or (\S$N.M$); and figure / table captions sit below their objects.

\chapter*{Notation and Symbols}
\addcontentsline{toc}{chapter}{Notation and Symbols}

\begin{center}
\begin{tabular}{@{}L{3.2cm} L{10.5cm}@{}}
\toprule
Symbol & Meaning \\
\midrule
$\x \in \R^d$ & Input or state vector, depending on context \\
$p(\x)$ & Policy/control associated with state $\x$ \\
$G(\x, p(\x))$ & Equilibrium or residual operator (e.g., Euler equation) \\
$\mathcal{N}_\rho$ & Neural network with parameters $\rho$ \\
$\ell$ & Loss function (supervised or residual-based) \\
$\eta$ & Learning rate \\
$\beta_1, \beta_2$ & Adam momentum coefficients \\
$\mathcal{D}$ & Dataset or collocation set \\
$\theta$ & Generic model or structural parameter vector; when structural parameters and neural-network weights must be distinguished, $\rho$ denotes network weights \\
$A$ & Number of OLG cohorts (Ch.~5) \\
$G_t(k_i, \varepsilon_j)$ & Histogram mass at grid point $k_i$, shock $\varepsilon_j$ (Ch.~6) \\
$\mu_t$ & Cross-sectional wealth distribution (Ch.~6--8) \\
$V(a,z)$ & Value function in continuous-time HJB (Ch.~7--8) \\
$g(a,z)$ & Stationary density from the KFE / Fokker--Planck equation (Ch.~7--8) \\
$\mathrm{SCC}_t$ & Social cost of carbon (Ch.~11) \\
\bottomrule
\end{tabular}
\end{center}

Where necessary, chapter-specific notation (e.g., HJB/PDE operators, kernel functions) is introduced locally to avoid ambiguity.  In a few places, the script intentionally reuses symbols such as $\eta$ when that is standard in the underlying literature; in those cases, the local chapter definition takes precedence.

\medskip
\noindent\textbf{Symbols with conflicting uses across chapters.}  Several symbols below are reused with different meanings depending on the chapter, because each chapter inherits the convention of its primary source.  This table collects the conflicts in one place; chapters that introduce a new local meaning also add a one-line warning at first use.

\begin{center}
\footnotesize
\renewcommand{\arraystretch}{1.15}
\begin{tabular}{l L{10cm}}
\toprule
Symbol & Meanings (by chapter) \\
\midrule
$\gamma$ & IES $=1/\text{CRRA}$ in Ch.~\ref{ch:irbc} (IRBC); CRRA in Ch.~\ref{ch:pinn} (cake-eating) and Ch.~\ref{ch:ct_theory} (continuous-time HA); reused as $\sigma_u$ in Ch.~\ref{ch:climate} (OLG-IAM) to free $\sigma_t$ for emissions intensity; RL discount factor $\gamma\in[0,1)$ and BatchNorm scale parameter in Ch.~\ref{ch:intro}; Hyperband / Successive-Halving reduction factor in Ch.~\ref{ch:nas}. \\
$\eta$ & Learning rate (Ch.~\ref{ch:intro}--\ref{ch:deqn}); TFP shock in OLG (Ch.~\ref{ch:olg}); idiosyncratic productivity in Krusell--Smith (Ch.~\ref{ch:young}); OU mean-reversion (Ch.~\ref{ch:ct_theory}); small numerical shift in I-spline basis; normalized temperature costate (Ch.~\ref{ch:climate}); normalized spatial coordinate in PINN bilinear BC construction (Ch.~\ref{ch:pinn}); functional-derivative test perturbation in KFE adjoint argument (Ch.~\ref{ch:ct_theory}). \\
$\alpha$ & Capital share in Cobb--Douglas production (Ch.~\ref{ch:deqn},~\ref{ch:olg},~\ref{ch:young},~\ref{ch:ct_theory},~\ref{ch:climate}); ReLoBRaLo smoothing parameter (Ch.~\ref{ch:nas}); boundary MPC head in I-spline (Ch.~\ref{ch:pinn}). \\
$\zeta$ vs.\ $\alpha$ & Capital share is denoted $\zeta$ in Ch.~\ref{ch:irbc} (Azinovic et al.\ convention) and $\alpha$ everywhere else. \\
$T$ & ReLoBRaLo softmax temperature (Ch.~\ref{ch:nas}); time horizon (Ch.~\ref{ch:pinn},~\ref{ch:climate}); atmospheric temperature $T^{\mathrm{AT}}_t$ in DICE (Ch.~\ref{ch:climate}); data sample size in SMM (Ch.~\ref{ch:estimation}). \\
$\delta$ & Capital depreciation rate (most chapters); Dirac measure in master equation (Ch.~\ref{ch:young}); Huber-loss threshold in robust regression (Ch.~\ref{ch:intro}). \\
$\psi$ & IES in Ch.~\ref{ch:climate} Epstein--Zin preferences (paired with $\gamma_u$ for risk aversion); Cobb--Douglas capital exponent in the GP surrogate model (Ch.~\ref{ch:gp}). \\
$\mu$ & Cross-sectional wealth distribution $\mu_t$ (Ch.~\ref{ch:young}--\ref{ch:ct_theory}); SGD momentum coefficient (Ch.~\ref{ch:intro}); Lagrange / KKT multiplier on investment irreversibility (Ch.~\ref{ch:irbc}); emissions abatement rate $\mu_t\in[0,1]$ (Ch.~\ref{ch:climate}). \\
$\rho$ & Network parameters $\mathcal{N}_\rho$ (most chapters); RMSprop decay coefficient and recurrence spectral radius (Ch.~\ref{ch:intro}); discount rate in continuous-time HJB (Ch.~\ref{ch:pinn},~\ref{ch:ct_theory}); ReLoBRaLo baseline-mix coefficient (Ch.~\ref{ch:nas}). The variant $\varrho$ is reserved for shock persistence in Ch.~\ref{ch:deqn}.  TFP persistence is denoted $\rho_z$ in Ch.~\ref{ch:irbc} (Azinovic et al.\ convention) and $\varrho$ elsewhere. \\
$\sigma$ & Logistic activation $\sigma(z)=1/(1+e^{-z})$ in Ch.~\ref{ch:intro} (single-output and final-layer use); the same symbol is used as a generic non-linearity in the RNN recurrence and as the LSTM gate non-linearity later in the same chapter, so the meaning is always logistic but the typographic role (specific vs.\ generic) shifts. Ch.~\ref{ch:climate} (climate) reserves $\sigma_t$ for emissions intensity, $\sigma_u$ for household CRRA. \\
$\tau$ & Bounded time index $\tau_t$ used as a network input in the deterministic CDICE-DEQN derivation (Ch.~\ref{ch:climate}, \S\ref{sec:deep_uq}); separately, the per-period carbon tax rate (also written $\tau_t$) later in the same chapter, in the OLG-IAM and Pareto-improving-tax discussion.  The notation is reset locally at each first use; readers should rely on the surrounding sentence rather than on the symbol alone. \\
\bottomrule
\end{tabular}
\end{center}

\medskip
\noindent\textbf{Default reading.}  When in doubt, default to the most common usage: $\rho$ is the neural-network parameter vector, $\eta$ is the learning rate, $\alpha$ is the Cobb--Douglas capital share, and $\sigma$ is the logistic activation.  Chapter-specific reuses always override locally, and the chapter introductions flag a divergent meaning at first use.  The conflict table above is a forward-reference for non-linear readers; a cover-to-cover reader will see each meaning introduced once and need not consult the table on a first pass.

\medskip
\noindent\textbf{Abbreviations and acronyms.}  The following acronyms appear throughout the script.  They are introduced in full at first use within each chapter; this list serves as a quick reference.

\begin{center}
\scriptsize
\renewcommand{\arraystretch}{1.05}
\setlength{\tabcolsep}{4pt}
\begin{tabular}{l p{5.4cm} @{\hspace{1.4em}} l p{5.4cm}}
\toprule
ABC    & Approximate Bayesian Computation         & KFE       & Kolmogorov forward (Fokker--Planck) Eq. \\
ACE    & Analytic Climate Economy (Traeger)        & KKT       & Karush--Kuhn--Tucker conditions \\
AD     & Automatic Differentiation                  & KS        & Krusell--Smith (1998) economy \\
AdamW  & Adam with decoupled weight decay           & LSTM      & Long Short-Term Memory net \\
AS     & Active Subspace                            & MAGICC    & Reduced-complexity climate emulator \\
BAL    & Bayesian Active Learning                   & MC        & Monte Carlo \\
BC     & Boundary Condition                         & MFG       & Mean Field Game \\
BSDE   & Backward Stochastic Differential Eq.       & ML        & Machine Learning \\
CDICE  & Calibrated DICE (Folini 2024)              & MLE       & Maximum Likelihood Estimator \\
CRN    & Common Random Numbers                      & MLP       & Multi-Layer Perceptron \\
CRRA   & Constant Relative Risk Aversion            & MMW       & Maliar--Maliar--Winant (2021) \\
DEQN   & Deep Equilibrium Net                       & MPC       & Marginal Propensity to Consume \\
DGM    & Deep Galerkin Method                       & NAS       & Neural Architecture Search \\
DICE   & Dyn.\ Integ.\ Climate-Econ.\ model         & NTK       & Neural Tangent Kernel \\
DKL    & Deep Kernel Learning                       & OLG       & Overlapping Generations \\
DL     & Deep Learning                              & PDE       & Partial Differential Equation \\
DNN    & Deep Neural Network                        & PINN      & Physics-Informed Neural Net \\
ECS    & Equilibrium Climate Sensitivity            & QMC       & Quasi-Monte Carlo \\
EGM    & Endogenous Grid Method                     & ReLoBRaLo & Relative Loss Balancing \\
ELU    & Exponential Linear Unit                    & RL        & Reinforcement Learning \\
EMINN  & Economic Model Informed NN                 & RNN       & Recurrent Neural Network \\
FaIR   & Reduced-complexity climate emulator        & SBI       & Simulation-Based Inference \\
FB     & Fischer--Burmeister loss                   & SCC       & Social Cost of Carbon \\
FD     & Finite Differences                         & SDE       & Stochastic Differential Equation \\
FNO    & Fourier Neural Operator                    & SGD       & Stochastic Gradient Descent \\
FOC    & First-Order Condition                      & SMM       & Simulated Method of Moments \\
GE     & General Equilibrium                        & TF / TF2  & TensorFlow / TensorFlow~2 \\
GMM    & Generalized Method of Moments              & UQ        & Uncertainty Quantification \\
GP     & Gaussian Process                           & VFI       & Value Function Iteration \\
HA     & Heterogeneous Agent                        & ZLB       & Zero Lower Bound \\
HJB    & Hamilton--Jacobi--Bellman Eq.              & XLA       & Accelerated Linear Algebra (TF/JAX) \\
IRBC   & Internat.\ Real Business Cycle             & DeepONet  & Deep Operator Network \\
JAX    & JAX autodiff library (Google)              & DeepHAM   & Deep Heterogeneous-Agent Model \\
\bottomrule
\end{tabular}
\end{center}

\chapter*{Execution Map}
\addcontentsline{toc}{chapter}{Execution Map}

The following table maps each manuscript chapter to its companion slide deck(s) and Jupyter notebooks.  All paths are relative to the repository root; the short names are intentionally compact so the table remains readable.

\begin{center}
\textbf{Execution map: manuscript chapters, slides, and notebooks}\par\smallskip
\scriptsize
\renewcommand{\arraystretch}{1.2}
\begin{tabular}{@{}c L{2.6cm} L{4.2cm} L{5.6cm}@{}}
\toprule
\textbf{Ch.} & \textbf{Topic} & \textbf{Lecture folder \& deck} & \textbf{Notebooks (role)} \\
\midrule
1  & Intro to ML \& DL                         & L02: \texttt{01\_Intro\_to\_DL}                          & L02 \texttt{01}--\texttt{09} (c$\times$9) \\
2  & Deep Equilibrium Nets                     & L03: \texttt{02\_DEQNs}; L07: \texttt{05b\_AutoDiff} & L03 \texttt{01}--\texttt{02} (c), \texttt{03}--\texttt{04} (e/s), \texttt{05} (c); L07 \texttt{01}--\texttt{04} (c) \\
3  & IRBC Model                                & L04: \texttt{03\_IRBC}                                  & L04 \texttt{01}--\texttt{02} (c) \\
4  & NAS \& Loss Norm.                         & L05: \texttt{04\_NAS}, \texttt{05\_Loss}                 & L05 \texttt{02}--\texttt{04} (c), \texttt{05} (e) \\
5  & OLG Models                                & L08: \texttt{08\_OLG}                                   & L08 \texttt{07}--\texttt{10} (c), \texttt{11} (e) \\
6  & HA, Young, Seq.\ Space                    & L09: \texttt{09\_HA\_Young}; L10: \texttt{10\_SeqSpace} & L09 \texttt{10}--\texttt{12} (c); L10 \texttt{05}, \texttt{05b}, \texttt{06} (c), \texttt{KrusellSmith\_Tutorial\_CPU} (x) \\
7  & PINNs                                     & L11: \texttt{06\_PINNs}                                 & L11 \texttt{01}--\texttt{05} (c) \\
8  & CT Het.\ Agents                           & L12: \texttt{07\_CT\_Theory}; L13: \texttt{08\_CT\_Num} & L13 \texttt{06}--\texttt{08} (c), \texttt{09} (e) \\
9  & Surrogates, GPs, DKL                      & L14: \texttt{07\_Surrogates\_GPs}                       & L14 \texttt{01}, \texttt{02}, \texttt{04}--\texttt{08} (c), \texttt{07}, \texttt{09}, \texttt{10} (x) \\
10 & Structural Estimation                     & L15: \texttt{08\_Struct\_Est}                          & L15 \texttt{03}, \texttt{03b} (c) \\
11 & Climate \& Deep UQ                        & L16: \texttt{08\_Climate}; L17: \texttt{09\_UQ}    & L16 \texttt{01}--\texttt{03} (c); L17 \texttt{09\_DICE\_2P\_UQ\_Analysis} (c) plus 4 \texttt{.py} pipeline drivers \\
12 & Synthesis \& Outlook                      & L18: \texttt{10\_Wrap\_Up}                              & --- \\
\bottomrule
\end{tabular}
\end{center}

\noindent{\scriptsize\textbf{Path conventions.} The repository organizes lectures by stable block id (\texttt{lectures/lecture\_$N$\_B$YY$\_*/}); the leading \texttt{L$N$} in the table is the student-facing lecture number and the parenthetical \texttt{B$YY$} is the canonical block id (which is stable across renumberings).  Slide and notebook names in the table are abbreviated; full paths follow \texttt{lectures/lecture\_$N$\_B$YY$\_*/\{slides,code\}/}.  Notebook role letters: \texttt{c}~=~core, \texttt{e}~=~exercise, \texttt{s}~=~solution (paired with an exercise notebook), \texttt{x}~=~extension/self-study.  See the README for complete file names and direct links.}

\medskip
\noindent{\scriptsize\textbf{Workshop material.} The live course includes a hands-on workshop on agentic programming (using AI agents as coding partners), delivered as L06.  Because this field is evolving quickly, it is presented through slides, two Python helper scripts, and exercises rather than as a fixed manuscript chapter.  The manuscript remains organized chapter by chapter, with the workshop material collected in the L06 slides and exercise prompts.}

\medskip
\noindent{\scriptsize\textbf{Reproducibility.} Random-seed conventions, the \tpath{RUN_MODE} budget split, hardware and software pins, and GPU-determinism flags used by every notebook in the table above are documented in Appendix~\ref{app:reproducibility}.  Worked solutions and guidance for the end-of-chapter exercises are collected in Appendix~\ref{app:solutions}.}

\chapter{Introduction to Machine Learning and Deep Learning}
\label{ch:intro}

The Preface argued that quantitative economics needs new tools because the models of interest, heterogeneous-agent economies, OLG models with aggregate risk, integrated assessment models, all share state spaces too large for traditional grid-based methods.  This chapter takes that argument as given and supplies the technical machinery the rest of the manuscript will build on.  Readers who want a fuller pitch for \emph{why} deep learning is the right response should re-read the Preface; readers who already accept the premise can dive directly into the machinery below.

We begin with a brief refresher on the motivation, mostly to fix vocabulary and citations, then survey the three broad paradigms of machine learning (supervised, unsupervised, and reinforcement learning), and then develop the core technical machinery: neural network architectures, loss functions, gradient-based optimization, backpropagation, weight initialization, activation functions, and the modern theory of generalization including the double descent phenomenon.  Readers already comfortable with these topics may skim this chapter and proceed to Chapter~\ref{ch:deqn}, where the economic applications begin.

The foundational references for the material in this chapter include \citet{mcculloch1943logical}, \citet{hebb1949organization} and \citet{rosenblatt1958perceptron} for the historical origins of artificial neurons and the first learning rules, \citet{cybenko1989approximation} and \citet{hornik1989multilayer} for universal approximation, \citet{rumelhart1986learning} for backpropagation, \citet{robbins1951stochastic} for the stochastic-approximation roots of SGD, \citet{geman1992biasvariance} for the bias/variance dilemma that underpins modern generalization theory, \citet{kingma2015adam} for the Adam optimizer, \citet{ioffe2015batch} and \citet{srivastava2014dropout} for batch normalization and dropout, and \citet{goodfellow2016deep} for a comprehensive textbook treatment, complemented by the historical survey of \citet{schmidhuber2015deep}.

\section{Motivation and Applications}

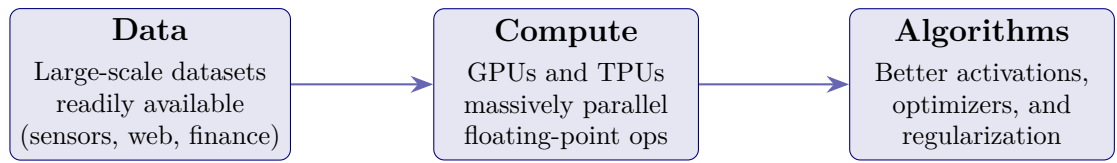
\begin{figure}[ht]
\centering
\begin{tikzpicture}[
    box/.style={rectangle, draw=uzhblue, fill=uzhblue!10, rounded corners=4pt,
                minimum width=3.5cm, minimum height=2cm, align=center, font=\small},
    arr/.style={-{Stealth[length=3mm]}, thick, uzhblue!60}
]
    \node[box] (data) at (0,0) {\textbf{\large Data}\\[3pt]Large-scale datasets\\readily available\\(sensors, web, finance)};
    \node[box] (hw) at (5.5,0) {\textbf{\large Compute}\\[3pt]GPUs and TPUs\\massively parallel\\floating-point ops};
    \node[box] (sw) at (11,0) {\textbf{\large Algorithms}\\[3pt]Better activations,\\optimizers, and\\regularization};
    \draw[arr] (data) -- (hw);
    \draw[arr] (hw) -- (sw);
\end{tikzpicture}
\caption{The three enablers of the deep-learning revolution: large-scale data, massively parallel compute, and algorithmic improvements.  None of the three alone is sufficient; their co-availability since the early 2010s is what has turned neural networks from a niche academic curiosity into a workhorse scientific and industrial tool.}
\label{fig:dl_enablers}
\end{figure}

The past decade has witnessed a remarkable convergence of three developments that have transformed machine learning from a niche academic pursuit into a practical tool of extraordinary power: the availability of large-scale datasets, the advent of massively parallel hardware (GPUs and TPUs), and algorithmic innovations in training deep neural networks (Figure~\ref{fig:dl_enablers}).  While much of the public attention has focused on applications such as image recognition, natural language processing, and game playing, the implications for economics and finance are equally profound.

Deep learning has already demonstrated its potential across a broad range of economic applications.  In macroeconomics, neural networks serve as global approximators of policy and value functions in high-dimensional equilibria that classical grid-based methods cannot reach \citep{maliar2021deep, azinovicDEEPEQUILIBRIUMNETS2022}; heterogeneous-agent extensions encode cross-sectional distributions via histograms or permutation-invariant moment networks \citep{young2010, han2023deepham, yang2025structural}.  Search-and-matching models with aggregate shocks \citep{payne2025deepsam} and continuous-time macro-finance settings requiring HJB approximation or deep-BSDE solvers \citep{gopalakrishna2024aliens, duarte2024ml, han2018solving} add further coverage, as do optimal monetary policy rules under persistent supply shocks \citep{nuno2024monetary}.  In climate economics, surrogate-based workflows solve integrated assessment models and derive Pareto-improving carbon-tax rules in OLG--IAMs with deep uncertainty \citep{kubler2025using, Folini_2021, fernandezvillaverde2025climate}.  In finance, surrogate models accelerate option pricing, sovereign-default computation, and portfolio optimization \citep{hutchinson1994nonparametric, scheideggertreccani_2018, arellano2008default, gaegauf2023portfolio, chen2025private}.  For structural estimation, neural-network surrogates make inference tractable and enable global uncertainty quantification for IAMs \citep{kase2022estimating, friedlDeep2023, chen2026Deep}.  Finally, these methods connect to an earlier generation of neural-network function approximation in economic computation: adaptive learning \citep{chenwhite1998adaptive}, derivative pricing \citep{hutchinson1994nonparametric}, and parameterized expectations \citep{duffy2001approximating}, with \citet{valaitisvilla2024} showing how the parameterized-expectations approach extends to contemporary deep architectures; structural discrete-choice estimation rounds out the historical picture \citep{norets2012structural}.

Two themes cut across these application areas and motivate the rest of the script.  First, every application area listed there involves a state space whose dimension grows with the number of agents, assets, shocks, or climate states; tensor-product grids become infeasible long before the modeling questions become uninteresting.  Second, neural networks are universal approximators with cost that scales with parameter count rather than with $N^d$, so they are the natural replacement function class once the problem becomes high-dimensional.  Subsequent chapters take this through-line and develop it: Chapter~\ref{ch:deqn} introduces the DEQN methodology that all macro / heterogeneous-agent / search applications above share; Chapter~\ref{ch:irbc} scales it to a 100+-country benchmark; Chapters~\ref{ch:pinn}--\ref{ch:ct_theory} develop the continuous-time analogue; Chapter~\ref{ch:gp} connects the deep-surrogate and Gaussian-process threads; and Chapters~\ref{ch:estimation}--\ref{ch:climate} return to structural estimation and integrated assessment.

In this course, we focus on the recent advances enabled by \emph{deep} neural networks, modern hardware, and algorithmic innovations in training.

For central banks, in particular, these tools address pressing practical needs.  Many of the models listed above, such as heterogeneous-agent New Keynesian (HANK) models, search and matching models with aggregate shocks, overlapping-generations (OLG) economies with aggregate risk, and integrated assessment models coupling climate and economic dynamics, involve state spaces of dimension $d \gg 5$, where traditional grid-based numerical methods are computationally infeasible: a tensor-product grid with $N$ points per dimension requires $N^{d}$ nodes, the canonical \emph{curse of dimensionality} of \citet{bellman1961adaptive}.  Deep learning provides a mesh-free function approximator.  For Barron-class functions, two-layer networks achieve dimension-independent rates in the number of hidden units (often stated as squared $L^2$ error of order $\mathcal{O}(1/n_1)$ under a bounded Barron norm), whereas tensor-product grids for generic smooth functions suffer rates such as $\mathcal{O}(M^{-2/d})$ in the total number $M=N^d$ of grid nodes.  The point is not that every economic object satisfies a Barron bound, but that neural approximators avoid the mechanical tensor-product explosion.  Chapter~\ref{ch:deqn} returns to this comparison with concrete numbers for a six-shock DSGE.

Specifically, this course focuses on using deep neural networks as computational tools for:  (i) solving high-dimensional dynamic stochastic general equilibrium (DSGE) models, (ii) approximating value and policy functions in continuous-time settings via physics-informed neural networks, (iii) constructing fast and accurate surrogate models for parameter estimation and uncertainty quantification, and (iv) leveraging Gaussian processes for sample-efficient Bayesian active learning \citep{azinovicDEEPEQUILIBRIUMNETS2022, friedlDeep2023, chen2026Deep}.  Supporting techniques include neural architecture search and adaptive multi-objective loss balancing (one option, ReLoBRaLo \citep{bischof2025relobralo}, is developed in the notebooks; alternatives include SoftAdapt and GradNorm).  For a comprehensive textbook treatment of the foundations, we refer the reader to \citet{goodfellow2016deep} and \citet{chollet2017deeplearning}; for concise surveys we recommend \citet{lecun2015deep}.

\section{Types of Machine Learning}

Before diving into the technical details, it is useful to recall the three broad paradigms of machine learning, each defined by the nature of the available data and the feedback signal provided to the algorithm.

\subsection{Supervised Learning}

Given a set of \emph{labeled} input--output pairs $\{(\x^{(i)}, y^{(i)})\}_{i=1}^{m}$, the goal is to learn a mapping $h:\mathcal{X}\to\mathcal{Y}$ that generalizes to unseen inputs.  The two main tasks are \emph{regression} and \emph{classification}.

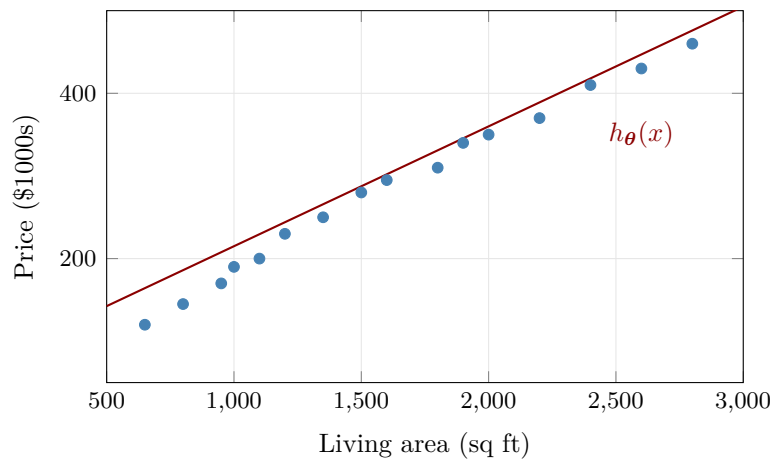
\begin{figure}[ht]
\centering
\begin{tikzpicture}
\begin{axis}[
    width=10cm, height=6.5cm,
    xlabel={Living area (sq ft)},
    ylabel={Price (\$1000s)},
    xlabel style={font=\small},
    ylabel style={font=\small},
    tick label style={font=\footnotesize},
    xmin=500, xmax=3000, ymin=50, ymax=500,
    grid=major, grid style={gray!20},
]
    \addplot[only marks, mark=*, mark size=2pt, softblue] coordinates {
        (650,120) (800,145) (950,170) (1000,190) (1100,200) (1200,230)
        (1350,250) (1500,280) (1600,295) (1800,310) (1900,340)
        (2000,350) (2200,370) (2400,410) (2600,430) (2800,460)
    };
    \addplot[thick, darkred, domain=500:3000, samples=2]{70 + 0.145*x};
    \node[font=\small, darkred] at (axis cs:2600,350) {$h_{\bm{\theta}}(x)$};
\end{axis}
\end{tikzpicture}
\caption{Supervised learning: regression.  The model $h_{\bm{\theta}}(x) = \theta_0 + \theta_1 x$ (red line) is fitted to observed house prices (blue dots).}
\label{fig:regression}
\end{figure}

\paragraph{Regression} ($y \in \R$): predict a continuous target from input features.  A simple linear model takes the form
\[
    h_{\bm{\theta}}(x) = \theta_0 + \theta_1 x,
\]
where the parameters $\bm{\theta} = (\theta_0, \theta_1)$ are learned from data.  Figure~\ref{fig:regression} illustrates regression on a house-price dataset: each dot is a training observation, and the line is the fitted model.

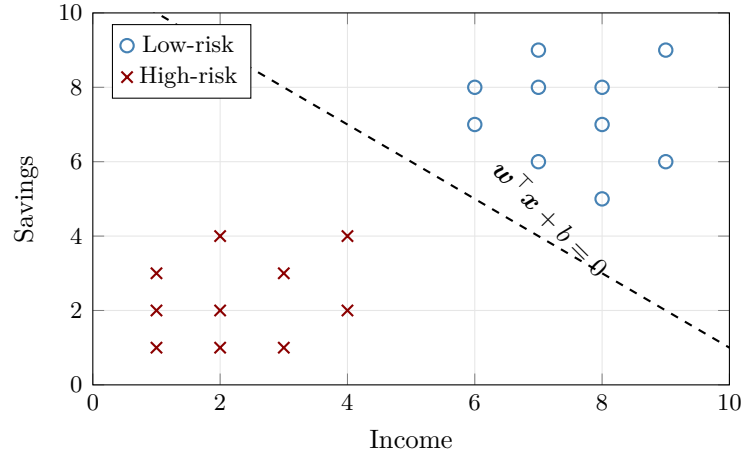
\begin{figure}[ht]
\centering
\begin{tikzpicture}
\begin{axis}[
    width=10cm, height=6.5cm,
    xlabel={Income},
    ylabel={Savings},
    xlabel style={font=\small},
    ylabel style={font=\small},
    tick label style={font=\footnotesize},
    xmin=0, xmax=10, ymin=0, ymax=10,
    grid=major, grid style={gray!20},
    legend style={at={(0.03,0.97)}, anchor=north west, font=\footnotesize},
]
    \addplot[only marks, mark=o, mark size=2.5pt, softblue, thick] coordinates {
        (6,7) (7,6) (8,5) (7,8) (8,7) (9,6) (6,8) (8,8) (9,9) (7,9)
    };
    \addlegendentry{Low-risk}
    \addplot[only marks, mark=x, mark size=3pt, darkred, thick] coordinates {
        (1,2) (2,1) (3,3) (2,4) (4,2) (1,3) (3,1) (2,2) (4,4) (1,1)
    };
    \addlegendentry{High-risk}
    \addplot[thick, dashed, black, domain=0:10, samples=2]{11 - x};
    \node[font=\small, rotate=-45] at (axis cs:7.2,4.5) {$\w^\top\x + b = 0$};
\end{axis}
\end{tikzpicture}
\caption{Supervised learning: classification.  A linear decision boundary separates low-risk (blue circles) from high-risk (red crosses) applicants in the income--savings feature space.}
\label{fig:classification}
\end{figure}

\medskip\noindent\textbf{Classification.} ($y \in \{0,1,\dots,K\}$): assign an input $\x$ to one of $K$ discrete categories.  A linear classifier predicts class~1 whenever $\w^\top \x + b > 0$ and class~0 otherwise.  Figure~\ref{fig:classification} shows a credit-scoring example: applicants are classified as low-risk or high-risk based on income and savings, and the dashed line is the learned decision boundary.

\phantomsection
\subsection{Unsupervised Learning}

Given only \emph{unlabeled} data $\{\x^{(i)}\}_{i=1}^{m}$, the goal is to discover hidden structure without any target signal.  Two common tasks are:
\begin{itemize}\itemsep2pt
    \item \textbf{Clustering:} partitioning data into groups of similar observations.  \emph{Example:} segmenting firms into peer groups based on financial characteristics such as size, leverage, and profitability.
    \item \textbf{Dimensionality reduction:} compressing features into fewer dimensions while preserving important variation.  \emph{Example:} principal component analysis of yield curves, where three factors (level, slope, curvature) capture most of the cross-sectional variation.
\end{itemize}

\begin{figure}[ht]
\centering
\begin{tikzpicture}
\begin{axis}[
    width=10cm, height=6.5cm,
    xlabel={Feature 1},
    ylabel={Feature 2},
    xlabel style={font=\small},
    ylabel style={font=\small},
    tick label style={font=\footnotesize},
    xmin=-1, xmax=11, ymin=-1, ymax=11,
    grid=major, grid style={gray!20},
    legend style={at={(0.03,0.97)}, anchor=north west, font=\footnotesize},
]
    \addplot[only marks, mark=*, mark size=2pt, softblue] coordinates {
        (1.2,1.5) (1.8,2.1) (2.0,1.0) (1.5,1.8) (2.3,2.5)
        (1.0,2.3) (2.5,1.5) (1.7,0.8) (0.8,1.2) (2.2,2.0)
    };
    \addlegendentry{Cluster A}
    \addplot[only marks, mark=*, mark size=2pt, softorange] coordinates {
        (7.0,2.0) (7.5,1.5) (8.0,2.5) (7.2,3.0) (8.5,2.0)
        (7.8,1.0) (8.2,2.8) (7.0,1.2) (8.8,1.8) (7.5,2.5)
    };
    \addlegendentry{Cluster B}
    \addplot[only marks, mark=*, mark size=2pt, softgreen] coordinates {
        (4.0,7.5) (4.5,8.0) (5.0,7.0) (3.8,8.2) (5.2,8.5)
        (4.2,7.8) (5.5,7.5) (3.5,7.0) (4.8,8.8) (5.0,8.0)
    };
    \addlegendentry{Cluster C}
    \addplot[only marks, mark=+, mark size=6pt, black, very thick] coordinates {
        (1.7,1.7) (7.75,2.03) (4.55,7.83)
    };
\end{axis}
\end{tikzpicture}
\caption{Unsupervised learning: clustering.  Unlabeled data points are grouped into three clusters; the $+$ markers indicate cluster centroids.  No target labels are used; the algorithm discovers the grouping from the data alone.}
\label{fig:clustering}
\end{figure}
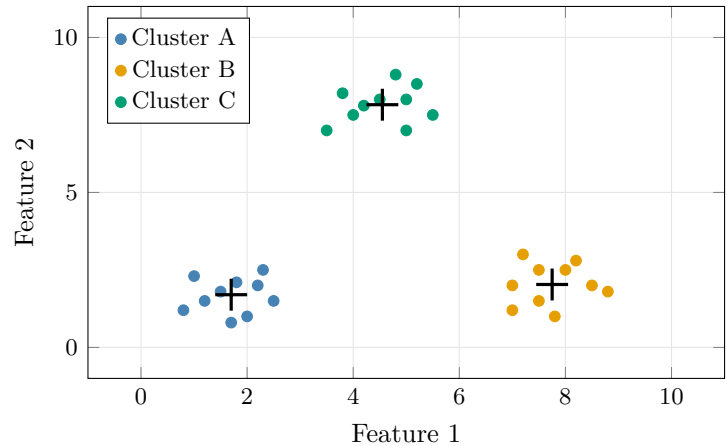

\noindent Figure~\ref{fig:clustering} illustrates a clustering task: unlabeled data points in two dimensions are partitioned into three clusters, each indicated by a different color and centroid marker.

\phantomsection
\subsection{Reinforcement Learning}

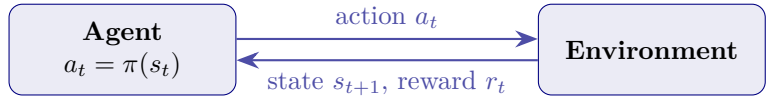
\begin{figure}[ht]
\centering
\begin{tikzpicture}[
    box/.style={rectangle, draw=uzhblue, fill=uzhblue!8, rounded corners=5pt,
                minimum width=3cm, minimum height=1.2cm, align=center, font=\small},
    arr/.style={-{Stealth[length=3mm]}, thick, uzhblue!70},
]
    \node[box] (agent) at (0,0) {\textbf{Agent}\\$a_t = \pi(s_t)$};
    \node[box] (env)   at (7,0) {\textbf{Environment}};
    \draw[arr] ([yshift=4pt]agent.east) -- node[above, font=\small]{action $a_t$} ([yshift=4pt]env.west);
    \draw[arr] ([yshift=-4pt]env.west) -- node[below, font=\small]{state $s_{t+1}$, reward $r_t$} ([yshift=-4pt]agent.east);
\end{tikzpicture}
\caption{Reinforcement learning: the agent--environment loop.  The agent observes a state, takes an action, and receives a reward signal.  Over time, it learns a policy $\pi$ that maximizes cumulative discounted reward.}
\label{fig:rl-loop}
\end{figure}

In reinforcement learning, an \emph{agent} interacts with an \emph{environment} over a sequence of time steps.  At each step $t$, the agent observes a state $s_t$, selects an action $a_t = \pi(s_t)$ according to its policy $\pi$, and receives a reward $r_t$ from the environment.  The goal is to learn a policy that maximizes the expected cumulative discounted return:
\[
    \max_{\pi}\; \mathbb{E}_{\pi}\!\left[\sum_{t=0}^{\infty} \gamma^t \, r_t\right], \qquad \gamma \in [0,1),
\]
where $\mathbb{E}_{\pi}[\,\cdot\,]$ is taken over the trajectory distribution induced jointly by the policy $\pi$ and the (possibly stochastic) environment dynamics, starting from a given initial-state distribution.
Figure~\ref{fig:rl-loop} illustrates this agent--environment interaction loop.

\noindent \emph{Example:} an algorithmic trader learning an execution strategy by optimizing realized profit over sequences of order placements; or a central bank learning an interest-rate rule by maximizing a welfare criterion over simulated macroeconomic trajectories.

\phantomsection
\section{Course Focus: Supervised vs. Unsupervised Learning}
\label{sec:supervised_vs_unsupervised}

This course begins with the \emph{supervised learning} paradigm, which provides the essential building blocks: choosing a parameterized model, defining a loss function, and minimizing it via gradient descent.

The core methods of this course, DEQNs and PINNs, are not supervised in the classical labeled-data sense.  More precisely, they are \emph{self-supervised residual methods}: the economic equilibrium conditions or governing equations generate the training signal.  To see why, recall the key distinction: in supervised learning, the loss function measures the discrepancy between the network's prediction $\hat{y}^{(i)}$ and a known target label $y^{(i)}$, for example, a mean squared error $\frac{1}{m}\sum_i (\hat{y}^{(i)} - y^{(i)})^2$.  This requires a dataset of correct input--output pairs.

In DEQNs and PINNs, \emph{no such labels exist}.  Consider the key differences:
\begin{itemize}\itemsep2pt
    \item \textbf{DEQNs:} A neural network approximates the unknown policy function of a dynamic economic model.  The loss function is the \emph{Euler equation residual}, which measures how much the network's output violates the model's optimality conditions at sampled state points.  The ``correct'' policy values are never provided; instead, the network discovers the equilibrium by driving these residuals to zero.
    \item \textbf{PINNs:} A neural network approximates the unknown solution to a partial differential equation (PDE).  The loss function is the \emph{PDE residual}, which evaluates how well the network satisfies the differential equation at randomly sampled collocation points.  Again, the true solution is not available as training data; the network learns it by enforcing the PDE constraint.
\end{itemize}
In both cases, the training data consists only of \emph{input locations} (sampled states or collocation points) with no associated output labels.  The loss is defined entirely by the structure of the economic model or the governing equation, not by example solutions.  They are therefore unsupervised in the narrow sense of using no labels, but the more informative term is equation-based self-supervision.

Despite this fundamental difference, the optimization machinery is shared: these approaches define a loss $J(\bm{\theta})$ over trainable parameters and minimize it via (stochastic) gradient descent.  This is why we introduce the supervised learning pipeline first in the next section: it establishes the model--loss--optimizer framework that DEQNs and PINNs then adapt by replacing the data-driven loss with a physics-based one.

\section{The Supervised Learning Pipeline}

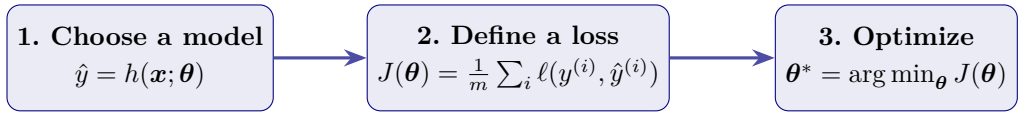
\begin{figure}[ht]
\centering
\begin{tikzpicture}[
    mlstep/.style={rectangle, draw=uzhblue, fill=uzhblue!8, rounded corners=5pt,
                 minimum width=3.2cm, minimum height=1.4cm, align=center, font=\small},
    arr/.style={-{Stealth[length=3mm]}, very thick, uzhblue!70}
]
    \node[mlstep] (m) at (0,0) {\textbf{1.\ Choose a model}\\[2pt]$\hat{y} = h(\x;\bm{\theta})$};
    \node[mlstep] (l) at (5,0) {\textbf{2.\ Define a loss}\\[2pt]$J(\bm{\theta}) = \frac{1}{m}\sum_{i}\ell(y^{(i)},\hat{y}^{(i)})$};
    \node[mlstep] (o) at (10,0) {\textbf{3.\ Optimize}\\[2pt]$\bm{\theta}^{*} = \argmin_{\bm{\theta}} J(\bm{\theta})$};
    \draw[arr] (m) -- (l);
    \draw[arr] (l) -- (o);
\end{tikzpicture}
\caption{The three-step supervised-learning recipe that underpins every model in this course.  Choose a parametric hypothesis $h(\x;\bm{\theta})$, measure its misfit on a labeled dataset via a loss $J(\bm{\theta})$, and minimize $J$ over the parameter vector $\bm{\theta}$.  DEQNs and PINNs modify step~2 (replacing the data-driven loss with an equilibrium or PDE residual) while keeping steps~1 and~3 identical.}
\label{fig:ml_recipe}
\end{figure}

Every supervised learning algorithm follows the same three-step recipe, regardless of whether the model is a linear regression, a random forest, or a deep neural network (Figure~\ref{fig:ml_recipe}).  Understanding this pipeline is essential because the DEQN and PINN methods in later chapters modify step~2 (replacing data-driven losses with physics-based residuals) while keeping steps~1 and~3 intact.

Given a training set $\{(\x^{(i)}, y^{(i)})\}_{i=1}^{m}$ of input--output pairs, we seek a hypothesis $h:\mathcal{X}\to\mathcal{Y}$ parameterized by $\bm{\theta}$ that minimizes the empirical risk.  For regression problems, the default choice is the mean squared error (MSE):
\begin{equation}
J(\bm{\theta}) = \frac{1}{m}\sum_{i=1}^{m}\big(h_{\bm{\theta}}(\x^{(i)}) - y^{(i)}\big)^{2}.
\label{eq:mse}
\end{equation}
This loss is not chosen arbitrarily.  If the data are generated by
\[
y^{(i)} = h_{\bm{\theta}}(\x^{(i)}) + \varepsilon^{(i)}, \qquad \varepsilon^{(i)} \sim \mathcal{N}(0,\sigma^2),
\]
then the log-likelihood of the sample is, up to constants, proportional to $-\sum_i (h_{\bm{\theta}}(\x^{(i)}) - y^{(i)})^2$.  Minimizing MSE is therefore equivalent to maximum likelihood under homoskedastic Gaussian observation noise.  This is one reason why squared error is the natural benchmark loss for regression; see \citet{bishop2006, goodfellow2016deep, deisenroth2020mathematics}.

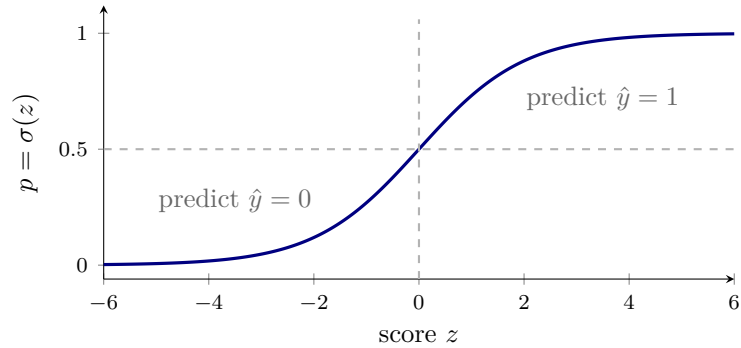
\begin{figure}[ht]
\centering
\begin{tikzpicture}
    \begin{axis}[
        width=0.62\textwidth,
        height=5.2cm,
        xmin=-6, xmax=6,
        ymin=-0.06, ymax=1.12,
        domain=-6:6,
        samples=250,
        xlabel={score $z$},
        ylabel={$p = \sigma(z)$},
        axis lines=left,
        enlargelimits=false,
        tick label style={font=\scriptsize},
        label style={font=\small},
        ytick={0, 0.5, 1},
        xtick={-6,-4,-2,0,2,4,6},
    ]
        \addplot[uzhblue, very thick] {1/(1+exp(-x))};
        \addplot[gray!60, dashed, thick] coordinates {(-6,0.5)(6,0.5)};
        \addplot[gray!60, dashed, thick] coordinates {(0,-0.06)(0,1.06)};
        \node[font=\small, gray!80!black, fill=white, inner sep=2pt] at (axis cs:-3.5,0.28) {predict $\hat{y}=0$};
        \node[font=\small, gray!80!black, fill=white, inner sep=2pt] at (axis cs: 3.5,0.72) {predict $\hat{y}=1$};
    \end{axis}
\end{tikzpicture}
\caption{Binary classification with a sigmoid output.  A scalar score $z$ (the model's raw output) is mapped to a probability $p = \sigma(z) \in (0,1)$.  The prediction rule assigns class~1 whenever $p > 0.5$, equivalently whenever $z > 0$.  The dashed lines mark the decision threshold; no neural-network architecture is assumed---any model that produces a real-valued score can be combined with this mapping and the binary cross-entropy loss.}
\label{fig:sigmoid_decision}
\end{figure}

For classification tasks, the model must output class probabilities.  In the binary case ($K=2$), the simplest approach passes a single scalar score $z$ through the \emph{sigmoid} function,
\[
p \;=\; \sigma(z) \;=\; \frac{1}{1+e^{-z}} \;\in\; (0,1),
\]
and assigns class~1 whenever $p > 0.5$, equivalently whenever $z > 0$ (Figure~\ref{fig:sigmoid_decision}).  For $K>2$ classes the natural generalization maps a raw score vector $\bm{z}\in\mathbb{R}^K$ onto the probability simplex via \emph{softmax}: $\hat{y}_k = e^{z_k}/\sum_j e^{z_j}$.  In both cases the misfit between predicted probabilities and true labels is measured by the cross-entropy loss:
\begin{equation}
J = -\frac{1}{m}\sum_{i=1}^{m}\sum_{k=1}^{K} y_k^{(i)}\log \hat{y}_k^{(i)},
\label{eq:cross_entropy}
\end{equation}
where $\hat{y}_k^{(i)}$ is the predicted probability that observation $i$ belongs to class $k$.  If the target is encoded as a one-hot vector, exactly one component of $(y_1^{(i)},\dots,y_K^{(i)})$ equals one and all others are zero, so for an observation whose true class is $c$ the loss contribution reduces to $-\log \hat{y}_c^{(i)}$.  The model is rewarded for assigning high probability to the correct class and penalized heavily when that probability is near zero.

The origin of cross-entropy is again likelihood theory.  In binary classification, with label $y \in \{0,1\}$ and predicted success probability $p$, the Bernoulli log-likelihood is
\begin{equation*}
\log L = y \log p + (1-y)\log(1-p).
\end{equation*}
Negating and averaging this expression gives the binary cross-entropy.  The $K$-class formula above is the corresponding negative log-likelihood for a categorical distribution with probabilities generated by sigmoid ($K=2$) or softmax ($K>2$).  Cross-entropy is therefore the statistically natural loss whenever the model output is meant to represent class probabilities; see again \citet{bishop2006, deisenroth2020mathematics}.

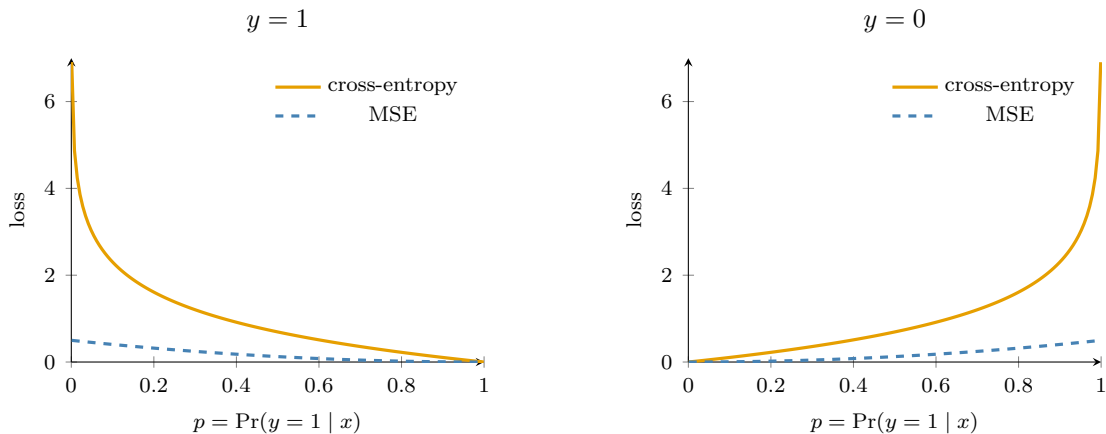
\begin{figure}[htbp]
\centering
\begin{tikzpicture}
    \begin{scope}
        \begin{axis}[
            width=0.44\textwidth,
            height=5.6cm,
            xmin=0, xmax=1,
            ymin=0, ymax=7,
            domain=0.001:0.999,
            samples=150,
            xlabel={$p = \Pr(y=1\mid x)$},
            ylabel={loss},
            title={$y=1$},
            axis lines=left,
            enlargelimits=false,
            legend style={draw=none, fill=none, font=\scriptsize, at={(0.97,0.97)}, anchor=north east},
            tick label style={font=\scriptsize},
            label style={font=\scriptsize},
            title style={font=\small}
        ]
            \addplot[softorange, very thick] {-ln(x)};
            \addlegendentry{cross-entropy}
            \addplot[softblue, very thick, dashed] {0.5*(1-x)^2};
            \addlegendentry{MSE}
        \end{axis}
    \end{scope}
    \begin{scope}[xshift=0.51\textwidth]
        \begin{axis}[
            width=0.44\textwidth,
            height=5.6cm,
            xmin=0, xmax=1,
            ymin=0, ymax=7,
            domain=0.001:0.999,
            samples=150,
            xlabel={$p = \Pr(y=1\mid x)$},
            ylabel={loss},
            title={$y=0$},
            axis lines=left,
            enlargelimits=false,
            legend style={draw=none, fill=none, font=\scriptsize, at={(0.97,0.97)}, anchor=north east},
            tick label style={font=\scriptsize},
            label style={font=\scriptsize},
            title style={font=\small}
        ]
            \addplot[softorange, very thick] {-ln(1-x)};
            \addlegendentry{cross-entropy}
            \addplot[softblue, very thick, dashed] {0.5*x^2};
            \addlegendentry{MSE}
        \end{axis}
    \end{scope}
\end{tikzpicture}
\caption{Binary cross-entropy and mean squared error as functions of the predicted class probability $p$.  Cross-entropy rises much more sharply near confident mistakes, which is why it is usually better aligned with probabilistic classification.}
\label{fig:classification_losses}
\end{figure}

Figure~\ref{fig:classification_losses} makes the practical difference visible.  If the true label is $y=1$ but the model predicts a very small $p$, then the cross-entropy loss explodes because $-\log p \to \infty$ as $p \downarrow 0$.  The same holds symmetrically when $y=0$ and $p \uparrow 1$.  Mean squared error does penalize mistakes, but it does so much more mildly near the boundaries.  For probabilistic classification, that weaker penalty is usually undesirable because it does not strongly discourage overconfident wrong predictions.

The optimization is performed via gradient descent or one of its stochastic variants, which we discuss in Section~\ref{sec:training}.

\subsection{Beyond MSE: Robust and Asymmetric Losses}
\label{sec:robust_losses}

MSE is optimal under Gaussian noise, but real-world economic and financial data often contain outliers and heavy tails that inflate the squared penalty disproportionately.  Two classical alternatives are useful in this course and beyond.

\paragraph{Huber loss.}  Introduced by \citet{huber1964robust} in the context of robust location estimation, the Huber loss behaves like MSE near the origin and like $L_1$ in the tails, capping the influence of any single observation:
\begin{equation}
\ell_\delta(r) = \begin{cases}
\tfrac{1}{2} r^2 & \text{if } |r| \leq \delta, \\
\delta\,(|r| - \tfrac{1}{2}\delta) & \text{otherwise,}
\end{cases}
\qquad r \equiv y - \hat{y}.
\label{eq:huber}
\end{equation}
The threshold $\delta$ controls the transition and is typically chosen to be a few times the noise scale.  Huber loss retains the smoothness needed for gradient-based optimization while reducing the weight of extreme residuals, which makes it the default choice for regression problems with suspected outliers.

\paragraph{Quantile (pinball) loss.}  \citet{koenker1978regression} proposed the \emph{check function}
\begin{equation}
\ell_\tau(r) = \max\!\bigl(\tau\, r,\; (\tau-1)\, r\bigr), \qquad \tau \in (0,1),
\end{equation}
whose minimizer is the conditional $\tau$-quantile of $y$ given $\x$ rather than the conditional mean.  Setting $\tau = 0.5$ recovers the median (absolute-error) regression; setting $\tau = 0.05$ or $\tau = 0.95$ targets the lower or upper tail.  In financial risk management this is precisely the statistic of interest: $\tau = 0.05$ yields a neural-network estimator of the lower-tail $5\%$-quantile of returns, which corresponds to Value-at-Risk (VaR) at the conventional $5\%$ level, and averaging the pinball loss over many quantiles traces out the full conditional distribution of returns, an approach known as quantile regression or distributional regression.

\begin{remarkbox}[Why this matters in economics and finance]
Tail risk is often more important than average performance.  The Huber and quantile losses let the network focus explicitly on robustness to outliers and on worst-case outcomes, respectively.  A single quantile loss gives a Value-at-Risk estimator at the chosen probability level; Expected Shortfall requires additional structure, such as averaging lower-tail quantiles, fitting several quantiles jointly, or using a dedicated joint VaR--ES scoring rule.  Notebook \texttt{07\_Genz\_Approximation\_and\_Loss\_Functions} compares MSE, Huber, and quantile losses on a common regression task.
\end{remarkbox}

\phantomsection
\section{From Perceptrons to Deep Networks}

\begin{figure}[ht]
\centering
\begin{tikzpicture}[scale=0.9, transform shape,
    neuron/.style={circle, draw=uzhblue, thick, minimum size=10mm, fill=uzhblue!8},
    arr/.style={-{Stealth[length=2.5mm]}, thick, uzhblue!70}]
    \node[font=\small] (x1) at (0,2) {$x_1$};
    \node[font=\small] (x2) at (0,0.8) {$x_2$};
    \node[font=\small] at (0,-0.1) {$\vdots$};
    \node[font=\small] (xm) at (0,-1) {$x_d$};
    \node[font=\scriptsize, above] at (1.5,2) {$w_1$};
    \node[font=\scriptsize, above] at (1.5,0.8) {$w_2$};
    \node[font=\scriptsize, below] at (1.5,-0.7) {$w_d$};
    \node[neuron, font=\large] (sum) at (3.2,0.5) {$\Sigma$};
    \node[rectangle, draw=uzhblue, thick, fill=softorange!15,
          minimum width=10mm, minimum height=10mm, font=\small] (act) at (5.5,0.5) {$g(\cdot)$};
    \node[font=\small] (out) at (7.5,0.5) {$\hat{y}$};
    \draw[arr] (x1) -- (sum);
    \draw[arr] (x2) -- (sum);
    \draw[arr] (xm) -- (sum);
    \draw[arr] (sum) -- node[above, font=\scriptsize]{$z$} (act);
    \draw[arr] (act) -- (out);
\end{tikzpicture}
\caption{An artificial neuron in the McCulloch--Pitts lineage.  Inputs $x_i$ are multiplied by synaptic weights $w_i$, summed into a pre-activation $z$, and passed through a nonlinear activation $g(\cdot)$ to yield the output $\hat{y}$.  The original \citet{mcculloch1943logical} unit used a binary threshold for $g$; the modern artificial neuron generalizes this to arbitrary smooth activations, and all deep networks are compositions of neurons of this form.}
\label{fig:artificial_neuron}
\end{figure}
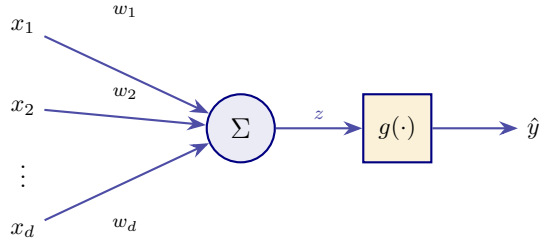

The building block of every neural network is the artificial neuron, first proposed by \citet{mcculloch1943logical}.  A companion question, how the synaptic weights $w_i$ themselves should adapt with experience, was first addressed by \citet{hebb1949organization}, whose rule ``neurons that fire together, wire together'' ($\Delta w_{ij} = \eta\, x_i y_j$) is the conceptual ancestor of all gradient-based learning rules discussed below.  \citet{rosenblatt1958perceptron} then introduced the Perceptron, the first trainable binary classifier built on these ideas.  A single neuron computes a weighted linear combination of its inputs, adds a bias term, and passes the result through a nonlinear activation function:
\begin{equation}
\hat{y} = g\!\big(w_0 + \x^\top \w\big),
\end{equation}
where $\w = (w_1, \dots, w_d)^\top$ are the synaptic weights, $w_0$ is the bias, and $g(\cdot)$ is the activation function (Figure~\ref{fig:artificial_neuron}).

Common choices for $g$ include the sigmoid $\sigma(z) = (1+e^{-z})^{-1}$, the hyperbolic tangent $\tanh(z)$, and the rectified linear unit $\mathrm{ReLU}(z) = \max(0,z)$ \citep{nair2010rectified, glorot2011deep}.  Without a nonlinear activation, any composition of linear layers collapses to a single affine map, a mathematical fact of fundamental importance: $\W_2(\W_1\x + \bb_1) + \bb_2 = \W_2\W_1\x + (\W_2\bb_1 + \bb_2)$.

\paragraph{From a single neuron to a layer.}  The single-neuron equation $\hat y = g(w_0 + \x^\top \w)$ produces a scalar output.  In practice we want vector-valued outputs (and, more importantly, vector-valued \emph{intermediate features}).  A \emph{layer} of $n$ parallel neurons, each with its own weights $\w^{(j)}$ and bias $w_0^{(j)}$, is a vector-valued generalization
\[
    \bm{a} \;=\; g\!\big(\W \x + \bb\big), \qquad \W \in \R^{n \times d},\ \bb \in \R^{n},\ \bm{a} \in \R^{n},
\]
where the nonlinearity $g$ is applied componentwise.  Each row of $\W$ is the weight vector of one neuron; stacking $n$ of them gives the matrix $\W$ at once, so the layer evaluates $n$ neurons in a single matrix--vector product.

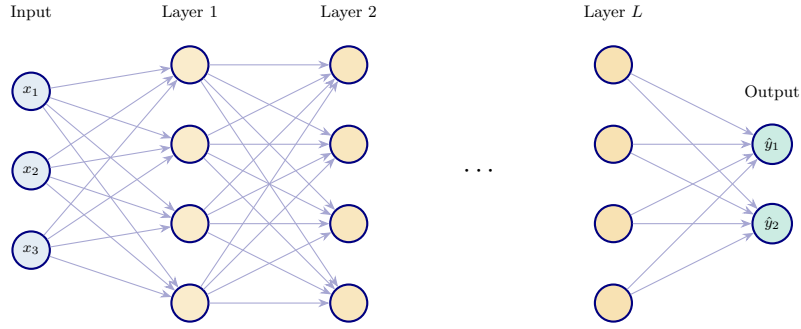
\begin{figure}[ht]
\centering
\begin{tikzpicture}[scale=0.7, transform shape,
    neuron/.style={circle, draw=uzhblue, thick, minimum size=7mm},
    arr/.style={-{Stealth[length=1.5mm]}, uzhblue!35}]
    \foreach \i/\y in {1/2,2/0.5,3/-1} {
        \node[neuron, fill=softblue!15, font=\scriptsize] (x\i) at (0,\y) {$x_\i$};
    }
    \foreach \i/\y in {1/2.5,2/1,3/-0.5,4/-2} {
        \node[neuron, fill=softorange!20, font=\tiny] (h1\i) at (3,\y) {};
    }
    \foreach \i/\y in {1/2.5,2/1,3/-0.5,4/-2} {
        \node[neuron, fill=softorange!25, font=\tiny] (h2\i) at (6,\y) {};
    }
    \foreach \i/\y in {1/2.5,2/1,3/-0.5,4/-2} {
        \node[neuron, fill=softorange!30, font=\tiny] (hL\i) at (11,\y) {};
    }
    \foreach \i/\y in {1/1,2/-0.5} {
        \node[neuron, fill=softgreen!20, font=\scriptsize] (y\i) at (14,\y) {$\hat{y}_\i$};
    }
    \node[font=\Large] at (8.5,0.5) {$\cdots$};
    \foreach \i in {1,2,3} { \foreach \j in {1,2,3,4} { \draw[arr] (x\i)--(h1\j); }}
    \foreach \i in {1,2,3,4} { \foreach \j in {1,2,3,4} { \draw[arr] (h1\i)--(h2\j); }}
    \foreach \i in {1,2,3,4} { \foreach \j in {1,2} { \draw[arr] (hL\i)--(y\j); }}
    \node[font=\footnotesize, above] at (0,3.2) {Input};
    \node[font=\footnotesize, above] at (3,3.2) {Layer 1};
    \node[font=\footnotesize, above] at (6,3.2) {Layer 2};
    \node[font=\footnotesize, above] at (11,3.2) {Layer $L$};
    \node[font=\footnotesize, above] at (14,1.7) {Output};
\end{tikzpicture}
\caption{An $L$-layer deep feedforward network.  Each layer applies an affine map followed by a pointwise nonlinearity; the composition realizes Eq.~\eqref{eq:dnn}.  Depth (rather than width) is what gives neural networks their efficient representational power for compositionally structured functions.}
\label{fig:deep_ff_net}
\end{figure}

\paragraph{From one layer to a deep composition.}  A single hidden layer is already a universal approximator (\citealt{cybenko1989approximation, hornik1989multilayer}; the universal-approximation theorem stated in the next subsection), but its hidden-layer width can grow exponentially in the input dimension to attain a target accuracy.  Stacking layers on top of one another reuses earlier features as inputs to later neurons; the resulting compositional representation is dramatically more efficient for many functions of interest \citep{telgarsky2016benefits, barron1993universal}.  A \emph{deep feedforward network} with $L$ layers is therefore a nested composition of layer maps:
\begin{equation}
f(\x;\bm{\theta}) = g_L\!\Big(\W^{(L)} g_{L-1}\!\big(\cdots g_1\!\big(\W^{(1)}\x + \bb^{(1)}\big)\cdots + \bb^{(L-1)}\big) + \bb^{(L)}\Big).
\label{eq:dnn}
\end{equation}
The architecture that implements~\eqref{eq:dnn} is sketched in Figure~\ref{fig:deep_ff_net}.

A useful geometric intuition, popularized by \citet{chollet2017deeplearning}, is that each layer of the network performs a nonlinear coordinate transformation, successively ``untangling'' the manifold on which the data lies.  In the input space, the data may be entangled in complex ways (e.g., two classes forming concentric spirals); each hidden layer warps the space so that the data become progressively more linearly separable.  By the final hidden layer, a simple linear readout suffices.  This perspective, formalized also by \citet{goodfellow2016deep}, explains why depth is so powerful: each layer adds an additional coordinate transformation, and the composition of many simple transformations can represent very complex mappings with far fewer parameters than a single, wide layer would require.

\paragraph{Other architectures.}  This course focuses almost entirely on feedforward networks of the form~\eqref{eq:dnn}, because DEQNs and PINNs operate on unstructured state vectors $\x \in \R^d$ for which feedforward maps are the natural choice.  For structured inputs other architecture families exist and are used widely elsewhere: convolutional networks \citep[e.g.,][]{lecun2015deep} for image data, graph neural networks for relational data, and Transformers (Section~\ref{sec:transformers}) for sequences.  We mention them so that readers who encounter these models in the empirical-finance or applied-ML literatures know where they fit; none are required for the methods developed in later chapters.

The \emph{universal approximation theorem} \citep{cybenko1989approximation, hornik1989multilayer} guarantees that even a single hidden layer with sufficiently many neurons can approximate any continuous function on a compact set to arbitrary precision.  However, in practice, deep (multi-layer) networks achieve the same accuracy with exponentially fewer parameters than wide (single-layer) ones, which motivates the use of depth; \citet{telgarsky2016benefits} makes this precise by exhibiting compositional functions that a depth-$L$ network can represent in $\mathrm{poly}(d,L)$ parameters but for which any depth-$(L-1)$ network requires width exponential in $L$.  \citet{barron1993universal} provides classical dimension-independent approximation rates for Barron-class targets, often stated as squared $L^2$ error of order $\mathcal{O}(1/n_1)$ in the hidden width, whereas tensor-product methods for generic smooth functions scale poorly in the total number of grid nodes.  This qualified comparison is the formal version of the ``deep learning can beat grids'' argument that motivates DEQNs in Chapter~\ref{ch:deqn}.

\begin{definitionbox}[Universal Approximation Theorem]
Let $g:\R\to\R$ be a bounded, non-constant, continuous activation function.  For any continuous function $f:\mathcal{K}\to\R$ on a compact set $\mathcal{K}\subset\R^d$ and any $\varepsilon>0$, there exists a single-hidden-layer network $F(\x) = \sum_{j=1}^{n_1} v_j\, g(\w_j^\top\x + b_j)$ such that $\sup_{\x\in\mathcal{K}} |F(\x) - f(\x)| < \varepsilon$.
\end{definitionbox}

\section{Training: Loss Functions, Gradient Descent, and Backpropagation}
\label{sec:training}

Given a loss function $J(\bm{\theta})$, training proceeds by iteratively updating the parameters in the direction of steepest descent:
\begin{equation}
\bm{\theta} \leftarrow \bm{\theta} - \eta\,\nabla_{\!\bm{\theta}} J(\bm{\theta}),
\label{eq:gd}
\end{equation}
where $\eta > 0$ is the learning rate.  In this introductory chapter $\bm{\theta}$ denotes generic trainable model parameters; in later chapters, when structural parameters and neural-network weights appear together, the script uses $\bm{\theta}$ for the structural parameters and $\rho$ for the network weights.  Computing the gradient $\nabla_{\!\bm{\theta}} J$ for a deep network is achieved through \emph{backpropagation} \citep{rumelhart1986learning}, an efficient application of the chain rule that propagates error signals from the output layer back to the input layer.  Appendix~\ref{app:ad} collects the matrix-calculus identities and the one-paragraph reverse-mode AD summary used throughout the script.

To see why the chain rule is central, consider a network with a single hidden layer.  Let $\z = \W^{(1)}\x + \bb^{(1)}$, $\a = g(\z)$, and $\hat{y} = \w^{(2)\top}\a + b^{(2)}$ with loss $J = \tfrac{1}{2}(\hat{y}-y)^2$.  Define the ``delta'' at the hidden layer:
\begin{equation}
\bm{\delta}^{(1)} = \frac{\partial J}{\partial \hat{y}} \cdot \frac{\partial \hat{y}}{\partial \a} \cdot \frac{\partial \a}{\partial \z} = (\hat{y}-y)\,\w^{(2)} \odot g'(\z).
\end{equation}
Then the weight gradient follows immediately:
\begin{equation}
\frac{\partial J}{\partial \W^{(1)}} = \bm{\delta}^{(1)}\,\x^\top.
\end{equation}
The key insight of \citet{rumelhart1986learning} is that this chain rule application can be organized as a single backward pass through the network, reusing intermediate quantities (the $\bm{\delta}$ vectors) from the forward pass.  The resulting algorithm has computational cost proportional to the forward pass; there is no need for finite differences or other expensive gradient approximations.

In practice, evaluating the full gradient over all $m$ training examples is expensive.  \emph{Stochastic gradient descent} (SGD), whose roots go back to the stochastic-approximation scheme of \citet{robbins1951stochastic}, replaces the full sum with a random mini-batch $\mathcal{B} \subset \{1, \dots, m\}$, yielding an unbiased estimate of the empirical-risk gradient at much lower cost per iteration:
\begin{equation}
\widehat{\nabla_{\!\bm{\theta}} J}_{\mathcal{B}}
= \frac{1}{|\mathcal{B}|}\sum_{i \in \mathcal{B}} \nabla_{\!\bm{\theta}}\,\ell\bigl(h_{\bm{\theta}}(\x^{(i)}),\, y^{(i)}\bigr).
\end{equation}
With mini-batch sizes of 32--256, SGD achieves both computational efficiency and an implicit regularization effect from gradient noise.  Two strands of theoretical work explain why this matters in deep nets specifically: the loss landscape of a deep network is dominated by saddle points rather than isolated bad local minima \citep{dauphin2014identifying}, and SGD's gradient noise tends to bias training toward \emph{flat} rather than \emph{sharp} regions of the loss surface, which often generalize better \citep{keskar2017large}.  Even on linearly separable data, gradient descent on the logistic loss converges to the maximum-margin solution, an instance of the broader principle that the optimizer itself imposes an \emph{implicit bias} that contributes to generalization \citep{soudry2018implicit}.  For a comprehensive modern review of stochastic optimization for large-scale learning (including convergence rates, adaptive methods, and variance reduction), see \citet{bottou2018optimization}.

\subsection{The Adam Optimizer}

Modern optimizers such as Adam \citep{kingma2015adam} adapt the learning rate for each parameter based on running averages of the first and second moments of the gradient:
\begin{align}
\bm{m}_t &= \beta_1\,\bm{m}_{t-1} + (1-\beta_1)\,\nabla J_t, \\
\bm{v}_t &= \beta_2\,\bm{v}_{t-1} + (1-\beta_2)\,(\nabla J_t)^2, \\
\hat{\bm{m}}_t &= \bm{m}_t/(1-\beta_1^t), \qquad \hat{\bm{v}}_t = \bm{v}_t/(1-\beta_2^t), \\
\bm{\theta}_{t+1} &= \bm{\theta}_t - \eta \cdot \hat{\bm{m}}_t / (\sqrt{\hat{\bm{v}}_t} + \varepsilon).
\end{align}
The bias-corrected first moment $\hat{\bm{m}}_t$ provides momentum (smoothing out gradient noise), while the second moment $\hat{\bm{v}}_t$ provides per-parameter adaptive learning rates (parameters with large gradients receive smaller effective steps).  The default hyperparameters $\beta_1=0.9$, $\beta_2=0.999$, $\varepsilon=10^{-8}$ work well across a wide range of problems, including all the economic applications in this course.

\subsection{The Optimizer Family Tree: Momentum, RMSprop, Adam, AdamW}
\label{sec:optimizer_zoo}

Adam did not appear out of thin air; it inherits from a family of refinements to plain SGD whose interactions are worth being explicit about for readers who will tune optimizers in practice.  Table~\ref{tab:optimizer_family} traces the lineage; each row is a one-line modification of the row above it.

\begin{table}[ht]
\centering
\footnotesize
\setlength{\tabcolsep}{4pt}
\begin{tabular}{@{} >{\bfseries}p{2.4cm} p{8.4cm} p{3.2cm} @{}}
\toprule
\textbf{Optimizer} & \textbf{Update rule (one parameter)} & \textbf{Reference} \\
\midrule
SGD             & $\theta \leftarrow \theta - \eta\,g_t$                                                                                       & \citet{robbins1951stochastic} \\
SGD + momentum  & $v_t = \mu v_{t-1} + g_t$;\ \ $\theta \leftarrow \theta - \eta\,v_t$                                                         & \citet{sutskever2013importance} \\
RMSprop         & $s_t = \rho s_{t-1} + (1-\rho)g_t^2$;\ \ $\theta \leftarrow \theta - \eta\,g_t/\sqrt{s_t+\varepsilon}$                       & \citet{tieleman2012rmsprop} \\
Adam            & momentum on $g_t$ and on $g_t^2$ with bias correction (Eqs.\ above)                                                          & \citet{kingma2015adam} \\
AdamW           & Adam plus \emph{decoupled} weight decay $\theta \leftarrow (1-\eta\lambda)\theta - \eta\,\hat m_t/(\sqrt{\hat v_t}+\varepsilon)$ & \citet{loshchilov2019decoupled} \\
\bottomrule
\end{tabular}
\caption{Lineage from plain SGD to AdamW.  Each row introduces exactly one new ingredient: momentum buffers gradient noise; RMSprop adds a per-parameter learning-rate scaling by the running second moment; Adam combines the two with bias correction; AdamW separates weight decay from the gradient step so that the implicit $L_2$ regularizer does not interact with the adaptive denominator.  PINN training in continuous-time chapters (Chapters~\ref{ch:pinn}--\ref{ch:ct_theory}) often uses Adam or AdamW; DEQNs in Chapters~\ref{ch:deqn}--\ref{ch:young} use plain Adam with the default $(\beta_1,\beta_2)=(0.9, 0.999)$ as in \citet{azinovicDEEPEQUILIBRIUMNETS2022}.}
\label{tab:optimizer_family}
\end{table}

The Adam-vs-AdamW distinction is sharper than the one-line table entry suggests, so it is worth writing out both rules side by side.  With $\hat m_t$, $\hat v_t$ the bias-corrected first and second moment of the gradient and $\lambda$ the weight-decay rate, Adam-with-$L_2$ (i.e.\ Adam applied to the loss $J + \tfrac{\lambda}{2}\|\bm\theta\|^2$) updates
\[
\bm\theta_{t+1} \;=\; \bm\theta_t \;-\; \eta\,\frac{\hat m_t + \lambda\,\bm\theta_t}{\sqrt{\hat v_t}+\varepsilon},
\]
so the implicit regularizer is itself rescaled by the adaptive denominator $\sqrt{\hat v_t}+\varepsilon$.  AdamW separates the two:
\[
\bm\theta_{t+1} \;=\; (1-\eta\lambda)\,\bm\theta_t \;-\; \eta\,\frac{\hat m_t}{\sqrt{\hat v_t}+\varepsilon},
\]
so the weight-decay term shrinks every parameter by the same proportional factor regardless of gradient magnitude.  This is why AdamW recovers the textbook intuition ``weight decay shrinks weights uniformly'' that Adam-with-$L_2$ loses.

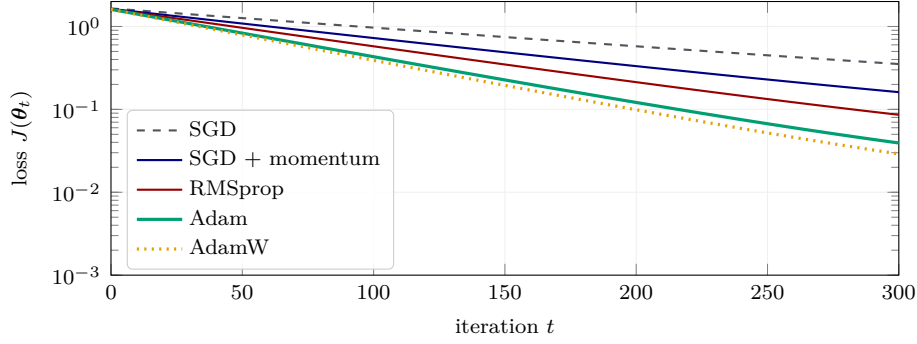
\begin{figure}[ht]
\centering
\begin{tikzpicture}
\begin{axis}[
    width=12cm, height=5.2cm,
    xlabel={iteration $t$}, ylabel={loss $J(\bm\theta_t)$},
    xmin=0, xmax=300, ymin=0.001, ymax=2,
    ymode=log,
    grid=major, grid style={gray!12},
    legend style={at={(0.02,0.02)}, anchor=south west, font=\scriptsize,
                  draw=gray!40, fill=white, fill opacity=0.9, text opacity=1, rounded corners=2pt},
    legend cell align=left,
    every axis plot/.append style={thick, no markers},
    label style={font=\scriptsize}, tick label style={font=\scriptsize}
]
\addplot[uzhgreydark, dashed, domain=0:300, samples=80] {1.6*exp(-x/180)+0.05};         \addlegendentry{SGD}
\addplot[uzhblue, domain=0:300, samples=80]              {1.6*exp(-x/120)+0.03};         \addlegendentry{SGD + momentum}
\addplot[harvardcrimson, domain=0:300, samples=80]       {1.6*exp(-x/95) +0.018};        \addlegendentry{RMSprop}
\addplot[softgreen, very thick, domain=0:300, samples=80]{1.6*exp(-x/75) +0.010};        \addlegendentry{Adam}
\addplot[softorange, dotted, very thick, domain=0:300, samples=80]
                                                        {1.6*exp(-x/70) +0.007};         \addlegendentry{AdamW}
\end{axis}
\end{tikzpicture}
\caption{Schematic loss-trajectory comparison on a moderately ill-conditioned objective.  Momentum and adaptive rescaling often improve early convergence relative to plain SGD, and Adam/AdamW are therefore useful defaults in the notebooks.  The ranking is illustrative rather than universal: on some objectives, carefully tuned SGD or RMSprop can match or beat Adam-family methods.}
\label{fig:optimizer_trajectories}
\end{figure}

Figure~\ref{fig:optimizer_trajectories} gives a schematic comparison of the qualitative convergence patterns behind this optimizer family tree.

\subsection{Learning Rate Schedules}

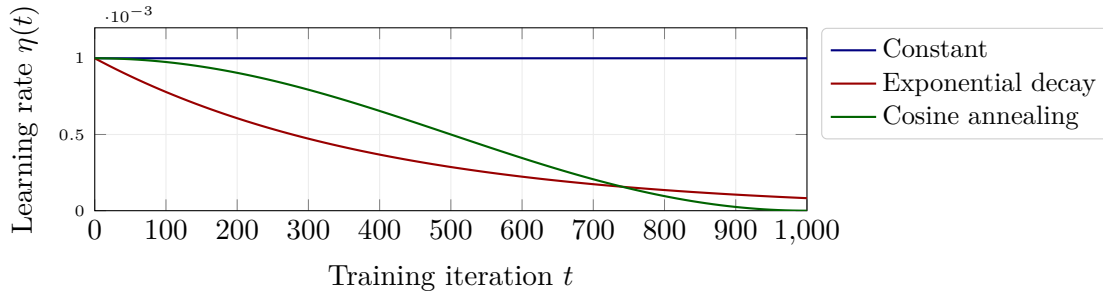
\begin{figure}[ht]
\centering
\begin{tikzpicture}
\begin{axis}[
    width=11cm, height=4cm,
    xlabel={Training iteration $t$},
    ylabel={Learning rate $\eta(t)$},
    xmin=0, xmax=1000, ymin=0, ymax=0.0012,
    grid=major, grid style={gray!15},
    legend style={at={(1.02,1.0)}, anchor=north west, font=\small,
                  draw=gray!40, fill=white, rounded corners=2pt},
    legend cell align=left,
    every axis plot/.append style={thick, no markers},
    yticklabel style={font=\tiny, /pgf/number format/fixed, /pgf/number format/precision=4},
]
\addplot[uzhblue, domain=0:1000, samples=100] {0.001};
\addlegendentry{Constant}
\addplot[harvardcrimson, domain=0:1000, samples=100] {0.001 * exp(-x/400)};
\addlegendentry{Exponential decay}
\addplot[darkgreen, domain=0:1000, samples=100]
    {0.001 * (1 + cos(180*x/1000))/2};
\addlegendentry{Cosine annealing}
\end{axis}
\end{tikzpicture}
\caption{Three common learning-rate schedules.  A constant rate is simple but often converges slowly in the fine-tuning phase; exponential decay shrinks monotonically; cosine annealing \citep{loshchilov2017sgdr} provides a smooth warm-to-cold transition that empirically performs well across a wide range of problems.  DEQNs and PINNs typically use exponential decay or cosine annealing to polish the solution after the initial coarse-grained phase.}
\label{fig:lr_schedules}
\end{figure}

The choice of learning rate $\eta$ is arguably the single most important hyperparameter in deep learning.  Too large, and the optimizer diverges; too small, and convergence is impractically slow.  A popular schedule is \emph{cosine annealing} \citep{loshchilov2017sgdr}, which smoothly decays the learning rate according to:
\begin{equation}
\eta(t) = \eta_{\min} + \tfrac{1}{2}(\eta_{\max} - \eta_{\min})\bigl(1 + \cos(\pi\, t / T)\bigr),
\end{equation}
where $T$ is the total number of training iterations.  Figure~\ref{fig:lr_schedules} compares the three learning-rate strategies used most often in practice.

In practice, decaying schedules such as exponential decay or cosine annealing tend to refine solutions in the later stages of training, once the optimizer has found a good basin of attraction.

\begin{keyinsightbox}[Loss function landscape]
The loss surface of a deep network is high-dimensional and non-convex, containing saddle points, plateaus, and sharp minima.  Stochastic optimization methods navigate this landscape effectively because the noise from mini-batch sampling helps escape shallow local minima and saddle points.  In economic applications, the loss function has direct economic interpretation, whether an Euler equation residual, a PDE residual, or a moment matching criterion, which provides a natural metric for assessing convergence quality.
\end{keyinsightbox}

\subsection{Backpropagation: The Chain Rule at Scale}

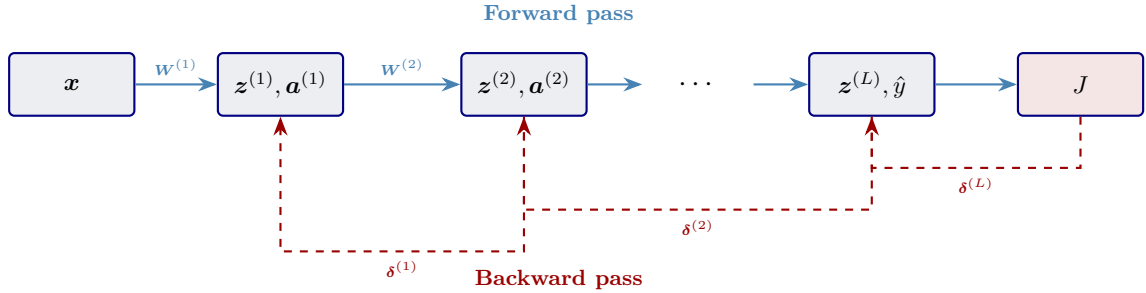
\begin{figure}[ht]
\centering
\begin{tikzpicture}[scale=0.92, transform shape,
    fwd/.style={-{Stealth[length=2.5mm]}, thick, softblue},
    bwd/.style={-{Stealth[length=2.5mm]}, thick, harvardcrimson, dashed},
    box/.style={rectangle, draw=uzhblue, thick, fill=uzhgreylight,
        minimum width=1.8cm, minimum height=0.9cm, font=\small, rounded corners=2pt}
]
    \node[box] (x) at (0,0) {$\x$};
    \node[box] (z1) at (3,0) {$\z^{(1)},\a^{(1)}$};
    \node[box] (z2) at (6.5,0) {$\z^{(2)},\a^{(2)}$};
    \node[font=\large] at (9,0) {$\cdots$};
    \node[box] (zL) at (11.5,0) {$\z^{(L)},\hat{y}$};
    \node[box, fill=harvardcrimson!10] (loss) at (14.5,0) {$J$};

    \draw[fwd] (x) -- node[above,font=\tiny]{$\W^{(1)}$} (z1);
    \draw[fwd] (z1) -- node[above,font=\tiny]{$\W^{(2)}$} (z2);
    \draw[fwd] (z2) -- (8.2,0);
    \draw[fwd] (9.8,0) -- (zL);
    \draw[fwd] (zL) -- (loss);

    \draw[bwd] (loss) -- ++(0,-1.2) -| node[below,font=\tiny,pos=0.25]{$\bm{\delta}^{(L)}$} (zL);
    \draw[bwd] (zL) -- ++(0,-1.8) -| node[below,font=\tiny,pos=0.25]{$\bm{\delta}^{(2)}$} (z2);
    \draw[bwd] (z2) -- ++(0,-2.4) -| node[below,font=\tiny,pos=0.25]{$\bm{\delta}^{(1)}$} (z1);

    \node[font=\footnotesize, softblue] at (7,1) {\textbf{Forward pass}};
    \node[font=\footnotesize, harvardcrimson] at (7,-2.8) {\textbf{Backward pass}};
\end{tikzpicture}
\caption{Backpropagation as forward and backward passes through the network.  The forward pass (blue) produces and caches every layer's pre-activations $\z^{(l)}$ and activations $\a^{(l)}$; the backward pass (red, dashed) propagates the ``delta'' vectors $\bm{\delta}^{(l)}$ from the loss back to the inputs, reusing the cached quantities to compute all parameter gradients in a single sweep.}
\label{fig:backprop_passes}
\end{figure}

For a network with $L$ layers, denote the pre-activation at layer $l$ as $\z^{(l)} = \W^{(l)}\a^{(l-1)} + \bb^{(l)}$ and the activation as $\a^{(l)} = g(\z^{(l)})$.  If the final layer is linear, set $g'(z)=1$ at $l=L$ and interpret $\a^{(L)}$ as the prediction $\hat y$.  The backpropagation algorithm computes $\partial J / \partial \W^{(l)}$ for all layers simultaneously by propagating a ``delta'' vector backward:
\begin{align}
\bm{\delta}^{(L)} &= \nabla_{\a^{(L)}} J \odot g'(\z^{(L)}), \\
\bm{\delta}^{(l)} &= \bigl((\W^{(l+1)})^\top \bm{\delta}^{(l+1)}\bigr) \odot g'(\z^{(l)}), \qquad l = L-1, \ldots, 1,
\end{align}
where $\odot$ denotes element-wise multiplication.  The parameter gradients are then $\partial J / \partial \W^{(l)} = \bm{\delta}^{(l)} (\a^{(l-1)})^\top$ and $\partial J / \partial \bb^{(l)} = \bm{\delta}^{(l)}$.  The computational cost is linear in the number of layers and the total number of parameters, a remarkable efficiency that enables training networks with millions of parameters.  Figure~\ref{fig:backprop_passes} shows the forward and backward passes side by side.

In modern deep learning frameworks such as TensorFlow and PyTorch, backpropagation is implemented automatically through computational graph tracing (``autograd'').  The user only needs to define the forward computation; the framework handles the differentiation.  This same automatic differentiation capability is what makes PINNs (Chapter~\ref{ch:pinn}) possible: the PDE residual requires derivatives of the network output with respect to its inputs, which autograd provides exactly and efficiently.

\section{Weight Initialization}
\label{sec:weight_init}

The initialization of network weights has a profound effect on training dynamics.  If weights are initialized too large, activations explode through the layers; if too small, they vanish to zero.  In both cases, the gradient signal degrades and training stalls.  The key principle is to choose initial weights so that the variance of activations remains approximately constant across layers.

\paragraph{Xavier/Glorot initialization.}  \citet{glorot2010understanding} derived the following rule for networks with symmetric activations (such as $\tanh$).  For a layer with $n_\mathrm{in}$ input neurons and $n_\mathrm{out}$ output neurons, initialize weights as:
\begin{equation}
W_{ij} \sim \mathcal{U}\!\left[-\sqrt{\frac{6}{n_\mathrm{in}+n_\mathrm{out}}},\; \sqrt{\frac{6}{n_\mathrm{in}+n_\mathrm{out}}}\right]
\qquad\text{or equivalently}\qquad
W_{ij} \sim \mathcal{N}\!\left(0,\; \frac{2}{n_\mathrm{in}+n_\mathrm{out}}\right).
\label{eq:xavier}
\end{equation}
This ensures $\mathrm{Var}[a^{(l)}] \approx \mathrm{Var}[a^{(l-1)}]$ under the assumption that activations are in the linear regime of $\tanh$.

\paragraph{He initialization.}  For ReLU activations, \citet{he2015delving} showed that the weight variance should be doubled relative to the forward fan-in rule:
\begin{equation}
W_{ij} \sim \mathcal{N}\!\left(0,\; \frac{2}{n_\mathrm{in}}\right).
\label{eq:he_init}
\end{equation}
The justification is a \emph{second-moment-preserving} calculation, not a variance one.  For a centered symmetric pre-activation $z$,
\[
    \E{\mathrm{ReLU}(z)^2}
    \;=\; \int_0^\infty z^2 \, p(z) \, dz
    \;=\; \tfrac{1}{2}\,\E{z^2},
\]
because $p(z)$ is symmetric about zero, so the negative half of the integrand is killed and the positive half is preserved.  Doubling the input weight variance therefore preserves the second moment $\E{(z^{(l)})^2}$ across layers under ReLU.  Strictly speaking $\E{\mathrm{ReLU}(z)} > 0$, so the \emph{variance} (centered second moment) is slightly smaller than the second moment, and the factor of 2 is an approximation rather than an identity; in practice the approximation is excellent and He initialization is the default for ReLU-family networks throughout this course.

\begin{remarkbox}[Practical guidance]
The applications in this course use different activations depending on the task:
\textbf{(i)}~the introductory DEQN examples (Brock--Mirman, Chapter~\ref{ch:deqn}, \S\ref{sec:bm}) use \emph{ReLU} for its simplicity and fast training;
\textbf{(ii)}~the multi-country IRBC model (Chapter~\ref{ch:irbc}) and the deep surrogate (Chapter~\ref{ch:gp}) use \emph{Swish} ($z\sigma(z)$) for its smooth gradients;
\textbf{(iii)}~all PINN examples (Chapter~\ref{ch:pinn}) use \emph{$\tanh$}, whose $C^\infty$ smoothness is essential when the loss involves second-order derivatives.
For ReLU-family networks, He initialization (\texttt{kaiming\_normal\_} in PyTorch, \texttt{he\_normal} in Keras) is the natural default.  For $\tanh$ and sigmoid networks, Xavier/Glorot initialization is usually the cleaner starting point; Swish and related smooth non-monotone activations often work well with either He-style or Xavier-style scaling, so the notebooks state the chosen initializer explicitly when it matters.
\end{remarkbox}

\section{Activation Functions in Depth}
\label{sec:activations}

Beyond the three classical choices (sigmoid, tanh, ReLU), several modern activation functions address specific shortcomings.  Table~\ref{tab:activations} summarizes the options used in this course.

\begin{table}[ht]
\centering
\small
\setlength{\tabcolsep}{5pt}
\begin{tabular}{@{}>{\bfseries}l l l l@{}}
\toprule
\textbf{Activation} & \textbf{Formula} & \textbf{Range} & \textbf{Key property} \\
\midrule
Sigmoid     & $\sigma(z) = (1+e^{-z})^{-1}$              & $(0,1)$                       & Smooth, saturates           \\
Tanh        & $\tanh(z)$                                 & $(-1,1)$                      & Zero-centered, saturates     \\
ReLU        & $\max(0,z)$                                & $[0,\infty)$                  & Non-saturating for $z>0$    \\
Leaky ReLU  & $\max(\alpha z, z)$, $\alpha\!=\!0.1$      & $(-\infty,\infty)$            & No dead neurons             \\
ELU         & $z\text{ if }z>0;\ \alpha(e^z-1)\text{ else}$ & $[-\alpha,\infty)$         & Negative saturation; $C^1$ if $\alpha=1$ \\
Swish       & $z \cdot \sigma(z)$                        & $\approx [-0.28,\infty)$      & Smooth, non-monotone        \\
GELU        & $z\cdot\Phi(z)$                            & $\approx [-0.17,\infty)$      & Smooth, default in BERT / GPT \\
Mish        & $z\cdot\tanh(\ln(1+e^z))$                  & $\approx [-0.31,\infty)$      & Smooth, used in YOLOv4        \\
Softplus    & $\ln(1+e^z)$                               & $(0,\infty)$                  & Smooth ReLU approximation   \\
\bottomrule
\end{tabular}
\caption{Activation functions used throughout the course.  Origin papers: ReLU \citep{nair2010rectified}, Leaky ReLU \citep{maas2013rectifier}, ELU \citep{clevert2016elu}, Swish \citep{ramachandran2017swish}, GELU \citep{hendrycks2016gelu}, Mish \citep{misra2019mish}.  \emph{Range} is the set of output values for $z \in \R$.  \emph{Smoothness} matters when derivatives of the network output are needed: sigmoid, tanh, Swish, GELU, Mish, and softplus are $C^\infty$; ReLU is only $C^0$; Leaky ReLU is piecewise linear; ELU is piecewise $C^\infty$ and is $C^1$ at the origin only for $\alpha=1$.  Smooth activations are required for PINN applications that involve second-order derivatives (Chapter~\ref{ch:pinn}).}
\label{tab:activations}
\end{table}

Leaky ReLU and ELU address the dying-neuron issue by providing a small but nonzero gradient for negative inputs.  The Swish activation $\mathrm{swish}(z) = z\sigma(z)$ \citep{ramachandran2017swish}, which is used extensively in the DEQN and IRBC implementations of this course, combines the benefits of ReLU (non-saturating for large $z$) with smoothness at the origin.  Its derivative $\mathrm{swish}'(z) = \sigma(z) + z\sigma(z)(1-\sigma(z))$ is smooth everywhere and bounded between approximately $-0.1$ and $1.1$, which can improve optimization stability.

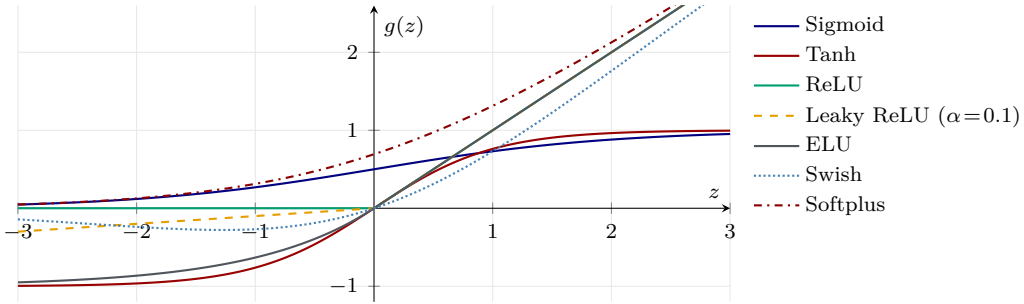
\begin{figure}[ht]
\centering
\begin{tikzpicture}
\begin{axis}[
    width=11cm, height=5.5cm,
    xlabel={$z$}, ylabel={$g(z)$},
    xmin=-3, xmax=3, ymin=-1.2, ymax=2.6,
    legend style={font=\scriptsize, at={(1.02,1)}, anchor=north west, draw=none, fill=none},
    legend cell align=left,
    grid=both, grid style={gray!18},
    axis lines=middle,
    samples=121, domain=-3:3,
    tick label style={font=\scriptsize},
    label style={font=\scriptsize}
]
\addplot[thick, uzhblue,        smooth] {1/(1+exp(-x))};                    \addlegendentry{Sigmoid}
\addplot[thick, harvardcrimson, smooth] {tanh(x)};                          \addlegendentry{Tanh}
\addplot[thick, softgreen,      smooth] {max(0,x)};                         \addlegendentry{ReLU}
\addplot[thick, softorange,     dashed, smooth] {max(0.1*x,x)};             \addlegendentry{Leaky ReLU ($\alpha\!=\!0.1$)}
\addplot[thick, uzhgreydark,    smooth]
        {(x>0)*x + (x<=0)*(exp(x)-1)};                                      \addlegendentry{ELU}
\addplot[thick, softblue,       densely dotted, smooth] {x/(1+exp(-x))};    \addlegendentry{Swish}
\addplot[thick, darkred,        dashdotted, smooth] {ln(1+exp(x))};         \addlegendentry{Softplus}
\end{axis}
\end{tikzpicture}
\caption{Seven representative activation functions from Table~\ref{tab:activations}.  Sigmoid and tanh saturate at large $|z|$ (vanishing gradients); ReLU is non-saturating but kinked at the origin; Leaky ReLU and ELU repair the dead-neuron problem with a small negative response; Swish and Softplus are everywhere $C^\infty$, which the PINN chapter (Chapter~\ref{ch:pinn}) requires.}
\label{fig:activations}
\end{figure}

For PDE applications (Chapter~\ref{ch:pinn}), the choice of activation function is particularly important because the PINN loss involves derivatives of the network output.  Since $\mathrm{ReLU}''(z) = 0$ almost everywhere, a ReLU network cannot represent second-order PDE residuals faithfully.  Smooth activations such as $\tanh$ ($C^\infty$) or Swish are therefore required for PINN applications involving second-order PDEs.  Figure~\ref{fig:activations} plots seven representative activations from Table~\ref{tab:activations}.

\section{Vanishing and Exploding Gradients}
\label{sec:vanishing}

A central obstacle to training deep networks is that the gradient signal reaching early layers can become either astronomically small or astronomically large as it is back-propagated through many layers.  The backward recursion derived above for the ``delta'' vector is
\begin{equation}
\bm{\delta}^{(l)} = \bigl((\W^{(l+1)})^\top \bm{\delta}^{(l+1)}\bigr) \odot g'(\z^{(l)}),
\end{equation}
so the magnitude of $\bm{\delta}^{(l)}$ is governed, roughly, by the product of derivatives $\prod_{k=l}^{L} g'(\z^{(k)})$ and the norms $\|\W^{(k)}\|$.  Two symmetric failure modes follow:
\begin{itemize}\itemsep2pt
\item \textbf{Vanishing gradients.}  For the sigmoid activation, $|g'(z)| \leq 1/4$; for $\tanh$, $|g'(z)| \leq 1$; and both derivatives approach zero when $|z|$ is large.  In the worst sigmoid case each layer shrinks the gradient by a factor close to $1/4$, so after $L=10$ layers the signal at the first layer can be attenuated by roughly $(1/4)^{10}\approx 10^{-6}$.  Early layers stop learning.
\item \textbf{Exploding gradients.}  If $|g'(z)| > 1$ or if $\|\W^{(k)}\|$ is large, the product grows instead of shrinking.  Updates become huge, parameters diverge, and the loss ``blows up''.
\end{itemize}
Three ingredients, each already introduced separately, combine to tame these problems:
\begin{enumerate}[itemsep=2pt]
\item \textbf{Non-saturating activations.}  ReLU has $g'(z) = 1$ for $z > 0$, eliminating the $(1/4)^L$ decay; Swish and tanh avoid vanishing when activations remain in a moderate range.
\item \textbf{Variance-preserving initialization.}  Xavier/Glorot \citep{glorot2010understanding} and He \citep{he2015delving} pick $\mathrm{Var}[W] \propto 1/n_\mathrm{in}$ precisely so that $\mathrm{Var}[\z^{(l)}]$ is constant across layers, keeping activations in the useful range of $g$ (Section~\ref{sec:weight_init}).
\item \textbf{Batch normalization} \citep{ioffe2015batch}.  Re-centering and re-scaling the pre-activations of each mini-batch prevents them from drifting toward the saturated tails of $g$ during training and allows much larger learning rates.  Its affine parameters $(\gamma, \beta)$ are learned.
\end{enumerate}
A practical complement is \emph{gradient clipping}: if $\|\nabla_{\bm{\theta}} J\|$ exceeds a threshold, rescale it to the threshold.  This eliminates the most damaging exploding-gradient events at negligible cost and is standard in RNN training; it is occasionally useful in DEQNs and PINNs when the residual magnitudes are highly unbalanced across collocation points.

\begin{keyinsightbox}[Course-wide implication]
Everywhere in this course where we train a network of non-trivial depth (Chapters~\ref{ch:deqn}--\ref{ch:pinn}), the combination of \emph{He/Xavier initialization}, a \emph{smooth non-saturating activation} (ReLU, Swish, tanh), and Adam's \emph{per-parameter adaptive step} keeps the gradient flow well conditioned.  Batch normalization is used when depth exceeds roughly ten layers or when the input distribution shifts substantially during training.
\end{keyinsightbox}

\subsection{Batch Normalization}
\label{sec:batchnorm}

Among the three mitigations listed above, batch normalization \citep{ioffe2015batch} (BN) deserves a closer look because it is simple to state, surprisingly effective in practice, and has become a default building block in supervised deep networks.  At its core BN is the standardization trick familiar from regression, re-centering each variable to mean zero and rescaling it to unit variance, applied separately to every layer's pre-activations and recomputed on every mini-batch of training data.

Let $z_1,\dots,z_B$ denote the pre-activations at one neuron over the $B$ examples in a mini-batch $\mathcal B$.  Batch normalization replaces them by
\begin{equation}
\mu_{\mathcal B} = \frac{1}{B}\sum_{i=1}^{B} z_i,
\quad
\sigma^2_{\mathcal B} = \frac{1}{B}\sum_{i=1}^{B} (z_i-\mu_{\mathcal B})^2,
\quad
\hat z_i = \frac{z_i-\mu_{\mathcal B}}{\sqrt{\sigma^2_{\mathcal B}+\varepsilon}},
\quad
y_i = \gamma\,\hat z_i + \beta,
\label{eq:batchnorm}
\end{equation}
where $\varepsilon$ is a small constant for numerical stability and $(\gamma,\beta)$ are \emph{learnable} scalar parameters specific to that neuron.  The transformed activation $y_i$ is what the next layer sees.  In the standard recipe BN is inserted between the linear map $\W^{(\ell)}\bm a^{(\ell-1)}+\bm b^{(\ell)}$ and the elementwise nonlinearity $g(\cdot)$.

\begin{figure}[ht]
\centering
\begin{tikzpicture}
\begin{axis}[
    name=plotNoBN,
    width=7cm, height=4.2cm,
    xmin=-4, xmax=6, ymin=0, ymax=0.75,
    xtick={-4,-2,0,2,4,6},
    xlabel={pre-activation $z$ at hidden layer $\ell$},
    ylabel={density},
    title={Without BatchNorm},
    title style={font=\small},
    label style={font=\footnotesize},
    tick label style={font=\footnotesize},
    axis lines=left, samples=120, no markers,
    legend style={font=\footnotesize, draw=none, fill=white, fill opacity=0.85,
                  at={(0.98,0.98)}, anchor=north east, row sep=-1pt}
]
    \addplot[very thick, softblue,   domain=-4:6] {1/sqrt(2*pi*0.4)*exp(-(x+1)^2/(2*0.4))};
    \addlegendentry{step 50}
    \addplot[very thick, softorange, domain=-4:6] {1/sqrt(2*pi*1.0)*exp(-(x-1)^2/(2*1.0))};
    \addlegendentry{step 500}
    \addplot[very thick, darkred,    domain=-4:6] {1/sqrt(2*pi*1.8)*exp(-(x-3)^2/(2*1.8))};
    \addlegendentry{step 5000}
\end{axis}
\begin{axis}[
    at={($(plotNoBN.east)+(1.0cm,0)$)}, anchor=west,
    width=7cm, height=4.2cm,
    xmin=-4, xmax=6, ymin=0, ymax=0.75,
    xtick={-4,-2,0,2,4,6},
    xlabel={normalized pre-activation $\hat z$},
    ylabel={density},
    title={With BatchNorm},
    title style={font=\small},
    label style={font=\footnotesize},
    tick label style={font=\footnotesize},
    axis lines=left, samples=120, no markers,
]
    \addplot[very thick, softgreen, domain=-4:6] {1/sqrt(2*pi)*exp(-x^2/2)};
    \node[font=\footnotesize, softgreen] at (axis cs:1.7,0.55) {all steps};
\end{axis}
\end{tikzpicture}
\caption{Distribution of pre-activations at one hidden neuron, sampled at three points during training.  \emph{Left:} without BatchNorm, the distribution drifts in mean and in variance as earlier layers update, each layer chases a moving target.  \emph{Right:} with BatchNorm, the affine pre-normalization transformation pins the inputs to $\mathcal{N}(0,1)$ at every training step, before the learned scale $\gamma$ and shift $\beta$ are applied.  The downstream layer always operates on inputs of the same scale, and the gradient signal flowing back is well conditioned.}
\label{fig:batchnorm_intuition}
\end{figure}
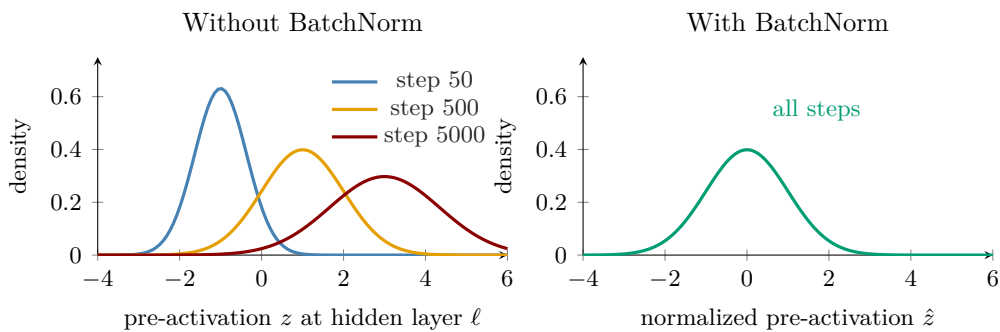

\paragraph{Why standardization, layer by layer.}  Without BN, the input distribution to a hidden layer $\ell$ depends on every weight in layers $1,\dots,\ell-1$.  As earlier weights update during gradient descent, the distribution faced by layer $\ell$ \emph{drifts} from one optimization step to the next: each layer therefore chases a moving target, a phenomenon \citet{ioffe2015batch} called \emph{internal covariate shift}.  BN pins the input distribution of every layer to mean zero and unit variance at every step (Figure~\ref{fig:batchnorm_intuition}).  Gradients become better conditioned, and substantially larger learning rates become safe.

\paragraph{The role of the affine parameters.}  At first glance the learnable shift and scale $(\gamma,\beta)$ seem to undo the normalization that BN just imposed.  This is exactly the point.  If a layer happens to prefer non-standard inputs, for example a tanh layer that needs slightly negative pre-activations to operate in its linear regime, the network is free to recover them via $(\gamma,\beta)$.  BN therefore never reduces the network's representational capacity; it merely shifts to a parameterization in which the optimization trajectory is easier to follow.

\paragraph{Why higher learning rates work.}  A precise Lipschitz bound depends on the operator norms of the surrounding weight matrices, the activation derivatives, and the learned affine scale $\gamma$.  The useful intuition is that BN reduces sensitivity to shifts and rescalings of intermediate activations, making the local optimization problem better conditioned and allowing step sizes that would otherwise cause divergence.  \citet{santurkar2018how} argue that loss-landscape smoothing, rather than the original ``internal covariate shift'' interpretation, better explains why BN helps optimization; the two views are complementary, but the smoothing perspective is the more directly testable one.

\paragraph{At inference time.}  During training, BN uses the current mini-batch to compute $\mu_{\mathcal B},\sigma^2_{\mathcal B}$.  At inference the mini-batch may be a single example, in which case those statistics would be ill-defined.  Implementations therefore maintain a running average of $\mu$ and $\sigma^2$ across training mini-batches and use these fixed estimates at test time, so the network's output is deterministic at deployment.

\subsection{Normalization Variants Beyond BatchNorm}
\label{sec:normalization_variants}

Batch normalization is the original normalization trick, but for several common deep-learning settings, small mini-batches, recurrent / sequence models, generative models, transformers, it is not the right one.  All variants share the form $\hat z = (z-\mu)/\sigma$ followed by a learned affine $\gamma\hat z + \beta$; they differ only in \emph{which axes} the statistics $(\mu,\sigma)$ are pooled over:

\begin{itemize}[itemsep=3pt]
\item \textbf{LayerNorm} \citep{ba2016layer}: pool across all features of a single example.  Decouples training from batch size and is the de-facto choice for RNNs and transformers; it can also be useful in small-batch residual methods, though PINNs more commonly rely on careful input/output scaling and smooth activations.
\item \textbf{GroupNorm} \citep{wu2018group}: pool over a group of channels of a single example.  Interpolates between LayerNorm ($G\!=\!1$) and InstanceNorm ($G$ = channels); the default in object detection and small-batch CNNs.
\item \textbf{WeightNorm} \citep{salimans2016weight}: reparameterize $\bm w = (g/\|\bm v\|)\,\bm v$ so the activation statistics never enter the gradient.  Avoids any batch-size dependence at the cost of slightly less representation power.
\end{itemize}

For the methods in this script, the practical guidance is: use BatchNorm for large supervised datasets (Chapter~\ref{ch:intro}), LayerNorm for recurrent or attention-based architectures and as an option when the effective batch size is small, and skip normalization entirely for small DEQN or PINN MLPs when input/output scaling plus Adam already condition the problem well.

\section{Generalization: Overfitting, Regularization, and Double Descent}
\label{sec:generalization}

\subsection{Train / Validation / Test Split}
\label{sec:traintest}

Before discussing overfitting formally, it is essential to fix the experimental protocol that every supervised-learning study should follow.  The available data are partitioned into three disjoint subsets:
\begin{itemize}\itemsep2pt
\item \textbf{Training set} (typically $\sim 70\%$): used to fit the model parameters $\bm{\theta}$ by minimizing the loss.
\item \textbf{Validation set} (typically $\sim 20\%$): used for \emph{model selection}, choosing hyperparameters, comparing architectures, deciding when to stop training.  The model's performance on this set guides these decisions but its parameters are never trained on it.
\item \textbf{Test set} (typically $\sim 10\%$): touched \emph{once}, at the very end, to report an unbiased estimate of generalization performance.
\end{itemize}
The key discipline is that no decision about the model (not hyperparameter tuning, not architecture, not early-stopping patience) may be informed by the test set.  Using the test set multiple times turns it into an implicit validation set and invalidates its role as a measure of out-of-sample error.  For small datasets, $k$-fold cross-validation replaces the fixed train/validation split: the training data are partitioned into $k$ equal folds; for each fold, the model is trained on the other $k-1$ folds and evaluated on the held-out fold; the $k$ resulting validation scores are averaged.  Common choices are $k = 5$ or $k = 10$; the test set is always held separately.  In DEQNs and PINNs (Chapters~\ref{ch:deqn} and~\ref{ch:pinn}), training and validation points are drawn from the same state distribution, and ``generalization'' is measured against an \emph{independently simulated test trajectory} rather than a held-out labeled set.

A model that memorizes the training data but fails on unseen examples is said to \emph{overfit}.  To understand overfitting precisely, consider the following thought experiment.  The decomposition below is the classical bias/variance analysis of \citet{geman1992biasvariance}, which provided the canonical framework for thinking about generalization in neural networks long before modern overparameterized regimes were studied.  Suppose we draw many independent training sets $\mathcal{D}$, each of size $n$, from the same data-generating process $y = f(\x) + \varepsilon$, where $\varepsilon$ is zero-mean noise with variance $\sigma^2$.  On each training set we fit our model, obtaining a predictor $\hat{f}_{\mathcal{D}}$.  Conditioning on a fixed test input $\x_0$ and averaging over both the training set and the new test noise $\varepsilon_0$, the squared prediction error decomposes into exactly three terms:
\begin{align}
\mathbb{E}_{\mathcal{D},\varepsilon_0}
\!\bigl[(y_0 - \hat{f}_{\mathcal{D}}(\x_0))^2\bigr]
&=
\underbrace{\bigl(f(\x_0) - \mathbb{E}_{\mathcal{D}}[\hat{f}_{\mathcal{D}}(\x_0)]\bigr)^2}_{\text{Bias}^2}
\notag\\
&\quad+
\underbrace{\mathbb{E}_{\mathcal{D}}\!\bigl[(\hat{f}_{\mathcal{D}}(\x_0) - \mathbb{E}_{\mathcal{D}}[\hat{f}_{\mathcal{D}}(\x_0)])^2\bigr]}_{\text{Variance}}
\;+\;
\underbrace{\sigma^2}_{\text{Irreducible noise}}.
\label{eq:bias-variance}
\end{align}
Each term captures a distinct source of error:
\begin{itemize}[itemsep=2pt]
\item \textbf{Bias$^2$} measures the systematic error: how far the \emph{average} prediction $\mathbb{E}_{\mathcal{D}}[\hat{f}_{\mathcal{D}}(\x_0)]$ is from the true function $f(\x_0)$.  A model that is too simple (e.g., a linear function fit to a nonlinear target) will have high bias regardless of how much data it sees.
\item \textbf{Variance} measures the sensitivity of the prediction to the particular training set drawn.  A highly flexible model (e.g., a large neural network) may fit each training set well, but the predictions can differ wildly across draws, which is overfitting.
\item \textbf{Irreducible noise} $\sigma^2$ is the error that no model can eliminate, because it stems from randomness in the data-generating process itself.
\end{itemize}
In classical statistics, there is a fundamental trade-off: reducing bias requires more flexible models, which increases variance.  However, modern deep networks challenge this picture.  \citet{zhang2017understanding} showed empirically that standard architectures can perfectly fit randomly labeled data, implying a VC-style capacity far beyond what classical bounds predict, yet still generalize well on real data.  \citet{belkin2019reconciling} subsequently demonstrated that with sufficient overparameterization, models exhibit a \emph{double descent} phenomenon: test error first increases as the model becomes more complex (classical regime), but then decreases again as the number of parameters greatly exceeds the number of data points (interpolation regime).  \citet{nakkiran2020deep} showed that this phenomenon extends to deep networks and occurs not only as a function of model size, but also as a function of training time (``epoch-wise double descent'') and dataset size.

The key techniques for preventing overfitting in neural networks are:
\begin{enumerate}[itemsep=2pt]
\item \textbf{Early stopping} \citep{prechelt1998early}: monitor validation loss and stop training when it begins to rise.
\item \textbf{Weight decay ($L_2$ regularization)} \citep{krogh1991simple}: add $\frac{\lambda}{2}\|\bm{\theta}\|^2$ to the loss.
\item \textbf{Dropout} \citep{srivastava2014dropout}: randomly drop a fraction $p$ of activations at every training step.  Two implementation conventions exist.  The original convention drops units during training and multiplies the outgoing weights or activations by the keep probability $1-p$ at test time, so that the expected activation matches the training-time expectation.  The now-standard \emph{inverted-dropout} convention divides the retained activations by $1-p$ during training, so no rescaling is needed at test time.  Either way, the mechanism is equivalent to training, on each mini-batch, a different sub-network drawn from an exponentially large ensemble that shares weights; the final network approximates the ensemble average.  Typical values are $p=0.5$ for hidden layers and $p=0.1$--$0.2$ for inputs.  Dropout is less commonly used in DEQN and PINN applications, where the loss is already noisy (stochastic collocation) and regularization is often supplied implicitly by the state-space sampling scheme.
\item \textbf{Data augmentation:} synthetically enlarge the training set via transformations.
\item \textbf{Batch normalization} \citep{ioffe2015batch}: normalize activations within each mini-batch to stabilize training; its mini-batch statistics also act as a mild regularizer.
\end{enumerate}

\begin{figure}[ht]
\centering
\begin{tikzpicture}
\begin{axis}[
    width=11cm, height=5.5cm,
    xlabel={Model complexity (number of parameters / $n$)},
    ylabel={Test error},
    xmin=0, xmax=5, ymin=0, ymax=3.5,
    xtick=\empty, ytick=\empty,
    grid=major, grid style={gray!15},
    axis lines=left,
    clip=false,
]
    \draw[dashed, darkred, thick] (axis cs:1,0) -- (axis cs:1,3.15);
    \node[font=\small\bfseries, darkred, anchor=south]
        at (axis cs:1, 3.22) {interpolation threshold ($p\!\approx\!n$)};

    \addplot[very thick, uzhblue, domain=0.05:1, samples=200]
        {1.3*exp(-1.8*x) + 0.9*x^2 + 2.0*exp(-(x-1)^2/0.05)};
    \addplot[very thick, uzhblue, domain=1:5, samples=200]
        {0.28 + 0.835*exp(-0.85*(x-1)) + 2.0*exp(-(x-1)^2/0.05)};

    \node[font=\footnotesize, darkred, align=center]
        at (axis cs:0.42, 2.2) {classical\\regime};
    \node[font=\footnotesize, softgreen, align=center]
        at (axis cs:3.5, 1.6) {modern /\\overparameterized regime};
\end{axis}
\end{tikzpicture}
\caption{Schematic of the double-descent phenomenon.  In the classical regime ($p < n$) test error follows the standard bias--variance U-curve; around the interpolation threshold $p \approx n$ test error can peak sharply because the fitted function is highly sensitive to noise; in the modern overparameterized regime ($p \gg n$) test error decreases again \citep{belkin2019reconciling, nakkiran2020deep}.  In some linearized, kernel, max-margin, or least-norm settings, gradient methods exhibit an implicit bias toward particular low-complexity interpolants; in nonlinear finite-width networks this bias depends on architecture, data, optimizer, initialization, and training protocol.  Axes are unitless; the qualitative shape, not the scale, is the point.  The curve is illustrative, not a measurement.}
\label{fig:double_descent}
\end{figure}

Figure~\ref{fig:double_descent} illustrates why classical bias--variance intuition breaks down for modern deep networks.  In the classical regime ($p < n$), increasing model capacity beyond a point leads to overfitting.  At the interpolation threshold ($p \approx n$), the model has just enough parameters to perfectly fit the training data, and the resulting solution can be extremely sensitive to noise.  In the modern regime ($p \gg n$), test error often decreases again because optimization and architecture bias select comparatively regular interpolating solutions rather than arbitrary ones \citep{belkin2019reconciling}.

This phenomenon has been documented across many architectures and datasets by \citet{nakkiran2020deep}, who showed that it persists even when controlling for effective model complexity.  The implications for computational economics are substantial but should not be overstated.  In DEQN and PINN applications, the practitioner controls both the network size (number of parameters $p$) and the amount of training data (number of collocation points $n$), and those collocation points are often resampled rather than fixed once and for all.  Overparameterized networks can therefore be useful and sometimes necessary, but their credibility must be checked by independent residual diagnostics, simulated trajectories, and benchmark comparisons rather than by parameter counting alone.

The double descent phenomenon can be explored interactively in the companion notebook \texttt{03\_Double\_Descent.ipynb}, which reproduces the spirit of the double-descent curve in a small kernel-regression / random-Fourier-features setting (\citealt{belkin2019reconciling}); see also \citet{nakkiran2020deep} for the deep-network / CIFAR-10 version of the experiment, which the notebook does \emph{not} attempt to replicate at scale.

\subsection{A Pointer to the Theory: Neural Tangent Kernel (NTK) and Benign Overfitting}
\label{sec:ntk_pointer}

Why \emph{does} an over-parameterized network with $p \gg n$ generalize instead of merely memorizing?  Two complementary lines of theory have emerged.

\paragraph{Neural Tangent Kernel (NTK).}  In the limit of infinite width and a particular initialization scaling, gradient-descent training of a deep network is described by kernel gradient descent with a fixed, deterministic kernel, the \emph{Neural Tangent Kernel} of \citet{jacot2018neural}.  In that lazy-training limit the network's function evolves almost linearly around its initialization, which explains why first-order optimization can be well behaved for very wide networks even though the finite-parameter objective is non-convex.

\paragraph{Benign overfitting.}  In linear and kernel settings, gradient methods often select a minimum-norm interpolating solution, which can behave much like a ridge-regularized least-squares estimator and inherit good generalization properties.  \citet{bartlett2020benign} make this precise for linear regression, showing that interpolation can be benign provided the spectrum of the input covariance has heavy enough tails.  Subsequent work has extended both stories beyond the simplest settings; for the practitioner the takeaway is that the NTK regime helps explain \emph{why training succeeds}, while benign-overfitting theory explains why interpolation need not imply poor test error under additional assumptions.

These two threads are not the final word: finite-width deviations from the NTK matter for feature learning, and benign overfitting requires conditions on the data covariance.  They nevertheless help explain why the rest of this script can use networks substantially wider than a classical degrees-of-freedom calculation would recommend, provided the numerical residuals are validated out of sample.

\section{Sequence Models: RNNs, LSTMs, and Attention}
\label{sec:sequence_models}

\begin{remarkbox}[Optional section]
This section and the in-context AR(1) aside (\S\ref{sec:incontext_ar1}) survey sequence architectures that the rest of the script does not use: Chapters~\ref{ch:deqn}--\ref{ch:climate} operate on unstructured state vectors and rely entirely on feedforward MLPs.  Readers focused on DEQN, PINN, and the structural-estimation chapters can skip directly to the Chapter Summary (page~\pageref{sec:ch1_summary}) without loss of continuity.  The material below is included for completeness and as a reference for readers who later encounter Transformers in empirical-finance or applied-ML work.
\end{remarkbox}

The chapters that follow rely almost entirely on feedforward networks, but many economic and financial datasets are intrinsically sequential.  Before closing this introduction, it is therefore useful to briefly survey the main architectures for sequence data.  We represent a sequence as \emph{tokens} $\x_1, \dots, \x_n$, where each token is a vector.  The word ``token'' is borrowed from natural-language processing, where a token is typically a word or sub-word piece; in the economic and financial context we use throughout this course a token is simply one element of the sequence, for example a scalar return, a vector of macroeconomic variables at one quarter, or a price-volume pair at one tick.  The algorithms below are agnostic to this choice; they only need the sequence elements to be real-valued vectors of a fixed dimension.

\subsection{Recurrent Neural Networks}

\begin{figure}[ht]
\centering
\begin{tikzpicture}[scale=0.9, transform shape,
    cell/.style={rectangle, draw=uzhblue, fill=uzhblue!5, minimum size=1cm, rounded corners=3pt},
    node/.style={circle, draw=gray, fill=gray!5, minimum size=0.6cm}]

    \foreach \t in {1,2,3} {
        \node[node] (x\t) at (2.5*\t-2.5, 0) {$\x_{\t}$};
        \node[cell] (h\t) at (2.5*\t-2.5, 1.8) {$\h_{\t}$};
        \node[node] (y\t) at (2.5*\t-2.5, 3.6) {$\hat{\y}_{\t}$};

        \draw[->, thick] (x\t) -- (h\t);
        \draw[->, thick] (h\t) -- (y\t);
    }

    \draw[->, thick] (h1) -- (h2);
    \draw[->, thick] (h2) -- (h3);

    \node at (-1.2, 1.8) {$\dots$};
    \draw[->, thick] (-0.8, 1.8) -- (h1);
    \node at (6.2, 1.8) {$\dots$};
    \draw[->, thick] (h3) -- (5.8, 1.8);

    \node[font=\small, uzhblue] at (1.25, 2.1) {$\Wh$};
    \node[font=\small, darkgreen] at (3.75, 2.1) {$\Wh$};
\end{tikzpicture}
\caption{An unrolled Recurrent Neural Network. The same parameters $\Wh, \Wx$ are reused at every time step, allowing the hidden state $\h_t$ to accumulate historical information.}
\label{fig:rnn}
\end{figure}
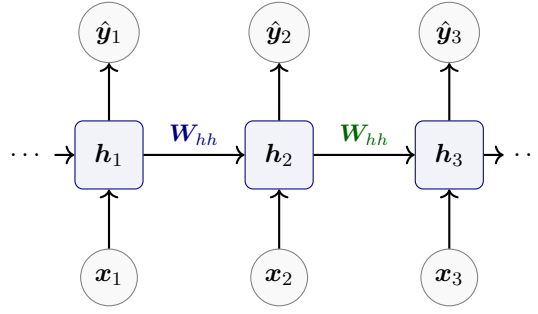

The traditional approach to sequence data is the \emph{Recurrent Neural Network} (RNN) \citep{elman1990finding}.  Unlike an MLP, which maps an input to an output in a single forward pass, an RNN maintains an internal \emph{hidden state} $\h_t$ that acts as a memory of past information.  At each time step $t$, the network updates this state using the current input $\x_t$ and the previous state $\h_{t-1}$:
\begin{equation}
\h_t = \sigma(\Wh \h_{t-1} + \Wx \x_t + \bb),
\end{equation}
where $\sigma$ is an activation function.  Concretely: for a scalar time series $\x_t$ is a scalar (e.g.\ log-return at date $t$), $\h_t \in \R^d$ is a $d$-dimensional hidden vector summarizing everything the network has seen so far, and $\Wh,\Wx,\bb$ are learnable parameters.  The same update is applied at every time step, so this recursive structure lets the network process sequences of arbitrary length with a fixed parameter budget.  Figure~\ref{fig:rnn} shows the resulting unrolled computation graph.

\paragraph{Training: Backpropagation Through Time (BPTT).}  Because the unrolled RNN \emph{is} a feedforward graph of depth $T$ with \emph{shared} weights, one can apply ordinary backpropagation and then \emph{sum} the weight-gradients across time.  Concretely, let $\mathcal{L}_T = \sum_{t=1}^{T}\ell(\hat{\y}_t, \y_t)$ denote the total loss.  For the recurrence above, the forward single-step Jacobian is
\[
J_t \equiv \frac{\partial \h_t}{\partial \h_{t-1}}
= \mathrm{diag}\!\bigl(\sigma'(\Wh\h_{t-1}+\Wx\x_t+\bb)\bigr)\Wh.
\]
With column gradients, the backward pass multiplies by $J_t^\top$.  Differentiating with respect to an early hidden state $\h_k$ therefore yields the schematic product
\begin{equation}\label{eq:bptt}
\nabla_{\h_k}\mathcal{L}_T
\;\approx\;
J_{k+1}^{\top}J_{k+2}^{\top}\cdots J_T^{\top}\,\nabla_{\h_T}\mathcal{L}_T,
\end{equation}
so the gradient picks up $T-k$ products of matrices containing the same recurrent weight matrix $\Wh$.  The relevant singular values or spectral radii of these factors determine the asymptotics.  If they are mostly below one, the gradient \emph{vanishes} exponentially in $T-k$ and the network cannot learn dependencies that span many steps; if they are mostly above one, it \emph{explodes}, producing NaNs during training.  This is the \emph{vanishing/exploding gradient problem}, originally analyzed by \citet{hochreiter1991untersuchungen} and developed formally in \citet{bengio1994learning, hochreiter2001gradient}, and revisited with a modern optimization-theoretic view by \citet{pascanu2013difficulty}.

Three practical remedies partially alleviate the pathology without changing the architecture.  \emph{Gradient clipping} \citep{pascanu2013difficulty} rescales the parameter-gradient whenever its norm exceeds a threshold; it eliminates the worst exploding events at negligible cost and is now a standard training default for any recurrent model.  \emph{Orthogonal or identity-like initialization} of $\Wh$ places its singular values near $1$ so that the Jacobian product preserves norms at the start of training.  \emph{Truncated BPTT} unrolls only $K$ steps back at each update, capping both the memory footprint and the effective gradient horizon at $K$.  These help but do not cure the problem: the structural fix is to change the recurrence itself, which is the move made by \emph{gated cells}.

\subsection{Long Short-Term Memory (LSTM) and GRUs}
\label{sec:lstm}

Before writing equations, it helps to keep a plain-language picture in mind.  A vanilla RNN asks one hidden state $\h_t$ to do three jobs at once: retain useful old information, absorb new information, and expose the relevant part to the next layer.  This is a fragile design.  The LSTM separates these tasks.  It keeps a dedicated \emph{memory lane} $\bm{C}_t$ flowing through time and at each date makes three soft decisions: what to keep, what to write, and what to reveal.  That is the intuition behind the gate equations below.

\begin{figure}[ht]
\centering
\begin{tikzpicture}[scale=1.0, transform shape,
    cell/.style={rectangle, draw=uzhblue, thick, rounded corners=5pt, fill=uzhgreylight,
                 minimum width=6.2cm, minimum height=1.6cm},
    g/.style={rectangle, draw=darkred, thick, fill=red!10,
              minimum size=0.7cm, font=\footnotesize, inner sep=2pt, align=center},
    plus/.style={circle, draw=darkgreen, thick, fill=white,
                 minimum size=0.48cm, inner sep=0pt, font=\footnotesize}]
\node[font=\footnotesize, darkgreen] at (0, 1.75) {\emph{protected memory lane}};
\node[cell] (c) at (0,0) {};
\node[g, fill=softorange!30]     (f) at (-1.9,0.05) {$\bm{f}_t$};
\node[g, fill=softgreen!30]      (i) at (-0.7,0.05) {$\bm{i}_t$};
\node[g, fill=softblue!30]       (u) at ( 0.55,0.05) {$\tilde{\bm{C}}_t$};
\node[g, fill=harvardcrimson!25] (o) at ( 1.8,0.05) {$\bm{o}_t$};
\node[plus] (add) at (0,1.15) {$+$};
\draw[->, very thick, darkgreen] (-3.9, 1.15) -- (-2.9, 1.15);
\draw[very thick, darkgreen] (-2.9, 1.15) -- (add.west);
\draw[very thick, darkgreen] (add.east) -- (2.9, 1.15);
\draw[->, very thick, darkgreen] (2.9, 1.15) -- (3.9, 1.15);
\node[darkgreen, font=\footnotesize, anchor=east] at (-3.95, 1.15) {$\bm{C}_{t-1}$};
\node[darkgreen, font=\footnotesize, anchor=west] at ( 3.95, 1.15) {$\bm{C}_t$};
\draw[->, very thick, uzhblue] (-3.9, -1.15) -- (-2.9, -1.15);
\draw[very thick, uzhblue] (-2.9, -1.15) -- (2.9, -1.15);
\draw[->, very thick, uzhblue] (2.9, -1.15) -- (3.9, -1.15);
\node[uzhblue, font=\footnotesize, anchor=east] at (-3.95, -1.15) {$\bm{h}_{t-1}$};
\node[uzhblue, font=\footnotesize, anchor=west] at ( 3.95, -1.15) {$\bm{h}_t$};
\node[font=\footnotesize, uzhblue] at (0, -1.75) {\emph{visible output}};
\draw[->, thick, softorange] (f.north) -- (-1.9,0.82);
\draw[->, thick, softgreen] (i.north) -- (-0.24,0.75);
\draw[->, thick, softblue] (u.north) -- (0.24,0.75);
\draw[->, thick, harvardcrimson] (o.south) -- (1.8,-0.82);
\node[below=0.04cm of f, font=\scriptsize, softorange] {keep old};
\node[below=0.04cm of i, font=\scriptsize, softgreen] {write};
\node[below=0.04cm of u, font=\scriptsize, softblue] {candidate};
\node[below=0.04cm of o, font=\scriptsize, harvardcrimson, fill=white, inner sep=1pt, rounded corners=0.5pt] {reveal};
\node[font=\footnotesize] at (0,-2.15) {$\x_t$};
\draw[->, thick] (0,-1.95) -- (0,-1.15);
\end{tikzpicture}
\caption{The LSTM cell.  The green top lane is the \emph{protected memory lane}: old memory can be kept, new information can be written additively, and the resulting state can later be revealed through the hidden output (\emph{visible output}, blue bottom lane).  This is the intuition behind the forget, input, candidate, and output components shown inside the cell.}
\label{fig:lstm_cell}
\end{figure}
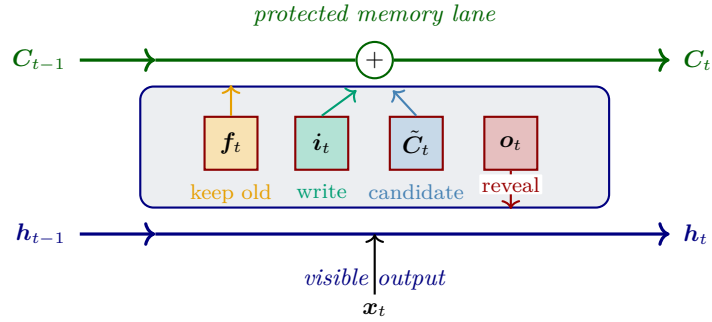

The LSTM cell \citep{hochreiter1997long} replaces the single-state recurrence $\h_t = \sigma(\Wh\h_{t-1} + \Wx\x_t)$ by a pair of states: a \emph{cell state} $\bm{C}_t$ that flows along the top of the cell with \emph{additive} updates, and a \emph{hidden state} $\h_t$ that is read off it.  Three learned sigmoid gates, each depending on the concatenation $[\h_{t-1}, \x_t]$, control what information flows through:
\begin{align}
\bm{f}_t       &= \sigma\!\big(\W_f\,[\h_{t-1}, \x_t] + \bb_f\big) && \text{(forget gate)} \label{eq:lstm_f}\\
\bm{i}_t       &= \sigma\!\big(\W_i\,[\h_{t-1}, \x_t] + \bb_i\big) && \text{(input gate)}  \\
\tilde{\bm{C}}_t &= \tanh\!\big(\W_C\,[\h_{t-1}, \x_t] + \bb_C\big) && \text{(candidate cell)} \\
\bm{C}_t       &= \bm{f}_t \odot \bm{C}_{t-1} + \bm{i}_t \odot \tilde{\bm{C}}_t && \text{(cell state update)} \label{eq:lstm_C}\\
\bm{o}_t       &= \sigma\!\big(\W_o\,[\h_{t-1}, \x_t] + \bb_o\big) && \text{(output gate)} \\
\h_t           &= \bm{o}_t \odot \tanh(\bm{C}_t)                   && \text{(hidden state).}
\end{align}
Each of $\bm{f}_t, \bm{i}_t, \bm{o}_t \in (0,1)^{d}$ acts as a soft switch applied element-wise.  The crucial structural change is in equation~\eqref{eq:lstm_C}: the cell state is \emph{additively} corrected rather than multiplicatively overwritten.  Along the direct memory path, differentiating $\bm{C}_t$ with respect to $\bm{C}_{t-1}$ contributes $\mathrm{diag}(\bm{f}_t)$ in place of a full recurrent matrix product.  The full derivative also contains indirect terms because the gates depend on $\h_{t-1}$ and hence on earlier cell states, but the direct path is the constant-error-carousel intuition: when the cell judges information worth keeping, it can open the forget gate ($\bm{f}_t \approx \bm{1}$) and allow gradients to flow through \emph{as if} the sequence were shorter.  Figure~\ref{fig:lstm_cell} sketches the resulting cell.

The \emph{Gated Recurrent Unit} (GRU) of \citet{cho2014gru} is a lighter sibling that merges the forget and input gates into a single \emph{update} gate $\bm{z}_t$ and drops the separate cell state in favor of the hidden state itself:
\begin{align}
\bm{z}_t       &= \sigma(\W_z\,[\h_{t-1}, \x_t] + \bb_z), \\
\bm{r}_t       &= \sigma(\W_r\,[\h_{t-1}, \x_t] + \bb_r), \\
\tilde{\h}_t   &= \tanh\!\big(\W_h\,[\bm{r}_t \odot \h_{t-1}, \x_t] + \bb_h\big), \\
\h_t           &= (1 - \bm{z}_t) \odot \h_{t-1} + \bm{z}_t \odot \tilde{\h}_t.
\end{align}
A GRU uses roughly $25\%$ fewer parameters than an LSTM of the same hidden size and performs comparably on many sequence tasks \citep{chung2014empirical}, at the cost of a slightly less expressive memory channel.

LSTMs remain particularly effective for economic time-series tasks where capturing specialized temporal patterns is more important than massive scalability.  For example, \citet{holt2024detecting} use LSTMs to detect \emph{Edgeworth cycles} in retail gasoline-price data.  These cycles are asymmetric, high-frequency price movements that are difficult to identify with traditional spectral analysis or simple rule-based methods.  In this setting the LSTM architecture excels at identifying the characteristic sawtooth patterns (sudden price jumps followed by slow decays) across thousands of retail stations, providing a robust tool for antitrust analysis and competition policy.

\subsection{Limits of Recurrent Models}
\label{sec:rnn_limits}

Even with gated cells, recurrent architectures retain two fundamental limitations that no amount of tuning can remove.
\begin{enumerate}[itemsep=2pt]
\item \textbf{Inherently sequential.}  Step $t$ cannot begin until step $t-1$ has finished, so the full computation takes $T$ serial wavefronts regardless of how many cores or tensor units are available.  On a modern GPU that can evaluate thousands of MLP rows in parallel, this forced serialization makes RNN training slow and wall-clock expensive.  Training a contemporary large language model on a trillion-token corpus with a plain LSTM would take years of wall-clock time rather than weeks.
\item \textbf{Path-length-dependent decay.}  Information from position $1$ must travel through all intermediate hidden states to influence position $T$, each hop applying a matrix and a nonlinearity.  Gating mitigates but does not eliminate this decay, and empirically LSTM performance saturates on sequences well below the theoretical limit implied by the cell's capacity.
\end{enumerate}
Both problems have the same structural cause: the recurrence forces a path of length $T$ between the two extreme positions of the sequence.  The remedy is to drop recurrence entirely and let every position read every other position \emph{directly}, in parallel.  This is the Transformer.

\phantomsection
\subsection{The Transformer Architecture}
\label{sec:transformers}

The \emph{Transformer} architecture \citep{vaswani2017attention} replaced recurrence entirely with \emph{self-attention}.  This allows the model to weigh the importance of all tokens in a sequence simultaneously, enabling massive parallelization and better capturing global dependencies.\footnote{The idea that a network can compute its own weights from its inputs has a long history: the \emph{Fast Weight Programmers} of \citet{schmidhuber1992learning} use one network to write the weights of another from context, which is widely viewed as a conceptual precursor of attention.  \citet{schlag2021linear} make the formal equivalence explicit, showing that linear-attention Transformers \emph{are} Fast Weight Programmers.}

For a first pass, four steps are enough.  Add position information so the model knows order; project each token into a \emph{query}, \emph{key}, and \emph{value}; compare each query to all keys; then take a weighted average of the values.  Multi-heads, LayerNorm, residual connections, and pointwise MLPs are refinements of this core idea, not a different idea.

\subsubsection{The Self-Attention Mechanism}
\label{sec:self_attention}

\paragraph{Intuition: search, then retrieval.}  Before writing the mechanism formally it is useful to see it as a \emph{soft library lookup}.  Picture a small library with $n$ shelves, one per position in the sequence.  Each shelf $i$ carries three vectors: a \emph{query} $q_i$ (``what am I looking for?''), a \emph{key} $k_i$ (``what is printed on my label?''), and a \emph{value} $v_i$ (``what do I actually contain?'').  All three are linear projections of the same input token $\x_i$, and the projection matrices $\W_Q,\W_K,\W_V$ are \emph{learned} during training; the librarian (queries and keys) and the books (values) are co-designed for whatever task the training objective encodes.

To produce the updated representation $\bm{o}_i$ at shelf $i$, the following four steps happen.
\begin{enumerate}[itemsep=2pt]
\item \textbf{Score.} Shelf $i$'s query $q_i$ is compared against every shelf's key $k_j$; the dot product $q_i^{\!\top} k_j$ is a \emph{similarity score}, high when shelf $j$'s label matches what shelf $i$ is looking for.
\item \textbf{Normalize.} The $n$ scores are divided by $\sqrt{d_k}$ (to keep them in a numerically sane range) and pushed through a softmax, producing probability weights $\alpha_{ij}\in[0,1]$ with $\sum_j \alpha_{ij}=1$.
\item \textbf{Retrieve.} The weights are applied to the values: $\bm{o}_i = \sum_j \alpha_{ij}\, v_j$ is a weighted average of the $n$ shelf contents.
\item \textbf{Repeat.} Steps 1--3 happen in parallel for every shelf $i$, producing the whole output sequence $\bm{o}_1,\dots,\bm{o}_n$ in a single layer.
\end{enumerate}
Shelves whose label matched the query contribute most to the output; shelves whose label was off-topic are almost ignored.

\paragraph{Why this is useful: a worked example.}  Consider the sentence \emph{``The cat sat on the mat.  It purred.''}  For the model to process ``it'' properly, it must first decide what ``it'' refers to, the cat or the mat.  Self-attention performs exactly this disambiguation: the query vector at the ``it'' position probes the key vectors at every earlier position, and the softmax converts the raw similarity scores into probability weights that concentrate most of the mass on the correct antecedent.  Figure~\ref{fig:attention} below illustrates the resulting pattern: the bulk of the weight lands on ``cat'', and the updated representation at the ``it'' position is formed as a weighted average of the values, driven mostly by ``cat''.  The same mechanism, run in parallel for every position, produces all of $\bm{o}_1,\dots,\bm{o}_n$ in a single layer.

\paragraph{A small concrete example.}  Suppose we have only three shelves and two-dimensional $q$'s and $k$'s.  Take the query at shelf~3 to be $q_3 = (1,0)$ and the three keys
$k_1 = (0.9,\,0.1),\ k_2 = (0.1,\,0.8),\ k_3 = (1,\,0).$
The raw similarity scores are $q_3^{\!\top}k_1 = 0.9$, $q_3^{\!\top}k_2 = 0.1$, $q_3^{\!\top}k_3 = 1.0$.  Ignoring the $1/\sqrt{d_k}$ scaling to keep the arithmetic clean, the softmax weights come out to roughly $(0.405,\,0.182,\,0.448)$.  Shelf~3's output $\bm{o}_3$ is therefore a blend of the three values in those proportions: most mass on shelf~3 itself and on shelf~1 (the closest match in label space); shelf~2, whose label $k_2$ points in an orthogonal direction, contributes about $18\%$.  The softmax is ``soft'' precisely in this sense, a smoothed nearest-neighbor retrieval rather than a hard argmax over the most similar shelf.

The ``self'' in \emph{self-}attention says that every shelf plays both roles simultaneously: every shelf's query probes every shelf's key, and every shelf receives an updated representation as output.  A single attention layer therefore pairs all $n$ positions with all $n$ positions in parallel.  Contrast this with an RNN or LSTM: to let position $t$ peek at information from position 1, the signal must be shuttled through every intermediate position via the hidden state, with each hop applying its own weight matrix and nonlinearity, so long-range information is either blurred or lost entirely.  Attention short-circuits that chain: position $t$ reads position $1$ directly, with no intermediate hops and no path-length-dependent attenuation.  This is the architectural source of the Transformer's advantage on long sequences, and the reason why attention-based models quickly displaced recurrent ones for language, code, and (increasingly) time-series forecasting.

Given a sequence of input vectors, stack the tokens as rows of $\X \in \R^{n\times d}$.  The attention mechanism computes three projections for each token: a \emph{Query} ($Q$), a \emph{Key} ($K$), and a \emph{Value} ($V$), using learned weight matrices $\W_Q, \W_K, \W_V$:
\begin{equation}
Q = \X \W_Q, \qquad K = \X \W_K, \qquad V = \X \W_V.
\end{equation}
The output of the attention layer is a weighted sum of the values, where the weights are determined by the compatibility (dot product) of the queries with the keys:
\begin{equation}
\mathrm{Attention}(Q, K, V) = \mathrm{softmax}\left(\frac{Q K^\top}{\sqrt{d_k}}\right) V.
\end{equation}
The scaling factor $\sqrt{d_k}$ (the dimensionality of the keys) prevents the dot products from growing too large in magnitude, which would otherwise make the softmax unstable.

\paragraph{An econometric lens.}  The attention layer is a \emph{data-dependent, learnable kernel smoother}.  Compare it to the Nadaraya--Watson estimator $\hat f(x) = \sum_i w_i(x)\,y_i$ with kernel-based weights $w_i(x) \propto k(x, x_i)$: attention has exactly this form, but the similarity $k(\cdot,\cdot)$ is the parametric bilinear form $(q,k)\mapsto q^{\!\top}k/\sqrt{d_k}$ and both $q$ and $k$ are themselves \emph{learned} projections of the input.  From this vantage point the Transformer's ``magic'' is less mysterious: it is a nonparametric smoother whose kernel the optimizer tunes to whatever task the training objective encodes.  Self-attention further recovers the classical recurrence-free property that every pair of positions interacts in a single parallel layer, with no signal decay along the sequence.

Figure~\ref{fig:attention} renders the attention pattern of the worked ``cat/it'' example on a compressed five-token version of the sentence.  The output $\bm{o}_{\textit{it}}$ is the new representation at the ``it'' position, formed as a weighted average of the values, with most weight coming from ``cat''.

\begin{figure}[ht]
\centering
\begin{minipage}[c]{0.64\textwidth}
\centering
\begin{tikzpicture}[scale=0.92, transform shape,
    tok/.style={rectangle, draw=uzhblue, thick, rounded corners=2pt, fill=uzhgreylight,
                minimum width=1.00cm, minimum height=0.48cm, inner sep=2pt, font=\footnotesize},
    wt/.style={rectangle, draw=white, rounded corners=1pt, minimum width=0.75cm,
               minimum height=0.36cm, inner sep=1pt, font=\scriptsize}]
\node[tok] (t1) at (0.0,0.0) {the};
\node[tok, fill=softgreen!14, draw=darkgreen] (t2) at (2.0,0.0) {\textbf{cat}};
\node[tok] (t3) at (4.0,0.0) {sat};
\node[tok, fill=softorange!12, draw=softorange] (t4) at (6.0,0.0) {mat};
\node[tok, fill=softblue!12, draw=softblue, very thick] (t5) at (8.0,0.0) {\textbf{it}};
\node[wt, fill=softblue!18] (w1) at (0.0,1.25) {0.05};
\node[wt, fill=darkgreen!75, text=white] (w2) at (2.0,1.25) {0.58};
\node[wt, fill=softblue!22] (w3) at (4.0,1.25) {0.08};
\node[wt, fill=softorange!70] (w4) at (6.0,1.25) {0.20};
\node[wt, fill=softblue!30] (w5) at (8.0,1.25) {0.09};
\node[draw=softblue, fill=softblue!10, rounded corners=3pt, inner sep=4pt, font=\footnotesize] (q) at (8.0,2.45) {$q_{\textit{it}}$};
\node[draw=darkgreen, fill=softgreen!10, rounded corners=3pt, inner sep=4pt, font=\footnotesize] (z) at (8.0,3.55) {$\bm{o}_{\textit{it}}$};
\draw[->, thick, softblue] (q.south) -- (t5.north);
\draw[softblue, thick] (0.0,1.78) -- (8.0,1.78);
\foreach \x in {0.0,2.0,4.0,6.0,8.0} {
    \draw[softblue, thick] (\x,1.78) -- (\x,1.54);
}
\draw[->, thick, softblue] (q.west) to[out=195,in=35] (4.0,1.78);
\draw[->, thick, darkgreen] (2.0,1.52) to[out=70,in=225] (z.south west);
\draw[->, thick, darkgreen] (w2.south) -- (t2.north);
\node[font=\footnotesize, text=uzhgreydark] at (4.0,-0.62) {keys and values at each token position};
\end{tikzpicture}
\end{minipage}\hfill
\begin{minipage}[c]{0.31\textwidth}
\footnotesize
\textbf{How to read the arrows}
\begin{enumerate}[leftmargin=*, itemsep=2pt]
\item Blue down arrow: build the query $q_{\textit{it}}$ from the current token ``it''.
\item Blue bent arrow: compare that query with all keys in the sequence.
\item Numbers above tokens: $0.05 + 0.58 + 0.08 + 0.20 + 0.09 = 1.00$, as softmax weights must.
\item The weight $0.58$ above ``cat'' is the largest one.
\item Green arrows: the ``cat'' value enters $\bm{o}_{\textit{it}}$.
\end{enumerate}
\end{minipage}
\caption{A worked self-attention pattern.  Both $q_{\textit{it}}$ and $\bm{o}_{\textit{it}}$ sit above the ``it'' position: $q_{\textit{it}}$ is the query projection of ``it'' (blue), and $\bm{o}_{\textit{it}}$ is the updated representation at that same position (green).  The blue arrows show how the query is built from ``it'' and then compared with every key.  The softmax weights above each token sum to one; here the largest weight lands on ``cat'', so the green aggregation arrow starts above ``cat'' and curves up to $\bm{o}_{\textit{it}}$, indicating that the update at ``it'' is driven mainly by the value at ``cat''.}
\label{fig:attention}
\end{figure}

\subsubsection{Multi-Head Attention}
\label{sec:mha}

A single attention layer implements \emph{one} similarity pattern between positions.  Linguistic and economic sequences, however, contain many relations that matter simultaneously: subject--verb agreement, coreference of pronouns, topic alignment, or, in a macro panel, sector co-movements, shock transmission lags, and autocorrelation structure.  A single head is forced to compress all of them into one kernel, and typically does a poor job.

The fix is \emph{multi-head attention}.  Run $H$ attention layers in parallel, each with its \emph{own} projection matrices $(\W_Q^{(h)}, \W_K^{(h)}, \W_V^{(h)})$ mapping the input to a lower-dimensional subspace of size $d_k = d/H$, compute $H$ attention outputs independently, then concatenate and linearly project back:
\begin{align}
\mathrm{head}_h
&= \mathrm{Attention}\bigl(\X\W_Q^{(h)},\, \X\W_K^{(h)},\, \X\W_V^{(h)}\bigr), \\
\mathrm{MHA}(\X)
&= \big[\mathrm{head}_1; \,\mathrm{head}_2; \,\ldots;\,\mathrm{head}_H\big]\,\W_O .
\end{align}
The total parameter count is essentially unchanged, because each head works on a $1/H$-dimensional slice, but the inductive bias is richer: different heads are free to specialize in different relations.  Interpretability studies of trained Transformers routinely find heads that focus on the previous token, on the closing bracket matching an open one, on the subject of the current clause, or, in time-series models, on the most recent analogue of the current calendar month.  Multi-head attention is therefore \emph{structurally analogous to a mixture of learned kernels} in a nonparametric regression; the weights $\W_O$ are the mixing coefficients, and the softmax-scored pairs are the kernels themselves.

\subsubsection{Positional Encoding}
\label{sec:pos_enc}

Self-attention treats its input as a \emph{set}: permuting the positions of the tokens permutes the rows of $\X$ and permutes the rows of the output, but changes nothing else.  This is problematic, because order matters in every sequence application (``dog bites man'' vs.\ ``man bites dog''; ``inflation before the shock'' vs.\ ``after'').  A model without a notion of position cannot distinguish them.

The fix that \citet{vaswani2017attention} propose is to \emph{add} a deterministic vector $\mathrm{PE}(t)\in\R^d$ to the input embedding at each position $t$, chosen so that the dot products $\mathrm{PE}(t)^\top \mathrm{PE}(t+k)$ encode the relative distance $k$.  The original sinusoidal scheme is
\begin{equation}
\mathrm{PE}(t, 2k) = \sin\!\big(t/10000^{2k/d}\big),
\qquad
\mathrm{PE}(t, 2k+1) = \cos\!\big(t/10000^{2k/d}\big),
\qquad k=0,\ldots,d/2-1.
\end{equation}
Three properties make this choice useful.  First, each coordinate $2k$ is a sine wave of wavelength $2\pi \cdot 10000^{2k/d}$, so the encoding spans wavelengths from roughly $2\pi$ up to $2\pi \cdot 10000$; the model sees both fine local positions and coarse global ones at once.  Second, for each sine/cosine pair, a fixed offset can be represented by a $2\times2$ rotation: $\mathrm{PE}(t+r)$ is a linear function of $\mathrm{PE}(t)$ with coefficients depending only on the relative offset $r$.  This gives attention a convenient way to represent relative positions.  Third, the encoding extrapolates: a model trained on sequences of length $512$ can be fed sequences of length $1024$ and the position code is still well-defined.  Modern alternatives, rotary position embeddings (RoPE) and ALiBi, refine these properties but keep the same philosophy.

\subsubsection{The Transformer Block}
\label{sec:transformer_block}

\begin{figure}[ht]
\centering
\begin{tikzpicture}[scale=0.95, transform shape,
    b/.style={draw=uzhblue, thick, rounded corners=3pt, fill=uzhgreylight,
              minimum width=5.2cm, minimum height=0.75cm, align=center, font=\small},
    attn/.style={draw=darkred, thick, rounded corners=3pt, fill=red!10,
              minimum width=5.2cm, minimum height=0.75cm, align=center, font=\small},
    ln/.style={draw=softblue, thick, rounded corners=3pt, fill=softblue!15,
              minimum width=5.2cm, minimum height=0.6cm, align=center, font=\small},
    plus/.style={circle, draw=harvardcrimson, thick, fill=white,
                 minimum size=0.44cm, inner sep=0pt, font=\small}]
\node[b] (in) at (0,0) {input tokens $\X$ $+$ positional encoding};
\node[ln, above=0.25cm of in] (ln1) {LayerNorm};
\node[attn, above=0.25cm of ln1] (mha) {\textbf{Multi-Head Self-Attention}\\[-1pt]{\footnotesize tokens exchange information}};
\node[plus, above=0.30cm of mha] (add1) {$+$};
\node[ln, above=0.30cm of add1] (ln2) {LayerNorm};
\node[b, above=0.25cm of ln2] (ff) {\textbf{Pointwise Feed-Forward MLP}\\[-1pt]{\footnotesize same MLP applied token by token}};
\node[plus, above=0.30cm of ff] (add2) {$+$};
\node[b, above=0.25cm of add2, fill=softgreen!15, draw=darkgreen] (out) {output tokens $\bm{Z}$};
\draw[->,thick] (in) -- (ln1);
\draw[->,thick] (ln1) -- (mha);
\draw[->,thick] (mha) -- (add1);
\draw[->,thick] (add1) -- (ln2);
\draw[->,thick] (ln2) -- (ff);
\draw[->,thick] (ff) -- (add2);
\draw[->,thick] (add2) -- (out);
\draw[->, thick, harvardcrimson, rounded corners=5pt]
    ($(in.east)+(0.10,0)$) -- ++(0.95,0) -- ++(0,3.45) -- (add1.east);
\draw[->, thick, harvardcrimson, rounded corners=5pt]
    ($(add1.east)+(0.10,0)$) -- ++(0.95,0) -- ++(0,3.05) -- (add2.east);
\end{tikzpicture}
\caption{One \emph{Transformer block} in pre-norm form.  Self-attention first mixes information across token positions, then the pointwise MLP transforms each token separately.  The red skip paths are the residual connections that let deep stacks train stably.  A full Transformer stacks $L$ such blocks; GPT-3, for instance, uses $L=96$.}
\label{fig:transformer_block}
\end{figure}
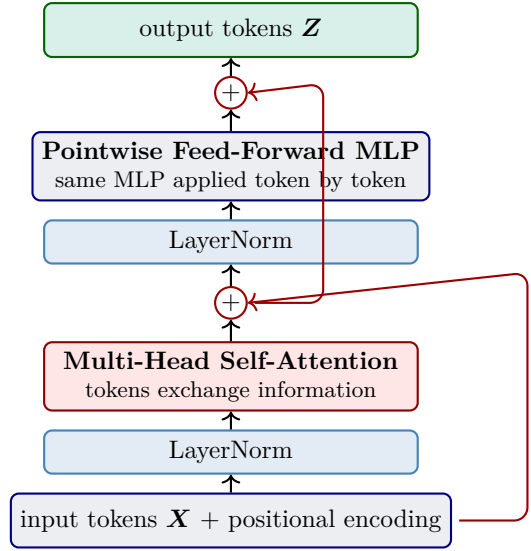

Single attention layers, even multi-headed, are not yet expressive enough: they are essentially linear in the values, with a nonlinear mixing pattern.  A full \emph{Transformer block} wraps one MHA layer and one pointwise MLP with residual connections and LayerNorm:
\begin{align}
\x^{+} &= \x + \mathrm{MHA}\!\big(\mathrm{LN}(\x)\big), \label{eq:tblock1}\\
\x^{\mathrm{out}}  &= \x^{+} + \mathrm{MLP}\!\big(\mathrm{LN}(\x^{+})\big). \label{eq:tblock2}
\end{align}
The LayerNorm steps \citep{ba2016layer} standardize across feature coordinates; together with the residual additions they stabilize training of very deep stacks.  Equations~\eqref{eq:tblock1}--\eqref{eq:tblock2} describe the modern \emph{pre-norm} variant (LN before each sub-block), which is easier to train than the original \emph{post-norm} variant of \citet{vaswani2017attention}.  Figure~\ref{fig:transformer_block} shows the architecture schematically.

\paragraph{Encoder, decoder, causal masking.}  The original Transformer of \citet{vaswani2017attention} pairs an \emph{encoder} stack (processes the source sequence) with a \emph{decoder} stack (generates the target sequence), linked by a cross-attention layer.  Most modern large language models are \emph{decoder-only}: they use the same block as above, but the attention softmax is applied to a masked score matrix that sets all entries above the diagonal to $-\infty$.  This \emph{causal mask} forbids a position from attending to future positions, turning the Transformer into a left-to-right autoregressive predictor suitable for language modeling.

\paragraph{Scaling and parallelism.}  For day~1 the key engineering fact is simpler than the modern LLM discussion: attention is parallel, recurrence is not.  Because every sub-block is pointwise in time and the only cross-position operation (attention) is fully parallelizable, a Transformer with $L$ blocks, $H$ heads, and hidden dimension $d$ can be trained on accelerators at roughly the theoretical peak throughput.  This is why modern foundation models, GPT-$n$, BERT, ViT, Claude, are Transformers rather than RNNs.  Empirical studies of \emph{scaling laws} \citep{kaplan2020scaling, hoffmann2022training} document that test loss decreases as a power law in parameters, data, and compute, giving rise to the compute-optimal prescriptions used by the largest labs.  For an economist, the relevant takeaway is that the marginal cost of additional capability is governed by a smooth, quantifiable, and \emph{very large} compute bill.

\subsubsection{At a glance: RNN vs LSTM vs Transformer}

Table~\ref{tab:seq_compare} summarizes the three architectures along the dimensions most relevant to a practitioner's choice.

\begin{table}[ht]
\centering
\small
\setlength{\tabcolsep}{5pt}
\renewcommand{\arraystretch}{1.15}
\begin{tabular}{@{}l
  >{\centering\arraybackslash}p{2.6cm}
  >{\centering\arraybackslash}p{3.0cm}
  >{\centering\arraybackslash}p{4.2cm}@{}}
\toprule
& \textbf{RNN} & \textbf{LSTM / GRU} & \textbf{Transformer} \\
\midrule
Hidden state                 & single $\h_t$         & $\h_t$ and $\bm{C}_t$       & none per step; the residual stream across layers is an implicit state \\
Path length $1 \to T$        & $\mathcal{O}(T)$      & $\mathcal{O}(T)$            & $\mathcal{O}(1)$ \\
Parallelism over $t$         & none                  & none                        & full (all positions at once) \\
Compute per layer            & $\mathcal{O}(T\,d^2)$ & $\mathcal{O}(T\,d^2)$       & $\mathcal{O}(T^2 d + T\,d^2)$ \\
Memory per layer             & $\mathcal{O}(T\,d)$   & $\mathcal{O}(T\,d)$         & $\mathcal{O}(T^2 + T\,d)$ \\
Training stability           & gradient decay/blow-up & much better, gated         & stable with LN + residuals \\
Sweet spot                   & short sequences       & mid-length, niche patterns  & long context, massive parallelism \\
\bottomrule
\end{tabular}
\caption{Comparison of the three sequence architectures.  Transformers trade a quadratic-in-$T$ attention cost for full parallelism and unit-length paths between any pair of positions, which is an excellent trade on modern accelerators and for the long sequences typical in language and high-frequency finance.}
\label{tab:seq_compare}
\end{table}

A practical rule of thumb follows immediately.  If the task is a specialized time-series problem with moderate history length and limited data, an LSTM remains a strong baseline.  If context is long and accelerator-friendly parallelism matters, one should usually start with a Transformer.

\subsection{Advanced Aside: In-Context Learning of an AR(1) Process}
\label{sec:incontext_ar1}

The material up to this point is the core Chapter~\ref{ch:intro} message; readers comfortable with the RNN $\to$ LSTM $\to$ Transformer summary may skip directly to the Chapter Summary on page~\pageref{sec:ch1_summary}.  This subsection is an optional but illuminating detour: it shows why economists often find attention intuitive once they see it through a regression lens, and it is the analytical companion to notebook~\texttt{09\_Transformer\_InContext\_AR1}.

A remarkable emergent property of large Transformers is \emph{in-context learning}: the ability to solve a task \emph{at inference time}, given only examples in the prompt, with no weight updates.  A model pretrained on a universe of sequences can be shown a fresh series it has never seen and produce sensible next-step forecasts.  For an economist this is striking: the Transformer behaves as though it \emph{runs a regression inside its forward pass}, with the prompt playing the role of the training sample and the final token playing the role of the test point.  \citet{vonoswald2023transformers} make this formal by showing that self-attention layers can implement gradient-based optimization internally, so a stack of such layers iteratively reduces an implicit loss.

Consider the simplest concrete setting, an autoregressive process of order~1:
\begin{equation}
x_t = \varrho \, x_{t-1} + \varepsilon_t, \qquad \varepsilon_t \sim \mathcal{N}(0, \sigma^2).
\end{equation}
If we provide a Transformer with a sequence $(x_1, \dots, x_t)$, it can predict $x_{t+1}$ by implicitly estimating $\varrho$ from the history.  There are two closely related but distinct perspectives.  First, under suitable linear-attention parameterizations, self-attention layers can implement gradient-descent-like updates on least-squares objectives \citep{vonoswald2023transformers}; specialized to the AR(1) target above, this reads as descent on $\min_\varrho \sum_{i=2}^{t} (x_i - \varrho\,x_{i-1})^2$.  Second, with ordinary softmax attention, \emph{one} natural Q/K/V assignment behaves like a kernel smoother:
\begin{itemize}[itemsep=2pt]
\item \textbf{Query:} the current state $Q_t = x_t$ (``where am I now?''),
\item \textbf{Key:} the lagged state $K_i = x_{i-1}$ (``past states that preceded each outcome''),
\item \textbf{Value:} the realized successor $V_i = x_i$ (``what came next'').
\end{itemize}
This is one particular parameterization that works because the AR(1) target is linear in the lagged state; with this choice the softmax-attention output becomes
\begin{equation}
\hat{x}_{t+1} = \sum_{i=2}^{t} \alpha_i\, x_i \approx \hat{\varrho} \, x_t,
\label{eq:incontext_kernel_smoother}
\end{equation}
where the softmax weights $\alpha_i \propto \exp(x_t\,x_{i-1})$ concentrate mass on those past states $x_{i-1}$ that \emph{look like} the current state $x_t$.

\paragraph{Why this approximates $\varrho\,x_t$, in three short steps.}  Equation~\eqref{eq:incontext_kernel_smoother} is exactly a Nadaraya--Watson kernel regression of $x_i$ on $x_{i-1}$, evaluated at $x_{i-1} = x_t$, with kernel $K(x_*, x_{i-1}) = \exp(x_*\,x_{i-1})$.  The intuition is then standard:
\begin{enumerate}[itemsep=2pt]
\item \emph{Population fact.}  The AR(1) data-generating process implies $\E{x_i \mid x_{i-1} = x_*} = \varrho\, x_*$ exactly.  So the population conditional mean we are trying to estimate at $x_* = x_t$ is just $\varrho\, x_t$.
\item \emph{Kernel smoother.}  The softmax attention output $\sum_i \alpha_i x_i$ with $\alpha_i \propto K(x_t, x_{i-1})$ is a kernel-weighted average of the values $x_i$ at past time steps $i$, with the weights peaked where the lagged state $x_{i-1}$ is closest to $x_t$.  Provided enough past observations land near $x_t$ (so the kernel concentrates around $x_t$), this is the empirical Nadaraya--Watson estimator of $\E{x \mid x_{i-1} = x_t}$.
\item \emph{Conclusion.}  Combining the two, $\sum_i \alpha_i x_i \approx \E{x_i \mid x_{i-1} = x_t} = \varrho\, x_t$.  The shock variance $\sigma^2$ controls how much the realized $x_i$'s scatter around the conditional mean and therefore the variance of the kernel estimate, but it does not enter the Nadaraya--Watson \emph{location} at first order.
\end{enumerate}
Note that the unscaled inner product $x_t\, x_{i-1}$ used in the kernel is dimension-1, so the standard $1/\sqrt{d_k}$ scaling of multi-head attention plays no role here: even without it, the softmax concentrates around the past $x_{i-1}$'s closest to $x_t$ as soon as the prompt is long enough.

For the econometrician's mental library the closest classical objects are the Nadaraya--Watson and local-linear estimators; the novelty is that the kernel is not hand-chosen but jointly learned with the data representation, over a large corpus of related tasks.  This is a form of \emph{meta-learning}: the network learned ``how to regress'' during pretraining, and at inference it runs that regression on a brand-new series.  Self-attention can therefore implement optimization-like or regression-like computations internally, not merely local pattern matching.

\paragraph{Code examples.}
The following Jupyter notebooks implement and extend the material in this chapter:
\begin{itemize}[itemsep=2pt]
\item \texttt{01\_BasicML\_intro}: linear regression, classification, and loss functions.
\item \texttt{02\_GradientDescent\_and\_SGD}: implementing and visualizing optimizers.
\item \texttt{03\_Double\_Descent}: the modern generalization regime.
\item \texttt{04\_Gentle\_DNN}: building a simple DNN from scratch.
\item \texttt{05\_Tensorboard}: monitoring training progress.
\item \texttt{06\_PyTorch\_intro}: introduction to PyTorch fundamentals.
\item \texttt{07\_Genz\_Approximation\_and\_Loss\_Functions}: high-dimensional integration using Genz test functions and robust losses.
\item \texttt{08\_MLP\_LSTM\_Transformer\_Edgeworth\_Cycles}: a three-way comparison on the same Edgeworth-cycle task.  An MLP that sees only $x_t$ collapses near the cycle mean (no memory); an LSTM with a 32-unit hidden state tracks the sawtooth almost exactly via its gated memory; a tiny encoder-only Transformer ($d_{\text{model}}\!=\!16$, two layers, four heads, $\sim\!4.7$k parameters) attends to the full window in parallel and beats the MLP by an order of magnitude, while remaining slightly behind the LSTM on this small, highly periodic, low-data signal -- a deliberate illustration that architectural inductive bias matters as much as flexibility on small problems.  Quantitatively, the test-set MAE ranking that the notebook prints in its summary cell is, in order, LSTM~$<$~Transformer~$\ll$~MLP, with the Transformer typically within a small multiple of the LSTM and the MLP an order of magnitude worse; readers should consult the notebook's printed table for the seed-specific numbers.
\item \texttt{09\_Transformer\_InContext\_AR1}: advanced / optional notebook.  A tiny 2-layer Transformer learns \emph{how to regress} across many AR(1) draws; at inference it recovers $\hat{\varrho}$ in-context without weight updates, reproducing the analytical prediction above.
\end{itemize}

\begin{keyinsightbox}[Chapter Summary]
\label{sec:ch1_summary}
\begin{itemize}[itemsep=2pt, leftmargin=*]
\item Deep networks compose simple nonlinear coordinate transformations: sufficiently wide shallow networks already attain universal approximation \citep{cybenko1989approximation, hornik1989multilayer}, but depth gives provably more efficient representations for compositional functions \citep{telgarsky2016benefits, barron1993universal}.
\item Training rests on three pillars: He/Xavier initialization, smooth non-saturating activations, and adaptive optimizers (SGD with momentum $\to$ RMSprop $\to$ Adam $\to$ AdamW); backpropagation makes the gradient cost essentially the same as the forward pass \citep{rumelhart1986learning, baydin2018automatic}.
\item Generalization in modern over-parameterized regimes is described by the double-descent curve and explained theoretically via the Neural Tangent Kernel and benign-overfitting results, not by classical bias--variance \citep{jacot2018neural, belkin2019reconciling, nakkiran2020deep, bartlett2020benign}.
\item Sequence architectures (RNN $\to$ LSTM $\to$ Transformer) are mentioned for completeness; the rest of the script uses feed-forward MLPs because DEQNs and PINNs operate on unstructured state vectors.
\end{itemize}
\end{keyinsightbox}

\section*{Further Reading}
\addcontentsline{toc}{section}{Further Reading}
\begin{itemize}[itemsep=2pt]
\item \citet{goodfellow2016deep}, the standard graduate textbook covering everything in this chapter at greater depth.
\item \citet{schmidhuber2015deep}, a historical survey tracing the deep-learning lineage; useful for context on LSTMs, Highway Networks, and Fast Weight Programmers.
\item \citet{bishop2006}, pattern recognition from a Bayesian/statistical viewpoint; the natural complement for econometric readers.
\item \citet{kingma2015adam, loshchilov2019decoupled}, the canonical references on Adam and AdamW; read together they explain modern optimizer tuning.
\item \citet{bottou2018optimization}, a comprehensive survey of stochastic optimization for large-scale learning, including convergence rates.
\end{itemize}

\section*{Exercises}
\addcontentsline{toc}{section}{Exercises}
\noindent Worked solutions and guidance for these exercises appear in Appendix~\ref{app:solutions}.

\noindent \textbf{Workload labels.}  Throughout the script, every exercise carries one of three workload tags inside its title.  \emph{[Core]} marks short analytical or pencil-and-paper questions suitable for a weekly problem set.  \emph{[Computational]} marks notebook-based exercises that involve running or modifying companion code; allow yourself a long evening or a weekend with verification gates and starter code in hand.  \emph{[Advanced/project]} marks longer, research-style assignments that may require a multi-day investment, a proper compute budget, or a small term-project plan.  The labels are advisory rather than prescriptive: students with prior exposure can promote a [Computational] exercise to a quick warm-up, while those new to the material can treat several [Advanced/project] entries as inspiration for term work.
\begin{enumerate}[itemsep=4pt, leftmargin=*]
\item\label{ex:ch1:1} \textbf{[Core] Backprop on a 2-layer net.}  Take a single hidden layer with ReLU activation: $\hat y = w_2\,\sigma(w_1 x + b_1)$ and squared loss $\ell = (\hat y - y)^2$.  Derive $\partial \ell / \partial w_1$ and $\partial \ell / \partial w_2$ by hand and compare with what \texttt{torch.autograd} returns on a worked numerical example ($x=2$, $y=1$, all weights $0.5$).
\item\label{ex:ch1:2} \textbf{[Core] MSE vs.\ MLE.}  Show that the maximum-likelihood estimator of the slope of $y_i = \beta x_i + \varepsilon_i$, $\varepsilon_i \sim \mathcal{N}(0,\sigma^2)$, coincides with the OLS estimator and with the minimizer of $\sum_i (y_i - \beta x_i)^2$.  Then repeat with $\varepsilon_i$ Laplace-distributed and discuss why the squared loss is no longer optimal.
\item\label{ex:ch1:3} \textbf{[Core] Activation choice for a PINN.}  Argue carefully why ReLU networks cannot be used to solve a second-order PDE in strong form.  Construct an explicit example where the second derivative of the network output is identically zero except on a measure-zero set.
\item\label{ex:ch1:4} \textbf{[Core] Adam vs.\ AdamW.}  In a regression with $L_2$ regularization, write down the per-parameter update rule for Adam (with $L_2$ added to the loss) and for AdamW (decoupled).  Show that the two updates do \emph{not} coincide and explain which one preserves the intuition that ``weight decay shrinks weights uniformly.''
\item\label{ex:ch1:5} \textbf{[Core] RNN forward pass by hand.}  Take a vanilla recurrent unit with hidden dimension $H=2$, scalar input, and update $h_t = \tanh(W_h h_{t-1} + W_x x_t + b)$, scalar readout $\hat y_t = W_y h_t$.  Use $W_h = 0.5\,I_2$, $W_x = (1,0)^\top$, $b = (0,0)^\top$, $W_y = (1,1)$, $h_0 = (0,0)^\top$, and the input sequence $x_{1:3} = (1, 0, 1)$.  (i) Compute $h_t$ and $\hat y_t$ for $t = 1, 2, 3$.  (ii) Derive a closed-form expression for $\partial \hat y_3 / \partial x_1$.  (iii) Show that with $\|W_h\|_2 < 1$ this gradient decays exponentially in the sequence length, and explain how this connects to the vanishing-gradient problem (\S\ref{sec:sequence_models}).
\item\label{ex:ch1:6} \textbf{[Core] Attention by hand.}  Consider a single attention head on a 3-token scalar sequence $x_{1:3} = (0.0,\, 1.0,\, 0.5)$ with identity projections $W_Q = W_K = W_V = 1$ (so $q_i = k_i = v_i = x_i$) and scaling $\sqrt{d_k} = 1$.  Compute the attention weights $a_{ij} = \mathrm{softmax}_j(q_i k_j)$ and the output $o_i = \sum_j a_{ij} v_j$ for $i = 1, 2, 3$.  Verify that each $o_i$ is a convex combination of $\{v_1, v_2, v_3\}$ and identify which token attends most to which.
\item\label{ex:ch1:7} \textbf{[Computational] Notebook.}  In \texttt{05\_Tensorboard}, plot the train and validation loss for three optimizer choices (SGD, Adam, AdamW) on a small classification task.  Comment on which converges fastest, which generalizes best, and where the curves diverge.
\end{enumerate}

\chapter{Deep Equilibrium Nets}
\label{ch:deqn}

This chapter introduces Deep Equilibrium Nets (DEQNs), the central computational framework of this script.  Rather than generating labeled training data, a DEQN embeds the equilibrium conditions of a dynamic stochastic model (Euler equations, budget constraints, complementarity conditions) directly into the loss function of a neural network.  The network then learns policy functions by minimizing the violation of these conditions via gradient descent.  We develop the methodology on the classical Brock--Mirman growth model, for which an analytical solution exists and serves as a benchmark.  The primary reference is \citet{azinovicDEEPEQUILIBRIUMNETS2022}.

\paragraph{Notation transition.}  Chapter~\ref{ch:intro} used $\bm{\theta}$ for the parameter vector of a generic supervised model, following the textbook convention.  From this chapter onward we switch to $\rho$ (so that the network is denoted $\mathcal{N}_\rho(\x)$), matching the notation of \citet{azinovicDEEPEQUILIBRIUMNETS2022} and most of the subsequent DEQN literature.  The two symbols denote the same object; we use $\rho$ for parameters wherever a network appears, and reserve $\bm{\theta}$ for structural economic parameters that the network does not see (Chs.~\ref{ch:estimation}--\ref{ch:climate}).

\section{The Curse of Dimensionality in Economics}
\label{sec:curse_of_dim}

A central challenge in computational economics is that many models of interest, such as overlapping-generations (OLG) economies, heterogeneous-agent models, and international business cycle models, feature state spaces of high dimension.  In an OLG model with $N$ cohorts and idiosyncratic risk, the state vector may include individual capital holdings, productivity levels, and wealth shares for each cohort, leading to state spaces of dimension $d \gg 10$.  Traditional grid-based solution methods, such as value function iteration on a Cartesian grid with $n$ nodes per dimension, require $n^d$ grid points; this is the \emph{curse of dimensionality}, the term coined by \citet{bellman1961adaptive} to describe the exponential growth of computational cost with the number of state variables.  A concrete back-of-the-envelope makes the bottleneck plain: at the modest resolution of $n=100$ nodes per dimension, a $d=5$ state space already requires $100^5 = 10^{10}$ grid points, which is infeasible under any reasonable compute budget; even more modest resolutions of $n=20$--$30$ become prohibitive once $d$ exceeds ten.

Table~\ref{tab:curse_of_dim} illustrates the curse concretely.  Even with a modest $n=10$ grid points per dimension, the total number of grid points grows astronomically.

\begin{table}[ht]
\centering
\small
\begin{tabular}{r r r l}
\toprule
$d$ & Grid points ($10^d$) & Memory (64-bit) & Feasibility \\
\midrule
1 & 10 & 80 B & trivial \\
2 & 100 & 800 B & trivial \\
5 & $10^5$ & 800 KB & feasible \\
10 & $10^{10}$ & 80 GB & borderline \\
20 & $10^{20}$ & $8 \times 10^{11}$ GB & impossible \\
50 & $10^{50}$ & --- & absurd \\
100 & $10^{100}$ & --- & a googol \\
\bottomrule
\end{tabular}
\caption{Size of an $n = 10$ Cartesian grid and the 64-bit memory required to store one floating-point value per grid point, as a function of state-space dimension $d$.  Grid-based methods are comfortable only at low dimension; by $d = 10$ even storing one scalar per grid point is already borderline before policies, interpolation objects, residuals, and simulation output are added.}
\label{tab:curse_of_dim}
\end{table}

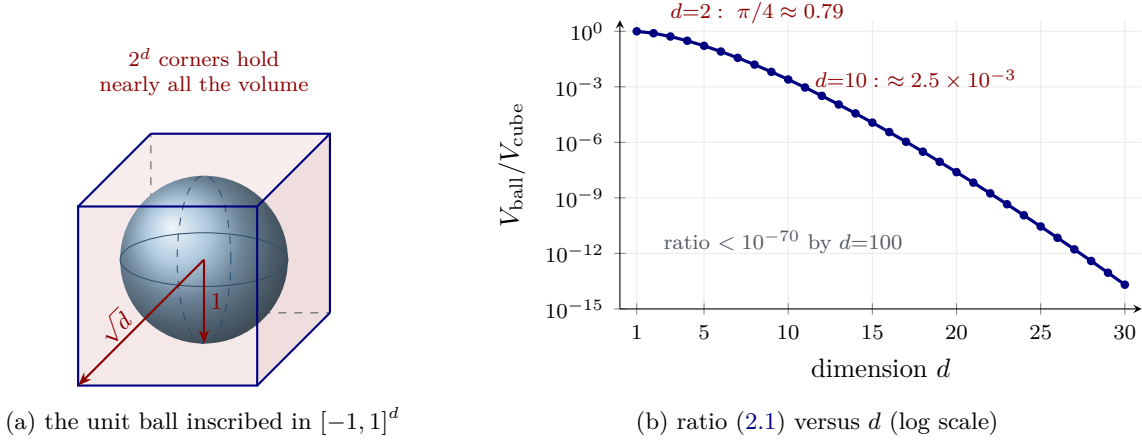
\begin{figure}[ht]
\centering
\begin{minipage}[b]{0.40\textwidth}
\centering
\begin{tikzpicture}[scale=0.74, every node/.style={font=\footnotesize}]
  \coordinate (F1) at (0,0);      \coordinate (F2) at (3.2,0);
  \coordinate (F3) at (3.2,3.2);  \coordinate (F4) at (0,3.2);
  \coordinate (B1) at (1.3,1.3);  \coordinate (B2) at (4.5,1.3);
  \coordinate (B3) at (4.5,4.5);  \coordinate (B4) at (1.3,4.5);
  \fill[darkred!9] (F1)--(F2)--(F3)--(F4)--cycle;          
  \fill[darkred!13] (F2)--(F3)--(B3)--(B2)--cycle;         
  \fill[darkred!7] (F4)--(F3)--(B3)--(B4)--cycle;          
  \draw[uzhgreydark, dashed, thin] (B1)--(B2) (B1)--(B4) (F1)--(B1);
  \shade[ball color=softblue!55] (2.25,2.25) circle (1.5);
  \draw[softblue!60!black, very thin] (2.25,2.25) ellipse (1.5 and 0.48);
  \draw[softblue!60!black, very thin, dashed] (2.25,2.25) ellipse (0.48 and 1.5);
  \draw[uzhblue, thick] (F1)--(F2)--(F3)--(F4)--cycle;
  \draw[uzhblue, thick] (B2)--(B3)--(B4) (F2)--(B2) (F3)--(B3) (F4)--(B4);
  \draw[-{Stealth[length=1.8mm]}, darkred, thick] (2.25,2.25) -- node[right, pos=0.5, inner sep=2pt]{$1$} (2.25,0.75);
  \draw[-{Stealth[length=1.8mm]}, darkred, thick] (2.25,2.25) -- node[sloped, above, pos=0.62, inner sep=1pt]{$\sqrt{d}$} (F1);
  \node[darkred, align=center, font=\scriptsize, anchor=south] at (2.25,5.0) {$2^d$ corners hold\\nearly all the volume};
\end{tikzpicture}\\[0.2em]
{\footnotesize (a) the unit ball inscribed in $[-1,1]^d$}
\end{minipage}\hfill
\begin{minipage}[b]{0.57\textwidth}
\centering
\begin{tikzpicture}
\begin{axis}[
  width=0.93\textwidth, height=5.4cm,
  xlabel={dimension $d$}, ylabel={$V_{\mathrm{ball}}/V_{\mathrm{cube}}$},
  ymode=log, xmin=0, xmax=31, ymin=1e-15, ymax=4,
  ytick={1e0,1e-3,1e-6,1e-9,1e-12,1e-15},
  xtick={1,5,10,15,20,25,30},
  tick label style={font=\scriptsize}, label style={font=\small},
  grid=both, grid style={gray!15}, axis lines=left, clip=false,
]
  \addplot[uzhblue, very thick, mark=*, mark size=1pt] coordinates {
    (1,1.0)(2,7.8540e-1)(3,5.2360e-1)(4,3.0843e-1)(5,1.6449e-1)(6,8.0746e-2)
    (7,3.6912e-2)(8,1.5854e-2)(9,6.4424e-3)(10,2.4904e-3)(11,9.1997e-4)
    (12,3.2599e-4)(13,1.1116e-4)(14,3.6576e-5)(15,1.1641e-5)(16,3.5909e-6)
    (17,1.0756e-6)(18,3.1336e-7)(19,8.8924e-8)(20,2.4611e-8)(21,6.6515e-9)
    (22,1.7572e-9)(23,4.5427e-10)(24,1.1501e-10)(25,2.8542e-11)(26,6.9485e-12)
    (27,1.6605e-12)(28,3.8981e-13)(29,8.9943e-14)(30,2.0410e-14)
  };
  \node[darkred, font=\scriptsize, anchor=south west] at (axis cs:2.4,7.85e-1) {$d{=}2:\ \pi/4\approx0.79$};
  \node[darkred, font=\scriptsize, anchor=west]       at (axis cs:11,2.49e-3)  {$d{=}10:\ \approx2.5\times10^{-3}$};
  \node[uzhgreydark, font=\scriptsize, anchor=south west] at (axis cs:2,3e-13) {ratio $<10^{-70}$ by $d{=}100$};
\end{axis}
\end{tikzpicture}\\[0.2em]
{\footnotesize (b) ratio~\eqref{eq:volume_ratio} versus $d$ (log scale)}
\end{minipage}
\caption{The volume paradox behind the curse of dimensionality.  \emph{(a)} The largest ball that fits inside the cube $[-1,1]^d$ touches the center of every face, at distance $1$ from the center, but its surface stays $\sqrt{d}-1$ away from each of the $2^d$ corners, which lie at distance $\sqrt{d}$.  As $d$ grows the corners recede while the ball does not, so almost all of the cube's volume ends up in the corners (tinted), far from the center.  \emph{(b)} The ball-to-cube volume ratio of equation~\eqref{eq:volume_ratio} on a logarithmic scale: $\pi/4$ at $d=2$, about $2.5\times10^{-3}$ at $d=10$, and below $10^{-70}$ at $d=100$.  A grid or quadrature rule built on the bounding hypercube therefore spends an exponentially growing share of its nodes in corners that the model's ergodic set never reaches.}
\label{fig:volume_paradox}
\end{figure}

\paragraph{The volume paradox.}  A complementary, and more geometric, perspective on the curse of dimensionality starts from a single ratio.  The volume of the unit $d$-ball relative to the smallest cube that contains it, $[-1,1]^d$, is
\begin{equation}
\frac{V_\mathrm{ball}(d)}{V_\mathrm{cube}(d)} = \frac{\pi^{d/2}}{2^d\,\Gamma(d/2+1)}.
\label{eq:volume_ratio}
\end{equation}
This ratio collapses with $d$ (Figure~\ref{fig:volume_paradox}): it is $\pi/4 \approx 0.785$ at $d=2$ and $\pi/6 \approx 0.524$ at $d=3$, has fallen to $\approx 2.5\times 10^{-3}$ by $d=10$, and is below $10^{-70}$ at $d=100$.  The geometry behind the numbers is sharp.  The inscribed ball touches the center of each of the $2d$ faces, at distance $1$ from the center, but its surface never gets within $\sqrt{d}-1$ of any of the $2^d$ corners, which sit at distance $\sqrt{d}$.  As $d$ grows the corners march off to infinity while the ball stays put, so essentially all of the cube's volume drains into its corners, far from the center.  A box-shaped object in high dimensions is, to a first approximation, all corners and no middle.

\begin{figure}[ht]
\centering
\begin{tikzpicture}[scale=1.0,
    dot/.style={circle, fill=#1, inner sep=0pt, minimum size=3pt},
    cube/.style={rectangle, draw=uzhblue, thick, minimum width=5cm, minimum height=5cm}
]
    \node[cube, label={above:\textbf{Cartesian grid}}] (g) at (0,0) {};
    \foreach \i in {-2,-1,0,1,2} {
        \foreach \j in {-2,-1,0,1,2} {
            \node[dot=softblue!70] at (\i,\j) {};
        }
    }
    \draw[darkred, ultra thick, opacity=0.55, line cap=round]
        plot[smooth, tension=0.7] coordinates {(-2.3,-0.8) (-1,-0.4) (0,0.1) (1,0.3) (2.3,0.7)};
    \node[font=\scriptsize, text=darkred, align=center] (ergtext) at (0,-1.95)
        {\textbf{ergodic set}\\{\tiny (states actually visited under the model's dynamics)}};
    \draw[->, darkred, thick, opacity=0.75]
        (ergtext.north) to[out=110, in=-80, looseness=1.0] (-0.55,-0.2);
    \node[font=\scriptsize, text=uzhgreydark] at (-1.6,1.6) {$n^d$ grid};
    \node[font=\scriptsize, text=uzhgreydark] at (-1.6,1.2) {points};

    \node[cube, label={above:\textbf{Simulation-based (DEQN)}}] (s) at (8,0) {};
    \begin{scope}[xshift=8cm]
        \draw[darkred, ultra thick, opacity=0.35, line cap=round]
            plot[smooth, tension=0.7] coordinates {(-2.3,-0.8) (-1,-0.4) (0,0.1) (1,0.3) (2.3,0.7)};
        \foreach \t in {-2.2,-2.0,-1.8,-1.5,-1.2,-1.0,-0.8,-0.5,-0.2,0,0.3,0.6,0.9,1.2,1.4,1.6,1.9,2.2} {
            \pgfmathsetmacro{\yy}{-0.8 + 1.5*(\t + 2.3)/4.6 + 0.1*sin(deg(\t*1.4))}
            \pgfmathsetmacro{\xx}{\t + 0.05*rand}
            \node[dot=darkred] at (\xx,\yy+0.03*rand) {};
        }
        \node[font=\scriptsize, text=uzhgreydark] at (-1.6,1.6) {simulated};
        \node[font=\scriptsize, text=uzhgreydark] at (-1.6,1.2) {trajectory};
    \end{scope}

    \draw[-{Stealth[length=2.5mm]}, thick, softgreen]
        (g.east) -- node[above, font=\scriptsize, text=softgreen]{\textit{targeted sampling}}
        (s.west);
\end{tikzpicture}
\caption{Grid-based vs.\ simulation-based state sampling.  A Cartesian grid (left) allocates effort uniformly over the hypercube and places most of its $n^d$ points far from the model's \emph{ergodic set}, the small subset of the state space that the economy actually visits under its own dynamics (red band).  The DEQN algorithm samples states along simulated trajectories (right), concentrating training points exactly on that set.  As $d$ grows, the fraction of the cube reached by the ergodic set shrinks rapidly, so the grid's relative waste grows exponentially.}
\label{fig:ergodic_vs_grid}
\end{figure}
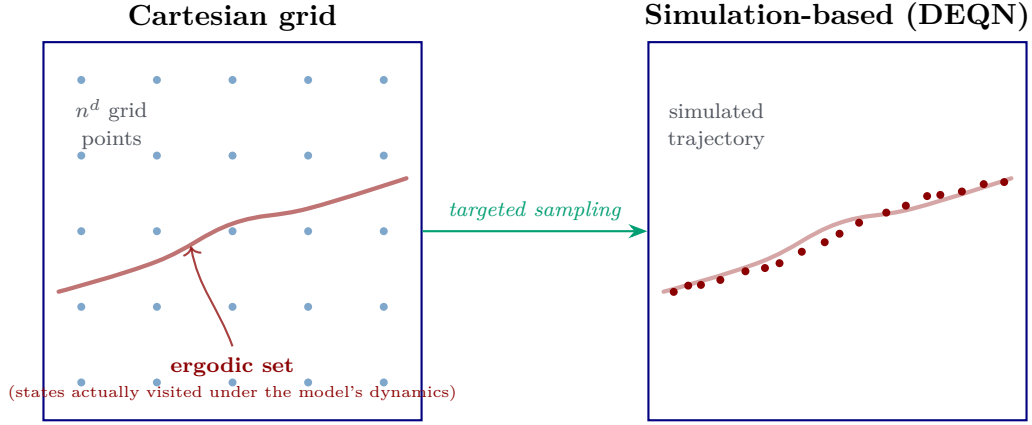

Why should an economist care about a fact about cubes and balls?  Because solving a dynamic stochastic model means producing a \emph{global} solution\footnote{In the sense of \citet{ECTA:ECTA1716}: an approximation of the equilibrium policy (or value) functions over the entire economically relevant region of the state space, in particular over the model's ergodic set, as opposed to a \emph{local} (perturbation) solution that is accurate only in a neighborhood of the deterministic steady state.}, and the region over which that solution actually has to be accurate is the model's \emph{ergodic set}, the states the economy visits under its own law of motion.  That set is typically a thin, curved, low-volume sliver of any hypercube $[\x_{\min},\x_{\max}]^d$ that one would draw around it, often well under one percent of the bounding box already at moderate $d$ \citep{judd2011numerically}.  Yet most of the high-dimensional approximation machinery on offer, tensor-product grids, adaptive sparse grids, tensor-product quadrature, is built on the hypercube; by the volume paradox an exponentially growing fraction of those nodes lands in corners the model never visits, so most of the computational effort is spent representing the function where it does not matter.  Figure~\ref{fig:ergodic_vs_grid} visualizes this mismatch.  This is exactly the waste the DEQN approach is designed to avoid: instead of tiling a box, it trains the network on states sampled from the model's own simulated trajectories, putting the approximation effort where the economics lives.

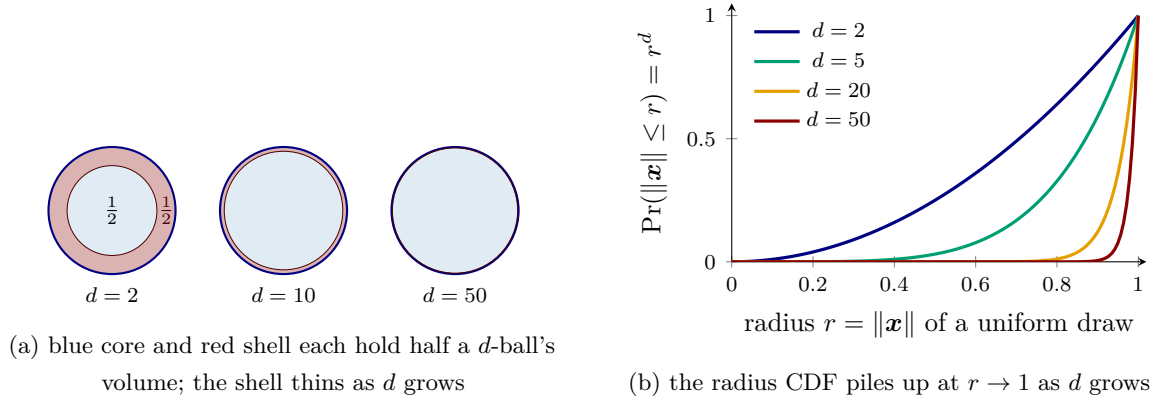
\begin{figure}[ht]
\centering
\begin{minipage}[b]{0.50\textwidth}
\centering
\begin{tikzpicture}[scale=0.84]
  \foreach \dd/\rh/\xc in {2/0.707/0, 10/0.933/2.7, 50/0.986/5.4}{
    \begin{scope}[shift={(\xc,0)}]
      \fill[darkred!30]  (0,0) circle (1);
      \fill[softblue!16] (0,0) circle (\rh);
      \draw[uzhblue, thick] (0,0) circle (1);
      \draw[darkred!55!black, very thin] (0,0) circle (\rh);
      \node[font=\scriptsize] at (0,-1.32) {$d=\dd$};
    \end{scope}
  }
  \node[font=\scriptsize] at (0,0) {$\tfrac12$};
  \node[font=\scriptsize, darkred!55!black] at (0.855,0) {$\tfrac12$};
\end{tikzpicture}\\[0.2em]
{\footnotesize (a) blue core and red shell each hold half a $d$-ball's volume; the shell thins as $d$ grows}
\end{minipage}\hfill
\begin{minipage}[b]{0.48\textwidth}
\centering
\begin{tikzpicture}
\begin{axis}[
  width=0.92\textwidth, height=5cm,
  xlabel={radius $r=\|\x\|$ of a uniform draw}, ylabel={$\Pr(\|\x\|\le r)=r^{d}$},
  domain=0:1, samples=140, xmin=0, xmax=1.02, ymin=0, ymax=1.05,
  legend pos=north west, legend style={font=\scriptsize, draw=none, fill=none},
  tick label style={font=\scriptsize}, label style={font=\small}, axis lines=left,
]
  \addplot[uzhblue, very thick]   {x^2};   \addlegendentry{$d=2$}
  \addplot[softgreen, very thick]  {x^5};   \addlegendentry{$d=5$}
  \addplot[softorange, very thick] {x^20};  \addlegendentry{$d=20$}
  \addplot[darkred, very thick]    {x^50};  \addlegendentry{$d=50$}
\end{axis}
\end{tikzpicture}\\[0.2em]
{\footnotesize (b) the radius CDF piles up at $r\to1$ as $d$ grows}
\end{minipage}
\caption{Why a ``random'' point in high dimensions lives on the shell, not in the core.  \emph{(a)} For each $d$, the blue inner disk and the red outer shell each hold half of the $d$-ball's volume; the shell's fractional thickness $1-(1/2)^{1/d}$ shrinks from $\approx0.29$ at $d=2$ to $\approx0.07$ at $d=10$ to $\approx0.014$ at $d=50$, so half of the mass crowds into an ever thinner rim near the surface.  \emph{(b)} Equivalently, the radius of a uniform draw has cumulative distribution $r^{d}$, which for large $d$ stays near zero until $r$ is almost $1$ and then jumps: the radius is essentially deterministic at $1$.  The companion fact is that pairwise Euclidean distances among random points concentrate, so a query point's nearest and farthest neighbors become nearly indistinguishable~\citep{aggarwal2001surprising}; this is why distance-based methods, including the Gaussian-process kernels of Chapter~\ref{ch:gp}, lose resolution in high dimensions.}
\label{fig:distance_concentration}
\end{figure}

Distance itself is the second casualty of high dimensions, and it is worth flagging now because it returns later with teeth.  Draw a point uniformly from the unit $d$-ball: its radius has cumulative distribution $r^d$ and density $d\,r^{d-1}$ on $[0,1]$, which for large $d$ piles essentially all of its mass against the shell $r\to 1$.  Random points sit on the pulp, not in the core, and they do so with overwhelming probability.  Relatedly, pairwise Euclidean distances among random points concentrate so tightly that the nearest and the farthest neighbor of a query point become almost equidistant, so ``distance'' loses most of its power to discriminate \citep{aggarwal2001surprising} (Figure~\ref{fig:distance_concentration}).  This is not a curio: any method that judges similarity through $\|\x-\x'\|$, from $k$-nearest-neighbors to kernel ridge regression to the Gaussian-process surrogates of Chapter~\ref{ch:gp}, whose RBF and Mat\'ern kernels are functions of $\|\x-\x'\|$ alone, loses resolution as $d$ grows.  It is one of the reasons Chapter~\ref{ch:gp} reaches for dimension reduction, active subspaces and deep kernels, before fitting a GP in a high-dimensional input space.

\section{From Supervised Learning to Self-Supervised Equilibrium Solving}

A word on terminology before we start, recalling the taxonomy of Section~\ref{sec:supervised_vs_unsupervised}.  A learning method is \emph{supervised} when its loss compares the network's output to an externally given target label; it is \emph{unsupervised} when there are no target labels at all (clustering, density estimation, dimension reduction); and it is \emph{self-supervised} when, although no human-provided labels exist, the method manufactures its own supervisory target out of the available structure.  Deep Equilibrium Nets are unsupervised in the narrow ``no labels'' sense, but the sharper and more useful description is \emph{equation-based self-supervision}: at every state where we evaluate the model, the equilibrium conditions tell us what the correct relationship between today's policy and tomorrow's must be, namely a residual of zero, and that residual plays exactly the role a label plays in supervised learning.  The DEQN objective introduced below is therefore a genuine regression loss, only against a target the model itself dictates rather than one handed to us as data.

\begin{figure}[ht]
\centering
\begin{tikzpicture}[
    mlstep/.style={rectangle, draw=softblue, thick, fill=blue!5,
        minimum width=3.5cm, minimum height=0.7cm, font=\small,
        rounded corners=3pt},
    deqnstep/.style={rectangle, draw=darkred, thick, fill=red!5,
        minimum width=3.5cm, minimum height=0.7cm, font=\small,
        rounded corners=3pt},
    arr/.style={-{Stealth[length=2.5mm]}, thick}
]
    \node[font=\small\bfseries, softblue] at (-4.2,3.2) {Supervised};
    \node[mlstep] (d) at (-4.2,2.4) {Labeled data $(\x_i, y_i)$};
    \node[mlstep] (l) at (-4.2,1.2) {$\ell = \|y_i - \mathcal{N}_\rho(\x_i)\|^2$};
    \node[mlstep] (s) at (-4.2,0) {SGD $\to$ $\rho^\star$};
    \draw[arr, softblue] (d) -- (l);
    \draw[arr, softblue] (l) -- (s);

    \node[font=\small\bfseries, darkred, align=center] at (4.2,3.2) {DEQN\\{\footnotesize self-supervised}};
    \node[deqnstep] (d2) at (4.2,2.4) {Equations $G(\x, p(\x))=0$};
    \node[deqnstep] (l2) at (4.2,1.2) {$\ell = \|G(\x_i, \mathcal{N}_\rho(\x_i))\|^2$};
    \node[deqnstep] (s2) at (4.2,0) {SGD $\to$ $\rho^\star$};
    \draw[arr, darkred] (d2) -- (l2);
    \draw[arr, darkred] (l2) -- (s2);
\end{tikzpicture}
\caption{Supervised learning (left) versus DEQN training (right).  Both paradigms train a parameter vector by minimizing a residual loss with SGD; the difference is the \emph{source} of the training signal, labeled data $(\x_i, y_i)$ in the supervised case, structural equilibrium equations $G(\x, p(\x)) = 0$ in the DEQN case.  No labeled solution data are required for DEQNs.  For Brock--Mirman the right-hand side specializes to $G(K, z) = 1 - \beta\,(C/C') \cdot \alpha z' K'^{\alpha-1}$ with $C = \mathcal{N}_\rho(K, z)$ and $K' = z K^\alpha - C$; the network learns the policy by driving the squared mean of this residual to zero on simulated trajectories.}
\label{fig:supervised_vs_deqn}
\end{figure}
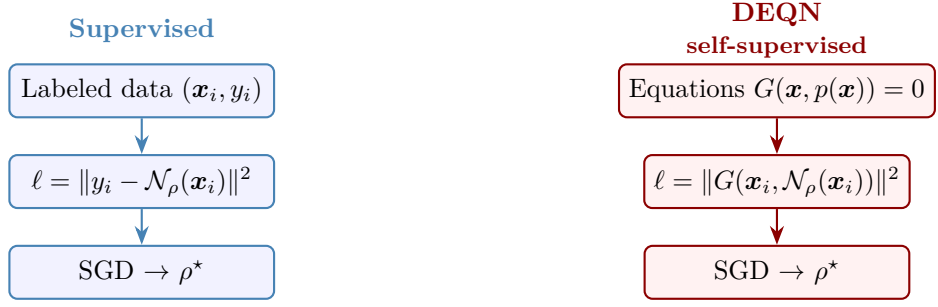

With that in mind, the key conceptual shift introduced by \citet{azinovicDEEPEQUILIBRIUMNETS2022} is to replace the labeled training data of standard supervised learning with the structural equations of the economic model.  In a generic dynamic stochastic economic model, the state of the economy at time $t$ is summarized by a vector $\x_t \in \R^d$, and agents choose policy (or decision) variables $p_t = p(\x_t) \in \R^m$.  The equilibrium is characterized by a system of functional equations
\begin{equation}
G\bigl(\x_t,\, p(\x_t),\, \mathbb{E}_t[H(\x_{t+1}, p(\x_{t+1}))]\bigr) = 0, \qquad \forall\, \x_t,
\end{equation}
where $G$ encodes optimality conditions (e.g., Euler equations) and $\mathbb{E}_t[\cdot]$ is the conditional expectation over next-period shocks via the transition law $\x_{t+1} = T(\x_t, p(\x_t), \varepsilon_{t+1})$.  The fundamental challenge is that this is a \emph{functional equation}: we seek functions $p:\R^d \to \R^m$ rather than finite-dimensional parameter vectors.  The DEQN approach parameterizes $p$ as a neural network and solves this functional equation via stochastic optimization (Figure~\ref{fig:supervised_vs_deqn}).

In the standard (supervised) paradigm, one minimizes the discrepancy between the network's predictions and known target values.  In the DEQN framework, the loss function is instead the squared norm of the equilibrium residual:
\begin{equation}
\boxed{
\ell_\rho = \frac{1}{N} \sum_{i=1}^{N} \bigl\| G\bigl(\x_i,\, \mathcal{N}_\rho(\x_i)\bigr) \bigr\|^2,
}
\label{eq:deqn_loss}
\end{equation}
where $\mathcal{N}_\rho$ denotes a deep neural network with parameters $\rho$ that maps states to policies.  If the residual loss is zero on a sufficiently rich set of states and the conditional expectations inside $G$ are evaluated accurately (rather than with a single shock realization), the network satisfies the equilibrium conditions on those states; global accuracy must always be assessed on independent validation trajectories and, where available, against benchmark solutions.  The continuous-time analogue is the residual-loss formulation of \citet{raissi2019physics}, which inspired the PINN literature developed in Chapter~\ref{ch:pinn}.  The training points $\x_i$ are drawn not from an exogenous dataset, but from the model's own simulated dynamics: the network learns on the ergodic distribution of the economy, concentrating computational effort on economically relevant regions of the state space.  The idea of using simulated paths as the support for projection goes back to the parameterized-expectations algorithm of \citet{marcet1988pea} and \citet{marcet_marshall:94}, which the DEQN literature reframed as ergodic-set sampling to focus computation on economically relevant states; the numerical stability of stochastic-simulation projection was further developed by \citet{judd2011numerically}.

\paragraph{Connection to the modern overparameterized regime.}
Note that DEQNs operate squarely in the ``modern regime'' of Section~\ref{sec:generalization}: the network often has $p \gg n$ parameters relative to the mini-batch size~$n$.  Unlike supervised learning with a fixed dataset, however, the training distribution is \emph{renewable}: each episode can generate fresh state samples from the model's own simulation.  This changes the fixed-dataset double-descent analogy, but it does not remove the need for validation: $p$ can still exceed the effective sample budget seen during training, and a flexible network can fit residuals on a narrow simulated region.  The practical lesson is to combine overparameterized networks and SGD with independent validation trajectories, held-out Euler residuals, and, where available, closed-form benchmarks.

\section{The DEQN Training Algorithm}
\label{sec:deqn_algo}

\begin{definitionbox}[Algorithm: Deep Equilibrium Net Training]
\begin{algorithmic}
\small
\STATE \textbf{Input:} Initial state $\x_0$, network $\mathcal{N}_\rho$, learning rate $\eta$, episodes $E$, simulation horizon $T_{\mathrm{sim}}$, training steps $T_{\mathrm{train}}$, expectation rule $\mathcal{Q}$ (path-average or quadrature)
\STATE \textbf{[NEW]} Burn-in: draw the first episode's states from a broad prior (uniform box around the deterministic steady state) and maintain a small replay buffer of early states.
\STATE \textbf{[NEW]} Independent validation trajectory $\x^{\mathrm{val}}_{0:T_{\mathrm{val}}}$ for out-of-sample residual diagnostics, simulated once with frozen $\rho$ at each checkpoint.
\FOR{episode $e = 1, \ldots, E$}
    \STATE \textbf{Simulate path:} $\x_0 \to \x_1 \to \cdots \to \x_{T_{\mathrm{sim}}}$ using $\mathcal{N}_\rho$ and transition law; guard infeasible states (e.g.\ clip $C\le 0$).
    \FOR{gradient step $t = 1, \ldots, T_{\mathrm{train}}$}
        \STATE Draw mini-batch $\mathcal{B} \subset \{\x_0, \ldots, \x_{T_{\mathrm{sim}}}\} \cup \mathrm{replay}$
        \STATE Compute loss:~$\ell_\rho = \frac{1}{|\mathcal{B}|}\sum_{\x_i \in \mathcal{B}} \|G(\x_i, \mathcal{N}_\rho(\x_i); \mathcal{Q})\|^2$ \textbf{[NEW: $\mathcal{Q}$ is the chosen path-average or Gauss--Hermite / monomial / QMC rule]}
        \STATE Update:~$\rho \leftarrow \rho - \eta \cdot \nabla_\rho \ell_\rho$ \textbf{[NEW: wrap the per-step kernel in \texttt{@tf.function} / \texttt{torch.compile} / \texttt{@jax.jit} for $5$--$50\times$ speed-up]}
    \ENDFOR
\ENDFOR
\STATE \textbf{Output:} Trained network $\mathcal{N}_{\rho^\star}$ approximating the policy function; report Euler residuals on $\x^{\mathrm{val}}$.
\end{algorithmic}
\end{definitionbox}

Several features of this algorithm deserve emphasis:

\begin{itemize}[itemsep=2pt]
\item \textbf{Endogenous training distribution.}  As the policy network improves, the simulated paths become more accurate, and the training points concentrate on the ergodic set of the true equilibrium.  This is fundamentally different from supervised learning, where the training distribution is fixed.  Early in training, however, the simulated distribution is generated by a poor and rapidly changing policy: it can be unstable, infeasible, or far too narrow.  Practical implementations therefore (i) burn in with broad sampling from the prior or a uniform box around the deterministic steady state, (ii) clip or guard against infeasible states (e.g.\ negative consumption), and (iii) often maintain a small replay buffer so that early states are revisited as the policy improves.  Out-of-sample validation on an independent trajectory remains essential.
\item \textbf{No labeled data.}  The algorithm requires no ``ground truth'' solutions, only the structural equations of the model.  The loss function has direct economic interpretation: it measures the violation of equilibrium conditions.
\item \textbf{Stochastic optimization.}  Mini-batch SGD naturally handles the stochasticity of the model: different mini-batches sample different states, providing implicit exploration of the state space.
\item \textbf{Scalability.}  The DEQN avoids explicit tensor-product state grids entirely: the per-iteration cost scales with network size, batch size, the number of equations, and the cost of integrating the conditional expectation; the input/output layer widths and the simulation step also depend on the state and shock dimensions, but only linearly.  This favorable empirical scaling does not eliminate all high-dimensional costs (sample complexity, expectation accuracy, optimization difficulty), but it does mitigate the practical grid-based curse of dimensionality.
\item \textbf{JIT-compile every gradient step.}  In any production or classroom-scale implementation, the per-batch gradient step (lines~5--6 inside the inner loop) should be wrapped in \texttt{@tf.function} (TensorFlow), \texttt{torch.compile} (PyTorch), or \texttt{@jax.jit} (JAX).  The speed-up over an un-traced Python loop is typically $5$--$50\times$ for the per-step kernel, and the trace cost is amortized after the first call.  All companion notebooks in this course follow this convention; un-traced Python loops are simply too slow for classroom use, let alone for research-scale runs.
\end{itemize}

\paragraph{Episode length $T_{\mathrm{sim}}$.}  The simulated-trajectory length~$T_{\mathrm{sim}}$ controls the effective size of the training set within an episode.  \citet{azinovicDEEPEQUILIBRIUMNETS2022} use $T_{\mathrm{sim}} = 10{,}000$ time steps (split into many short mini-batches) in their 113-dimensional 56-agent OLG benchmark.  Shorter episodes (e.g., $T_{\mathrm{sim}} = 1000$) speed up early training because the distribution is re-drawn more frequently, at the cost of higher variance; longer episodes produce smoother gradient estimates but run the risk of stale data if the policy has drifted during training.  A pragmatic rule is to start short, lengthen $T_{\mathrm{sim}}$ as the loss plateaus, and check out-of-sample accuracy on an \emph{independent} simulated trajectory.

\paragraph{Connection to continuous-time methods.}  The DEQN residual loss is the discrete-time analogue of the Deep Galerkin Method (DGM) of \citet{sirignano2018dgm}, which minimizes a PDE residual on neural-network-parameterized PDE solutions in continuous time, and of the deep BSDE solver of \citet{han2018solving} (Han--Jentzen--E), which formulates the same problem as a backward stochastic differential equation.  Within the discrete-time deep-learning toolkit, \citet{maliar2021deep} unify lifetime-reward, Bellman-equation, and Euler-equation training into a single framework, paired with an ``all-in-one'' integration operator that estimates all conditional expectations from a single Monte Carlo realization; DEQN as developed here is the Euler-equation branch of that taxonomy.  Chapters~\ref{ch:pinn} and~\ref{ch:ct_theory} develop the continuous-time analogues explicitly; the present chapter should be read as introducing the discrete-time machinery on which those continuous-time methods are built.

\section{The Brock--Mirman Benchmark}
\label{sec:bm}

Having laid out the generic training loop, we now put it to work on a concrete model.  To validate the DEQN methodology, \citet{azinovicDEEPEQUILIBRIUMNETS2022} begin with the stochastic growth model of \citet{brock1972optimal}, which admits a closed-form solution and therefore serves as an ideal test case.

The social planner solves:
\begin{equation}
\max_{\{C_t,\, K_{t+1}\}_{t=0}^{\infty}} \E{\sum_{t=0}^{\infty} \beta^t \ln(C_t)}
\quad\text{s.t.}\quad
K_{t+1} + C_t = z_t K_t^\alpha, \quad \ln z_{t+1} = \varrho \ln z_t + \sigma_z \epsilon_{t+1},
\label{eq:bm_problem}
\end{equation}
where $\beta \in (0,1)$ is the discount factor, $\alpha \in (0,1)$ the capital share, $\varrho \in [0,1)$ the shock persistence, $\sigma_z > 0$ the shock volatility, and $\epsilon_{t+1} \sim \mathcal{N}(0,1)$ i.i.d.\ innovations.  Depreciation is full ($\delta = 1$).

\paragraph{Derivation of the Euler equation.}
To derive the optimality conditions, form the Lagrangian by attaching a multiplier $\beta^t \lambda_t$ to the resource constraint at each date $t$:
\begin{equation}
\mathcal{L} = \E{\sum_{t=0}^{\infty} \beta^t \Bigl[\ln(C_t) + \lambda_t \bigl(z_t K_t^\alpha - C_t - K_{t+1}\bigr)\Bigr]}.
\label{eq:bm_lagrangian}
\end{equation}
Since the planner chooses $C_t$ and $K_{t+1}$ at each date, we take partial derivatives with respect to each:

\medskip
\noindent\textbf{FOC w.r.t.\ $C_t$:}
\begin{equation}
\frac{\partial \mathcal{L}}{\partial C_t} = \beta^t\!\left(\frac{1}{C_t} - \lambda_t\right) = 0
\qquad\Longrightarrow\qquad
\lambda_t = \frac{1}{C_t}.
\label{eq:bm_foc_c}
\end{equation}

\noindent\textbf{FOC w.r.t.\ $K_{t+1}$:}\; The variable $K_{t+1}$ appears in the date-$t$ constraint (with coefficient $-1$) and in the date-$(t\!+\!1)$ constraint via output $z_{t+1}K_{t+1}^\alpha$.  Differentiating:
\begin{equation}
\frac{\partial \mathcal{L}}{\partial K_{t+1}}
= \beta^t\bigl(-\lambda_t\bigr) + \beta^{t+1}\,\mathbb{E}_t\!\bigl[\lambda_{t+1}\,\alpha\, z_{t+1}\, K_{t+1}^{\alpha-1}\bigr] = 0.
\label{eq:bm_foc_k}
\end{equation}
Dividing both sides by $\beta^t$ yields:
\begin{equation}
\lambda_t = \beta\,\mathbb{E}_t\!\bigl[\lambda_{t+1}\,\alpha\, z_{t+1}\, K_{t+1}^{\alpha-1}\bigr].
\end{equation}
Finally, substituting $\lambda_t = 1/C_t$ from~\eqref{eq:bm_foc_c} into this expression gives the Euler equation:
\begin{equation}
\frac{1}{C_t}
= \beta\,\mathbb{E}_t\!\left[
\frac{\alpha\, z_{t+1}\, K_{t+1}^{\alpha-1}}{C_{t+1}}
\right].
\label{eq:bm_euler}
\end{equation}
The economic intuition is transparent: the left-hand side is the marginal utility cost of saving one additional unit today; the right-hand side is the expected discounted marginal utility gain from the extra output $\alpha\, z_{t+1}\, K_{t+1}^{\alpha-1}$ that unit produces tomorrow.  At the optimum, the planner is indifferent between consuming and saving at the margin.

\paragraph{Analytical solution.}  When depreciation is full ($\delta = 1$), this model admits a closed-form solution.  One can verify by direct substitution into the Euler equation~\eqref{eq:bm_euler} that the optimal consumption policy is:
\begin{equation}
C_t^\star = (1 - \beta\alpha)\, z_t K_t^\alpha.
\label{eq:bm_analytical}
\end{equation}
The derivation proceeds by guessing that the value function takes the form $V(K,z) = V_0 + B\ln K + D\ln z$ (with the constant $V_0$ written as $V_0$ rather than $A$ to avoid clashing with the TFP / cohort-count uses of $A$ elsewhere in the script) and using the envelope theorem.  Substituting this guess into the Bellman equation $V(K,z) = \max_C \{\ln C + \beta \E{V(K',z')}\}$ with $K' = zK^\alpha - C$ yields a system of equations for $V_0$, $B$, and $D$.  Matching coefficients on $\ln K$ pins down $B = \alpha/(1-\alpha\beta)$; matching on $\ln z$ pins down $D = 1/[(1-\alpha\beta)(1-\beta\varrho)]$ (so $D$, unlike $B$, depends on the shock persistence); and the constant $V_0$ then absorbs the remaining terms.  The first-order condition $1/C_t = \beta\,B/K_{t+1}$ together with $C_t + K_{t+1} = z_t K_t^\alpha$ delivers the linear savings rule $K_{t+1} = \beta\alpha\, z_t K_t^\alpha$, from which~\eqref{eq:bm_analytical} follows immediately via the resource constraint.  Notably, this closed-form solution holds regardless of the shock persistence $\varrho$ for the savings rule itself, because $z_{t+1}$ cancels in the ratio $z_{t+1}/C_{t+1}$ under the linear consumption rule: $C_{t+1} \propto z_{t+1}$, so $z_{t+1}/C_{t+1} = 1/[(1-\beta\alpha) K_{t+1}^\alpha]$ no longer depends on the next-period shock.

\paragraph{Persistent shocks.}  When $\varrho > 0$, productivity shocks are autocorrelated and the state of the economy becomes the pair $(K_t, z_t)$, since the current shock level $z_t$ now carries information about future productivity.  Under log utility with Cobb--Douglas production \emph{and full depreciation ($\delta = 1$)}, the analytical solution~\eqref{eq:bm_analytical} continues to hold regardless of the persistence $\varrho$.  The reason fits in one line: under the linear consumption rule $C_t = (1-\beta\alpha)\,z_t K_t^\alpha$ we have $z_{t+1}/C_{t+1} = 1/[(1-\beta\alpha)K_{t+1}^\alpha]$, so $z_{t+1}$ cancels and the conditional expectation in the Euler equation~\eqref{eq:bm_euler} no longer depends on the next-period shock; the constant $D$ in the value function still depends on $\varrho$, but $D$ enters $V$ and cancels in the FOC for the savings rule.  However, for more general preferences (e.g., CRRA with $\gamma \neq 1$), partial depreciation ($\delta < 1$), or non-Cobb--Douglas production, the closed-form solution breaks down and numerical methods, and the DEQN approach in particular, become essential: the policy function $C(K_t, z_t)$ must be approximated rather than derived analytically.  (Concretely: the deterministic companion notebook uses $\delta = 1$ so that the closed form applies and can be used as a benchmark; the stochastic companion notebook switches to $\delta = 0.1$, where it does not, and the policy is genuinely learned.  Only the loss-kernel notebook of \S\ref{sec:loss_kernels} returns to $\delta = 1$, precisely so that the closed-form savings rate $s^\star = \alpha\beta$ is available as ground truth.)

\begin{figure}[ht]
\centering
\begin{tikzpicture}
\begin{axis}[
    width=11cm, height=4.4cm,
    xlabel={Training episode},
    ylabel={Mean abs.\ Euler residual (log)},
    xmin=0, xmax=300, ymin=1e-5, ymax=1,
    ymode=log,
    grid=major, grid style={gray!15},
    legend style={at={(0.98,0.97)}, anchor=north east, font=\footnotesize, draw=gray!40},
    every axis plot/.append style={very thick},
    label style={font=\small}, tick label style={font=\footnotesize},
]
\addplot[uzhblue, dashed, domain=0:50, samples=50] {0.5*exp(-x/20) + 0.05};
\addlegendentry{early phase}
\addplot[harvardcrimson, domain=50:200, samples=80] {0.06*exp(-(x-50)/45) + 8e-4};
\addlegendentry{mid phase}
\addplot[softgreen, domain=200:300, samples=50] {1.1e-3*exp(-(x-200)/120) + 1.5e-4};
\addlegendentry{late phase}
\addplot[black, dotted, very thick, domain=0:300] {1e-4};
\addlegendentry{analytical benchmark}
\end{axis}
\end{tikzpicture}
\caption{Schematic, not measured: the qualitative convergence behavior typical of a successful Brock--Mirman DEQN run.  An early high-residual phase reflects an untrained network feeling out the state space; a mid phase descends roughly exponentially as the policy locks onto the equilibrium structure; a late phase plateaus near the irreducible quadrature/training-noise floor.  The dotted line marks the analytical benchmark below which the residual cannot be driven without higher-precision quadrature.  For the actual numbers on a specific seed, consult the companion notebook.}
\label{fig:bm_convergence_schematic}
\end{figure}
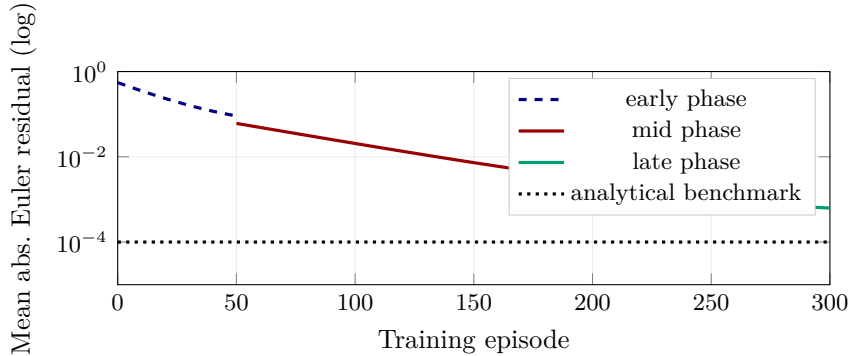

\paragraph{DEQN formulation.}  We parametrize the consumption policy as $C_t = \mathcal{N}_\rho(K_t, z_t)$ and define the residual
\begin{equation}
G(K_t, z_t) = 1 - \beta \, \frac{C_t}{C_{t+1}} \cdot \alpha z_{t+1} K_{t+1}^{\alpha-1},
\end{equation}
where $z_{t+1}$ denotes a single realization of the next-period shock and $K_{t+1} = z_t K_t^\alpha - \mathcal{N}_\rho(K_t, z_t)$.  In the code listing below the network actually emits a \emph{savings share} $s_t\in(0,1)$ through a sigmoid head rather than $C_t$ directly, which guarantees $C_t>0$ \emph{and} $K_{t+1}>0$ at every training step; Section~\ref{sec:deqn_hard_soft} explains the hard-vs-soft constraint split this exemplifies.  Note that the Euler equation~\eqref{eq:bm_euler} involves the conditional expectation $\mathbb{E}_t[\cdot]$, whereas the residual $G$ above is written for a single realization of $z_{t+1}$.  Two approaches are common for handling this expectation:
\begin{enumerate}[itemsep=2pt]
\item \textbf{Path averaging:} draw a single $z_{t+1}$ per state, compute $G$ for that draw, and average $G^2$ over many states along the simulated path.  This minimizes $\mathbb{E}_{(K_t,z_t,z_{t+1})}[G^2]$ -- the squared \emph{pathwise} residual.  It is an unbiased Monte Carlo objective for that stronger loss, and Jensen's inequality gives $\bigl(\mathbb{E}_t[G]\bigr)^2 \leq \mathbb{E}_t[G^2]$.  Thus $G=0$ almost surely implies the conditional Euler equation, but the converse need not hold: a policy can have $\mathbb{E}_t[G]=0$ while $G$ varies with the shock.  In the log/full-depreciation Brock--Mirman benchmark the closed-form policy happens to drive the pathwise residual itself to zero, so this stronger target is harmless; in richer models it is a modeling choice.  This is the approach used in the Brock--Mirman code listing below.  Note that the pathwise-residual-zero coincidence is special to $\delta = 1$: for $\delta < 1$ (the calibration of the stochastic notebook) a converged policy has $\mathbb{E}_\varepsilon[G\mid x]=0$ but $G$ itself is nonzero realization-by-realization, so path averaging then minimizes a strictly stronger objective than the conditional Euler residual.
\item \textbf{Quadrature:} for each state $(K_t, z_t)$, approximate $\mathbb{E}_t[\cdot]$ explicitly via deterministic nodes and weights (Section~\ref{sec:quadrature_rules}), form the residual from the estimated expectation, and then square.  This targets $\bigl(\mathbb{E}_t[G]\bigr)^2$ directly and is the approach used for the IRBC model of Chapter~\ref{ch:irbc}, where accurate expectations are critical for convergence.
\end{enumerate}
Both approaches recover the analytical solution~\eqref{eq:bm_analytical} to high accuracy, providing a rigorous validation of the methodology.  Because convergence curves depend on the exact training run, random seed, and solver configuration, this manuscript does \emph{not} include a hand-drawn convergence plot.  In practice, one should report diagnostics from the \emph{actual notebook run}: residual trajectories, held-out Euler errors, and the gap between the learned policy and the analytical benchmark.  Figure~\ref{fig:bm_convergence_schematic} sketches the qualitative shape one should expect to see.

\paragraph{Relative Euler equation error.}  Following \citet{judd1998numerical} and \citet{azinovicDEEPEQUILIBRIUMNETS2022} (Eq.~43), the standard accuracy diagnostic for a DEQN is the \emph{relative} Euler error, which measures the percentage consumption error implied by a violation of the Euler equation.  For Brock--Mirman with log utility, the relative error at state $\x_j = (K_j, z_j)$ is
\begin{equation}
e^{\mathrm{REE}}_{\x_j}(\rho)
\;=\;
\frac{
  \bigl(\beta\,\mathbb{E}[\alpha\, z' (K_j')^{\alpha-1}/C'_\rho(\x'_j)\mid \x_j]\bigr)^{-1}
}{C_\rho(\x_j)}
\;-\; 1,
\label{eq:ree_bm}
\end{equation}
where $C_\rho = \mathcal{N}_\rho$, $K'_j$ is next-period capital under the network policy, the expectation conditions on $\x_j$, and the $^{-1}$ inverts the marginal-utility relation $u'(c) = 1/c$.  The value $e^{\mathrm{REE}} = 10^{-4}$ means the agent's optimal consumption is mispriced by $0.01\%$, independent of units or utility scale.  This is the metric reported in Table~3 of \citet{azinovicDEEPEQUILIBRIUMNETS2022}, where the trained DEQN achieves mean relative Euler errors of order $10^{-4}$ on the 113-dimensional 56-agent OLG benchmark.

\begin{remarkbox}[Why does this work so well?]
The success of the DEQN approach rests on three pillars.  First, neural networks are universal function approximators that can represent the smooth policy functions arising in most economic models.  Second, the training distribution is endogenous: the network learns on the model's own ergodic distribution, concentrating computational effort precisely where it matters.  Third, stochastic gradient descent operates directly on the economic equilibrium conditions, so the loss function has a clear economic interpretation: a pointwise relative Euler error of $10^{-4}$ means the consumption level implied by the Euler equation differs from the network's consumption by about $0.01\%$.
\end{remarkbox}

\begin{lstlisting}[caption={Representative DEQN loss for Brock--Mirman with path averaging.  The network outputs a savings share $s \in (0,1)$ via a sigmoid, which jointly enforces $C > 0$ \emph{and} $K' > 0$; softplus on $C$ alone would not, since $C > z K^\alpha$ would yield $K' < 0$ and an undefined $K'^{\,\alpha-1}$.}]
def deqn_loss(model, K, z, beta, alpha, z_next):
    Y       = z * K**alpha                   # output today
    s       = model(K, z)                    # savings share in (0,1) via sigmoid
    C       = (1.0 - s) * Y                  # consumption today  (>0)
    K_next  = s * Y                          # capital tomorrow   (>0)
    Y_next  = z_next * K_next**alpha
    s_next  = model(K_next, z_next)
    C_next  = (1.0 - s_next) * Y_next        # consumption tomorrow
    # z_next: single draw; expectation via path averaging
    G = 1.0 - beta * (C / C_next) * alpha * z_next * K_next**(alpha-1)
    return tf.reduce_mean(G**2)
\end{lstlisting}

A Gauss--Hermite variant (developed formally in \S\ref{sec:gh_tensor_product}) replaces the single shock draw \texttt{z\_next} by a $Q$-node weighted sum
$\mathrm{E}[\,\alpha z' K'^{\alpha-1}/C'\mid z\,] \approx \sum_{q=1}^{Q} w_q\,\alpha z_q' K'^{\alpha-1}/C'(K',z_q')$
with $z_q' = z^\varrho\exp(\sigma_z\varepsilon_q)$ at the rescaled Hermite nodes $\varepsilon_q$; in practice $Q=5$ already drives the quadrature error below the training error.  The autodiff companion notebook \tpath{03_Brock_Mirman_Uncertainty_AutoDiff_DEQN.ipynb} implements this variant explicitly.

\section{Encoding Equilibrium Conditions: Hard vs.\ Soft Constraints}
\label{sec:deqn_hard_soft}

The Brock--Mirman DEQN loss above already used one design choice without dwelling on it: the network emits a savings share through a sigmoid rather than consumption directly.  This is an instance of a general principle.  \citet{azinovicDEEPEQUILIBRIUMNETS2022}~\S4.2.2 do not treat every equilibrium equation symmetrically in the loss~\eqref{eq:deqn_loss}; they split the conditions into two groups.  We make the split explicit on the Brock--Mirman model of the previous section and then indicate how it generalizes.

\begin{description}[itemsep=3pt, leftmargin=1.4em]
\item[Hard constraints, encoded in the architecture (never in the loss).]  Some equations can be satisfied exactly for every state $\x = (K,z)$:
  \begin{itemize}\itemsep1pt
  \item The \emph{resource / state-transition} equation $K_{t+1} = z_t K_t^\alpha - C_t$ defines next-period capital from the network's consumption policy.  It is \emph{not} minimized; it is evaluated as a closed-form function of $\x_t$ and $C_t$.
  \item The economic requirements $C_t > 0$ \emph{and} $K_{t+1} > 0$ are jointly imposed by parameterizing the network's output as a \emph{savings share} $s_t \in (0,1)$ via a \emph{sigmoid} activation, $\mathrm{sigmoid}(z) = 1/(1+e^{-z})$, and recovering both quantities in closed form from the resource constraint, $C_t = (1-s_t)\,z_t K_t^\alpha$ and $K_{t+1} = s_t\,z_t K_t^\alpha$.  A softplus head on $C_t$ alone would guarantee $C_t>0$ but not $K_{t+1}>0$ (the network could output $C_t > z_t K_t^\alpha$).  This sigmoid-savings parameterization removes an entire class of infeasible candidate policies before training begins; see the code listing in Section~\ref{sec:bm} above.
  \end{itemize}
\item[Soft constraint, minimized in the loss.]
  The only equilibrium condition that cannot be enforced analytically is the \emph{Euler equation}~\eqref{eq:bm_euler}.  The squared relative Euler error~\eqref{eq:ree_bm} is averaged over the mini-batch and driven toward zero by stochastic gradient descent.
\end{description}

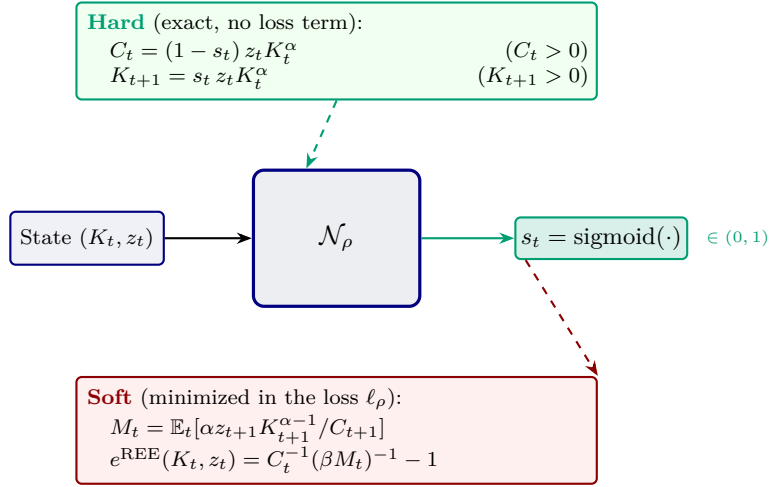
\begin{figure}[ht]
\centering
\begin{tikzpicture}[
    io/.style={rectangle, draw=uzhblue, thick, rounded corners=2pt, fill=uzhblue!6,
               minimum width=1.6cm, minimum height=0.7cm, font=\scriptsize, align=center},
    dnn/.style={rectangle, draw=uzhblue, very thick, fill=uzhgreylight, rounded corners=4pt,
                minimum width=2.2cm, minimum height=1.8cm, font=\small\bfseries, align=center},
    softbox/.style={rectangle, draw=darkred, thick, fill=red!6, rounded corners=2pt,
                 font=\scriptsize, align=left, inner sep=4pt},
    hardbox/.style={rectangle, draw=softgreen, thick, fill=green!6, rounded corners=2pt,
                 font=\scriptsize, align=left, inner sep=4pt},
    sp/.style={rectangle, draw=softgreen, thick, fill=softgreen!15, rounded corners=2pt,
               minimum width=2.2cm, minimum height=0.55cm, font=\footnotesize, align=center, inner sep=2pt},
    arr/.style={-{Stealth[length=2mm]}, thick}
]
    \node[io] (x) at (0,0) {State $(K_t,z_t)$};
    \node[dnn] (nn) at (3.3,0) {$\mathcal{N}_\rho$};
    \node[sp] (s) at (6.8,0) {$s_t = \mathrm{sigmoid}(\cdot)$};
    \node[font=\tiny, text=softgreen, anchor=west] at ($(s.east)+(0.1,0)$) {$\in(0,1)$};

    \draw[arr] (x) -- (nn);
    \draw[arr, softgreen] (nn.east) -- (s.west);

    \node[hardbox, above=0.9cm of nn, text width=6.6cm] (hb) {%
        \textbf{\textcolor{softgreen}{Hard}} (exact, no loss term):\\[1pt]
        \quad$C_t = (1-s_t)\,z_t K_t^\alpha$ \hfill ($C_t>0$)\\
        \quad$K_{t+1} = s_t\,z_t K_t^\alpha$ \hfill ($K_{t+1}>0$)};

    \node[softbox, below=0.9cm of nn, text width=6.6cm] (sb) {%
        \textbf{\textcolor{darkred}{Soft}} (minimized in the loss $\ell_\rho$):\\[1pt]
        \quad $M_t=\mathbb{E}_t[\alpha z_{t+1}K_{t+1}^{\alpha-1}/C_{t+1}]$\\
        \quad $e^{\mathrm{REE}}(K_t,z_t)=C_t^{-1}(\beta M_t)^{-1}-1$};

    \draw[dashed, softgreen, arr] (hb.south) -- ($(nn.north)+(-0.4,0)$);
    \draw[dashed, darkred,  arr] ($(s.south west)+(0.15,0)$) -- (sb.north east);
\end{tikzpicture}
\caption{Hard vs.\ soft constraints in the DEQN architecture for Brock--Mirman.  The network $\mathcal{N}_\rho$ reads the state $(K_t,z_t)$ and emits a \emph{savings share} $s_t \in (0,1)$ via a \emph{sigmoid} head.  Both consumption $C_t = (1-s_t)\,z_t K_t^\alpha$ and next-period capital $K_{t+1} = s_t\,z_t K_t^\alpha$ are then defined in closed form (green, top), guaranteeing $C_t>0$ \emph{and} $K_{t+1}>0$ simultaneously; a softplus head on $C_t$ alone could not, since $C_t > z_t K_t^\alpha$ would push $K_{t+1}<0$ and make $K_{t+1}^{\alpha-1}$ undefined.  Only the relative Euler-equation residual (red, bottom) is squared, averaged over the mini-batch, and minimized by SGD.  This figure matches the code listing in Section~\ref{sec:bm} above.  In richer models (OLG, Chapter~\ref{ch:olg}) the same split extends to firm first-order conditions, household budget constraints, and KKT multipliers with softplus heads and Fischer--Burmeister complementarity residuals.}
\label{fig:hard_soft}
\end{figure}

This split is pedagogically important for three reasons: (i)~only the genuinely non-closed-form conditions enter the loss, which speeds up training; (ii)~it eliminates a family of bad local minima in which the network produces, e.g., slightly negative consumption or infeasible states; and (iii)~it explains why the loss typically converges to a small but nonzero value even at the optimum, since the Euler residual is intrinsic and cannot be removed by re-parameterization.  Figure~\ref{fig:hard_soft} summarizes this construction for Brock--Mirman.

\paragraph{How the split generalizes.}  In models with more equilibrium objects (the OLG benchmark of \citet{azinovicDEEPEQUILIBRIUMNETS2022}, the IRBC model of Chapter~\ref{ch:irbc}, the HA economies of Chapter~\ref{ch:young}), the same logic extends: firm first-order conditions for prices $(w_t,r_t)$ are closed-form in the aggregate state, the household budget constraint algebraically determines consumption given savings, and each inequality constraint contributes a softplus-headed KKT multiplier paired with a Fischer--Burmeister complementarity residual in the soft part of the loss (Chapter~\ref{ch:irbc}).  The Brock--Mirman case above is the minimal instance of the pattern.

\paragraph{Market-clearing layers.}  One can push this design philosophy further by encoding the \emph{market-clearing condition itself} into the network as a dedicated output layer rather than minimizing a market-clearing residual in the loss: the network outputs unnormalized cohort savings (or shares) and a final layer rescales them so that aggregate clearing holds by construction.  \citet{azinoviczemlicka_2024} use such a ``market-clearing layer'' in an OLG economy with rare disasters; this is the discrete-time counterpart of the PINN-style practice of baking conservation laws directly into the network output (Chapter~\ref{ch:pinn}).  Sigmoid heads, softplus heads, Fischer--Burmeister residuals, and market-clearing rescaling layers thus form a small toolbox of architecture-level encodings that each move part of the equilibrium structure from the soft loss into the hard part of the network.

\section{Quadrature Rules for Conditional Expectations}
\label{sec:quadrature_rules}

\subsection{Why We End Up Integrating, and What Numerical Integration Is}
\label{sec:integration_primer}

\paragraph{Why an integral always shows up.}  Sooner or later, every dynamic stochastic model in this script presents the same algorithmic bottleneck: a forward-looking agent has to form an expectation over future shocks before any decision can be made.  The Euler equation~\eqref{eq:bm_euler} hinges on $\mathbb{E}_t[\,\alpha z_{t+1} K_{t+1}^{\alpha-1}/C_{t+1}\,]$; the Bellman operator behind a value-function iteration evaluates $\mathbb{E}[V(\bm{s}')\mid\bm{s},a]$ at every state and action; the relative Euler error~\eqref{eq:ree_bm} embeds the same expectation in the diagnostic; the integrated-assessment models of Chapter~\ref{ch:climate} require expected discounted utility under climate-shock distributions; and the structural-estimation moments of Chapter~\ref{ch:estimation} are themselves expectations over simulated paths.  Outside a handful of conjugate or fully-linear-quadratic-Gaussian special cases, none of these expectations admit closed-form expressions, so the integral $\mathbb{E}[g(\bm\varepsilon)] = \int g(\bm\varepsilon)\mu(\bm\varepsilon)\,d\bm\varepsilon$ has to be approximated numerically by a finite, cheap-to-evaluate, ideally differentiable, weighted sum $\sum_q w_q\, g(\bm\varepsilon_q)$.  Choosing the nodes $\bm\varepsilon_q$ and weights $w_q$ well is the entire content of \emph{quadrature theory}.  The path-averaging recipe used in the Brock--Mirman code above is one such choice (a Monte Carlo sample average with one node per state along a simulated trajectory); the rest of this section makes the deterministic and quasi-random alternatives explicit and explains when each is preferable.

\begin{figure}[ht]
\centering
\begin{minipage}[c]{0.46\textwidth}
\centering
{\scriptsize\textbf{Midpoint sum, $N=8$ rectangles}}\\[2pt]
\begin{tikzpicture}[scale=1.0, font=\small]
    \foreach \i in {0,1,2,3,4,5,6,7} {
        \pgfmathsetmacro{\mx}{0.5*\i + 0.25}
        \pgfmathsetmacro{\hh}{1.6 + 0.6*sin(deg(\mx*0.85)) + 0.18*\mx}
        \draw[fill=harvardcrimson!18, draw=harvardcrimson!70, very thin]
            (0.5*\i, 0) rectangle (0.5*\i + 0.5, \hh);
        \fill[harvardcrimson] (\mx, \hh) circle (0.9pt);
    }
    \draw[thick, uzhblue, smooth, samples=80, domain=0:4]
        plot(\x, {1.6 + 0.6*sin(deg(\x*0.85)) + 0.18*\x});
    \draw[->, gray!80] (-0.05,0) -- (4.5,0) node[right, font=\scriptsize] {$x$};
    \draw[->, gray!80] (0,-0.05) -- (0,3.0) node[above, font=\scriptsize] {$f(x)$};
    \node[below, font=\scriptsize] at (0,-0.05) {$a$};
    \node[below, font=\scriptsize] at (4,-0.05) {$b$};
\end{tikzpicture}
\\[2pt]
{\scriptsize\textbf{(a)} Riemann/midpoint approximation: tile the area under $f$ by $N$ rectangles whose heights equal $f$ at the midpoints (red dots).  Error $\mathcal{O}(N^{-2})$ in 1D; with $M$ total Cartesian nodes in $d$ dimensions the rate is $\mathcal{O}(M^{-2/d})$.}
\end{minipage}
\hfill
\begin{minipage}[c]{0.46\textwidth}
\centering
{\scriptsize\textbf{Quarter-disc $x^2+y^2 \le 1$ inside $[0,1]^2$}}\\[2pt]
\begin{tikzpicture}[scale=2.4, font=\small]
    \draw[thick, uzhblue] (0,0) rectangle (1,1);
    \fill[harvardcrimson!10] (0,0) -- (1,0) arc(0:90:1) -- cycle;
    \draw[thick, harvardcrimson] (1,0) arc (0:90:1);
    \pgfmathsetseed{42}
    \foreach \i in {1,...,120} {
        \pgfmathsetmacro{\xx}{rnd}
        \pgfmathsetmacro{\yy}{rnd}
        \pgfmathsetmacro{\rsq}{\xx*\xx + \yy*\yy}
        \pgfmathparse{\rsq < 1 ? 1 : 0}
        \edef\inside{\pgfmathresult}
        \ifnum\inside=1
          \fill[harvardcrimson!85] (\xx,\yy) circle (0.012);
        \else
          \fill[uzhblue!85] (\xx,\yy) circle (0.012);
        \fi
    }
    \node[below, font=\scriptsize] at (0,-0.02) {$0$};
    \node[below, font=\scriptsize] at (1,-0.02) {$1$};
    \node[left,  font=\scriptsize] at (-0.02,1) {$1$};
\end{tikzpicture}
\\[2pt]
{\scriptsize\textbf{(b)} Monte Carlo estimate of $\pi$: throw $N$ uniform darts at $[0,1]^2$, count $N_{\mathrm{in}}$ inside the quarter-disc (shaded red), and report $\widehat{\pi}_N = 4N_{\mathrm{in}}/N$.  Error $\mathcal{O}(N^{-1/2})$ regardless of dimension.}
\end{minipage}
\caption{Two paradigms for numerical integration that underlie every rule in this section.  Deterministic tiling (left) is exact and highly accurate at low dimension but suffers exponential cost growth in $d$.  Random sampling (right) is dimension-independent in its error rate but slow to converge in any single dimension.  The Gauss--Hermite, monomial, and QMC rules of \S\S\ref{sec:gh_tensor_product}--\ref{sec:qmc_cdf} are sophisticated descendants of the two ideas, designed to combine deterministic accuracy with manageable cost in $d$.}
\label{fig:integration_primer}
\end{figure}
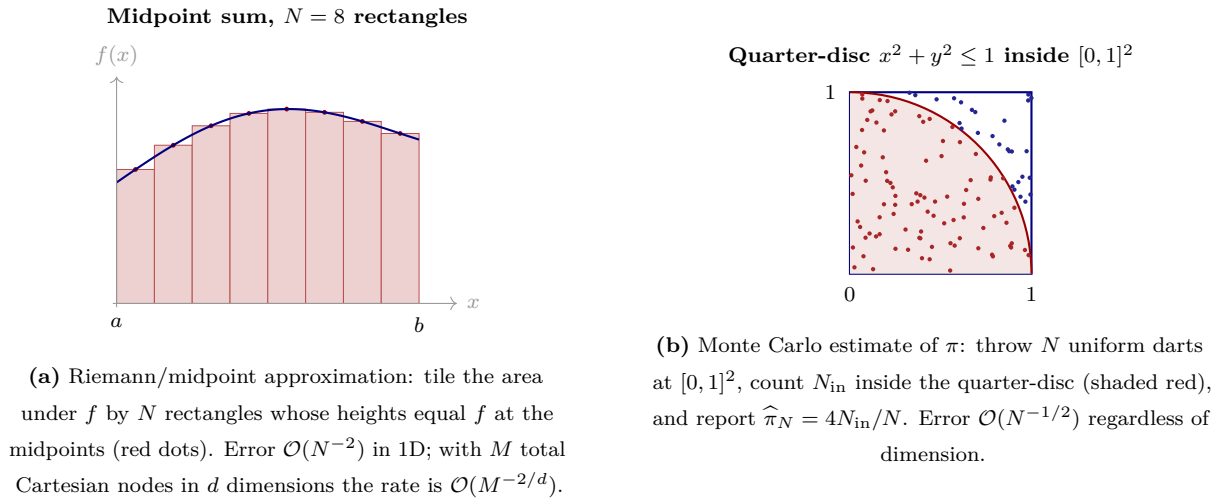

\paragraph{Picture~1: a definite integral as area, and the Riemann/midpoint sum.}  Strip away the economics.  A definite integral $\int_a^b f(x)\,dx$ is the (signed) area between the graph of $f$ and the $x$-axis on $[a,b]$.  The simplest deterministic numerical rule, the \emph{midpoint rule}, replaces this area by a stack of $N$ rectangles of equal width $\Delta x = (b-a)/N$, each one as tall as $f$ at the midpoint of its base (Figure~\ref{fig:integration_primer}, left).  Adding the rectangle areas yields
\begin{equation}
I_N = \Delta x \sum_{i=1}^{N} f\bigl(x_i^{\text{mid}}\bigr)
\;\xrightarrow[N \to \infty]{}\; \int_a^b f(x)\,dx,
\qquad \text{with error } |I_N - I| = \mathcal{O}(N^{-2})
\end{equation}
for smooth $f$.  Trapezoid and Simpson rules refine the same idea by replacing each rectangle with a trapezoid or a parabolic arc and reach $\mathcal{O}(N^{-2})$ and $\mathcal{O}(N^{-4})$ respectively; Gauss--Hermite, the workhorse of \S\ref{sec:gh_tensor_product}, is the optimal rule when the integrand is multiplied by the Gaussian weight $e^{-x^2}$ and reaches degree-$(2Q-1)$ exactness using only $Q$ nodes.  The same tile-the-area idea generalizes to higher dimensions by laying out a Cartesian grid.  With $N$ nodes per coordinate the cost is $N^d$; equivalently, with $M=N^d$ total nodes the midpoint rate becomes $\mathcal{O}(M^{-2/d})$.  Notice that it is the \emph{exponent} of the rate, not just the constant, that degrades from $-2$ in 1D to $-2/d$ in $d$-D Cartesian, the curse of dimensionality made quantitative.  This is the explicit form of the curse of dimensionality flagged in Section~\ref{sec:curse_of_dim}, and it motivates both the Stroud monomial rules of \S\ref{sec:monomial_cubature} and the randomized methods of \S\ref{sec:qmc_cdf}.

\paragraph{Picture~2: throwing darts to estimate $\pi$.}  An entirely different idea is to abandon the grid and approximate the integral by a \emph{sample average}: draw $N$ random points $\bm{u}_1, \ldots, \bm{u}_N$ uniformly from the integration domain $\Omega$ and report $I_N = \mathrm{vol}(\Omega)\cdot \tfrac{1}{N}\sum_{i=1}^{N} f(\bm{u}_i)$.  This is plain Monte Carlo (MC), and its error decays as $\mathcal{O}(N^{-1/2})$ with a rate that is \emph{independent of the dimension $d$}, although the variance constant can still worsen with dimension.  This is why MC dominates deterministic grids when $d$ is large.  The cleanest illustration uses the unit square $[0,1]^2$ and the indicator function of a quarter-disc:
\begin{equation}
\frac{\pi}{4} \;=\; \int_0^1\!\!\int_0^1 \mathbf{1}\bigl[x^2 + y^2 \leq 1\bigr]\, dx\, dy
\;\approx\; \frac{N_{\mathrm{in}}}{N},
\qquad
\widehat{\pi}_N \;=\; 4\cdot \frac{N_{\mathrm{in}}}{N},
\label{eq:pi_mc}
\end{equation}
where $N_{\mathrm{in}}$ counts how many of the $N$ uniform ``darts'' land inside the quarter-circle $x^2+y^2 \leq 1$ (Figure~\ref{fig:integration_primer}, right).  With $N=100$ a typical run gives $\widehat{\pi}_{100} \approx 3.04$ (about 3\% off); with $N=10^6$ a typical run gives $\widehat{\pi}_{10^6} \approx 3.1417$ (about $10^{-4}$ off).  The error shrinks as $\mathcal{O}(1/\sqrt{N})$, requiring a hundredfold increase in $N$ to gain one extra decimal of accuracy, which is glacially slow compared to $\mathcal{O}(N^{-4})$ for Simpson's rule on a smooth 1D integrand.  But the MC \emph{rate} has no dependence on the dimension of the domain: replacing the quarter-disc by a $d$-dimensional unit ball would leave the rate untouched, while the deterministic grid would suffer the $N^d$ cost explosion of the curse of dimensionality.  This is what makes MC and its quasi-random refinement (QMC, \S\ref{sec:qmc_cdf}) the natural tools for the conditional expectations encountered in DEQNs at $d \gtrsim 10$.

\paragraph{Where the rest of this section is going.}  With this picture in hand, the design of every quadrature rule in the literature can be read as an answer to two questions: (i)~where do we place the nodes $\bm\varepsilon_q$, and (ii)~what weights $w_q$ do we attach to them?  Tensor-product Gauss--Hermite (\S\ref{sec:gh_tensor_product}) places nodes deterministically on a Cartesian grid of Hermite roots; the Stroud-3 monomial rule of \S\ref{sec:monomial_cubature} places only $2d$ nodes on the principal axes and accepts a controlled bias on fourth-order moments in exchange for linear-in-$d$ scaling; the QMC construction of \S\ref{sec:qmc_cdf} places nodes from a low-discrepancy sequence in the unit cube and pulls them back through the inverse CDF.  The cost--accuracy trade-offs differ dramatically with $d$, as Table~\ref{tab:quadrature_costs} (page~\pageref{tab:quadrature_costs}) makes concrete.  Chapter~\ref{ch:irbc} returns to these numbers when scaling DEQNs to multi-country economies.

\subsection{Tensor-Product Gauss--Hermite}
\label{sec:gh_tensor_product}

The classical Gauss--Hermite rule approximates integrals against the weight function $e^{-x^2}$:
\begin{equation}
\int_{-\infty}^{\infty} f(x)\, e^{-x^2}\, dx \approx \sum_{q=1}^{Q} \tilde{w}_q\, f(x_q),
\end{equation}
where the $Q$ nodes $\{x_q\}$ are the roots of the $Q$-th Hermite polynomial $H_Q(x)$ and the weights are $\tilde{w}_q = 2^{Q-1} Q!\sqrt{\pi}/(Q^2 [H_{Q-1}(x_q)]^2)$.  For expectations under the standard normal distribution, we apply the change of variables $\varepsilon = \sqrt{2}\, x$ to obtain:
\begin{equation}
\E{h(\varepsilon)} = \int_{-\infty}^{\infty} h(\varepsilon)\frac{e^{-\varepsilon^2/2}}{\sqrt{2\pi}}\,d\varepsilon \approx \sum_{q=1}^{Q} w_q\, h(\varepsilon_q),
\qquad w_q = \frac{\tilde{w}_q}{\sqrt{\pi}},\quad \varepsilon_q = \sqrt{2}\,x_q.
\label{eq:gh_1d}
\end{equation}
The 1D rule~\eqref{eq:gh_1d} is exact for univariate polynomials in $\varepsilon$ of degree at most $2Q-1$, an attractive accuracy guarantee for smooth integrands.

\paragraph{Worked example: Brock--Mirman Euler expectation.}  In the Brock--Mirman model, the conditional expectation in the Euler equation~\eqref{eq:bm_euler} is one-dimensional:
\begin{equation}
\mathbb{E}_t\!\left[
\frac{\alpha\, z_{t+1}\,K_{t+1}^{\alpha-1}}{C_{t+1}}
\right],
\qquad
z_{t+1}=z_t^{\varrho}\exp(\sigma_z\varepsilon_{t+1}),
\quad
\varepsilon_{t+1}\sim\mathcal{N}(0,1).
\end{equation}
Replacing path averaging by Gauss--Hermite with $Q=5$ nodes turns the residual~\eqref{eq:bm_euler} into a deterministic function of $(K_t, z_t)$ and the network parameters, eliminating the shock noise inside the loss while costing only five extra forward passes per state.  This is the simplest concrete instance of the general construction below and a useful sanity check: at $Q=5$ the quadrature error is negligible relative to the training error in this benchmark, so a well-trained network should match the analytical Brock--Mirman policy to numerical tolerance.

For shock dimension $d > 1$, the multivariate expectation is computed via a tensor product of one-dimensional rules.  Each multi-dimensional quadrature node is a $d$-tuple $(\varepsilon_1^{q_1}, \ldots, \varepsilon_d^{q_d})$ with weight $\prod_{j=1}^{d} w_{q_j}$.  The total number of quadrature nodes grows as $Q^{d}$, which becomes prohibitive once $d$ is large \citep[see][Section~7.7, for a textbook treatment of product rules and their cost]{judd1998numerical}.  For $d \geq 10$, the monomial rule of \S\ref{sec:monomial_cubature} or \emph{sparse-grid} (Smolyak) alternatives can substantially reduce the computational cost: \citet{gerstner1998numerical} introduce the construction in numerical analysis, \citet{heiss2008likelihood} apply it to econometric likelihoods, and \citet{ECTA:ECTA1716} develop adaptive sparse grids tailored to dynamic-programming and equilibrium computations.  The polynomial-versus-exponential gap depends on smoothness assumptions and the effective dimension of the integrand: sparse grids attain polynomial rates for sufficiently regular functions; QMC achieves close-to-$\mathcal{O}(1/M)$ rates under randomized constructions and finite-variation / low-effective-dimension conditions \citep{sobol1967distribution, niederreiter1992random, owen1995randomly, caflisch1998monte, novak2008tractability}, but neither method is universally polynomial.

\subsection{Monomial (Stroud-3) Cubature with Linear Scaling}
\label{sec:monomial_cubature}

The exponential cost $Q^{d}$ of the previous subsection is the price paid for being exact in the highest-degree univariate polynomial along \emph{each} coordinate.  In equilibrium-residual applications, however, the integrand $h(\bm\varepsilon')$ obtained by composing the policy network with model primitives is typically only mildly nonlinear in the shock vector, and reproducing every cross-coordinate monomial of total degree up to $2Q-1$ is overkill.  Once one is willing to give up that high degree, dramatically cheaper rules become available.

The simplest such rule is the \emph{Stroud degree-3 monomial cubature} \citep[formula E$_n^{r^2}$ 3-1]{stroud1971approximate}, a cornerstone of the toolkit reviewed in \citet[\S 7.5]{judd1998numerical} and \citet[\S 5.3]{Maliar2014325} for large-scale dynamic models.  For a $d$-dimensional standard normal expectation it uses only $2d$ nodes, all on the principal axes:
\begin{equation}
\E{h(\bm\varepsilon')} \;\approx\; \frac{1}{2d} \sum_{k=1}^{d}\Bigl[\, h\bigl(+\sqrt{d}\,\bm{e}_k\bigr) \;+\; h\bigl(-\sqrt{d}\,\bm{e}_k\bigr)\,\Bigr],
\label{eq:stroud3}
\end{equation}
where $\bm{e}_k$ is the $k$-th standard basis vector in $\R^{d}$.  All weights are equal to $1/(2d)$, and the nodes lie at the common radius $\sqrt{d}$ from the origin (so they expand outward as the dimension grows, consistent with the fact that the bulk of a high-dimensional Gaussian sits in a thin spherical shell, recalling the dimension-of-volumes discussion in Section~\ref{sec:curse_of_dim}).  At $d=1$, the rule collapses to the two-point cubature $\{+1, -1\}$ with weights $\{\tfrac12, \tfrac12\}$, exact for polynomials in $\varepsilon$ up to degree three.

\paragraph{Why it works (and where it stops).}  Rule~\eqref{eq:stroud3} integrates every monomial in $\bm\varepsilon'$ of total degree at most three exactly: the first moments vanish by $\pm$-symmetry, the variances $\E{\varepsilon_i'^{\,2}}$ all evaluate to $1$ (each axis contributes $\tfrac{1}{2d} \cdot 2 \cdot d = 1$), the cross-moments $\E{\varepsilon_i' \varepsilon_j'}$ ($i \neq j$) vanish because each node has only one nonzero coordinate, and the third moments vanish again by symmetry.  Higher moments are not exact: $\E{\varepsilon_i'^{\,4}}$ is reproduced as $d$ rather than the truth $3$, so the rule introduces a controlled bias proportional to the integrand's fourth-order shock content.  In the IRBC class of Chapter~\ref{ch:irbc} this bias is empirically negligible at the Euler-equation tolerance one cares about ($\sim 10^{-3}$); see \citet{pichler2011} for a careful demonstration on a multi-country RBC and \citet[Table~5]{Maliar2014325} for a broader cost--accuracy benchmark.  When higher accuracy is required, Stroud's degree-5 monomial rule with $2d^2 + 1$ nodes still scales polynomially, not exponentially \citep[\S 7.5]{judd1998numerical}.

\paragraph{Cost comparison.}  At $Q=3$ nodes per dimension, the tensor-product rule of \S\ref{sec:gh_tensor_product} costs $3^{d}$ evaluations per state, while~\eqref{eq:stroud3} costs $2d$.  Table~\ref{tab:quadrature_costs} contrasts the two for the dimensionalities encountered later in the script.  The monomial rule is the default integration scheme behind the production DEQN code for IRBC-type models in Chapter~\ref{ch:irbc} once the shock dimension exceeds about five, dropped in by replacing a single quadrature kernel.

\begin{table}[ht]
\centering
\small
\begin{tabular}{r l r r l@{}}
\toprule
shock dim.\ $d$ & where it appears & tensor-product $3^{d}$ & monomial $2d$ & speed-up \\
\midrule
$1$  & Brock--Mirman (Ch.~\ref{ch:deqn})        & $3$              & $2$   & $\sim 1.5\times$\\
$3$  & 2-country IRBC (Ch.~\ref{ch:irbc})       & $27$             & $6$   & $\sim 5\times$\\
$6$  & 5-country IRBC                           & $729$            & $12$  & $\sim 60\times$\\
$11$ & 10-country IRBC                          & $177{,}147$      & $22$  & $\sim 8{,}000\times$\\
$21$ & 20-country IRBC                          & $\sim 10^{10}$   & $42$  & $\sim 2 \times 10^8\times$\\
$51$ & 50-country IRBC                          & $\sim 10^{24}$   & $102$ & $\sim 10^{22}\times$\\
\bottomrule
\end{tabular}
\caption{Quadrature cost per residual evaluation, tensor-product Gauss--Hermite ($Q=3$) versus the Stroud-3 monomial rule of equation~\eqref{eq:stroud3}, as a function of the shock dimension $d$.  The tensor-product cost grows exponentially in $d$; the monomial cost grows linearly.  Brock--Mirman ($d=1$) is the simplest case where the two rules agree to within a factor of $1.5$; the gap explodes once the shock vector is multi-dimensional.}
\label{tab:quadrature_costs}
\end{table}

\subsection{Inverse-CDF Transform and Quasi--Monte Carlo}
\label{sec:qmc_cdf}

For very high-dimensional expectations ($d \geq 10$), even sparse-grid rules can become expensive, and, when the integrand contains material fourth-order shock content, the bias of the monomial rule~\eqref{eq:stroud3} starts to dominate.  An effective alternative is Quasi-Monte Carlo (QMC) integration combined with the inverse-CDF transform \citep[see][for a textbook treatment]{judd1998numerical}.  This method maps the unbounded domain of normal shocks $\R^{d}$ to the unit hypercube $[0,1]^{d}$ using the inverse normal CDF:
\begin{equation}
\E{f(\bm{\varepsilon})} \approx \frac{1}{M} \sum_{m=1}^{M} f\bigl(\Phi^{-1}(\bm{u}_m)\bigr),
\end{equation}
where $\bm{\varepsilon} \in \R^{d}$ is the vector of i.i.d.\ standard normal shocks, $\Phi^{-1}$ is the element-wise inverse standard normal CDF, and $\bm{u}_m$ are points from a low-discrepancy sequence (e.g., Sobol) on $[0,1]^{d}$.  For sufficiently smooth integrands, randomized QMC often achieves error rates close to $\mathcal{O}(1/M)$ up to logarithmic factors, compared with the $\mathcal{O}(1/\sqrt{M})$ rate of standard Monte Carlo.  This makes QMC highly effective in many high-dimensional applications, including the multi-country IRBC of Chapter~\ref{ch:irbc}.  The companion notebook \texttt{07\_Genz\_Approximation\_and\_Loss\_Functions.ipynb} explores QMC integration using the Genz test functions, a standard suite of integrands for benchmarking high-dimensional quadrature methods.

\section{Automatic Differentiation for DEQNs}
\label{sec:autodiff}

The quadrature rules of the previous section answer the question ``how do we evaluate the expectation in the Euler equation?''  This section answers the companion question that every DEQN application faces: ``how do we evaluate the \emph{derivatives} in the Euler equation, the marginal utility $u'(C_t)$, the marginal product $\partial Y / \partial K$, and the derivative of the value function $V'(K)$?''  For the textbook Brock--Mirman model of \S\ref{sec:bm} we derived all of these by hand.  But every variation of the model, whether CRRA utility in place of log, CES production in place of Cobb--Douglas, a capital adjustment cost, or a borrowing constraint, would force us to redo pages of algebra before writing the loss.  Automatic differentiation (AD) removes that tax: the user writes only the \emph{period payoff} $\Pi$ of the model once, and the first-order conditions, the envelope theorem, and all higher-order sensitivities are computed exactly from the code that evaluates $\Pi$.  The result is not a numerical approximation of the Euler equation: it is the textbook Euler equation, produced by a different mechanism.

\subsection{Three ways to take a derivative}
\label{sec:ad_three_methods}

There are three generic ways to compute a derivative of a programmable function $f:\R^n \to \R$ at a point $x_0$: symbolic, finite differences, and autodiff.  Each has a distinct profile.

\paragraph{Symbolic differentiation} uses a computer algebra system (Mathematica, SymPy) to manipulate the algebraic expression for $f$ into an algebraic expression for $f'$.  Its output is exact and human-readable, and for clean analytic formulas it is ideal.  Its two weaknesses are (i) \emph{expression swell}: the chain rule applied to a deeply composed expression can produce a result with many more terms than the original, sometimes enough to defeat further symbolic manipulation; and (ii) it cannot handle arbitrary code with loops and conditional branches.

\paragraph{Finite differences (FD)} approximates the derivative by the definition itself.  For the common central-difference variant,
\begin{equation}
\widehat{f'}(x_0) \;=\; \frac{f(x_0 + h) - f(x_0 - h)}{2\,h}
\;=\; f'(x_0) \;+\; \tfrac{1}{6}\,f'''(x_0)\,h^2 \;+\; \mathcal{O}(h^4).
\label{eq:fd_central}
\end{equation}
The geometric content of this formula is the mundane fact that the slope of a smooth curve at a point can be approximated by the slope of a nearby \emph{secant line}.  Figure~\ref{fig:fd_geometry} makes this concrete: the true tangent at $x_0$ (red) is approximated by the secant connecting $(x_0 - h, f(x_0-h))$ to $(x_0+h, f(x_0+h))$ (blue dashed).  As $h$ shrinks the secant rotates toward the tangent, and its slope converges to $f'(x_0)$; taking $h$ too small, however, forces the numerator $f(x_0+h) - f(x_0-h)$ to become the difference of two nearly equal floating-point numbers, at which point catastrophic cancellation dominates.

\begin{figure}[ht]
\centering
\begin{tikzpicture}
\begin{axis}[
    width=0.78\linewidth, height=6.2cm,
    axis lines=left,
    xlabel={$x$}, ylabel={$f(x)$},
    xmin=-0.1, xmax=2.6, ymin=-0.3, ymax=6.0,
    xtick={0.6, 1.2, 1.8}, xticklabels={$x_0 - h$, $x_0$, $x_0 + h$},
    ytick=\empty,
    label style={font=\small}, tick label style={font=\small},
    samples=100, thick,
    legend style={at={(0.98,0.98)}, anchor=north east, font=\small,
                  fill=white, draw=black!30},
]
\addplot[domain=-0.1:2.6, thick, black!75] {(x - 0.3)^2 + 0.5};
\addlegendentry{$f(x)$}
\addplot[domain=0.5:2.1, thick, red!80!black] {1.8*(x - 1.2) + 1.31};
\addlegendentry{tangent at $x_0$ (true $f'(x_0)$)}
\addplot[domain=0.2:2.4, dashed, thick, blue!70!black] {(2.75 - 0.59)/1.2 * (x - 0.6) + 0.59};
\addlegendentry{secant through $(x_0 \pm h, f(x_0 \pm h))$}
\fill[blue!70!black] (axis cs:0.6,0.59) circle (2pt);
\fill[blue!70!black] (axis cs:1.8,2.75) circle (2pt);
\fill[red!80!black] (axis cs:1.2,1.31) circle (2pt);
\draw[dotted, black!50] (axis cs:0.6,0.59) -- (axis cs:0.6,0);
\draw[dotted, black!50] (axis cs:1.2,1.31) -- (axis cs:1.2,0);
\draw[dotted, black!50] (axis cs:1.8,2.75) -- (axis cs:1.8,0);
\draw[decorate, decoration={brace, amplitude=5pt, mirror}, black!60]
    (axis cs:0.6,-0.12) -- (axis cs:1.8,-0.12) node[midway, below=6pt, font=\small] {$2h$};
\end{axis}
\end{tikzpicture}
\caption{Geometric meaning of a central finite difference.  The derivative $f'(x_0)$ is the slope of the tangent to $f$ at $x_0$ (red).  Finite differences replace this slope by the slope of the secant through the two nearby points $(x_0 - h, f(x_0-h))$ and $(x_0 + h, f(x_0+h))$ (blue, dashed).  For small $h$, the secant approaches the tangent at rate $\mathcal{O}(h^2)$; for $h$ too small, the subtraction in the numerator becomes unreliable in finite-precision arithmetic and the approximation degrades.  Figure~\ref{fig:fd_ucurve} plots the resulting error against $h$.}
\label{fig:fd_geometry}
\end{figure}
\begin{figure}[ht]
\centering
\begin{tikzpicture}
\begin{axis}[
    width=0.78\linewidth, height=6.2cm,
    xlabel={$\log_{10}(h)$},
    ylabel={$\log_{10}\bigl|\widehat{f'}(x_0) - f'(x_0)\bigr|$},
    xmin=-16, xmax=0, ymin=-14, ymax=4,
    grid=both, minor grid style={draw=black!8}, major grid style={draw=black!20},
    samples=200, thick,
    legend style={at={(0.5,1.02)}, anchor=south, legend columns=-1, font=\small,
                  fill=white, draw=black!30,
                  /tikz/every even column/.append style={column sep=0.5cm}},
    label style={font=\small}, tick label style={font=\small},
]
\addplot[domain=-16:0, thick, blue!65!black]   {log10(10^(-16)/10^x + 10^(2*x)/6)};
\addplot[dashed, red!75!black, thick, domain=-16:0] {log10(10^(2*x)/6)};
\addplot[dashed, green!55!black, thick, domain=-16:0] {log10(10^(-16)/10^x)};
\fill[red!80!black] (axis cs:-5.33,-10.8) circle (2pt);
\node[font=\footnotesize, red!80!black, anchor=west]
    at (axis cs:-5.0,-10.8) {optimal $h^\star \approx 5\cdot 10^{-6}$};
\legend{total error, truncation $\sim h^2$, roundoff $\sim \epsilon/h$}
\end{axis}
\end{tikzpicture}
\caption{The classic U-curve of central finite differences, here for $f(x) = \mathrm{e}^x$ at $x_0 = 1$.  Truncation error (red, dashed) falls as $h^2$; roundoff error (green, dashed) grows as $\epsilon / h$.  Their sum (blue) has a minimum at $h^\star \approx 5 \cdot 10^{-6}$, where the achievable precision is around $10^{-11}$, roughly five digits worse than the input precision.  Autodiff, by contrast, attains machine precision at no step-size tuning.}
\label{fig:fd_ucurve}
\end{figure}
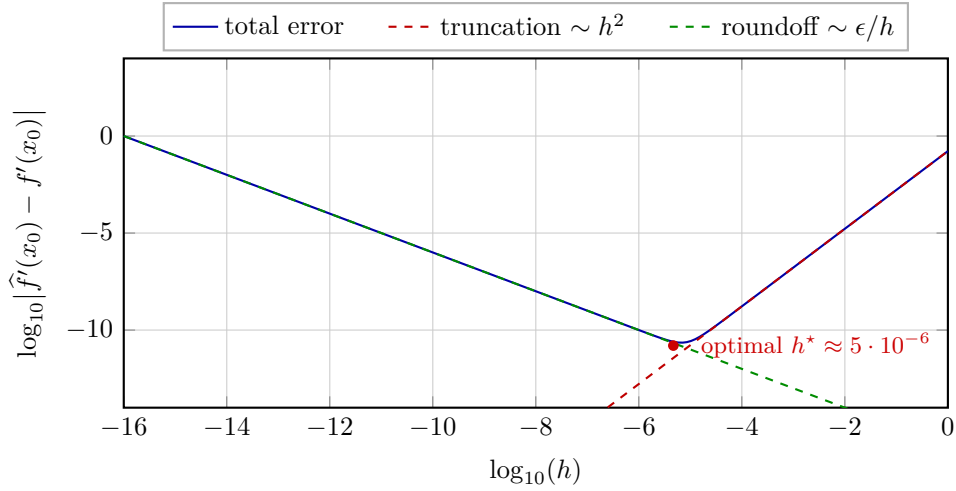

Two sources of error compete as $h$ changes.  The \emph{truncation error} $\tfrac{1}{6}f'''(x_0)\,h^2$ shrinks as $h$ decreases; the \emph{rounding error} associated with computing the near-zero subtraction $f(x_0+h)-f(x_0-h)$ in floating-point arithmetic grows as $\epsilon/h$, where $\epsilon \approx 2.2\cdot 10^{-16}$ is machine precision in double precision.  The total error is minimized at the step size that balances the two,
\begin{equation}
h^\star \sim \epsilon^{1/3},\qquad\text{minimum error}\quad \sim \epsilon^{2/3} \approx 10^{-11}.
\end{equation}
The practical consequence is that central FD loses roughly \emph{five digits of precision} relative to the input precision.  For second derivatives the loss is worse: Hessians obtained by FD are usually unusable for precision-critical work.  FD has one undeniable advantage: it treats $f$ as a black box, requiring only the ability to evaluate $f$.  For debugging or for a one-off comparative-statics check it is fine.  For \emph{training} a DEQN, where the gradient is used millions of times, it is badly suboptimal.  Figure~\ref{fig:fd_ucurve} makes the trade-off visible.

\paragraph{Automatic differentiation} exploits the fact that every piece of Python (or any other imperative language) that evaluates a numerical function is, at run-time, a composition of \emph{elementary operations} ($+$, $-$, $\times$, $/$, $\exp$, $\log$, $\sin$, $\cos$, $\ldots$) whose derivatives are known in closed form.  The chain rule composes these local derivatives into the derivative of the whole program, exactly.  The cost is a small multiple of one function evaluation, typically $2$--$4\times$ in reverse mode, and is \emph{independent of the number of inputs}.  There is no step-size $h$ to tune, and the result is exact to machine precision.  For ML-scale problems with $10^6$ parameters and a scalar loss, AD is the \emph{only} feasible method.  The classic computer-science references are \citet{baydin2018automatic} and \citet{margossian2019review}; both motivate AD from machine-learning workloads and lay out the forward/reverse-mode taxonomy used below.

\paragraph{Where do local derivatives come from?  A tiny built-in table.}  It is worth pausing on a question that beginners reasonably ask: \emph{``How does the computer know that the derivative of $\sin(x)$ is $\cos(x)$?''}  The answer is deflating but honest: it does not compute this from first principles.  Every autodiff framework ships with a short, hand-written table of derivatives for the elementary operations it supports, a few dozen entries such as those in Table~\ref{tab:ad_primitives}.  When your program executes $y = \sin(x)$, the framework looks up the rule ``gradient of $\sin$ is $\cos$'' in that table and records $\cos(x)$ on the edge of the computational graph.  The chain rule then composes those table entries.  \emph{Autodiff is not a theorem prover}; it is a library of primitive rules plus an automatic application of the chain rule.  This explains both its power (every program expressible in terms of these primitives is differentiable) and its limits (genuinely new functions require extending the table).

\begin{table}[ht]
\centering
\small
\renewcommand{\arraystretch}{1.15}
\begin{tabular}{l l l}
\toprule
\textbf{Primitive} & \textbf{Local derivative (built in)} & \textbf{Where to find it} \\
\midrule
$y = x + z$ & $\partial y/\partial x = 1,\ \partial y/\partial z = 1$ & \texttt{tf.math.add} \\
$y = x \cdot z$ & $\partial y/\partial x = z,\ \partial y/\partial z = x$ & \texttt{tf.math.multiply} \\
$y = x^{\alpha}$ & $\partial y/\partial x = \alpha\,x^{\alpha-1}$ & \texttt{tf.math.pow} \\
$y = \exp(x)$ & $\partial y/\partial x = \exp(x)$ & \texttt{tf.math.exp} \\
$y = \log(x)$ & $\partial y/\partial x = 1/x$ & \texttt{tf.math.log} \\
$y = \sin(x)$ & $\partial y/\partial x = \cos(x)$ & \texttt{tf.math.sin} \\
$y = \cos(x)$ & $\partial y/\partial x = -\sin(x)$ & \texttt{tf.math.cos} \\
$y = \mathrm{ReLU}(x)$ & $\partial y/\partial x = \mathbf{1}\{x > 0\}$ & \texttt{tf.nn.relu} \\
\bottomrule
\end{tabular}
\caption{A representative slice of the primitive-derivative table that every autodiff framework ships with.  TensorFlow, PyTorch, and JAX each include a few hundred such entries: one per elementary operation they support.  User code composes these primitives; the chain rule composes their derivatives.  When an economist writes the Brock--Mirman period payoff $\Pi(K_t, K_{t+1}) = \log(K_t^{\alpha} + (1-\delta)K_t - K_{t+1})$, only entries from this table are invoked; no calculus is performed at runtime.}
\label{tab:ad_primitives}
\end{table}

This explanation also makes the \emph{limits} of AD concrete.  Operations absent from the table (a black-box solver call, a non-differentiable step like sorting or \texttt{argmax}) require either a custom rule (via \texttt{@tf.custom\_gradient}) or a redesign of the code.  The pitfalls in \S\ref{sec:ad_pitfalls} are essentially cases where either the table's entry is degenerate at a point (kinks, $\partial|x|/\partial x$ at $x=0$) or the composed gradient is numerically unstable even though each local rule is correct.

\subsection{Computational graph; forward and reverse modes}
\label{sec:ad_modes}

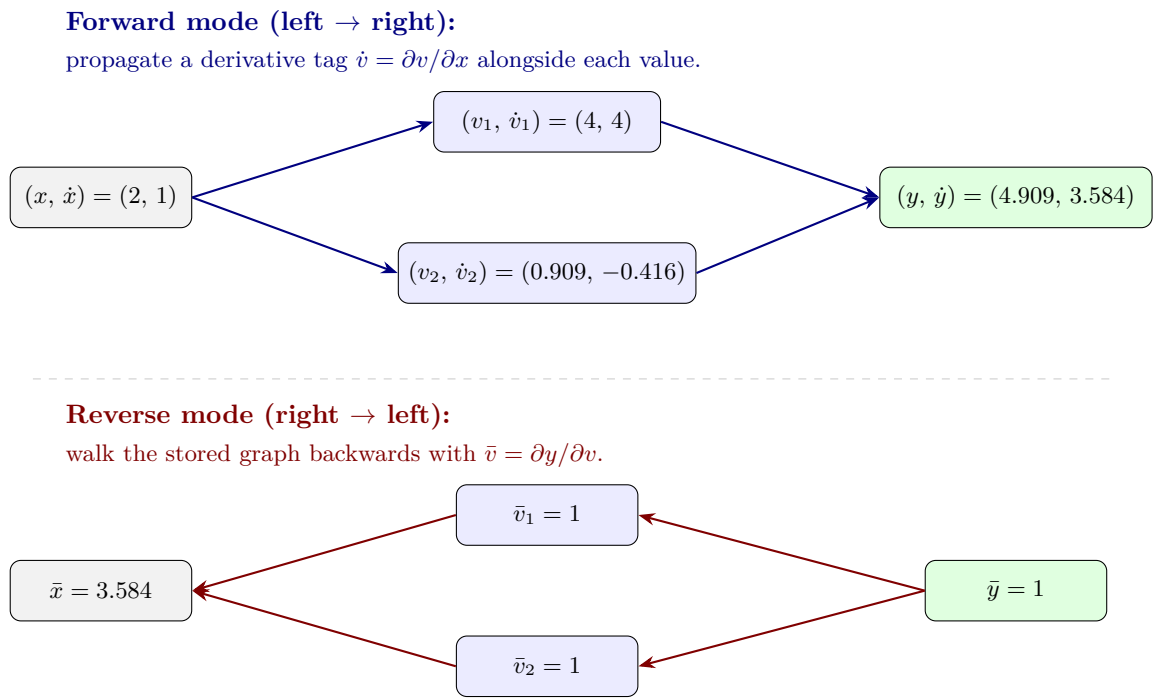
\begin{figure}[ht]
\centering
\begin{tikzpicture}[every node/.style={font=\footnotesize}]
\node[anchor=west, font=\small\bfseries, blue!50!black]
    at (0, 4.6) {Forward mode (left $\to$ right):};
\node[anchor=west, font=\footnotesize, blue!50!black]
    at (0, 4.1) {propagate a derivative tag $\dot v = \partial v / \partial x$ alongside each value.};
\node[draw, rounded corners, fill=black!5, minimum width=2.4cm, minimum height=8mm]
    (fx)  at (0.6, 2.3)  {$(x,\,\dot x) = (2,\,1)$};
\node[draw, rounded corners, fill=blue!8, minimum width=3.0cm, minimum height=8mm]
    (fv1) at (6.5, 3.3) {$(v_1,\,\dot v_1) = (4,\,4)$};
\node[draw, rounded corners, fill=blue!8, minimum width=3.8cm, minimum height=8mm]
    (fv2) at (6.5, 1.3) {$(v_2,\,\dot v_2) = (0.909,\,-0.416)$};
\node[draw, rounded corners, fill=green!12, minimum width=3.6cm, minimum height=8mm]
    (fy)  at (12.7, 2.3)  {$(y,\,\dot y) = (4.909,\,3.584)$};
\draw[-{Stealth[length=2.2mm]}, blue!50!black, thick] (fx.east) -- (fv1.west);
\draw[-{Stealth[length=2.2mm]}, blue!50!black, thick] (fx.east) -- (fv2.west);
\draw[-{Stealth[length=2.2mm]}, blue!50!black, thick] (fv1.east) -- (fy.west);
\draw[-{Stealth[length=2.2mm]}, blue!50!black, thick] (fv2.east) -- (fy.west);

\draw[black!20, dashed] (-0.3, -0.1) -- (14.0, -0.1);

\node[anchor=west, font=\small\bfseries, red!50!black]
    at (0, -0.6) {Reverse mode (right $\to$ left):};
\node[anchor=west, font=\footnotesize, red!50!black]
    at (0, -1.1) {walk the stored graph backwards with $\bar v = \partial y / \partial v$.};
\node[draw, rounded corners, fill=black!5, minimum width=2.4cm, minimum height=8mm]
    (rx)  at (0.6, -2.9)  {$\bar x = 3.584$};
\node[draw, rounded corners, fill=blue!8, minimum width=2.4cm, minimum height=8mm]
    (rv1) at (6.5, -1.9) {$\bar v_1 = 1$};
\node[draw, rounded corners, fill=blue!8, minimum width=2.4cm, minimum height=8mm]
    (rv2) at (6.5, -3.9) {$\bar v_2 = 1$};
\node[draw, rounded corners, fill=green!12, minimum width=2.4cm, minimum height=8mm]
    (ry)  at (12.7, -2.9)  {$\bar y = 1$};
\draw[-{Stealth[length=2.2mm]}, red!50!black, thick] (ry.west) -- (rv1.east);
\draw[-{Stealth[length=2.2mm]}, red!50!black, thick] (ry.west) -- (rv2.east);
\draw[-{Stealth[length=2.2mm]}, red!50!black, thick] (rv1.west) -- (rx.east);
\draw[-{Stealth[length=2.2mm]}, red!50!black, thick] (rv2.west) -- (rx.east);
\end{tikzpicture}
\caption{The two modes of autodiff on $y = x^2 + \sin(x)$ at $x = 2$.  \emph{Top:} forward mode carries a derivative tag $\dot v = \partial v / \partial x$ alongside each value and reads $\dot y = dy/dx$ at the output.  \emph{Bottom:} reverse mode evaluates $f$ forward, stores the graph, then walks backwards with $\bar v = \partial y / \partial v$ and reads $\bar x = dy/dx$ at the input.  Both deliver $3.584$, equal to $f'(2) = 2 \cdot 2 + \cos(2)$ at machine precision.  Forward mode scales linearly with the number of \emph{inputs}; reverse mode scales linearly with the number of \emph{outputs}, which is why it wins for DEQN training (scalar loss, many parameters).}
\label{fig:ad_graph}
\end{figure}

Every numerical function, once evaluated, corresponds to a directed acyclic \emph{computational graph} whose nodes are elementary operations and whose edges carry intermediate values.  Take the toy example $y = f(x) = x^2 + \sin(x)$ evaluated at $x_0=2$:
\[
x \;\to\; v_1 = x^2 \;\to\; y = v_1 + v_2,\qquad
x \;\to\; v_2 = \sin(x) \;\to\; y = v_1 + v_2.
\]
The values along the graph are $v_1=4$, $v_2=\sin(2) \approx 0.909$, $y \approx 4.909$.  The \emph{edges} carry local derivatives: $\partial v_1/\partial x = 2x = 4$, $\partial v_2/\partial x = \cos(x) \approx -0.416$, $\partial y/\partial v_1 = \partial y/\partial v_2 = 1$.  AD combines the edge derivatives by the chain rule in one of two traversal orders.  Figure~\ref{fig:ad_graph} shows the graph together with both traversals.

\paragraph{Forward mode} carries a derivative tag $\dot v$ next to every value $v$ and updates both in a single forward pass through the graph.  One seeds $\dot x = 1$ at the input; by the chain rule, $\dot v_1 = 4$, $\dot v_2 = -0.416$, and $\dot y = \dot v_1 + \dot v_2 = 3.584$, which is exactly $f'(2) = 2x + \cos x$ evaluated at $x=2$.  The cost is one extra float per variable and one extra pass through the graph.  For a function with $n$ inputs and $m$ outputs, the cost of the full Jacobian is $\mathcal{O}(n)$ forward passes; forward mode is therefore attractive when \emph{the number of inputs is small}.

\paragraph{Reverse mode} first evaluates $f$ forward, storing the computational graph and all intermediate values.  It then walks the graph backwards, carrying a sensitivity $\bar v = \partial y / \partial v$ along each edge.  Seeded with $\bar y = 1$ at the output, the backward pass accumulates
\[
\bar v_i \;=\; \sum_{j:\, v_i \to v_j} \bar v_j \cdot \frac{\partial v_j}{\partial v_i},
\]
yielding at the input node $\bar x = \bar v_1 \cdot 4 + \bar v_2 \cdot (-0.416) = 1\cdot 4 + 1\cdot(-0.416) = 3.584$.  The computational cost is $2$--$4\times$ a single forward pass, and, crucially, \emph{one reverse pass produces the full gradient with respect to all inputs simultaneously}.  Reverse mode is what \texttt{tf.GradientTape}, \texttt{torch.autograd}, and \texttt{jax.grad} use by default.  The price paid is memory: the entire forward graph must be stored for the backward pass.  For extremely long simulations this can be a binding constraint; we return to the point in the pitfalls below.

\paragraph{Which mode for DEQNs.}  The DEQN loss is a \emph{scalar} mean-squared Euler residual, computed from a network with $n \sim 10^3$ to $10^5$ parameters.  Reverse mode produces the full gradient with respect to all network parameters in $2$--$4\times$ one forward pass, a speedup of $n/4$ over forward mode.  This is the same reason reverse mode is the workhorse of deep-learning training: it exactly fits the ``few outputs, many inputs'' regime.

\subsection{The autodiff Euler residual}
\label{sec:ad_euler}

We now apply autodiff to the very object we spent \S\ref{sec:bm} deriving by hand: the Euler equation of the Brock--Mirman planner.  Define the \emph{period payoff} as a function of the current state and the current choice:
\begin{equation}
\Pi(K_{\text{in}}, K_{\text{out}}, z_{\text{in}}) \;=\; u\!\left(\,z_{\text{in}} K_{\text{in}}^{\alpha} + (1-\delta)K_{\text{in}} \;-\; K_{\text{out}}\,\right),
\label{eq:ad_pi}
\end{equation}
where the three argument \emph{slots} are the state $K_{\text{in}}$, the choice $K_{\text{out}}$, and the exogenous shock $z_{\text{in}}$ (the same slot names the code listing below uses); applying $\Pi$ at a date $t$ means plugging $K_t$ into slot~1, $K_{t+1}$ into slot~2, and $z_t$ into slot~3.  This is the only place the model enters.  Change $u$, $Y$, or $\delta$ and nothing else in what follows needs to move.  The Bellman equation of \S\ref{sec:bm} then reads
\begin{equation}
V(K_t, z_t) \;=\; \max_{K_{t+1}}\;\Pi(K_t, K_{t+1}, z_t) + \beta\,\mathbb{E}\!\left[V(K_{t+1}, z_{t+1})\,\big|\,z_t\right].
\end{equation}

\paragraph{Notation: what do $\partial_1\Pi$ and $\partial_2\Pi$ mean?}  Throughout this section the subscript names the \emph{slot} of $\Pi$ being differentiated, not a time index.  With the three-slot definition $\Pi(K_{\text{in}},K_{\text{out}},z_{\text{in}})$ of~\eqref{eq:ad_pi}, we write
\[
\partial_1\Pi \;\equiv\; \frac{\partial\Pi}{\partial K_{\text{in}}}, \qquad \partial_2\Pi \;\equiv\; \frac{\partial\Pi}{\partial K_{\text{out}}}.
\]
The first slot $K_{\text{in}}$ is the \emph{state}, the second slot $K_{\text{out}}$ is the \emph{choice}, the third slot $z_{\text{in}}$ is an exogenous parameter and is not differentiated.  An expression like $\partial_2\Pi(K_t,K_{t+1},z_t)$ therefore denotes the derivative of $\Pi$ in its second slot, evaluated with $K_t$ plugged into slot~1, $K_{t+1}$ into slot~2, and $z_t$ into slot~3, so it equals $\partial\Pi/\partial K_{t+1}$.  The expression $\partial_1\Pi(K_{t+1},K_{t+2},z_{t+1})$ denotes the derivative in the \emph{first} slot, evaluated at the time-$(t+1)$ state pair, so it also equals $\partial\Pi/\partial K_{t+1}$ but for a different physical reason (because $K_{t+1}$ now sits in the state slot of the period-$t+1$ problem).  This overloading is the whole point: the same partial expression handles the FOC in one period and the envelope term in the next.

Two facts, both derived in \S\ref{sec:bm}, are now worth restating in terms of $\Pi$:
\begin{description}
\item[FOC w.r.t.\ the choice $K_{t+1}$:]
\[
\partial_2 \Pi(K_t,K_{t+1},z_t)
\;+\;
\beta\,\mathbb{E}_{z_{t+1}\mid z_t}[\,V'(K_{t+1},z_{t+1})\,]
\;=\;0.
\]
\item[Envelope at the optimum, evaluated at the state $K_t$:]
\[
V'(K_t, z_t)
\;=\;
\partial_1 \Pi(K_t, g(K_t, z_t), z_t),
\]
where $g$ is the optimal policy.
\end{description}
Substituting the envelope (evaluated one period ahead, so that $K_{t+1}$ sits in the \emph{state} slot) into the FOC, the familiar Euler equation becomes
\begin{equation}
\boxed{
\begin{aligned}
0
= {}& \partial_2\,\Pi(K_t, K_{t+1}, z_t)\\
&+ \beta\,\mathbb{E}_{z_{t+1}\mid z_t}\!\left[
\partial_1\,\Pi(K_{t+1}, K_{t+2}, z_{t+1})
\right].
\end{aligned}}
\label{eq:ad_euler}
\end{equation}
Here $K_{t+2}=g(K_{t+1},z_{t+1})$.  Both terms in~\eqref{eq:ad_euler} are derivatives with respect to the same physical variable $K_{t+1}$, but one treats $K_{t+1}$ as the \emph{choice} of period~$t$ (slot~2) and the other treats $K_{t+1}$ as the \emph{state} of period~$t+1$ (slot~1).  Every term is therefore a \emph{partial derivative of the same function $\Pi$}.  Neither $u'$ nor the marginal product of capital, nor the envelope formula $V'(K) = u'(C)(\alpha K^{\alpha-1} + 1 - \delta)$ need to be written out by hand.  \texttt{tf.GradientTape} records the forward evaluation of $\Pi$ and produces both partials on demand; the expectation is approximated by any of the quadrature rules of \S\ref{sec:quadrature_rules}.  The code is shorter than the pen-and-paper counterpart and, more importantly, entirely \emph{model-agnostic}: swapping log for CRRA utility means editing the body of \texttt{Pi} and nothing else.

\begin{lstlisting}[caption={Autodiff Euler residual for Brock--Mirman.  The function \texttt{Pi} is the only model-specific code; the rest is generic.  A full implementation lives in the autodiff chapter's code folder, notebook \protect\tpath{02_Brock_Mirman_AutoDiff_DEQN.ipynb}.}, label=lst:autodiff_euler]
def Pi(K_in, K_out, z_in):
    Y = z_in * K_in ** alpha
    C = Y + (1.0 - delta) * K_in - K_out
    return tf.math.log(C)

def euler_residual(K_t, z_t, K_tp1, K_tp2, z_tp1):
    # FOC term:       d Pi / d K_{t+1} at (K_t, K_{t+1}, z_t)
    with tf.GradientTape() as t1:
        t1.watch(K_tp1)
        pi_t = Pi(K_t, K_tp1, z_t)
    d2 = t1.gradient(pi_t, K_tp1)
    # Envelope term:  d Pi / d K_t at (K_{t+1}, K_{t+2}, z_{t+1})
    with tf.GradientTape() as t2:
        t2.watch(K_tp1)
        pi_tp1 = Pi(K_tp1, K_tp2, z_tp1)
    d1 = t2.gradient(pi_tp1, K_tp1)
    return d2 + beta * d1
\end{lstlisting}

\paragraph{Cross-checking the autodiff loss.}  An important pedagogical point is that the autodiff residual~\eqref{eq:ad_euler} and the hand-derived residual of \S\ref{sec:bm} are the same mathematical object.  The companion autodiff notebooks implement both expressions on the same network and report the maximum absolute difference, which in our (seeded) runs is of order $10^{-6}$--$10^{-7}$ in the deterministic case and $10^{-6}$--$10^{-5}$ in the stochastic case with Gauss--Hermite quadrature (float32 arithmetic; float64 tightens this by roughly seven orders of magnitude).  In every case the residual is consistent with finite-precision arithmetic, graph ordering, and quadrature accumulation rather than with any difference in the underlying mathematics.  Under full depreciation ($\delta = 1$) the trained policy from the autodiff loss matches the analytical closed-form $K_{t+1} = \alpha\beta K_t^{\alpha}$ (respectively $\alpha\beta z_t K_t^{\alpha}$) to mean relative error $\sim 10^{-4}$ in the deterministic case and $\sim 10^{-3}$ on a coarse classroom grid in the stochastic case (the residual rises with the quadrature footprint), confirming that minimizing the autodiff residual recovers the true policy when one is available.

\subsection{Pitfalls of autodiff}
\label{sec:ad_pitfalls}

AD is exact and cheap but it is not magical.  Five pitfalls are worth naming; each has a concrete workaround.

\paragraph{Non-differentiable kinks.}  Operations such as $\max(0,x)$ (ReLU), $|x|$, $\min$, $\max$, \texttt{argmax}, $\mathrm{sort}$, and indicators are non-differentiable at isolated points.  Frameworks return a \emph{subgradient} at the kink -- in TensorFlow and PyTorch, $\partial \mathrm{ReLU}/\partial x\,|_{x=0} = 0$ by convention.  If the loss repeatedly lands on such a kink, SGD can receive an uninformative gradient and stall.  The cure is to either \emph{smooth} the kink ($\mathrm{softplus}$ for ReLU, $\sqrt{x^2 + \varepsilon^2}$ for $|x|$, log-sum-exp for $\max$) or, in complementarity problems, to replace the non-smooth indicator by a Fischer--Burmeister residual (Chapter~\ref{ch:irbc}, \S\ref{sec:irbc_fischer_burmeister}).

\paragraph{Boundary singularities.}  The Brock--Mirman loss contains $\log C_t$ and hence $1/C_t$.  If the network's output happens to drive $C_t$ close to zero during training, AD dutifully returns a very large gradient; SGD then takes a very large step and the network explodes or produces \texttt{NaN}.  Two cures of different quality are in wide use:
\begin{itemize}[itemsep=2pt]
\item \emph{Reparameterize so the domain is respected by architecture.}  In the Brock--Mirman notebooks, the network outputs the \emph{savings share} $s \in (0,1)$ through a sigmoid; this guarantees $C_t > 0$ \emph{and} $K_{t+1} > 0$ simultaneously, at every training iteration, with no penalty term.  This is the hard/soft constraint split of Figure~\ref{fig:hard_soft}.
\item \emph{Use numerically stable primitives.}  Prefer \tpath{tf.math.log1p(x)} to \tpath{log(1+x)}, \tpath{tf.math.softplus} to a hand-coded $\log(1 + e^x)$, and \tpath{tf.math.xlogy(a,b)} to $a\log b$ when the $a=0$ convention matters.  AD does not see the cancellations these functions implement; the human must.
\end{itemize}

\paragraph{Reverse-mode memory.}  Reverse mode stores the entire forward graph so it can walk it backwards.  Unrolling a $10{,}000$-step simulation and back-propagating through all of it has memory cost $\mathcal{O}(T \times \mathrm{dim}\,\text{state})$ and can exhaust GPU memory on non-trivial models.  Standard mitigations are (i) \emph{gradient checkpointing} (\texttt{tf.recompute\_grad}), which recomputes parts of the graph on the backward pass in exchange for memory, (ii) \emph{truncated back-propagation through time} for long recurrences, and (iii) path-averaging mini-batches of trajectory segments rather than full trajectories (notebook 02 of Day~2 already does this).

\paragraph{Loss of structural insight.}  AD returns a number, not an algebraic expression.  Useful structural facts -- the cancellation $u'(C_t) - \beta u'(C_{t+1}) R$ has the sign of the Euler wedge, the Euler residual is homogeneous of a known degree in productivity, and so on -- are invisible to AD and come from the hand derivation.  In practice one should retain the ability to derive the FOC on a toy version of the model, precisely as we did in \S\ref{sec:bm}: AD scales that derivation; it does not replace the understanding it provides.

\paragraph{\texttt{stop\_gradient} and the envelope theorem.}  A subtle point worth emphasizing.  The envelope identity $V'(K) = \partial_1 \Pi(K, g(K))$ holds at the optimum because the FOC kills the term $(-u'(C) + \beta V'(g(K)))\cdot g'(K)$ that otherwise contributes to the total derivative of the composed continuation value.  During training we are not yet at the optimum.  Two natural codings of the next-period term therefore give different numbers: one differentiates \emph{through} the policy $g$ and computes a total derivative of a rollout, while the other treats $K'=g(K)$ as a fixed choice and computes the partial derivative used in the Euler FOC.  The autodiff residual~\eqref{eq:ad_euler} uses the latter, which is what falls out of \texttt{t2.gradient(pi\_tp1, K\_tp1)} in Listing~\ref{lst:autodiff_euler}.  This is the correct residual for checking the Euler equation; differentiating through the policy is a different object.  The two codings agree as the FOC residual vanishes, but off the optimum they can disagree.

\paragraph{Companion notebooks.}  Three notebooks (in the autodiff chapter's code folder) put the above into practice, with a fourth provided as self-study:
\begin{itemize}[itemsep=2pt]
\item \texttt{01\_AutoDiff\_Analytical\_Examples}: warm-up ($y=x^2 + \sin x$), the FD U-curve regenerated numerically, CRRA utility's first and second derivative, Cobb--Douglas production's 2-D gradient field, the capital adjustment cost $\Gamma(K,K')$ with AD vs hand partials side by side, and the Hessian of Cobb--Douglas via a nested \texttt{GradientTape}.
\item \texttt{02\_Brock\_Mirman\_AutoDiff\_DEQN}: the loss of~\eqref{eq:ad_euler} implemented on the deterministic model of \S\ref{sec:bm}; cross-check against the hand residual at float32 tolerance; cross-check of the trained policy against $K_{t+1} = \alpha\beta K_t^\alpha$.
\item \texttt{03\_Brock\_Mirman\_Uncertainty\_AutoDiff\_DEQN}: the same, with AR(1) productivity and Gauss--Hermite quadrature; cross-check against the stochastic hand residual and the closed-form $K_{t+1} = \alpha\beta z_t K_t^\alpha$ in a full-depreciation side-experiment.
\item \texttt{04\_IRBC\_AutoDiff\_DEQN} (self-study): the same $\partial_2\Pi + \beta\,\mathbb{E}[\partial_1\Pi]$ template scaled up to the multi-country IRBC model of Chapter~\ref{ch:irbc} (per-country planner Lagrangian, Fischer--Burmeister irreversibility, tensor-product Gauss--Hermite), with a machine-precision cross-check against the Chapter~\ref{ch:irbc} hand-derived residual.
\end{itemize}

\section{Data Parallelization with MPI}

For very large-scale applications (e.g., $N \geq 50$ countries in the IRBC model of Chapter~\ref{ch:irbc}), training can be accelerated by distributing the gradient computation across multiple GPUs or compute nodes.  The standard approach uses synchronous data parallelism via MPI \texttt{Allreduce}; in the DEQN paper the corresponding implementation is built on \textit{Horovod} \citep{sergeev2018horovod}, which wraps \texttt{Allreduce} into a drop-in replacement for the training optimizer.

\begin{definitionbox}[Data-Parallel DEQN Training]
\begin{algorithmic}
\small
\STATE \textbf{Input:} $P$ workers, each with local copy of $\mathcal{N}_\rho$
\FOR{each training iteration}
    \STATE Each worker $p$ draws local mini-batch $\mathcal{B}_p$ from simulation
    \STATE Each worker computes local gradient: $\bm{g}_p = \nabla_\rho \ell(\mathcal{B}_p)$
    \STATE \textbf{MPI\_Allreduce:} $\bar{\bm{g}} = \frac{1}{P}\sum_{p=1}^P \bm{g}_p$
    \STATE Each worker updates: $\rho \leftarrow \rho - \eta\,\bar{\bm{g}}$
\ENDFOR
\end{algorithmic}
\end{definitionbox}

Since each worker processes an independent mini-batch and the \texttt{Allreduce} operation averages the gradients, the effective batch size scales linearly with the number of workers.  In practice, this yields near-linear speedup for moderate numbers of workers ($P \leq 32$), with communication overhead becoming significant only for very large clusters.  The key advantage for economics applications is that each worker can simulate its own trajectory of the economy, naturally exploring different regions of the state space and improving the diversity of the training distribution.

\section{Choice of Loss Kernel: Beyond Mean-Squared Loss}
\label{sec:loss_kernels}

Every DEQN derivation in this script ends with the same step: take the equilibrium residual $r(\bm{x})$, square it, average over a mini-batch, and minimize.  The squared average, mean-squared error (MSE), is the canonical default, but it is only one of many reasonable reductions of a residual vector to a scalar.  Different reductions emphasize different parts of the residual distribution and have visibly different convergence properties, even on the same model with the same network and data.  This section makes the trade-off concrete on the stochastic Brock--Mirman model with full depreciation, where the optimal savings rate is known in closed form ($s^\star = \alpha\beta$) so that the deviation between learned and optimal policy can be measured exactly.

\paragraph{Six kernels.}  We compare:

\begin{itemize}[itemsep=2pt]
\item \textbf{MSE}: $\ell = \tfrac1B \sum r^2$.  Quadratic everywhere; large residuals dominate the gradient.
\item \textbf{MAE}: $\ell = \tfrac1B \sum |r|$.  Robust to outliers, but the gradient has \emph{constant magnitude}, so the optimizer takes the same step size regardless of how close to zero a residual already is.  This is fine for learning a coarse fit, but it means MAE training plateaus at a finite ``noise floor'' that is several orders of magnitude above what MSE attains at the same training budget.
\item \textbf{Huber}: quadratic for $|r| \le \delta$, linear for $|r| > \delta$.  Smooth hybrid that combines MSE's fine convergence with MAE's tail robustness, but the transition at $|r|=\delta$ is a kink in the gradient, which can introduce small training-loop oscillations as residuals migrate across the threshold.
\item \textbf{Quantile pinball} at level $\tau$: $\ell = \tfrac1B \sum \max(\tau r, (\tau-1)r)$.  Targets the signed $\tau$-quantile of the residual distribution; high $\tau$ emphasizes positive residual tails, low $\tau$ emphasizes negative tails, and all observations contribute with asymmetric weights.
\item \textbf{CVaR-style} at confidence $\alpha$: $\ell = $ mean of the top $(1{-}\alpha)$ fraction of $|r|$.  Explicitly trains the worst-case states.
\item \textbf{LogCosh}: $\ell = \tfrac1B \sum \log\cosh(r)$.  This kernel is $C^\infty$ everywhere and combines MSE-like behavior near $r=0$ ($\log\cosh r \approx \tfrac12 r^2$) with MAE-like saturation in the tails ($\log\cosh r \approx |r| - \log 2$, gradient saturates at $\pm 1$).  Crucially, the transition between the two regimes is smooth: there is no kink at any threshold.
\end{itemize}

\begin{figure}[ht]
\centering
\includegraphics[width=\linewidth]{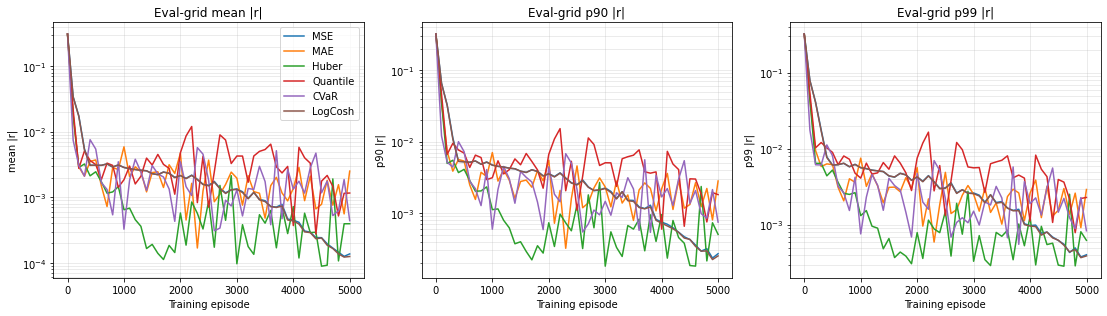}
\caption{Convergence of the relative Euler-error distribution under six different loss kernels on the stochastic Brock--Mirman model with full depreciation.  Same network (2$\times$32 swish, sigmoid head), same Adam optimizer, same CRN training stream; only the loss kernel changes.  Left: mean $|r|$.  Middle: $p_{90}\,|r|$.  Right: $p_{99}\,|r|$.  All on a log scale.  Closed-form benchmark: $s^\star = \alpha\beta$.}
\label{fig:loss_kernels}
\end{figure}

\paragraph{Convergence behavior.}  Figure~\ref{fig:loss_kernels} shows the mean, $p_{90}$, and $p_{99}$ of the absolute relative Euler error on a fixed evaluation grid as a function of the training episode, for each of the six kernels.  Three patterns are immediate.

\noindent\emph{First}, MAE clearly stalls at a noise floor an order of magnitude above the other kernels, exactly the constant-gradient pathology described above; the curve is monotone but flattens out and refuses to drop further.  \emph{Second}, MSE and log-cosh achieve essentially the same final mean residual ($\sim 1.4\!\times\!10^{-4}$), and both pull the $p_{99}$ down with comparable efficacy.  \emph{Third}, the tail-aware kernels (Huber, Quantile, CVaR) typically produce the narrowest spread between mean and max residual but pay a small constant cost in the bulk; the practical pay-off shows up in problems where rare-but-consequential states matter.

\paragraph{Why log-cosh tends to converge most smoothly.}  Of the six kernels, log-cosh is the one whose convergence curve is essentially monotone with no spikes.  This is not an accident: $\log\cosh(r)$ is a convenient smooth interpolant between $\tfrac12 r^2$ near zero and $|r|$ in the tails, with no kink at any threshold.  Near the optimum its gradient is $\tanh(r) \approx r$ (linear, like MSE), so it inherits MSE's fine-grained convergence.  Far from the optimum its gradient saturates at $\pm 1$ (like MAE), so an unusually large residual cannot dominate the mini-batch gradient.  In effect, log-cosh is what one would design if asked for ``MSE near zero and MAE in the tails, smoothly''.  For practical DEQN settings where the residual distribution is unknown in advance, log-cosh is therefore a useful default to compare against MSE rather than a mere robustness afterthought.

\paragraph{Why this is an \emph{economic} comparison.}  The relative Euler-equation residual already has a direct economic interpretation: a $1\%$ relative Euler error translates approximately into a $1\%$ per-period consumption error along the simulated path \citep[\S4.2]{judd1998numerical}.  The mean / $p_{90}$ / $p_{99}$ panels of Figure~\ref{fig:loss_kernels} therefore measure, in interpretable units, the average and tail consumption mistakes that each loss kernel leaves behind.  The point of the experiment is that \emph{the training-loss ranking is not the same as the ranking on this economic metric}: a kernel that achieves a slightly higher \emph{mean} residual but a much narrower \emph{tail} produces smaller worst-case consumption errors, which is what matters when rare states are economically consequential (occasionally binding constraints, fat-tailed shocks, tipping risks).  The training loss is the instrument; the relative Euler error along simulated paths is the criterion.  Choosing the kernel with that hierarchy in mind is part of the modeling decision.  The companion notebook pushes this one step further by collapsing the per-period error distribution into a single consumption-equivalent welfare loss against $s^\star$; readers who want a single-number summary of the trade-off should consult that table directly.

The full experiment, including a path-residual histogram, a policy-error heatmap on the $(z, \log K)$ plane, and a CE-welfare-loss summary table, is in the companion notebook \tpath{05_StochasticBM_LossComparison.ipynb}.

\paragraph{Extensions of the basic DEQN template.}
The Brock--Mirman model establishes the core DEQN recipe, but it is only the starting point.  Later chapters modify the same template in several directions: Chapter~\ref{ch:olg} adds lifecycle structure and complementarity constraints, Chapter~\ref{ch:young} replaces a finite vector of states by a histogram representation of the cross-sectional distribution, Section~\ref{sec:sequence_space} then shows that one can also feed \emph{shock histories} to the network rather than the current endogenous aggregate state, and Chapter~\ref{ch:climate} extends the method to nonstationary climate-economy models with pseudo-states.  The equilibrium logic is the same in all cases: choose a network parameterization, simulate the model forward, evaluate equilibrium residuals, and update the network by gradient descent.

\paragraph{Code examples.}
The following Jupyter notebooks implement and extend the material in this chapter.  Notebooks~01 and~02 illustrate the \emph{sampling progression} that is pedagogically central to the method: 01 uses uniform random states to isolate the loss-and-architecture mechanics, while 02 replaces the exogenous grid by simulated trajectories and learns on the model's ergodic set (§\ref{sec:deqn_algo}); both derive the FOC and apply the envelope theorem on paper before writing the loss.  Notebooks~03 and~04 introduce additional techniques (endogenous labor, a KKT-constrained labor-time ceiling encoded via a \emph{Fischer--Burmeister} complementarity function, and a six-period OLG extension) that anticipate material developed formally in Chapters~\ref{ch:irbc} and~\ref{ch:olg}.  Three further notebooks that re-solve the same two Brock--Mirman models with an \emph{autodiff} loss (illustrating the template of \S\ref{sec:autodiff}) are placed in the autodiff-chapter code folder as the autodiff primer: \tpath{01_AutoDiff_Analytical_Examples.ipynb}, \tpath{02_Brock_Mirman_AutoDiff_DEQN.ipynb}, \tpath{03_Brock_Mirman_Uncertainty_AutoDiff_DEQN.ipynb}.
\begin{itemize}[itemsep=2pt]
\item \tpath{01_Brock_Mirman_1972_DEQN.ipynb}: deterministic Brock--Mirman; exogenous uniform sampling; hand-derived FOC + envelope.
\item \tpath{02_Brock_Mirman_Uncertainty_DEQN.ipynb}: stochastic Brock--Mirman ($\varrho > 0$, $\sigma_z > 0$); simulation-based sampling on the ergodic set; hand-derived FOC + envelope.
\item \tpath{03_DEQN_Exercises_Blanks.ipynb}: four guided exercises (endogenous labor, KKT + Fischer--Burmeister, simple OLG extension); blank version.
\item \tpath{04_DEQN_Exercises_Solutions.ipynb}: the same four exercises with complete solutions.
\item \tpath{05_StochasticBM_LossComparison.ipynb}: the stochastic Brock--Mirman is re-solved six times with identical network, optimizer, and CRN training data, and only the loss kernel changes (MSE, MAE, Huber, quantile pinball, CVaR, log-cosh).  The notebook switches to full depreciation $\delta=1$ so that the closed-form optimal savings rate $s^\star=\alpha\beta$ is available, then evaluates each trained policy on the \emph{economic} metric: the relative Euler-equation error along a simulated path, plus the consumption-equivalent welfare loss against $s^\star$.  The exercise makes concrete that the choice of training loss and the convergence of the metric we ultimately care about are not the same thing, and that tail-aware kernels (Huber, CVaR, quantile pinball) trade a small loss in the bulk for a much cleaner $p_{99}$.
\end{itemize}

\begin{keyinsightbox}[Chapter Summary]
\begin{itemize}[itemsep=2pt, leftmargin=*]
\item DEQNs solve dynamic-equilibrium models by treating the equilibrium conditions $G(\x_t, p(\x_t), \mathbb{E}_t[H(\cdot)]) = 0$ as a residual loss; SGD on this loss replaces the traditional fixed-point iteration \citep{azinovicDEEPEQUILIBRIUMNETS2022}.
\item The hard/soft split is the key design pattern: state transitions and budget constraints are encoded \emph{exactly} in the network architecture (e.g.\ a sigmoid savings-share head); only the Euler residual is minimized in the loss.  This eliminates infeasible policies and accelerates training.
\item The Brock--Mirman benchmark with closed-form solution validates the methodology and is the reference example reused throughout the script.
\item Pathwise and conditional-expectation residuals target different objectives: path averaging is cheap and can work well in benchmarks such as Brock--Mirman, while explicit quadrature directly targets the conditional Euler equation and is preferred when expectation accuracy is central.
\end{itemize}
\end{keyinsightbox}

\section*{Further Reading}
\addcontentsline{toc}{section}{Further Reading}
\begin{itemize}[itemsep=2pt]
\item \citet{azinovicDEEPEQUILIBRIUMNETS2022}, the foundational DEQN paper; required reading.
\item \citet{maliar2021deep}, ``all-in-one'' deep learning, an alternative formulation discussed in Chapter~\ref{ch:young}.
\item \citet{fernandezvillaverde2024taming}, broad survey of deep learning in macro, situating DEQNs against PINNs and other approaches.
\item \citet{judd1998numerical}, the classical numerical-methods reference whose toolkit DEQNs supplement (rather than replace); especially Chapters 7 and 7.5 on Gauss--Hermite quadrature and monomial rules underlying \S\ref{sec:quadrature_rules}.
\item \citet{stroud1971approximate}, the canonical reference for the monomial cubature formulas of \S\ref{sec:monomial_cubature}.
\item \citet{pichler2011} and \citet{Maliar2014325}, monomial integration in large-scale dynamic economic models.
\end{itemize}

\section*{Exercises}
\addcontentsline{toc}{section}{Exercises}
\noindent Worked solutions and guidance for these exercises appear in Appendix~\ref{app:solutions}.
\begin{enumerate}[itemsep=4pt, leftmargin=*]
\item\label{ex:ch2:1} \textbf{[Core] Closed-form Brock--Mirman.}  Verify that for log utility, $\delta = 1$, and AR(1) productivity $\ln z_{t+1} = \varrho \ln z_t + \sigma_z \varepsilon_{t+1}$, the optimal savings share is the constant $s^\star = \alpha\beta$.  Use this to check that a converged DEQN's average sigmoid output should equal $\alpha\beta$.
\item\label{ex:ch2:2} \textbf{[Core] Hard vs.\ soft constraints.}  Consider a softplus head on $C_t$ alone (so $C_t > 0$ but $K_{t+1}$ unconstrained).  Construct an explicit input $(K_t, z_t)$ for which a randomly initialized network would predict $K_{t+1} < 0$.  Explain why the sigmoid-savings parameterization eliminates this failure mode.
\item\label{ex:ch2:3} \textbf{[Core] Path averaging vs.\ conditional expectation.}  Let $G_\rho(x_t,\varepsilon_{t+1})$ denote the one-shock Euler residual under network parameters $\rho$.  Show that, under ergodicity, the path average of $G_\rho^2$ converges almost surely to $\mathbb{E}_{\mu,\varepsilon}[G_\rho(x,\varepsilon)^2]$ as $T_{\text{sim}} \to \infty$.  Compare this target with the conditional residual objective $\mathbb{E}_{\mu}[(\mathbb{E}[G_\rho(x,\varepsilon)\mid x])^2]$.  Use Jensen's inequality to explain why the pathwise squared objective is generally stronger, and discuss the finite-sample variance trade-off.
\item\label{ex:ch2:4} \textbf{[Core] Brock--Mirman with Gauss--Hermite.}  In the Brock--Mirman Euler equation~\eqref{eq:bm_euler}, replace the single-shock pathwise residual by a $Q=5$ Gauss--Hermite expectation (\S\ref{sec:gh_tensor_product}).  Write out the resulting deterministic loss in closed form, and check on a few representative $(K_t, z_t)$ that the value matches a Monte Carlo estimate with $10^4$ shock draws to four significant digits.
\item\label{ex:ch2:5} \textbf{[Core] Monomial rule by hand.}  Take the Stroud-3 rule of equation~\eqref{eq:stroud3} for shock dimension $d=4$ (so $8$ nodes).  Verify by direct computation that the rule reproduces $\E{\varepsilon_i'}=0$, $\E{\varepsilon_i'^{\,2}}=1$, $\E{\varepsilon_i'\varepsilon_j'}=0$ for $i\neq j$, and $\E{\varepsilon_i'^{\,3}}=0$ exactly, but gives $\E{\varepsilon_i'^{\,4}} = d = 4$ instead of the true value $3$.  Show that the relative bias on the fourth moment grows linearly in $d$, and discuss when this matters for an Euler-equation residual.
\item\label{ex:ch2:6} \textbf{[Core] Loss-kernel selection.}  Three application scenarios are described below; match each to the most appropriate loss kernel from the menu \{MSE, MAE, Huber($\delta$), pinball loss at $\tau$, CVaR at $\alpha$, log-cosh\} and justify the choice in two or three sentences.  (a)~``Quadrature noise occasionally produces a few large Euler residuals; we want a smooth loss that behaves quadratically near zero so gradients vanish at the optimum, but only linearly in the tails so that a single noisy point cannot dominate the gradient.''  (b)~``A regulator will inspect the worst $1\%$ of residuals; we want the optimizer to drive the conditional mean above the $99$th percentile down to tolerance, not the average.''  (c)~``We care that the \emph{median} Euler residual is small.  Tails are an artefact of badly conditioned states near the borrowing constraint and should not pull the gradient.''
\item\label{ex:ch2:7} \textbf{[Computational] Multivariate shock scaling.}  Extend notebook \tpath{lecture_03_02_Brock_Mirman_Uncertainty_DEQN.ipynb} to $d$ independent productivity shocks, $d \in \{1, 2, 4, 8\}$ (e.g., add country-level TFP terms to the production technology, all i.i.d.\ standard normal).  For each $d$, train the network twice: once with tensor-product Gauss--Hermite at $Q=3$ ($3^d$ nodes per residual), once with the Stroud-3 rule of \eqref{eq:stroud3} ($2d$ nodes).  Plot training time per epoch and final relative Euler error against $d$ on the same axes.  Confirm that the Stroud-3 cost grows linearly while Gauss--Hermite is exponential, and identify the $d$ at which Gauss--Hermite becomes impractical on a single GPU.
\item\label{ex:ch2:8} \textbf{[Computational] Implementation.}  Modify notebook \tpath{lecture_03_02_Brock_Mirman_Uncertainty_DEQN.ipynb} to use a tanh activation instead of Swish.  Does training still converge?  How does the time-to-converge change?
\end{enumerate}

\chapter{The International Real Business Cycle Model}
\label{ch:irbc}

Having established the DEQN framework on the one-dimensional Brock--Mirman model in Chapter~\ref{ch:deqn}, we now scale it to the multi-country international real business cycle (IRBC) model of \citet{backus1992international}.  This model features $N$ countries with heterogeneous productivity, complete markets, irreversible investment, and convex capital adjustment costs.  It is the standard testbed for high-dimensional solution methods in macroeconomics, and applying DEQNs to it illustrates how the framework handles high-dimensional state spaces, multiple equilibrium conditions, and complementarity constraints.

\section{Why IRBC for Macro-Finance Research?}
\label{sec:irbc_motivation}

Beyond its computational-testbed role, the IRBC model is the workhorse framework for \emph{open-economy asset pricing} and \emph{international risk sharing}.  Several first-order questions in macro-finance can be posed sharply within it:

\begin{itemize}[itemsep=2pt]
\item \textbf{International risk sharing.}  Under complete markets and homogeneous preferences, the planner's allocation implies perfectly correlated consumption growth across countries, whereas the data show consumption correlations that are lower than output correlations (the \emph{consumption-correlation puzzle} of \citealt{backus1992international}): the benchmark BKK model predicts a cross-country consumption correlation close to 1, while the empirical correlation between the US and a typical industrial country is in the range 0.3--0.5 and is systematically below the corresponding output correlation.  IRBC extensions with incomplete markets, frictions, or heterogeneous preferences are designed precisely to close this gap.
\item \textbf{Asset-market structure and welfare.}  \citet{heathcote2002financial} use an IRBC-style setup to quantify the welfare cost of moving from complete markets to one-bond economies (``financial autarky''), obtaining a welfare cost of the same order of magnitude as the business cycle itself.  More generally, variants of this model are the standard laboratory for comparing complete vs.\ incomplete market structures.
\item \textbf{Capital flows and current-account dynamics.}  With intertemporal savings and heterogeneous productivity, the IRBC delivers persistent current-account imbalances as an equilibrium outcome rather than as a reduced-form residual.  This is the starting point for the modern open-economy DSGE literature.
\item \textbf{Home bias.}  The frictionless benchmark of near-unit consumption correlation is the anchor against which observed portfolio home bias must be explained; \citet{heathcote2013international} show that accounting for nontraded goods and labor-income hedging substantially narrows the gap between theory and observed portfolios.
\item \textbf{Macro-financial transmission.}  Adjustment costs, borrowing constraints, and Pareto weights become levers for studying how financial frictions propagate across borders.  This is the class of extensions that motivates the DEQN treatment: once frictions are added, the policy functions acquire kinks and nonlinearities that are hard to handle with traditional grid-based methods.
\end{itemize}

The IRBC model is therefore an interesting substantive object, not merely a scaling test.  Its combination of a clean complete-markets benchmark and rich, realistic frictions makes it a natural next step after the one-country Brock--Mirman benchmark of Chapter~\ref{ch:deqn}.

\paragraph{A calibration caveat for the puzzles above.}  The shock decomposition of \S\ref{sec:irbc_setup} below, $z^{j\prime} = \rho_z z^j + \sigma_e(\varepsilon^j + \varepsilon^{\mathrm{agg}})$, hard-wires a cross-country innovation correlation of exactly $1/2$ for any number of countries~$N$.  The consumption-correlation and Backus--Smith puzzles cited in the bullets above should therefore be read as statements about this specific calibration: richer correlation structures, country-specific factor loadings, or fewer aggregate factors would change the quantitative bite of the puzzles in this model.  This is calibration, not theory.

\section{Model Setup}
\label{sec:irbc_setup}

\begin{table}[ht]
\centering
\small
\setlength{\tabcolsep}{4pt}
\begin{tabular}{@{}>{$}l<{$} l l l@{}}
\toprule
\textbf{Symbol} & \textbf{Role} & \textbf{Range / sign} & \textbf{Calibration} \\
\midrule
\gamma_j         & IES of country $j$ (\emph{not} CRRA)                                & $>0$                & $[0.25, 1.0]$ linearly spaced \\
\tau^j           & Pareto weight on country $j$                                        & $>0$                & $(A_{\mathrm{tfp}}-\delta)^{1/\gamma_j}$ \\
\lambda_t        & Aggregate resource-constraint multiplier                            & $>0$                & $\lambda_{\mathrm{ss}} = 1$ \\
\mu_t^j          & Irreversibility KKT multiplier on $I^j \ge 0$                       & $\ge 0$             & $0$ in slack regime \\
A_{\mathrm{tfp}} & TFP normalization constant                                          & $>0$                & $\approx 0.0559$ \\
\zeta            & Capital share in Cobb--Douglas                                      & $\in (0,1)$         & $0.36$ \\
\Gamma^j         & Quadratic adjustment-cost level                                     & $\ge 0$             & $\kappa=0.50$ \\
\rho_z           & TFP persistence                                                     & $\in [0,1)$         & $0.95$ \\
\sigma_e         & Innovation s.d.\ per component                                      & $>0$                & $0.01$ \\
\varepsilon^j    & Idiosyncratic innovation                                            & $\mathcal N(0,1)$   & i.i.d.\ across $j,t$ \\
\varepsilon^{\mathrm{agg}} & Aggregate innovation                                       & $\mathcal N(0,1)$   & common factor \\
\kappa           & Adjustment-cost intensity                                           & $\ge 0$             & $0.50$ \\
\bottomrule
\end{tabular}
\caption{Symbol cheat-sheet for the IRBC model.  Note the IES-vs-CRRA convention: here $\gamma_j$ is the intertemporal elasticity, and the implied risk aversion is $1/\gamma_j$; later chapters on continuous-time HA models and climate use $\gamma$ for CRRA and $\psi$ for IES.}
\label{tab:irbc_symbols}
\end{table}

The international real business cycle (IRBC) model, introduced by \citet{backus1992international}, extends the single-country growth model to $N$ heterogeneous countries, each endowed with country-specific capital $k^j$ and total factor productivity $z^j$.  The model features complete markets, irreversible investment, and convex capital adjustment costs, and serves as the workhorse test case for high-dimensional solution methods \citep[see, e.g.,][]{ECTA:ECTA1716}.  Here, we apply the DEQN methodology of \citet{azinovicDEEPEQUILIBRIUMNETS2022} to this setting.

\paragraph{Preferences.}  Each country $j$ has CRRA utility
\begin{equation}
u^j(c) \;=\; \begin{cases}
\dfrac{c^{1-1/\gamma_j} - 1}{1 - 1/\gamma_j}, & \gamma_j \neq 1,\\[6pt]
\ln c, & \gamma_j = 1,
\end{cases}
\end{equation}
where the intertemporal elasticity of substitution (IES) $\gamma_j$ is heterogeneous across countries; risk aversion under this CRRA specification equals $1/\gamma_j$.  \textbf{Notation warning:} this chapter uses $\gamma$ for the IES, while later chapters on continuous-time HA models and climate use $\gamma$ for CRRA risk aversion and $\psi$ for the IES.  The convention is stated explicitly at the start of each chapter.  A social planner maximizes
\begin{equation}
\max \; \sum_{t=0}^{\infty} \beta^t \, \E{\sum_{j=1}^{N} \tau^j \, u^j(c_t^j)}
\label{eq:irbc_utility}
\end{equation}
with Pareto weights $\tau^j > 0$.

\paragraph{Production.}  Country $j$ produces $Y^j = A_\mathrm{tfp} \exp(z^j)(k^j)^\zeta$, where the total factor productivity constant $A_\mathrm{tfp}$ is calibrated to normalize the steady-state capital stock to unity.  In steady state (where $z^j = 0$, $k^j = 1$, and $k^{j\prime} = 1$), the Euler equation implies:
\begin{equation}
A_\mathrm{tfp} = \frac{1/\beta - 1 + \delta}{\zeta}.
\label{eq:atfp}
\end{equation}
This normalization ensures that the deterministic steady state lies at $(k^\star, z^\star) = (1, 0)$ for all countries, which simplifies the network's learning task and provides a natural center for the training distribution.

\paragraph{TFP process.}  Log productivity follows an AR(1) with common and idiosyncratic shocks:
\begin{equation}
z^{j\prime} = \rho_z z^j + \sigma_e(\varepsilon^j + \varepsilon^\mathrm{agg}), \qquad \varepsilon^j, \varepsilon^\mathrm{agg} \sim \mathcal{N}(0,1)\text{ i.i.d.}, \qquad |\rho_z| < 1
\label{eq:irbc_tfp}
\end{equation}
The persistence restriction $|\rho_z|<1$ guarantees stationarity of the TFP process, which in turn underlies the existence of an ergodic distribution on which DEQN training samples (Section~\ref{sec:deqn_algo}).  Here $\sigma_e$ is the per-component standard deviation, so the marginal innovation variance for country $j$ is $2\sigma_e^2$ and the cross-country innovation covariance is $\sigma_e^2$.  These two facts imply a fixed cross-country innovation correlation of $1/2$ regardless of $N$, a direct consequence of the equal-weighted aggregate-shock decomposition $\varepsilon^j + \varepsilon^\mathrm{agg}$.  Asset-pricing implications (in particular the international consumption-correlation puzzle and the cyclicality of trade balances) inherit this hard-wired common-factor structure: results below should be interpreted with that calibration choice in mind.  If a desired total innovation scale $\bar\sigma$ is targeted instead, set $\sigma_e = \bar\sigma/\sqrt{2}$.

\paragraph{Adjustment costs and irreversibility.}  Changing the capital stock incurs a quadratic adjustment cost:
\begin{equation}
\Gamma^j = \frac{\kappa}{2}\, k^j \left(\frac{k^{j\prime}}{k^j} - 1\right)^{\!2},
\label{eq:irbc_adjcost}
\end{equation}
with marginal derivatives that appear in the Euler equations:
\begin{align}
\frac{\partial \Gamma^j}{\partial k^{j\prime}} &= \kappa\left(\frac{k^{j\prime}}{k^j} - 1\right), &
\frac{\partial \Gamma^j}{\partial k^j} &= \frac{\kappa}{2}\left(1 - \left(\frac{k^{j\prime}}{k^j}\right)^{\!2}\right).
\label{eq:irbc_adjcost_derivs}
\end{align}
Note that $\partial\Gamma^j/\partial k^j$ is \emph{negative} whenever $k^{j\prime} > k^j$, i.e.\ in expanding states.  Consequently the term $-\partial\Gamma^j/\partial k^j$ that appears in the marginal product of capital below~\eqref{eq:irbc_mpk} \emph{raises} MPK in expansion phases; a reader who plugs in $|\partial\Gamma/\partial k|$ here will introduce a sign error.  Investment is irreversible: $I^j = k^{j\prime} - (1-\delta)k^j \geq 0$.

\paragraph{Pareto-weight calibration.}  With heterogeneous IES $\gamma_j$, a symmetric deterministic steady state is most easily obtained by choosing the Pareto weights as
\begin{equation}
\tau^j \;=\; \bigl(A_\mathrm{tfp} - \delta\bigr)^{1/\gamma_j}, \qquad j=1,\ldots,N.
\label{eq:pareto_calibration}
\end{equation}
The derivation is a two-step inversion of the planner's first-order condition.  The consumption-sharing condition~\eqref{eq:irbc_consumption} (derived in the next section from the FOC for $c^j_t$) reads $\tau^j (c^j_t)^{-1/\gamma_j} = \lambda_t$, so $c^j_t = (\lambda_t / \tau^j)^{-\gamma_j}$.  In the deterministic steady state with the normalizations $\lambda_\mathrm{ss} = 1$ and $k^j_\mathrm{ss} = 1$ we want every country to consume the same amount $c^j_\mathrm{ss} = A_\mathrm{tfp} - \delta$ implied by the resource constraint $c^j_\mathrm{ss} = Y^j_\mathrm{ss} - I^j_\mathrm{ss}$.  Setting $(1/\tau^j)^{-\gamma_j} = A_\mathrm{tfp} - \delta$ and solving for $\tau^j$ gives Eq.~\eqref{eq:pareto_calibration}.  The symmetric steady state thus serves as a natural anchor for training: the network's initial predictions need only match this point to avoid infeasible economies during the early simulated trajectories.

\paragraph{Reference calibration.}  Throughout the companion notebooks \tpath{lecture_04_01_IRBC_DEQN_smooth.ipynb} and \tpath{lecture_04_02_IRBC_DEQN_irreversible.ipynb}, we use the quarterly calibration summarized in Table~\ref{tab:irbc_params}.  The implied total factor productivity and deterministic steady-state quantities can then be computed analytically.

\begin{table}[ht]
\centering
\small
\begin{tabular}{llrl}
\toprule
\textbf{Symbol} & \textbf{Name} & \textbf{Value} & \textbf{Description} \\
\midrule
$\beta$ & Discount factor & 0.99 & Quarterly \\
$\zeta$ & Capital share & 0.36 & Cobb--Douglas \\
$\delta$ & Depreciation & 0.01 & Low quarterly rate \\
$\rho_z$ & TFP persistence & 0.95 & Highly persistent \\
$\sigma_e$ & Shock std.\ dev. & 0.01 & Small innovations \\
$\kappa$ & Adjustment-cost intensity & 0.50 & Moderate frictions \\
$\gamma_{\min}$ & Min IES & 0.25 & Risk aversion $=4$ \\
$\gamma_{\max}$ & Max IES & 1.00 & Log utility \\
$k^\star$ & Steady-state capital & 1.00 & Normalization \\
\bottomrule
\end{tabular}
\caption{Reference IRBC calibration used in the companion notebook.  Countries' IES values $\gamma_j$ are linearly spaced in $[\gamma_{\min}, \gamma_{\max}]$.  Pareto weights are computed from~\eqref{eq:pareto_calibration}.}
\label{tab:irbc_params}
\end{table}

\paragraph{Worked steady state.}  Equation~\eqref{eq:atfp} is most compactly written as $A_\mathrm{tfp}=(1/\beta - 1 + \delta)/\zeta$; multiplying numerator and denominator by $\beta$ gives the algebraically equivalent form $A_\mathrm{tfp}=(1-\beta(1-\delta))/(\zeta\beta)$ used below.  Substituting the reference values:
\begin{align}
A_\mathrm{tfp} &= \frac{1-\beta(1-\delta)}{\zeta\,\beta}
               = \frac{1 - 0.99 \cdot 0.99}{0.36 \cdot 0.99}
               \;\approx\; 0.0559, \\
Y^\star_j      &= A_\mathrm{tfp}\,(k^\star)^\zeta \approx 0.0559, \qquad
I^\star_j      = \delta\,k^\star = 0.01, \qquad
c^\star_j      = Y^\star_j - I^\star_j \approx 0.0459.
\end{align}
The aggregate resource constraint~\eqref{eq:irbc_arc} is then satisfied country by country, $Y^\star_j - I^\star_j - c^\star_j = 0$, as a trivial check.  These numbers provide a baseline against which the trained network's predictions on an out-of-sample simulation can be compared.

\section{The Planner's Problem and Equilibrium Conditions}
\label{sec:irbc_planner_problem}

\paragraph{The planner's problem.}  The social planner maximizes the weighted sum of utilities across all $N$ countries, subject to the aggregate resource constraint~\eqref{eq:irbc_arc}, the irreversibility constraints, the production technology, and the TFP process~\eqref{eq:irbc_tfp}:
\begin{equation}
\max_{\{c_t^j,\, k_{t+1}^j\}_{j,t}} \; \sum_{t=0}^{\infty} \beta^t \, \E{\sum_{j=1}^{N} \tau^j \, u^j(c_t^j)}
\label{eq:irbc_planner}
\end{equation}
with Pareto weights $\tau^j > 0$ and discount factor $\beta \in (0,1)$.

\paragraph{The Lagrangian.}  Following the same approach as in Section~\ref{sec:bm} for the Brock--Mirman model, we form the Lagrangian by attaching discounted multipliers to each constraint.  Let $\beta^t \lambda_t$ be the multiplier on the aggregate resource constraint at date $t$, and $\beta^t \mu_t^j$ the multiplier on the irreversibility constraint for country $j$ at date $t$.  The Lagrangian is:
\begin{equation}
\begin{split}
\mathcal{L} = \mathbb{E}\Biggl[\sum_{t=0}^{\infty} \beta^t \Biggl(
&\sum_{j=1}^{N} \tau^j \, \frac{(c_t^j)^{1-1/\gamma_j} - 1}{1-1/\gamma_j}
+ \lambda_t \sum_{j=1}^{N} \bigl(Y_t^j + (1-\delta)k_t^j - k_{t+1}^j - \Gamma_t^j - c_t^j\bigr) \\
&+ \sum_{j=1}^{N} \mu_t^j \bigl(k_{t+1}^j - (1-\delta)k_t^j\bigr)
\Biggr)\Biggr].
\end{split}
\label{eq:irbc_lagrangian}
\end{equation}
The planner chooses $c_t^j$ and $k_{t+1}^j$ for each country $j$ and each date $t$.  The complementary slackness conditions require $\mu_t^j \geq 0$, $I_t^j \geq 0$, and $\mu_t^j \cdot I_t^j = 0$.  Two notation reminders before we differentiate.  First, the irreversibility multiplier is $\mu_t^j$, not the resource-constraint multiplier $\lambda_t$; the two play different roles ($\lambda_t$ shadow-prices the aggregate goods market; $\mu_t^j$ shadow-prices country $j$'s individual investment floor) and they enter the FOCs through entirely different channels.  Second, $\mu_t^j \geq 0$ is the standard KKT sign: the multiplier on a $\geq$-constraint is non-negative at the optimum, and the Fischer--Burmeister residual constructed below packages this sign restriction together with the slackness condition into a single smooth squared term that is compatible with SGD.

\paragraph{FOC w.r.t.\ $c_t^j$:}  Differentiating the Lagrangian with respect to $c_t^j$:
\begin{equation}
\frac{\partial \mathcal{L}}{\partial c_t^j}
= \beta^t \bigl[\tau^j (c_t^j)^{-1/\gamma_j} - \lambda_t\bigr] = 0
\qquad\Longrightarrow\qquad
\tau^j (c_t^j)^{-1/\gamma_j} = \lambda_t.
\label{eq:irbc_foc_c}
\end{equation}
This is the \emph{consumption-sharing condition}: the planner equates the Pareto-weighted marginal utility of consumption across all countries to a common shadow price $\lambda_t$.  Solving~\eqref{eq:irbc_foc_c} for $c_t^j$:
\begin{equation}
c_t^j = \left(\frac{\lambda_t}{\tau^j}\right)^{-\gamma_j}.
\label{eq:irbc_consumption}
\end{equation}
This shows that all $N$ consumption levels are determined by the single variable $\lambda_t$: a higher shadow price (resources are scarcer) lowers consumption in every country.  Countries with a higher IES $\gamma_j$ respond more elastically to changes in $\lambda_t$.

\paragraph{FOC w.r.t.\ $k_{t+1}^j$:}  The variable $k_{t+1}^j$ appears in three places in the Lagrangian: (i)~the date-$t$ resource constraint with coefficient $-\lambda_t(1 + \partial\Gamma_t^j/\partial k_{t+1}^j)$, (ii)~the date-$t$ irreversibility constraint with coefficient $+\mu_t^j$, and (iii)~the date-$(t\!+\!1)$ terms via output $Y_{t+1}^j$, depreciated capital $(1-\delta)k_{t+1}^j$, adjustment costs $\Gamma_{t+1}^j$, and the irreversibility constraint.  Differentiating and collecting terms:
\begin{multline}
\frac{\partial \mathcal{L}}{\partial k_{t+1}^j}
= \beta^t \!\left[-\lambda_t\!\left(1 + \frac{\partial \Gamma_t^j}{\partial k_{t+1}^j}\right) + \mu_t^j\right] \\
+ \beta^{t+1}\,\mathbb{E}_t\!\left[\lambda_{t+1}\!\left(\frac{\partial Y_{t+1}^j}{\partial k_{t+1}^j} + (1-\delta) - \frac{\partial \Gamma_{t+1}^j}{\partial k_{t+1}^j}\right) - \mu_{t+1}^j(1-\delta)\right] = 0.
\label{eq:irbc_foc_k_raw}
\end{multline}
Now define the \emph{marginal product of capital} (inclusive of depreciation and adjustment cost effects):
\begin{equation}
\mathrm{MPK}^j \;\equiv\; 1-\delta + \zeta A_\mathrm{tfp}\exp(z^j)(k^j)^{\zeta-1} - \frac{\partial \Gamma^j}{\partial k^j},
\label{eq:irbc_mpk}
\end{equation}
and note from~\eqref{eq:irbc_adjcost_derivs} that $\partial \Gamma_t^j / \partial k_{t+1}^j = \kappa(k_{t+1}^j/k_t^j - 1)$.  Dividing~\eqref{eq:irbc_foc_k_raw} by $\beta^t$ and substituting the MPK definition:
\begin{equation}
\lambda_t\!\left(1 + \frac{\partial \Gamma_t^j}{\partial k_{t+1}^j}\right) - \mu_t^j
= \beta\,\mathbb{E}_t\!\bigl[\lambda_{t+1}\,\mathrm{MPK}_{t+1}^j - (1-\delta)\,\mu_{t+1}^j\bigr].
\label{eq:irbc_euler_level}
\end{equation}
This is the \emph{Euler equation} for country $j$.  The left-hand side is the cost of investing one more unit in country $j$'s capital: the shadow price $\lambda_t$ of the resources used (scaled by the marginal adjustment cost) minus the value $\mu_t^j$ of relaxing the irreversibility constraint.  The right-hand side is the expected discounted benefit: next period's shadow price times the marginal product of capital, minus the option-value loss from tightening next period's irreversibility constraint.

\paragraph{Relative error form.}  For numerical purposes, regroup~\eqref{eq:irbc_euler_level} so that the cost-of-investment term $\lambda_t(1+\partial\Gamma_t^j/\partial k_{t+1}^j)$ stands alone on the left, $\lambda_t(1+\partial\Gamma_t^j/\partial k_{t+1}^j) = \beta\,\mathbb{E}_t[\lambda_{t+1}\,\mathrm{MPK}_{t+1}^j - (1-\delta)\mu_{t+1}^j] + \mu_t^j$, and divide through by it.  This gives a scale-free formulation:
\begin{equation}
\frac{\beta\,\mathbb{E}_t\!\left[\lambda' \cdot \mathrm{MPK}^{j\prime} - (1-\delta)\mu^{j\prime}\right] + \mu^j}{\lambda(1+\partial\Gamma^j/\partial k^{j\prime})} - 1 = 0, \qquad j=1,\ldots,N.
\label{eq:irbc_euler_relerr}
\end{equation}
This ensures that all $N$ Euler equations are dimensionless and residuals can be interpreted directly as percentage deviations from optimality.

\paragraph{Aggregate resource constraint.}  All output is allocated to consumption, investment, and adjustment costs:
\begin{equation}
\sum_{j=1}^{N}\bigl[Y^j + (1-\delta)k^j - k^{j\prime} - \Gamma^j - c^j\bigr] = 0.
\label{eq:irbc_arc}
\end{equation}

\paragraph{Summary of equilibrium conditions.}  The complete system consists of three blocks:
\begin{enumerate}[itemsep=2pt]
\item \textbf{Consumption sharing}~\eqref{eq:irbc_consumption}: determines all $N$ consumption levels from $\lambda_t$.
\item \textbf{Euler equations}~\eqref{eq:irbc_euler_relerr}: $N$ intertemporal optimality conditions, one per country.
\item \textbf{Aggregate resource constraint}~\eqref{eq:irbc_arc}: closes the model by equating world supply and demand.
\end{enumerate}
In addition, the $N$ irreversibility constraints are enforced via complementary slackness ($\mu^j \geq 0$, $I^j \geq 0$, $\mu^j I^j = 0$).

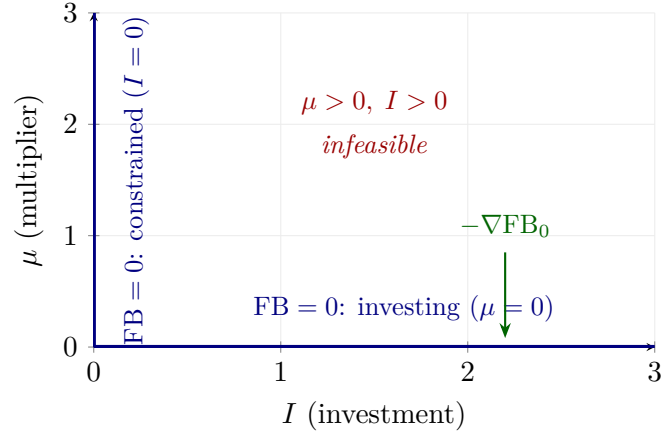
\begin{figure}[ht]
\centering
\begin{tikzpicture}
\begin{axis}[
    width=9cm, height=6cm,
    xlabel={$I$ (investment)}, ylabel={$\mu$ (multiplier)},
    xmin=0, xmax=3, ymin=0, ymax=3,
    xtick={0,1,2,3}, ytick={0,1,2,3},
    grid=major, grid style={gray!15},
    axis lines=left,
    every axis plot/.append style={ultra thick, no markers},
]
\addplot[uzhblue, domain=0:3, samples=2] {0};
\addplot[uzhblue, samples=2] coordinates {(0,0) (0,3)};
\node[font=\small, text=harvardcrimson] at (axis cs:1.5,2.2) {$\mu > 0,\; I > 0$};
\node[font=\small, text=harvardcrimson] at (axis cs:1.5,1.8) {\textit{infeasible}};
\node[font=\small, text=uzhblue, anchor=west] at (axis cs:0.8,0.35)
    {$\mathrm{FB}=0$: investing ($\mu=0$)};
\node[font=\small, text=uzhblue, rotate=90, anchor=south] at (axis cs:0.35,1.5)
    {$\mathrm{FB}=0$: constrained ($I=0$)};
\draw[-{Stealth[length=2mm]}, thick, darkgreen] (axis cs:2.2,0.85) -- (axis cs:2.2,0.08);
\node[font=\small, darkgreen, anchor=south] at (axis cs:2.2,0.9) {$-\nabla\mathrm{FB}_0$};
\end{axis}
\end{tikzpicture}
\caption{The Fischer--Burmeister complementarity function, drawn in the investment--multiplier plane: investment $I^j$ on the horizontal axis, the irreversibility multiplier $\mu^j$ on the vertical axis.  The exact map $\mathrm{FB}_0(\mu,I)=\mu+I-\sqrt{\mu^2+I^2}$ packs the three Karush--Kuhn--Tucker conditions $\mu\ge 0$, $I\ge 0$, $\mu I=0$ into a single smooth equation: $\mathrm{FB}_0=0$ holds \emph{exactly} on the two heavy blue half-axes and nowhere else.  The horizontal half-axis ($\mu=0$, $I>0$) is the \emph{investing} regime, where the country invests a strictly positive amount, the irreversibility constraint is slack, and its shadow price $\mu$ is therefore zero.  The vertical half-axis ($I=0$, $\mu>0$) is the \emph{constrained} regime, where the constraint binds, investment is pinned at zero, and $\mu>0$ measures how much the planner would pay to relax it; the origin is the knife-edge where both hold with equality.  The open interior of the first quadrant ($\mu>0$ \emph{and} $I>0$ together) is \emph{infeasible} because it violates complementarity, and there $\mathrm{FB}_0>0$ strictly (since $\mu+I>\sqrt{\mu^2+I^2}$ whenever both are positive).  This is exactly what makes the function useful as a loss term: when the network's predicted $(\mu^j,I^j)$ lands in that forbidden region, the squared residual $\mathrm{FB}_\varepsilon^2$ is positive and its negative gradient $-\nabla\mathrm{FB}_0$ (green arrow) pushes the iterate back toward the nearest feasible half-axis, so the network learns which regime applies at each state without any explicit regime switch.  The exact map has a single kink, at the origin; the smoothed version $\mathrm{FB}_\varepsilon(\mu,I)=\mu+I-\sqrt{\mu^2+I^2+\varepsilon^2}$ actually used in the code rounds that corner, restoring differentiability everywhere at the price of an $\mathcal{O}(\varepsilon)$ relaxation of exact complementarity.}
\label{fig:fb_zeroset}
\end{figure}

\paragraph{Fischer--Burmeister complementarity.}\label{sec:irbc_fischer_burmeister}  The irreversibility constraint is enforced via a smoothed Fischer--Burmeister residual:
\begin{equation}
\mathrm{FB}_\varepsilon(\mu^j, I^j) = \mu^j + I^j - \sqrt{(\mu^j)^2 + (I^j)^2 + \varepsilon^2} = 0.
\end{equation}
The exact Fischer--Burmeister map is the limiting case $\mathrm{FB}_0(\mu,I)=\mu+I-\sqrt{\mu^2+I^2}$.  Its zero set coincides with the positive axes in the $(\mu, I)$-plane, ensuring $\mu^j \geq 0$, $I^j \geq 0$, and $\mu^j \cdot I^j = 0$ (Figure~\ref{fig:fb_zeroset}).  The smoothed version with $\varepsilon > 0$ rounds the corner at the origin and is differentiable there, improving numerical conditioning at the cost of a slight relaxation of exact complementarity.  The companion notebooks use $\varepsilon = 10^{-4}$ as the default; tighter values ($10^{-6}$--$10^{-5}$) are sometimes preferred when complementarity must hold to higher accuracy, at the cost of stiffer gradients near the origin.

The complementarity conditions $\mu^j \geq 0$, $I^j \geq 0$, $\mu^j \cdot I^j = 0$ have a natural economic interpretation: when investment is strictly positive ($I^j > 0$), the irreversibility constraint is slack and the multiplier is zero ($\mu^j = 0$); conversely, when the constraint binds ($I^j = 0$), the multiplier is positive, reflecting the shadow value of the binding constraint.  The FB function smoothly encodes both regimes, allowing the neural network to learn which regime applies for each state without explicit regime switching.

\begin{remarkbox}[Why Fischer--Burmeister works so well in DEQNs]
The squared FB residual converts a discrete regime-switching problem (constraint slack vs binding) into a smooth gradient field that SGD can navigate.  Three properties matter: (i) the zero set of $\mathrm{FB}_0$ \emph{exactly} coincides with the KKT complementarity axes, so a converged network satisfies the constraint structure to whatever tolerance the loss is driven; (ii) the residual is smooth everywhere away from the origin, so backpropagation through it is well behaved; and (iii) the $\varepsilon^2$ smoothing rounds the single remaining kink at the origin, restoring differentiability there at the price of an $\mathcal{O}(\varepsilon)$ relaxation of exact complementarity.  In the IRBC context, the network learns which states fall on the ``investing'' axis and which on the ``constrained'' axis without ever being told which regime applies, a major saving over methods that require manual regime indicators.
\end{remarkbox}

\section{DEQN Formulation}

\paragraph{From Brock--Mirman to IRBC.}  It is useful to see the IRBC as the natural extension of the one-country benchmark of Chapter~\ref{ch:deqn}.  Table~\ref{tab:bm_vs_irbc} summarizes what changes.

\begin{table}[ht]
\centering
\small
\begin{tabular}{@{}>{\raggedright\arraybackslash}p{0.19\linewidth}
                >{\raggedright\arraybackslash}p{0.35\linewidth}
                >{\raggedright\arraybackslash}p{0.36\linewidth}@{}}
\toprule
& \textbf{Brock--Mirman (Ch.~\ref{ch:deqn})} & \textbf{IRBC (this chapter)} \\
\midrule
Countries      & 1 & $N$ \\
States         & $(K, z)$ & $(k^1,\ldots,k^N, z^1,\ldots,z^N)$ \\
Policies       & $C$ & $(k^{1\prime},\ldots,k^{N\prime}, \lambda, \mu^1,\ldots,\mu^N)$ \\
Loss terms     & 1 Euler & $N$ Euler $+$ 1 ARC $+$ $N$ Fischer--Burmeister \\
Constraints    & none & irreversibility, convex adjustment costs \\
Shocks per period & 1 & $N+1$ (one idiosyncratic per country + one aggregate) \\
Output activation & softplus or sigmoid & softplus \\
Analytical solution & yes (log utility, $\delta=1$) & no \\
\bottomrule
\end{tabular}
\caption{The DEQN template is the same in both cases; only the input/output dimensions, the number of loss terms, and the presence of complementarity constraints change.}
\label{tab:bm_vs_irbc}
\end{table}

The full system of equations comprises $N$ Euler equations, $N$ Fischer--Burmeister conditions, and 1 aggregate resource constraint, totaling $2N+1$ equations.  Table~\ref{tab:irbc_scalability} summarizes how the problem dimensions scale with $N$.

\begin{table}[ht]
\centering
\scriptsize
\begin{tabular}{@{}r r r r r r r@{}}
\toprule
$N$ & States & Policies & Equations & Shock dim. & GH nodes ($Q=3$) & Stroud-3 nodes \\
\midrule
2 & 4 & 5 & 5 & 3 & $27$ & $6$ \\
5 & 10 & 11 & 11 & 6 & $729$ & $12$ \\
10 & 20 & 21 & 21 & 11 & $1.8\times 10^5$ & $22$ \\
50 & 100 & 101 & 101 & 51 & $\sim 2.2\times 10^{24}$ & $102$ \\
100 & 200 & 201 & 201 & 101 & $\sim 1.5\times 10^{48}$ & $202$ \\
\bottomrule
\end{tabular}
\caption{Scaling of the IRBC state, policy, equation, and quadrature dimensions with the number of countries $N$.  The state, policy, and equation counts grow linearly.  Tensor-product Gauss--Hermite quadrature grows as $Q^{N+1}$, while the Stroud-3 monomial rule uses only $2(N+1)$ nodes; this is why the notebook uses Gauss--Hermite only for the two-country classroom case and switches to monomial or QMC rules in larger IRBC applications.}
\label{tab:irbc_scalability}
\end{table}

\begin{figure}[ht]
\centering
\begin{tikzpicture}
\begin{axis}[
    width=12cm, height=5.5cm,
    xlabel={Number of countries $N$},
    ylabel={Quadrature nodes per residual evaluation (log)},
    xmin=2, xmax=100, ymin=1, ymax=1e50,
    ymode=log,
    grid=major, grid style={gray!15},
    legend style={at={(0.03,0.97)}, anchor=north west, font=\small, draw=gray!40},
    every axis plot/.append style={very thick},
    label style={font=\small}, tick label style={font=\footnotesize},
]
\addplot[harvardcrimson, mark=*, mark size=2pt] coordinates {
    (2,27) (5,729) (10,177147) (50,2.2e24) (100,1.5e48)
};
\addlegendentry{Tensor-product GH ($Q=3$)}
\addplot[uzhblue, mark=square*, mark size=2pt] coordinates {
    (2,6) (5,12) (10,22) (50,102) (100,202)
};
\addlegendentry{Stroud-3 monomial ($2(N+1)$)}
\end{axis}
\end{tikzpicture}
\caption{Quadrature-cost crossover for the IRBC model as a function of the number of countries $N$.  Tensor-product Gauss--Hermite (red) grows exponentially in $N$ and becomes infeasible by $N=10$; the Stroud-3 monomial rule (blue) grows linearly and stays well under $10^3$ nodes even at $N=100$.  This is the operational reason every IRBC application beyond the classroom $N=2$ case uses monomial or QMC integration.}
\label{fig:irbc_quad_cost}
\end{figure}
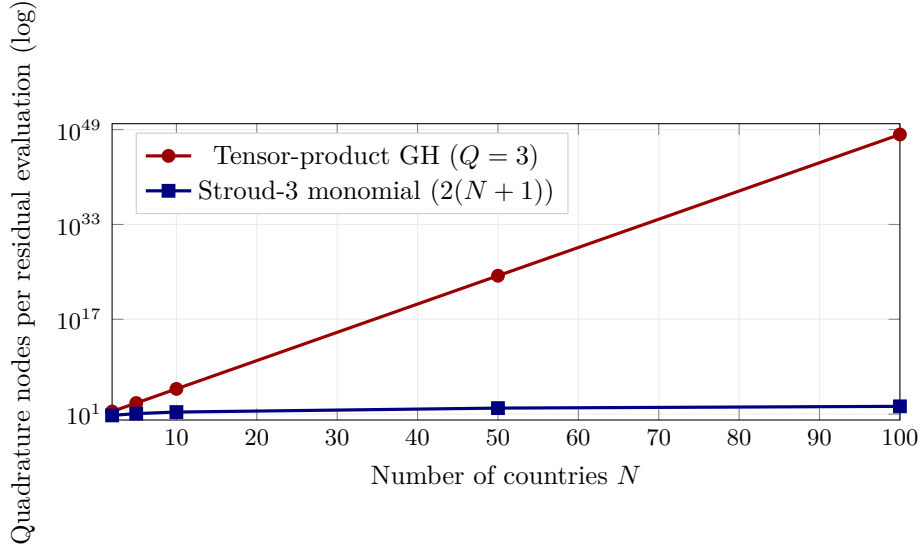

\begin{figure}[ht]
\centering
\begin{tikzpicture}[
    layer/.style={rectangle, draw=uzhblue, thick, fill=uzhgreylight,
        minimum width=2.0cm, minimum height=0.7cm, font=\footnotesize,
        rounded corners=3pt, inner sep=3pt},
    lossbox/.style={rectangle, draw=darkred, thick, fill=red!6,
        minimum width=2.4cm, minimum height=0.55cm, font=\scriptsize,
        rounded corners=3pt, inner sep=2pt, align=center},
    arr/.style={-{Stealth[length=2.5mm]}, thick, uzhblue},
    larr/.style={-{Stealth[length=2mm]}, thick, darkred}
]
    \node[layer, fill=blue!8] (in) at (0,0) {Input: $2N$};
    \node[layer] (h1) at (2.8,0) {Dense 64};
    \node[layer] (h2) at (5.6,0) {Dense 64};
    \node[layer, fill=red!8] (out) at (8.4,0) {Output: $2N\!+\!1$};
    \draw[arr] (in) -- (h1);
    \draw[arr] (h1) -- (h2);
    \draw[arr] (h2) -- (out);
    \node[below=0.15cm, font=\tiny] at (h1.south) {Swish};
    \node[below=0.15cm, font=\tiny] at (h2.south) {Swish};
    \node[below=0.15cm, font=\tiny, darkred] at (out.south) {Softplus};
    \node[lossbox] (eul) at (11.6, 0.85) {Euler$^j$};
    \node[lossbox] (arc) at (11.6, 0.05) {ARC};
    \node[lossbox] (fb)  at (11.6,-0.75) {FB$^j$};
    \node[lossbox, fill=red!12, minimum width=1.4cm] (sum) at (13.7, 0.05) {$\ell_\rho$};
    \draw[larr] (out.east) -- (eul.west);
    \draw[larr] (out.east) -- (arc.west);
    \draw[larr] (out.east) -- (fb.west);
    \draw[larr] (eul.east) -- (sum.west);
    \draw[larr] (arc.east) -- (sum.west);
    \draw[larr] (fb.east)  -- (sum.west);
\end{tikzpicture}
\caption{Reference network architecture used for the $N$-country IRBC model.  The diagram shows the irreversible companion notebook (\texttt{lecture\_04\_02\_IRBC\_DEQN\_irreversible.ipynb}): two hidden layers of 64 Swish units mapping the $2N$-dimensional state to a $2N+1$-dimensional output ($N$ capital choices, the resource-constraint multiplier $\lambda$, and the $N$ irreversibility multipliers $\mu^j$); softplus on the $\lambda$ and $\mu^j$ heads enforces non-negativity, and capital choices use the bounded growth head described below.  The smooth-benchmark companion (\texttt{lecture\_04\_01\_IRBC\_DEQN\_smooth.ipynb}) drops the $\mu^j$ block, leaving an $N+1$-dimensional output head and no Fischer--Burmeister residual; in both notebooks the capital head is parameterized as the bounded log-growth $k_{t+1}^j = k_t^j\exp\{\bar g\,\tanh r_j(\bm s)\}$ (smooth) or the additive form $k_{t+1}^j = (1-\delta)k_t^j + \mathrm{softplus}(r_j)$ (irreversible), both of which keep $k_{t+1}^j > 0$ by construction.}
\label{fig:irbc_nn_arch}
\end{figure}
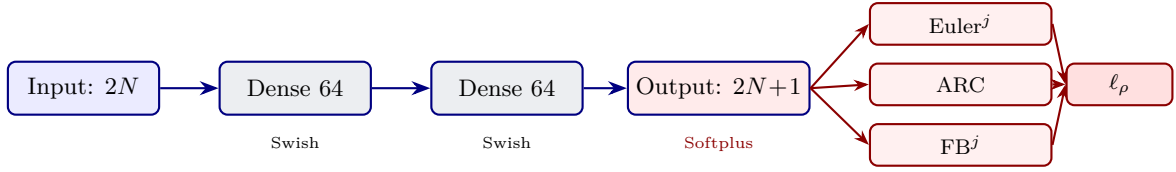

The neural network maps the full state vector $\bm{s} = (k^1,\ldots,k^N, z^1,\ldots,z^N) \in \R^{2N}$ to all $2N+1$ policy variables $(k^{1\prime},\ldots,k^{N\prime}, \lambda, \mu^1,\ldots,\mu^N)$ simultaneously through the small Swish--softplus network in Figure~\ref{fig:irbc_nn_arch}.

The hidden layers use the Swish activation $\mathrm{swish}(x) = x \cdot \sigma(x)$, while the output layer employs the softplus function $\ln(1+e^x)$ to keep the multipliers and capital choice positive.  Two approximation caveats deserve emphasis.  First, $\mathrm{softplus}(x) > 0$ for all $x$, so the multipliers $\mu^j$ are strictly positive rather than exactly zero when the constraint is slack; complementarity is enforced only approximately.  Second, irreversibility requires $I^j = k^{j\prime} - (1-\delta)k^j \geq 0$; a softplus on $k^{j\prime}$ alone does \emph{not} enforce this, since the network can output a positive $k^{j\prime}$ that nonetheless implies negative investment.  A cleaner alternative is to output investment directly via $I^j = \mathrm{softplus}(r^j)$ and set $k^{j\prime} = (1-\delta)k^j + I^j$, which hard-enforces the constraint by construction.

The total DEQN loss aggregates the equilibrium conditions.  In the smooth benchmark (companion notebook \tpath{lecture_04_01_IRBC_DEQN_smooth.ipynb}) only the Euler and aggregate-resource-constraint residuals appear:
\begin{equation}
\ell^{\mathrm{smooth}}_\rho = \frac{1}{N_s} \sum_{i=1}^{N_s} \left[
\sum_{j=1}^{N} \bigl(\mathrm{Euler}^j(\bm{s}_i)\bigr)^2
+ \bigl(\mathrm{ARC}(\bm{s}_i)\bigr)^2
\right].
\label{eq:irbc_loss_smooth}
\end{equation}
The irreversibility extension (companion notebook \tpath{lecture_04_02_IRBC_DEQN_irreversible.ipynb}) augments \eqref{eq:irbc_loss_smooth} with the Fischer--Burmeister complementarity block:
\begin{equation}
\ell^{\mathrm{irrev}}_\rho = \ell^{\mathrm{smooth}}_\rho \;+\; \frac{1}{N_s}\sum_{i=1}^{N_s} \sum_{j=1}^{N} \bigl(\mathrm{FB}^j(\bm{s}_i)\bigr)^2,
\label{eq:irbc_loss}
\end{equation}
where $N_s$ is the number of training states.  When the individual loss components differ in magnitude across countries (which is typical when countries differ in size or calibration), an adaptive loss-balancing scheme from Chapter~\ref{ch:nas} (e.g., ReLoBRaLo, SoftAdapt, GradNorm) can be applied to reweight the components during training.

\paragraph{Representative implementation.}  The architecture is a 2-hidden-layer Swish network with a softplus output head.  In the smooth benchmark the head has dimension $N + 1$ (the $N$ capital choices and the resource-constraint multiplier $\lambda$); in the irreversible extension the head expands to $2N + 1$, adding the irreversibility multipliers $\mu^j \ge 0$ (softplus enforces non-negativity by construction).  Only the irreversible loss carries a non-textbook line, the Fischer--Burmeister smoothing of the complementarity $0 \le \mu^j \perp I^j \ge 0$:
\begin{lstlisting}[caption={Fischer--Burmeister smoothing of $\mu \perp I$ (irreversible companion notebook only).}, label=lst:irbc_fb]
def fischer_burmeister(mu, I, eps=1e-4):
    return mu + I - tf.sqrt(mu**2 + I**2 + eps**2)
\end{lstlisting}
This residual is then squared elementwise and averaged across the mini-batch and across the $N$ countries, in line with the squared-residual treatment of the Euler and ARC blocks; that elementwise square is what makes the gradient field push iterates toward the complementarity axes (see Figure~\ref{fig:fb_zeroset}).
Inside the per-batch cost function of the irreversible notebook, this residual is squared and averaged alongside the Euler-equation residual (whose conditional expectation is handled by the Stroud-3 monomial rule of \S\ref{sec:monomial_cubature} -- $2(N+1)$ nodes for the $N$ idiosyncratic and one aggregate shock) and the aggregate-resource-constraint residual.  The smooth companion implements the same \tpath{compute_cost} pipeline with the $\mu^j$ outputs and the FB block removed.

\section{Persistent-Simulation Training}
\label{sec:irbc_persistent_simulation}

The companion notebooks train the IRBC DEQN with a single training pipeline: a continuing ensemble of stochastic trajectories that evolves alongside the policy network.  There is no Phase~1 / Phase~2 switch and no reset to the steady state between training segments.

\begin{definitionbox}[Persistent-Simulation Training]
Maintain a vector of $M$ stochastic trajectory heads $\bm{X}^{(1)}_t,\ldots,\bm{X}^{(M)}_t$.  Each \emph{training segment} simulates these heads forward for $T$ stochastic periods under the \emph{current} policy network, flattens the simulated states into a training cloud of size $M\cdot T$, performs a fixed number of SGD passes on that cloud, and then continues from the segment's terminal states $\bm{X}^{(m)}_{t+T}$.  The trajectory ensemble therefore co-evolves with the policy and is never reset to the steady state.
\end{definitionbox}

What makes the single-pipeline approach feasible is that both companion notebooks parameterize the policy so that capital cannot leave the feasible set, even at random initialization.  In the smooth notebook the network outputs a bounded log-growth term, $k_{t+1}^j = k_t^j\exp\{\bar g\,\tanh r_j(\bm{s})\}$, which keeps $k_{t+1}^j$ strictly positive and per-period capital growth bounded by $\exp\{\pm\bar g\}$.  In the irreversible notebook the policy network outputs an investment fraction shaped by a sigmoid head and the law of motion $k_{t+1}^j = (1-\delta)k_t^j + I^j$ is hard-coded with $I^j \ge 0$.  Either choice removes the reason historical implementations needed a uniform-sampling burn-in: the simulation cannot diverge.

A \tpath{SAMPLING_MODE} switch (\tpath{simulation} vs \tpath{exogenous}) is exposed for ablation studies and debugging, exogenous sampling on a wide box can be useful to confirm that a finding is not an artefact of the ergodic set, but the default \tpath{simulation} mode runs for the entire training horizon without a phase change.

A typical schedule on the two-country benchmark uses $M = 10$ trajectories of length $T = 256$ per segment, a batch size of $256$, and one or a small number of optimizer passes per segment, with Adam at learning rate $\eta \sim 10^{-3}$ and a cosine decay; convergence is read off the diagnostics of the next section rather than off a phase-transition criterion.  As a budgeting reference, the companion notebooks typically run on the order of $200$--$500$ training segments before mean Euler errors drop below $10^{-3}$ on a held-out trajectory.

\begin{remarkbox}[Why persistent-simulation training works]
Three properties make the single-pipeline approach robust.  First, the bounded capital-growth heads of the previous section keep the simulation feasible at every weight setting, so there is no need for a separate uniform-sampling warm-up phase to prevent the trajectories from diverging.  Second, because the training cloud co-evolves with the policy, the network is always trained on states drawn from the current policy's ergodic distribution, which is the same distribution out-of-sample evaluation will face; there is no train/test distributional shift.  Third, the lack of a phase-transition criterion makes the protocol model-agnostic: scaling from $N=2$ to $N=10$ requires only changing the network's input dimension and the number of equations in the loss, not redesigning the training schedule.  The trade-off is that early-training states reflect a poor and rapidly changing policy, so a small replay buffer or generous mini-batch size is helpful to keep the gradient signal stable.
\end{remarkbox}

\section{Results and Scalability}

The DEQN approach has been successfully applied to IRBC models with up to $N=100$ countries (200 state variables, 201 policy outputs), producing equilibrium errors below $10^{-3}$ in all Euler equations, a level comparable to the best existing solution methods at a fraction of the computational cost, while substantially mitigating curse-of-dimensionality effects in practice.

\paragraph{Convergence diagnostics.}  The quality of the DEQN solution is assessed using several complementary diagnostics:
\begin{enumerate}[itemsep=2pt]
\item \textbf{Euler equation errors:} For each country $j$, compute $\max_{\bm{s} \in \mathcal{S}_\mathrm{test}} |\mathrm{Euler}^j(\bm{s})|$.  Errors below $10^{-3}$ indicate that the optimality condition is violated by less than 0.1\% of consumption, an acceptable tolerance for most applications.
\item \textbf{Resource constraint residual:} Verify that $|\mathrm{ARC}(\bm{s})| < 10^{-4}$ on the test set.
\item \textbf{Complementarity check (irreversible companion only):} Confirm that $\mathrm{FB}^j \approx 0$ and that the multiplier $\mu^j$ is positive only when investment is at its lower bound.
\item \textbf{Economic diagnostics:} Verify that the ergodic distribution of capital, output, and consumption has sensible properties (e.g., positive trade balances for productive countries, capital flowing to high-productivity states).
\item \textbf{Policy-drift / time-invariance check:} Evaluate the policy on a fixed anchor cloud \tpath{X_anchor} after each monitoring interval and report \tpath{policy_drift_rms} and \tpath{policy_drift_max}.  The architecture has no calendar-time input, so any fixed weight vector is a stationary recursive policy by construction; the empirical question is whether SGD has stopped moving the policy function.  The run is treated as time-invariant once both drift statistics fall below the prescribed tolerances \tpath{TIME_INVARIANCE_TOL_RMS} and \tpath{TIME_INVARIANCE_TOL_MAX}.
\item \textbf{Zero-shock stochastic steady state (SSS):}  Iterate the learned policy from \tpath{ZERO_SHOCK_N_STARTS} dispersed feasible starts with all shocks set to zero.  A well-trained policy converges to a common point with $I^j \approx \delta\,k^j$ and (in the irreversible case) $\mu^j \approx 0$; the SSS is a fixed point of the learned stochastic policy that is not imposed during training.
\end{enumerate}

\subsection*{Validation Protocol}
To keep the manuscript self-contained, we summarize here the validation diagnostics used for the IRBC model:
\begin{enumerate}[itemsep=2pt]
\item \textbf{Held-out residual table.}  Evaluate mean and max absolute residuals on an out-of-sample test set for each equation block (Euler and ARC always; FB only in the irreversible companion).  In the two-country benchmark, typical values are mean $\sim 10^{-4}$ and max $\sim 10^{-3}$ for Euler/ARC, with smaller FB residuals.
\item \textbf{Euler-side comparison.}  Compare left and right sides of the Euler equation directly on the test set (scatter around the 45-degree line).  Target thresholds are mean relative error below $10^{-3}$ and max relative error below $10^{-2}$.
\item \textbf{Constraint diagnostics (irreversible companion only).}  Verify $I^j \ge 0$ everywhere and that $(\mu^j, I^j)$ lies close to the complementarity axes ($\mu^j \approx 0$ when $I^j > 0$).
\item \textbf{Economic sanity checks.}  Confirm market-wide accounting identities (e.g., trade balances summing to zero), sensible consumption-sharing behavior, and stable ergodic state distributions around economically plausible regions.
\item \textbf{Policy-drift / time-invariance check.}  Track \tpath{policy_drift_rms} and \tpath{policy_drift_max} on a fixed anchor cloud across training segments; flag the run as time-invariant once both drop below the prescribed tolerances.  This check distinguishes ``the policy has stabilized'' from ``the residuals are small''; both are needed for a trustworthy recursive solution.
\item \textbf{Zero-shock stochastic steady state.}  Simulate the learned policy with all shocks set to zero from several dispersed feasible initial states.  Convergence to a single point with $I^j \approx \delta\,k^j$ (and $\mu^j \approx 0$ in the irreversible case) is a coordinate-free sanity check that complements the held-out residual table.
\end{enumerate}
This protocol makes solution quality auditable and comparable across model sizes and network configurations.

\paragraph{Policy function properties.}
The learned policy functions exhibit the expected economic properties.  Consumption sharing follows the Pareto-weight and IES structure in~\eqref{eq:irbc_consumption}: holding the common shadow price $\lambda_t$ fixed, a higher Pareto weight raises country $j$'s consumption, and with heterogeneous IES the consumption ratio varies with $\lambda_t$.  Productivity affects consumption only indirectly through the equilibrium shadow price and the resource constraint, not through a mechanical bilateral ratio $z^j/z^k$.  This is the textbook complete-markets prediction: the cross-country consumption ratio depends on the Pareto weights $\tau^j/\tau^k$ and the IES gap, not on the productivity differential.  The empirical failure of this prediction is the consumption-correlation puzzle introduced in §\ref{sec:irbc_motivation}; a closely related but distinct failure is the Backus--Smith puzzle, which concerns the correlation between relative consumption growth and the real exchange rate, predicted to be near one under complete markets but empirically near zero or even negative.  Any model that aims to reproduce either puzzle has to break some of the assumptions used here (e.g.\ by restricting the asset menu, \citet{heathcote2002financial}, or adding non-traded goods, \citet{heathcote2013international}).  Investment responds procyclically to productivity shocks: a high realization of $z^j$ raises the marginal product of capital in country $j$, triggering increased investment.  When the irreversibility constraint binds ($I^j = 0$), capital cannot be disinvested and the multiplier $\mu^j$ becomes positive; the network learns this regime-switching behavior smoothly through the Fischer--Burmeister loss.  Trade balances adjust to channel resources toward productive countries: positive trade balances (net exports of goods) correspond to countries whose current productivity exceeds the average, and the implied capital flows are consistent with standard international macroeconomic theory.

The key advantage of the DEQN approach is its scaling behavior: while traditional Cartesian grid-based methods \citep{judd1998numerical} exhibit exponential growth in computation time as $N$ increases, and even adaptive sparse grid methods \citep{ECTA:ECTA1716}, which significantly mitigate the curse of dimensionality, become computationally demanding for $N > 10$, DEQN runtimes in our implementations are reported close to linear in $N$ over a broad range of model sizes (see \citet{azinovicDEEPEQUILIBRIUMNETS2022}, Table~2 and surrounding discussion, for timings across $N \in \{2, \ldots, 100\}$).  This favorable empirical scaling arises because the network's parameter count grows roughly linearly (more input/output neurons), while each SGD step avoids state-space grids.  The companion notebooks (\tpath{lecture_04_01_IRBC_DEQN_smooth.ipynb} and \tpath{lecture_04_02_IRBC_DEQN_irreversible.ipynb}) only run the $N=2$ case, so the linear-scaling claim cannot be reproduced from the in-class material; readers who wish to verify it directly should consult the published timings or replicate the larger-$N$ runs from the Azinovic et al.\ codebase.

\paragraph{Comparison with adaptive sparse grids.}  The approach of \citet{ECTA:ECTA1716} handles kinks in the policy function (e.g., those induced by the irreversibility constraint) by \emph{refining} the grid locally around the kink using hierarchical surplus indicators.  This keeps the method accurate but the grid remains anchored to a hypercube, so computation still scales poorly once the number of active kinks or the dimensionality grows.  DEQNs do not represent kinks by grid refinement; instead, they fit a smooth approximator (Swish/softplus network) to the Fischer--Burmeister-regularized problem, which produces a globally smooth policy that tracks the true piecewise structure without needing localized grid points.  The two methods are therefore complementary: adaptive sparse grids give deterministic error bounds on a hypercube; DEQNs give simulation-based error bounds on the ergodic set with no grid at all.  From a theoretical perspective, \citet{montanelli2019deep} establish error bounds showing that deep ReLU networks can approximate functions on sparse grids without the exponential growth in parameters that afflicts classical polynomial methods, providing formal underpinning for why deep learning can mitigate (though not eliminate) the high-dimensional approximation cost.  Exact runtimes depend on architectural choices, quadrature design, and hardware; the robust finding is that the DEQN formulation avoids explicit tensor-product state grids and remains computationally viable in dimensions where standard methods become prohibitively expensive.

Beyond the IRBC setting, closely related neural-equilibrium methods have been applied to other policy-relevant problems.  \citet{nuno2024monetary} use DEQNs to compute optimal \emph{monetary policy rules} under persistent supply shocks, replacing the linearization step around steady state with a globally trained policy network.  \citet{bretscherRicardianBusinessCycles2022} apply DEQN to multi-country international real business cycles with comparative advantage.  Most recently, \citet{azinovicyangzemlicka2025sequencespace} replace the endogenous cross-sectional state with a \emph{truncated history of exogenous aggregate shocks} (the sequence-space representation), so that the network's input dimension scales with the truncation horizon rather than with the number of agents, which is the heterogeneous-agent extension developed in Chapter~\ref{ch:young}.

\begin{keyinsightbox}[Chapter Summary]
\begin{itemize}[itemsep=2pt, leftmargin=*]
\item The IRBC model is the standard testbed for high-dimensional global solution methods: $N$ countries each with capital and TFP push the state dimension to $2N$ in the planner formulation used here, and the irreversibility constraint introduces non-smooth kinks via Karush--Kuhn--Tucker complementarity.
\item Fischer--Burmeister smoothing converts the non-differentiable irreversibility complementarity $0 \le \mu^j \perp I^j \ge 0$ into a smooth squared residual $\Phi_\varepsilon^2$ that is compatible with SGD.
\item Persistent-simulation training, a single continuing ensemble of stochastic trajectories, never reset to steady state, is the recipe used in the companion notebooks; bounded policy parameterizations ($k_{t+1}^j = k_t^j\exp\{\bar g\tanh r_j(\bm{s})\}$ in the smooth notebook, $k_{t+1}^j = (1-\delta)k_t^j + I^j$ with $I^j \ge 0$ in the irreversible notebook) keep the simulation feasible without a separate burn-in phase.
\item Time-invariance via policy drift on a fixed anchor cloud, and a zero-shock stochastic-steady-state check from dispersed feasible starts, are the two convergence diagnostics specific to recursive DEQN training; both are run inside the companion notebooks.
\item Gauss--Hermite tensor-product quadrature (\S\ref{sec:gh_tensor_product}, introduced in the previous chapter) handles expectations in low-dimensional shock spaces; once $N \gtrsim 5$ the linear-scaling Stroud-3 monomial rule (\S\ref{sec:monomial_cubature}) is the workhorse for IRBC, and is also what the companion notebooks use for $N=2$, with QMC (\S\ref{sec:qmc_cdf}) and sparse grids reserved for high-accuracy or very high-dimensional cases.
\end{itemize}
\end{keyinsightbox}

\section*{Further Reading}
\addcontentsline{toc}{section}{Further Reading}
\begin{itemize}[itemsep=2pt]
\item \citet{ECTA:ECTA1716}, adaptive sparse grids for IRBC, the classical-method benchmark this chapter contrasts with.
\item \citet{pichler2011}, an IRBC-specific application of the monomial rule of \S\ref{sec:monomial_cubature}, useful as a sanity check for the multi-country setting.
\item \citet{niederreiter1992random}, the standard reference for quasi-Monte Carlo and low-discrepancy sequences for the high-dimensional integrals encountered at large $N$.
\item \citet{nuno2024monetary}, a recent DEQN application to optimal monetary policy.
\end{itemize}

\section*{Exercises}
\addcontentsline{toc}{section}{Exercises}
\noindent Worked solutions and guidance for these exercises appear in Appendix~\ref{app:solutions}.
\begin{enumerate}[itemsep=4pt, leftmargin=*]
\item\label{ex:ch3:1} \textbf{[Core] Fischer--Burmeister.}  (a)~Show that $\Phi(a,b) = a + b - \sqrt{a^2+b^2} = 0 \iff a\ge 0,\, b\ge 0,\, ab=0$.  (b)~Plot the level set $\Phi(a,b)=0$ and a smoothed variant $\Phi_\varepsilon(a,b) = a + b - \sqrt{a^2+b^2+\varepsilon^2}$ for $\varepsilon=10^{-3}$ in the $(a,b)$ plane.  (c)~Compute the gradient $\nabla\Phi(a,b) = (1 - a/\sqrt{a^2+b^2},\, 1 - b/\sqrt{a^2+b^2})$ and evaluate it at $(a,b) = (1,1)$.  Note that the raw gradient $\nabla\Phi$ at $(1,1)$ equals $(1-1/\sqrt 2,\, 1-1/\sqrt 2) > 0$ and therefore points \emph{away} from the complementarity axes (northeast).  Now show that gradient descent on the \emph{squared} residual $\Phi^2$ instead points toward the zero set: in the open positive quadrant $\Phi(a,b) > 0$, so $-\nabla\Phi^2 = -2\Phi\,\nabla\Phi$ points southwest, back to the L-shape.  (d)~Explain in two sentences why replacing $\Phi$ by $-\Phi$ has no effect on the squared loss: $\Phi^2 = (-\Phi)^2$, so the gradient field of $\Phi^2$ is unchanged.  In particular, the sign convention $a+b-\sqrt{a^2+b^2}$ versus $\sqrt{a^2+b^2}-a-b$ is irrelevant once the residual is squared, and the operative quantity for SGD is $-\nabla(\Phi^2)$ rather than $-\nabla\Phi$.
\item\label{ex:ch3:2} \textbf{[Core] State-space scaling.}  For an IRBC with $N$ symmetric countries, write down the state, policy, and equation dimensions.  At what $N$ does a tensor-product Gauss--Hermite rule with $Q=3$ nodes per dimension exceed $10^4$ evaluations per Euler equation?  At what $N$ does the Stroud-3 monomial rule of \S\ref{sec:monomial_cubature} stay below $100$ nodes?
\item\label{ex:ch3:3} \textbf{[Core] Persistent-simulation feasibility.}  The companion notebooks parameterize next-period capital as $k_{t+1}^j = k_t^j\exp\{\bar g\,\tanh r_j(\bm{s}_t)\}$ in the smooth model and $k_{t+1}^j = (1-\delta)k_t^j + I^j$ with $I^j\ge 0$ in the irreversible model.  (a)~Show that under either parameterization the simulated capital stock cannot leave the feasible set $\{k>0\}$ even when $r_j$ is generated by a randomly initialized network.  (b)~Contrast with the naive head $k_{t+1}^j = \mathrm{softplus}(r_j)$: construct a $\bm{s}_t$ at which a random initialization can drive the simulation arbitrarily far from any economically sensible region in one step.  (c)~Explain in two sentences why feasibility-by-construction is what allows the notebooks to dispense with a separate uniform-sampling burn-in phase.
\item\label{ex:ch3:4} \textbf{[Core] Adjustment-cost partials and Tobin's Q.}  Starting from the quadratic adjustment cost~\eqref{eq:irbc_adjcost}, $\Gamma^j = (\kappa/2)\,k^j (k^{j\prime}/k^j - 1)^2$, derive the partial derivatives $\partial\Gamma^j/\partial k^{j\prime}$ and $\partial\Gamma^j/\partial k^j$ shown in~\eqref{eq:irbc_adjcost_derivs}.  Show that both partials vanish at the steady state $k^{j\prime} = k^j$, so the deterministic steady-state allocation is identical to the one of the frictionless model.  Define the \emph{net investment rate} $g^j = k^{j\prime}/k^j - 1$ and re-express the marginal adjustment cost $\partial\Gamma^j/\partial k^{j\prime}$ as $\kappa\,g^j$; this is the Tobin's-Q-style wedge that the planner must balance against the marginal product of capital in the Euler equation.  Discuss qualitatively how raising $\kappa$ changes the speed of convergence to the steady state under a positive productivity shock.
\item\label{ex:ch3:5} \textbf{[Core] Complete-markets risk sharing.}  Use the consumption-sharing condition~\eqref{eq:irbc_consumption} to derive the cross-country consumption ratio $c_t^i / c_t^j$ as a function of the Pareto weights $(\tau^i, \tau^j)$, the IES parameters $(\gamma_i, \gamma_j)$, and the shadow price $\lambda_t$.  (i)~Show that with homogeneous preferences ($\gamma_i = \gamma_j = \gamma$), the ratio $c_t^i/c_t^j$ is \emph{constant in time}, reproducing the perfect-risk-sharing result of \citet{backus1992international}.  (ii)~Show that with heterogeneous IES, the ratio fluctuates with $\lambda_t$, but consumption growth rates are still scalar multiples of the same aggregate shadow-price growth when all $\gamma_j>0$.  What does this imply for cross-country consumption-growth correlations in this planner allocation?  (iii)~Explain in two sentences why heterogeneous IES alone does not resolve the empirical consumption-correlation puzzle.
\item\label{ex:ch3:6} \textbf{[Core] Notebook: steady-state comparative statics.}  Open \tpath{lecture_05_05_IRBC_Exercise.ipynb} (it lives in the Lecture-05 code folder, since it doubles as the entry point to the NAS / loss-normalization material of Chapter~\ref{ch:nas}).  Working from the closed-form deterministic steady state, in which the Euler condition pins $\mathrm{MPK} = 1/\beta$ and hence $k_\mathrm{ss} = \bigl[(1/\beta - 1 + \delta)/(\zeta A_\mathrm{tfp})\bigr]^{1/(\zeta-1)}$, compute $k_\mathrm{ss}$ and $c_\mathrm{ss}$ for (a)~higher depreciation $\delta = 0.05$, (b)~a more impatient household $\beta = 0.95$, and (c)~a lower capital share $\zeta = 0.30$, each relative to the baseline.  Predict the sign of each change before computing, and explain why $k_\mathrm{ss}$ falls in every case.  (The notebook deliberately uses a standalone calibration $A_\mathrm{tfp} = 1$; the IRBC training notebook \tpath{lecture_04_01_IRBC_DEQN_smooth.ipynb} instead pins $A_\mathrm{tfp}$ so that $k_\mathrm{ss} = 1$, a rescaling that does not change any of the qualitative conclusions.)
\item\label{ex:ch3:7} \textbf{[Computational] Notebook: loss-component weighting.}  In the same notebook, take a snapshot of the IRBC training loss with five components (two Euler residuals, the aggregate resource constraint, and two Fischer--Burmeister complementarity residuals) whose magnitudes span several orders ($\ell_\mathrm{ARC} \sim 5$, $\ell_\mathrm{FB} \sim 10^{-4}$).  Replace the equal weights $w_i = 1$ with inverse-loss weights $w_i = (1/\ell_i)/\sum_j (1/\ell_j)$, verify that every weighted contribution $w_i \ell_i$ is then equal, and identify which component receives the most weight (and why).  Finally, on the supplied pair of (synthetic) equal-weight vs.\ inverse-weight training curves, plot both on a log axis and quantify the convergence speed-up.  In which regime do you expect inverse-loss weighting to help most, and when can it hurt; think of a component that is small because it is genuinely satisfied by construction, e.g., a hard-coded resource constraint?
\end{enumerate}

\chapter{Neural Architecture Search and Loss Normalization}
\label{ch:nas}

The DEQN models of Chapters~\ref{ch:deqn}--\ref{ch:irbc} involve several hyperparameters (network depth, width, activation functions, learning rate) and multi-component loss functions whose relative scales can differ by orders of magnitude.  To fix ideas, even a modest sweep over depth $\in\{1,\ldots,10\}$ and width $\in\{16, 32, 64, 128, 256, 512\}$ on the companion NAS regression task already spans $10\times 6 = 60$ configurations, and the production sweep below (six axes) reaches $\sim 3{,}000$; on that task (illustrative numbers), the best mean absolute error ($\approx 3\times 10^{-3}$) is attained at a $5\times 256$ network, while $10\times 512$ overfits by almost an order of magnitude ($\approx 2\times 10^{-2}$).  Hand-tuning at this scale is infeasible.  This chapter addresses both challenges.  We first survey hyperparameter-search methods (random search, Bayesian optimization, Hyperband, and BOHB, which combines TPE with Hyperband, \citep{bergstra2012random, snoek2012practical, jamieson2016nonstochastic, li2018hyperband, falkner2018bohb, garnett2023bayesian}, and then develop a classroom-friendly version of the ReLoBRaLo algorithm \citep{bischof2025relobralo} for adaptive multi-objective loss balancing.  In the notebooks, this is implemented as a deterministic convex blend of step-wise and baseline loss comparisons, which keeps the code compact while preserving the balancing intuition.

A terminology note before we begin: in this chapter we use ``NAS'' loosely to cover both \emph{hyperparameter optimization} (HPO; choosing widths, depths, activations, learning rates from a fixed parameterization) and ``true'' NAS in the sense of \citet{elsken2019neural}, where the network's wiring graph itself is searched (e.g.\ regularized evolution \citep{real2019regularized}).  The two literatures share methodology (Random / Bayesian / Hyperband-style search) but differ in scope.  All four methods discussed here are HPO; for graph-level NAS, the textbook reference is \citet{hutter2019automl}.  \citet{elsken2019neural} provide the canonical survey; the local copy in \texttt{readings/} is recommended as the first deep-dive reference.

\paragraph{Hands-on notebooks for this chapter.}
Two NAS walkthroughs are provided alongside the ReLoBRaLo notebook, plus the IRBC exercise notebook that doubles as the entry point to this chapter.  All four live in the NAS chapter's code folder:
\begin{itemize}
    \item \tpath{02_NAS_Random_Search_10D.ipynb}: a library-free Random Search loop (model in TF/Keras) on a 10-dimensional analytical regression task, used to illustrate the projection argument of \citet{bergstra2012random} in its cleanest form.
    \item \tpath{03_NAS_RandomSearch_Hyperband.ipynb}: from-scratch Random Search and Successive Halving (Hyperband's inner loop) on a two-dimensional Genz Gaussian, written in $\sim$25 lines of plain Python so the algorithms in this chapter are visible without library abstraction; after the first run, the cached records in \tpath{nas_results/} short-circuit re-runs for instant re-inspection.
    \item \tpath{04_Loss_Normalization.ipynb}: the classroom ReLoBRaLo implementation, matched to the notation below.
    \item \tpath{05_IRBC_Exercise.ipynb}: the IRBC exercise notebook (closed-form steady-state comparative statics and inverse-loss weighting on a multi-component IRBC residual); it is the notebook referenced by Chapter~\ref{ch:irbc} Exercises~\ref{ex:ch3:6}--\ref{ex:ch3:7}, and it reuses the loss-balancing ideas of this chapter on a deliberately small, library-free example.
\end{itemize}

\section{The Hyperparameter Space}

The performance of a neural network depends sensitively on its architecture (number of layers, neurons per layer, activation functions) and training configuration (learning rate, batch size, optimizer, regularization).  Choosing these hyperparameters by hand is tedious and often suboptimal.

To appreciate the scale of the search problem, consider as a stylized example a typical DEQN setup where we must select: the number of hidden layers ($L \in \{2, 3, 4, 5\}$), neurons per layer ($n \in \{32, 64, 128, 256\}$), activation function (ReLU, Swish, Tanh), learning rate ($\eta \in \{10^{-4}, 5\times 10^{-4}, 10^{-3}, 5\times 10^{-3}\}$), batch size ($B \in \{64, 128, 256, 512\}$), and weight decay ($\lambda \in \{0, 10^{-5}, 10^{-4}, 10^{-3}\}$).  The total number of configurations is $4 \times 4 \times 3 \times 4 \times 4 \times 4 = 3{,}072$.  If each configuration requires 30 minutes to train, exhaustive evaluation would take 64 days on a single GPU.  With additional choices (optimizer type, learning rate schedule, dropout rate), the space easily exceeds $10^4$ configurations.  (The slide deck uses a slightly larger illustrative space, $5 \times 8 \times 3 \times 20 \times 4 = 9{,}600$ configurations, for the same point.)

\section{Grid Search}

The simplest approach is to define a grid of values for each hyperparameter and evaluate all combinations.  For $d$ hyperparameters, each taking $k$ values, the cost is $\mathcal{O}(k^d)$, the same exponential scaling that plagues grid-based PDE solvers.  Grid search is deterministic and easy to implement, but it has a fundamental flaw: it allocates the same density of evaluations to all hyperparameters, including those to which performance is insensitive.  If only 2 out of 6 hyperparameters matter (which is typical in practice), the remaining 4 dimensions contribute only wasted computation.

\section{Random Search}
\label{sec:nas_random_search}

\citet{bergstra2012random} demonstrated that random sampling of hyperparameter configurations often outperforms grid search, particularly when only a few hyperparameters are important.  The key insight is a \emph{projection argument}: when a random configuration is projected onto any single hyperparameter axis, the marginal distribution covers the entire range densely, regardless of how many other hyperparameters exist.  In contrast, a grid with the same total number of evaluations provides only $k = N^{1/d}$ distinct values per axis, which can be very coarse in high dimensions.

\begin{figure}[ht]
\centering
\begin{tikzpicture}[scale=1.0]
    \draw[uzhblue, thick] (0,0) rectangle (3.6,3.6);
    \node[above, font=\small\bfseries] at (1.8,3.7) {Grid: $3\times 3$ (9 points)};
    \foreach \x in {0.6,1.8,3.0}
      \foreach \y in {0.6,1.8,3.0}
        \fill[softblue] (\x,\y) circle (2.5pt);
    \draw[uzhgreydark, thick] (0,-0.4) -- (3.6,-0.4);
    \foreach \x in {0.6,1.8,3.0}
        \fill[softblue] (\x,-0.4) circle (2pt);
    \node[below, font=\scriptsize] at (1.8,-0.7) {only \textbf{3 distinct values} of the relevant axis};
    \node[below, font=\scriptsize] at (1.8,-1.1) {\emph{important} hyperparameter $\longrightarrow$};
    \node[rotate=90, font=\scriptsize] at (-0.4,1.8) {\emph{unimportant} hyperparameter};

    \begin{scope}[xshift=6cm]
    \draw[uzhblue, thick] (0,0) rectangle (3.6,3.6);
    \node[above, font=\small\bfseries] at (1.8,3.7) {Random: 9 points};
    \foreach \p in {(0.35,2.65),(0.95,0.55),(1.4,2.95),(2.1,1.55),(2.55,0.85),(2.95,3.15),(0.75,1.9),(3.2,2.3),(1.75,0.25)}
        \fill[darkred] \p circle (2.5pt);
    \draw[uzhgreydark, thick] (0,-0.4) -- (3.6,-0.4);
    \foreach \x in {0.35,0.75,0.95,1.4,1.75,2.1,2.55,2.95,3.2}
        \fill[darkred] (\x,-0.4) circle (2pt);
    \node[below, font=\scriptsize] at (1.8,-0.7) {\textbf{9 distinct values} of the relevant axis};
    \node[below, font=\scriptsize] at (1.8,-1.1) {\emph{important} hyperparameter $\longrightarrow$};
    \end{scope}
\end{tikzpicture}
\caption{Why random search beats grid search when only one hyperparameter matters.  Both designs spend the same budget of nine evaluations on a two-dimensional space in which only the horizontal axis affects performance; the vertical axis is a ``nuisance'' hyperparameter.  Project each design onto that important axis (the strip below each panel): the $3\times 3$ grid stacks its nine points into only three columns, so it probes just three distinct values of the parameter that matters, whereas the random design lands on nine distinct values.  Equivalently, with a budget of $N$ points in $d$ dimensions a grid resolves only $N^{1/d}$ values per axis no matter which axes matter, while a random design resolves $N$ values per axis in the marginal sense.  Since in practice only two or three of many hyperparameters typically drive performance \citep{bergstra2012random}, the random design extracts far more information about the dimensions that count for the same compute.}
\label{fig:grid_vs_random}
\end{figure}
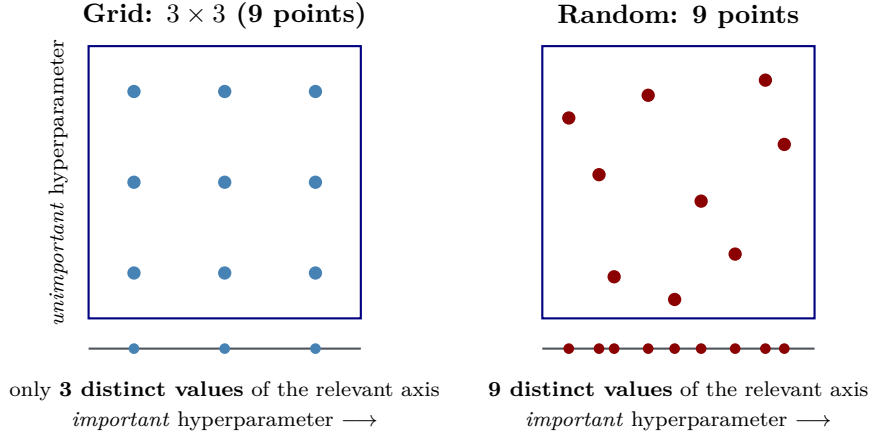

For example, with a budget of $N = 60$ evaluations in $d = 6$ dimensions, a grid provides only $60^{1/6} \approx 2$ values per hyperparameter, while random search provides 60 distinct values per hyperparameter (in the marginal sense).  This makes random search much more likely to find good values for the hyperparameters that matter most.  Figure~\ref{fig:grid_vs_random} shows the same projection argument in two dimensions.

\section{Bayesian Optimization}

Bayesian optimization treats the validation loss $\ell(\bm{h})$ as an expensive black-box function of the hyperparameter vector $\bm{h}$, and uses a probabilistic surrogate model, typically a Gaussian process (see Chapter~\ref{ch:gp}) \citep{snoek2012practical, garnett2023bayesian}, to guide the search.  After evaluating $n$ configurations $\{(\bm{h}_i, \ell_i)\}_{i=1}^n$, the GP posterior provides both a prediction $\mu(\bm{h})$ and an uncertainty estimate $\sigma(\bm{h})$ at any untried configuration.

\begin{figure}[ht]
\centering
\begin{tikzpicture}
    \begin{axis}[
        name=gp,
        width=12cm, height=5cm,
        xmin=0, xmax=10, ymin=-1.3, ymax=1.0,
        xtick=\empty, ytick=\empty,
        ylabel={validation loss $\ell(h)$},
        axis lines=left,
        legend style={font=\scriptsize, at={(0.5,1.05)}, anchor=south,
                      draw=none, fill=none, legend columns=-1,
                      /tikz/every even column/.append style={column sep=0.5cm}},
        clip=true,
    ]
    \addplot[name path=UP, draw=none] coordinates {
      (0.000,+0.7045) (0.125,+0.6214) (0.250,+0.5309) (0.375,+0.4346) (0.500,+0.3347) (0.625,+0.2679) (0.750,+0.3315) (0.875,+0.3816) (1.000,+0.4166) (1.125,+0.4350) (1.250,+0.4359) (1.375,+0.4190) (1.500,+0.3846) (1.625,+0.3337) (1.750,+0.2677) (1.875,+0.1885) (2.000,+0.0984) (2.125,+0.0001) (2.250,-0.1036) (2.375,-0.1048) (2.500,-0.0299) (2.625,+0.0506) (2.750,+0.1338) (2.875,+0.2172) (3.000,+0.2983) (3.125,+0.3747) (3.250,+0.4445) (3.375,+0.5058) (3.500,+0.5573) (3.625,+0.5977) (3.750,+0.6262) (3.875,+0.6421) (4.000,+0.6451) (4.125,+0.6350) (4.250,+0.6120) (4.375,+0.5763) (4.500,+0.5286) (4.625,+0.4697) (4.750,+0.4008) (4.875,+0.3233) (5.000,+0.2390) (5.125,+0.1500) (5.250,+0.0584) (5.375,-0.0331) (5.500,-0.1221) (5.625,-0.0245) (5.750,+0.0683) (5.875,+0.1538) (6.000,+0.2297) (6.125,+0.2939) (6.250,+0.3446) (6.375,+0.3802) (6.500,+0.4000) (6.625,+0.4035) (6.750,+0.3908) (6.875,+0.3627) (7.000,+0.3206) (7.125,+0.2662) (7.250,+0.2019) (7.375,+0.1304) (7.500,+0.0546) (7.625,+0.1401) (7.750,+0.2200) (7.875,+0.2914) (8.000,+0.3518) (8.125,+0.3988) (8.250,+0.4307) (8.375,+0.4464) (8.500,+0.4452) (8.625,+0.4271) (8.750,+0.3927) (8.875,+0.3429) (9.000,+0.2795) (9.125,+0.2043) (9.250,+0.1195) (9.375,+0.0277) (9.500,+0.0797) (9.625,+0.1712) (9.750,+0.2634) (9.875,+0.3542) (10.000,+0.4417)
    };
    \addplot[name path=DN, draw=none] coordinates {
      (0.000,-0.1673) (0.125,-0.0730) (0.250,+0.0202) (0.375,+0.1098) (0.500,+0.1932) (0.625,+0.2336) (0.750,+0.1339) (0.875,+0.0383) (1.000,-0.0503) (1.125,-0.1294) (1.250,-0.1964) (1.375,-0.2495) (1.500,-0.2871) (1.625,-0.3081) (1.750,-0.3122) (1.875,-0.2995) (2.000,-0.2709) (2.125,-0.2276) (2.250,-0.1715) (2.375,-0.2099) (2.500,-0.3161) (2.625,-0.4195) (2.750,-0.5179) (2.875,-0.6093) (3.000,-0.6921) (3.125,-0.7649) (3.250,-0.8270) (3.375,-0.8776) (3.500,-0.9166) (3.625,-0.9437) (3.750,-0.9591) (3.875,-0.9627) (4.000,-0.9549) (4.125,-0.9357) (4.250,-0.9055) (4.375,-0.8643) (4.500,-0.8126) (4.625,-0.7508) (4.750,-0.6796) (4.875,-0.5998) (5.000,-0.5125) (5.125,-0.4193) (5.250,-0.3218) (5.375,-0.2220) (5.500,-0.1221) (5.625,-0.2059) (5.750,-0.2819) (5.875,-0.3479) (6.000,-0.4016) (6.125,-0.4412) (6.250,-0.4655) (6.375,-0.4735) (6.500,-0.4650) (6.625,-0.4402) (6.750,-0.4000) (6.875,-0.3457) (7.000,-0.2794) (7.125,-0.2033) (7.250,-0.1203) (7.375,-0.0332) (7.500,+0.0546) (7.625,-0.0223) (7.750,-0.0971) (7.875,-0.1669) (8.000,-0.2286) (8.125,-0.2797) (8.250,-0.3181) (8.375,-0.3422) (8.500,-0.3508) (8.625,-0.3436) (8.750,-0.3205) (8.875,-0.2823) (9.000,-0.2302) (9.125,-0.1659) (9.250,-0.0912) (9.375,-0.0085) (9.500,-0.0686) (9.625,-0.1669) (9.750,-0.2648) (9.875,-0.3602) (10.000,-0.4512)
    };
    \addplot[softblue!25] fill between[of=UP and DN];
    \addplot[very thick, softblue] coordinates {
      (0.000,+0.2686) (0.125,+0.2742) (0.250,+0.2755) (0.375,+0.2722) (0.500,+0.2640) (0.600,+0.2538) (0.750,+0.2327) (0.875,+0.2100) (1.000,+0.1832) (1.125,+0.1528) (1.250,+0.1197) (1.375,+0.0847) (1.500,+0.0488) (1.625,+0.0128) (1.750,-0.0223) (1.875,-0.0555) (2.000,-0.0862) (2.125,-0.1137) (2.250,-0.1375) (2.300,-0.1459) (2.500,-0.1730) (2.625,-0.1845) (2.750,-0.1921) (2.875,-0.1961) (3.000,-0.1969) (3.125,-0.1951) (3.250,-0.1912) (3.375,-0.1859) (3.500,-0.1796) (3.625,-0.1730) (3.750,-0.1664) (3.875,-0.1603) (4.000,-0.1549) (4.125,-0.1504) (4.250,-0.1467) (4.375,-0.1440) (4.500,-0.1420) (4.625,-0.1406) (4.750,-0.1394) (4.875,-0.1382) (5.000,-0.1368) (5.125,-0.1347) (5.250,-0.1317) (5.375,-0.1275) (5.500,-0.1221) (5.625,-0.1152) (5.750,-0.1068) (5.875,-0.0970) (6.000,-0.0859) (6.125,-0.0736) (6.250,-0.0605) (6.375,-0.0466) (6.500,-0.0325) (6.625,-0.0184) (6.750,-0.0046) (6.875,+0.0085) (7.000,+0.0206) (7.125,+0.0314) (7.250,+0.0408) (7.375,+0.0486) (7.500,+0.0546) (7.625,+0.0589) (7.750,+0.0614) (7.875,+0.0623) (8.000,+0.0616) (8.125,+0.0596) (8.250,+0.0563) (8.375,+0.0521) (8.500,+0.0472) (8.625,+0.0418) (8.750,+0.0361) (8.875,+0.0303) (9.000,+0.0246) (9.125,+0.0192) (9.250,+0.0142) (9.375,+0.0096) (9.400,+0.0087) (9.500,+0.0056) (9.625,+0.0021) (9.750,-0.0007) (9.875,-0.0030) (10.000,-0.0047)
    };
    \addlegendentry{GP mean $\mu(h)$}
    \addplot[thick, dashed, darkred] coordinates {
      (0.000,+0.1238) (0.500,+0.2382) (1.000,+0.2786) (1.500,+0.2082) (2.000,+0.0189) (2.500,-0.2730) (3.000,-0.6221) (3.500,-0.9293) (3.750,-1.0217) (4.000,-1.0505) (4.250,-1.0071) (4.500,-0.8939) (5.000,-0.5230) (5.500,-0.1221) (6.000,+0.1471) (6.500,+0.2334) (7.000,+0.1771) (7.500,+0.0546) (8.000,-0.0577) (8.500,-0.1050) (9.000,-0.0681) (9.500,+0.0315) (10.000,+0.1427)
    };
    \addlegendentry{(unknown) truth}
    \addplot[only marks, mark=*, mark size=2.8pt, black]
      coordinates {(0.60,+0.254) (2.30,-0.146) (5.50,-0.122) (7.50,+0.055) (9.40,+0.009)};
    \addlegendentry{observed $(h_i, \ell_i)$}
    \draw[dashed, gray, thin] (axis cs:0,-0.146) -- (axis cs:10,-0.146);
    \node[font=\scriptsize, gray, anchor=west] at (axis cs:0.2,-0.28) {$\ell^\star$ so far};
    \end{axis}

    \begin{axis}[
        at={($(gp.south)+(0cm,-1.1cm)$)}, anchor=north,
        width=12cm, height=3cm,
        xmin=0, xmax=10, ymin=0, ymax=1.15,
        xtick=\empty, ytick=\empty,
        xlabel={hyperparameter $h$},
        ylabel={$\mathrm{EI}(h)$},
        axis lines=left,
        clip=false,
    ]
    \addplot[thick, softgreen, fill=softgreen!25] coordinates {
      (0.000,+0.013) (0.125,+0.002) (0.250,+0.000) (0.500,+0.000) (0.750,+0.000) (1.000,+0.000) (1.125,+0.004) (1.250,+0.016) (1.375,+0.034) (1.500,+0.055) (1.625,+0.074) (1.750,+0.086) (1.875,+0.086) (2.000,+0.073) (2.125,+0.046) (2.250,+0.007) (2.375,+0.069) (2.500,+0.232) (2.625,+0.383) (2.750,+0.518) (2.875,+0.637) (3.000,+0.738) (3.125,+0.822) (3.250,+0.889) (3.375,+0.939) (3.500,+0.974) (3.625,+0.994) (3.750,+1.000) (3.875,+0.993) (4.000,+0.973) (4.125,+0.942) (4.250,+0.898) (4.375,+0.843) (4.500,+0.777) (4.625,+0.699) (4.750,+0.610) (4.875,+0.511) (5.000,+0.402) (5.125,+0.286) (5.250,+0.165) (5.375,+0.049) (5.500,+0.000) (5.625,+0.028) (5.750,+0.096) (5.875,+0.159) (6.000,+0.208) (6.125,+0.241) (6.250,+0.256) (6.375,+0.253) (6.500,+0.234) (6.625,+0.200) (6.750,+0.154) (6.875,+0.103) (7.000,+0.054) (7.125,+0.017) (7.250,+0.001) (7.500,+0.000) (7.750,+0.000) (7.875,+0.008) (8.000,+0.026) (8.125,+0.050) (8.250,+0.074) (8.375,+0.090) (8.500,+0.098) (8.625,+0.093) (8.750,+0.078) (8.875,+0.055) (9.000,+0.027) (9.125,+0.006) (9.250,+0.000) (9.500,+0.000) (9.625,+0.006) (9.750,+0.048) (9.875,+0.119) (10.000,+0.203)
    };
    \node[font=\footnotesize, darkred, anchor=west] (nextLbl) at (axis cs:5.1,0.82)
          {next: $h_{n+1} = \arg\max \mathrm{EI}$};
    \draw[-{Stealth[length=2mm]}, thick, darkred]
         (nextLbl.west) to[out=180, in=-30] (axis cs:3.85,0.97);
    \end{axis}
\end{tikzpicture}
\caption{Bayesian optimization in one dimension.  All curves come from a genuine Gaussian-process fit (RBF kernel, length scale $\ell = 1.2$, signal variance $\sigma_f^2 = 0.25$); the same setup is reproduced in the companion notebook.  \emph{Top:} the validation loss $\ell(h)$ is treated as an expensive black box.  After five evaluations (black dots) the GP posterior is summarized by its mean $\mu(h)$ (solid blue), which \emph{interpolates} every observation exactly, and by a $\pm 2$ s.d.\ credible band (shaded), which pinches to nearly zero at each observation and balloons in the wide gaps where no data constrain it.  The dashed red curve is the (in practice unknown) true loss; it has a pronounced minimum near $h \approx 4$ that these five points completely miss, and there it even slips below the $\pm 2$ s.d.\ band, an honest reminder that a Gaussian process with a long length scale can be over-confident between observations.  The horizontal grey line marks $\ell^\star$, the best loss observed so far (the point near $h \approx 2.3$).  \emph{Bottom:} the Expected-Improvement acquisition function $\mathrm{EI}(h) = \E{\max\!\bigl(\ell^\star - \ell(h),\, 0\bigr)}$ scores each untried $h$ by how much it is expected to beat $\ell^\star$ under the posterior.  It is essentially zero at the existing observations, where there is nothing to learn, rises in the unexplored gaps, and peaks at $h \approx 3.75$, the place that combines a predicted mean already below $\ell^\star$ with substantial residual uncertainty; that maximizer is selected as the next configuration to evaluate (red arrow).  EI therefore balances \emph{exploitation} (low predicted mean) against \emph{exploration} (high predicted variance), and here it steers the search straight at the neighborhood of the hidden true minimum.}
\label{fig:bayesopt}
\end{figure}
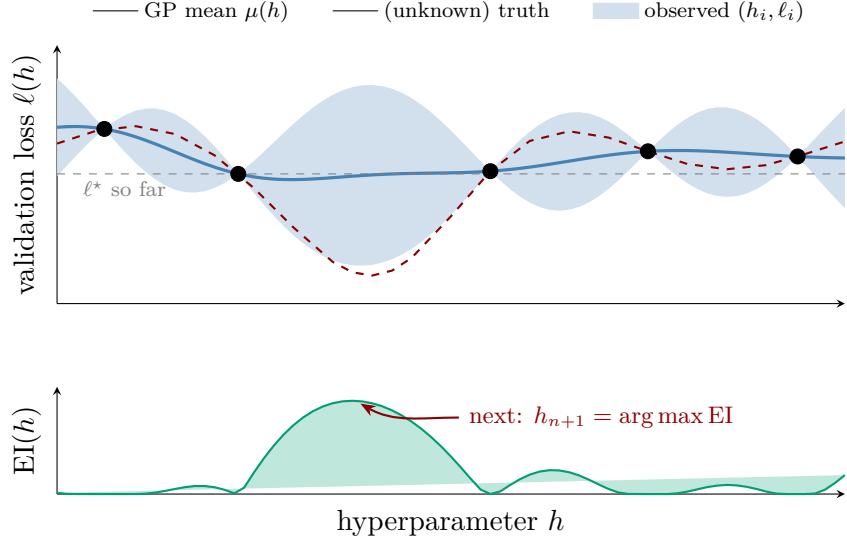

The next configuration to evaluate is selected by maximizing an \emph{acquisition function}.  The most common choice is Expected Improvement (EI):
\begin{equation}
\mathrm{EI}(\bm{h}) = \E{\max\bigl(\ell^\star - \ell(\bm{h}),\; 0\bigr)},
\end{equation}
where $\ell^\star = \min_i \ell_i$ is the best loss observed so far.  Under the GP posterior, this has a closed-form expression:
\begin{equation}
\mathrm{EI}(\bm{h}) = (\ell^\star - \mu(\bm{h}))\,\Phi(Z) + \sigma(\bm{h})\,\phi(Z),
\qquad Z = \frac{\ell^\star - \mu(\bm{h})}{\sigma(\bm{h})},
\end{equation}
where $\Phi$ and $\phi$ are the standard normal CDF and PDF, respectively.  For the closed-form derivation under the GP posterior, see \citet[\S 5]{garnett2023bayesian}.  EI naturally balances exploitation ($\mu(\bm{h})$ is small, i.e., predicted to be good) and exploration ($\sigma(\bm{h})$ is large, i.e., uncertain).  Bayesian optimization is particularly effective when the number of hyperparameters is moderate ($d \leq 20$) and each evaluation is expensive, which describes many economic applications well.  Figure~\ref{fig:bayesopt} illustrates the GP posterior and EI acquisition rule in one dimension.

\section{Hyperband and Successive Halving}
\label{sec:hyperband}

\citet{li2018hyperband} proposed an entirely different approach based on \emph{adaptive resource allocation}, building on the Successive Halving Algorithm popularized by \citet{jamieson2016nonstochastic}.  The key observation is that poor configurations can often be identified early in training, without running them to completion.  The Successive Halving Algorithm (SHA) formalizes this:

\begin{definitionbox}[Successive Halving Algorithm]
\begin{algorithmic}
\small
\STATE \textbf{Input:} Budget $B$, initial candidates $n$, reduction factor $\eta = 3$
\STATE Allocate each candidate a budget of $r = B/(n \lceil\log_\eta n\rceil)$
\FOR{round $s = 0, 1, \ldots, \lceil\log_\eta n\rceil - 1$}
    \STATE Train all remaining candidates for $r$ additional resources (epochs)
    \STATE Keep the top $1/\eta$ fraction; discard the rest
    \STATE $r \leftarrow r \cdot \eta$
\ENDFOR
\STATE \textbf{Output:} Best surviving configuration
\end{algorithmic}
\end{definitionbox}

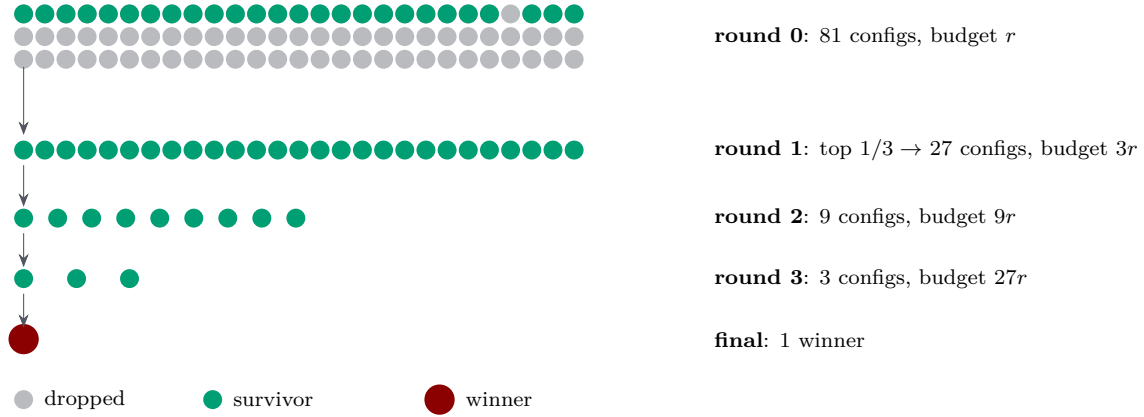
\begin{figure}[ht]
\centering
\begin{tikzpicture}[
    surv/.style={circle, fill=softgreen, minimum size=2.5mm, inner sep=0pt},
    drop/.style={circle, fill=uzhgreydark!40, minimum size=2.5mm, inner sep=0pt},
    win/.style ={circle, fill=darkred,  minimum size=4mm,   inner sep=0pt},
    lbl/.style={font=\scriptsize, anchor=west}
]
\foreach \i in {0,...,80} {
    \pgfmathsetmacro{\col}{mod(\i,27)}
    \pgfmathsetmacro{\row}{int(\i/27)}
    \node[drop] at (\col*0.28, -\row*0.30) {};
}
\foreach \i in {0,4,8,12,16,20,24,1,5,9,13,17,21,25,2,6,10,14,18,22,26,3,7,11,15,19} {
    \pgfmathsetmacro{\col}{mod(\i,27)}
    \pgfmathsetmacro{\row}{int(\i/27)}
    \node[surv] at (\col*0.28, -\row*0.30) {};
}
\node[lbl] at (9.0, -0.30) {\textbf{round 0}: 81 configs, budget $r$};

\foreach \i in {0,...,26} {
    \node[surv] at (\i*0.28, -1.8) {};
}
\node[lbl] at (9.0, -1.8) {\textbf{round 1}: top $1/3 \to$ 27 configs, budget $3r$};

\foreach \i in {0,...,8} {
    \node[surv] at (\i*0.45, -2.7) {};
}
\node[lbl] at (9.0, -2.7) {\textbf{round 2}: 9 configs, budget $9r$};

\foreach \i in {0,1,2} {
    \node[surv] at (\i*0.70, -3.5) {};
}
\node[lbl] at (9.0, -3.5) {\textbf{round 3}: 3 configs, budget $27r$};

\node[win] at (0, -4.3) {};
\node[lbl] at (9.0, -4.3) {\textbf{final}: 1 winner};

\draw[-{Stealth[length=1.8mm]}, uzhgreydark] (0.0,-0.7) -- (0.0,-1.6);
\draw[-{Stealth[length=1.8mm]}, uzhgreydark] (0.0,-2.0) -- (0.0,-2.55);
\draw[-{Stealth[length=1.8mm]}, uzhgreydark] (0.0,-2.9) -- (0.0,-3.35);
\draw[-{Stealth[length=1.8mm]}, uzhgreydark] (0.0,-3.7) -- (0.0,-4.15);

\node[drop, label={[font=\scriptsize, anchor=west, inner sep=1pt]right:\ dropped}] at (0.0,-5.1) {};
\node[surv, label={[font=\scriptsize, anchor=west, inner sep=1pt]right:\ survivor}] at (2.5,-5.1) {};
\node[win,  label={[font=\scriptsize, anchor=west, inner sep=1pt]right:\ winner}]   at (5.5,-5.1) {};
\end{tikzpicture}
\caption{Successive Halving with 81 initial candidates and reduction factor $\eta = 3$.  Each round trains the surviving configurations for $\eta$ times the previous budget, then discards the bottom $(1-1/\eta)$ fraction.  Total compute per bracket is only $\mathcal{O}(B)$ rather than $\mathcal{O}(nB)$ for training every candidate to completion.  Hyperband runs several such brackets in parallel with different $(n,r)$ trade-offs to hedge against unknown early-vs-late performance correlations \citep{li2018hyperband}.}
\label{fig:hyperband}
\end{figure}

For example, with $n = 81$ initial candidates and $\eta = 3$: round~0 trains all 81 for $r$ epochs and keeps the top 27; round~1 trains these 27 for $3r$ epochs and keeps the top 9; round~2 trains 9 for $9r$ epochs and keeps 3; round~3 trains 3 for $27r$ and selects the winner.  The total budget is $81r + 27\cdot 3r + 9\cdot 9r + 3\cdot 27r = 4 \cdot 81 r = 324\,r$, equivalent to about a dozen full $R = 27r$ trainings rather than the $81$ that naive parallel evaluation would require.  Figure~\ref{fig:hyperband} visualizes this resource-allocation cascade.

Hyperband extends SHA by running multiple SHA brackets with different trade-offs between the number of candidates $n$ and the per-candidate budget $r$, ensuring robustness to the unknown early-stopping behavior of different hyperparameter configurations.  The displayed sequence $(81,1) \to (27,3) \to (9,9) \to (3,27) \to (1,81)$ used in Algorithm~\ref{fig:hyperband} is the \emph{stage schedule} of the most exploratory SHA bracket, $s = s_{\max} = 4$, not Hyperband's full bracket set.  With maximum resource $R = 81$, reduction factor $\eta = 3$, and budget $B = (s_{\max}+1)R = 405$, Hyperband itself cycles through five SHA brackets indexed by $s = 0,\ldots,4$: each bracket starts with $n_s = \lceil (B/R)\,\eta^s/(s+1) \rceil$ configurations at initial resource $r_s = R\,\eta^{-s}$, then runs the SHA halving rule to convergence, so more aggressive brackets (large $s$) start with many cheap configurations while more conservative brackets (small $s$) start with fewer, longer-trained configurations.  The companion notebook implements the SHA inner loop only; running the full Hyperband schedule is a straightforward outer loop that iterates the inner loop across these five $(n_s, r_s)$ starting points.

\section{Method Comparison}

Table~\ref{tab:nas_methods} contrasts the four hyperparameter-search strategies covered above on three dimensions that matter in practice: the cost of $N$ objective evaluations, the degree to which the evaluations can be parallelised, and the sample efficiency (how much of the budget actually improves the best-so-far value).

\begin{table}[ht]
\centering
\small
\begin{tabular}{l c c c c}
\toprule
\textbf{Method} & \textbf{Cost} & \textbf{Parallelizable} & \textbf{Sample efficiency} & \textbf{Best for} \\
\midrule
Grid search & $\mathcal{O}(k^d)$ evals & fully & low & $d \leq 3$ \\
Random search & $N$ evals & fully & moderate & general use \\
Bayesian opt. & $N$ evals + GP fit & limited & high & expensive evals \\
Hyperband & $N$ resource units & within brackets & moderate & cheap evals \\
\bottomrule
\end{tabular}
\caption{Comparison of hyperparameter-search methods.  Grid search scales exponentially in the number of hyperparameters $d$; random search and Hyperband scale linearly in the chosen evaluation/resource budget and parallelise well; Bayesian optimization has the highest per-evaluation information gain but adds surrogate-fitting overhead and is partly sequential.}
\label{tab:nas_methods}
\end{table}

For the DEQN and PINN applications in this course, random search or Bayesian optimization are typically the most practical choices.  Hyperband is attractive when training is relatively cheap and many configurations need to be screened quickly.

\section{Implementing the Search in Practice}
\label{sec:nas-implementation}

To keep the algorithms transparent, the companion notebook \tpath{03_NAS_RandomSearch_Hyperband.ipynb} implements both Random Search (\S\,\ref{sec:nas_random_search}) and the Successive Halving Algorithm (\S\,\ref{sec:hyperband}) directly in plain Python, with no hyperparameter-search library involved.  The search space is encoded as an ordinary dict (number of hidden layers $\in \{1,\ldots,5\}$, units per layer $\in \{32, 64, \ldots, 256\}$, activation function $\in \{\texttt{relu}, \texttt{tanh}, \texttt{swish}\}$, and learning rate log-uniform in $[10^{-4}, 10^{-2}]$), and a single \texttt{sample\_config(rng)} function draws candidates from it.  Random Search is then a $30$-iteration loop that builds, trains, and scores each candidate; Successive Halving is the same loop wrapped in a halving schedule ($n_0 = 27$ candidates at $r_0 = 8$ epochs $\to$ $9$ at $24$ $\to$ $3$ at $72$ $\to$ winner, with $\eta = 3$).  Both implementations fit on a single slide and reproduce the qualitative finding of \citet{li2018hyperband} that Successive Halving reaches comparable accuracy to Random Search at substantially lower compute: in the notebook run, the same MAE is recovered with $\sim 2.3\times$ less compute (648 SHA config-epochs vs.\ 1500 for 30 Random Search trials at 50 epochs each) at a comparable number of architectures (27 vs.\ 30).  The precise multipliers are notebook-specific; the magnitudes reported in Li et al.\ vary by benchmark.

\paragraph{Production tooling (footnote).}  Real projects rarely hand-roll the search loop.  Several established libraries wrap (and parallelise) the same algorithms behind uniform APIs: \texttt{KerasTuner}\footnote{\url{https://keras.io/keras_tuner/}} (Random, Bayesian, Hyperband; tight Keras integration), \texttt{Optuna}\footnote{\url{https://optuna.org/}} (TPE, CMA-ES, Hyperband, NSGA-II; framework-agnostic), \texttt{Ray Tune}\footnote{\url{https://docs.ray.io/en/latest/tune/}} (all of the above plus ASHA and population-based training, distributed by design), \texttt{Hyperopt}\footnote{\url{http://hyperopt.github.io/hyperopt/}} (the original TPE reference), \texttt{Ax} / \texttt{BoTorch}\footnote{\url{https://ax.dev/}} (PyTorch-native multi-objective Bayesian optimization), \texttt{NNI}\footnote{\url{https://nni.readthedocs.io/}} (Microsoft; full graph-NAS support), and \texttt{AutoKeras}\footnote{\url{https://autokeras.com/}} (full AutoML pipeline).  We deliberately teach the algorithms rather than the wrappers because library APIs change every few years; the underlying search procedures (Random, SHA / Hyperband, GP+EI, TPE) do not.  The notebook additionally compares the best NAS-found architecture to a hand-tuned baseline, which makes the pedagogical value of automated search concrete.

\section{Multi-Component Losses: The Scale Problem}

In many applications, including DEQNs and PINNs, the loss function is a weighted sum of several components:
\begin{equation}
\ell = \sum_{k=1}^{K} w_k \, \ell_k.
\end{equation}
From a multi-objective-optimization standpoint, the vector $(\ell_1, \dots, \ell_K)$ is the object of interest: the equilibrium is defined by all $K$ residuals being zero, and any nonzero weight vector $\bm{w}$ picks a particular scalarization of the same underlying Pareto problem.  When the components are on comparable scales, uniform weights work; when they are not, the scalarized problem is dominated by a single component and the optimizer effectively ignores the others.  Adaptive weighting methods (discussed below) can be seen as online strategies for traversing the Pareto front rather than committing to a single scalarization up front.
If the individual components $\ell_k$ differ in magnitude by several orders of magnitude, training can become unstable or converge slowly.  Consider a concrete example from the IRBC model with $N=10$ countries: at initialization, the Euler equation residual for country~1 might be $\ell_1 \approx 50$, while for country~10 it might be $\ell_{10} \approx 0.05$, a ratio of $10^3$.  With uniform weights, the gradient is dominated by $\nabla \ell_1$, and the optimizer essentially ignores $\ell_{10}$ until $\ell_1$ is nearly converged.  This sequential convergence pattern can be $5$--$10\times$ slower than balanced convergence.

The fundamental difficulty is that the gradient of the total loss $\nabla \ell = \sum_k w_k \nabla \ell_k$ is dominated by the components with the largest $|w_k \nabla \ell_k|$.  Even if all components are equally important for the economic solution, the optimizer cannot ``see'' the small components through the noise of the large ones.

\subsection{Inverse-Loss Weighting}

A simple first approach is to set $w_k = 1/\bar{\ell}_k$, where $\bar{\ell}_k$ is an exponential moving average of $\ell_k$.  This normalizes each component to have approximately unit magnitude.  While straightforward, this method can be unstable when loss components change rapidly.

\subsection{SoftAdapt}

\citet{heydari2019softadapt} proposed a more principled approach based on the \emph{rates of change} of the loss components.  Define $\Delta_k^{(t)} = \ell_k^{(t)} - \ell_k^{(t-1)}$.\footnote{The raw $\Delta_k^{(t)}$ is dimensional, so a component at scale $10^3$ produces fluctuations that swamp one at scale $10^{-3}$; in practice one rescales before the softmax, e.g.\ $\tilde\Delta_k^{(t)} = \Delta_k^{(t)} / (\ell_k^{(t)} + \varepsilon)$, so the rule reacts to \emph{relative} progress, the same idea that underlies ReLoBRaLo's loss \emph{ratios} below.}  The SoftAdapt weights are:
\begin{equation}
w_k^{(t)} = \frac{\exp(\Delta_k^{(t)}/\tau)}{\sum_{j=1}^K \exp(\Delta_j^{(t)}/\tau)},
\end{equation}
where $\tau > 0$ is a temperature parameter.  Components that are decreasing slowly (or increasing) receive higher weight, directing the optimizer's attention to the lagging components.  In practice, SoftAdapt uses smoothed rates (averaged over a window of recent iterations) for stability.  SoftAdapt is discussed here for context; the companion notebook \texttt{04\_Loss\_Normalization.ipynb} implements equal-, inverse-loss-, and ReLoBRaLo-weighting, and Exercise~\ref{ex:ch4:6} asks you to add a GradNorm balancer to the same testbed.

\subsection{ReLoBRaLo: Adaptive Loss Balancing}
\label{sec:relobralo}

The \emph{Relative Loss Balancing with Random Lookback} (ReLoBRaLo) algorithm of \citet{bischof2025relobralo} motivates the deterministic classroom implementation used here.  In the notebooks, we use a convex blend of the same ingredients, which is easier to follow while preserving the balancing logic.

\paragraph{High-level intuition.}  ReLoBRaLo combines two complementary signals into a single weight per loss component.  The \emph{step-wise} signal asks ``which component lagged the most since the last iteration?'' and rewards it with more weight; this is fast and reactive but noisy.  The \emph{baseline} signal asks ``which component lagged the most since the start of training?'' and is slow but globally aware.  The two are then averaged with a one-step smoother to dampen oscillations.  Concretely the algorithm stacks three pieces (Components 1--3 below); only the temperature $T$ usually needs tuning, while the smoothing parameters $\alpha,\rho$ work at their textbook defaults.

\paragraph{Component 1: Relative balancing.}  At iteration $t$, compute relative losses with respect to the previous iteration:
\begin{equation}
\begin{aligned}
r_{k,\mathrm{step}}^{(t)}
&= \frac{\ell_k^{(t)}}{T\,\ell_k^{(t-1)}+\varepsilon_{\mathrm{num}}},\\
\hat{w}_{k,\mathrm{step}}^{(t)}
&= K \cdot
\frac{\exp\!\bigl(r_{k,\mathrm{step}}^{(t)}\bigr)}
{\sum_{j=1}^{K}\exp\!\bigl(r_{j,\mathrm{step}}^{(t)}\bigr)}.
\end{aligned}
\end{equation}
This upweights components whose relative loss increased (lagging behind) and downweights those that improved.  The small $\varepsilon_{\mathrm{num}}$ prevents division by zero; in code the softmax is evaluated after subtracting the largest ratio for numerical stability.

\paragraph{Component 2: Baseline comparison.}  To maintain a global perspective, compare the current losses to their initial values at $t=0$:
\begin{equation}
\begin{aligned}
r_{k,\mathrm{base}}^{(t)}
&= \frac{\ell_k^{(t)}}{T\,\ell_k^{(0)}+\varepsilon_{\mathrm{num}}},\\
\hat{w}_{k,\mathrm{base}}^{(t)}
&= K \cdot
\frac{\exp\!\bigl(r_{k,\mathrm{base}}^{(t)}\bigr)}
{\sum_{j=1}^{K}\exp\!\bigl(r_{j,\mathrm{base}}^{(t)}\bigr)}.
\end{aligned}
\end{equation}
This baseline comparison provides robustness to non-monotone loss trajectories and prevents the algorithm from losing sight of overall training progress.

\paragraph{Component 3: Smoothed combination.}  The final weight blends historical weights, baseline weights, and step-wise weights:
\begin{equation}
w_k^{(t)} =
\alpha\bigl[\rho\, w_k^{(t-1)} + (1-\rho)\, \hat{w}_{k,\mathrm{base}}^{(t)}\bigr]
+ (1-\alpha)\, \hat{w}_{k,\mathrm{step}}^{(t)},
\label{eq:relobralo_full}
\end{equation}
where $\alpha \in [0,1]$ is a smoothing parameter controlling how much to trust historical weights versus the current step-wise signal, and $\rho \in [0,1]$ is a baseline-mix coefficient controlling the relative importance of the previous weights versus the initial-loss comparison.  Equivalently, $w_k^{(t)}$ is a convex combination of $\{w_k^{(t-1)}, \hat{w}_{k,\mathrm{base}}^{(t)}, \hat{w}_{k,\mathrm{step}}^{(t)}\}$ with mixture weights $(\alpha\rho,\, \alpha(1-\rho),\, 1-\alpha)$, which sum to 1.  (In the original ReLoBRaLo formulation, $\rho$ governs a stochastic Bernoulli lookback mechanism; here and in the notebooks we use a deterministic convex blend, which is simpler and easier to reproduce.)  Typical values are collected in Table~\ref{tab:relobralo_hp}.

\begin{table}[ht]
\centering
\small
\begin{tabular}{l l l}
\toprule
\textbf{Hyperparameter} & \textbf{Role} & \textbf{Typical value} \\
\midrule
$T$ (temperature) & Controls weight concentration (softmax sharpness) & $0.5$--$2.0$ \\
$\alpha$ (smoothing) & History vs.\ new information & $0.99$--$0.999$ \\
$\rho$ (mixing coefficient) & Initial-loss baseline vs.\ historical weight & $0.99$--$0.999$ \\
\bottomrule
\end{tabular}
\caption{ReLoBRaLo hyperparameters used in the companion notebook.  Default ranges follow \citet{bischof2025relobralo}.  $T$ is the only one that usually needs tuning; $\alpha$ and $\rho$ at their defaults give slow, stable adaptation that works across a wide range of multi-component problems.}
\label{tab:relobralo_hp}
\end{table}

\subsection{GradNorm}

An alternative approach proposed by \citet{chen2018gradnorm} directly normalizes the gradient magnitudes rather than the loss values.  GradNorm adjusts the weights so that $\|w_k \nabla \ell_k\|$ is approximately equal across all components, using the ratio of each component's training rate to the average training rate as a signal.  While more computationally expensive than ReLoBRaLo (it requires computing per-component gradient norms), GradNorm can be effective when gradient magnitudes are a better proxy for training difficulty than loss magnitudes.

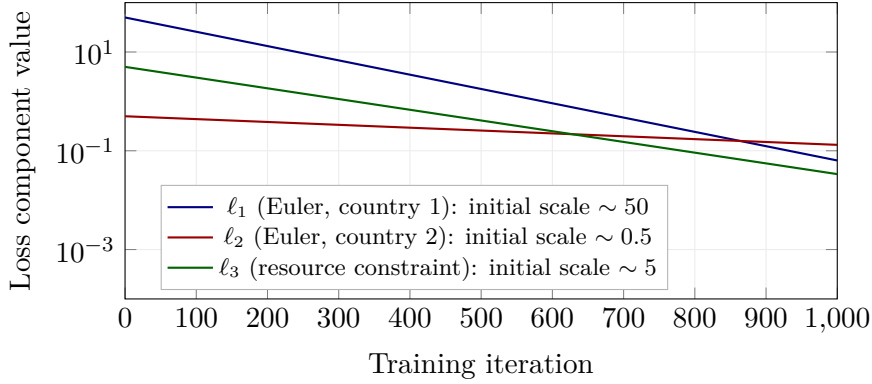
\begin{figure}[ht]
\centering
\begin{tikzpicture}
\begin{axis}[
    width=11cm, height=5.5cm,
    xlabel={Training iteration},
    ylabel={Loss component value},
    xmin=0, xmax=1000, ymin=0.0001, ymax=100,
    ymode=log,
    grid=major, grid style={gray!15},
    legend style={at={(0.05,0.03)}, anchor=south west, font=\footnotesize, fill=white, draw=black!30, row sep=-1pt, inner sep=2pt},
    every axis plot/.append style={thick, no markers},
]
\addplot[uzhblue, domain=0:1000, samples=100] {50*exp(-x/150)};
\addlegendentry{$\ell_1$ (Euler, country 1): initial scale $\sim 50$}
\addplot[harvardcrimson, domain=0:1000, samples=100] {0.5*exp(-x/750)};
\addlegendentry{$\ell_2$ (Euler, country 2): initial scale $\sim 0.5$}
\addplot[darkgreen, domain=0:1000, samples=100] {5*exp(-x/200)};
\addlegendentry{$\ell_3$ (resource constraint): initial scale $\sim 5$}
\end{axis}
\end{tikzpicture}
\caption{\emph{Stylized} sketch of the multi-component loss-scale problem, drawn to mimic what one typically sees early in a two-country IRBC training run; this is \emph{not} measured data.  The three curves are hand-picked exponentials $a_k\,e^{-t/\tau_k}$ (with $a_1{=}50,\tau_1{=}150$; $a_2{=}0.5,\tau_2{=}750$; $a_3{=}5,\tau_3{=}200$), chosen only to make the mechanism visible: at initialization the residuals differ by about two orders of magnitude, and under \emph{uniform} weighting the optimizer drives the largest component $\ell_1$ (blue) down fastest because it dominates the summed gradient, while the smaller-scale but equally important country-2 Euler residual $\ell_2$ (red) decays roughly five times more slowly and is left all but flat next to the others.  Adaptive loss balancing such as ReLoBRaLo re-weights the components so that all three decrease at comparable rates.  For the \emph{actual} recorded trajectories on this problem, see the companion notebook \texttt{04\_Loss\_Normalization.ipynb}.}
\label{fig:multi_component_loss}
\end{figure}

Figure~\ref{fig:multi_component_loss} illustrates the typical behavior: without adaptive reweighting, the optimizer focuses almost exclusively on $\ell_1$ (the largest component), allowing $\ell_2$ to stagnate; with adaptive loss balancing (e.g., ReLoBRaLo, GradNorm), all components converge at comparable rates.  As a concrete reference, an unweighted run of the two-country IRBC training loop in the companion notebook typically prints something like the trace below (numbers indicative, seed-dependent):

\begin{lstlisting}[caption={Indicative residual log from an unweighted two-country IRBC run; the largest component falls quickly, the smaller component stalls.}, label=lst:irbc_residual_trace, basicstyle=\ttfamily\footnotesize]
epoch    0:  ell_1=49.700  ell_2=0.510  ell_arc=4.820
epoch  200:  ell_1=0.0123  ell_2=0.494  ell_arc=0.041
epoch  500:  ell_1=8.2e-4  ell_2=0.470  ell_arc=3.5e-3
\end{lstlisting}

The pathology is immediate: $\ell_1$ drops four orders of magnitude while $\ell_2$ barely moves.  Replacing the equal weights with ReLoBRaLo (\S\ref{sec:relobralo}) typically produces a trace in which all three components decay together; see the companion notebook for the actual ReLoBRaLo trace on the same seed.  Reported convergence-speed improvements vary across schemes and benchmarks; multi-physics PINN benchmarks have shown substantial gains with ReLoBRaLo \citep{bischof2025relobralo}, while gains on DEQN-style Euler-equation losses tend to be smaller and problem-specific (the multi-component scale gap there is usually one to two orders of magnitude rather than the four to six common in PINN systems).  A complementary line uses Neural-Tangent-Kernel diagnostics to choose the weights \citep{wang2022when}, and an older multi-task baseline weights losses by their predictive uncertainty \citep{kendall2018multi}.

\subsection{Summary of Balancing Methods}

Table~\ref{tab:balancing_methods} compares the four balancing strategies on the two dimensions that reliably matter in practice: runtime overhead per step and the number of hyperparameters the user must set.

\begin{table}[ht]
\centering
\small
\begin{tabular}{l c c}
\toprule
\textbf{Method} & \textbf{Overhead} & \textbf{Hyperparameters} \\
\midrule
Uniform weights & none & 0 \\
Inverse-loss & negligible & 1 (smoothing) \\
Uncertainty weighting \citep{kendall2018multi} & negligible & $K$ (one log-variance per loss) \\
SoftAdapt \citep{heydari2019softadapt} & negligible & 2 ($\tau$, window) \\
ReLoBRaLo \citep{bischof2025relobralo} & negligible & 3 ($T$, $\alpha$, $\rho$) \\
GradNorm \citep{chen2018gradnorm} & moderate & 1 ($\alpha$) \\
NTK-based \citep{wang2022when} & moderate--high & 0 (data-driven) \\
\bottomrule
\end{tabular}
\caption{Summary of adaptive loss-balancing methods.  Overhead is a per-step wall-clock cost (additional softmaxes for SoftAdapt/ReLoBRaLo; per-component gradient norms for GradNorm).  Quantitative speedups depend strongly on the problem; see the companion notebook \texttt{04\_Loss\_Normalization.ipynb} for problem-specific measurements.}
\label{tab:balancing_methods}
\end{table}

Quantitative speedup claims depend on the specific problem (PDE vs.\ Euler residual, number of components, imbalance ratio), the baseline (uniform vs.\ manually tuned), and the success criterion.  The companion notebook \texttt{04\_Loss\_Normalization.ipynb} runs the four methods on a shared multi-scale regression task so that the reader can generate problem-specific numbers rather than rely on headline speedup factors from unrelated benchmarks.

\begin{remarkbox}[Practical guidance]
When implementing ReLoBRaLo:
\begin{itemize}[itemsep=2pt]
\item Set $T \in [0.5, 2.0]$; higher values yield more uniform weighting.  In the limit, $T \to 0$ approximates a winner-take-all scheme that concentrates all weight on the single most-lagging component, while $T \to \infty$ recovers uniform weighting regardless of loss dynamics.
\item Start with $\alpha = \rho = 0.999$ and reduce if weights change too slowly.
\item ReLoBRaLo adds negligible computational overhead (one softmax per iteration) but can dramatically improve convergence for multi-component losses; GradNorm and SoftAdapt make analogous trade-offs.
\item For PINN applications (Chapter~\ref{ch:pinn}), adaptive loss balancing in general (ReLoBRaLo, GradNorm, NTK-based schemes) is particularly effective at balancing PDE residual terms against boundary condition penalties.
\item In DSGE applications, multi-component losses arise naturally: a model with $N$ countries has $N$ Euler equations, an aggregate resource constraint, and $N$ complementarity conditions, often differing by several orders of magnitude.  Without loss balancing, the optimizer focuses on the largest component and ignores smaller but economically important residuals (see Chapter~\ref{ch:irbc}).
\end{itemize}
\end{remarkbox}

\begin{keyinsightbox}[Chapter Summary]
\begin{itemize}[itemsep=2pt, leftmargin=*]
\item Architecture and hyperparameter choice are first-order: random search dominates grid search whenever only a few hyperparameters matter, and Bayesian optimization dominates random search whenever each evaluation is expensive.
\item Hyperband-style multi-fidelity scheduling is the modern compromise: cheap evaluations of many candidates, with budget concentrated on the survivors.
\item Multi-component DEQN/PINN losses suffer from scale imbalance; ReLoBRaLo and related adaptive schemes restore balance by reweighting components on the fly.
\item All three families (NAS, Bayesian optimization, adaptive loss balancing) are implementation details, but they make the difference between a DEQN that converges on a coffee break and one that diverges overnight.
\end{itemize}
\end{keyinsightbox}

\section*{Further Reading}
\addcontentsline{toc}{section}{Further Reading}
\begin{itemize}[itemsep=2pt]
\item \citet{bergstra2012random}, the original case for random over grid search.
\item \citet{snoek2012practical}, foundational reference for Bayesian optimization in ML.
\item \citet{li2018hyperband}, Hyperband and successive halving.
\item \citet{bischof2025relobralo}, ReLoBRaLo loss-balancing scheme used throughout the PINN chapter.
\end{itemize}

\section*{Exercises}
\addcontentsline{toc}{section}{Exercises}
\noindent Worked solutions and guidance for these exercises appear in Appendix~\ref{app:solutions}.
\begin{enumerate}[itemsep=4pt, leftmargin=*]
\item\label{ex:ch4:1} \textbf{[Core] Random vs.\ grid.}  Reproduce the random-vs-grid figure of \citet{bergstra2012random}.  With $9$ evaluations and only one ``important'' axis out of two, suppose the near-optimal region occupies a fraction $p$ of that axis and its location relative to the grid is unknown.  What is the hit probability for a $3\times 3$ grid, and what is the hit probability for random search?
\item\label{ex:ch4:2} \textbf{[Computational] Bayesian optimization toy problem.}  Implement a Gaussian-process-based BO loop on the 1D function $f(x) = -\sin(3x) - x^2 + 0.7x$ over $x \in [-1,2]$.  How many BO evaluations are typically needed before it matches the best value found by a grid with step size $0.01$?
\item\label{ex:ch4:3} \textbf{[Core] Hyperband budget allocation.}  For a Hyperband run with maximum resource $R=81$ and $\eta=3$, list the brackets $(n_i, r_i)$ used and the total resource consumed.  Compare with a naive ``train each candidate to $R$'' strategy at fixed candidate count $n_0 = 27$.
\item\label{ex:ch4:4} \textbf{[Core] Loss balancing.}  In a multi-component PINN loss with three terms of magnitude $10^0$, $10^{-2}$, $10^{-4}$, write down what fixed weights $\lambda_i$ would be needed to equalise their gradient contributions.  Why does this become impractical when the gradients are correlated?
\item\label{ex:ch4:5} \textbf{[Core] Pareto front geometry.}  Consider the toy two-component loss $\mathcal{L}(\theta;\lambda) = \lambda\,(\theta - a)^2 + (1-\lambda)(\theta - b)^2$ with $\theta \in \mathbb{R}$, $a < b$, and $\lambda \in [0,1]$.  (i)~Solve for the minimizer $\theta^\star(\lambda)$ in closed form.  (ii)~Compute the per-component residuals $\ell_1^\star(\lambda) = (\theta^\star - a)^2$ and $\ell_2^\star(\lambda) = (\theta^\star - b)^2$.  (iii)~Eliminate $\lambda$ to express the Pareto frontier in the $(\ell_1, \ell_2)$ plane and show it is the curve $\sqrt{\ell_1} + \sqrt{\ell_2} = b - a$ for $\ell_1, \ell_2 \ge 0$, hence convex.  (iv)~Sketch the front and identify which point on it corresponds to the equal-weight choice $\lambda = 1/2$.  (v)~In higher-dimensional parameter spaces, explain why nonzero gradient inner products $\langle\nabla \ell_1, \nabla \ell_2\rangle$ make fixed scalar weights fragile.  Contrast ReLoBRaLo's relative-loss progress rule with GradNorm's direct gradient-norm balancing.
\item\label{ex:ch4:6} \textbf{[Computational] ReLoBRaLo vs.\ GradNorm.}  In notebook \texttt{04\_Loss\_Normalization}, swap the ReLoBRaLo balancer for a GradNorm balancer (\citealt{chen2018gradnorm}; see the chapter for the per-component gradient-norm equalization rule).  Run both on the same multi-scale regression target with components of magnitude $10^0, 10^{-2}, 10^{-4}$.  Report (i)~wall-clock training time per epoch, (ii)~final residual on each component, (iii)~the gradient-balance ratio $\|\nabla\ell_k\|/\sum_j \|\nabla\ell_j\|$ at convergence.  Comment on the cost--benefit trade-off: under what circumstances is the extra per-step cost of GradNorm worth it?
\item\label{ex:ch4:7} \textbf{[Core] HPO vs.\ full NAS decision.}  You have access to either (a)~a single consumer GPU (RTX 3060, $\sim 12$ GB) or (b)~one A100 ($80$ GB), for one week of compute.  Your search problem is either (i)~a fixed-topology MLP with unknown depth $\in \{1,\ldots,5\}$, width $\in \{32,\ldots,512\}$, activation $\in \{\mathrm{ReLU}, \mathrm{Swish}, \tanh\}$, learning rate (log-uniform), or (ii)~a flexible network that can be MLP / RNN / shallow Transformer with unknown layer connectivity (graph-level NAS).  For each of the four (hardware $\times$ problem) cells, recommend in three to five sentences whether to use Random Search with Successive Halving, Bayesian Optimization, or full graph-level NAS.  Justify by referencing the per-method overhead and search-space size.
\end{enumerate}

\chapter{Overlapping Generations Models with DEQNs}
\label{ch:olg}

In Chapters~\ref{ch:deqn}--\ref{ch:irbc} all agents were infinitely lived.  We now extend the DEQN framework to \emph{overlapping generations} (OLG) models \citep{diamond1965national}, where $A$ finitely-lived cohorts coexist in every period.  OLG models introduce lifecycle savings, intergenerational transfers, age-dependent heterogeneity, and inequality constraints on portfolio choices, phenomena that are central to fiscal policy analysis, pension reform, and demographic modeling.  We proceed in two stages.  We first solve a deliberately small 6-agent OLG that admits a \emph{closed-form solution} \citep{Krueger20041411}, which gives a clean ground truth against which to validate the neural-network solver.  We then scale up to the 56-agent research benchmark of \citet{azinovicDEEPEQUILIBRIUMNETS2022}, where the no-short-sale-of-capital constraint $k'^h\ge 0$ binds on a non-trivial slice of the ergodic set; that constraint introduces a kink, the main new computational challenge of the benchmark, and we handle it by combining softplus output activations (for non-negativity) with squared product residuals for the orthogonality conditions in the loss.  The model also carries a collateral constraint $k'^h+\kappa\,b'^h\ge 0$ that the current notebook parameterization of $\hat q^h$ keeps slack on the learned ergodic set; we develop both constraints below so that the architecture is in place when a future calibration makes the collateral side bind.

\section{Why Overlapping Generations?}
\label{sec:olg_why}

In the Brock--Mirman and IRBC models of Chapters~\ref{ch:deqn}--\ref{ch:irbc}, all agents are infinitely lived.  Picture instead a photograph of the economy taken at a single instant: it contains a twenty-something just entering the workforce with no savings, a forty-something at peak earnings putting money aside, and a retiree drawing down a lifetime of accumulated wealth, all making decisions in the same period and all linked through the prices that their collective saving determines.  The infinitely-lived-agent assumption collapses this picture and rules out several economically important phenomena:
\begin{itemize}[itemsep=2pt]
\item \textbf{Lifecycle savings.}  Agents accumulate wealth when young, draw it down in old age.
\item \textbf{Intergenerational transfers.}  Pensions, social security, and bequests cannot be studied without age structure.
\item \textbf{Age-dependent heterogeneity.}  Labor endowments, risk preferences, and portfolio composition vary systematically over the lifecycle.
\end{itemize}

\begin{figure}[ht]
\centering
\begin{tikzpicture}
\begin{axis}[
    width=11cm, height=5.4cm,
    xlabel={Age $h$}, ylabel={Level (stylized)},
    xmin=22, xmax=83, ymin=0, ymax=1.42,
    xtick={25,35,45,55,65,75,80},
    ytick=\empty,
    grid=major, grid style={gray!16},
    axis lines=left,
    label style={font=\small}, tick label style={font=\small},
    legend style={font=\footnotesize, draw=none, fill=none, at={(0.5,1.03)}, anchor=south, legend columns=-1, /tikz/every even column/.append style={column sep=12pt}},
    every axis plot/.append style={very thick, smooth},
    domain=25:80, samples=180,
]
\addplot[softblue]                       {0.66 + 0.60*exp(-((x-50)/17)^2)};                          \addlegendentry{labor income $w\,\ell^h$}
\addplot[darkgreen]                      {0.75};                                                     \addlegendentry{consumption $c^h$}
\addplot[harvardcrimson, densely dashed] {(x-25)^3 * (80-x) / 1.86e6};                               \addlegendentry{assets $k^h$}
\end{axis}
\end{tikzpicture}
\caption{Stylized lifecycle profiles in an OLG economy (schematic, \emph{not} a solution of the model).  Labor income (blue) is hump-shaped, peaking in mid-career, while agents prefer a roughly flat consumption path (green); so they accumulate assets out of income during their high-earning years and run them down near the end of life.  The asset profile (red, dashed) is therefore a hump that starts near zero for the newborn cohort, peaks toward the end of working life, and returns to zero for the oldest cohort, which consumes everything.  The 6-agent analytic model of \S\ref{sec:olg_analytic} is a stripped-down version of this picture (only the youngest cohort earns labor income); the 56-agent benchmark of \S\ref{sec:olg_56} reproduces the full hump.}
\label{fig:olg_lifecycle}
\end{figure}
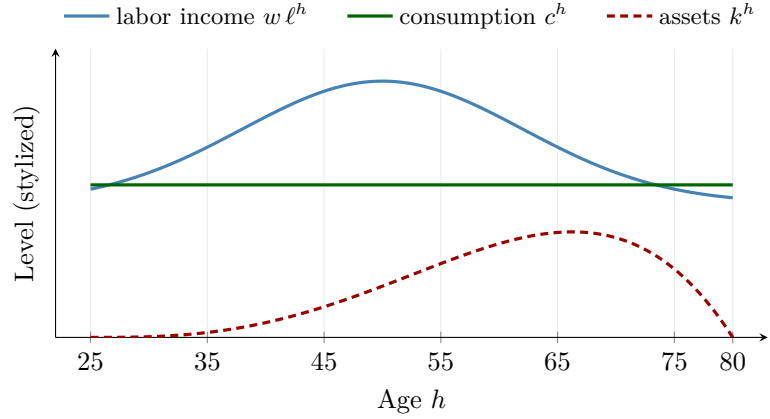

An OLG economy consists of $A$ cohorts that coexist in each period: a new cohort of age~1 is born, the oldest cohort of age~$A$ dies, and everyone else ages by one period.  Crucially, the number of agent types is \emph{finite}, so the cross-sectional distribution has only $A$ entries and the state space remains finite-dimensional, in contrast to the continuum-of-agents models treated in Chapter~\ref{ch:young}.  The mechanism that ties the three phenomena above together is consumption smoothing over a hump-shaped earnings path (Figure~\ref{fig:olg_lifecycle}): because labor income rises and then falls over the lifecycle while agents prefer a steady consumption stream, they accumulate assets in their high-earning years and run them down afterwards, and the equilibrium interest rate is whatever clears the resulting demand for savings against the economy's capital stock.

We develop the OLG framework in two stages.  Section~\ref{sec:olg_analytic} works through the 6-agent model with a closed-form solution, maps it to a DEQN (\S\ref{sec:olg_deqn}), and validates the trained network against the analytical savings rates; \S\ref{sec:olg_fb} then explains how binding borrowing and collateral constraints are encoded, and \S\ref{sec:olg_56} solves the 56-agent research benchmark with exactly the same training loop.

\section{The 6-Agent Analytic OLG Model}
\label{sec:olg_analytic}
\label{sec:olg_setup}

\begin{remarkbox}[End-to-end vignette: a 6-agent OLG in five steps]
Before stepping through the formal model, it is worth seeing the full DEQN pipeline in one breath.  The same five steps apply to every model in this script.
\begin{enumerate}[itemsep=2pt, leftmargin=*]
\item \textbf{State and policy.}  Stack the 6 cohort capital holdings, the aggregate $K_t$, prices $(r_t, w_t)$, the TFP shock $\eta_t$, and the depreciation shock $\delta_t$ into a state vector $\x_t$.  A single MLP $\mathcal{N}_\theta(\x_t) \to \R^{5}$ outputs the savings of cohorts 1--5 (cohort 6 saves nothing).
\item \textbf{Loss.}  Stack 5 Euler-equation residuals into a single mean-square loss; a sigmoid savings-fraction head makes each $k'^h \in [0,\mathrm{inc}^h]$, so $k'^h \ge 0$ and $c^h \ge 0$ both hold exactly, and capital-market clearing is satisfied by construction (aggregate $K_{t+1}$ is read off as the sum of cohort savings).  Small additive penalties on negative $c$ and $K$ remain in the loss as numerical backstops but are inactive with this head.
\item \textbf{Training distribution.}  At each segment, construct a state cloud: in the persistent notebooks the cloud is generated by rolling parallel trajectories forward under the current policy; in the exogenous companions it is drawn from broad feasible boxes.  Mini-batches are sampled from the current cloud.
\item \textbf{Optimization.}  Adam with the standard moment coefficients $(0.9,0.999)$ (avoiding the $\beta$ symbol of the household discount factor) and learning rate $\sim 10^{-4}$ in the short presets, decaying to $10^{-5}$ in the production preset of the analytic-OLG notebook.  In the 56-agent benchmark of \S\ref{sec:olg_56}, the production schedule is $10^{-5}$ for the first $60{,}000$ episodes followed by $10^{-6}$ for the remaining episodes.
\item \textbf{Diagnostics.}  At convergence, check (i) the average savings rate by cohort against the closed-form $\beta_h$ in Eq.~\eqref{eq:olg_savings_rate} below, (ii) Euler residuals across the simulated ergodic set, (iii) (in the 56-agent benchmark only) the bond-market-clearing residual, and (iv) policy-drift / time-invariance on a fixed anchor cloud: evaluate the policy on \tpath{X_anchor} after each monitoring interval and flag the run as \emph{time-invariant} once \tpath{policy_drift_rms} and \tpath{policy_drift_max} fall below \tpath{TIME_INVARIANCE_TOL_RMS} and \tpath{TIME_INVARIANCE_TOL_MAX}.  In the 6-agent analytic model, capital-market clearing is satisfied by construction, since aggregate next-period capital is taken as the sum of cohort savings.
\end{enumerate}
The notebook \tpath{lecture_08_08_OLG_Analytic_DEQN_persistent.ipynb} ships with a \texttt{RUN\_MODE} switch: \texttt{"smoke"} ($\sim$30~s, sanity check), \texttt{"teaching"} ($\sim$5~min, savings close to the closed form and mean Euler residuals $\sim 10^{-3}$ on the simulated cloud), and \texttt{"production"} (longer run, Table~3-level Euler accuracy on the full ergodic set); a feedback-free exogenous-sampling companion is available as \tpath{lecture_08_07_OLG_Analytic_DEQN_exogenous.ipynb}.  A fifth notebook, \tpath{lecture_08_11_OLG_Exercise.ipynb}, is a self-contained warm-up (closed-form savings rates, a single-cohort lifecycle simulation, and a discount-factor comparison) on the same analytic model.  The 56-agent benchmark of \S\ref{sec:olg_56} (notebook \tpath{lecture_08_10_OLG_Benchmark_DEQN_persistent.ipynb}, with the exogenous-sampling companion \tpath{lecture_08_09_OLG_Benchmark_DEQN_exogenous.ipynb}) runs in a few hours on a GPU with the \emph{same} code template.
\end{remarkbox}

\citet{Krueger20041411} proposed a deliberately simple OLG model with a \emph{closed-form solution}, making it an ideal validation benchmark for the DEQN approach.  We develop it here as the first of the two OLG instances of this chapter; \S\ref{sec:olg_deqn} maps it to a DEQN and validates the trained network against the closed form derived below.

We instantiate the OLG environment with $A=6$ overlapping cohorts, indexed by age $h \in \{1,\ldots,6\}$.  Time is discrete and infinite.  The model equations below are written for general $A$ and specialized to $A=6$ in the calibration that follows.

\paragraph{Household problem.}
An agent of age $h$ at time $t$ maximizes expected lifetime utility:
\begin{equation}
\max_{\{c_{t+j}^{h+j},\, k_{t+j+1}^{h+j+1}\}_{j=0}^{A-h}} \;\mathbb{E}_t\!\left[\sum_{j=0}^{A-h} \beta^j\, u(c_{t+j}^{h+j})\right],
\label{eq:olg_hh}
\end{equation}
subject to the period budget constraint
\begin{equation}
c_t^h + k_{t+1}^{h+1} = r_t \, k_t^h + w_t \, \ell^h \equiv \mathrm{inc}_t^h,
\label{eq:olg_budget}
\end{equation}
where $k_t^h$ denotes capital holdings, $r_t$ is the gross return on capital, $w_t$ is the wage, $\ell^h$ is an age-dependent labor endowment, and $\mathrm{inc}_t^h$ is total income.

\paragraph{Boundary conditions.}
\begin{itemize}[itemsep=2pt]
\item Newborns have no initial wealth: $k_t^1 = 0$.
\item The oldest cohort consumes everything: $k_{t+1}^{A+1} = 0$.
\item Borrowing is not permitted: $k_{t+1}^{h+1} \geq 0$ for all $h$.
\end{itemize}

\paragraph{Euler equations.}
The first-order conditions yield $A-1$ Euler equations (for ages $h=1,\ldots,A-1$):
\begin{equation}
u'(c_t^h) = \beta\,\mathbb{E}_t\!\left[r_{t+1}\, u'(c_{t+1}^{h+1})\right].
\label{eq:olg_euler}
\end{equation}

\paragraph{Firm problem and market clearing.}
A representative firm operates a Cobb--Douglas technology with value added $F_t = \eta_t K_t^\alpha L_t^{1-\alpha}$, where $\eta_t$ is a TFP shock and $L_t = \sum_{h=1}^A \ell^h$; the gross resource available to households is $Y_t = F_t + (1-\delta_t)K_t = r_t K_t + w_t L_t$ (it is $Y_t$, not $F_t$, that the notebook passes as an engineered feature).  Competitive factor markets imply:
\begin{equation}
r_t = \alpha\,\eta_t\, K_t^{\alpha-1}L_t^{1-\alpha} + (1-\delta_t), \qquad
w_t = (1-\alpha)\,\eta_t\, K_t^\alpha L_t^{-\alpha},
\label{eq:olg_prices}
\end{equation}
where $\delta_t$ is the depreciation rate (potentially stochastic).  Market clearing requires that aggregate capital at $t+1$ is the sum of holdings across cohorts:
\begin{equation}
\sum_{h=2}^{A} k_{t+1}^{h} = K_{t+1},
\label{eq:olg_mc}
\end{equation}
with $k_{t+1}^{1}=0$ as a newborn boundary condition (cohort~1 enters life with no assets), and where $k_{t+1}^{h}$ for $h=2,\dots,A$ is the savings of cohort $h-1$ at date~$t$ (which becomes the date-$(t+1)$ holdings of the cohort once it has aged by one period).

\paragraph{Calibration.}
The model has $A=6$ agents with log utility ($\gamma = 1$), Cobb--Douglas production ($\alpha = 0.3$), and discount factor $\beta = 0.7$.  Only agent~1 works ($\ell = (1,0,0,0,0,0)$); this stripped-down labor profile is what gives the closed form below, not a realistic lifecycle assumption, and the 56-agent benchmark of \S\ref{sec:olg_56} restores a hump-shaped endowment.  Four exogenous shock states combine TFP $\eta \in \{0.95,1.05\}$ and depreciation $\delta \in \{0.5,0.9\}$, with i.i.d.\ transitions ($\pi_{ss'} = 0.25$).

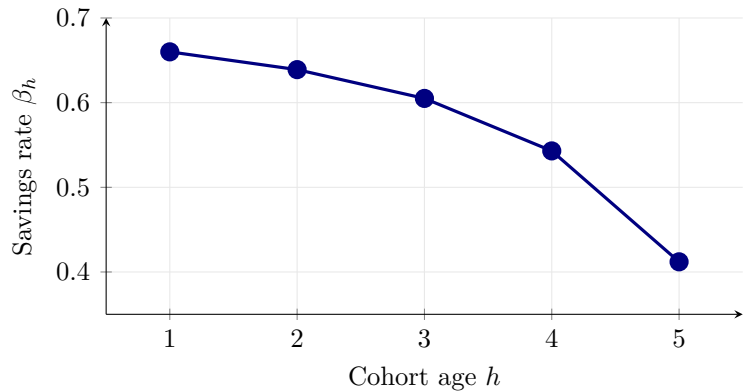
\begin{figure}[ht]
\centering
\begin{tikzpicture}
\begin{axis}[
    width=10cm, height=5.5cm,
    xlabel={Cohort age $h$},
    ylabel={Savings rate $\beta_h$},
    xmin=0.5, xmax=5.5, ymin=0.35, ymax=0.70,
    xtick={1,2,3,4,5},
    grid=major, grid style={gray!18},
    axis lines=left,
    label style={font=\small}, tick label style={font=\small},
    every axis plot/.append style={very thick, mark=*, mark size=3pt},
]
\addplot[uzhblue] coordinates {(1, 0.660) (2, 0.639) (3, 0.605) (4, 0.543) (5, 0.412)};
\end{axis}
\end{tikzpicture}
\caption{Closed-form savings rates $\beta_h$ from Table~\ref{tab:olg6_savings_rates} for the 6-agent analytic OLG ($\beta=0.7$, log utility).  The monotone decline with age reflects the shrinking forward horizon: cohort~$h$ has only $A-h$ remaining periods over which to consume future income, so the marginal incentive to save weakens as $h$ grows.  This is the validation target the trained DEQN's average sigmoid output should match cohort by cohort.}
\label{fig:olg6_savings}
\end{figure}

\paragraph{Analytical solution.}
With log utility and i.i.d.\ shocks, the optimal savings rate has a closed form.  Define the age-dependent savings rate:
\begin{equation}
\beta_h = \beta \cdot \frac{1 - \beta^{A-h}}{1 - \beta^{A-h+1}}, \qquad h = 1, \ldots, A-1.
\label{eq:olg_savings_rate}
\end{equation}
The optimal policy is then $k'^h = \beta_h \cdot \mathrm{inc}^h$: each agent saves a \emph{fixed fraction} of total income, regardless of the shock.  Two features of the calibration drive this clean form.  First, under log utility the income and substitution effects of a return shock exactly cancel, so the savings \emph{rate} is invariant to $(r_t, w_t)$.  Second, because the shocks are i.i.d.\ there is nothing about the future to forecast, so the rate does not depend on the current shock either; only the horizon matters.  The fraction $\beta_h$ therefore declines with age: cohort $h$ has only $A-h$ remaining periods over which to spread its future income, so the marginal incentive to carry resources forward weakens as $h$ grows.  For $A=6$, $\beta=0.7$, Table~\ref{tab:olg6_savings_rates} reports the resulting savings rates.
\begin{table}[ht]
\centering
\small
\begin{tabular}{cccccc}
\toprule
Age $h$ & 1 & 2 & 3 & 4 & 5 \\
\midrule
$\beta_h$ & 0.660 & 0.639 & 0.605 & 0.543 & 0.412 \\
\bottomrule
\end{tabular}
\caption{Closed-form age-specific savings rates in the 6-agent analytic OLG with log utility and $\beta=0.7$.}
\label{tab:olg6_savings_rates}
\end{table}
Young agents save more (more periods ahead); old agents save less; Figure~\ref{fig:olg6_savings} plots the same numbers across $h$.  This vector is the validation target: at convergence, the trained network's average sigmoid output should reproduce $\beta_h$ cohort by cohort.

\section{Mapping the Analytic OLG to a DEQN}
\label{sec:olg_deqn}

The mapping follows the same ``states $\to$ network $\to$ loss'' structure as Brock--Mirman (Chapter~\ref{ch:deqn}).  We now write each block explicitly for the 6-agent analytic model just set up; this is exactly what slides II.7--II.9 of \tpath{lectures/lecture_08_olg_models_deqns/slides/lecture_08_olg_models_deqns.tex} render in pictures.  The 56-agent benchmark of \S\ref{sec:olg_56} extends the same template with two extra policy blocks (multipliers, bond price) and an additional market-clearing residual; we write that version out there.

\paragraph{State $\x_t$ entering the network.}
What does the network actually need to know?  The \emph{informational} state of the analytic model is just the pair
\begin{equation}
\bigl(z_t,\,\bm{k}_t\bigr) \;\in\; \{1,\ldots,4\}\times\R^A
\qquad\text{where}\qquad \bm{k}_t = (k_t^1,\ldots,k_t^A),
\label{eq:olg_state_min}
\end{equation}
the current shock index plus the cross-sectional capital distribution.  This is the minimal vector that pins down the equilibrium, and it is what slide II.8 displays in the FREE signature.  Everything else, the aggregate capital $K_t=\sum_h k_t^h$, the prices $(r_t,w_t)$, output $Y_t$, each cohort's income, the row of next-period transition probabilities, is a deterministic function of $(z_t,\bm{k}_t)$.  The network could in principle re-derive all of it from the raw pair, but there is no reason to make it: we hand the network those quantities pre-computed, which is a pure change of input coordinates that leaves the equilibrium map untouched and frees the network's capacity for the one genuinely hard thing it has to learn, the savings policy.  Concretely the notebook feeds an \emph{extended state} of dimension $16+4A$,
\begin{equation}
\x_t \;=\; \bigl(\,\underbrace{z_t,\,\mathbf{1}\{z_t\},\,\eta_t,\delta_t,K_t,L_t,r_t,w_t,Y_t}_{12\text{ aggregate}},\;
\underbrace{k_t^{1:A},\,\mathrm{fw}_t^{1:A},\,\mathrm{linc}_t^{1:A},\,\mathrm{inc}_t^{1:A}}_{4A\text{ per-agent}},\;
\underbrace{\pi(z_t,\cdot)}_{4\text{ transition probs}}\bigr) \;\in\; \R^{16+4A},
\label{eq:olg_state}
\end{equation}
with $\mathbf{1}\{z_t\}$ the 4-state one-hot of the current shock, $K_t=\sum_h k_t^h$, $L_t=\sum_h \ell^h$, $(r_t,w_t)$ from~\eqref{eq:olg_prices}, $Y_t$ the gross resource $\eta_t K_t^\alpha L_t^{1-\alpha} + (1-\delta_t)K_t$, and the per-agent blocks $\mathrm{fw}_t^h = r_t k_t^h$ (capital income), $\mathrm{linc}_t^h = w_t\,\ell^h$ (labor income), $\mathrm{inc}_t^h = \mathrm{fw}_t^h + \mathrm{linc}_t^h$ (total income).  Since the map $(z_t,\bm{k}_t)\mapsto\x_t$ is deterministic, \eqref{eq:olg_state_min} and~\eqref{eq:olg_state} carry exactly the same information.  For $A=6$ this is $16+4\cdot 6 = 40$ inputs (the notebook constant \texttt{FEATURE\_DIM}).

\paragraph{Policies approximated by the network.}
A single multilayer perceptron with a \emph{sigmoid savings-fraction} output head approximates the equilibrium policy as a function of the state.  (Throughout this OLG chapter we use $\theta$ for the network parameters rather than the $\rho$ of Chapters~\ref{ch:deqn}--\ref{ch:irbc}; both refer to the same object, and the switch follows the convention of the public OLG reference implementation.)
\begin{equation}
\boxed{\;\mathcal{N}_\theta:\;\R^{16+4A} \;\longrightarrow\; \R^{A-1},\qquad
\mathcal{N}_\theta(\x_t) \;=\; \bigl(\hat\beta^1(\x_t),\ldots,\hat\beta^{A-1}(\x_t)\bigr),\qquad
\hat a^h(\x_t) := \hat\beta^h(\x_t)\,\mathrm{inc}_t^h\;}
\label{eq:olg_policy}
\end{equation}
where the network output $\hat\beta^h(\x_t)\in(0,1)$ is cohort $h$'s \emph{savings rate} and $\hat a^h(\x_t)$ its savings level (slide II.9, output column).  This parameterization mirrors the closed-form solution's structure (each cohort saves a fixed fraction of income, Eq.~\eqref{eq:olg_savings_rate}).  Cohort $A$ saves nothing by terminal boundary, so the network has $A-1$ outputs rather than $A$.  Three by-construction guarantees follow:
\begin{itemize}[itemsep=2pt]
    \item Non-negativity of savings $\hat a^h \ge 0$ holds at every iteration, so the borrowing constraint~\eqref{eq:olg_kkt} is satisfied without an explicit Lagrange multiplier (in this calibration $\lambda^h \equiv 0$ on the ergodic set; see \S\ref{sec:olg_fb} for the multiplier-based variant used in the 56-agent benchmark).
    \item Non-negativity of consumption $\hat c^h = (1-\hat\beta^h)\,\mathrm{inc}_t^h \ge 0$ also holds by construction, so the soft penalty on $\max(-\hat c^h,0)$ in the loss (next paragraph) is a dead backstop.
    \item Capital-market clearing $K_{t+1} = \sum_{h=2}^{A} k_{t+1}^h$ also holds by construction, since aggregate next-period capital is read off as the sum of the network's savings outputs together with the newborn boundary $k_{t+1}^1 = 0$.
\end{itemize}

\paragraph{Equilibrium residual.}
Each cohort $h\in\{1,\ldots,A-1\}$ contributes one \emph{relative} Euler-equation residual, built from three quantities.  First, the implied current consumption, read off from the budget~\eqref{eq:olg_budget} as $\hat c^h(\x_t) := \mathrm{inc}_t^h - \hat a^h(\x_t)$.  Second, the next-state map $\Phi$, which combines the current policy with a fresh shock $z_{t+1}$ to produce next period's extended state $\hat\x_{t,+}=\Phi(\x_t,z_{t+1};\theta)$ (the construction of $\Phi$ is spelled out in the next paragraph).  Third, the implied next-period consumption $\hat c^{h+1}(\hat\x_{t,+})$ of the cohort that has just aged from $h$ to $h+1$.  The relative Euler-equation residual is then
\begin{equation}
e_{\mathrm{REE}}^h(\x_t)
\;:=\;
\frac{(u')^{-1}\!\Bigl(\beta\,\E{\,r(\hat\x_{t,+})\,u'\!\bigl(\hat c^{h+1}(\hat\x_{t,+})\bigr)\,}\Bigr)}{\hat c^h(\x_t)} \;-\; 1,
\qquad h=1,\ldots,A-1,
\label{eq:olg_ree}
\end{equation}
with $u(c)=\ln c$ in the analytic model so $(u')^{-1}(y) = 1/y$.  Equation~\eqref{eq:olg_ree} is the unit-free residual of the standard Euler equation~\eqref{eq:olg_euler}: a value of $10^{-3}$ means cohort $h$'s implied consumption is mispriced by $0.1\%$ relative to the conditional certainty equivalent.  This is the residual displayed in slide II.7.

\paragraph{Sampling the conditional expectation.}
The expectation in~\eqref{eq:olg_ree} is over the next-period shock $z_{t+1}$.  Because the analytic-model shock has only four states with i.i.d.\ transition $\pi_{ss'} = 1/4$, the expectation is computed \emph{exactly} (no Monte Carlo) by summing over the four next-period shocks:
\(
\E{\,r(\hat\x_{t,+})\,u'(\hat c^{h+1})\,} = \tfrac14\,\sum_{s'=1}^{4} r(\Phi(\x_t,s';\theta))\,u'\!\bigl(\hat c^{h+1}(\Phi(\x_t,s';\theta))\bigr).
\)
For each candidate $s'$ the next-state map $\Phi$ ages the cross-section by one period, sets the newborn to $k^1=0$, evaluates the firm prices~\eqref{eq:olg_prices} at $K_{t+1}$, and produces the next-period extended state $\hat\x_{t,+}$ on which the network is evaluated again to obtain $\hat a^h(\hat\x_{t,+})$ and hence $\hat c^{h+1}(\hat\x_{t,+})$.  When the shock has more states or is continuous, the same construction is replaced by a sample of $z_{t+1}\sim\pi(\cdot|z_t)$ inside the mini-batch (see \S\ref{sec:olg_56}).

\paragraph{The DEQN loss for the analytic OLG.}
Given a mini-batch $D_{\mathrm{train}}\subset\{\x_j\}_{j=1}^{N}$ sampled from the ergodic set of the current policy, the loss is the mean-squared relative Euler residual averaged across cohorts and states:
\begin{equation}
\boxed{\;\mathcal{L}_{D_{\mathrm{train}}}(\theta)
\;=\;
\frac{1}{|D_{\mathrm{train}}|}\;\frac{1}{A-1}\,
\sum_{\x_j\in D_{\mathrm{train}}}\;\sum_{h=1}^{A-1}
\bigl(e_{\mathrm{REE}}^h(\x_j)\bigr)^2\;}
\label{eq:olg_loss}
\end{equation}
\emph{(matching slide II.7).}  Two small barrier-style additive penalties on rescaled negative-consumption and negative-aggregate-capital hinges are summed in alongside~\eqref{eq:olg_loss} to keep training numerically robust away from convergence; in the notebook they carry the weight \texttt{PENALTY\_WEIGHT}~$=10$ and act on terms such as $\max(-\hat c^h,0)/(1+|\hat c^h|)$ rather than the raw squared hinge.  With the sigmoid savings-fraction head described above these hinges are in fact identically zero (savings stay in $[0,\mathrm{inc}^h]$, so $\hat c^h\ge 0$ and $K_{t+1}\ge 0$ always), so the penalties are pure backstops and do not bias the solution.\footnote{The 56-agent benchmark of \S\ref{sec:olg_56} adds two genuine extras to~\eqref{eq:olg_loss}: KKT product residuals (because the borrowing and collateral constraints actually bind) and an explicit bond-market-clearing residual (because the network outputs each agent's bond holding independently).  An orthogonal extension is to encode capital-market clearing \emph{exactly} via a dedicated output layer that rescales unnormalised cohort savings so that $\sum_{h=2}^{A} k_{t+1}^{h} = K_{t+1}$ holds by construction; \citet{azinoviczemlicka_2024} adopt this design in an OLG economy with rare disasters.}

\paragraph{DEQN architecture and training.}
The network takes a 40-dimensional input (the extended state~\eqref{eq:olg_state}, $16 + 4 \times 6$) and outputs 5 savings \emph{rates} $\hat\beta^h$ via a $40 \to 100 \to 50 \to 5$ architecture with ReLU hidden layers and a sigmoid savings-fraction output ($\approx 9{,}400$ parameters).  Training uses the episode-based procedure from Chapter~\ref{ch:deqn}: the current network generates a capital path (episode), equilibrium residuals are computed and used for SGD updates, and a new episode is simulated periodically.  The companion notebook exposes a \texttt{RUN\_MODE} switch with three calibrated budgets: \texttt{"smoke"} ($\sim$25 training segments, $\sim$30~s on CPU; a code-path sanity check, well short of convergence), \texttt{"teaching"} ($\sim$500 segments, $\sim$5~min on CPU; savings rates match the closed form to a few parts in $10^{4}$ and mean relative Euler errors are already $\sim 10^{-3}$ on the simulated cloud, though larger off-trajectory), and \texttt{"production"} ($\sim$10{,}000 segments with longer trajectories, several hours on CPU; mean Euler errors $\sim 10^{-3}$ or below, matching Table~3 of \citet{azinovicDEEPEQUILIBRIUMNETS2022}).  Adam is used throughout (learning rate $\sim 3\times 10^{-4}$ in the short presets, $10^{-5}$ in the production preset); the analogous decay to $10^{-6}$ used by the 56-agent benchmark (\S\ref{sec:olg_56}) is not needed at the analytic model's scale.

\begin{keyinsightbox}[Validation principle]
The 6-agent analytic OLG provides a known ground truth against which the DEQN can be validated exactly.  This validation step gives confidence that the same framework will produce reliable results for models without closed-form solutions.
\end{keyinsightbox}

\section{Inequality Constraints and KKT Complementarity}
\label{sec:olg_fb}

The 6-agent calibration above is deliberately frictionless: the no-short-sale-of-capital constraint never binds on its ergodic set, so we could solve it with a plain sigmoid-savings head and no multipliers.  Realistic OLG economies are not so kind.  The 56-agent benchmark of the next section carries a no-short-sale-of-capital constraint that binds on a non-trivial slice of states and a collateral constraint that the current notebook parameterization keeps slack on the learned ergodic set; binding inequality constraints in general bring in Karush--Kuhn--Tucker (KKT) complementarity, with its characteristic non-smooth orthogonality condition.  This section sets out how the DEQN framework encodes that complementarity; the next section puts it to work.

The no-short-sale-of-capital constraint $k'^h \geq 0$ introduces a complementarity condition via the Karush--Kuhn--Tucker (KKT) system:
\begin{equation}
k'^h \geq 0, \qquad \lambda^h \geq 0, \qquad k'^h \cdot \lambda^h = 0,
\label{eq:olg_kkt}
\end{equation}
where $\lambda^h$ is the KKT multiplier on the constraint.  In a generic non-linear program, the orthogonality condition $k'^h \cdot \lambda^h = 0$ is non-smooth at the origin and cannot be differentiated through naively.

The DEQN setup of \citet{azinovicDEEPEQUILIBRIUMNETS2022} sidesteps the kink by splitting enforcement across the architecture and the loss:
\begin{itemize}[itemsep=2pt]
\item \textbf{Hard side (architecture).}  The savings $k'^h$ and the multiplier $\lambda^h$ are both produced by the network through a softplus activation, so the inequalities $k'^h \geq 0$ and $\lambda^h \geq 0$ hold by construction at every iteration.
\item \textbf{Soft side (loss).}  With non-negativity already guaranteed, the orthogonality $k'^h \cdot \lambda^h = 0$ is enforced by adding the squared product residual $(k'^h \lambda^h)^2$ to the loss.
\end{itemize}
This product form is what the public reference implementation accompanying \citet{azinovicDEEPEQUILIBRIUMNETS2022} uses, and what we adopt in the \emph{56-agent benchmark} of \S\ref{sec:olg_56} (Notebook~\tpath{lecture_08_10_OLG_Benchmark_DEQN_persistent.ipynb}).  As noted above, in the 6-agent analytic calibration of \S\ref{sec:olg_analytic} the no-short-sale-of-capital constraint is non-binding everywhere on the ergodic set, so $\lambda^h \equiv 0$ and the multipliers (and the KKT residual) drop out of both the network output and the loss; that is why the mapping there was the simpler $\mathcal{N}_\theta: \mathbb{R}^{16+4A} \to \mathbb{R}^{A-1}$ of \S\ref{sec:olg_deqn} above, with no multiplier outputs.  The smoother \emph{Fischer--Burmeister} (FB) reformulation, $\Phi(a,b) = a + b - \sqrt{a^2 + b^2}$, is an alternative used in the IRBC notebook of Chapter~\ref{ch:irbc} for the investment-irreversibility constraint.

\paragraph{When to choose product form vs.\ Fischer--Burmeister.}  The product form $(k'^h \lambda^h)^2$ is simpler, gradient-cheaper, and sufficient whenever the constraint is rarely active on the ergodic set, since the optimizer just needs to verify slackness in expectation.  The Fischer--Burmeister residual $\Phi(a,b)^2$ keeps gradient information \emph{on both sides} of the active set: when the constraint is frequently binding (e.g.\ the IRBC irreversibility constraint on a non-trivial fraction of states), product-form gradients vanish whenever the constraint is locally inactive, which can stall training; FB does not have this pathology.  As a rule of thumb: product form for occasionally-binding KKT, FB for frequently-binding KKT.  In the OLG benchmark of \S\ref{sec:olg_56} the no-short-sale-of-capital constraint binds on a thin slice of the ergodic set, so the product form was sufficient; the IRBC application of the previous chapter binds more often and benefits from FB.

\paragraph{The two OLG models we solve, side by side.}
We have now built and solved the first of the two OLG instances that anchor the rest of the manuscript: the 6-agent analytic model used to validate the DEQN against a closed form (Sections~\ref{sec:olg_analytic}--\ref{sec:olg_deqn}).  The second is the 56-agent benchmark of \citet{azinovicDEEPEQUILIBRIUMNETS2022}, developed in the next section.  Table~\ref{tab:olg_6_vs_56} summarizes the structural and computational gap between them before we turn to it.

\begin{table}[ht]
\centering
\footnotesize
\setlength{\tabcolsep}{4pt}
\renewcommand{\arraystretch}{1.25}
\begin{tabular}{@{}>{\bfseries\raggedright\arraybackslash}p{2.8cm} >{\raggedright\arraybackslash}p{4.7cm} >{\raggedright\arraybackslash}p{7.2cm}@{}}
\toprule
                              & \textbf{6-agent analytic (\S\ref{sec:olg_analytic})}    & \textbf{56-agent benchmark (\S\ref{sec:olg_56})} \\
\midrule
Cohorts $A$                   & 6 (childhood-style)                                     & 56 (ages 25--80, one period = one year)         \\
Utility                       & Log ($\gamma=1$)                                        & CRRA ($\gamma=2$)                               \\
Shocks                        & i.i.d.\ TFP \& depreciation, 4 states                   & Persistent Markov on $(\eta,\delta)$            \\
Labor profile                 & Only youngest cohort works                              & Hump-shaped lifecycle endowment $\ell^h$        \\
Assets                        & Capital only                                            & Capital $+$ bonds                               \\
Constraints                   & None binding in calibration                             & No-short-sale of capital $k'^h\!\ge\!0$ binds; collateral $k'^h\!+\!\kappa b'^h\!\ge\!0$ kept slack by the $\hat q^h$ parameterization \\
Adjustment cost               & None                                                    & Quadratic $\tfrac{\zeta}{2}(k'^h\!-\!rk^h)^2$   \\
Network input dim             & 40 (extended; minimal 7)                                & 240 (extended; minimal 113)                     \\
Output dim                    & 5 (savings rates of cohorts 1--5)                       & 221 ($4(A-1)+1$: policies, multipliers, price)  \\
Loss terms                    & 5 Euler $+$ market clearing by construction             & 221: $4(A-1)$ Euler/KKT $+$ 1 bond clearing     \\
Network                       & Input(40) $\to$ 100 $\to$ 50 $\to$ 5                  & Input(240) $\to$ 128 $\to$ 128 $\to$ 221 (teaching) / Input(240) $\to$ 1000 $\to$ 1000 $\to$ 221 (production) \\
Validation target             & Closed-form $\beta_h$ of \citet{Krueger20041411}        & Mean Euler residual on simulated trajectory     \\
Notebook                      & \tpath{lecture_08_08_OLG_Analytic_DEQN_persistent.ipynb}  & \tpath{lecture_08_10_OLG_Benchmark_DEQN_persistent.ipynb}  \\
\bottomrule
\end{tabular}
\caption{The two OLG models solved in this chapter, side by side.  The economic richness of the 56-agent benchmark adds two assets, an effectively binding no-short-sale-of-capital constraint (the collateral constraint is kept slack by the $\hat q^h$ parameterization), persistent shocks, lifecycle labor, and adjustment costs, raising the network input dimension from 40 to 240 and the output dimension from 5 to 221.  The DEQN training loop is structurally identical in both cases.  Each variant additionally ships with a feedback-free exogenous-sampling companion notebook (\texttt{lecture\_08\_07\_OLG\_Analytic\_DEQN\_exogenous.ipynb}, \texttt{lecture\_08\_09\_OLG\_Benchmark\_DEQN\_exogenous.ipynb}) that exercises the same model under a non-co-evolving training cloud.}
\label{tab:olg_6_vs_56}
\end{table}

\section{The 56-Agent Benchmark}
\label{sec:olg_56}
\label{sec:olg_benchmark}

Table~\ref{tab:olg_6_vs_56} above previewed the gap; we now develop the second model in full.  The benchmark of \citet{azinovicDEEPEQUILIBRIUMNETS2022} scales the OLG framework to $A = 56$ agents (ages 25--80) with several realistic features:

\begin{itemize}[itemsep=2pt]
\item \textbf{CRRA utility} with $\gamma = 2$ (replacing log utility).
\item \textbf{Two assets:} capital $k$ and one-period bonds $b$, with bond price $p$ determined in equilibrium.
\item \textbf{Hump-shaped labor endowment} $e^h$ peaking in the early 50s.
\item \textbf{No-short-sale of capital:} $k'^h \geq 0$ (the constraint historically labelled the ``borrowing constraint'' in this literature; we keep the more precise name to free ``borrowing'' for the bond side).
\item \textbf{Collateral constraint:} $k'^h + \kappa\, b'^h \geq 0$, where $\kappa = 1/(1-\delta_{\max})$.
\item \textbf{Capital adjustment costs:} $\Psi^h = \frac{\zeta}{2}(k'^h - r\cdot k^h)^2$.
\item \textbf{Persistent shocks:} a 4-state Markov chain for TFP $\times$ depreciation (contrast with i.i.d.\ in the analytic model).
\end{itemize}

\paragraph{Lifecycle labor endowments.}
The labor endowment profile $e^h$ follows \citet{BKS1}.  In the implementation used here, $e^h$ is a quadratic in age that rises from $0.60$ at age $25$, peaks at $\approx 1.36$ around age $53$, then decays linearly between ages $\sim 62$ and $\sim 70$ to a flat post-retirement floor of $\approx 0.64$.  Table~\ref{tab:olg56_labor_profile} lists the values produced by the notebook formula at a few representative ages.
\begin{table}[ht]
\centering
\small
\begin{tabular}{l*{7}{c}}
\toprule
Age & 25 & 30 & 40 & 48 & 53 & 65 & 80 \\
\midrule
$e^h$ & 0.60 & 0.85 & 1.20 & 1.34 & 1.36 & 1.04 & 0.64 \\
\bottomrule
\end{tabular}
\caption{Representative points on the lifecycle labor-endowment profile in the 56-agent benchmark.}
\label{tab:olg56_labor_profile}
\end{table}
This hump-shaped profile ensures realistic savings heterogeneity: young agents with low labor income and no initial wealth are borrowing-constrained, mid-career agents with high earnings accumulate both capital and bonds, and older agents gradually decumulate toward the end of life.

\paragraph{Persistent aggregate shocks.}
The 4-state Markov chain combines TFP $\eta$ and depreciation $\delta$ into the pairs $(\eta,\delta) \in \{(0.978, 0.08),\, (1.022, 0.08),\, (0.978, 0.11),\, (1.022, 0.11)\}$.  The transition matrix is persistent (diagonal entries $\sim$0.63--0.88), in contrast to the i.i.d.\ shocks in the analytic model.  This persistence creates richer dynamics in capital accumulation: a sequence of bad TFP draws can push young agents deep into their borrowing constraint, producing endogenous amplification that a single-period shock would not generate.

\paragraph{Budget constraint.}
Each agent of age $h$ faces:
\begin{equation}
c^h + k'^h + p\cdot b'^h + \Psi^h = r\cdot k^h + b^h + w\cdot e^h.
\end{equation}
The collateral constraint $k'^h + \kappa\,b'^h \geq 0$ acts as a margin requirement: it limits bond borrowing ($b'^h < 0$) relative to capital holdings.  Since $\kappa = 1/(1-\delta_{\max})$, the constraint tightens when depreciation is high, precisely when agents are most likely to seek insurance through borrowing.

\paragraph{State $\x_t$ entering the network.}
The informational state of the benchmark is the triple
\(
(z_t,\,\bm{k}_t,\,\bm{b}_t) \in \{1,\ldots,4\}\times\R^A\times\R^A
\),
where $\bm{k}_t = (k_t^1,\ldots,k_t^A)$ and $\bm{b}_t = (b_t^1,\ldots,b_t^A)$ are the cross-sectional capital and bond distributions, so the minimal state has dimension $1+2A = 113$.  As in the analytic case, the notebook feeds the network an extended state of the same $16+4A$ form -- twelve aggregate scalars (shock index and its one-hot, $\eta_t$, $\delta_t$, $K_t$, $L_t$, $r_t$, $w_t$, and the gross resource $Y_t = \eta_t K_t^\alpha L_t^{1-\alpha} + (1-\delta_t)K_t$), four per-agent blocks ($k_t^h$, financial income $r_t k_t^h + b_t^h$, labor income $w_t e^h$, and cash $r_t k_t^h + b_t^h + w_t e^h$ -- the bond holdings $b_t^h$ are recoverable from financial income and are \emph{not} passed as a separate block, and the bond price $\hat p_t$ is an output, not an input), and the row of next-period transition probabilities $\pi(z_t,\cdot)$ (used by the conditional-expectation block of the loss); concretely $240 = 12 + 4\times 56 + 4$ (the notebook constant \texttt{FEATURE\_DIM}).  This is the analogue of slide III.8.

\paragraph{Policies approximated by the network.}
A single network $\mathcal{N}_\theta$ with softplus output produces a $4(A-1)+1$-dimensional vector that is sliced into five economic blocks (slide III.9):
\begin{equation}
\boxed{\;\mathcal{N}_\theta:\;\R^{240} \;\longrightarrow\; \R^{4(A-1)+1},\qquad
\mathcal{N}_\theta(\x_t) \;=\; \bigl(\hat k'^{1:A-1},\;\hat \lambda_b^{1:A-1},\;\hat q^{1:A-1},\;\hat \mu^{1:A-1},\;\hat p\bigr)(\x_t)\;}
\label{eq:olg56_policy}
\end{equation}
where $\hat k'^h$ are capital savings, $\hat\lambda_b^h$ the no-short-sale-of-capital multipliers, $\hat q^h \equiv \hat k'^h + \kappa\,\hat b'^h$ the collateral requirement (from which bond holdings are recovered as $\hat b'^h = (\hat q^h - \hat k'^h)/\kappa$), $\hat\mu^h$ the collateral-constraint multipliers, and $\hat p$ the equilibrium bond price.  Each raw output is mapped to an admissible value: softplus for the multipliers, and a bounded-exponential map around a baseline for the positive levels.  Concretely, writing $z^k_h$, $z^q_h$, and $z^p$ for the raw network outputs, the heads are
\begin{equation}
\hat k'^h \;=\; k^h_{\mathrm{baseline}}\,\exp(\tanh z^k_h), \quad
\hat q^h \;=\; q^h_{\mathrm{baseline}}\,\exp(\tanh z^q_h), \quad
\hat p \;=\; p_{\mathrm{baseline}}\,\exp(\tanh z^p),
\label{eq:olg56_heads}
\end{equation}
so the four non-negativity inequalities $\hat k'^h\ge 0$, $\hat\lambda_b^h\ge 0$, $\hat q^h\ge 0$, $\hat\mu^h\ge 0$ hold \emph{by construction}, leaving the orthogonality conditions of the KKT systems to be enforced softly in the loss (next paragraph).\footnote{In the current notebook implementation $\hat q^h$ is parameterized \emph{relative} to $\hat k'^h$, so it cannot fall to zero while $\hat k'^h>0$; the collateral-complementarity residual is then satisfied by $\hat\mu^h\to 0$, and the collateral constraint is effectively non-binding on the learned ergodic set, consistent with the chapter-opening note.  Allowing it to bind exactly requires a free positive slack output (a softplus head on $\hat q^h$); the architecture above already accommodates this swap.}  The production network uses $1000\times 1000$ hidden units ($\sim\!1.5$M parameters); the teaching version uses $128\times 128$.

\paragraph{Equilibrium residuals.}
Each cohort $h\in\{1,\ldots,A-1\}$ contributes \emph{four} residuals, one per equilibrium condition (slide III.6).  To keep the displayed form compact, introduce \emph{numerator/denominator shorthands} for the two Euler conditions:
\begin{equation*}
\begin{aligned}
\mathcal{N}^h_k(\x_t) &:= \beta\,\E{r_{t+1}\,\mathcal{D}^{h+1}_k(\hat\x_{t,+})\,u'(\hat c^{h+1}_{t+1})} + \hat\lambda_b^h + \hat\mu^h, \quad &
\mathcal{D}^h_k(\x_t) &:= 1 + \zeta\bigl(\hat k'^h - r_t k_t^h\bigr), \\[2pt]
\mathcal{N}^h_b(\x_t) &:= \beta\,\E{u'(\hat c^{h+1}_{t+1})} + \kappa\,\hat\mu^h, &
\mathcal{D}^h_b(\x_t) &:= \hat p.
\end{aligned}
\end{equation*}
Here $\mathcal{D}^h_k$ is the marginal-adjustment-cost wedge from $\Psi^h = \tfrac{\zeta}{2}(k'^h - r_t k^h)^2$: the capital Euler equation in envelope form reads $u'(c_t^h)\,\mathcal{D}^h_k = \beta\,\E{r_{t+1}\,\mathcal{D}^{h+1}_k\,u'(c_{t+1}^{h+1})} + \lambda_b^h + \mu^h$, so the same wedge appears next period on the marginal return to capital (this is the factor \texttt{adj\_factor\_next} in the notebook).  With $\zeta = 0$ it collapses to the textbook Euler equation.
The bond Euler reduces to the textbook stochastic-discount-factor form $\hat p = \beta\,\mathbb{E}[u'(c')]/u'(c)$ \emph{only when the collateral constraint is slack} ($\hat\mu^h = 0$); whenever $\hat\mu^h > 0$, the bond price carries an additional shadow-value term $\kappa\hat\mu^h/u'(\hat c^h)$ that captures the value of relaxing the collateral constraint.
The four per-cohort residuals are then
\begin{equation}
\begin{aligned}
e_{\mathrm{REE},k}^h(\x_t) &:= \frac{(u')^{-1}\bigl(\mathcal{N}^h_k(\x_t)\,/\,\mathcal{D}^h_k(\x_t)\bigr)}{\hat c^h(\x_t)} - 1, & \text{(Euler, $k$)}\\[2pt]
e_{\mathrm{REE},b}^h(\x_t) &:= \frac{(u')^{-1}\bigl(\mathcal{N}^h_b(\x_t)\,/\,\mathcal{D}^h_b(\x_t)\bigr)}{\hat c^h(\x_t)} - 1, & \text{(Euler, $b$)}\\[2pt]
e_{\mathrm{KKT},b}^h(\x_t) &:= \hat\lambda_b^h \cdot \hat k'^h, & \text{(borrowing complementarity)}\\
e_{\mathrm{KKT},c}^h(\x_t) &:= \hat\mu^h \cdot \bigl(\hat k'^h + \kappa\,\hat b'^h\bigr) = \hat\mu^h \cdot \hat q^h. & \text{(collateral complementarity)}
\end{aligned}
\label{eq:olg56_residuals}
\end{equation}
On top of these per-agent residuals the bond market must clear: bonds are in zero net supply, so the residual is the cross-sectional sum of bond holdings against the target $\bar B = 0$,
\begin{equation}
e_{\mathrm{MC},b}(\x_t) \;:=\; \sum_{h=1}^{A} \hat b'^h(\x_t) \;-\; \bar B \;=\; \frac{1}{\kappa}\sum_{h=1}^{A-1}\bigl(\hat q^h - \hat k'^h\bigr), \qquad \bar B = 0.
\label{eq:olg56_mc}
\end{equation}
Capital-market clearing $K_{t+1} = \sum_{h=2}^{A} k_{t+1}^h$ is once again satisfied by construction and does \emph{not} appear as a residual.  The conditional expectation in the two Euler equations is computed exactly as in~\eqref{eq:olg_ree}: by summing over the four next-period shocks weighted by the persistent-Markov transition probabilities $\pi(z_t,\cdot)$.

\paragraph{The DEQN loss for the 56-agent benchmark.}
Stack the four per-cohort residuals into one squared-cohort term
\(
R^h(\x)^2 := (e_{\mathrm{REE},k}^h)^2 + (e_{\mathrm{REE},b}^h)^2 + (e_{\mathrm{KKT},b}^h)^2 + (e_{\mathrm{KKT},c}^h)^2,
\)
then add the bond-market-clearing residual.  The mini-batch loss is
\begin{equation}
\boxed{\;\mathcal{L}_{D_{\mathrm{train}}}(\theta)
\;=\;
\frac{1}{|D_{\mathrm{train}}|}\;\frac{1}{4(A-1)+1}\,
\sum_{\x_j\in D_{\mathrm{train}}}\!\Biggl[\;\sum_{h=1}^{A-1} R^h(\x_j)^2 \;+\; \bigl(e_{\mathrm{MC},b}(\x_j)\bigr)^2\;\Biggr]\;}
\label{eq:olg56_loss}
\end{equation}
\emph{(matching slide III.6).}  With $A=56$ this is $4\times 55 + 1 = 221$ squared residuals per training state.  Each residual enters with weight one: no adaptive loss balancing (cf.\ Chapter~\ref{ch:nas}) is applied because the relative-Euler convention~\eqref{eq:olg_ree} already homogenizes the per-cohort Euler scales, and the product-form KKT residuals are unit-free under the softplus head; ReLoBRaLo or GradNorm would be the natural next step if a future calibration broke this homogeneity.  Comparison with~\eqref{eq:olg_loss}: the analytic case is the special instance of~\eqref{eq:olg56_loss} in which the no-short-sale-of-capital constraint never binds (so $\lambda_b^h\equiv 0$), there are no bonds (so all $b$- and collateral-related blocks drop out), and $4(A-1)+1$ collapses to $A-1$.  The two losses are the same template instantiated at different complexity.  Table~\ref{tab:olg56_residual_count} unpacks the residual blocks.

\begin{table}[ht]
\centering
\footnotesize
\begin{tabular}{llc}
\toprule
\textbf{Component} & \textbf{Symbol} & \textbf{Count} \\
\midrule
Euler (capital) & $e_{\mathrm{REE},k}^h$ & 55 \\
Euler (bonds) & $e_{\mathrm{REE},b}^h$ & 55 \\
KKT (borrowing) & $e_{\mathrm{KKT},b}^h = \hat\lambda_b^h\,\hat k'^h$ & 55 \\
KKT (collateral) & $e_{\mathrm{KKT},c}^h = \hat\mu^h\,\hat q^h$ & 55 \\
Market clearing (bonds) & $e_{\mathrm{MC},b} = \sum_h \hat b'^h$ & 1 \\
\midrule
\textbf{Total residuals} & & \textbf{221} \\
\bottomrule
\end{tabular}
\caption{Residual blocks entering the 56-agent benchmark loss for one training state.}
\label{tab:olg56_residual_count}
\end{table}

\paragraph{Training and results.}
Production training uses 60,000 episodes at lr $= 10^{-5}$ followed by 140,000 episodes at lr $= 10^{-6}$, with runtime of several hours on GPU.  The teaching version ($\sim$200 segments, 128-128 hidden units) runs in a few minutes on CPU and is meant to show the mechanics and qualitative lifecycle patterns, not final accuracy.  The loss trajectory typically exhibits oscillations, caused by re-simulation of the capital path at each episode, but the overall trend is steadily downward.

\paragraph{Lifecycle diagnostics.}
The trained model produces economically plausible lifecycle patterns.  Capital savings $k'^h$ follow a hump shape that mirrors the labor income profile: young agents save little (borrowing constraint binds), mid-career agents accumulate rapidly, and older agents decumulate.  Bond holdings $b'^h$ are initially negative (young agents borrow against future income) and increase with age as agents shift from illiquid capital to liquid bonds.  Bond prices vary across shock states, with higher prices in high-TFP states reflecting stronger demand for savings.  In the teaching run the Euler residuals are still large enough to treat the output as diagnostic; in production runs the mean Euler equation errors are of order $10^{-4}$--$10^{-3}$ for both capital and bond equations (matching Table~3 of \citet{azinovicDEEPEQUILIBRIUMNETS2022}), corresponding to a $\sim$0.01\%--0.1\% deviation in consumption.  Market clearing residuals are comparably small.  Convergence is also confirmed by the policy-drift check on the fixed anchor cloud: the run is treated as time-invariant once \tpath{policy_drift_rms} and \tpath{policy_drift_max} fall below their prescribed tolerances.

\begin{keyinsightbox}[Scalability of the DEQN framework]
The \emph{same algorithm} that solved the 6-agent model scales to 56 agents with two assets and inequality constraints.  Only the network size and training duration change; the DEQN framework itself is unchanged.  The 56-agent benchmark has a 113-dimensional minimal state and a 240-dimensional engineered network input, both infeasible for traditional grid-based methods.
\end{keyinsightbox}

\begin{keyinsightbox}[Chapter Summary]
OLG models discretise the cross-section into a finite number of cohorts $A$, so the minimal state vector contains the aggregate shock together with the cohort asset holdings and has dimension $\mathcal{O}(A)$.  Market clearing $\sum_h k_{t+1}^h = K_{t+1}$ closes the model and is the natural place to encode aggregation exactly via a market-clearing output layer (Chapter~\ref{ch:deqn}, footnote on Azinovic--Yang \& \v{Z}emli\v{c}ka 2024).  No-short-sale-of-capital and collateral constraints introduce KKT complementarity, which we enforce by combining softplus output activations (for non-negativity) with squared product residuals $(a\cdot b)^2$ in the loss.  Across both instances, the 6-agent analytic OLG and the 56-agent IER benchmark, the \emph{same} training loop covers the spectrum from textbook closed-form validation to research-scale scalability.
\end{keyinsightbox}

\section*{Further Reading}
\addcontentsline{toc}{section}{Further Reading}
\begin{itemize}[itemsep=2pt]
\item \citet{azinovicDEEPEQUILIBRIUMNETS2022}, the IER paper that established the 56-agent benchmark.
\item \citet{auerbach1987dynamic}, the classical OLG reference.
\item \citet{azinoviczemlicka_2024}, market-clearing output layer in OLG with rare disasters.
\end{itemize}

\section*{Exercises}
\addcontentsline{toc}{section}{Exercises}
\noindent Worked solutions and guidance for these exercises appear in Appendix~\ref{app:solutions}.
\begin{enumerate}[itemsep=4pt, leftmargin=*]
\item\label{ex:ch5:1} \textbf{[Core] OLG market clearing for $A=3$.}  Write the budget constraint and Euler equation for each of three cohorts (young, middle, old) and the market-clearing condition that closes the model.  Verify that the count of equations matches the count of unknown policies.
\item\label{ex:ch5:2} \textbf{[Core] KKT residuals: product form vs.\ Fischer--Burmeister.}  For a borrowing-constrained OLG agent, write the KKT conditions and rewrite them as \emph{both} (i) a product-form residual $(k'^h\cdot\lambda^h)^2$ and (ii) a Fischer--Burmeister residual $\Phi(\lambda^h, k'^h)^2$ with $\Phi(a,b)=a+b-\sqrt{a^2+b^2}$.  In each case, show that the loss vanishes \emph{exactly} at any KKT point, not just approximately.  Using the rule of thumb in \S\ref{sec:olg_fb}, argue which formulation you would choose for the \emph{6-agent analytic OLG} (\S\ref{sec:olg_analytic}, where the borrowing constraint is non-binding on the ergodic set) versus the \emph{investment-irreversibility constraint} of Chapter~\ref{ch:irbc} (which binds on a non-trivial fraction of states).
\item\label{ex:ch5:3} \textbf{[Computational] Hump-shaped lifecycle.}  Argue why a hump-shaped labor-income profile (rising through working years, falling after retirement) implies a hump in the savings policy $k'^h$.  Sketch the expected shape and check it against the trained-network output in notebook \tpath{lecture_08_10_OLG_Benchmark_DEQN_persistent.ipynb}.
\item\label{ex:ch5:4} \textbf{[Computational] Hard aggregation layer.}  The analytic notebook currently predicts cohort savings and then defines aggregate next-period capital as their sum, so capital-market clearing is already exact.  Implement an alternative architecture in notebook \tpath{lecture_08_08_OLG_Analytic_DEQN_persistent.ipynb} in which the network outputs a scalar $\widehat K_{t+1}>0$ and unnormalised cohort scores $(z^2,\ldots,z^A)$, then sets $s^h=\mathrm{softmax}(z^h)$ and $k_{t+1}^h=\widehat K_{t+1}s^h$.  Verify that $\sum_{h=2}^{A} k_{t+1}^h=\widehat K_{t+1}$ is at machine precision (below $10^{-12}$) at every training step.  Compare the Euler residual and runtime against the current ``sum-of-savings'' implementation, and explain why the hard layer is useful mainly when aggregate $K_{t+1}$ is a separate policy head.
\item\label{ex:ch5:5} \textbf{[Core] Bond pricing in equilibrium.}  Assume the borrowing/collateral constraint is slack for cohort $h$ throughout this exercise.  In the 56-agent OLG (\S\ref{sec:olg_56}) cohort $h$ holds capital $k^h$ and one-period riskless bonds $b^h$; capital pays the stochastic gross return $R_{t+1}$, bonds pay one unit of consumption next period and trade at price $p_t$.  Write the agent's Euler equations for capital and for bonds separately.  Eliminate the marginal utilities to derive the equilibrium bond-pricing equation
\[
   p_t \;=\; \frac{\beta\,\mathbb{E}_t\!\bigl[u'(c^{h+1}_{t+1})\bigr]}{u'(c^h_t)},
\]
and show that the same expression equals $\mathbb{E}_t[M_{t,t+1}]$ for the stochastic discount factor $M_{t,t+1}:=\beta u'(c^{h+1}_{t+1})/u'(c^h_t)$.  Show that absence of arbitrage between bonds and capital implies $1/p_t = \mathbb{E}_t[R_{t+1}] + \mathrm{Cov}_t(M_{t,t+1}, R_{t+1})/p_t$, equivalently $\mathbb{E}_t[R_{t+1}] - 1/p_t = -\mathrm{Cov}_t(M_{t,t+1}, R_{t+1})/p_t$, the standard risk-premium decomposition.  Then explain qualitatively how the bond-pricing equation changes when the collateral multiplier $\mu^h_t > 0$ is positive (i.e., the constraint binds), and identify which cohorts are most likely to be affected.  Briefly explain why the 6-agent analytic OLG of \S\ref{sec:olg_analytic} does not need an explicit bond-market residual in its DEQN loss.
\item\label{ex:ch5:6} \textbf{[Advanced/project] Collateral sensitivity.}  In the 56-agent benchmark notebook \tpath{lecture_08_10_OLG_Benchmark_DEQN_persistent.ipynb}, the collateral constraint is $k'^h + \kappa b'^h \ge 0$ with reference $\kappa = 1/(1-\delta_{\max}) \approx 1.1236$.  Sweep $\kappa \in \{0.8, 1.0, 1.1236, 1.3, 1.5\}$, retraining the network for each.  Report (i)~the fraction of ergodic states at which the collateral constraint binds, measured by $\hat q^h < 10^{-4}$ together with a positive multiplier $\hat\mu^h > 10^{-4}$, (ii)~the cross-cohort dispersion of bond holdings $b^h$ at the deterministic steady state, (iii)~the equilibrium bond price $p_\mathrm{ss}$.  Plot all three against $\kappa$ and explain qualitatively how tightening the collateral requirement (larger $\kappa$, hence less negative admissible bond positions for a fixed $k'^h$) changes portfolio choice.
\item\label{ex:ch5:7} \textbf{[Advanced/project] KKT-binding frequency under aggregate volatility.}  Still in the 56-agent benchmark, vary the standard deviation of the aggregate productivity shock $\sigma_z \in \{0.005, 0.01, 0.02, 0.04\}$ (the reference calibration uses $\sigma_z \approx 0.01$).  For each, retrain and report the fraction of ergodic-set draws on which (a)~the borrowing constraint $k'^h \ge 0$ binds and (b)~the collateral constraint $k'^h + \kappa b'^h \ge 0$ binds, broken out by cohort age.  Use the same small-slack/positive-multiplier convention as in Exercise~\ref{ex:ch5:6}.  Show that the binding fraction grows roughly linearly in $\sigma_z$ for the youngest cohorts but stays near zero for older cohorts, and connect this to the chapter's recommendation (\S\ref{sec:olg_fb}) that product-form KKT residuals suffice when the constraint binds rarely.
\end{enumerate}

\chapter{Heterogeneous Agents and Young's Method}
\label{ch:young}

The OLG models of Chapter~\ref{ch:olg} featured a finite number of agent types, so the cross-sectional state was simply a vector $(k^1,\ldots,k^A)$.  Many important macroeconomic applications instead require a \emph{continuum} of agents subject to idiosyncratic shocks.  In the \citet{krusell1998income} economy, an aggregate productivity shock additionally moves the cross section in a stochastic way, and the entire wealth distribution $\mu_t$ then becomes part of the aggregate state.  The \citet{aiyagari1994uninsured} model is the special case without aggregate risk: $\mu_t$ is fixed in stationary equilibrium and evolves deterministically along transitions, so it is a parameter of the equilibrium rather than a stochastic state variable.  Incomplete markets prevent full insurance in both, making explicit distributional tracking essential, but the ``master-equation'' challenge of treating $\mu_t$ as a high-dimensional aggregate state arises only once aggregate risk is added on top of Aiyagari.  Why represent $\mu_t$ as a histogram on a discrete grid rather than as a Monte Carlo panel?  Two reasons motivate the choice up front: a histogram is \emph{deterministic}, so re-running the same equilibrium gives identical aggregates and the loss is a smooth function of the network weights, and it is \emph{noise-free}, so Euler-equation residuals reflect approximation error rather than $\mathcal{O}(1/\sqrt N)$ Monte Carlo sampling noise (Figure~\ref{fig:young_vs_mc} makes this contrast quantitative).  This chapter develops Young's non-stochastic simulation method \citep{young2010} for representing $\mu_t$ as a histogram and shows how to embed that method within the DEQN framework of \citet{azinovicDEEPEQUILIBRIUMNETS2022} to solve heterogeneous-agent economies with neural network policy functions.

\paragraph{How this chapter maps onto the slides and notebooks.}
The heterogeneous-agent material of this chapter, together with the companion deck in \tpath{lecture_09_heterogeneous_agents_youngs_method/slides/}, can be read independently of the sequence-space material in \S\ref{sec:sequence_space}; readers already comfortable with Krusell--Smith may skip straight there on a first pass.  Two notebooks in \tpath{lecture_09_heterogeneous_agents_youngs_method/code/} accompany Sections~\ref{sec:young_method}--\ref{sec:young_deqn}: \tpath{lecture_09_10_Youngs_Method_Examples.ipynb} isolates Young's redistribution operator on toy examples, and \tpath{lecture_09_11_Continuum_of_Agents_DEQN.ipynb} embeds the same operator inside the Appendix~A.5 endowment-economy DEQN.  Section~\ref{sec:ks_alternatives} on alternative deep-learning approaches is paired with \tpath{lecture_09_12_KrusellSmith_DeepLearning.ipynb}, a classroom-scale all-in-one DL solver in the spirit of \citet{maliar2021deep}.

\paragraph{The Bewley--Huggett--Aiyagari lineage.}  The continuum-agent framework that this chapter operationalises has three foundational sources.  \citet{bewley1986stationary} introduced stationary monetary equilibrium with a continuum of agents subject to iid endowment shocks; the explicit self-insurance-through-borrowing-constraints mechanism that defines the modern incomplete-markets workhorse is due to \citet{imrohoroglu1989cost}, \citet{huggett1993riskfree}, and \citet{aiyagari1994uninsured}.  \citet{huggett1993riskfree} cast the idea as a tractable endowment economy with a single risk-free asset, focusing on the equilibrium interest rate.  \citet{aiyagari1994uninsured} added neoclassical production, closing the model in general equilibrium with capital accumulation; this is the canonical incomplete-markets economy.  \citet{krusell1998income} layered aggregate productivity shocks on top, producing the modern Krusell--Smith economy that this chapter targets, and its continuous-time reformulation is developed by \citet{achdou2022income} and revisited in Chapter~\ref{ch:ct_theory}.

\begin{figure}[ht]
\centering
\begin{tikzpicture}[
    node/.style={rectangle, draw=uzhblue, very thick, fill=uzhblue!8,
                 rounded corners=4pt, minimum width=2.6cm, minimum height=1.0cm,
                 align=center, font=\small},
    arr/.style={-{Stealth[length=2.5mm]}, thick, uzhblue!75},
    label/.style={font=\scriptsize\itshape, gray}
]
\node[node]                          (bewley)  at (0,0)   {Bewley (1986)\\\scriptsize stationary continuum};
\node[node]                          (huggett) at (4,1)   {Huggett (1993)\\\scriptsize endowment, risk-free};
\node[node]                          (aiy)     at (4,-1)  {Aiyagari (1994)\\\scriptsize production, GE};
\node[node, fill=harvardcrimson!10]  (ks)      at (8,-1)  {Krusell--Smith\\\scriptsize aggregate shocks};
\node[node, fill=softgreen!12]       (achdou)  at (8,1)   {Achdou et al.\ (2022)\\\scriptsize continuous time};
\node[node, fill=softorange!18]      (deqnY)   at (12,0)  {DEQN $+$ Young\\\scriptsize this chapter};

\draw[arr] (bewley) -- (huggett);
\draw[arr] (bewley) -- (aiy);
\draw[arr] (huggett) -- (achdou);
\draw[arr] (aiy) -- (achdou);
\draw[arr] (aiy) -- (ks);
\draw[arr] (huggett) -- (ks);
\draw[arr] (ks) -- (deqnY);
\draw[arr] (achdou) -- (deqnY);
\end{tikzpicture}
\caption{Genealogy of the heterogeneous-agent models treated in this script.  This chapter targets the Krusell--Smith branch (right) by combining a DEQN policy with Young's histogram update; the continuous-time branch (Achdou--Han--Lasry--Lions--Moll) is taken up in Chapter~\ref{ch:ct_theory}.}
\label{fig:bewley_lineage}
\end{figure}
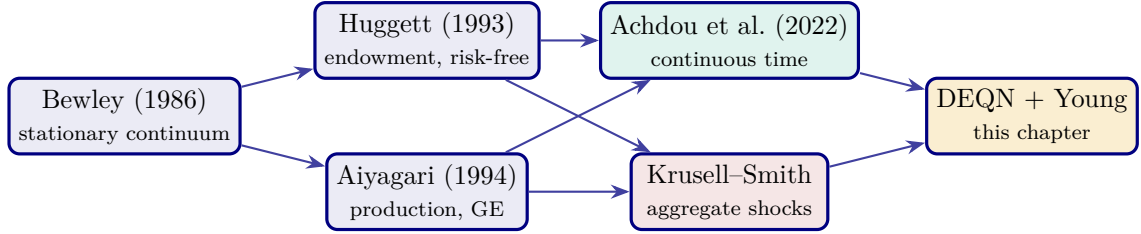

\section{From Representative to Heterogeneous Agents}
\label{sec:young_motivation}

In the representative-agent models of Chapters~\ref{ch:deqn}--\ref{ch:irbc}, aggregate capital $K_t$ is a sufficient statistic for the economy's state.  In reality, agents differ in wealth, income, and employment status, and incomplete markets prevent full insurance against idiosyncratic risk.  This heterogeneity matters for policy analysis:

\begin{itemize}[itemsep=2pt]
\item Fiscal stimulus: agents near the borrowing constraint have high marginal propensities to consume; wealthy agents save most of a windfall.
\item Monetary policy affects borrowers and savers differently.
\item The shape of the wealth distribution feeds back into aggregate demand and equilibrium prices.
\end{itemize}

The mathematical challenge depends on whether the economy carries aggregate risk.  Without it (the Aiyagari case), the cross-section $\mu_t$ is fixed in stationary equilibrium and evolves deterministically along transitions, so it parameterizes the equilibrium rather than serving as a stochastic state variable.  With aggregate risk (the Krusell--Smith case that this chapter targets), an aggregate shock moves the cross-section in a stochastic way, and the entire distribution $\mu_t$ then becomes part of the aggregate state.  Since $\mu_t$ is in either case a measure (an infinite-dimensional object), it cannot be placed on a grid without encountering the curse of dimensionality.

\paragraph{Two approaches.}
There are two main strategies for handling heterogeneous agents in discrete time.  \emph{Finite-agent methods} keep a high-dimensional but finite vector $(k_1,\ldots,k_N)$ of individual states, which is exact in $\mu_t$ but pays a permutation-symmetry cost that scales poorly with $N$ (the OLG models of Chapter~\ref{ch:olg}, and the finite-agent Monte-Carlo \emph{implementation} of the continuum-agent Krusell--Smith model by \citet{maliarmaliarvalli2010}, fall in this category).  \emph{Continuum-agent methods} replace the agent labels with the distributional state $\mu_t(\cdot)$, which is infinite-dimensional but eliminates sampling noise and the permutation problem, at the price of needing to approximate $\mu_t$ (\citet{krusell1998income} and \citet{young2010} are the canonical references).  This chapter pursues the second approach: tracking the distribution via a histogram, with Young's histogram method \citep{young2010} providing the distribution update.

\section{The Krusell--Smith (1998) Economy}
\label{sec:ks_economy}

The canonical heterogeneous-agent model with aggregate uncertainty is due to \citet{krusell1998income}.  Its key features are:

\begin{itemize}[itemsep=2pt]
\item A \emph{continuum} of ex ante identical, infinitely-lived agents.
\item \emph{Incomplete markets:} no insurance against idiosyncratic labor income risk.
\item \emph{Aggregate uncertainty:} productivity shocks that affect all agents.
\item A single asset (capital/bonds) with a borrowing constraint.
\end{itemize}

Each agent must forecast future prices to make optimal savings decisions.  But prices depend on the entire wealth distribution, which is infinite-dimensional.  The key insight of \citet{krusell1998income} is that the \emph{mean} of the wealth distribution is a nearly sufficient statistic for forecasting: a log-linear forecasting rule
\begin{equation}
\hat{H}(m_1, a) = \exp\!\bigl(A(a) + B(a)\log m_1\bigr),
\label{eq:ks_forecast}
\end{equation}
where $\hat{H}(m_1,a)$ is the approximate next-period aggregate capital (the price-forecasting function), $A(a)$ and $B(a)$ are the OLS intercept and slope coefficients (each a function of the aggregate shock state $a$), and $m_1 = \int k\,d\mu$ is mean capital; this rule achieves $R^2 > 0.9999$ in practice.

\begin{remarkbox}[Approximate aggregation: what it means and what it does \emph{not} mean]
The empirical observation that a single moment $m_1 = \int k\,d\mu$ is sufficient for price forecasting is known as \emph{approximate aggregation}.  Conceptually, it says that the equilibrium price mapping $r_{t+1}, w_{t+1}$ depends on the cross-sectional distribution \emph{almost only through its mean}, so the high-dimensional state $\mu_t$ collapses to a one-dimensional summary for forecasting purposes.

Importantly, \citet{krusell1998income} did not restrict themselves to the one-moment specification.  Their original paper explicitly reports results for two- and three-moment forecasting rules (adding the variance, and further adding the skewness of the wealth distribution) and finds that additional moments provide only marginal improvement: $R^2$ barely changes.  The one-moment benchmark is therefore an \emph{empirical} finding, not a modeling assumption.

Two cautions are essential.  First, this is a property of the \emph{Krusell--Smith model class} (one risky asset, modest wealth dispersion, moderate income risk) and is not a general theorem; in models with multiple assets, large MPC heterogeneity, or occasionally binding constraints that bite on a non-trivial mass of agents, the forecasting rule typically requires higher moments or a more flexible function approximator (Chapters~\ref{ch:gp},~\ref{ch:climate}).  Second, $R^2 > 0.9999$ on simulated mean dynamics does \emph{not} imply that the distribution itself is well summarized by its mean: tails, percentiles, and the borrowing-constrained mass continue to matter for welfare and policy analysis.  Approximate aggregation is therefore best read as a \emph{computational} result, and Sections~\ref{sec:young_method}--\ref{sec:young_deqn} accordingly track the full histogram and use the mean only as one of the network's input features rather than as a state-space substitute.
\end{remarkbox}

\paragraph{Equilibrium.}
A recursive competitive equilibrium consists of:
\begin{enumerate}[itemsep=2pt]
\item Individual optimization: each agent chooses savings $k'$ to maximize utility given the forecasting rule.
\item Market clearing: aggregate demand equals supply.
\item Consistent distribution evolution: $\mu_{t+1} = T(\mu_t, a_t)$ given optimal policies.
\item Rational expectations: agents' forecasts are self-confirming.
\end{enumerate}

The question is: how do we simulate the distribution forward to compute the realized mean and check the forecasting rule?

\paragraph{The traditional Krusell--Smith algorithm in detail.}
Before turning to the distribution-update step, it is useful to write the canonical KS algorithm explicitly.  It is a nested fixed-point iteration with an \emph{outer} loop over forecasting-rule coefficients and an \emph{inner} loop that solves the individual household's Bellman equation given those coefficients.

\begin{definitionbox}[Krusell--Smith (1998), Traditional Algorithm]
\begin{algorithmic}
\small
\REQUIRE Initial forecast coefficients $(A^{(0)}(a), B^{(0)}(a))$ from Eq.~\eqref{eq:ks_forecast}; capital grid $\mathcal{K}$; tolerances $\varepsilon_{\mathrm{inner}},\varepsilon_{\mathrm{outer}}$; simulation length $T_{\mathrm{sim}}$.
\STATE \textit{Key notation:} $a$: aggregate TFP shock; $\varepsilon$: idiosyncratic employment shock; $m_1 = \int k\,d\mu$: mean capital; $\beta$: household discount factor; $u(c)$: period utility; $c = w(\hat H)\,y(\varepsilon) + (1+r(\hat H))\,k - k'$: consumption (wages $w$ and return $r$ come from the forecasting rule $\hat H$); $k'$: end-of-period savings (the policy choice); $\varepsilon', a'$: next-period shocks; $R^2$: OLS coefficient of determination.
\FOR{outer iteration $\ell = 0, 1, 2, \ldots$ until $\|A^{(\ell+1)}-A^{(\ell)}\| + \|B^{(\ell+1)}-B^{(\ell)}\| < \varepsilon_{\mathrm{outer}}$}
    \STATE \textbf{(a) Solve individual Bellman equation} given $\hat H(m_1, a) = \exp\!\bigl(A^{(\ell)}(a) + B^{(\ell)}(a)\log m_1\bigr)$.
    \STATE \quad Iterate $V^{(n+1)}(k, \varepsilon, m_1, a) = u(c) + \beta\, \E{V^{(n)}(k', \varepsilon', \hat H(m_1,a), a')}$ until $\|V^{(n+1)}-V^{(n)}\| < \varepsilon_{\mathrm{inner}}$.
    \STATE \quad Return policy $k'(k, \varepsilon, m_1, a)$.
    \STATE \textbf{(b) Simulate the cross-section forward} for $T_{\mathrm{sim}}$ periods with (say) $10{,}000$ agents, starting from an initial distribution and drawing idiosyncratic shocks independently.
    \STATE \textbf{(c) Update forecasting rule} by OLS on the simulated path: regress $\log m_1^{t+1}$ on $\log m_1^t$ within each aggregate-shock state to obtain $(A^{(\ell+1)}(a), B^{(\ell+1)}(a))$.
    \STATE \textbf{(d) Convergence check:} retain the rule if $R^2 \geq 0.9999$ and the coefficient update $\|A^{(\ell+1)}-A^{(\ell)}\| + \|B^{(\ell+1)}-B^{(\ell)}\|$ is below $\varepsilon_{\mathrm{outer}}$.
\ENDFOR
\end{algorithmic}
\end{definitionbox}

The remarkable empirical finding of \citet{krusell1998income} is that a log-linear rule in $m_1$ alone achieves $R^2 > 0.9999$ and very small forecast errors: the ``approximate aggregation'' result.  Textbook implementations (e.g., the \texttt{econ-ark/KrusellSmith} REMARK) typically converge in $\sim 20$ outer iterations with standard parameters $\beta = 0.99$, $\alpha = 0.36$, log utility, aggregate shocks on $\{\text{good},\text{bad}\}$ with persistence $\approx 0.875$, and unemployment rates $4\%$ (good) vs.\ $10\%$ (bad).

\paragraph{Bottleneck.}
The inner Bellman solve on a two-dimensional $(k, m_1)$ grid is the expensive step.  It scales exponentially if one tries to add additional moments (e.g., tracking variance or skewness), which is why the KS algorithm as stated caps out at one moment.  Extensions that need more moments must use more powerful function approximators.  This is where the Young-histogram DEQN of §\ref{sec:young_deqn} (and the alternatives in §\ref{sec:ks_alternatives}) come in.

\section{Young's (2010) Non-Stochastic Simulation}
\label{sec:young_method}

\citet{young2010} proposed a deterministic method for propagating the wealth distribution that avoids Monte Carlo sampling entirely.

\paragraph{Core idea.}
Represent the wealth distribution at time $t$ as a \emph{histogram} over discrete bins, using two ingredients:
\begin{itemize}[itemsep=1pt, leftmargin=*]
\item a fixed capital grid $\{k_1, k_2, \ldots, k_{N_k}\}$, where $N_k$ is the number of bins and $i \in \{1,\ldots,N_k\}$ indexes individual grid points $k_i$;
\item a finite set of idiosyncratic-shock states $\{\varepsilon_1,\ldots,\varepsilon_{N_\varepsilon}\}$, indexed by $j \in \{1,\ldots,N_\varepsilon\}$.
\end{itemize}
The histogram value $G_t(k_i, \varepsilon_j) \in [0,1]$ is the probability mass of agents sitting at capital bin $k_i$ in shock state $\varepsilon_j$ at time $t$; by construction, $\sum_{i,j} G_t(k_i, \varepsilon_j) = 1$.

\begin{figure}[ht]
\centering
\begin{tikzpicture}[scale=0.95]
    \draw[thick, gray] (0,0) -- (12.5,0);

    \foreach \x in {1.5, 4, 8, 10.5} {
        \draw[thick] (\x,0.15) -- (\x,-0.15);
    }
    \draw[thick, darkred] (6.2,0.15) -- (6.2,-0.15);

    \node[below, font=\small] at (1.5,-0.2) {$k_{n-1}$};
    \node[below, font=\small] at (4,-0.2) {$k_n$};
    \node[below, font=\small, darkred] at (6.2,-0.2) {$k'$};
    \node[below, font=\small] at (8,-0.2) {$k_{n+1}$};
    \node[below, font=\small] at (10.5,-0.2) {$k_{n+2}$};

    \draw[thick, darkred, fill=darkred!25] (5.85,0) rectangle (6.55,2.8);
    \node[above, font=\small, darkred] at (6.2,2.8) {mass $m$};

    \draw[-{Stealth[length=3mm]}, thick, softblue, line width=1.5pt]
        (5.85,2.0) -- (4.35,1.1) node[midway, above, font=\small, sloped] {$\omega \cdot m$};
    \draw[-{Stealth[length=3mm]}, thick, softorange, line width=1.5pt]
        (6.55,2.0) -- (7.65,1.1) node[midway, above, font=\small, sloped] {$(1\!-\!\omega) \cdot m$};

    \draw[thick, softblue, fill=softblue!30] (3.65,0) rectangle (4.35,1.1);
    \draw[thick, softorange, fill=softorange!30] (7.65,0) rectangle (8.35,1.1);

    \node[draw=gray, rounded corners=3pt, fill=uzhgreylight, inner sep=5pt,
          font=\small, right] at (9.5,2.0) {$\displaystyle \omega = 1 - \frac{k' - k_n}{k_{n+1} - k_n}$};
\end{tikzpicture}
\caption{Linear interpolation in Young's method.  Mass~$m$ at off-grid point~$k'$ is redistributed to the two bracketing grid points~$k_n$ and~$k_{n+1}$ using weights~$\omega$ and~$1-\omega$.  Closer proximity to a grid point yields a larger share of the mass.}
\label{fig:young_interp}
\end{figure}
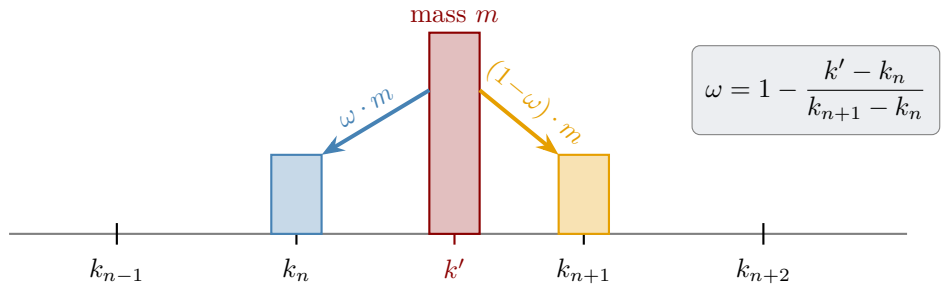

Given the household policy function $k' = g(k, \varepsilon, m_1, a)$, where:
\begin{itemize}[itemsep=1pt, leftmargin=*]
\item $k$ is the individual's current capital and $\varepsilon$ their current idiosyncratic shock;
\item $m_1 = \int k\,d\mu$ is aggregate mean capital (the summary statistic from Krusell--Smith);
\item $a$ is the aggregate productivity shock (the same $a$ that drives prices in the KS setup above);
\end{itemize}
the key operation is \emph{mass redistribution}.  Evaluating $g$ at a bin $(k_i,\varepsilon_j)$ produces the savings target $k' = g(k_i, \varepsilon_j, m_1, a)$, which generically lies \emph{off} the grid.  Let $n = n(i,j) \in \{1,\ldots,N_k-1\}$ denote the bracketing index defined by $k_n \leq k' < k_{n+1}$, and let
\[ m \equiv G_t(k_i,\varepsilon_j) \]
be the probability mass currently sitting at bin $(k_i,\varepsilon_j)$.  This mass $m$ is then split between the two bracketing grid points $k_n$ and $k_{n+1}$ using linear-interpolation weights:
\begin{equation}
\omega = 1 - \frac{k' - k_n}{k_{n+1} - k_n}, \qquad
\text{mass } \omega\cdot m \text{ to } k_n, \quad (1-\omega)\cdot m \text{ to } k_{n+1}.
\label{eq:young_weights}
\end{equation}
Figure~\ref{fig:young_interp} illustrates this operation.  A mass $m$ sitting at the off-grid savings target $k'$ is split between the two bracketing grid points $k_n$ and $k_{n+1}$, with the weight $\omega$ proportional to the proximity of $k'$ to $k_n$ (so $\omega \to 1$ when $k' \to k_n$ and $\omega \to 0$ when $k' \to k_{n+1}$).

\medskip\noindent\textbf{Why exactly this weight?}
The lottery weight $\omega$ in Eq.~\eqref{eq:young_weights} is not an arbitrary choice: it is the \emph{unique} value for which the two-point split has conditional mean equal to the off-grid policy choice $k'$.  Solving the equation $\omega\cdot k_n + (1-\omega)\cdot k_{n+1} = k'$ for $\omega$ recovers Eq.~\eqref{eq:young_weights} in one line:
\begin{equation}
\omega\cdot k_n + (1-\omega)\cdot k_{n+1} = k_n + (1-\omega)(k_{n+1} - k_n) = k_n + (k' - k_n) = k'.
\label{eq:young_meanpreserve}
\end{equation}
By linearity of expectation, this conditional mean equality at every grid bin extends to the full distribution: the unconditional mean of $G_{t+1}$ equals the unconditional mean of the policy-implied next-period capital.  Higher moments (variance, percentiles) are only approximated; the leading error is of order $(\Delta k)^2$ on smooth densities, so a finer grid improves higher-moment fidelity at no cost to mean preservation.

\begin{remarkbox}[Lottery interpretation: what makes this ``non-stochastic'']
Equation~\eqref{eq:young_weights} can equivalently be read as a \emph{fair lottery}: every agent at $(k, \varepsilon)$ whose policy choice lands at off-grid $k'$ is reassigned to $k_n$ with probability $\omega$ and to $k_{n+1}$ with probability $1-\omega$.  In a Monte Carlo panel of size $N$, each agent draws this lottery once and the empirical histogram converges to its expectation at rate $\mathcal{O}(N^{-1/2})$.  Young's method instead computes that expectation in closed form, which is equivalent to ``running'' an infinite number of agents through the lottery and integrating out the result.  This is why the procedure is called non-stochastic: the lottery is still there, but its outcome is computed analytically rather than sampled.  The same logic, applied to the discrete shock transitions $\pi_{\varepsilon'\mid\varepsilon}$, sends each piece of mass into all reachable next-period $\varepsilon'$ in proportion to $\pi$ rather than drawing one realisation.
\end{remarkbox}

\paragraph{Worked example.}
We illustrate the full histogram update with a small grid of $N_k=4$ capital levels $\{1.0,\, 2.0,\, 3.0,\, 4.0\}$ and two idiosyncratic states $\varepsilon \in \{\text{low},\, \text{high}\}$.

\medskip\noindent
\textbf{Step 1, Setup.}  The initial histogram $G_0$ (masses summing to 1):
\begin{center}\small
\begin{tabular}{l cccc c}
\toprule
& $k=1.0$ & $k=2.0$ & $k=3.0$ & $k=4.0$ & Row sum \\
\midrule
$\varepsilon = \text{low}$  & 0.10 & 0.20 & 0.10 & 0.05 & 0.45 \\
$\varepsilon = \text{high}$ & 0.05 & 0.15 & 0.20 & 0.15 & 0.55 \\
\bottomrule
\end{tabular}
\end{center}
The mean capital is $\bar{k}_0 = \sum_{i,j} G_0(k_i,\varepsilon_j)\cdot k_i = 0.10(1) + 0.20(2) + 0.10(3) + 0.05(4) + 0.05(1) + 0.15(2) + 0.20(3) + 0.15(4) = 2.55$.

\medskip\noindent
\textbf{Step 2, Policy evaluation.}  Let $y(\varepsilon)$ denote the dollar productivity associated with shock state $\varepsilon$, with $y_{\text{low}}=1$ and $y_{\text{high}}=3$ (here $\varepsilon$ is the \emph{state index}, $y$ is its numerical value).  Using a simple linear savings rule $k' = 0.4 k + 0.5\,y(\varepsilon)$:
\begin{center}\small
\begin{tabular}{l cccc}
\toprule
& $k=1.0$ & $k=2.0$ & $k=3.0$ & $k=4.0$ \\
\midrule
$\varepsilon = \text{low}$:  $k'$ & 0.9 & 1.3 & 1.7 & 2.1 \\
$\varepsilon = \text{high}$: $k'$ & 1.9 & 2.3 & 2.7 & 3.1 \\
\bottomrule
\end{tabular}
\end{center}

\medskip\noindent
\textbf{Step 3, Interpolation weights.}  Since the grid spacing is $\Delta k = 1.0$, each off-grid $k'$ is bracketed by $[k_n, k_{n+1}]$ with weight $\omega = 1 - (k' - k_n)/\Delta k$:
\begin{center}\small
\begin{tabular}{l cccc}
\toprule
& $k=1.0$ & $k=2.0$ & $k=3.0$ & $k=4.0$ \\
\midrule
$\varepsilon = \text{low}$:  bracket, $\omega$ & $[1,1]$, clip & $[1,2]$, 0.7 & $[1,2]$, 0.3 & $[2,3]$, 0.9 \\
$\varepsilon = \text{high}$: bracket, $\omega$ & $[1,2]$, 0.1 & $[2,3]$, 0.7 & $[2,3]$, 0.3 & $[3,4]$, 0.9 \\
\bottomrule
\end{tabular}
\end{center}
When $k' < k_1 = 1.0$ (here $k' = 0.9$ for the low-state, lowest-capital agents), all mass is assigned to $k_1$ (clipped to the boundary).

\medskip\noindent
\textbf{Step 4, Mass redistribution.}  For simplicity, assume the shock transition is the identity ($\varepsilon' = \varepsilon$), so mass stays in its current $\varepsilon$-state.  Building $G_1$ by accumulating the redistributed mass from each source bin (mass is conserved bin-by-bin: e.g.\ the $\varepsilon=\text{low}$, $k=2$ source of mass $0.20$ contributes $0.7\cdot0.20 = 0.14$ to $k=1$ and $0.3\cdot0.20 = 0.06$ to $k=2$):
\begin{center}\small
\begin{tabular}{l cccc c}
\toprule
& $k=1.0$ & $k=2.0$ & $k=3.0$ & $k=4.0$ & Row sum \\
\midrule
$\varepsilon = \text{low}$  & 0.270 & 0.175 & 0.005 & 0.000 & 0.450 \\
$\varepsilon = \text{high}$ & 0.005 & 0.210 & 0.320 & 0.015 & 0.550 \\
\midrule
total                       & 0.275 & 0.385 & 0.325 & 0.015 & 1.000 \\
\bottomrule
\end{tabular}
\end{center}

\medskip\noindent
\textbf{Step 5, Mean verification.}  The mean of $G_1$ is $\bar{k}_1 = 0.275(1)+0.385(2)+0.325(3)+0.015(4) = 2.08$.  The unclipped, policy-implied mean is $\sum_{i,j} G_0(k_i,\varepsilon_j)\,k'(k_i,\varepsilon_j) = 2.07$, so boundary clipping at $k' = 0.9$ raises the mean by only $0.01$.  Mean preservation is exact for source bins whose policy choice $k'$ lies strictly inside the grid; clipping at the boundary slightly biases the mean, here upward, because mass that would have landed at $k'=0.9$ is forced to $k=1$.  In general, boundary clipping biases the mean in the \emph{direction of the violated boundary}: clipping at $k_{\min}$ biases the mean upward (mass is pulled in from below the grid), clipping at $k_{\max}$ biases it downward.  With a wider grid ($k_{\min} < 0.9$) the mean would be preserved exactly.

\begin{remarkbox}[Practical take-away]
This small example shows why the grid must extend beyond the range of the policy function.  In production code, a safeguard checks that boundary bins contain negligible mass (typically $<10^{-6}$).
\end{remarkbox}

\medskip\noindent\textbf{The full picture: one cell splits into four.}
The worked example above used identity shock transitions to keep the arithmetic visible.  The general case combines the capital lottery with a \emph{shock fork}: each piece of mass split between $k_J$ and $k_{J+1}$ is then split again across the reachable next-period $\varepsilon'$ values according to $\pi_{\varepsilon'\mid\varepsilon, a}$.  With two shocks $\varepsilon \in \{L, H\}$ this produces \emph{four} destination bins for every source bin, with weights given by the product $\{\omega, 1-\omega\} \times \{\pi_{\varepsilon L}, \pi_{\varepsilon H}\}$.  Figure~\ref{fig:young_cascade} reproduces this two-stage cascade (essentially Fig.~1 of \citet{young2010}), annotated with a concrete numerical case.

\begin{figure}[ht]
\centering
\begin{tikzpicture}[scale=0.95, every node/.style={transform shape},
    src/.style={circle, draw=uzhblue, thick, fill=softblue!15,
        minimum size=0.95cm, font=\small, inner sep=0pt},
    mid/.style={circle, draw=harvardcrimson, thick, fill=red!8,
        minimum size=0.85cm, font=\footnotesize, inner sep=0pt},
    leaf/.style={circle, draw=darkgreen, thick, fill=green!8,
        minimum size=0.85cm, font=\footnotesize, inner sep=0pt},
    arr/.style={-{Stealth[length=2.5mm]}, thick},
    lab/.style={font=\scriptsize, midway, fill=white, inner sep=1pt}
]
    \node[src] (S) at (0,3.6) {$(k,\varepsilon{=}L)$};
    \node[font=\scriptsize, right=0.15cm] at (S.east) {mass $m = 0.05$};

    \node[mid] (M1) at (-3.4,1.5) {$(k_J,\,\cdot\,)$};
    \node[mid] (M2) at ( 3.4,1.5) {$(k_{J+1},\,\cdot\,)$};

    \node[leaf] (L1) at (-5.1,-0.9) {$(k_J,L)$};
    \node[leaf] (L2) at (-1.7,-0.9) {$(k_J,H)$};
    \node[leaf] (L3) at ( 1.7,-0.9) {$(k_{J+1},L)$};
    \node[leaf] (L4) at ( 5.1,-0.9) {$(k_{J+1},H)$};

    \draw[arr, harvardcrimson] (S) -- (M1) node[lab, above, sloped] {$\omega = 0.6$};
    \draw[arr, harvardcrimson] (S) -- (M2) node[lab, above, sloped] {$1-\omega = 0.4$};

    \draw[arr, darkgreen] (M1) -- (L1) node[lab, above, sloped] {$\pi_{LL}{=}0.9$};
    \draw[arr, darkgreen] (M1) -- (L2) node[lab, above, sloped] {$\pi_{LH}{=}0.1$};
    \draw[arr, darkgreen] (M2) -- (L3) node[lab, above, sloped] {$\pi_{LL}{=}0.9$};
    \draw[arr, darkgreen] (M2) -- (L4) node[lab, above, sloped] {$\pi_{LH}{=}0.1$};

    \node[font=\scriptsize, below=0.05cm] at (L1.south) {$0.027$};
    \node[font=\scriptsize, below=0.05cm] at (L2.south) {$0.003$};
    \node[font=\scriptsize, below=0.05cm] at (L3.south) {$0.018$};
    \node[font=\scriptsize, below=0.05cm] at (L4.south) {$0.002$};

    \node[font=\footnotesize\itshape, harvardcrimson, anchor=west] at (5.7,2.5) {Stage~1:};
    \node[font=\footnotesize\itshape, harvardcrimson, anchor=west] at (5.7,2.05) {capital lottery};
    \node[font=\footnotesize\itshape, darkgreen,    anchor=west] at (5.7,0.3) {Stage~2:};
    \node[font=\footnotesize\itshape, darkgreen,    anchor=west] at (5.7,-0.15) {shock fork};
\end{tikzpicture}

\smallskip
{\footnotesize Verification: $0.027 + 0.003 + 0.018 + 0.002 = 0.05 = m$ $\checkmark$.\quad The conditional mean of next-period capital is $\omega k_J + (1-\omega) k_{J+1} = k'$, by Eq.~\eqref{eq:young_meanpreserve}.}
\caption{Young's cascade for one source bin (essentially Fig.~1 of \citet{young2010}).  Mass $m$ at $(k, \varepsilon{=}L)$ flows in two stages: the capital lottery ($\omega$ vs.~$1-\omega$) sends it to the bracketing grid points $k_J$ and $k_{J+1}$, and the shock fork ($\pi_{LL}$ vs.~$\pi_{LH}$) splits each piece across the reachable next-period idiosyncratic states.  Each of the four leaves receives the product of its stage-1 and stage-2 weights times the source mass; the four leaf masses sum back to $m$.  Repeating the cascade for every active source bin and accumulating the leaves yields the new histogram $G_{t+1}$.}
\label{fig:young_cascade}
\end{figure}

A reader implementing the method should recognize three properties from the figure: (i)~mass is conserved bin-by-bin because $\omega + (1-\omega) = 1$ and $\sum_{\varepsilon'} \pi_{\varepsilon\varepsilon'} = 1$; (ii)~the capital lottery's expected next-period $k$ equals the policy choice $k'$ by Eq.~\eqref{eq:young_meanpreserve}; (iii)~the entire update is \emph{linear} in $G_t$, so it is a sparse matrix-vector product $G_{t+1} = T(\rho)\,G_t$ where the transition operator $T$ depends on the current policy.  This linearity is what makes the histogram update differentiable in the policy values almost everywhere, conditional on the interpolation brackets, when we embed it inside a neural-network training loop in \S\ref{sec:young_deqn}.  Index changes at bin boundaries and clipping at domain edges are nondifferentiable; standard implementations either ignore these measure-zero events, smooth the assignment, or stop gradients through the index-selection step, and in practice none of these choices materially affects training in the calibrations covered in this chapter.

\paragraph{The histogram update algorithm.}
At each time step, the histogram is updated deterministically:

\begin{algorithm}[H]
\caption{Young's histogram update at time $t$}
\label{alg:young}
\begin{algorithmic}
\REQUIRE Current histogram $G_t$, policy $g(\cdot)$, transition matrix $\pi_{\varepsilon'|\varepsilon,a}$, grid $\{k_j\}$
\STATE Initialize $G_{t+1}(k_j, \varepsilon') = 0$ for all $(j, \varepsilon')$
\FOR{each $(k_i, \varepsilon_j)$ with $G_t(k_i, \varepsilon_j) > 0$}
    \STATE Compute optimal savings: $k' = g(k_i, \varepsilon_j, m_1, a_t)$
    \STATE Find bracketing grid points: $k' \in [k_J, k_{J+1}]$
    \STATE Compute weight: $\omega = 1 - (k' - k_J)/(k_{J+1} - k_J)$
    \FOR{each next-period shock $\varepsilon'$}
        \STATE $G_{t+1}(k_J, \varepsilon') \mathrel{+}= \omega \cdot \pi_{\varepsilon'|\varepsilon_j, a_t}\cdot G_t(k_i, \varepsilon_j)$
        \STATE $G_{t+1}(k_{J+1}, \varepsilon') \mathrel{+}= (1-\omega) \cdot \pi_{\varepsilon'|\varepsilon_j, a_t}\cdot G_t(k_i, \varepsilon_j)$
    \ENDFOR
\ENDFOR
\RETURN Updated histogram $G_{t+1}$
\end{algorithmic}
\end{algorithm}

\medskip\noindent\textbf{Implementation cheatsheet.}
The pseudocode above translates into a few lines of \texttt{NumPy}; the inner double loop can also be vectorised with \texttt{np.add.at} for production use.

\begin{lstlisting}[language=Python, basicstyle=\footnotesize\ttfamily, frame=single, backgroundcolor=\color{backcolor}, numbers=none, escapeinside={(*}{*)}]
import numpy as np

def young_step(G, kp, pi_eps, k_grid):
    """One Young (2010) histogram update on a uniform k-grid.
    G[i,j]      mass at (k_grid[i], eps_j),       sums to 1
    kp[i,j]     policy choice k'(k_i, eps_j),     possibly off-grid
    pi_eps[j,jp] = Pr(eps_{t+1}=jp | eps_t=j)
    """
    G_next = np.zeros_like(G)
    dk     = k_grid[1] - k_grid[0]                       # uniform spacing
    n_k    = len(k_grid)
    for j in range(G.shape[1]):                          # current shock
        for i in range(G.shape[0]):                      # current capital bin
            x = kp[i, j]
            # Boundary handling: clip BOTH the bracket index J AND the
            # lottery weight omega.  Otherwise an off-grid x produces
            # omega < 0 or omega > 1 and hence negative or super-unit mass.
            if x <= k_grid[0]:
                J, omega = 0, 1.0                        # all mass to the floor
            elif x >= k_grid[-1]:
                J, omega = n_k - 2, 0.0                  # all mass to the cap
            else:
                J     = int((x - k_grid[0]) // dk)
                J     = min(max(J, 0), n_k - 2)
                omega = (k_grid[J + 1] - x) / dk         # in [0, 1]
            for jp in range(G.shape[1]):                 # next-period shock
                w = pi_eps[j, jp] * G[i, j]              # shock fork x source mass
                G_next[J,     jp] += omega       * w
                G_next[J + 1, jp] += (1 - omega) * w
    return G_next
\end{lstlisting}

The four scatter-add lines correspond exactly to the four leaves of Figure~\ref{fig:young_cascade}: each leaf receives \texttt{omega} or \texttt{(1-omega)} from the capital lottery, multiplied by \texttt{pi\_eps[j, jp]} from the shock fork, multiplied by the source mass.  The Krusell--Smith JAX tutorial in \tpath{lectures/lecture_10_sequence_space_deqns/code/lecture_10_KrusellSmith_Tutorial_CPU.ipynb} implements this same operation as \texttt{distribution\_step}, vectorised over the grid via \texttt{jax.vmap} and accumulated with \texttt{.at[\,].add(\,)}.

\begin{remarkbox}[Closed-form bracketing on GPUs, and how to keep it on log-spaced grids]
The expensive sub-step of Young's update (and of any piecewise-linear interpolation) is the \emph{bracketing index}
\[
  J(k') \;=\; \max\{\,n : k_n \le k'\,\},\qquad k' \in [k_J,\, k_{J+1}).
\]
The textbook implementation uses a binary search (e.g.\ \texttt{numpy.searchsorted} or \texttt{jnp.searchsorted}): $\mathcal{O}(\log N_k)$ per query, with data-dependent branches.  On a CPU this is essentially free; on a GPU under \texttt{XLA}/\texttt{CUDA} it is one of the worst patterns one can write, because (i)~SIMT threads of the same warp take different branches (warp divergence), (ii)~the resulting gathers are irregular, and (iii)~the opaque search op breaks operator fusion with the surrounding arithmetic, forcing extra kernel launches.

For an \emph{equidistant} grid the search collapses to a single fused multiply--add and a cast,
\[
  J(k') \;=\; \mathrm{clip}\!\left(\Big\lfloor \tfrac{k' - k_0}{\Delta k} \Big\rfloor,\;0,\;N_k-2\right),
\]
which is exactly what the line \verb|J = int((x - k_grid[0]) // dk)| in the cheatsheet does.  This is branch-free, uniform across threads, and fuses with the lottery weight $\omega = (k_{J+1}-k')/\Delta k$ into a single GPU kernel.

\paragraph{Log-spaced grids without losing $\mathcal{O}(1)$ lookup.}  Uniform $k$-grids waste resolution where the consumption policy is flat (the right tail) and starve resolution where it is steepest (near the borrowing constraint $k\!\to\!0$).  A simple fix preserves the closed-form bracketing: place equidistant knots in a \emph{transformed} coordinate $x = \phi(k)$ that maps the desired refinement region uniformly.  For a log-spaced grid with shift $c>0$,
\[
  x_n \,=\, x_0 + n\,\Delta x,\qquad k_n \,=\, e^{x_n} - c,\qquad n=0,\dots,N_k,
\]
the bracketing index becomes
\[
  J(k') \;=\; \mathrm{clip}\!\left(\Big\lfloor \tfrac{\log(c+k') - x_0}{\Delta x}\Big\rfloor,\;0,\;N_k-2\right),
\]
again $\mathcal{O}(1)$ per query and branch-free.  Crucially, the \emph{index} is computed in $x$-space but the lottery \emph{weights} are taken in the original $k$-space (using $k_J$, $k_{J+1}$ from the table), so the interpolated policy remains piecewise linear in $k$, consistent with the on-grid consumption values and with the mean-preserving lottery of Eq.~\eqref{eq:young_meanpreserve}.  More generally, any bijective $\phi$ for which $\phi^{-1}$ is cheap admits the same trick.  See \texttt{interpolate()} and \texttt{distribution\_step()} in the Krusell--Smith JAX tutorial of \citet{azinovicyangzemlicka2025sequencespace}~for a production implementation.
\end{remarkbox}

\begin{figure}[ht]
\centering
\begin{tikzpicture}[scale=0.88, every node/.style={transform shape},
    box/.style={rectangle, draw=uzhblue, thick, fill=uzhgreylight,
        minimum width=2.6cm, minimum height=0.85cm, font=\small,
        rounded corners=3pt, inner sep=4pt, text width=2.5cm, align=center},
    arr/.style={-{Stealth[length=3mm]}, thick, uzhblue}
]
    \node[box, fill=softblue!12] (hist) at (0,0) {Current histogram\\$G_t(k,\varepsilon)$};
    \node[box] (pol) at (3.6,0) {Evaluate policy\\$k' = g(k,\varepsilon,m_1,a)$};
    \node[box] (interp) at (7.2,0) {Interpolate\\to grid: $\omega$};
    \node[box, fill=softorange!15] (trans) at (10.8,0) {Apply shock\\transitions $\pi$};
    \node[box, fill=softgreen!12] (next) at (7.2,-1.8) {Next histogram\\$G_{t+1}(k,\varepsilon')$};

    \draw[arr] (hist) -- (pol);
    \draw[arr] (pol) -- (interp);
    \draw[arr] (interp) -- (trans);
    \draw[arr] (trans) |- (next);

    \node[font=\tiny, gray, above=0.08cm] at (1.8,0.55) {for each $(k_i,\varepsilon_j)$};
    \node[font=\tiny, gray, below=0.08cm] at (3.6,-0.55) {off-grid $k'$};
    \node[font=\tiny, gray, below=0.08cm] at (7.2,-0.55) {weights $\omega,\,1\!-\!\omega$};
    \node[font=\tiny, gray, right=0.08cm] at (10.8,-0.9) {$\varepsilon\!\to\!\varepsilon'$};
\end{tikzpicture}
\caption{Flow diagram for one forward step of Young's histogram update (Algorithm~\ref{alg:young}).  Starting from~$G_t$, the policy function is evaluated at every active bin, the resulting off-grid savings are interpolated back onto the grid, and idiosyncratic shock transitions redistribute mass across $\varepsilon$-states to produce~$G_{t+1}$.}
\label{fig:young_forward}
\end{figure}
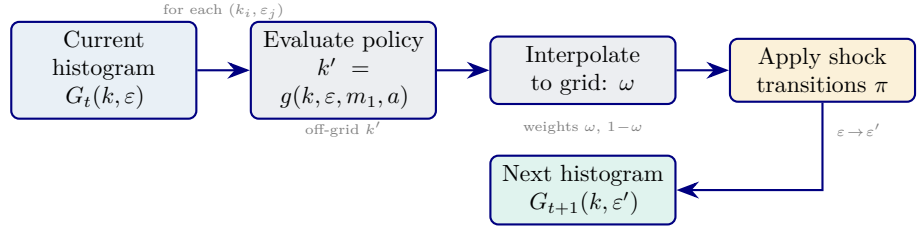

\noindent
Figure~\ref{fig:young_forward} visualizes the five stages of a single forward step.

\begin{figure}[ht]
\centering
\begin{tikzpicture}
\begin{axis}[
    name=leftpanel,
    width=6.5cm, height=4.5cm,
    ybar, bar width=6pt,
    xlabel={Capital $k$}, ylabel={Mass},
    xlabel style={font=\small}, ylabel style={font=\small},
    tick label style={font=\tiny},
    title={\small Young's method ($N_k=100$)},
    xmin=0, xmax=5, ymin=0, ymax=0.065,
    axis lines=left,
    grid=major, grid style={gray!12},
    legend pos=north west,
    legend style={font=\tiny, fill=white, fill opacity=0.9, draw=gray!40, text opacity=1, inner sep=2pt},
]
    \addplot[fill=softblue!50, draw=softblue!80, forget plot] coordinates {
        (0.5,0.005) (0.8,0.010) (1.1,0.018) (1.4,0.028) (1.7,0.038)
        (2.0,0.048) (2.3,0.055) (2.6,0.060) (2.9,0.058) (3.2,0.050)
        (3.5,0.040) (3.8,0.028) (4.1,0.016) (4.4,0.008) (4.7,0.003)
    };
    \addplot[thick, darkred, dashed, smooth, domain=0.2:4.9, samples=40]
        {0.062*exp(-0.5*((x-2.7)/0.8)^2)};
    \addlegendentry{True density}
\end{axis}

\begin{axis}[
    at={(leftpanel.east)}, anchor=west, xshift=1.2cm,
    width=6.5cm, height=4.5cm,
    ybar, bar width=6pt,
    xlabel={Capital $k$}, ylabel={},
    xlabel style={font=\small},
    tick label style={font=\tiny},
    title={\small Monte Carlo ($N=1000$ agents)},
    xmin=0, xmax=5, ymin=0, ymax=0.065,
    axis lines=left,
    grid=major, grid style={gray!12},
]
    \addplot[fill=softorange!45, draw=softorange!70, forget plot] coordinates {
        (0.5,0.007) (0.8,0.013) (1.1,0.015) (1.4,0.032) (1.7,0.042)
        (2.0,0.044) (2.3,0.058) (2.6,0.053) (2.9,0.062) (3.2,0.046)
        (3.5,0.043) (3.8,0.024) (4.1,0.019) (4.4,0.006) (4.7,0.005)
    };
    \addplot[thick, darkred, dashed, smooth, domain=0.2:4.9, samples=40]
        {0.062*exp(-0.5*((x-2.7)/0.8)^2)};
\end{axis}
\end{tikzpicture}
\caption{Young's histogram (left) versus Monte Carlo panel simulation (right).  Both approximate the same underlying wealth density (dashed).  The histogram method is deterministic and smooth; the Monte Carlo panel exhibits $\mathcal{O}(1/\sqrt{N})$ sampling noise that contaminates downstream OLS regressions in the Krusell--Smith algorithm.  \emph{The bars in this figure are a TikZ schematic illustrating the two regimes; for the actual histograms produced by the algorithm see notebook \texttt{lecture\_09\_10\_Youngs\_Method\_Examples} in the Lecture-09 \texttt{code/} folder.}}
\label{fig:young_vs_mc}
\end{figure}
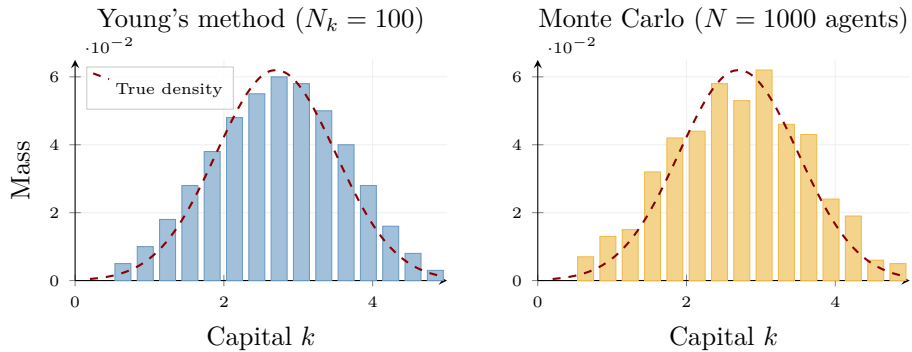

\medskip\noindent\textbf{Comparison with Monte Carlo.}
Young's method produces \emph{zero sampling noise} (deterministic), preserves the mean \emph{exactly}, requires only $\sim$100--5,000 grid points (versus $>$50,000 agents for Monte Carlo), and is fully reproducible.  The trade-off is that it approximates higher moments and requires a grid that is wide enough to contain all mass.  The following table summarizes the comparison:
\begin{center}
\small
\begin{tabular}{lcc}
\toprule
& \textbf{Young's method} & \textbf{Panel simulation} \\
\midrule
Sampling noise & None & $\mathcal{O}(1/\sqrt{N})$ \\
Mean preservation & Exact & Approximate \\
Typical size & 100--5,000 grid points & $>$50,000 agents \\
Reproducibility & Deterministic & Seed-dependent \\
Higher moments & Approximated & Approximated \\
\bottomrule
\end{tabular}
\end{center}
Figure~\ref{fig:young_vs_mc} contrasts the two approaches visually: the histogram method yields a smooth, noise-free distribution, while a Monte Carlo panel of comparable size exhibits visible sampling noise.

\noindent
The absence of sampling noise matters for the Krusell--Smith algorithm: Monte Carlo noise in the realized mean contaminates the OLS regression that updates the forecasting rule, potentially destabilizing convergence.

\medskip\noindent\textbf{Grid design.}
Two separate grids are used in practice.  The \emph{value function grid} (typically $\sim$150 irregularly spaced points) is finer near the borrowing constraint where the policy function has high curvature and coarser for large $k$ where behavior is smooth.  The \emph{simulation grid} for Young's histogram (typically $\sim$1,000--5,000 uniformly spaced points) uses uniform spacing to produce smooth histograms without artifacts.  The upper bound $k_{\max}$ must be chosen large enough that no mass reaches the boundary; if $k' > k_{\max}$ for any agent, all mass is assigned to the last grid point, which violates mean preservation.  A practical safeguard is to run a preliminary simulation and verify that the boundary bins contain negligible mass.

\medskip\noindent\textbf{The full Krusell--Smith algorithm.}
Combining value function iteration (VFI) with Young's simulation yields:
\begin{enumerate}[itemsep=2pt]
\item Initialize forecasting coefficients $A(a), B(a)$.
\item Solve the household problem via VFI given the forecasting rule $\hat{H}$.
\item Forward-simulate the distribution for $T$ periods via Young's method, recording realized means $m_1^t$.
\item Re-estimate $A(a), B(a)$ by OLS on simulated $(m_1^t, m_1^{t+1}, a_t)$.
\item Check convergence; if not converged, return to step~2.
\end{enumerate}
Young's non-stochastic simulation makes step~3 essentially noise-free and fast relative to a large Monte Carlo panel.  The full outer-loop iteration count and wall-clock time, however, remain implementation- and tolerance-dependent because the VFI solve and forecasting-rule fixed point are still present; the speedup applies to the simulation step, not as a generic wall-clock guarantee for the traditional KS workflow.

\paragraph{Convergence and accuracy.}
A remarkable empirical finding is that the log-linear forecasting rule~\eqref{eq:ks_forecast} achieves $R^2 > 0.9999$ in the standard Krusell--Smith economy: the first moment of the wealth distribution is a nearly sufficient statistic for predicting next-period prices.  Adding higher moments (variance, skewness) to the forecasting rule barely improves the fit.  This ``approximate aggregation'' result does not hold universally; it relies on the specific features of the Krusell--Smith calibration (small aggregate shocks, moderate borrowing constraint), but it is a useful benchmark against which richer models can be compared.

\section{DEQN with a Continuum of Agents}
\label{sec:young_deqn}

The traditional Krusell--Smith algorithm provides the benchmark logic for this chapter, but the classroom DEQN notebook used in the course is the simpler Bewley endowment economy of Appendix~A.5 in \citet{azinovicDEEPEQUILIBRIUMNETS2022}.  This distinction is pedagogically useful.  Krusell--Smith explains \emph{why} distribution tracking matters and \emph{why} Young's method is valuable inside an outer forecasting-rule loop.  Appendix~A.5 then shows \emph{how} the same histogram machinery enters a neural-equilibrium implementation once one replaces the forecasting rule by a price network.  By combining Young's histogram method with neural network policies, the DEQN approach overcomes both main limitations of the traditional KS workflow: the network can condition on the \emph{full histogram}, and there is no need for a separate forecasting rule because the price network directly takes the distribution as input.

\paragraph{What \S\ref{sec:young_deqn} inherits from Appendix~A.5.}
Five features of the Appendix-A.5 teaching model anchor the rest of this section and the companion notebook \tpath{11_Continuum_of_Agents_DEQN.ipynb}; they are the distinguishing departures from the canonical Krusell--Smith calibration of \S\S\ref{sec:ks_economy}--\ref{sec:young_method}:
\begin{itemize}[itemsep=2pt, leftmargin=*]
\item \textbf{Endowment economy, not production.}  Aggregate output is exogenous, $Y_t = w(a_t)$, instead of $Y_t = z_t K_t^\alpha L_t^{1-\alpha}$; there is no capital and no firm problem.
\item \textbf{Bonds in unit net supply.}  Households trade a single one-period bond at endogenous price $p_t$, and the market-clearing condition is $\int b_{t+1}\,d\mu_t = \bar B = 1$.
\item \textbf{Epstein--Zin recursive utility.}  Risk aversion ($\sigma=8$) and inverse IES ($\rho=2$) are separated, with discount factor $\beta=0.95$.  The KS benchmark in \S\ref{sec:ks_economy} instead used log utility with $\beta=0.99$.
\item \textbf{Six-state aggregate shock.}  $a_t \in \{0,\ldots,5\}$ encodes a $2$-state uncertainty regime crossed with a $3$-state income level, replacing the $2$-state TFP shock of canonical KS.
\item \textbf{Two idiosyncratic productivity types.}  Labour endowment $\eta_t \in \{0.8, 1.2\}$ on a transition matrix $\Pi_\eta$, in place of the employed/unemployed two-state Markov chain of canonical KS.
\end{itemize}

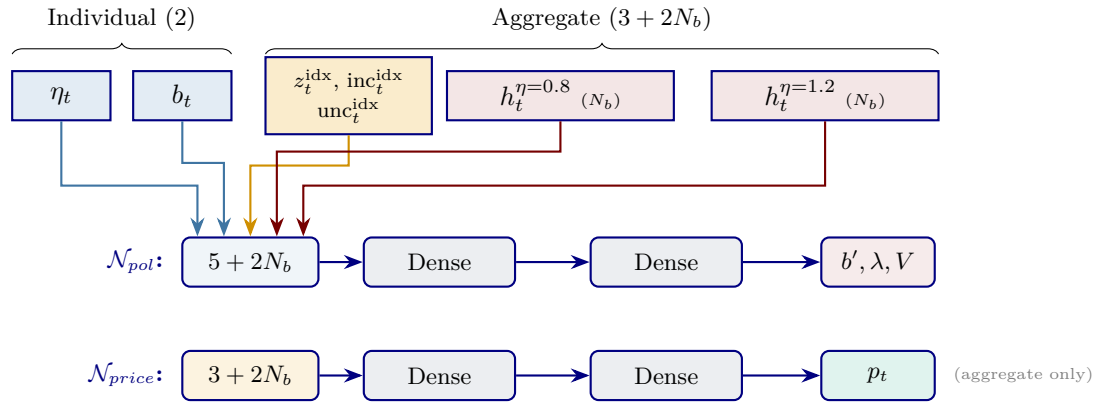
\begin{figure}[ht]
\centering
\begin{tikzpicture}[
    seg/.style={rectangle, draw=uzhblue, thick, minimum height=0.65cm,
        font=\small, inner sep=3pt},
    layer/.style={rectangle, draw=uzhblue, thick, fill=uzhgreylight,
        minimum width=2.0cm, minimum height=0.65cm, font=\footnotesize,
        rounded corners=3pt, inner sep=3pt},
    arr/.style={-{Stealth[length=2.5mm]}, thick, uzhblue},
    arrind/.style ={-{Stealth[length=2.5mm]}, thick, softblue!90!black},
    arragg/.style ={-{Stealth[length=2.5mm]}, thick, softorange!90!black},
    arrhist/.style={-{Stealth[length=2.5mm]}, thick, darkred!85!black}
]
    \node[seg, fill=softblue!15, minimum width=1.3cm] (eta) at (0,0) {$\eta_t$};
    \node[seg, fill=softblue!15, minimum width=1.3cm] (b) at (1.6,0) {$b_t$};

    \node[seg, fill=softorange!20, minimum width=2.2cm] (a) at (3.8,0)
        {\begin{tabular}{c}\scriptsize $z_t^{\mathrm{idx}},\,\mathrm{inc}_t^{\mathrm{idx}}$\\[-1pt]\scriptsize $\mathrm{unc}_t^{\mathrm{idx}}$\end{tabular}};
    \node[seg, fill=darkred!10, minimum width=3.0cm] (h1) at (6.6,0)
        {$h_t^{\eta=0.8}$ {\tiny ($N_b$)}};
    \node[seg, fill=darkred!10, minimum width=3.0cm] (h2) at (10.1,0)
        {$h_t^{\eta=1.2}$ {\tiny ($N_b$)}};

    \draw[decorate, decoration={brace, amplitude=4pt}]
        (-0.65,0.55) -- (2.25,0.55)
        node[midway, above=5pt, font=\footnotesize] {Individual (2)};
    \draw[decorate, decoration={brace, amplitude=4pt}]
        (2.7,0.55) -- (11.6,0.55)
        node[midway, above=5pt, font=\footnotesize] {Aggregate ($3+2N_b$)};

    \node[layer, fill=softblue!8, minimum width=1.8cm] (pin) at (2.5,-2.2) {\footnotesize $5+2N_b$};
    \node[layer] (ph1) at (5.0,-2.2) {Dense};
    \node[layer] (ph2) at (8.0,-2.2) {Dense};
    \node[layer, fill=darkred!8, minimum width=1.5cm] (pout) at (10.8,-2.2) {$b',\lambda,V$};

    \node[font=\footnotesize\bfseries, uzhblue, left=0.1cm] at (pin.west) {$\mathcal{N}_{pol}$:};
    \draw[arr] (pin) -- (ph1);
    \draw[arr] (ph1) -- (ph2);
    \draw[arr] (ph2) -- (pout);

    \draw[arrind]  (eta.south) -- ++(0,-0.85) -| ([xshift=-0.70cm]pin.north);
    \draw[arrind]  (b.south)   -- ++(0,-0.55) -| ([xshift=-0.35cm]pin.north);
    \draw[arragg]  (a.south)   -- ++(0,-0.40) -| ([xshift= 0.00cm]pin.north);
    \draw[arrhist] (h1.south)  -- ++(0,-0.40) -| ([xshift= 0.35cm]pin.north);
    \draw[arrhist] (h2.south)  -- ++(0,-0.85) -| ([xshift= 0.70cm]pin.north);

    \node[layer, fill=softorange!12, minimum width=1.8cm] (qin) at (2.5,-3.7) {\footnotesize $3+2N_b$};
    \node[layer] (qh1) at (5.0,-3.7) {Dense};
    \node[layer] (qh2) at (8.0,-3.7) {Dense};
    \node[layer, fill=softgreen!12, minimum width=1.5cm] (qout) at (10.8,-3.7) {$p_t$};

    \node[font=\footnotesize\bfseries, uzhblue, left=0.1cm] at (qin.west) {$\mathcal{N}_{price}$:};
    \draw[arr] (qin) -- (qh1);
    \draw[arr] (qh1) -- (qh2);
    \draw[arr] (qh2) -- (qout);

    \node[font=\tiny, gray, right=0.1cm] at (qout.east) {(aggregate only)};
\end{tikzpicture}
\caption{Histogram encoding and neural network architecture.  The individual state $(\eta_t, b_t)$ (blue boxes / blue arrows) and the aggregate state $(z_t^{\mathrm{idx}},\mathrm{inc}_t^{\mathrm{idx}},\mathrm{unc}_t^{\mathrm{idx}}, h_t^{\eta=0.8}, h_t^{\eta=1.2})$ (orange box for the three shock-index entries, red boxes for the two histograms) are concatenated as input to the policy network~$\mathcal{N}_{pol}$; each input arrow is colored to match its source box and enters the policy-input layer at a distinct horizontal offset, so the five arrows are uniquely identifiable at a glance.  The price network~$\mathcal{N}_{price}$ receives only the aggregate state.  Both networks use softplus output activations.}
\label{fig:young_encoding}
\end{figure}

\paragraph{Histogram encoding.}
The aggregate state is encoded as a vector containing three aggregate-shock index entries plus $N_b$ histogram values for each idiosyncratic shock type.  In the Appendix~A.5 notebook these three entries are the combined six-state aggregate index, the income-level index, and the uncertainty-regime index:
\begin{equation}
x_t^{agg} = \bigl(\underbrace{z_t^{\mathrm{idx}},\,\mathrm{inc}_t^{\mathrm{idx}},\,\mathrm{unc}_t^{\mathrm{idx}}}_{3\text{ shock-index entries}},\; \underbrace{h_t^{\eta=0.8}(b_1),\ldots,h_t^{\eta=0.8}(b_{N_b})}_{N_b\text{ values}},\;\underbrace{h_t^{\eta=1.2}(b_1),\ldots,h_t^{\eta=1.2}(b_{N_b})}_{N_b\text{ values}}\bigr).
\label{eq:young_hist_state}
\end{equation}
For $N_b = 100$ grid points and 2 idiosyncratic states, the aggregate state has dimension $3 + 200 = 203$.  The full input to the policy network adds the individual state $(b_t, \eta_t)$, giving total dimension $205$.  This is the \texttt{production} setting; the checked-in \texttt{smoke} and \texttt{teaching} runs use $N_b = 50$, so the aggregate state has dimension $103$ and the policy input $105$.  Figure~\ref{fig:young_encoding} shows how the histogram vector and individual state are assembled and fed into the two networks.

\paragraph{How the notebooks fit together.}
The companion notebook sequence mirrors this decomposition.  \tpath{10_Youngs_Method_Examples.ipynb} isolates Young's redistribution operator on toy examples, first in one dimension and then with idiosyncratic shocks.  \tpath{11_Continuum_of_Agents_DEQN.ipynb} then reuses the same logic inside the full aggregate state vector~\eqref{eq:young_hist_state}.  In other words, the second notebook is not a new distributional device; it is the first notebook's histogram update embedded in a larger equilibrium-learning loop.

\medskip\noindent\textbf{Neural network architecture.}
Two networks are trained jointly:
\begin{itemize}[itemsep=2pt]
\item \textbf{Policy network} $\mathcal{N}_{pol}$: takes individual + aggregate state ($\mathbb{R}^{5+2N_b}$) $\to$ outputs savings $b'$, borrowing multiplier $\lambda$, and value $V$ (3 outputs with softplus activation).
\item \textbf{Price network} $\mathcal{N}_{price}$: takes only aggregate state ($\mathbb{R}^{3+2N_b}$) $\to$ outputs the bond price $p$ (1 output with softplus).
\end{itemize}
Both networks use two hidden layers with 500 units (\texttt{production}) or 128 (the checked-in \texttt{smoke}/\texttt{teaching} runs).

\medskip\noindent\textbf{Equilibrium conditions as loss.}
The loss function comprises five terms, all of which should be zero in equilibrium:
\begin{equation}
\mathcal{L} = \frac{1}{N}\sum_{i=1}^{N}\left[\mathrm{EE}_i^2 + \mathrm{BE}_i^2 + \left(n_Z\,\mathrm{MC}_i\right)^2 + \mathrm{KKT}_i^2 + \mathrm{CB}_i^2\right],
\label{eq:young_loss}
\end{equation}
where EE is the Euler equation residual, BE is the Bellman equation consistency condition, MC is the bond market clearing condition, KKT is the borrowing complementarity, and CB penalizes negative consumption.  The market-clearing residual is rescaled by $n_Z$ (the number of aggregate-shock states) before squaring, which puts the single market-clearing residual on the same scale as the per-state Euler, Bellman, and complementarity residuals (themselves evaluated in relative terms, divided through by consumption).

\paragraph{Market clearing via histogram.}
A key advantage of the histogram representation is that market clearing can be computed \emph{exactly} (no Monte Carlo).  The equilibrium condition is $\sum_{\eta,j} h_t(\eta,b_j)\,b'(\eta,b_j,x_t^{agg}) = \bar B$; the residual that enters the loss is its deviation from zero,
\begin{equation}
\mathrm{MC} \;=\; \sum_{\eta}\sum_{j=1}^{N_b} h_t(\eta, b_j)\cdot b'(\eta, b_j, x_t^{agg}) \;-\; \bar B \;=\; 0 \text{ at equilibrium},
\label{eq:young_mc}
\end{equation}
where $h_t(\eta, b_j)$ is the histogram mass, $b'(\cdot)$ is the policy network output, and $\bar B$ is aggregate net bond supply.  We write the \emph{signed} residual here because it enters the loss~\eqref{eq:young_loss} squared, which is also how the notebook computes it.  The Appendix~A.5 notebook normalizes $\bar B=1$, which is the source of the ``$-1$'' term sometimes seen in code and slides.

\paragraph{Young's method inside the training loop.}
During each training episode, the histogram is propagated forward using Young's method (Algorithm~\ref{alg:young}) with the current neural network providing the policy function.  This creates a sequence of aggregate states $(x_0^{agg}, x_1^{agg}, \ldots, x_T^{agg})$ on which the equilibrium residuals~\eqref{eq:young_loss} are evaluated.  Young's redistribution operator is differentiable almost everywhere (linear interpolation conditional on fixed brackets), so the \emph{one-step} histogram update that enters the market-clearing residual at each sampled state carries gradients back to the policy network.  The longer simulated path of aggregate states is treated as data: it is regenerated each episode from the current network and held fixed inside the gradient tape, rather than backpropagated through end to end.  Even so the distribution co-evolves with the network: early in training, when the policy network is inaccurate, the simulated path will be ``wrong,'' but the market-clearing residuals evaluated along it provide corrective feedback through the loss, and as the network improves the distribution converges toward the ergodic steady state.

\section{Results and Discussion}
\label{sec:young_results}

In the Appendix~A.5 teaching model, the DEQN with histogram encoding achieves competitive accuracy: Euler equation errors are of order $10^{-3}$, and market-clearing residuals are comparably small.  These figures come from the checked-in \texttt{teaching}/\texttt{smoke} configuration of the companion notebook (a small network trained for a modest number of episodes); the \texttt{production} configuration tightens them further.  The broader lesson for the Krusell--Smith benchmark is conceptual rather than model-specific.  Compared to the traditional KS algorithm, the DEQN approach has two advantages: (i) the neural network can condition on the \emph{full} distribution rather than just its mean, providing a richer approximation that can capture situations where higher moments of the distribution matter for prices, and (ii) there is no need for a separate outer loop to update forecasting coefficients, since equilibrium conditions are enforced directly through the loss function.

\begin{remarkbox}[Connection to continuous time]
The discrete-time histogram approach presented here has a natural continuous-time counterpart.  In Chapter~\ref{ch:ct_theory}, we will see that the Kolmogorov forward equation (KFE; Fokker--Planck) describes the evolution of the wealth \emph{density} in continuous-time heterogeneous-agent models such as \citet{achdou2022income}.  While the mathematical formulation differs (histogram update vs.\ PDE), the economic question is the same: how does the wealth distribution evolve, and how does it feed back into equilibrium prices?
\end{remarkbox}

\section{Alternative Deep-Learning Approaches to Krusell--Smith}
\label{sec:ks_alternatives}

Before turning to the deep-learning alternatives, it is useful to set the histogram-DEQN method in the broader landscape of solution techniques for heterogeneous-agent equilibria with aggregate shocks.  Table~\ref{tab:ha_methods_landscape} compares classical and modern approaches along four dimensions that drive method choice in practice.

\begin{table}[ht]
\centering
\footnotesize
\setlength{\tabcolsep}{4pt}
\begin{tabular}{@{} >{\bfseries}p{2.8cm} p{2.6cm} p{2.6cm} p{2.6cm} p{2.6cm} @{}}
\toprule
\textbf{Method} & \textbf{Distribution rep.} & \textbf{Aggregate state} & \textbf{Solution principle} & \textbf{Best when} \\
\midrule
Classical KS \citep{krusell1998income}
 & Panel of $N \sim 10^4$ agents
 & First moment(s)
 & Bounded-rationality fixed point of forecasting rule
 & Standard incomplete-markets, low-dim aggregate state \\
\addlinespace
Reiter (back-loaded) \citep{reiter2009}
 & Histogram on a fixed grid
 & First-order perturbation around stationary
 & Linearize after solving the steady state
 & Small aggregate shocks, smooth policies \\
\addlinespace
Continuous-time Achdou et al.\ \citep{achdou2022income}
 & Density solving a KFE PDE
 & In the limit, the entire density
 & Coupled HJB+KFE finite differences
 & PDE-friendly model, smooth densities \\
\addlinespace
Histogram-DEQN \citep{azinovicDEEPEQUILIBRIUMNETS2022}
 & Young histogram on grid
 & Histogram entries (or moments thereof)
 & SGD on equilibrium residuals
 & Strong nonlinearities, occasionally binding constraints \\
\addlinespace
All-in-one DL \citep{maliar2021deep}
 & Panel of $N \gtrsim 10^3$ agents
 & Agent-level states, policy and aggregate together
 & SGD on stacked Euler+market-clearing residuals
 & Many states per agent, GPU available \\
\addlinespace
DeepHAM \citep{han2023deepham}
 & Permutation-invariant set encoder (DeepSets)
 & Learned $M\!\ll\!N$ generalized moments
 & Cumulative utility along simulated paths (policy-gradient with structural individual dynamics)
 & Want a low-dim aggregate state without committing to a moment \emph{a priori} \\
\bottomrule
\end{tabular}
\caption{Heterogeneous-agent solution methods at a glance.  The first three rows are classical or finite-difference; the last three are the modern deep-learning families compared in detail in Table~\ref{tab:ks_dl_comparison} and the rest of this section.}
\label{tab:ha_methods_landscape}
\end{table}

Whereas Table~\ref{tab:ha_methods_landscape} is panoramic (classical and DL methods on common axes), Table~\ref{tab:ks_dl_comparison} drills into the DL trio along the axes that matter when choosing among them.  Histogram-DEQN is not the only deep-learning approach to heterogeneous-agent equilibria, and it is pedagogically useful to see how the space of deep-learning strategies decomposes.  Three broad families have emerged in the literature.  Two informative axes organize them: how the cross-sectional distribution is represented as input to the network, and what objective is optimized.  Histogram-DEQN and All-in-One DL minimize residuals of the structural equilibrium equations; DeepHAM instead maximizes cumulative utility along simulated paths and uses Bellman residuals as a validation diagnostic.

\begin{table}[ht]
\centering
\small
\begin{tabular}{p{3.1cm} p{3.1cm} p{3.1cm} p{3.1cm}}
\toprule
 & \textbf{Histogram DEQN} & \textbf{All-in-one DL} & \textbf{DeepHAM} \\
 & \citep{azinovicDEEPEQUILIBRIUMNETS2022} & \citep{maliar2021deep} & \citep{han2023deepham} \\
\midrule
State distribution representation
 & Explicit histogram vector on a fixed grid
 & Explicit panel of $N$ agents' states
 & Learned generalized moments via a permutation-invariant encoder \\
\addlinespace
Dimension of aggregate state seen by the network
 & $\mathcal{O}(N_b)$ histogram entries
 & $\mathcal{O}(N)$ agent states
 & $M \ll N_b$ learned scalars \\
\addlinespace
Interpretability of distributional state
 & High (histogram readable)
 & Low (permutation-dependent)
 & High in the economic sense (the learned basis is interpretable, e.g.\ concave in assets, linking the moment to MPCs and redistribution) \\
\addlinespace
Permutation invariance
 & Automatic (histogram is invariant)
 & Requires careful architecture / data augmentation
 & Baked into the encoder by DeepSets structure \\
\addlinespace
Training objective
 & Euler + Bellman + market clearing + FB residuals (squared)
 & Euler + Bellman + market clearing residuals (squared)
 & Cumulative utility along simulated paths plus value-function error; Bellman residual is a validation diagnostic only \\
\addlinespace
Reported accuracy (baseline KS)
 & Euler errors $\sim 10^{-3}$--$10^{-4}$
 & Approximation errors $< 1\%$ with $>10^3$ agents
 & Bellman residual (used as a validation diagnostic) reduced by $68.9\%$ vs.\ KS with one learned generalized moment \\
\bottomrule
\end{tabular}
\caption{Three deep-learning approaches to heterogeneous-agent equilibria, all applied to the Krusell--Smith benchmark.  They differ on two axes: how the cross-sectional distribution is encoded as input to the network, and what objective is optimized.  Histogram-DEQN and All-in-One DL minimize squared structural residuals (Euler, Bellman, market clearing); DeepHAM instead maximizes cumulative utility along simulated paths, with the individual law of motion embedded in the computational graph and Bellman residuals tracked as a validation diagnostic.}
\label{tab:ks_dl_comparison}
\end{table}

\subsection{All-in-One Deep Learning (Maliar, Maliar \& Winant, 2021)}
\label{sec:mmw_ks}

\citet{maliar2021deep} propose what they call \emph{all-in-one deep learning}.  The key observation is that every dynamic economic model can be cast, at its core, as a collection of \emph{expectation conditions} (optimality, feasibility, market clearing) that vanish at the true solution.  They rewrite these conditions as non-linear regression equations with zero dependent variable, parameterize the policy and value functions by deep neural networks, and minimize the expected-squared residual by stochastic gradient descent on simulated paths.

For the Krusell--Smith benchmark with aggregate shocks and a continuum of agents, \citet{maliar2021deep} formulate the following components of the loss.  An alternative deep-learning route to the same problem is the symmetry-exploiting parameterisation of \citet{kahou2021exploiting}, which uses a permutation-invariant aggregation of agent-level features; the modern perturbation route is the refined Reiter implementation of \citet{bayer2020solving}.  Both are useful complements to the Young/DEQN combination developed below, with different scaling profiles and code-complexity tradeoffs.
\begin{itemize}[itemsep=2pt]
\item \textbf{Euler residual.}  The household's consumption-saving first-order condition:
\begin{equation}
\mathrm{EE}(s) \;=\; u'(c_t(s)) - \beta\,\mathbb{E}_{t}\!\left[u'(c_{t+1}(s'))\,R_{t+1}\right],
\end{equation}
where $s = (k_t,\varepsilon_t; \bm{S}_t)$ is the individual-plus-aggregate state and $\bm{S}_t$ is an $N$-agent panel of capital holdings.  The policy network outputs $c_t = \pi_\rho(s)$; the Euler residual is evaluated at simulated states and the expectation by Monte Carlo over next-period idiosyncratic and aggregate shocks.  With the borrowing constraint $k'\ge 0$, the plain Euler residual is insufficient: at a binding constraint the equation can be slack, so the loss must combine EE with a complementarity condition via Fischer--Burmeister or KKT (see below).
\item \textbf{Bellman residual / value-function error} for off-policy learning of a value function, used in the ``lifetime reward'' formulation.
\item \textbf{Market-clearing residual:} $\sum_i k_{t+1}^i = \bar K_{t+1}$ summed across the $N$ agents in the panel.
\item \textbf{Borrowing constraint} non-negativity enforced architecturally (softplus on savings); the complementarity optimality condition still enters the loss via a Fischer--Burmeister term regardless.
\end{itemize}
A central computational contribution is the \emph{all-in-one integration operator}: a single Monte Carlo realization of the next-period state is used to estimate the combined expectation across all residual terms simultaneously, rather than evaluating separate quadrature nodes for each conditional expectation.  This reduces the per-iteration cost from $\mathcal{O}(Q^m)$ (with $Q$ nodes per shock and $m$ shocks) to $\mathcal{O}(1)$ in expectation.

\paragraph{Scale reported by the authors.}
\citet{maliar2021deep} demonstrate the approach across several increasingly demanding setups: a Krusell--Smith variant with $N \geq 1{,}000$ explicit agents and $2{,}001$ state variables (one per agent plus one aggregate) in the baseline parameterization, scaled up to $N = 10{,}000$ agents in a moments-reduced variant where the cross section is summarized by $10$--$20$ aggregate moments; and on a one-agent consumption-savings problem with kinked policies they report approximation errors of at most a fraction of one percentage point.  A Python/TensorFlow replication is available at \url{https://github.com/marcmaliar/deep-learning-euler-method-krusell-smith} \citep{marcmaliar2022dlks}; it is the basis for the companion notebook \tpath{lectures/lecture_09_heterogeneous_agents_youngs_method/code/lecture_09_12_KrusellSmith_DeepLearning.ipynb} (described below).

\paragraph{Why this approach?}
All-in-one DL is the closest cousin of the DEQN framework of Chapter~\ref{ch:deqn}.  The only conceptual difference is that the aggregate state is represented by an \emph{explicit panel} of $N$ agents (a large vector $\bm{S}_t \in \R^N$) rather than by a histogram on a fixed grid.  Stochastic mini-batches draw both individual state and aggregate state, and the law of large numbers delivers permutation-invariance in the limit $N \to \infty$ without requiring a permutation-invariant architecture.  The appeal is conceptual simplicity; the cost is that the input dimension grows with~$N$ and the learner must rediscover permutation symmetry from the data.

\subsection{DeepHAM: Generalized Moments via DeepSets (Han, Yang \& E, 2024)}
\label{sec:deepham_ks}

\citet{han2023deepham} ask a more ambitious question: can the network \emph{learn} which summary statistics of the cross-sectional distribution are relevant for individual decisions, rather than having the researcher commit to tracking a histogram or the first moment?  They propose replacing the distribution~$\mu_t$ with a small set of \emph{generalized moments} obtained by averaging a flexible neural feature over the cross-section,
\begin{equation}
m_t^\ell \;=\; \frac{1}{N}\sum_{i=1}^{N} \phi_\theta^{\,\ell}(s_t^i) \quad\xrightarrow{N\to\infty}\quad \int \phi_\theta^{\,\ell}(s)\, d\mu_t(s), \qquad \ell = 1,\ldots,M,
\label{eq:deepham_moments}
\end{equation}
with $\bm{m}_t = (m_t^1,\ldots,m_t^M)$ and $\phi_\theta^{\,\ell}: \R^{d} \to \R$ a neural feature encoder trained jointly with the policy and value networks.  Equation~\eqref{eq:deepham_moments} is the canonical DeepSets architecture of permutation-invariant set functions \citep{zaheer2017deep}: averaging (or, in the continuum limit, integration against $\mu_t$) makes $\bm{m}_t$ invariant to permutations of the agents, and the $\phi_\theta^{\,\ell}$ encoders are flexible enough (by universal approximation on the permutation-invariant functions) to represent any fixed-arity moment.

The individual's value and policy functions then take the form
\begin{equation}
V_t = V_\rho(s_t^i; \bm{m}_t, a_t), \qquad k_{t+1}^i = \pi_\rho(s_t^i; \bm{m}_t, a_t),
\end{equation}
and all networks $(V_\rho, \pi_\rho, \phi_\theta)$ are trained jointly.  The primitive training objective is to \emph{maximize} cumulative utility along simulated paths,
\begin{equation}
J(\theta,\rho) \;=\; \mathbb{E}\!\left[\sum_{t=0}^{T-1} \beta^{t}\,u\!\bigl(c_\rho(s_t^i; \bm{m}_t, a_t)\bigr) \;+\; \beta^{T}\, V_\psi(s_T^i; \bm{m}_T, a_T)\right],
\label{eq:deepham_objective}
\end{equation}
where the expectation is taken over simulated idiosyncratic and aggregate histories generated under the current policy.  Squared Bellman residuals are reported as a validation diagnostic, not as the optimization target.  Because the individual law of motion, the budget constraint, and the transition structure are known economic objects, they are written directly into the computational graph: gradients of $J$ flow through these structural dynamics, in contrast to model-free reinforcement learning where transitions are observed only as samples \citep{yang2025structural}.  Because the generalized moments are themselves parameters of the optimization (not hyperparameters), the method automatically discovers the minimal set of distributional statistics required for equilibrium pricing.

A practical consequence of the policy-gradient formulation is that DeepHAM is well suited to problems where first-order conditions are difficult to write down or inconvenient to use, including constrained-efficiency problems with aggregate shocks, optimal-policy design, and behavioral macro questions; the headline application in \citet{han2023deepham} is precisely a constrained-efficiency problem solved by simulating the economy under candidate allocation rules and updating the rules to improve social welfare.

\paragraph{Reported accuracy (validation diagnostics).}
The numbers that follow are taken from \citet{han2023deepham}; they have not been independently replicated by the companion notebooks of this chapter, and should be read as reported results rather than as figures we can vouch for from our own runs.  On the baseline Krusell--Smith model, the authors report the following Bellman-error reductions, computed \emph{ex post} as validation measures of the converged solution rather than as the training objective:
\begin{itemize}[itemsep=2pt]
\item DeepHAM with \emph{only} the first moment in the state vector reduces the Bellman residual by \textbf{61.6\%} compared to the classical KS algorithm.
\item DeepHAM with \emph{one} learned generalized moment (on top of, or replacing, the first moment) reduces the residual by \textbf{68.9\%}.
\item The method extends to HA models with multiple shocks, multiple endogenous states, large shocks (risky steady state), and nonlinearities (e.g., ZLB) where both KS and the local perturbation method of \citet{reiter2009} break down.
\end{itemize}
Conceptually, DeepHAM bridges two traditions: the approximate-aggregation insight of \citet{krusell1998income} (a small number of moments can suffice) and the deep-function-approximator philosophy of \citet{maliar2021deep} and \citet{azinovicDEEPEQUILIBRIUMNETS2022} (the NN need not be restricted to pre-specified basis functions).  The official reference implementation, including replication code for the Krusell--Smith benchmarks above, is available at \url{https://github.com/frankhan91/DeepHAM}.

\paragraph{Interpretability of the learned moments.}
Although the encoders $\phi_\theta^{\,\ell}$ are flexible neural networks, the moments they produce are interpretable in the economic sense relevant to heterogeneous-agent macro: they summarize how the cross-section affects welfare and policy.  In the Krusell--Smith application of \citet{han2023deepham}, the learned basis function is concave in assets, so a marginal asset held by a poor household contributes more to the moment than a marginal asset held by a rich household.  This links the learned distributional representation directly to marginal propensities to save, redistribution, and welfare effects, which is the natural object of interest in HA models.  Generalized moments are therefore not just a flexible compression of $\mu_t$; they double as a device for reading economic content out of the trained equilibrium.

\paragraph{From learned moments to learned state spaces.}
Once equilibrium computation is written as policy-gradient optimization over simulated paths, one can ask a sharper question: do agents need the full distribution as a state variable at all?  In Walrasian heterogeneous-agent models, agents care about $\mu_t$ only indirectly, through equilibrium prices.  \citet{yang2025structural} exploit this in their \emph{structural reinforcement learning} (SRL) framework: agents' policy functions take low-dimensional prices, or short price histories, as the aggregate state, while $\mu_t$ remains part of the simulated environment used to clear markets.  This sidesteps the master equation entirely for a substantial class of HA economies and produces a natural conceptual arc: KS chooses moments \emph{a priori} (Section~\ref{sec:young_method}); DeepHAM \emph{learns} moments from the equilibrium objective (this section); SRL replaces moments with prices when the model permits, and lets ML help \emph{define} a tractable equilibrium concept rather than only solving a fixed one.

\subsection{Beyond Walrasian Markets: DeepSAM and Search Frictions}
\label{sec:deepsam_ks}

A natural follow-up question is whether the DeepHAM machinery extends to economies in which the cross-sectional distribution affects decisions through channels other than aggregate prices.  In standard heterogeneous-agent models the distribution is felt only through equilibrium prices: $\mu_t$ enters individual problems by setting $r_t = r(K_t)$ and $w_t = w(K_t)$.  In labor markets, search-and-matching, and other non-Walrasian settings, the distribution enters more directly: through the matching technology, the type composition on each side of the market, outside options, and bargained transfers.  This makes the equilibrium mapping intrinsically non-Walrasian and forecloses simple price-only summaries of the cross-section.

\citet{payne2025deepsam} address this case with \emph{DeepSAM}, a deep-learning solver for search-and-matching models with two-sided heterogeneity and aggregate shocks.  The architecture inherits the DeepHAM idea of a permutation-invariant set encoder, applied to each side of the market separately, and feeds the resulting type-composition summaries into networks that approximate value functions, matching surplus, and policy.  The training objective is again policy-improvement on simulated paths, with the matching technology and bargaining rule embedded structurally.  In Walrasian HAM the cross-sectional asset/wealth/income distribution enters individual problems only \emph{indirectly}, through equilibrium prices $r_t,w_t$; in SAM the type composition on each side of the matching market enters \emph{directly}, through the matching technology, outside options, and bargained transfers.  That is why DeepHAM and DeepSAM, despite sharing a permutation-invariant set-encoder architecture, treat the distribution differently.

\subsection{Which Method, When?}

The three deep-learning approaches (Histogram DEQN, §\ref{sec:young_deqn}; All-in-One DL, §\ref{sec:mmw_ks}; and DeepHAM, §\ref{sec:deepham_ks}) are complements rather than substitutes.  Table~\ref{tab:ks_dl_comparison} summarizes the practical trade-offs:
\begin{itemize}[itemsep=2pt]
\item For \emph{teaching} purposes, the Histogram DEQN is the cleanest: the network input is an interpretable distribution vector, and the training loop directly mirrors the DEQN template introduced in Chapter~\ref{ch:deqn}.
\item For \emph{research} problems where the number of agents is the natural state dimension (e.g., overlapping generations with many cohorts, or finite-agent social-planner problems), All-in-One DL is often more convenient because it requires no grid design.
\item For \emph{policy analysis in richer HA environments} (risky steady states, multiple endogenous states, ZLB), DeepHAM's learned-moment representation pays off both in accuracy and in interpretability, because the learned moments can be plotted and analyzed as functions of the distribution.
\end{itemize}

Table~\ref{tab:ks_dl_chooser} distils the same trade-offs into a quick decision aid: when the model fits the row's "When it shines" column, the matching method is the first one to try.

\begin{table}[ht]
\centering
\footnotesize
\setlength{\tabcolsep}{4pt}
\begin{tabular}{@{} >{\bfseries}p{3.0cm} p{4.4cm} p{6.0cm} @{}}
\toprule
\textbf{Method} & \textbf{Pros} & \textbf{When it shines} \\
\midrule
Histogram DEQN \citep{azinovicDEEPEQUILIBRIUMNETS2022}
 & Interpretable state; exact market clearing; reuses the DEQN template
 & Teaching; $N_b$ moderate; smooth policies \\
\addlinespace
All-in-One DL \citep{maliar2021deep}
 & No grid design; large-$N$ panels; single optimizer for all residuals
 & Large-$N$ research problems; OLG-many-cohort extensions \\
\addlinespace
DeepHAM \citep{han2023deepham}
 & Learned moments; cumulative-utility objective; risky steady state; ZLB
 & Rich HA macro-finance; constrained-efficiency / optimal-policy design \\
\bottomrule
\end{tabular}
\caption{Practical chooser for the three DL approaches to Krusell--Smith.  When several rows look applicable, the recommended ordering is: start with Histogram DEQN if a clean teaching narrative is the goal; switch to All-in-One DL if grid design is awkward; switch to DeepHAM if the cumulative-utility objective or learned moments are first-order to the question.  Adapted from the L09 deck's \emph{Head-to-Head} slide.}
\label{tab:ks_dl_chooser}
\end{table}

\paragraph{Notebook: a classroom-scale all-in-one KS solver.}
The accompanying Jupyter notebook \tpath{lecture_09_12_KrusellSmith_DeepLearning.ipynb} (introduced in §\ref{sec:young_results} and described in detail in the README) implements a classroom-scale version of the all-in-one DL approach of \citet{maliar2021deep} on the Krusell--Smith benchmark with the parameters of \citet{krusell1998income}.  It uses a single policy network $\pi_\rho(k, \varepsilon, K, a)$ parameterized by TensorFlow/Keras and an explicit panel of $N$ agents as the distributional input; the loss is the squared Euler residual, augmented with a Fischer--Burmeister complementarity term at the borrowing constraint, and aggregate consistency is imposed by construction (next-period capital is the cross-sectional mean of the panel's savings choices) rather than through a separate market-clearing penalty.  This is a deliberate simplification of the full all-in-one formulation of \S\ref{sec:mmw_ks}, which also carries a value network and an explicit market-clearing residual.  The notebook is annotated cell-by-cell with the correspondence to the equations in this chapter, and is designed to converge in under ten minutes on a standard CPU.  For the production-scale counterpart (up to $N = 10^4$ agents, GPU acceleration, richer shock structure), we refer the reader to the replication repository of \citet{marcmaliar2022dlks}.

\paragraph{Benchmarks and replication pointers.}  For readers who want to benchmark any of the deep-learning approaches against the traditional KS algorithm, the canonical reference implementation is \texttt{econ-ark/KrusellSmith} \citep{econark2020ks}, which implements the forecasting-rule-update algorithm of \citet{krusell1998income} and reports $R^2 > 0.9999$ within about 20 outer iterations under standard parameters ($\beta = 0.99$, $\alpha = 0.36$, log utility, two aggregate states).  Any deep-learning method must at least match that accuracy on the baseline model; the advantages are supposed to appear when the model is extended in directions KS cannot handle cleanly (more moments, many endogenous states, risky steady state).

\section{Extension: Deep Learning in the Sequence Space}
\label{sec:sequence_space}

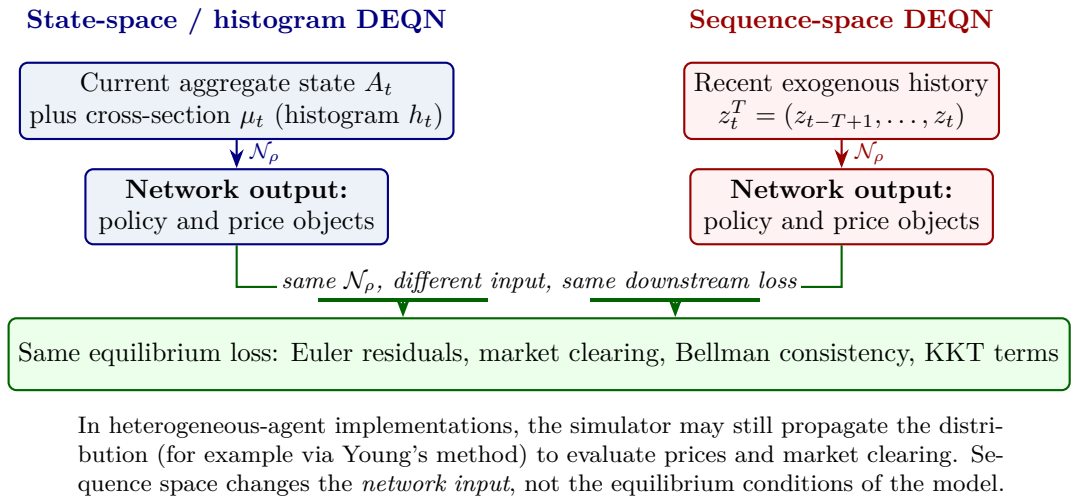
\begin{figure}[ht]
\centering
\begin{tikzpicture}[
    repstate/.style={rectangle, rounded corners=3pt, draw=uzhblue, thick,
        fill=softblue!10, minimum width=4.0cm, minimum height=0.95cm,
        align=center, font=\small},
    repseq/.style={rectangle, rounded corners=3pt, draw=harvardcrimson, thick,
        fill=red!5, minimum width=4.0cm, minimum height=0.95cm,
        align=center, font=\small},
    same/.style={rectangle, rounded corners=3pt, draw=darkgreen, thick,
        fill=green!8, minimum width=12.2cm, minimum height=0.95cm,
        align=center, font=\small},
    arrstate/.style={-{Stealth[length=2.5mm]}, thick, uzhblue},
    arrseq/.style={-{Stealth[length=2.5mm]}, thick, harvardcrimson},
    arrsame/.style={-{Stealth[length=2.5mm]}, thick, darkgreen}
]
    \node[font=\bfseries\small, uzhblue] at (-4.0,1.9) {State-space / histogram DEQN};
    \node[repstate] (s1) at (-4.0,0.85) {Current aggregate state $A_t$\\plus cross-section $\mu_t$ (histogram $h_t$)};
    \node[repstate] (s2) at (-4.0,-0.55) {\textbf{Network output:}\\policy and price objects};
    \draw[arrstate] (s1) -- node[right=1pt, font=\scriptsize, uzhblue] {$\mathcal{N}_\rho$} (s2);

    \node[font=\bfseries\small, harvardcrimson] at (4.0,1.9) {Sequence-space DEQN};
    \node[repseq] (q1) at (4.0,0.85) {Recent exogenous history\\ $z_t^T = (z_{t-T+1}, \ldots, z_t)$};
    \node[repseq] (q2) at (4.0,-0.55) {\textbf{Network output:}\\policy and price objects};
    \draw[arrseq] (q1) -- node[right=1pt, font=\scriptsize, harvardcrimson] {$\mathcal{N}_\rho$} (q2);

    \node[same] (loss) at (0,-2.5) {Same equilibrium loss: Euler residuals, market clearing, Bellman consistency, KKT terms};

    \draw[arrsame] (s2.south) -- ++(0,-0.55) -|
        node[pos=0.5, fill=white, inner sep=1.5pt, font=\footnotesize, darkgreen]
            {policy + prices}
        ([xshift=-1.8cm]loss.north);
    \draw[arrsame] (q2.south) -- ++(0,-0.55) -|
        node[pos=0.5, fill=white, inner sep=1.5pt, font=\footnotesize, darkgreen]
            {policy + prices}
        ([xshift=1.8cm]loss.north);

    \node[font=\footnotesize\itshape, align=center, text=black,
          fill=white, inner sep=1.5pt] at (0,-1.55)
        {same $\mathcal{N}_\rho$, different input, same downstream loss};

    \node[font=\footnotesize, align=center, text width=12.2cm] at (0,-3.85) {In heterogeneous-agent implementations, the simulator may still propagate the distribution (for example via Young's method) to evaluate prices and market clearing.  Sequence space changes the \emph{network input}, not the equilibrium conditions of the model.};
\end{tikzpicture}
\caption{Two ways to encode the aggregate state in deep equilibrium learning.  Each pipeline reads top-to-bottom: the upper (colored) box is the \emph{input} the user gives to the same neural network $\mathcal{N}_\rho$, the middle (colored) box is the network's \emph{output} (policy and price objects), and the green box is the equilibrium loss that consumes those outputs.  Histogram DEQNs (left, blue) feed an endogenous state representation $(A_t, \mu_t)$; sequence-space DEQNs (right, red) feed a truncated exogenous shock history $z_t^T$.  Crucially, the network and the residual-based training loss are identical across the two pipelines, only the input changes.}
\label{fig:sequence_space_compare}
\end{figure}

The histogram-based DEQN above is transparent because it feeds a direct approximation of the endogenous cross-sectional distribution into the neural network.  The price of that transparency is dimensionality: in richer heterogeneous-agent economies, the aggregate state can contain hundreds of histogram entries.  \citet{azinovicyangzemlicka2025sequencespace} propose a different representation of the aggregate state.  Instead of feeding the \emph{current endogenous state} to the network, they feed a \emph{truncated history of exogenous aggregate shocks}.  The equilibrium logic does not change: one still enforces Euler equations, market clearing, and occasionally binding constraints inside the loss.  What changes is the object that summarizes the aggregate state for the network.  Figure~\ref{fig:sequence_space_compare} contrasts the two views.

\medskip\noindent\textbf{The sequence-space representation.}
Let $z_t^T := (z_{t-T+1}, \ldots, z_t) \in \R^T$ denote the last $T$ realizations of the exogenous aggregate shock.  The key claim is that, in an ergodic economy, this history is an \emph{approximate sufficient statistic} for the endogenous aggregate state.  In the Brock--Mirman warm-up notebook, the network maps the shock history to a bounded savings rate, from which next-period capital follows by the resource constraint,
\[
s_t = \sigma\!\bigl(\mathcal{N}_\rho(z_t^T)\bigr) \in (0,1), \qquad K_{t+1} = s_t\, z_t K_t^\alpha,
\]
where $\sigma$ is the logistic squashing that keeps $K_{t+1}$ feasible.
In the richer heterogeneous-agent version, the network instead maps the same history to higher-level equilibrium objects such as policy-function coefficients or pricing objects.  This connects the method to the MIT-shock and sequence-space Jacobian literature of \citet{boppart2018exploiting} and \citet{auclert2021using}, but replaces local linear approximations with a global residual-based neural approximation.

\begin{definitionbox}[State-space vs sequence-space: the equilibrium operator]
\small
The two formulations can be written symmetrically.  Let $y_t$ denote the equilibrium objects of interest (policies, prices) at date $t$, $x_t$ the endogenous aggregate state, and $\varepsilon_t$ the exogenous shock.

\medskip\noindent\textbf{State-space recursion.}  The decision rule is a function $f$ of the current state, the state evolves through a known transition $H$, and equilibrium is the functional equation
\[
y_t = f(x_t), \qquad x_{t+1} = H(x_t, y_t, \varepsilon_{t+1}), \qquad G(f, x) = 0 \;\;\forall x.
\]

\noindent\textbf{Sequence-space formulation.}  Let $\mathcal{E}_t = (\varepsilon_t, \varepsilon_{t-1}, \ldots)$ denote the full shock history.  The decision rule is now a function $\Psi$ of the history (with initial condition $x_0$), and the state $x_t$ is recovered by iterating the same law of motion under $\Psi$:
\[
y_t = \Psi(\mathcal{E}_t \mid x_0), \qquad x_t = \mathcal{H}(\mathcal{E}_t, x_0 \mid \Psi), \qquad G(\Psi, \mathcal{E}, x_0) = 0 \;\;\forall \mathcal{E}, \forall x_0.
\]

\noindent Both formulations describe the \emph{same} equilibrium.  What differs is the \emph{domain of approximation}: $f$ lives on the (potentially infinite-dimensional) endogenous state space, while $\Psi$ lives on the (also infinite-dimensional but exogenously driven) space of shock histories.  In an ergodic economy the partial derivative $\partial\Psi/\partial\varepsilon_{t-\tau}$ vanishes as $\tau\to\infty$, so $\Psi$ admits a finite-history truncation $\widehat\Psi(z_t^T)$ with controllable error.  This truncation step, developed in the next paragraph, is what makes the sequence-space formulation computable.
\end{definitionbox}

\medskip\noindent\textbf{Intuition first.}
The easiest way to think about the method is as a \emph{memory compression} device.  A positive aggregate shock today raises output and therefore raises tomorrow's capital.  That extra capital still matters the period after, but only through the production elasticity $\alpha$, so its influence is smaller.  One more period later it is smaller again.  In other words, the current aggregate state stores a decaying memory of past shocks.  The sequence-space idea is to feed that shock history directly to the network rather than feeding the current endogenous state itself.

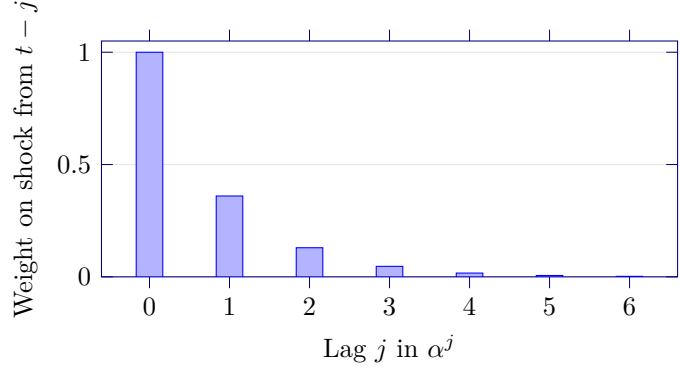
\begin{figure}[ht]
\centering
\begin{tikzpicture}
\begin{axis}[
    width=9.2cm,
    height=4.7cm,
    ybar,
    bar width=10pt,
    ymin=0,
    ymax=1.05,
    xlabel={Lag $j$ in $\alpha^j$},
    ylabel={Weight on shock from $t-j$},
    symbolic x coords={0,1,2,3,4,5,6},
    xtick=data,
    ymajorgrids=true,
    grid style={gray!20},
    axis line style={draw=uzhblue},
    tick style={draw=uzhblue},
    ticklabel style={font=\small},
    label style={font=\small},
    every axis plot/.append style={fill=softblue!60, draw=uzhblue}
]
\addplot coordinates {(0,1.0000) (1,0.3600) (2,0.1296) (3,0.0467) (4,0.0168) (5,0.0060) (6,0.0022)};
\end{axis}
\end{tikzpicture}
\caption{Intuition for sequence space in Brock--Mirman.  $\log K_t$ depends on past shocks with weights that decay like $\alpha^j$ in the lag $j$ (here $\alpha = 0.36$, the standard capital share).  Already at $j = 3$ the weight has fallen to $\sim\!0.05$, so a finite history of recent shocks summarizes the relevant aggregate information; very old shocks matter little.}
\label{fig:sequence_space_decay}
\end{figure}

\medskip\noindent\textbf{Brock--Mirman: what changes relative to Chapter~\ref{ch:deqn}?}
The Brock--Mirman warm-up is useful because the change can be written down exactly.  In Chapter~\ref{ch:deqn}, the state-space DEQN uses the current state as input,
\[
x_t^{\mathrm{state}} = (K_t, z_t), \qquad C_t = \mathcal{N}_\rho(K_t, z_t), \qquad K_{t+1} = z_t K_t^\alpha - C_t.
\]
In the sequence-space version, the \emph{economic model} is unchanged, but the network sees a different input:
\[
x_t^{\mathrm{seq}} = z_t^T = (z_{t-T+1}, \ldots, z_t), \qquad s_t = \sigma\!\bigl(\mathcal{N}_\rho(z_t^T)\bigr), \qquad K_{t+1} = s_t\, z_t K_t^\alpha, \qquad C_t = (1-s_t)\, z_t K_t^\alpha.
\]
The Euler residual is the same object as before,
\[
G_t = 1 - \beta \,\frac{C_t}{C_{t+1}} \,\alpha z_{t+1} K_{t+1}^{\alpha-1},
\]
so the economics are unchanged.  What changes is the computational representation:
\begin{itemize}[itemsep=2pt]
\item the \textbf{network input} changes from the current state $(K_t, z_t)$ to the recent history $z_t^T$;
\item the \textbf{network output} changes from current consumption $C_t$ to a bounded savings rate $s_t\in(0,1)$ in the warm-up notebook, so that $K_{t+1}=s_t z_t K_t^\alpha$ is feasible by construction;
\item the \textbf{current capital stock} is no longer fed directly into the network, but is generated recursively from the initial condition and previously predicted capital choices;
\item the \textbf{training samples} are overlapping shock histories rather than pointwise states $(K_t, z_t)$.
\end{itemize}

\medskip
\noindent This distinction is important conceptually.  For Brock--Mirman, sequence space is \emph{not} a dimensionality reduction, since $(K_t, z_t)$ is only two-dimensional whereas a history of length $T=25$ is larger.  The Brock--Mirman notebook is therefore a pedagogical demonstration of the idea that histories can stand in for endogenous states.  The dimensionality gain appears only in richer heterogeneous-agent models, where the relevant alternative is a large histogram or other high-dimensional distributional summary.

\medskip
\noindent\textbf{Intermediate bridge: sequence-space IRBC.}  Between the one-shock Brock--Mirman warm-up and the infinite-dimensional Krusell--Smith state, the companion notebook \tpath{lectures/lecture_10_sequence_space_deqns/code/lecture_10_05b_SequenceSpace_IRBC.ipynb} re-trains the two-country IRBC model of Chapter~\ref{ch:irbc} under sequence-space inputs: the policy network reads the last $T=80$ shock vectors (a $240$-dimensional history with $\rho_z^T \approx 1.7\times 10^{-2}$ truncation error) instead of the four-dimensional current state.  The $2N+1$ equilibrium residuals (Euler, ARC, Fischer--Burmeister), the Gauss--Hermite quadrature, and the cloud-method sampler are literally unchanged from nb~01; only the input domain changes.  Because the current capital stock is no longer an input, we parametrize the output head around the steady state, $k'_j = k_{ss}\exp(\tanh z^k_j)$ and $\lambda = \lambda_{ss}\exp(\tanh z^\lambda)$, which keeps gradients lively at the target policy and prevents the cold-start divergence that plagued a naive softplus head.  This notebook is a \emph{pedagogical bridge} rather than a computational win, at a four-dimensional state the history is much larger, not smaller, but it shows that the same template handles a multi-equation system with multiple independent shock channels before we hand the method over to Krusell--Smith, where the dimensionality gain is real.

\begin{figure}[ht]
\centering
\begin{tikzpicture}[
    flow/.style={rectangle, rounded corners=3pt, draw=uzhblue, thick,
        fill=uzhgreylight, text width=2.70cm, minimum height=1.00cm,
        align=center, font=\small},
    flowred/.style={rectangle, rounded corners=3pt, draw=harvardcrimson, thick,
        fill=red!5, text width=2.70cm, minimum height=1.00cm,
        align=center, font=\small},
    flowgreen/.style={rectangle, rounded corners=3pt, draw=darkgreen, thick,
        fill=green!8, text width=2.70cm, minimum height=1.00cm,
        align=center, font=\small},
    arr/.style={-{Stealth[length=2.5mm]}, thick, uzhblue}
]
    \node[flow] (z) at (0,0) {Sample exogenous\\ shock path $\{z_t\}_{t=0}^S$};
    \node[flowred] (h) at (4.5,0) {Build history windows\\ $z_t^T$};
    \node[flowred] (n) at (9.1,0) {Network output\\ $K_{t+1}$ or policy coefficients};

    \node[flow] (sim) at (10.1,-3.4) {Forward simulator\\ prices, aggregates, Young update if HA};
    \node[flowgreen] (loss) at (4.5,-3.4) {Residual loss\\ Euler, MC, Bellman, KKT};
    \node[flowgreen] (sgd) at (-1.1,-3.4) {Gradient step\\ update $\rho$};

    \draw[arr] (z) -- (h);
    \draw[arr] (h) -- (n);
    \draw[arr] (n.south) -- ++(0,-0.70) -| (sim.north);
    \draw[arr] (sim.west) -- (loss.east);
    \draw[arr] (loss.west) -- (sgd.east);
    \draw[arr] (sgd.north) to[out=110,in=250] node[left=2pt, font=\footnotesize] {repeat} (z.south);
\end{tikzpicture}
\caption{Training flow for sequence-space DEQNs.  The exogenous shock history is the network input, but the forward simulator still produces endogenous objects such as prices, aggregate capital, or cross-sectional distributions needed for residual evaluation.}
\label{fig:sequence_space_training}
\end{figure}
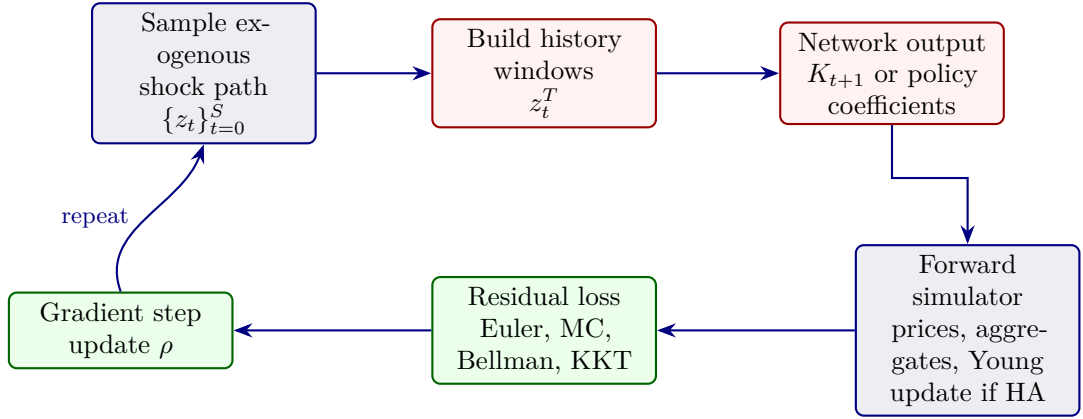

\medskip\noindent\textbf{Training logic.}
The computational pattern is also close to the rest of this chapter.  One samples an exogenous shock path, constructs overlapping history windows $z_t^T$, evaluates the network on those windows, and then uses the resulting decisions to simulate the endogenous economy forward.  In the Brock--Mirman warm-up this produces the capital sequence directly; in the Krusell--Smith tutorial it produces policy-function objects, while Young's method still propagates the cross-sectional distribution inside the simulator.  Residuals are then evaluated on the simulated path and backpropagated through the full pipeline.  Figure~\ref{fig:sequence_space_training} summarizes this workflow.

\begin{remarkbox}[Worked example: what the network input looks like]
In the companion notebook \tpath{KrusellSmith_Tutorial_CPU.ipynb}, the helper function \texttt{encode\_Z\_history} represents a discrete shock history as a one-hot block concatenated with the corresponding realized levels.  Suppose $N_Z = 2$ (so $Z_t \in \{Z_L, Z_H\}$ with $Z_L = 0.93$, $Z_H = 1.07$) and the truncated history of length $H = 3$ is $(Z_L, Z_H, Z_L)$.  \texttt{encode\_Z\_history} then returns
\[
\underbrace{\bigl[\,\underbrace{1,0}_{Z_L},\ \underbrace{0,1}_{Z_H},\ \underbrace{1,0}_{Z_L}\,\bigr]}_{\text{one-hot block, length }H \cdot N_Z}
\;\bigm\Vert\;
\underbrace{\bigl[\,0.93,\ 1.07,\ 0.93\,\bigr]}_{\text{level block, length }H},
\]
a single vector of length $H \cdot (N_Z + 1) = 9$ that is fed to the MLP.  In the Krusell--Smith tutorial, $H = 50$ and $N_Z = 2$, giving an input of length $150$.  The corresponding histogram-based input, by contrast, would have hundreds of bins from the wealth distribution alone.
\end{remarkbox}

\medskip\noindent\textbf{Why truncated histories can work.}
The Brock--Mirman warm-up makes the logic especially transparent.  With full depreciation ($\delta = 1$) and log utility, recursive substitution shows that the capital stock depends on the last $T$ shocks up to an error of order $\alpha^T \log(K_{t-T})$.  Since $\alpha < 1$ (typically $\alpha \approx 0.36$), this error vanishes exponentially: for $T = 25$, the truncation error is of order $10^{-11}$.  More generally, in ergodic economies with persistent aggregate shocks, the approximation error decays at roughly $\max\{|\varrho|, |\alpha|\}^T$.  In richer heterogeneous-agent models this is no longer an exact algebraic statement, so the history length $T$ becomes an empirical accuracy choice rather than a theorem.

\medskip\noindent\textbf{Why this is useful in heterogeneous-agent models.}
Two advantages are worth separating.  First, \emph{as a network input}, a history of $T \approx 25$ shocks can be much smaller than a histogram with hundreds of bins.  Second, exogenous shock histories are sampled from a fixed distribution.  This removes one source of instability in residual-based training: the set of network inputs is anchored by model primitives even though the endogenous simulator still evolves with the current policy network.  In the Krusell--Smith tutorial, this means that the network is conditioned on shock histories, while Young's method remains responsible for propagating the distribution used in market-clearing calculations.

\begin{keyinsightbox}[Why sequence-space training is more stable: the feedback loop]
\small
The second advantage above deserves to be unpacked, because it is empirically the single biggest source of stability gains reported in \citet{azinovicyangzemlicka2025sequencespace}.  In a state-space deep-learning HA solver the network reads the endogenous distribution $\mu_t$ as part of its input, and the training set of $\mu$'s is generated by simulating the economy under the \emph{current} policy network.  This creates a self-amplifying loop:
\[
\begin{aligned}
\rho^{(k)} \;\longrightarrow\; \pi_{\rho^{(k)}} \;\longrightarrow\; \{\mu_t\}_{\rho^{(k)}} \;\longrightarrow\;& \text{input distribution shifts}\\[-2pt]
\;\longrightarrow\; \text{out-of-distribution evaluations} \;\longrightarrow\;& \text{large residual gradient} \;\longrightarrow\; \rho^{(k+1)}\;\text{overshoots},
\end{aligned}
\]
and the next outer iteration starts from inputs the network has never seen before, often producing even larger shifts.  In sequence space the network input is the truncated shock history $z_t^T$, drawn from the \emph{exogenous} law of motion of the aggregate shock.  That distribution is fixed by model primitives and \emph{does not move} with the policy update.  The feedback loop is broken at its first link: training inputs are stationary even when the policy is far from optimal.  Empirically this often turns a calibration that fails to converge in the state-space formulation (across random seeds and learning rates) into one that converges robustly in sequence space.
\end{keyinsightbox}

\medskip\noindent\textbf{Shape-preserving operator learning.}
A second contribution of \citet{azinovicyangzemlicka2025sequencespace} is to let the network output \emph{policy-function objects} rather than a single scalar choice.  In particular, they construct architectures that guarantee monotonicity and concavity of the predicted consumption rule by representing it with an I-spline basis and non-negative coefficients.  In the Krusell--Smith tutorial, the network maps the shock history to these coefficients; the resulting policy can then be evaluated at all idiosyncratic states on the wealth grid.  This operator-learning view pairs naturally with the endogenous grid method (EGM) of \citet{carroll2006method} and avoids ad hoc penalties for monotonicity or concavity.

\medskip\noindent\textbf{Explicit I-spline MPC parameterization.}
Having seen \emph{why} a shape-preserving output head matters (above), we now write the construction down concretely; this is the most technical paragraph of the section and a reader who already accepts the monotonicity/concavity guarantees can skip to ``Fischer--Burmeister KKT loss'' below.  Let $\{k_n\}_{n=0}^N$ be a fixed log-spaced wealth grid and let $B \in \mathbb{R}^{J \times (N+1)}$ be a precomputed I-spline basis evaluated on it,
\[
B_{j,n} \;=\; I_j\!\bigl(\log(\eta + k_n)\bigr), \qquad j = 1,\ldots,J,\; n=0,\ldots,N,
\]
where $\eta>0$ is a small numerical shift (the \texttt{BASIS\_SHIFT} constant in the notebook) and each $I_j$ is an integrated B-spline that is monotonically increasing from $0$ to $1$.  For each idiosyncratic state $\varepsilon$, the network outputs two objects: a boundary marginal propensity to consume $\alpha(\varepsilon)\in(0,1)$ (sigmoid head) and non-negative weights $\widetilde w_j(\varepsilon)\ge 0$ with $\sum_j \widetilde w_j(\varepsilon) < 1$ (a ``phantom-zero'' softmax head).  The grid MPC is
\begin{equation}
\mathrm{MPC}_{\varepsilon,n} \;=\; \alpha(\varepsilon)\Bigl(1 - \sum_{j=1}^J \widetilde w_j(\varepsilon)\, B_{j,n}\Bigr),
\label{eq:ispline_mpc}
\end{equation}
which is decreasing in $n$ by construction (positive weights times an increasing basis, subtracted off a constant), bounded in $[0, \alpha(\varepsilon)] \subset [0,1]$, and continuous in the network parameters.  Consumption is then recovered on the grid by cumulation of the MPC schedule along cash-on-hand $m = w\varepsilon + Rk$,
\begin{equation}
c(\varepsilon, k_0) \;=\; \mathrm{MPC}_{\varepsilon,0}\, m(\varepsilon, k_0), \qquad
c(\varepsilon, k_n) \;=\; c(\varepsilon, k_{n-1}) + \mathrm{MPC}_{\varepsilon,n}\, R\, (k_n - k_{n-1}),
\label{eq:ispline_cumulation}
\end{equation}
and off-grid evaluation uses piecewise-linear interpolation.  Equations~\eqref{eq:ispline_mpc}--\eqref{eq:ispline_cumulation} guarantee, by construction and without any auxiliary penalty, that the consumption rule is non-negative, monotonically increasing in $k$, concave in $k$, and feasible ($c \le m$).  In code, $B$ is the matrix \texttt{ispline\_basis}, $\alpha$ and $\widetilde w$ come from the two heads of \texttt{actor\_c\_grid}, and the cumulation is the closing block of that same function.

\medskip\noindent\textbf{Fischer--Burmeister KKT loss.}
Households face a borrowing constraint $k_{t+1} \ge 0$.  The Karush--Kuhn--Tucker conditions of the household problem split into two regimes: at an interior optimum the Euler equation holds with equality, while at a binding constraint the Euler equation can be slack but next-period capital is zero.  Define the (relative) Euler residual and the (relative) savings slack (in this section $g$ is reused as the Euler residual, matching the tutorial code's variable name; it is \emph{not} the household policy function $g(k,\varepsilon,\ldots)$ of \S\ref{sec:young_method})
\[
g \;=\; \frac{c_{\text{Euler}} - c}{c}, \qquad s \;=\; \frac{k'}{c},
\]
where $c_{\text{Euler}} = (u')^{-1}\!\bigl(\beta\,\mathbb{E}_t [R'\,u'(c')]\bigr)$ is the consumption level implied by the Euler equation given the network's continuation policy.  The KKT pair is then
\[
g = 0,\ s \ge 0 \quad\text{(interior)} \qquad\text{or}\qquad g \ge 0,\ s = 0 \quad\text{(constrained)},
\]
which is a complementarity condition.  The Fischer--Burmeister envelope, in the same sign convention used in Ch.~\ref{ch:irbc} and Ch.~\ref{ch:olg},
\begin{equation}
\mathrm{FB}(g, s) \;=\; g \;+\; s \;-\; \sqrt{g^2 + s^2 + \epsilon_{\text{fb}}}
\label{eq:fb}
\end{equation}
is smooth and satisfies $\mathrm{FB}(g,s) = 0$ if and only if $\min(g, s) = 0$ with both non-negative; the small constant $\epsilon_{\text{fb}}$ (set to $10^{-12}$ in the notebooks) is a numerical stabilizer for the square root.  The upstream JAX tutorial code uses the negative-sign variant $\sqrt{g^2+s^2+\epsilon}-g-s$, which has the same zero set when squared.  The training loss is the buffer-and-grid average of $\mathrm{FB}^2$,
\[
\mathcal{L}(\rho) \;=\; \mathbb{E}_{(z^H,\mu)\sim\mathcal{B}}\biggl[\, \frac{1}{N_\varepsilon (N+1)} \sum_{\varepsilon,n} \mathrm{FB}\!\bigl(g_{\varepsilon,n}(\rho), s_{\varepsilon,n}(\rho)\bigr)^2 \,\biggr],
\]
so that one differentiable scalar simultaneously enforces the Euler equation in the interior region and the complementarity condition at the borrowing constraint, without case splits or shadow-price augmentation.  This reuses the smooth complementarity construction of \citet{fischer1992special} that is standard in nonlinear programming, applied here to a heterogeneous-agent equilibrium loss.

\medskip\noindent\textbf{Putting the pieces together: the HA training loop.}
The Krusell--Smith tutorial assembles the encoder, the I-spline policy head, Young's distribution step, and the Fischer--Burmeister loss into a single replay-buffer training loop.  Algorithm~\ref{algo:ks_seqspace} states it explicitly.

\begin{definitionbox}[Algorithm: Sequence-Space DEQN with Young's method (KS tutorial)]
\refstepcounter{algorithm}\label{algo:ks_seqspace}
\begin{algorithmic}
\small
\STATE \textbf{Input:} network $\mathcal{N}_\rho$ (I-spline MPC heads), I-spline basis $B$, transition matrices $\Pi_\varepsilon, \Pi_Z$, prices $R(K,L,Z), w(K,L,Z)$
\STATE \textbf{Hyperparameters:} history length $H{=}50$, roll-out $T{=}100$, buffer cap $C{=}128$, $N_{\text{agg}}{=}8$, FB steps per epoch $S{=}10$, mini-batch $B_{\text{fb}}{=}16$, learning rate $\alpha{=}5\!\times\!10^{-5}$, gradient clip $\|\cdot\|_2 \le 1$
\STATE \textbf{Initialize} replay buffer $\mathcal{B}$ with copies of $(\mathbf{0}_H,\, \mu_0)$, where $\mu_0$ is centered at the deterministic-RA reference $K_{ss}$
\FOR{epoch $e = 1, 2, \ldots$}
    \STATE Draw $N_{\text{agg}}$ replay states $\{(z^H_i, \mu_i)\}$ from $\mathcal{B}$
    \STATE Draw fresh shock paths $\{Z_{i,1:T}\}$ with $\Pi_Z$ from each terminal $z^H_i[-1]$
    \FOR{$i = 1, \ldots, N_{\text{agg}}$}
        \STATE \textbf{Forward roll (no grad):} for $t=1,\ldots,T$ compute $K_t,L_t$ from $\mu_{i,t-1}$, prices $R_t,w_t$, MPC and $c_t$ from $\mathcal{N}_\rho$ on the running history, advance $\mu_{i,t}$ via the Young step
        \STATE Append $(z^H_{i,T}, \mu_{i,T})$ to $\mathcal{B}$; evict oldest if $|\mathcal{B}| > C$
    \ENDFOR
    \FOR{$s = 1, \ldots, S$}
        \STATE Sample mini-batch $\mathcal{M}\subset\mathcal{B}$ of size $B_{\text{fb}}$
        \STATE Compute Fischer--Burmeister loss $\mathcal{L}(\rho)$ on $\mathcal{M}$ using \eqref{eq:fb}
        \STATE Update $\rho \leftarrow \mathrm{Adam}\bigl(\rho,\, \nabla_\rho \mathcal{L}(\rho)\bigr)$ with global-norm clipping
    \ENDFOR
\ENDFOR
\STATE \textbf{Output:} trained $\mathcal{N}_{\rho^\star}$ that maps a shock history to grid-MPC coefficients
\end{algorithmic}
\end{definitionbox}

Three implementation choices in Algorithm~\ref{algo:ks_seqspace} are worth flagging.  First, the forward roll is wrapped in \texttt{stop\_gradient} so that gradients only flow through the FB residual evaluated on the buffer, not through long simulation chains; this is what makes training tractable for $T = 100$ horizons.  Second, because Young's step gives \emph{exact} aggregate sums, the only stochasticity in the gradient comes from buffer mini-batching, which acts as standard SGD noise rather than a Monte-Carlo aggregation noise floor.  Third, the buffer simultaneously decouples training-state drawing from the current policy and lets the network see distributions $\mu$ generated by earlier policies, which improves coverage of the ergodic set during early training.

\medskip\noindent\textbf{Two training algorithms: residual minimization vs.\ time iteration.}
Algorithm~\ref{algo:ks_seqspace} is one of two families of training schemes used in the sequence-space DL literature.  In \emph{direct residual minimization} (the version above and in our notebook), the network is trained by gradient descent on the squared equilibrium residual itself.  In \emph{time iteration with EGM}, the network is trained on a sequence of supervised regression problems: at each outer iteration, one (i)~uses the current network to construct next-period policies, (ii)~backs out implied current-period policies via the endogenous-grid method of \citet{carroll2006method}, and (iii)~updates the network by minimizing the squared error against those EGM targets.  Time iteration is more involved and requires a per-batch root-finding step, but it is more flexible: it tolerates non-trivial market clearing and, crucially, handles \emph{non-convex} choices (e.g., a discrete retirement decision) and non-monotone Laffer curves where the Euler equation has multiple roots that direct residual minimization cannot disambiguate.  In practice, residual minimization is the simpler entry point on smooth, convex problems; switch to time iteration when convergence stalls, when the model contains discrete choices, or when the continuation value has convex regions that admit multiple optimal savings.

\begin{remarkbox}[Practical heuristics for sequence-space DEQNs]
\small
Four operational rules of thumb, distilled from \citet{azinovicyangzemlicka2025sequencespace}:

\begin{itemize}[itemsep=2pt]
\item \textbf{Choosing the truncation length $T$.}  Three sensible heuristics, in increasing order of conservatism: (i)~for OLG models, set $T$ to roughly two life-cycles, so that all shocks experienced by any household alive today are inside the window; (ii)~start short and iteratively increase $T$, monitoring the equilibrium residual; (iii)~``overkill'', set $T$ such that $\varrho_{\text{shock}}^T \le \text{tol}$ with $\text{tol} \in [10^{-8}, 10^{-6}]$, since long histories are cheap to feed.

\item \textbf{Innovations or realisations?}  For an AR(1) shock $z_t = \varrho z_{t-1} + \varepsilon_t$, feeding the history of \emph{realisations} $z_t^T$ to the network typically allows shorter truncation than feeding the history of \emph{innovations} $\varepsilon_t^T$, because the levels already integrate the persistence.  Feeding both is the most accurate option in practice.

\item \textbf{What to approximate.}  Smooth, bounded equilibrium objects (savings rates, MPCs) are easier to learn and yield better Euler accuracy than raw policy levels (consumption, savings).  This is the deeper reason the I-spline MPC parameterization in \eqref{eq:ispline_mpc}--\eqref{eq:ispline_cumulation} pays off so much.

\item \textbf{When the method breaks.}  The truncation argument relies on \emph{ergodicity} of the underlying dynamics, $\partial\Psi/\partial\varepsilon_{t-\tau} \to 0$ as $\tau \to \infty$.  Models with stable limit cycles or deterministic chaos violate this assumption; in those exotic settings the sequence-space approach is not guaranteed to converge to the equilibrium policy.
\end{itemize}
\end{remarkbox}

\medskip\noindent\textbf{Applications and notebooks.}
The paper applies this framework to three demanding models: (i)~an OLG economy with portfolio choice and aggregate risk, (ii)~an economy with a continuum of heterogeneous firms and households featuring idiosyncratic and aggregate shocks, and (iii)~a lifecycle model with a discrete early-retirement choice that introduces local convexities.  Mean Euler equation errors are below 0.2\% in all cases.  The two TensorFlow~2 companion notebooks are intentionally complementary: \tpath{05_SequenceSpace_BrockMirman.ipynb} is a Brock--Mirman warm-up that isolates the history-to-policy logic in the simplest possible environment, while \tpath{06_SequenceSpace_KrusellSmith.ipynb} is a compact teaching implementation that combines sequence-space inputs, I-spline policies, and Young's method in a heterogeneous-agent setting.

\medskip
A third notebook, \tpath{KrusellSmith_Tutorial_CPU.ipynb}, is a JAX/optax port of the upstream pedagogical tutorial released by the paper's authors.  It exposes the same shape-preserving I-spline MPC parameterization, the same Young step inside the simulator, and the same Fischer--Burmeister KKT loss as the TensorFlow notebook, but in the original JAX form.  It is adapted from the upstream tutorial \tpath{01_KrusellSmith_Tutorial_CPU.ipynb} in the companion code repository \citep{azinovicyangzemlicka2025sequencespacecode}, available at \url{https://github.com/azinoma/DeepLearningInTheSequenceSpace}; the local adaptation adds an explicit shape-guarantee diagnostic and additional inline commentary, leaving the algorithm unchanged.

\begin{definitionbox}[Math-to-code glossary for the KS sequence-space tutorial]
\footnotesize
\renewcommand{\arraystretch}{1.15}
\begin{tabular}{@{}L{2.6cm} L{6.2cm} L{4.4cm}@{}}
\toprule
\textbf{Symbol} & \textbf{Meaning} & \textbf{Notebook name(s)} \\
\midrule
$\rho$                                  & Policy-network parameters                                 & \texttt{psi} \\
$z_t^T$ (or $z^H$)                      & Truncated shock history of length $H$                     & \texttt{z\_history}, \texttt{z\_hist} \\
encoded $z^H$                           & One-hot $\Vert$ value vector, length $H(N_Z{+}1)$         & output of \texttt{encode\_Z\_history} \\
$\mathcal{N}_\rho$                      & MLP mapping history to MPC heads                          & \texttt{actor\_c\_grid} \\
$B_{j,n}$                               & Precomputed I-spline basis matrix                         & \texttt{ispline\_basis} \\
$\alpha(\varepsilon)$                   & Boundary MPC head (sigmoid output)                        & \texttt{alpha} \\
$\widetilde w_j(\varepsilon)$           & Non-negative I-spline weights (phantom-zero softmax)      & \texttt{w\_tilde} \\
$\mathrm{MPC}_{\varepsilon,n}$, $c$     & Decreasing MPC and cumulated grid consumption             & \texttt{mpc}, \texttt{c\_grid} \\
$\mu(\varepsilon, k)$                   & Cross-sectional distribution on the $(\varepsilon,k)$ grid & \texttt{mu} \\
$K_t, L_t$                              & Exact aggregates from $\mu$                               & \texttt{distribution\_aggregates} \\
Young step                              & Non-stochastic distribution update $\mu \mapsto \mu'$     & \texttt{distribution\_step} \\
$g$, $s$                                & Relative Euler residual and savings slack                 & \texttt{g}, \texttt{s} \\
$\mathrm{FB}(g,s)$                      & Smooth complementarity envelope                           & \texttt{fb} in \texttt{fb\_loss\_one\_state} \\
$\mathcal{B}$                           & Replay buffer of $(z^H, \mu)$ pairs                       & \texttt{buffer\_z}, \texttt{buffer\_mu} \\
\bottomrule
\end{tabular}
\end{definitionbox}

\begin{keyinsightbox}[Chapter Summary]
Continuum-agent equilibria require explicit distribution tracking; Young's (2010) histogram is a deterministic mass-redistribution scheme on a fixed grid that converges to the ergodic distribution exactly as the grid is refined.  Embedding Young's update inside a DEQN training loop yields fully differentiable heterogeneous-agent solutions, with gradients flowing through the histogram and the policy network simultaneously.  The sequence-space DEQN of \citet{azinovicyangzemlicka2025sequencespace} is the natural alternative when the entire path of aggregate variables, rather than the cross-section, is the state of interest.  Bewley--Huggett--Aiyagari--Krusell--Smith is the foundational lineage; \citet{achdou2022income} supply the continuous-time counterpart taken up in Chapter~\ref{ch:ct_theory}.
\end{keyinsightbox}

\section*{Further Reading}
\addcontentsline{toc}{section}{Further Reading}
\begin{itemize}[itemsep=2pt]
\item \citet{young2010}, the original non-stochastic histogram paper.
\item \citet{krusell1998income}, the canonical heterogeneous-agent benchmark with aggregate shocks.
\item \citet{azinovicyangzemlicka2025sequencespace}, sequence-space DEQNs, the natural extension.
\item \citet{achdou2022income}, the continuous-time treatment that Chapter~\ref{ch:ct_theory} builds on.
\item \citet{maliar2021deep, han2023deepham}, alternative deep-learning approaches to KS, contrasted in \S\ref{sec:ks_alternatives}.
\end{itemize}

\section*{Exercises}
\addcontentsline{toc}{section}{Exercises}
\noindent Worked solutions and guidance for these exercises appear in Appendix~\ref{app:solutions}.
\begin{enumerate}[itemsep=4pt, leftmargin=*]
\item\label{ex:ch6:1} \textbf{[Core] Mean-preserving lottery.}  Show that the unique two-point split that places probability $\omega$ at $k_n$ and $1-\omega$ at $k_{n+1}$ such that $\omega k_n + (1-\omega) k_{n+1} = k'$ is given by $\omega = (k_{n+1} - k')/(k_{n+1} - k_n)$.  Verify mass conservation.
\item\label{ex:ch6:2} \textbf{[Computational] Closed-form bracketing on log-spaced grids.}  Implement the $\mathcal{O}(1)$ bracketing index for a log-spaced grid $k_n = e^{x_0 + n\Delta x} - c$ in five lines of NumPy.  Verify against \texttt{numpy.searchsorted} on a random batch of queries; measure the relative speed-up.
\item\label{ex:ch6:3} \textbf{[Core] Approximate aggregation, scope.}  Construct an example economy in which the KS log-linear forecasting rule fails (e.g.\ multiple assets with switching liquidity).  Why does adding higher moments not always rescue the rule?
\item\label{ex:ch6:4} \textbf{[Computational] Sequence-space vs.\ histogram DEQN.}  Use Notebook~11 as the histogram-state baseline and compare it with the sequence-space implementation in Notebook~06 or the JAX tutorial.  In the sequence-space notebook, vary the history length $T$ (or $H$ in the code) and compare the residual after training.  Comment on which formulation generalizes better to a much longer test horizon and why.
\item\label{ex:ch6:5} \textbf{[Core] DeepSets permutation invariance.}  Consider the DeepHAM aggregator from~\eqref{eq:deepham_moments}, $\bm{m}_t = \bigl(\sum_i g_\theta^1(s_t^i), \ldots, \sum_i g_\theta^M(s_t^i)\bigr)$.  (i)~Show that for any permutation $\pi$ of the agents, $\bm{m}_t(\pi \cdot s) = \bm{m}_t(s)$, so the encoder is permutation-invariant by construction.  (ii)~Show that the policy $\pi_\rho(s_t^i; \bm{m}_t, a_t)$ is consequently equivariant in the agent index $i$: if we permute the agents, each agent's policy moves with its own index but the dependence on the population through $\bm{m}_t$ is unchanged.  (iii)~State the converse (\citet{zaheer2017deep}): any continuous permutation-invariant function on $\mathbb{R}^d$-valued sets of fixed cardinality can be written in the form $\rho(\sum_i g(s^i))$.  Use this to argue why the DeepHAM architecture can in principle replace tracking the entire histogram with a finite vector $\bm{m}_t$ of learned moments, provided the dimension $M$ is large enough.
\item\label{ex:ch6:6} \textbf{[Computational] Histogram noise vs.\ Monte Carlo sampling.}  In the Krusell--Smith tutorial notebook \tpath{KrusellSmith_Tutorial_CPU.ipynb}, fix one aggregate shock path and one initial distribution.  Run Young's histogram with $N_\mathrm{grid} = 100$ wealth grid points once, and run $R$ repeated Monte Carlo panels under the same aggregate path for $N \in \{100, 1000, 10\,000\}$ agents.  At each $t = 1, \dots, 200$, record aggregate capital $K_t^{(r)}$ for every Monte Carlo replication $r$ and the corresponding Young aggregate $K_t^Y$.  Plot the cross-replication standard deviation $\mathrm{sd}_r(K_t^{(r)})$ over time, or its time average, rather than the time-series standard deviation of $K_t$.  Verify (i)~Young's method has zero sampling variance conditional on the aggregate path, (ii)~MC sampling variance scales as $1/\sqrt{N}$, (iii)~at $N \approx N_\mathrm{grid}^2 = 10^4$, MC's stochastic noise becomes comparable to Young's discretization error on a smooth statistic like $K_t$.  Comment on why the discretization-error-vs-MC-error trade-off depends on which functional of $\mu_t$ is targeted (mean, variance, tail mass).
\item\label{ex:ch6:7} \textbf{[Computational] Two-moment forecasting rule.}  In the same notebook, replace the Krusell--Smith log-linear forecasting rule $\log K_{t+1} = a_0 + a_1 \log K_t + a_2 \mathbb{1}[z_t = z_g]$ with a two-moment extension that also conditions on cross-sectional dispersion: $\log K_{t+1} = a_0 + a_1 \log K_t + a_2 \mathbb{1}[z_t = z_g] + a_3 \log V_t$, where $V_t = \mathrm{Var}_{\mu_t}(k)$.  Re-fit the rule and the network jointly.  Report (i)~the forecasting $R^2$ for $\log K_{t+1}$ before and after, (ii)~the maximum Euler residual after training, (iii)~by how much the second moment ``buys'' you in absolute residual terms.  Connect the result to the chapter's discussion of \emph{approximate aggregation}: in the standard Krusell--Smith calibration the marginal information in the second moment is small, but it becomes material once you introduce features (e.g., binding borrowing constraints in a non-trivial fraction of the population, multiple assets) that break aggregation.
\end{enumerate}

\chapter{Physics-Informed Neural Networks}
\label{ch:pinn}

Chapters~\ref{ch:deqn}--\ref{ch:young} operated in discrete time, where the equilibrium conditions take the form of expectational (Euler) equations.  We now shift to \emph{continuous time}, where the optimality conditions are partial differential equations -- Hamilton--Jacobi--Bellman (HJB) equations for optimal control and Kolmogorov forward equations for wealth distributions.  \emph{Physics-Informed Neural Networks} (PINNs) approximate the solution of such PDEs by minimizing the PDE residual at collocation points, using automatic differentiation to compute the required derivatives \citep{sirignano2018dgm, raissi2019physics}.  The PINN loss is structurally analogous to the DEQN loss of Chapter~\ref{ch:deqn}, but instead of time-stepping a fixed-point iteration we solve a single PDE globally over the state space.  This chapter develops the PINN methodology from simple ODEs through boundary-condition strategies to the HJB equation for consumption--savings problems, with applications to macro-finance.

\section{From DEQNs to PINNs: Discrete vs.\ Continuous Time}

Many frontier models in macroeconomics and finance are written in continuous time.  Examples include Hamilton--Jacobi--Bellman (HJB) equations for optimal control, Kolmogorov forward equations for wealth distributions, and Black--Scholes PDEs for asset pricing.  The DEQN methodology from Chapters~\ref{ch:deqn}--\ref{ch:irbc} handles discrete-time equilibrium conditions.  Its continuous-time analogue relies on derivatives of the network with respect to its inputs, which automatic differentiation provides directly.

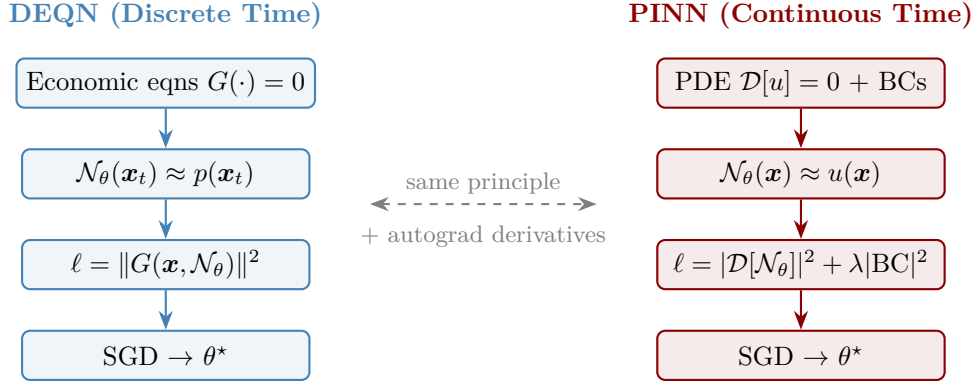
\begin{figure}[ht]
\centering
\begin{tikzpicture}[
    deqnbox/.style={rectangle, draw=softblue, thick, fill=softblue!8,
        minimum width=3.8cm, minimum height=0.65cm, font=\small, rounded corners=3pt},
    pinnbox/.style={rectangle, draw=darkred, thick, fill=darkred!8,
        minimum width=3.8cm, minimum height=0.65cm, font=\small, rounded corners=3pt},
    deqnarr/.style={-{Stealth[length=2.5mm]}, thick, softblue},
    pinnarr/.style={-{Stealth[length=2.5mm]}, thick, darkred}
]
    \node[font=\small\bfseries, softblue] at (-4.2,3.5) {DEQN (Discrete Time)};
    \node[deqnbox] (d1) at (-4.2,2.6) {Economic eqns $G(\cdot)=0$};
    \node[deqnbox] (d2) at (-4.2,1.4) {$\mathcal{N}_\theta(\x_t) \approx p(\x_t)$};
    \node[deqnbox] (d3) at (-4.2,0.2) {$\ell = \|G(\x, \mathcal{N}_\theta)\|^2$};
    \node[deqnbox] (d4) at (-4.2,-1.0) {SGD $\to$ $\theta^\star$};
    \draw[deqnarr] (d1) -- (d2);
    \draw[deqnarr] (d2) -- (d3);
    \draw[deqnarr] (d3) -- (d4);

    \node[font=\small\bfseries, darkred] at (4.2,3.5) {PINN (Continuous Time)};
    \node[pinnbox] (p1) at (4.2,2.6) {PDE $\mathcal{D}[u]=0$ + BCs};
    \node[pinnbox] (p2) at (4.2,1.4) {$\mathcal{N}_\theta(\x) \approx u(\x)$};
    \node[pinnbox] (p3) at (4.2,0.2) {$\ell = |\mathcal{D}[\mathcal{N}_\theta]|^2 + \lambda|\mathrm{BC}|^2$};
    \node[pinnbox] (p4) at (4.2,-1.0) {SGD $\to$ $\theta^\star$};
    \draw[pinnarr] (p1) -- (p2);
    \draw[pinnarr] (p2) -- (p3);
    \draw[pinnarr] (p3) -- (p4);

	    \draw[dashed, thick, gray, {Stealth[length=2.5mm]}-{Stealth[length=2.5mm]}]
	        (-1.5,1.0) -- (1.5,1.0)
	        node[midway, above, font=\footnotesize, text=gray] {same principle};
	    \node[below, font=\footnotesize, text=gray] at (0,0.85) {+ autograd derivatives};
\end{tikzpicture}
\caption{Discrete-time DEQNs and continuous-time PINNs use the same residual-minimization principle with different mathematical residuals.  DEQNs minimize algebraic equilibrium conditions such as Euler-equation errors evaluated along simulated states, while PINNs minimize PDE and boundary residuals evaluated at collocation points using derivatives of the network with respect to its inputs.}
\label{fig:deqn_pinn_comparison}
\end{figure}

The resulting framework, known as \emph{Physics-Informed Neural Networks} (PINNs), was introduced by \citet{raissi2019physics}.  In economics and finance, continuous-time models governed by HJB and related PDEs arise naturally in asset pricing under climate uncertainty \citep{barnett2020pricing}, portfolio choice \citep{duarte2024ml}, and heterogeneous-agent economies with aggregate shocks \citep{gopalakrishna2024aliens}.  \citet{duarte2024ml} apply machine-learning methods to continuous-time finance problems, while \citet{gopalakrishna2024aliens} develops the ALIENs framework for solving continuous-time economies with deep learning.  PINNs provide a natural and scalable computational framework for such problems.  Figure~\ref{fig:deqn_pinn_comparison} summarizes the discrete-time DEQN template and its continuous-time PINN analogue.

\section{The PINN Loss and Automatic Differentiation for PDEs}

\paragraph{Prerequisites in one paragraph.}  A few terms from PDE numerics recur throughout the chapter and deserve to be fixed in advance.  \emph{Collocation points} are interior evaluation points at which the PDE residual is enforced; they do not need to lie on a Cartesian grid, and in high dimensions they are typically drawn at random or from a low-discrepancy sequence.  The \emph{strong form} minimizes the residual point-wise and requires the network's activation to be at least $C^k$ if the differential operator is of order $k$.  The \emph{weak form} integrates the residual against test functions, which tolerates rougher solutions but is rarely needed for the PDEs in this chapter.  We assume throughout that each problem is \emph{well-posed} (existence, uniqueness, and continuous dependence on data); proving well-posedness for a particular HJB or KFE is part of the PDE literature, not of this course.

Given a PDE $\mathcal{D}[u](\x) = 0$ on a domain $\Omega$ with boundary conditions $\mathcal{B}[u](\x) = 0$ on $\partial\Omega$, the PINN loss is:
\begin{equation}
\boxed{
\ell_\theta^\mathrm{PINN} =
\underbrace{\frac{1}{N_r}\sum_{i=1}^{N_r} \big|\mathcal{D}[\mathcal{N}_\theta](\x_i^r)\big|^2}_{\text{PDE residual}}
+ \underbrace{\frac{\lambda}{N_b}\sum_{j=1}^{N_b} \big|\mathcal{B}[\mathcal{N}_\theta](\x_j^b)\big|^2}_{\text{boundary conditions}}.
}
\label{eq:pinn_loss}
\end{equation}
The derivatives appearing in $\mathcal{D}[\mathcal{N}_\theta]$ are computed algorithmically via automatic differentiation\footnote{Automatic differentiation is the algorithmic backbone of every deep-learning framework discussed in this script; \citet{baydin2018automatic} survey forward- and reverse-mode AD, the dual-number and tape-based implementations, and the practical trade-offs that determine why reverse mode is the standard for neural-network training (one backward pass costs roughly the same as one forward pass, irrespective of the parameter count).} (e.g., \texttt{torch.autograd.grad} in PyTorch), up to floating-point precision and without any finite-difference approximation.  This removes finite-difference truncation error in derivative terms and is a major practical advantage in high dimensions because no state-space grid is required.  Remaining error sources are function-approximation error, optimization error, and collocation sampling error.

\paragraph{Activation function requirements.}  The choice of activation function is particularly important for classical \emph{strong-form} PINNs that compute high-order derivatives via automatic differentiation.  If the PDE operator $\mathcal{D}$ involves $k$-th order derivatives, the network's activation function must be at least $C^k$ for the strong residual to be well-defined.  $\mathrm{ReLU}(z) = \max(0,z)$ is only $C^0$ and its second derivative is zero almost everywhere (and undefined at the kink), so a ReLU network is not suitable for the strong form of a second-order PDE.  For second-order PDEs (HJB, Black--Scholes, Poisson), one should use $C^\infty$ activations such as $\tanh$, Swish, or softplus.  This requirement stands in contrast to the supervised setting, where ReLU is the default.  The trade-off is real: smooth saturating activations propagate gradients more slowly than ReLU on the same architecture (the upper plateau of $\tanh$ has near-zero slope), so PINNs typically need wider networks or more iterations to reach the same training loss as a comparable supervised ReLU network.  In practice this is the price of having well-defined second derivatives.  Note that the limitation is specific to strong-form auto-differentiation PINNs: weak/variational formulations and methods that handle nonsmooth solutions explicitly can still use ReLU activations, since high-order derivatives never need to be differentiated point-wise.  More broadly, PINN optimization can suffer from gradient pathologies across loss terms, which is why adaptive loss balancing is often beneficial \citep{wang2021understanding, bischof2025relobralo}.

\begin{remarkbox}[Framework choice: PyTorch vs.\ TensorFlow]
The PINN notebooks use PyTorch, whereas the DEQN code (Chapters~\ref{ch:deqn}--\ref{ch:irbc}) uses TensorFlow.  This is a practical implementation choice rather than a mathematical one: PINNs repeatedly differentiate the network with respect to its \emph{inputs} (e.g., $V_{SS}$ in the Black--Scholes PDE), and PyTorch's eager-mode \texttt{autograd} makes those higher-order derivatives easy to compose.  TensorFlow remains a natural fit for the DEQN training loop already used earlier in the course.  The underlying mathematics is identical in both frameworks: a PINN written here in PyTorch can be ported to TensorFlow line-for-line by replacing \texttt{torch.autograd.grad} with \texttt{tf.GradientTape} (one tape per derivative order, since TF tapes are by default not persistent), and \texttt{torch.compile} with \texttt{tf.function}.
\end{remarkbox}

\subsection{A Concrete Example: Solving a 1D ODE with a PINN}

To build intuition, consider the boundary value problem
\begin{equation}
y''(x) = -1, \qquad x \in (0,1), \qquad y(0) = y(1) = 0,
\label{eq:1d_ode_bvp}
\end{equation}
whose analytical solution is $y^\star(x) = \tfrac{1}{2}\,x(1-x)$.  Approximate the unknown function by a raw neural network $\mathcal{N}_\theta(x)$, with no boundary information built in.  The strong-form residual is
\begin{equation}
R(x;\theta) = \partial_{xx}\mathcal{N}_\theta(x) + 1,
\end{equation}
where $\partial_{xx}\mathcal{N}_\theta$ is obtained by two applications of \texttt{torch.autograd.grad}.  The boundary conditions are then added to the loss as a penalty term -- the \emph{soft} enforcement strategy:
\begin{equation}
\ell_\theta = \frac{1}{N_r}\sum_{i=1}^{N_r} R(x_i;\theta)^2
            \;+\; \lambda\bigl(\mathcal{N}_\theta(0)^2 + \mathcal{N}_\theta(1)^2\bigr),
\label{eq:1d_ode_soft_loss}
\end{equation}
with collocation points $x_i$ drawn uniformly from $(0,1)$ and a weight $\lambda>0$ on the boundary term.  Building the boundary conditions directly into the network output instead -- the \emph{hard} enforcement alternative, which removes the $\lambda$ term entirely -- is taken up in \S\ref{sec:bc_soft_hard}.

\begin{remarkbox}[Collocation strategies]
Uniform random sampling is the simplest collocation strategy but not necessarily the most efficient.  Alternatives include low-discrepancy Sobol sequences \citep{sobol1967distribution}; Latin Hypercube Sampling \citep[LHS;][]{mckay1979comparison}, which provides better space coverage; other quasi-Monte Carlo points \citep{niederreiter1992random, owen1995randomly}, which fill the domain more evenly; and residual-based adaptive refinement \citep[RAR;][]{lu2021deepxde}, which concentrates points in regions where the current PDE residual is largest.  For most applications in this course, uniform or LHS sampling is sufficient, but adaptive refinement can be beneficial for PDEs with sharp gradients, boundary layers, and near payoff discontinuities (e.g., the kink at $S = K$ in Black--Scholes).  \emph{Mini-batch sizes.}  Per-step collocation batches are typically $32$--$256$ points in low-dimensional problems and $512$--$4096$ when the input dimension or interaction-order grows; the right number is the largest that fits comfortably in GPU memory together with the AD graph for the highest-order derivative needed.  All companion notebooks make this an explicit hyperparameter near the top of the training cell.
\end{remarkbox}

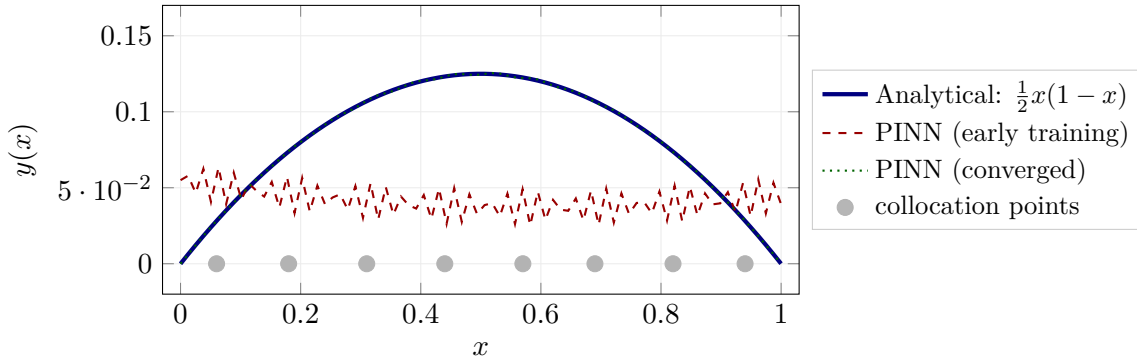
\begin{figure}[ht]
\centering
\begin{tikzpicture}
\begin{axis}[
    width=10cm, height=5.4cm,
    xlabel={$x$}, ylabel={$y(x)$},
    xmin=0, xmax=1, ymin=-0.02, ymax=0.17,
    grid=major, grid style={gray!15},
    legend style={
        at={(1.02,0.5)}, anchor=west, font=\small,
        draw=gray!40, fill=white, fill opacity=0.95, text opacity=1,
        cells={anchor=west}, row sep=1pt,
    },
    every axis plot/.append style={no markers},
    enlarge x limits=0.03,
]
\addplot[uzhblue, ultra thick, domain=0:1, samples=80] {0.5*x*(1-x)};
\addlegendentry{Analytical: $\tfrac12 x(1-x)$}
\addplot[harvardcrimson, thick, dashed, domain=0:1, samples=80]
    {0.055 - 0.06*x + 0.05*x*x + 0.012*sin(deg(720*x))};
\addlegendentry{PINN (early training)}
\addplot[darkgreen, thick, dotted, domain=0:1, samples=80]
    {0.5*x*(1-x) + 0.003*x*(1-x)*sin(deg(900*x))};
\addlegendentry{PINN (converged)}
\addplot[only marks, mark=|, mark size=3pt, gray!60] coordinates
    {(0.06,0) (0.18,0) (0.31,0) (0.44,0) (0.57,0) (0.69,0) (0.82,0) (0.94,0)};
\addlegendentry{collocation points}
\end{axis}
\end{tikzpicture}
\caption{PINN solution of the 1D ODE $y''=-1$ on $(0,1)$ with $y(0)=y(1)=0$ and the soft-penalty loss~\eqref{eq:1d_ode_soft_loss}. The analytical solution $\tfrac12 x(1-x)$ (solid blue) is recovered to plotting accuracy by the converged network (dotted green); the dashed red curve illustrates a typical early-training iterate, which still misses the endpoints because the boundary penalty is enforced only approximately. Tick marks on the $x$-axis are the uniformly drawn collocation points.  The curves above are TikZ illustrations rather than direct exports.  Notebook \texttt{lecture\_11\_01\_ODE\_PINN\_ZeroBCs} runs exactly this experiment; notebook \texttt{lecture\_11\_02\_ODE\_PINN\_SoftVsHardBCs} then contrasts the soft penalty against the hard trial-solution construction of \S\ref{sec:bc_soft_hard} on a non-zero-BC variant.}
\label{fig:pinn_1d_ode_soft}
\end{figure}

Figure~\ref{fig:pinn_1d_ode_soft} shows this calculation graphically.  This simple example illustrates the key ingredients of every PINN: (i)~a neural network that approximates the unknown function, (ii)~automatic differentiation to compute derivatives, (iii)~a loss function built from the PDE residual (plus, here, a boundary penalty), and (iv)~collocation points sampled from the domain interior.  The same machinery extends directly to PDEs in two or more dimensions, as we demonstrate in the following applications.

\section{Boundary Conditions: Soft vs.\ Hard Enforcement}
\label{sec:bc_soft_hard}

The treatment of boundary conditions is a critical design choice in PINNs.

\begin{figure}[ht]
\centering
\begin{tikzpicture}
    \begin{scope}[shift={(-5,0)}]
        \node[font=\small\bfseries, softorange] at (1.8,3.1) {$\lambda$ too small};
        \draw[thick, uzhblue] (0,0) -- (3.6,0);
        \fill[darkred] (0,0) circle (4pt);
        \fill[darkred] (3.6,0) circle (4pt);
        \node[above right=-1pt and 2pt, font=\tiny, darkred] at (0,0.05) {$y=1$};
        \node[above, font=\tiny, darkred] at (3.6,0.1) {$y=2$};
        \draw[thick, softorange, domain=0:3.6, samples=60]
            plot (\x, {-0.14*\x*\x + 0.62*\x + 0.7});
        \node[font=\tiny, softorange] at (1.8,-0.5) {PDE OK, BCs violated};
        \draw[thick, darkred, <->] (0,0) -- (0,0.7) node[pos=0.65, left=1pt, font=\tiny] {gap};
    \end{scope}

    \begin{scope}[shift={(-0.5,0)}]
        \node[font=\small\bfseries, softgreen] at (1.8,3.1) {$\lambda$ balanced};
        \draw[thick, uzhblue] (0,0) -- (3.6,0);
        \fill[darkred] (0,0) circle (4pt);
        \fill[darkred] (3.6,0) circle (4pt);
        \node[above, font=\tiny, darkred] at (0,0.1) {$y=1$};
        \node[above, font=\tiny, darkred] at (3.6,0.1) {$y=2$};
        \draw[thick, softgreen, domain=0:3.6, samples=60]
            plot (\x, {-0.04*\x*\x + 0.42*\x + 1.0});
        \draw[thick, dashed, gray, domain=0:3.6, samples=60]
            plot (\x, {-0.04*\x*\x + 0.42*\x + 1.0 + 0.03*sin(deg(3.14*\x/3.6))});
        \node[font=\tiny, softgreen] at (1.8,-0.5) {Compromise (hard to find)};
    \end{scope}

    \begin{scope}[shift={(4,0)}]
        \node[font=\small\bfseries, darkred] at (1.8,3.1) {$\lambda$ too large};
        \draw[thick, uzhblue] (0,0) -- (3.6,0);
        \fill[darkred] (0,0) circle (4pt);
        \fill[darkred] (3.6,0) circle (4pt);
        \node[above, font=\tiny, darkred] at (0,0.1) {$y=1$};
        \node[above, font=\tiny, darkred] at (3.6,0.1) {$y=2$};
        \draw[thick, dashed, gray!60] (0,1.0) -- (3.6,2.0);
        \draw[thick, darkred, domain=0:3.6, samples=60]
            plot (\x, {1.0 + \x/3.6 + 0.15*sin(deg(2*3.14*\x/3.6))});
        \node[font=\tiny, darkred] at (1.8,-0.5) {BCs OK, PDE violated};
        \draw[thick, gray, <->] (2.7,1.6) -- (2.7,1.75);
        \node[font=\tiny, gray, anchor=south west] at (2.78,1.95) {PDE err};
    \end{scope}
\end{tikzpicture}
\caption{Failure modes of soft boundary-condition enforcement on a Dirichlet problem with $y(0)=1$, $y(1)=2$.  \emph{Left:} the BC penalty weight $\lambda$ is too small, so the optimizer minimizes the interior PDE residual but lets the candidate solution miss both endpoints (visible as the ``gap'' at $x=0$).  \emph{Right:} $\lambda$ is too large, so the network nails the boundary values but distorts the interior, producing a wiggly profile with PDE residual error against the affine reference (dashed grey).  \emph{Centre:} a balanced $\lambda$ approximately satisfies both objectives, but the right-shaped value depends on the network, the PDE, and the geometry, and is not known a priori.  This trade-off motivates the hard-enforcement construction below, which removes the boundary loss term entirely.}
\label{fig:soft_bc_failure_modes}
\end{figure}
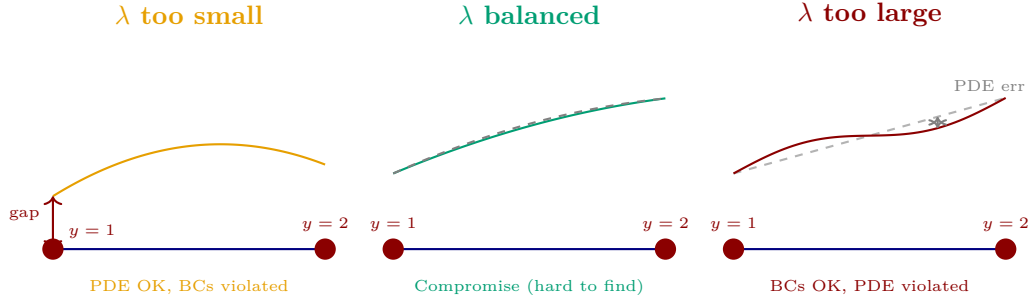

\paragraph{Soft enforcement.}  Boundary conditions are penalized as an additional loss term, weighted by $\lambda$.  The difficulty is that $\lambda$ must be tuned: too small and the BCs are violated; too large and the optimizer ignores the PDE interior.  Figure~\ref{fig:soft_bc_failure_modes} sketches the two failure modes and the compromise regime in between.

\begin{figure}[ht]
\centering
\begin{tikzpicture}
    \begin{axis}[
        name=plotA,
        width=4.0cm, height=4.0cm,
        title={\small $A(x)$},
        title style={font=\small\bfseries, text=darkred, yshift=-2pt},
        xlabel={$x$}, xlabel style={font=\small},
        tick label style={font=\footnotesize},
        xmin=0, xmax=1, ymin=0.6, ymax=2.6,
        grid=major, grid style={gray!15}, axis lines=left,
    ]
        \addplot[ultra thick, darkred, domain=0:1, samples=2]{1+x};
        \addplot[only marks, mark=*, mark size=2.5pt, darkred] coordinates {(0,1) (1,2)};
    \end{axis}
    \node[font=\Large\bfseries] at ($(plotA.east)+(0.9,0)$) {$+$};
    \begin{axis}[
        at={($(plotA.east)+(1.8cm,0)$)}, anchor=west,
        name=plotBN,
        width=4.0cm, height=4.0cm,
        title={\small $B(x)\cdot\mathcal{N}_\theta(x)$},
        title style={font=\small\bfseries, text=softblue, yshift=-2pt},
        xlabel={$x$}, xlabel style={font=\small},
        tick label style={font=\footnotesize},
        xmin=0, xmax=1, ymin=-0.45, ymax=0.45,
        grid=major, grid style={gray!15}, axis lines=left,
    ]
        \addplot[ultra thick, softblue, domain=0:1, samples=80]
            {x*(1-x)*(0.8*sin(deg(3.14*x)) - 0.3*cos(deg(2*3.14*x)) + 0.5)};
        \addplot[only marks, mark=*, mark size=2.5pt, softblue] coordinates {(0,0) (1,0)};
    \end{axis}
    \node[font=\Large\bfseries] at ($(plotBN.east)+(0.9,0)$) {$=$};
    \begin{axis}[
        at={($(plotBN.east)+(1.8cm,0)$)}, anchor=west,
        name=plotY,
        width=4.0cm, height=4.0cm,
        title={\small $\hat{y}(x)$},
        title style={font=\small\bfseries, text=softgreen, yshift=-2pt},
        xlabel={$x$}, xlabel style={font=\small},
        tick label style={font=\footnotesize},
        xmin=0, xmax=1, ymin=0.6, ymax=2.6,
        grid=major, grid style={gray!15}, axis lines=left,
    ]
        \addplot[thick, dashed, darkred, domain=0:1, samples=2]{1+x};
        \addplot[ultra thick, softgreen, domain=0:1, samples=80]
            {1 + x + x*(1-x)*(0.8*sin(deg(3.14*x)) - 0.3*cos(deg(2*3.14*x)) + 0.5)};
        \addplot[only marks, mark=*, mark size=2.5pt, softgreen] coordinates {(0,1) (1,2)};
    \end{axis}
\end{tikzpicture}
\caption{Hard boundary-condition decomposition for the trial solution $\hat y(x) = A(x) + B(x)\cdot\mathcal{N}_\theta(x)$ with Dirichlet data $\hat y(0)=1$, $\hat y(1)=2$. \emph{Left:} anchor $A(x)=1+x$ matches the boundary values exactly. \emph{Centre:} mask times network, $B(x)\cdot\mathcal{N}_\theta(x)$ with mask $B(x)=x(1-x)$, which vanishes at $x\in\{0,1\}$ regardless of $\mathcal{N}_\theta$. \emph{Right:} their sum $\hat y(x) = A(x) + B(x)\cdot\mathcal{N}_\theta(x)$ (solid green) is a candidate solution that satisfies both BCs by construction; the dashed red line is the affine anchor $A(x)=1+x$ for visual reference. Training loss reduces to the interior PDE residual alone.}
\label{fig:trial_solution_decomposition}
\end{figure}
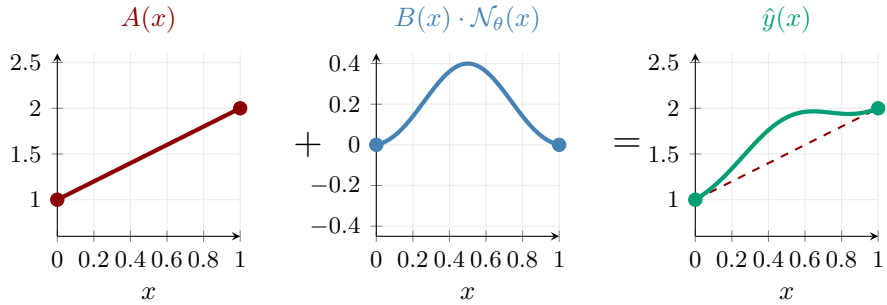

\paragraph{Hard enforcement.}  A \emph{trial solution} is constructed that satisfies the boundary conditions \emph{by construction}:
\begin{equation}
\hat{y}(x) = A(x) + B(x) \cdot \mathcal{N}_\theta(x),
\label{eq:trial}
\end{equation}
where $A(x)$ is an anchor function satisfying the BCs exactly, and $B(x)$ is a mask function that vanishes at the boundary.  For Dirichlet BCs $\hat{y}(0) = a$, $\hat{y}(1) = b$, one may choose $A(x) = a + (b-a)x$ and $B(x) = x(1-x)$.  Figure~\ref{fig:trial_solution_decomposition} visualizes this anchor-plus-mask decomposition for non-zero endpoint data.

Hard enforcement eliminates the boundary loss term entirely, reducing the number of hyperparameters and improving accuracy near the boundaries.  The trade-off is real, however: because $\hat y = A + B\cdot\mathcal{N}_\theta$ multiplies the network output by a mask $B$ that vanishes on $\partial\Omega$, the input-gradient $\partial\hat y/\partial x$ inherits a factor of $B$ near the boundary and is thus damped by construction.  When the true solution exhibits a steep boundary layer (e.g., a Hamilton--Jacobi--Bellman equation at the borrowing constraint, see \S\ref{sec:hjb_theory}), the network must compensate by making $\mathcal{N}_\theta$ itself locally large, which can slow convergence.  In short: hard BCs trade boundary-loss tunability for vanishing input gradients at $\partial\Omega$, and the trade is favorable for smooth Dirichlet problems but less obviously so for problems with sharp boundary features.

\paragraph{A worked 1D instance.}
As a concrete instantiation, return to the simple ODE $y''(x)+y(x)=0$ on $[0,\pi/2]$ with $y(0)=0$, $y(\pi/2)=1$, whose analytical solution is $\sin x$.  A trial solution that builds in both endpoints is
\begin{equation}
\hat{y}(x) = \underbrace{\frac{2x}{\pi}}_{A(x)} + \underbrace{x\!\left(\frac{\pi}{2} - x\right)}_{B(x)} \cdot \mathcal{N}_\theta(x),
\label{eq:1d_ode_trial}
\end{equation}
which has the anchor-plus-mask form~\eqref{eq:trial}: the anchor $A$ matches the boundary data ($A(0)=0$, $A(\pi/2)=1$) and the mask $B$ vanishes at both endpoints, so $\hat{y}(0)=0$ and $\hat{y}(\pi/2)=1$ hold exactly for any network output, and the loss reduces to the interior PDE residual alone,
\[
\ell_\theta = \frac{1}{N_r}\sum_{i=1}^{N_r}\big(\hat{y}''(x_i) + \hat{y}(x_i)\big)^2,
\]
with $\hat{y}''$ obtained by two applications of \texttt{torch.autograd.grad} and collocation points $x_i$ drawn uniformly from $[0,\pi/2]$.

\begin{figure}[ht]
\centering
\begin{tikzpicture}
\begin{axis}[
    width=10cm, height=5.4cm,
    xlabel={$x$}, ylabel={$y(x)$},
    xmin=0, xmax=1.75, ymin=0, ymax=1.15,
    grid=major, grid style={gray!15},
    legend style={
        at={(1.02,0.5)}, anchor=west, font=\small,
        draw=gray!40, fill=white, fill opacity=0.95, text opacity=1,
        cells={anchor=west}, row sep=1pt,
    },
    every axis plot/.append style={no markers},
    enlarge x limits=0.05,
]
\addplot[uzhblue, ultra thick, domain=0:1.5708, samples=80] {sin(deg(x))};
\addlegendentry{Analytical: $\sin(x)$}
\addplot[harvardcrimson, thick, dashed, domain=0:1.5708, samples=80]
    {2*x/3.1416 + x*(1.5708-x)*0.18*sin(deg(2*x))};
\addlegendentry{PINN (early training)}
\addplot[darkgreen, thick, dotted, domain=0:1.5708, samples=80]
    {sin(deg(x)) + 0.005*sin(deg(5*x))*x*(1.5708-x)};
\addlegendentry{PINN (converged)}
\addplot[only marks, mark=|, mark size=3pt, gray!60] coordinates
    {(0.1,0) (0.3,0) (0.5,0) (0.7,0) (0.9,0) (1.1,0) (1.3,0) (1.5,0)};
\addlegendentry{collocation points}
\end{axis}
\end{tikzpicture}
\caption{PINN solution of the 1D ODE $y''+y=0$ on $[0,\pi/2]$ with the hard-BC trial solution~\eqref{eq:1d_ode_trial}. The analytical solution $\sin(x)$ (solid blue) is recovered to plotting accuracy by the converged network (dotted green); the dashed red curve illustrates a typical early-training iterate. Tick marks on the $x$-axis are the uniformly drawn collocation points.  The curves above are TikZ illustrations rather than direct exports.  Notebook \texttt{lecture\_11\_02\_ODE\_PINN\_SoftVsHardBCs} contrasts this hard-trial-solution construction with the soft-penalty alternative on a non-zero-BC variant.}
\label{fig:pinn_1d_ode}
\end{figure}
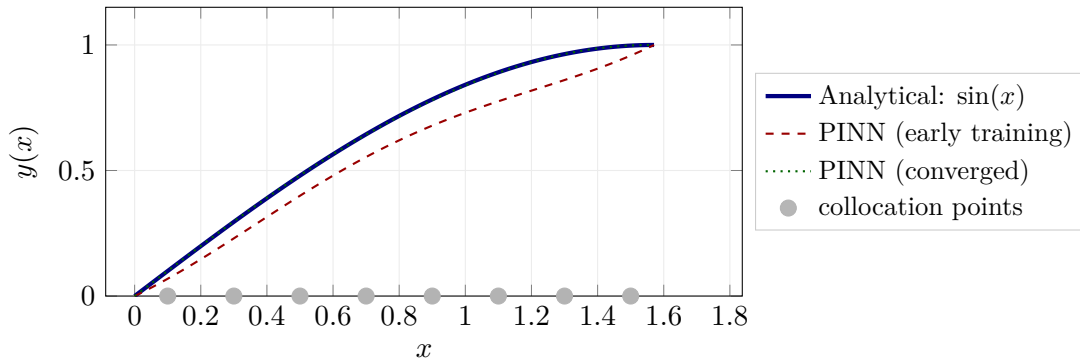

Figure~\ref{fig:pinn_1d_ode} confirms that the converged trial solution recovers $\sin x$ to plotting accuracy, with the boundary values exact by construction; contrast the early-training iterate of Figure~\ref{fig:pinn_1d_ode_soft}, which still misses the endpoints under the soft penalty.

\paragraph{Transfinite interpolation for 2D domains.}
The 1D hard-enforcement idea ($\hat{y} = A + B \cdot \mathcal{N}_\theta$) extends naturally to rectangular 2D domains, but the anchor function $A$ must now interpolate prescribed boundary data along entire \emph{edges}, not just at two endpoints.  The classical technique for this is transfinite interpolation (also known as Gordon--Coons blending): it constructs a function over the rectangle that exactly matches given boundary curves, in much the same way that bilinear interpolation matches four corner values, except that here we match four continuous edge functions rather than four discrete values.

Consider a rectangular domain $[a,b] \times [c,d]$ with Dirichlet boundary data $f_L(y)$ (left edge, $x=a$), $f_R(y)$ (right edge, $x=b$), $f_B(x)$ (bottom edge, $y=c$), and $f_T(x)$ (top edge, $y=d$).  Introduce the normalized coordinates $\xi = (x-a)/(b-a) \in [0,1]$ and $\eta = (y-c)/(d-c) \in [0,1]$.  The idea is to blend the four edge functions using linear weights, then correct for the corners that get counted twice.

\smallskip\noindent\textbf{Step 1, Interpolate in $x$:}  The function $(1-\xi)\,f_L(y) + \xi\,f_R(y)$ matches the left and right edges exactly for every $y$, but says nothing about the top and bottom edges.

\smallskip\noindent\textbf{Step 2, Interpolate in $y$:}  Similarly, $(1-\eta)\,f_B(x) + \eta\,f_T(x)$ matches the bottom and top edges exactly for every $x$.

\smallskip\noindent\textbf{Step 3, Add and correct.}  Summing these two would double-count the corner values (e.g., $f_L(c)$ and $f_B(a)$ both contribute at $(a,c)$).  We subtract a bilinear interpolant through the four corner values to compensate.  The result is the blending (anchor) function:
\begin{equation}
\begin{split}
A(x,y) &= \underbrace{(1-\xi)\,f_L(y) + \xi\,f_R(y)}_{\text{left/right interpolation}}
       \;+\; \underbrace{(1-\eta)\,f_B(x) + \eta\,f_T(x)}_{\text{bottom/top interpolation}} \\[4pt]
       &\quad -\; \underbrace{\bigl[(1-\xi)(1-\eta)\,f_{BL} + \xi(1-\eta)\,f_{BR} + (1-\xi)\eta\,f_{TL} + \xi\,\eta\,f_{TR}\bigr]}_{\text{bilinear corner interpolant}},
\end{split}
\end{equation}
where $f_{BL} = f_B(a) = f_L(c)$, etc., are the four corner values.  These values must be \emph{consistent} across the two edges that meet at each corner (e.g., the value of $f_L$ at $y=c$ must equal the value of $f_B$ at $x=a$); otherwise the corner-correction bilinear cancels imperfectly and $A$ does not match any of the four edges exactly at the offending corner.  Under consistent corner data, $A$ matches all four edge functions exactly by construction.

The mask function must vanish on all four edges so that the network can modify the interior without disturbing the boundaries:
\begin{equation}
B(x,y) = \xi\,(1-\xi)\,\eta\,(1-\eta).
\end{equation}
Note that $B = 0$ whenever $\xi \in \{0,1\}$ or $\eta \in \{0,1\}$, i.e., on every edge of the rectangle.  The hard-enforced trial solution is then:
\begin{equation}
\hat{u}(x,y) = A(x,y) + B(x,y) \cdot \mathcal{N}_\theta(x,y).
\end{equation}
On any boundary edge, $B=0$ and $\hat{u}$ reduces to $A$, which matches the prescribed data.  In the interior, $B > 0$ and the network $\mathcal{N}_\theta$ is free to learn whatever shape is needed to satisfy the PDE.  This construction satisfies all four Dirichlet conditions exactly for \emph{any} network output $\mathcal{N}_\theta$.

\begin{keyinsightbox}[When to use hard vs.\ soft BCs]
\textbf{Hard enforcement} is preferred whenever a valid trial solution can be constructed analytically, which is the case for most standard boundary value problems in economics (e.g., Dirichlet conditions on finite domains).  \textbf{Soft enforcement} is necessary when the boundary conditions are complicated (e.g., free-boundary problems, state-dependent constraints) or when the domain geometry makes it difficult to construct the mask function $B(x)$.  In practice, hard enforcement typically improves accuracy by 1--2 orders of magnitude near the boundaries.
\end{keyinsightbox}

\subsection{2D Poisson Benchmark}

As a pedagogical 2D benchmark, the accompanying notebook \tpath{lecture_11_03_PDE_PINN_Poisson2D} solves a Poisson equation on the unit square,
\begin{equation}
\nabla^2 u(x,y) = f(x,y), \qquad (x,y) \in [0,1]^2,
\end{equation}
with manufactured solution $u^\star(x,y) = x^2 + y + \sin(\pi x)\sin(\pi y)$, yielding $f(x,y) = 2 - 2\pi^2\sin(\pi x)\sin(\pi y)$ and non-homogeneous Dirichlet data $u\big|_{y=0} = x^2$, $u\big|_{y=1} = x^2+1$, $u\big|_{x=0} = y$, $u\big|_{x=1} = 1+y$.  (Some PDE references prefer the standard-elliptic convention $-\nabla^2 u = f$; switching to that form flips the sign of $f$ but leaves the boundary data and the network architecture unchanged.)  Because $\sin(\pi x)\sin(\pi y)$ vanishes on $\partial\Omega$, the polynomial part $A(x,y) = x^2 + y$ already matches all four edges (and their corner values) exactly, so it serves as the transfinite-interpolation anchor (the corner-consistency requirement of \S\ref{sec:bc_soft_hard} is automatic here).  The mask $B(x,y) = x(1-x)y(1-y)$ vanishes on every edge, and the hard-enforced trial solution $\hat u(x,y) = A(x,y) + B(x,y)\,\mathcal{N}_\theta(x,y)$ satisfies all Dirichlet conditions by construction; the network is trained on the interior PDE residual alone.  In the reference notebook configuration, the verification gate is a mode-dependent relative-$L_2$ error against the manufactured solution on a $100 \times 100$ test grid (disabled in the \texttt{smoke} CI run, with progressively tighter bounds in \texttt{teaching} and \texttt{production}), computed for a modest network and a few thousand collocation points; the exact numbers depend on seed, hardware, optimizer settings, and the collocation sample.  This benchmark is the clean 2D extension of the 1D ODE example and a useful bridge before turning to economic applications, and it exercises the full hard-BC transfinite-interpolation machinery on non-zero edge data rather than collapsing to the trivial $u\big|_{\partial\Omega}=0$ case.

\section{Common PINN Failure Modes and Remedies}

In practice, PINN training often fails for optimization reasons rather than approximation capacity.  Table~\ref{tab:pinn_failure_modes} lists the most common pathologies and practical remedies.
\begin{table}[ht]
\centering
\small
\begin{tabular}{@{} L{3.0cm} L{4.0cm} L{5.5cm} @{}}
\toprule
\textbf{Symptom} & \textbf{Typical cause} & \textbf{Practical remedy} \\
\midrule
Large boundary violations & Soft BC penalties underweighted or unstable & Prefer hard BC trial solutions when possible; otherwise use adaptive loss balancing (e.g., ReLoBRaLo). \\
Good BC fit, poor interior PDE fit & Boundary penalties overweighted & Decrease BC weights or rebalance losses dynamically. \\
Residual spikes in narrow regions & Collocation undercoverage (boundary layers / kinks) & Use residual-based adaptive resampling, Sobol/LHS points, and local point refinement. \\
Slow or stalled optimization & Gradient imbalance across loss terms; or initialization in a basin disconnected from the true solution & Normalize loss components and monitor per-term gradients; use adaptive balancing schemes \citep{wang2021understanding, bischof2025relobralo}; restart with different seeds or warm-start from a coarser solver if the loss plateaus far from any plausible solution. \\
Optimizer stuck on a symmetric / structureless ansatz & Initialization too symmetric to represent an asymmetric solution; gradient updates preserve the symmetry & Break the symmetry explicitly: asymmetric weight initialization, an auxiliary symmetry-breaking input feature, or a short pre-training pass that nudges the ansatz toward the correct shape. \\
Oscillatory solution near payoff kink & Spectral bias and non-smooth target geometry & Increase sampling density near kink and split domain if needed; use problem-specific transforms. \\
\bottomrule
\end{tabular}
\caption{Common PINN failure modes and practical remedies.  The main implementation lesson is to monitor loss components separately: a small total loss can hide boundary violations, interior residual spikes, or gradient imbalance across terms.}
\label{tab:pinn_failure_modes}
\end{table}

The penultimate row above is folklore in computational quantum chemistry, where solvers routinely ship dedicated symmetry-breaking modules because a symmetric initial guess is preserved exactly under gradient training; the analogous trap shows up in PINNs whenever the true solution lacks a symmetry that the architecture happens to enforce.

For the course applications, a robust default stack is: smooth activation ($\tanh$ or Swish), hard BCs when analytically available, adaptive loss balancing, and adaptive collocation near high-residual regions.

\section{The Deep Galerkin Method (DGM) Architecture}

The plain MLP used in the examples so far -- and, below, for the cake-eating HJB -- is sufficient for low-dimensional PDEs with smooth solutions, but its expressive bottleneck shows up in two regimes: (i)~genuinely high-dimensional state spaces (where curse-of-dimensionality effects compound across layers), and (ii)~PDEs with sharp internal features such as boundary layers, wave fronts, or kinks at policy switches.  The Deep Galerkin Method addresses both by adding LSTM-style gates and skip connections from the input to every layer, so that the network can carry input information forward unchanged through depth and route it past intermediate transformations.  In the 1D problems studied so far, an MLP and a DGM block train to comparable accuracy and the MLP is preferred for transparency; the architectural complexity pays off as dimension grows or sharp features appear.  Table~\ref{tab:mlp_vs_dgm} gives a qualitative comparison across the problem classes treated in this chapter; the chapter's exercises invite a concrete benchmark of the two architectures on the same PDE.

\begin{table}[ht]
\centering
\footnotesize
\setlength{\tabcolsep}{5pt}
\begin{tabular}{@{} >{\bfseries}p{3.0cm} c c p{4.5cm} @{}}
\toprule
\textbf{Problem class}                          & \textbf{MLP}            & \textbf{DGM}              & \textbf{Where DGM helps} \\
\midrule
1D ODE (\S\ref{sec:bc_soft_hard})                & comparable              & comparable                & not in 1D smooth problems; MLP preferred for transparency \\
2D Poisson (\S\ref{sec:bc_soft_hard})            & comparable              & comparable                & marginally, on stiff sources \\
Cake-eating HJB (\S\ref{sec:cake_eating_hjb})    & preferred               & comparable                & MLP wins when hard BCs already kill the boundary loss; DGM wins if kinks at policy-switching points dominate \\
Black--Scholes (\S\ref{sec:bs_pinn})             & adequate                & cleaner near kink         & sharp payoff kink at $S=K$; DGM gates absorb local geometry better \\
High-dim HJB ($d\gtrsim 4$)                       & curse of dim. visible   & preferred                 & input-skip connections retain raw coordinates at every depth, mitigating expressivity loss across layers \\
\bottomrule
\end{tabular}
\caption{Qualitative MLP vs.\ DGM comparison across the problem classes of this chapter.  ``Comparable'' means the two architectures train to similar final residual at similar wall time; ``preferred'' indicates which one a practitioner should reach for first.  A quantitative benchmark on a fixed pair (residual, wall time, parameters) is the natural project extension and is not run here.}
\label{tab:mlp_vs_dgm}
\end{table}

For high-dimensional PDEs, \citet{sirignano2018dgm} introduced the Deep Galerkin Method (DGM), an architecture with LSTM-style gating \citep{hochreiter1997long} and skip connections from the input layer to every hidden layer reminiscent of Highway Networks \citep{srivastava2015highway} (the immediate precursor of ResNets, in which a learned gate controls how much of the previous representation is carried forward unchanged versus transformed).\footnote{Notation reuse: the gate $G^{(l)}$ in the DGM block below is unrelated to the cross-sectional density $g$ that appears in heterogeneous-agent models (Chapters~\ref{ch:young}, \ref{ch:ct_theory}); the symbols share a letter only.}  A related deep BSDE-based formulation was introduced by \citet{e2017deep} (E--Han--Jentzen) and developed in the companion paper of \citet{han2018solving} (Han--Jentzen--E).  An accessible exposition of DGM together with several PDE applications is given by \citet{al2018solving}, which also introduces the gate naming convention adopted below.  The original DGM architecture of \citet{sirignano2018dgm} uses four gates at each layer $l$:
\begin{align}
Z^{(l)} &= \sigma\!\big(\W_z^{(l)} \a^{(l-1)} + \bm{U}_z^{(l)} \x + \bb_z^{(l)}\big), && \text{(update gate)} \\
G^{(l)} &= \sigma\!\big(\W_g^{(l)} \a^{(l-1)} + \bm{U}_g^{(l)} \x + \bb_g^{(l)}\big), && \text{(forget gate)} \\
R^{(l)} &= \sigma\!\big(\W_r^{(l)} \a^{(l-1)} + \bm{U}_r^{(l)} \x + \bb_r^{(l)}\big), && \text{(relevance gate)} \\
H^{(l)} &= \tanh\!\big(\W_h^{(l)} (R^{(l)} \odot \a^{(l-1)}) + \bm{U}_h^{(l)} \x + \bb_h^{(l)}\big), && \text{(candidate state)}
\end{align}
and the state update is:
\begin{equation}
\a^{(l)} = (1 - G^{(l)}) \odot H^{(l)} + Z^{(l)} \odot \a^{(l-1)}.
\end{equation}
Note that this formulation uses two separate gates: $Z^{(l)}$ controls how much of the previous state $\a^{(l-1)}$ to retain, while $(1 - G^{(l)})$ scales the new candidate $H^{(l)}$.  Because $Z$ and $G$ are independent in the original Sirignano--Spiliopoulos formulation, the two coefficients need not sum to one.  A simpler GRU-style variant \citep{cho2014gru} collapses the two gates into a single convex combination $\a^{(l)} = (1-G) \odot \a^{(l-1)} + G \odot H$, where the coefficients sum to one; in practice, both variants perform comparably on the PDE benchmarks in this course.  The gate $R^{(l)}$ controls which parts of the previous state are relevant for computing the candidate $H^{(l)}$.  The gate naming convention $(Z,G,R,H)$ used here follows \citet{al2018solving}; in GRU terminology, $Z$ corresponds to the update gate and $R$ to the reset gate.  The input $\x$ enters every layer through the $\bm{U}$ matrices, providing skip connections that ensure input information is available at every depth.  This gating mechanism, directly analogous to LSTM recurrent networks, helps the DGM architecture learn functions that depend sensitively on the input coordinates, a common feature of PDE solutions near boundary layers.  Figure~\ref{fig:dgm_architecture} shows the resulting feed-forward architecture.

\begin{figure}[ht]
\centering
\begin{tikzpicture}[scale=0.88, transform shape,
    layer/.style={rectangle, draw=uzhblue, thick, fill=uzhgreylight,
        minimum width=2.0cm, minimum height=0.6cm, font=\footnotesize, rounded corners=3pt},
    arr/.style={-{Stealth[length=2mm]}, thick, uzhblue},
    skip/.style={-{Stealth[length=2mm]}, thick, softorange, dashed}
]
    \node[layer, fill=softblue!15] (input) at (0,0) {Input $\x$};
    \node[layer] (h1) at (3.5,0) {DGM Layer 1};
    \node[layer] (h2) at (7,0) {DGM Layer 2};
    \node[font=\large] at (9.2,0) {$\cdots$};
    \node[layer] (hL) at (11.2,0) {DGM Layer $L$};
    \node[layer, fill=softgreen!15] (out) at (14.5,0) {Output $\hat{u}$};

    \draw[arr] (input) -- (h1);
    \draw[arr] (h1) -- (h2);
    \draw[arr] (h2) -- (8.5,0);
    \draw[arr] (9.9,0) -- (hL);
    \draw[arr] (hL) -- (out);

    \draw[skip] (input) -- ++(0,-1.2) -| (h2);
    \draw[skip] (input) -- ++(0,-1.8) -| (hL);

    \node[font=\small, softorange, below] at (7.25,-2.05) {skip connections from $\x$};
\end{tikzpicture}
\caption{The DGM (Deep Galerkin Method) architecture of \citet{sirignano2018dgm}. The input $\x$ feeds the first layer through a standard forward path (solid arrows) and is, in addition, routed to every subsequent DGM block via skip connections (dashed arrows), so each layer can see $\x$ directly as well as the running hidden state. Each DGM block combines $\x$ with the hidden state through update, forget, and relevance gates (see body text), in the spirit of LSTM/GRU recurrences applied across depth rather than time.}
\label{fig:dgm_architecture}
\end{figure}

\paragraph{Stationary vs.\ evolutionary PDEs.}
The PINN framework applies uniformly to both \emph{stationary} PDEs, where the unknown depends only on state variables ($u = u(\mathbf{x})$), and \emph{evolutionary} PDEs, where time enters as an additional input ($u = u(\mathbf{x}, t)$).  The network architecture is identical in both cases: only the input dimension changes.  The HJB equation below is stationary: the value function $V(a)$ depends on a single state variable.  The Black--Scholes PDE of Section~\ref{sec:bs_pinn} is evolutionary: $V(S,t)$ depends on both the asset price and time to maturity.

\section{Application: HJB Equation and the Cake-Eating Problem}
\label{sec:cake_eating_hjb}

\begin{remarkbox}[Notation: $\gamma$ as CRRA]
In this section, $\gamma$ denotes the coefficient of relative risk aversion (CRRA), with utility $c^{1-\gamma}/(1-\gamma)$.  This convention differs from Chapter~\ref{ch:irbc}, where $\gamma_j$ denotes the intertemporal elasticity of substitution (IES $= 1/\text{CRRA}$).  Both conventions are standard in their respective literatures.
\end{remarkbox}

Consider a household with wealth $a(t) > 0$ that chooses a consumption stream $c(t) \geq 0$ to maximize discounted lifetime utility:
\begin{equation}
\max_{\{c(t)\}_{t \geq 0}} \int_0^\infty e^{-\rho t}\,\frac{c(t)^{1-\gamma}}{1-\gamma}\,dt
\quad\text{s.t.}\quad \dot{a}(t) = r\,a(t) - c(t),
\label{eq:cake_problem}
\end{equation}
where $\rho > 0$ is the subjective discount rate, $\gamma > 0$ ($\gamma \neq 1$) is the coefficient of relative risk aversion (CRRA), and $r$ is the interest rate on wealth.  The budget constraint says that wealth grows at rate $r$ minus consumption; we call this the ``cake-eating'' problem in a loose sense because the household consumes out of a single finite stock of wealth.  Strictly speaking, the textbook cake-eating problem has $r = 0$ (the cake does not grow); the $r > 0$ variant solved here is the consumption--savings problem with deterministic returns, and Exercise~\ref{ex:ch7:3} traces the $r = 0$ limit explicitly.
The dynamic-programming perspective is classical in economics; see \citet{stokeylucas1989} for the standard textbook treatment.  Conceptually, this is the same recursive optimization that produced the discrete-time Bellman equations of Chapters~\ref{ch:deqn}--\ref{ch:irbc}; the only difference is that the household now optimizes over a continuous time grid, so the resulting optimality condition is a partial differential equation (the HJB) rather than an algebraic Euler equation.  The derivation below makes this $\Delta t \to 0$ link precise.

\paragraph{Deriving the HJB equation.}
Define the \emph{value function} $V(a)$ as the maximum attainable lifetime utility starting from wealth $a$:
\begin{equation}
V(a) = \max_{\{c(t)\}} \int_0^\infty e^{-\rho t}\,\frac{c(t)^{1-\gamma}}{1-\gamma}\,dt.
\label{eq:cake_value}
\end{equation}
Since the problem is stationary (no explicit time dependence in the return or the law of motion), $V$ depends only on the current state $a$, not on calendar time.

To derive the HJB equation, apply the \emph{principle of optimality}: over a short interval $[0, \Delta t]$, the household chooses $c$ optimally and then continues optimally from the resulting state $a + \dot{a}\,\Delta t$.  This gives:
\begin{equation}
V(a) = \max_c \left\{\frac{c^{1-\gamma}}{1-\gamma}\,\Delta t + e^{-\rho\,\Delta t}\,V\!\bigl(a + (ra - c)\,\Delta t\bigr)\right\} + \mathcal{O}(\Delta t^2).
\label{eq:cake_bellman_dt}
\end{equation}
Now expand both the discount factor and the value function to first order in $\Delta t$:
\begin{align}
e^{-\rho\,\Delta t} &\approx 1 - \rho\,\Delta t, \label{eq:cake_expand_disc} \\
V\!\bigl(a + (ra - c)\,\Delta t\bigr) &\approx V(a) + V'(a)\,(ra - c)\,\Delta t. \label{eq:cake_expand_V}
\end{align}

\begin{remarkbox}[Where the upwind scheme comes from]
The expansion~\eqref{eq:cake_expand_V} silently assumes that $V$ is differentiable at $a$.  When $V$ has a kink, a common feature of HJB solutions, e.g.\ at borrowing constraints or policy-switching points, the expansion still holds, but $V'(a)$ must be replaced by the appropriate \emph{one-sided} derivative chosen by the sign of the drift $(ra-c)$.  This is exactly the origin of the upwind scheme used in finite-difference HJB solvers: the side of the derivative is selected to follow information flow.  PINNs that minimize the strong-form residual implicitly demand two-sided differentiability and therefore tend to over-smooth genuine kinks, one of the mechanisms behind the oscillatory failure mode in Table~\ref{tab:pinn_failure_modes}.
\end{remarkbox}

Substituting~\eqref{eq:cake_expand_disc} and~\eqref{eq:cake_expand_V} into~\eqref{eq:cake_bellman_dt}:
\begin{equation}
V(a) = \max_c \left\{\frac{c^{1-\gamma}}{1-\gamma}\,\Delta t + (1 - \rho\,\Delta t)\Bigl[V(a) + V'(a)(ra-c)\,\Delta t\Bigr]\right\} + \mathcal{O}(\Delta t^2).
\end{equation}
Expanding the product and dropping terms of order $(\Delta t)^2$:
\begin{equation}
V(a) = \max_c \left\{\frac{c^{1-\gamma}}{1-\gamma}\,\Delta t + V(a) + V'(a)(ra-c)\,\Delta t - \rho\,V(a)\,\Delta t\right\} + \mathcal{O}(\Delta t^2).
\end{equation}
Cancel $V(a)$ from both sides, divide by $\Delta t$, and let $\Delta t \to 0$:
\begin{equation}
\boxed{\rho\, V(a) = \max_c \left\{\frac{c^{1-\gamma}}{1-\gamma} + V'(a)\,(ra - c)\right\}.}
\label{eq:cake_hjb}
\end{equation}
This is the \emph{Hamilton--Jacobi--Bellman (HJB) equation}.  The left-hand side is the ``cost of holding wealth $a$'' (the required return $\rho V$).  The right-hand side is the ``benefit'': instantaneous utility from consuming $c$, plus the capital gain $V'(a)\,\dot{a}$ from the change in wealth.

\paragraph{First-order condition and optimal consumption.}
The maximization in~\eqref{eq:cake_hjb} is over $c$ with the objective being concave in $c$ (since $\gamma > 0$).  Differentiating the term inside the braces with respect to $c$ and setting to zero:
\begin{equation}
\frac{\partial}{\partial c}\left[\frac{c^{1-\gamma}}{1-\gamma} + V'(a)(ra - c)\right]
= c^{-\gamma} - V'(a) = 0
\qquad\Longrightarrow\qquad
c^\star(a) = \bigl(V'(a)\bigr)^{-1/\gamma}.
\label{eq:cake_foc}
\end{equation}
The economic content is intuitive: the marginal utility of consumption $c^{-\gamma}$ equals the marginal value of wealth $V'(a)$.  A higher $V'(a)$ (wealth is more valuable) implies lower optimal consumption.

\paragraph{Analytical solution.}  The HJB~\eqref{eq:cake_hjb} with the FOC~\eqref{eq:cake_foc} substituted in becomes a nonlinear ODE in $V(a)$.  To solve it, conjecture $V(a) = \Lambda\, a^{1-\gamma}/(1-\gamma)$ for some constant $\Lambda > 0$.  Then $V'(a) = \Lambda\, a^{-\gamma}$, and from the FOC~\eqref{eq:cake_foc}:
\begin{equation}
c^\star = (\Lambda\, a^{-\gamma})^{-1/\gamma} = \Lambda^{-1/\gamma}\, a.
\end{equation}
Substituting into the HJB:
\begin{equation}
\rho\,\Lambda\,\frac{a^{1-\gamma}}{1-\gamma}
= \frac{(\Lambda^{-1/\gamma}\,a)^{1-\gamma}}{1-\gamma} + \Lambda\, a^{-\gamma}\bigl(ra - \Lambda^{-1/\gamma}\,a\bigr).
\end{equation}
Dividing through by $a^{1-\gamma}/(1-\gamma)$ and solving for $\Lambda$ gives $\Lambda = \kappa^{-\gamma}$ with $\kappa = (\rho - (1-\gamma)r)/\gamma$, so $V^\star(a) = \kappa^{-\gamma}\,a^{1-\gamma}/(1-\gamma)$ and the optimal consumption rule is $c^\star(a) = \kappa\, a$.  This solution is well-defined provided $\kappa > 0$, i.e., $\rho > (1-\gamma)r$, a standard transversality condition ensuring that the agent discounts future utility sufficiently relative to wealth growth.  This closed form is the natural validation target for the PINN in notebook \texttt{lecture\_11\_04\_Cake\_Eating\_HJB\_PINN}: the trained network should recover $V(a)$ to mean relative error well below $10^{-3}$, and any larger discrepancy signals an under-trained network or a malformed loss.

\paragraph{PINN formulation and PDE residual.}
To solve the HJB equation~\eqref{eq:cake_hjb} with a PINN (rather than using the closed-form above), we proceed as follows.  A neural network $\hat{V}(a) = \mathcal{N}_\theta(a)$ approximates the value function.  Its derivative $\hat{V}'(a)$ is computed by automatic differentiation up to floating-point precision, so no finite differences are needed.  From this derivative, we reconstruct the optimal consumption via the FOC~\eqref{eq:cake_foc}:
\begin{equation}
\hat{c}(a) = \bigl(\hat{V}'(a)\bigr)^{-1/\gamma}.
\end{equation}
Substituting $\hat{V}$, $\hat{V}'$, and $\hat{c}$ into the HJB~\eqref{eq:cake_hjb} and moving everything to one side defines the \emph{PDE residual}.  In practice both the FOC inversion and the residual evaluation use a positivity-guarded derivative $\widetilde V_a := \operatorname{softplus}(\hat V'(a)) + \varepsilon$ rather than the raw $\hat V'(a)$, so that consumption is well-defined even where the network's raw derivative is negative during early training (see also the listing below):
\begin{equation}
\hat{c}(a) = \widetilde V_a^{-1/\gamma}, \qquad
\mathcal{R}(a) = \rho\,\hat{V}(a) - \left[\frac{\hat{c}(a)^{1-\gamma}}{1-\gamma} + \widetilde V_a\,(ra - \hat{c}(a))\right].
\label{eq:cake_residual}
\end{equation}
Once training has converged with $\hat V'(a) \gg 0$ on the support, $\operatorname{softplus}(\hat V'(a)) \approx \hat V'(a)$ to high precision (the exponential tail decays rapidly), so the safeguarded residual is asymptotically equivalent to the original HJB residual; using $\widetilde V_a$ throughout keeps the FOC inversion and the HJB residual mutually consistent during training, when $\hat V'(a)$ may transiently be small or negative.
If the network has perfectly learned the true value function, $\mathcal{R}(a) = 0$ for all $a$.  The PINN loss is the mean squared residual over a set of collocation points $\{a_i\}_{i=1}^M$ sampled in the interior of the domain:
\begin{equation}
\ell_{\mathrm{HJB}} = \frac{1}{M}\sum_{i=1}^{M} \mathcal{R}(a_i)^2.
\end{equation}
The DGM (Deep Galerkin Method) architecture of \citet{sirignano2018dgm}, which provides LSTM-style gating and skip connections from the input to every hidden layer, may be used in place of the plain MLP below to improve expressivity for this type of PDE problem; the listing keeps a plain MLP for clarity.  In low dimension the trial-solution MLP of notebook \texttt{lecture\_11\_04\_Cake\_Eating\_HJB\_PINN} typically outperforms a soft-BC DGM, because the boundary loss no longer competes with the interior residual; DGM becomes worthwhile primarily in higher dimensions or for PDEs with sharp internal features.

\begin{remarkbox}[Residual loss alone is not enough]
A small HJB residual is necessary but not sufficient for an economically meaningful solution.  The same residual minimum can correspond to wildly different value-function levels when the boundary anchor is weak, and to spurious non-monotonic policies when $\hat V'(a) \le 0$ is left unpenalized.  Practical safeguards used in notebook \texttt{lecture\_11\_04\_Cake\_Eating\_HJB\_PINN}: (i)~hard-BC trial solution that fixes both endpoints exactly, removing the boundary--interior trade-off; (ii)~a softplus-transformed derivative $\widetilde V_a=\operatorname{softplus}(\hat V_a)+\varepsilon$ inside the FOC inversion, eliminating the negative-derivative pathology in training; (iii)~checking the implied consumption policy, the raw derivative $\hat V_a$, and the HJB residual against the closed-form benchmark.  When the residual, boundary conditions, and policy diagnostics agree, the solution is also economically sensible; when only the HJB residual is small, it usually is not.
\end{remarkbox}

\begin{remarkbox}[Optimizer pipeline and double precision]
The PINN notebooks of this chapter use a two-stage pipeline: \emph{Adam} on resampled collocation points to find a basin of attraction, then a \emph{deterministic L-BFGS} polish on a fixed grid in float64.  Two practical points are not cosmetic.  First, switching collocation points each L-BFGS evaluation breaks the strong Wolfe line search; the polish requires a deterministic objective.  Second, second derivatives of a $\tanh$ MLP lose substantial precision in float32, which is exactly the scale L-BFGS probes when comparing successive line-search points.  FP64 is therefore a stability device, not a guarantee of machine-precision residuals: in a longer \texttt{teaching}/\texttt{production} run, the one-dimensional cake HJB reaches final HJB loss about $3\times 10^{-4}$ (the checked-in \texttt{RUN\_MODE="smoke"} notebook stops well short of that), while the heterogeneous-agent examples still require residual, policy, density, and aggregate diagnostics.  Recent quantitative discussions of PINN precision and floating-point limits are reviewed in the further-reading list at the end of the chapter.
\end{remarkbox}

\begin{lstlisting}[caption={PINN residual for the HJB equation (PyTorch).}]
def pde_residual(model, a, gamma, rho, r):
    a.requires_grad_(True)
    V = model(a)
    V_a = torch.autograd.grad(V.sum(), a, create_graph=True)[0]
    safe_Va = F.softplus(V_a) + 1e-6    # positivity guard; +1e-6 avoids division by zero in safe_Va.pow(-1/gamma)
    c = safe_Va.pow(-1.0 / gamma)        # FOC
    u_c = c.pow(1 - gamma) / (1 - gamma)
    R = rho * V - (u_c + safe_Va * (r * a - c))
    return R
\end{lstlisting}

\section{Continuous-Time Heterogeneous Agent Models}
\label{sec:ct_hank}

Many frontier models in macroeconomics feature a continuum of agents who face idiosyncratic risk and interact through equilibrium prices.  Chapter~\ref{ch:young} studied the discrete-time version with Young's histogram method inside a DEQN.  In continuous time the same economic question becomes a \emph{coupled PDE system}: a Hamilton--Jacobi--Bellman equation for individual optimization and a Kolmogorov forward (Fokker--Planck) equation for the stationary cross-sectional density, closed by a market-clearing condition that pins down prices.  This is the canonical example of a PINN applied to a \emph{system} of equilibrium PDEs, in contrast to the single HJB of the cake-eating problem (\S\ref{sec:cake_eating_hjb}) and the single Black--Scholes PDE (\S\ref{sec:bs_pinn}): a PINN parameterizes the value function and the density by two networks $\hat V_\theta(a,z)$, $\hat g_\psi(a,z)$ and minimizes a four-term loss -- the HJB residual, the KFE residual, the no-flux boundary residual at the borrowing constraint, and a mass-normalization residual -- with the loss-balancing and boundary-layer issues familiar from the rest of this chapter.

The full development -- the stochastic-calculus background, the formal HJB and KFE derivations, the Huggett and Aiyagari equilibria, the PINN solver for the stationary coupled system, the master equation that handles aggregate shocks, and the EMINN method -- is the subject of Chapter~\ref{ch:ct_theory}: see \S\ref{sec:ct_equilibrium} and \S\ref{sec:ct_pinn} for the stationary problem and \S\ref{sec:master_eq}--\S\ref{sec:eminn} for the aggregate-shock case.  Following \citet{achdou2022income} for the model setup (whose own numerical solution uses finite differences), that treatment replaces the traditional fixed-point iteration over $r$ with joint training of all components.

\section{Application: Black--Scholes PDE}
\label{sec:bs_pinn}

The Black--Scholes PDE has a closed-form solution, so a PINN run on it is not motivated by the absence of an analytical answer.  The pedagogical purpose is exactly the opposite: it is a known-answer benchmark.  We verify that the same PINN recipe (smooth activations, hard or soft BCs, autodiff for $V_S$ and $V_{SS}$, Adam-then-L-BFGS) reproduces a textbook formula on a clean domain before applying it to PDEs without closed forms (American options, jump diffusions, multi-asset pricing, HJBs with multiple state variables).  If the network cannot recover Black--Scholes to plotting accuracy, no further trust should be placed in its output on harder problems.

The Black--Scholes PDE for a European call option with strike $K$, maturity $T$, risk-free rate $r$, and volatility $\sigma$ is the canonical option-pricing benchmark of \citet{black1973pricing}:
\begin{equation}
\frac{\partial V}{\partial t} + \frac{1}{2}\sigma^2 S^2 \frac{\partial^2 V}{\partial S^2} + rS\frac{\partial V}{\partial S} - rV = 0,
\end{equation}
with terminal condition $V(S,T) = \max(S-K, 0)$ and boundary conditions $V(0,t)=0$, $V(S_\mathrm{max}, t) = S_\mathrm{max} - Ke^{-r(T-t)}$.

The PINN approach approximates $V(S,t) \approx \mathcal{N}_\theta(S,t)$ and computes the partial derivatives $V_t$, $V_S$, and $V_{SS}$ via automatic differentiation.  The total PINN loss for the Black--Scholes problem has four terms:
\begin{equation}
\begin{aligned}
\ell ={}& \underbrace{\frac{1}{N_r}\sum_i
\left|V_t(S_i^r,t_i^r) + \tfrac{1}{2}\sigma^2 (S_i^r)^2 V_{SS}(S_i^r,t_i^r)
+ rS_i^r V_S(S_i^r,t_i^r) - rV(S_i^r,t_i^r)\right|^2}_{\text{PDE residual}} \\
&+ \underbrace{\frac{\lambda_\mathrm{TC}}{N_T}\sum_j \left|V(S_j,T) - \max(S_j-K,0)\right|^2}_{\text{terminal condition}}
+ \underbrace{\frac{\lambda_0}{N_0}\sum_k \left|V(0,t_k)\right|^2}_{\text{lower BC}} \\
&+ \underbrace{\frac{\lambda_\infty}{N_\infty}\sum_l
\left|V(S_\mathrm{max},t_l) - \bigl(S_\mathrm{max}-Ke^{-r(T-t_l)}\bigr)\right|^2}_{\text{upper BC}} .
\end{aligned}
\end{equation}
The loss has four terms (PDE residual, terminal condition, lower BC, upper BC) and three relative penalty weights $(\lambda_\mathrm{TC}, \lambda_0, \lambda_\infty)$; the PDE residual coefficient is normalized to one.  These weights must be tuned or adaptively balanced via ReLoBRaLo from Chapter~\ref{ch:nas}.  As a practical diagnostic of what goes wrong if the weights are mis-set: underweighting $\lambda_{\mathrm{TC}}$ produces a smooth surface that mis-prices the payoff at maturity (the network ignores the kink at $S=K$); overweighting $\lambda_{\mathrm{TC}}$ pins the network at $t=T$ but the interior PDE residual then fails to balance the $V_t$ term, producing visible drift in the early-time slice.  In notebook \texttt{lecture\_11\_05\_Black\_Scholes\_PINN}, the benchmark parameters are $K=50$, $T=1$, $r=0.05$, $\sigma=0.2$, and $S_\mathrm{max}=100$, for which the analytical at-the-money call value is roughly $4.8$.  A longer \texttt{teaching}/\texttt{production} run reports a max absolute error around $0.13$ and a mean absolute error around $0.04$ against the analytical Black--Scholes formula (the checked-in \texttt{smoke} notebook prints larger errors); at the ATM value scale this is roughly a $2.7\%$ peak and $0.8\%$ average relative error.  Accuracy is therefore a diagnostic to check after training, not a guaranteed property of the architecture.  The choice of $\tanh$ is natural here: it is $C^\infty$ (so $V_{SS}$ is well-defined everywhere), and its bounded range $(-1,1)$ provides a stable starting point for learning the option price surface.

\section{From PINNs to Operator Learning: One Network, Many Problems}
\label{sec:operator_learning_bridge}

\noindent\emph{This section is script-only outlook material; it has no companion slide and no notebook.  It can be skipped on a first read without loss of continuity, and is intended for readers preparing to scale a PINN-based pipeline across many parameter configurations.}

\medskip
A PINN learns \emph{one solution} to one PDE.  Each new boundary condition, parameter set, or coefficient field forces a fresh training run.  In economic and financial applications this is often exactly the bottleneck: we want option prices for many strike--maturity pairs, value functions for many discount factors, or HJB solutions across an entire parameter sweep.  \emph{Operator learning} flips the question: instead of learning a function $u : \mathbb{R}^d \to \mathbb{R}$ that solves the PDE for a single instance, one learns the \emph{solution operator}
\[
   \mathcal{G}: \;\;\text{(input field, BCs, parameters)} \;\longmapsto\; \text{solution function } u,
\]
i.e.\ a map between two function spaces.

Two mature architectures dominate the literature.

\paragraph{DeepONet \citep{lu2021learning}.}  Inspired by the universal-approximation theorem for nonlinear operators, DeepONet uses two sub-networks: a \emph{branch net} encodes the input field at sensor locations into a latent vector, and a \emph{trunk net} encodes the query point at which the output is requested; their inner product is the predicted operator output.  The architecture is generic and trains entirely on input--output pairs of an offline solver.

\paragraph{Fourier Neural Operator (FNO) \citep{li2021fourier}.}  FNO parameterizes a kernel integral operator by truncating it in Fourier space and applying a learned linear transformation to the low-frequency modes per layer.  This captures global interactions cheaply ($\mathcal{O}(n\log n)$ via FFT) and exhibits resolution invariance: a network trained on one grid can be evaluated on a finer grid at test time.

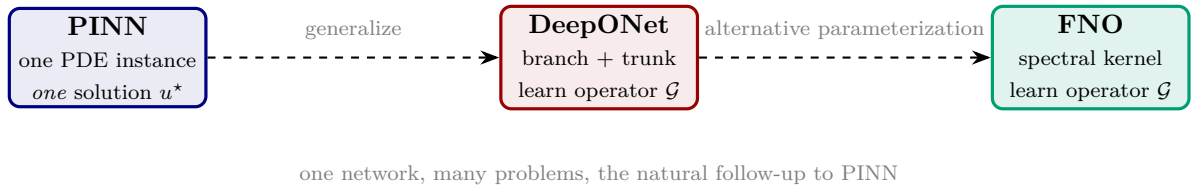
\begin{figure}[ht]
\centering
\begin{tikzpicture}[
    nbox/.style={rectangle, draw=uzhblue, very thick, fill=uzhblue!8, rounded corners=4pt,
                 minimum width=2.6cm, minimum height=1.05cm, align=center, font=\small,
                 inner sep=3pt},
    pbox/.style={rectangle, draw=harvardcrimson, very thick, fill=harvardcrimson!8, rounded corners=4pt,
                 minimum width=2.6cm, minimum height=1.05cm, align=center, font=\small,
                 inner sep=3pt},
    obox/.style={rectangle, draw=softgreen, very thick, fill=softgreen!12, rounded corners=4pt,
                 minimum width=2.6cm, minimum height=1.05cm, align=center, font=\small,
                 inner sep=3pt},
    arr/.style={-{Stealth[length=2.5mm]}, thick}
]
\node[nbox] (pinn) at (0,0)    {\textbf{PINN}\\\scriptsize one PDE instance \\\scriptsize \emph{one} solution $u^\star$};
\node[pbox] (don)  at (6.5,0)  {\textbf{DeepONet}\\\scriptsize branch + trunk \\\scriptsize learn operator $\mathcal{G}$};
\node[obox] (fno)  at (13,0)   {\textbf{FNO}\\\scriptsize spectral kernel \\\scriptsize learn operator $\mathcal{G}$};
\draw[arr, dashed] (pinn) -- node[midway, above=3pt, font=\scriptsize, gray] {generalize} (don);
\draw[arr, dashed] (don)  -- node[midway, above=3pt, font=\scriptsize, gray] {alternative parameterization} (fno);
\node[font=\scriptsize, gray, align=center] at (6.5,-1.55) {one network, many problems, the natural follow-up to PINN};
\end{tikzpicture}
\caption{Operator learning generalizes PINNs by amortising over an entire parametric family of PDEs.  For economic applications such as option-price surfaces over $(K,T)$, value functions across a parameter range, or HJB sweeps for sensitivity analysis (Chapter~\ref{ch:gp}), training the operator once gives instant predictions at test time.}
\label{fig:operator_learning_bridge}
\end{figure}

Figure~\ref{fig:operator_learning_bridge} summarizes this progression from one-instance PINNs to operator-learning architectures.  In the rest of this script, we mostly stay with PINNs because the focus is on solving one model carefully; we revisit operator learning briefly in Chapter~\ref{ch:outlook} and point readers who want to amortise across an entire parametric family of PDEs to \citet{lu2021learning, li2021fourier}.

\begin{keyinsightbox}[Chapter Summary]
PINNs approximate PDE solutions by minimizing the residual at collocation points, with automatic differentiation supplying the required derivatives algorithmically up to floating-point precision \citep{raissi2019physics}.  Hard versus soft enforcement of boundary conditions is the central design lever: trial-function constructions of the form $\hat y = A(x) + B(x)\mathcal{N}_\theta(x)$ enforce BCs by construction, while soft penalties cover cases where hard enforcement is intractable.  For second-order PDEs (HJB, Black--Scholes, Poisson) the activation must be at least $C^2$, which excludes ReLU and makes $\tanh$ and Swish the standard choices.  Operator learning (DeepONet, FNO) is the natural generalization when one wants to amortise across a parametric family rather than solve one instance.
\end{keyinsightbox}

\section*{Further Reading}
\addcontentsline{toc}{section}{Further Reading}
\begin{itemize}[itemsep=2pt]
\item \citet{raissi2019physics}, the foundational PINN paper.
\item \citet{sirignano2018dgm}, the Deep Galerkin Method, the original deep-PDE solver in finance.
\item \citet{e2017deep, han2018solving}, deep BSDE methods, the SDE counterpart.
\item \citet{wang2021understanding, bischof2025relobralo}, adaptive loss balancing for PINNs.
\item \citet{lu2021deepxde}, the DeepXDE software ecosystem.
\end{itemize}

\section*{Exercises}
\addcontentsline{toc}{section}{Exercises}
\noindent Worked solutions and guidance for these exercises appear in Appendix~\ref{app:solutions}.
\begin{enumerate}[itemsep=4pt, leftmargin=*]
\item\label{ex:ch7:1} \textbf{[Core] Trial-function BC enforcement.}  For the BVP $y'' + y = 0$ on $[0,\pi/2]$ with $y(0)=0$, $y(\pi/2)=1$, verify that $\hat y(x) = 2x/\pi + x(\pi/2-x)\,\mathcal{N}_\theta(x)$ satisfies both BCs for any network output.  Why is this preferable to a soft penalty?
\item\label{ex:ch7:2} \textbf{[Core] ReLU pathology.}  Explain why a ReLU network is unsuitable for the strong-form Black--Scholes residual.  Then sketch a weak-form formulation that would work with ReLU activations.
\item\label{ex:ch7:3} \textbf{[Core] Discrete $\to$ continuous bridge.}  Take the discrete-time Bellman operator
\[
V(a) = \max_c \bigl[u(c)\,\Delta t + \beta_{\Delta t}\,\E V(a')\bigr],
\qquad
a' = a - c\,\Delta t,
\qquad
\beta_{\Delta t}=e^{-\rho\Delta t}.
\]
Show that as $\Delta t \to 0$ it formally yields the HJB $\rho V = \max_c [u(c) - V'(a)\,c]$.  (This is the pure cake-eating problem with $r=0$; including a return $r$ on wealth, $\dot a = ra - c$, recovers~\eqref{eq:cake_hjb}.)
\item\label{ex:ch7:4} \textbf{[Computational] BC penalty weight $\lambda$ tuning.}  In notebook \tpath{lecture_11_02_ODE_PINN_SoftVsHardBCs}, run the soft-BC variant with PDE residual $\ell_\mathrm{int}$ and BC residual $\ell_\mathrm{BC}$ combined as $\mathcal{L} = \ell_\mathrm{int} + \lambda\,\ell_\mathrm{BC}$ (the notebook exposes $\lambda$ as the \texttt{bc\_weight} hyperparameter dispatched from \texttt{RUN\_MODE}) for $\lambda \in \{10^{-1}, 10^0, 10^1, 10^2, 10^3,10^4\}$.  For each, train to a fixed budget of $5{,}000$ iterations and record (i)~the BC violation $|\hat y(0) - y_0|$ at convergence, (ii)~the maximum interior residual, and (iii)~training wall time.  Plot all three on log--log axes against $\lambda$, identify the elbow at which boundary fit stops improving cheaply, and compare against the hard-BC variant.
\item\label{ex:ch7:5} \textbf{[Advanced/project] Collocation strategy comparison.}  In notebook \tpath{lecture_11_03_PDE_PINN_Poisson2D} solving $\Delta u = f$ on $[0,1]^2$ with Dirichlet BCs, replace the default uniform-random collocation with (a)~Latin Hypercube sampling, (b)~a Sobol low-discrepancy sequence, (c)~residual-based adaptive sampling that draws each batch of collocation points proportional to the residual magnitude at the current iterate.  Train each variant to the same residual tolerance ($10^{-4}$) and report (i)~the number of collocation points needed, (ii)~the wall time, (iii)~the spatial distribution of final residuals (compute $\sup_x |\Delta u_\theta - f|$ on a fine $200\times 200$ test grid).  Discuss when each strategy is preferred: uniform for smooth problems, low-discrepancy for moderately rough, adaptive for problems with localized features (e.g., near boundary layers).
\item\label{ex:ch7:6} \textbf{[Advanced/project] Strong vs.\ weak forms.}  Extend the Black--Scholes notebook.  First compare the current strong-form $\tanh$ PINN with a ReLU version and explain why the second derivative term is problematic for ReLU.  Then derive a weak-form variant: multiply the residual by a smooth test function $\varphi$ supported inside $[0,S_\mathrm{max}]$, integrate by parts to move the second derivative onto $\varphi$, and state which derivatives of $V$ remain.  If you implement the weak-form ReLU variant, compare its held-out pricing error with the strong-form $\tanh$ baseline.  Connect the result to Exercise~\ref{ex:ch7:2}.
\item\label{ex:ch7:7} \textbf{[Core] Operator learning vs.\ PINN.}  For an option pricing problem where the strike $K$ varies across $\{45,46,\ldots,55\}$, count the training cost of (a)~eleven independent PINN runs vs.\ (b)~one operator-learning or parametric-PINN run that takes $K$ as an input; see Section~\ref{sec:operator_learning_bridge} and Chapter~\ref{ch:outlook}.  At what cost ratio does amortized training win?
\end{enumerate}

\chapter[Continuous-Time Heterogeneous Agent Models]{Heterogeneous Agent Models in Continuous Time}
\label{ch:ct_theory}

Building on the PINN foundations of Chapter~\ref{ch:pinn}, this chapter develops the full continuous-time heterogeneous-agent framework: the coupled system of the Hamilton--Jacobi--Bellman equation (for individual optimization) and the Kolmogorov forward equation (for the stationary wealth distribution), closed by a market-clearing condition.  This is the continuous-time analogue of the discrete-time Krusell--Smith economy of Chapter~\ref{ch:young}.  The primary reference is \citet{achdou2022income}; for pedagogical background on the continuous-time methods, see also Moll's lecture notes.\footnote{\url{https://benjaminmoll.com/lectures/}}

In the previous chapter, we applied PINNs to individual PDEs (ODEs, the Poisson equation, the HJB for cake-eating, and the Black--Scholes equation).  This chapter makes the leap to \emph{equilibrium systems}: coupled PDEs that arise when a continuum of heterogeneous agents interact through prices.  The theoretical framework draws on the Bewley--Huggett--Aiyagari tradition, formulated in continuous time following \citet{achdou2022income}.  We derive the two core PDEs, the Hamilton--Jacobi--Bellman (HJB) equation for individual optimization and the Kolmogorov forward equation (KFE) for the cross-sectional density, and show how they are coupled through market clearing.  The chapter culminates with the \emph{master equation}, a single (infinite-dimensional) PDE that encapsulates the full equilibrium, and with EMINNs, introduced by \citet{gu2024masterequations}, which solve it using deep learning.

\paragraph{Companion notebook.}  One notebook accompanies this chapter and the Lecture~13 numerical deck: \tpath{lecture_13_08_Aiyagari_Continuous_Time_FD_and_PINN_PyTorch.ipynb}.  It first computes the stationary Aiyagari equilibrium with an upwind finite-difference solver, then freezes those equilibrium prices and trains a PINN for the coupled HJB--KFE system at them (with normalization, boundary-flux, state-constraint, and aggregate-capital diagnostics) -- so the PINN stage is an equilibrium-consistency and representation exercise, not a price-discovery algorithm.  For transparency the notebook specializes the general CRRA/switching equations below to log utility ($\gamma=1$) and symmetric two-state Poisson switching; the formulas in this chapter are the general case.  The detailed upwind finite-difference scheme and PINN losses are developed in the accompanying Lecture~13 deck; this chapter focuses on the continuous-time equilibrium equations and the conceptual bridge to master-equation methods.  (A partial-equilibrium HJB on its own is just a single-PDE PINN problem of the kind treated in Chapter~\ref{ch:pinn}; here we go straight to the coupled equilibrium system.)

\paragraph{Where does $g$ come from?}  Before turning to the PDEs themselves, it is worth fixing the economic picture.  The model has a continuum of agents, each indexed by an idiosyncratic state $(a,z)$ comprising wealth $a$ and a labor or productivity component $z$.  Each agent solves its own HJB equation taking the prices $(r,w)$ as given, which yields an optimal savings policy that pushes mass through wealth space.  The Kolmogorov forward equation tracks how this aggregate mass evolves, and its stationary solution $g^\star(a,z)$ is the cross-sectional density that the general-equilibrium clearing equations integrate against to obtain aggregate capital, $K = \int a\, g^\star(a,z)\,da\,dz$, and aggregate labor, $L = \int z\, g^\star(a,z)\,da\,dz$.

\section{Why Continuous Time?}
\label{sec:ct_why}

Chapters~\ref{ch:deqn}--\ref{ch:young} formulated heterogeneous-agent models in discrete time.  Working in continuous time offers several complementary advantages:
\begin{itemize}[itemsep=2pt]
\item \textbf{Analytical tractability.}  It\^o calculus provides clean first-order conditions, and the separation between the backward HJB and the forward KFE is sharper than in discrete time.
\item \textbf{No expectations operator.}  Conditional expectations are replaced by differential operators, avoiding numerical integration over shock distributions.
\item \textbf{Powerful numerical methods.}  Finite differences, PINNs, and deep learning methods (EMINNs) can be applied directly to the PDE system.
\item \textbf{Connection to mean field games.}  The coupled HJB--KFE system is precisely a \emph{mean field game} (MFG) in the sense of \citet{lasry2007mean}: each atomistic agent solves an HJB taking the cross-sectional density as given, while the density itself evolves via a KFE driven by those individual best responses.  Equilibrium is the fixed point of this two-way coupling.  Recasting the problem in MFG language gives access to a large mathematical literature on existence, uniqueness, and numerical analysis \citep{carmona2018probabilistic, cardaliaguet2019master}.
\end{itemize}

\paragraph{Historical context.}  The models in this chapter build on a long tradition: \citet{bewley1986stationary} introduced precautionary savings with borrowing constraints; \citet{huggett1993riskfree} studied endowment economies with incomplete markets; \citet{aiyagari1994uninsured} added production and general equilibrium; and \citet{krusell1998income} incorporated aggregate uncertainty.  \citet{achdou2022income} reformulated these models in continuous time and demonstrated that finite-difference methods can solve the coupled HJB--KFE system efficiently.  More recently, \citet{gu2024masterequations} introduced EMINNs to solve the master equation globally, enabling treatment of aggregate shocks without moment-based approximations.

\section{Stochastic Calculus Refresher}
\label{sec:stoch_calc}

We briefly review the stochastic calculus tools needed for continuous-time models; for a standard finance-oriented textbook treatment, see \citet{shreve2004stochasticii}.

\paragraph{Quick reference.}  Appendix~\ref{app:stoch} contains a one-page summary of the same material (Brownian motion, It\^o's lemma, ergodicity in one paragraph) for readers who want a compact reminder rather than the full exposition below.

\subsection{Brownian Motion}

\begin{definitionbox}[Standard Brownian Motion]
A stochastic process $\{B_t\}_{t \geq 0}$ is a \emph{standard Brownian motion} (Wiener process) if: (i) $B_0 = 0$; (ii) it has independent increments: $B_t - B_s \perp B_s - B_r$ for $r < s < t$; (iii) increments are Gaussian: $B_t - B_s \sim \mathcal{N}(0, t-s)$; and (iv) paths $t \mapsto B_t$ are continuous almost surely.
\end{definitionbox}

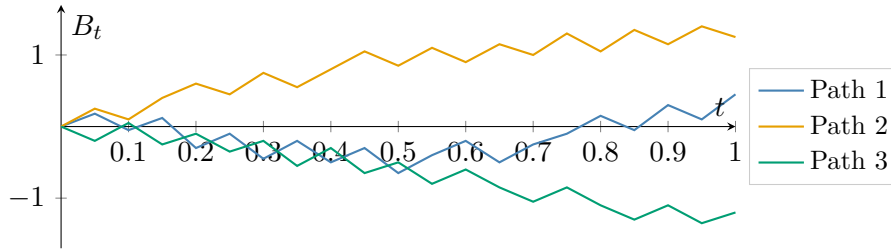
\begin{figure}[ht]
\centering
\begin{tikzpicture}
\begin{axis}[
    width=10.5cm, height=4.8cm,
    xlabel={$t$}, ylabel={$B_t$},
    xmin=0, xmax=1, ymin=-1.7, ymax=1.7,
    axis lines=middle,
    legend style={
        at={(1.02,0.5)}, anchor=west, font=\small,
        draw=gray!40, fill=white, fill opacity=0.95, text opacity=1,
        cells={anchor=west}, row sep=1pt,
    },
    every axis plot/.append style={thin, mark=none},
]
\addplot[softblue, thick] coordinates {
(0,0) (0.05,0.18) (0.10,-0.05) (0.15,0.12) (0.20,-0.30) (0.25,-0.10)
(0.30,-0.45) (0.35,-0.20) (0.40,-0.50) (0.45,-0.30) (0.50,-0.65)
(0.55,-0.40) (0.60,-0.20) (0.65,-0.50) (0.70,-0.25) (0.75,-0.10)
(0.80,0.15) (0.85,-0.05) (0.90,0.30) (0.95,0.10) (1.00,0.45)};
\addlegendentry{Path 1}
\addplot[softorange, thick] coordinates {
(0,0) (0.05,0.25) (0.10,0.10) (0.15,0.40) (0.20,0.60) (0.25,0.45)
(0.30,0.75) (0.35,0.55) (0.40,0.80) (0.45,1.05) (0.50,0.85)
(0.55,1.10) (0.60,0.90) (0.65,1.15) (0.70,1.00) (0.75,1.30)
(0.80,1.05) (0.85,1.35) (0.90,1.15) (0.95,1.40) (1.00,1.25)};
\addlegendentry{Path 2}
\addplot[softgreen, thick] coordinates {
(0,0) (0.05,-0.20) (0.10,0.05) (0.15,-0.25) (0.20,-0.10) (0.25,-0.35)
(0.30,-0.20) (0.35,-0.55) (0.40,-0.30) (0.45,-0.65) (0.50,-0.50)
(0.55,-0.80) (0.60,-0.60) (0.65,-0.85) (0.70,-1.05) (0.75,-0.85)
(0.80,-1.10) (0.85,-1.30) (0.90,-1.10) (0.95,-1.35) (1.00,-1.20)};
\addlegendentry{Path 3}
\end{axis}
\end{tikzpicture}
\caption{Three simulated standard Brownian sample paths $\{B_t\}_{t\in[0,1]}$, generated with discretization step $\Delta t = 0.05$ and Gaussian increments $B_{t+\Delta t} = B_t + \sqrt{\Delta t}\,\varepsilon_t$, $\varepsilon_t \sim \mathcal{N}(0,1)$. Paths are jagged at the chosen $\Delta t$; limiting Brownian paths are continuous almost surely but nowhere differentiable, with $\mathrm{Var}(B_t) = t$.}
\label{fig:brownian_paths}
\end{figure}

Key properties include $\E{B_t} = 0$, $\mathrm{Var}(B_t) = t$, nowhere-differentiable paths, and quadratic variation $\langle B \rangle_t = t$.  Brownian motion arises as the scaling limit of a random walk: if $X_{t+\Delta t} = X_t + \sqrt{\Delta t}\,\varepsilon_t$ with $\varepsilon_t \in \{-1,+1\}$ equiprobably, then $X^{\Delta t} \xrightarrow{d} B_t$ as $\Delta t \to 0$ (Donsker's theorem).  The $\sqrt{\Delta t}$ scaling ensures that $\mathrm{Var}(X_t) = t$ in the limit.  Figure~\ref{fig:brownian_paths} shows three discretized sample paths with the same variance scaling.

\subsection{It\^o Processes and SDEs}

An \emph{It\^o process} $X_t$ satisfies the stochastic differential equation (SDE):
\begin{equation}
dX_t = \mu(X_t, t)\,dt + \sigma(X_t, t)\,dB_t,
\label{eq:ito_sde}
\end{equation}
where $\mu(\cdot)$ is the \emph{drift} (deterministic trend) and $\sigma(\cdot)$ is the \emph{diffusion coefficient} (volatility).  Readers unfamiliar with It\^o calculus will benefit from \citet{shreve2004stochasticii} as a prerequisite; the key fact $(dB_t)^2 = dt$ is what forces the second-order term in It\^o's lemma below.  Two important special cases recur throughout this chapter:

\begin{itemize}[itemsep=2pt]
\item \textbf{Geometric Brownian motion (GBM):} $dS_t = \mu S_t\,dt + \sigma S_t\,dB_t$ (stock prices, GDP).
\item \textbf{Ornstein--Uhlenbeck (OU) process:} $dZ_t = \eta(\bar{Z} - Z_t)\,dt + \sigma\,dB_t$ (productivity, interest rates).  The OU process is mean-reverting with stationary distribution $Z_\infty \sim \mathcal{N}(\bar{Z}, \sigma^2/(2\eta))$.  It will model aggregate TFP in the Krusell--Smith economy below.
\end{itemize}

For discrete income switching, we use a \emph{continuous-time Markov chain}: a labor state $n_t \in \{n_1, n_2\}$ that switches with Poisson intensities $\lambda_1$ (from $n_1$ to $n_2$) and $\lambda_2$ (from $n_2$ to $n_1$), yielding ergodic probabilities $\pi_1 = \lambda_2/(\lambda_1 + \lambda_2)$ and $\pi_2 = \lambda_1/(\lambda_1 + \lambda_2)$.

\subsection{It\^o's Lemma}

\begin{definitionbox}[It\^o's Lemma]
Let $X_t$ follow $dX_t = \mu\,dt + \sigma\,dB_t$ and let $f \in C^2$.  Then:
\begin{equation}
df(X_t) = \left[f'(X_t)\,\mu + \tfrac{1}{2}f''(X_t)\,\sigma^2\right]dt + f'(X_t)\,\sigma\,dB_t.
\label{eq:ito_lemma}
\end{equation}
\end{definitionbox}

The key difference from ordinary calculus is the second-order correction $\tfrac{1}{2}f''\sigma^2\,dt$, which arises because $(dB_t)^2 = dt \neq 0$.  The differential algebra rules are: $dt \cdot dt = 0$, $dt \cdot dB_t = 0$, $dB_t \cdot dB_t = dt$.

\paragraph{Worked example.}  For GBM $dX_t = \mu X_t\,dt + \sigma X_t\,dB_t$, applying It\^o's lemma to $f(x) = \ln x$ gives $d(\ln X_t) = (\mu - \tfrac{1}{2}\sigma^2)\,dt + \sigma\,dB_t$, so $X_t = X_0\exp[(\mu - \tfrac{1}{2}\sigma^2)t + \sigma B_t]$.  The $-\tfrac{1}{2}\sigma^2$ It\^o correction explains why the expected log return differs from the expected return for volatile assets.

\paragraph{Time-dependent version.}  For $V(X_t, t)$ where $dX_t = \mu\,dt + \sigma\,dB_t$:
\begin{equation}
dV = \left[\partial_t V + \mu\,\partial_x V + \tfrac{1}{2}\sigma^2\,\partial_{xx} V\right]dt + \sigma\,\partial_x V\,dB_t.
\label{eq:ito_multi}
\end{equation}
With Poisson jumps of intensity $\lambda$ and jump size $\Delta X$, an additional term $[V(X_{t^-} + \Delta X, t) - V(X_{t^-}, t)]\,dN_t$ appears.

\section{The Kolmogorov Forward Equation}
\label{sec:kfe_theory}

The \emph{Kolmogorov forward equation} (KFE), also known as the \emph{Fokker--Planck equation}, describes how the probability density of a stochastic process evolves over time.  It was introduced by Kolmogorov (1931) and, independently in the physics literature, by Fokker (1914) and Planck (1917) to describe diffusion of particles.  In stationary equilibrium, $\partial_t g = 0$ and the KFE reduces to an elliptic PDE for the cross-sectional density (used throughout \S\ref{sec:ct_equilibrium}--\S\ref{sec:ct_pinn}); when aggregate shocks make prices time-varying, the parabolic time-dependent form returns and motivates the master-equation reformulation of \S\ref{sec:master_eq}.

\subsection{Derivation from First Principles}

Consider a population of particles, each independently following $dX_t = \mu\,dt + \sigma\,dB_t$ with constant coefficients.  If the initial density is $g_0(x)$, what is the density $g(x,t)$ at time $t > 0$?

The derivation proceeds in four steps.  (i) For any smooth test function $\varphi(x)$, $\E{\varphi(X_t)} = \int \varphi(x)\,g(x,t)\,dx$.  (ii) Differentiate with respect to $t$ and apply It\^o's lemma: $\tfrac{d}{dt}\int \varphi\,g\,dx = \int [\mu\,\varphi' + \tfrac{\sigma^2}{2}\varphi'']\,g\,dx$.  (iii) Integrate by parts.  Take $\varphi$ to be \emph{compactly supported}: this kills the boundary terms cleanly and costs nothing here, since the identity must hold for every such $\varphi$.  We obtain $\int \mu\,\varphi'\,g\,dx = -\int \varphi\,\mu\,\partial_x g\,dx$ and $\int \tfrac{\sigma^2}{2}\varphi''\,g\,dx = \int \varphi\,\tfrac{\sigma^2}{2}\partial_{xx}g\,dx$.  (The natural-looking assumption $g \to 0$ at $\pm\infty$ is automatic for any continuous density, but it is not by itself enough: what is needed is that the product $g\,\varphi$ vanishes at infinity, which compact support of $\varphi$ delivers for free.  Spatial boundaries, e.g.\ a borrowing constraint, require separate treatment in the next subsection.)  (iv) Since the identity holds for all test functions $\varphi$, we obtain:

\begin{equation}
\boxed{\frac{\partial g}{\partial t}(x,t) = -\mu\frac{\partial g}{\partial x}(x,t) + \frac{\sigma^2}{2}\frac{\partial^2 g}{\partial x^2}(x,t).}
\label{eq:kfe_const}
\end{equation}

The two terms on the right encode competing effects: $-\mu\,\partial_x g$ is \emph{advection} (drift transports the density), and $\tfrac{\sigma^2}{2}\partial_{xx}g$ is \emph{diffusion} (noise spreads the density).

\paragraph{General form.}  For state-dependent coefficients $dX_t = \mu(X_t)\,dt + \sigma(X_t)\,dB_t$, the KFE in \emph{divergence form} is:
\begin{equation}
\partial_t g = -\partial_x\!\left[\mu(x)\,g(x,t)\right] + \tfrac{1}{2}\partial_{xx}\!\left[\sigma(x)^2\,g(x,t)\right] = -\partial_x J(x,t),
\label{eq:kfe_general}
\end{equation}
where $J = \mu g - \tfrac{1}{2}\partial_x(\sigma^2 g)$ is the probability flux.  The identity $\partial_t g + \partial_x J = 0$ is a conservation law: probability is neither created nor destroyed.  When $\sigma$ depends on $x$, the diffusion coefficient must be written \emph{inside} the second derivative; in particular, $\tfrac{1}{2}\partial_{xx}[\sigma(x)^2\,g] \neq \tfrac{\sigma(x)^2}{2}\partial_{xx}g$, with the two expressions differing by $(\sigma^2)'\,\partial_x g + \tfrac{1}{2}(\sigma^2)''\,g$.  This distinction is invisible in the constant-coefficient case but is what keeps the operator self-adjoint and probability-conserving when $\sigma$ varies.

\subsection{Examples}

\paragraph{Heat equation.}  With $\mu = 0$ and constant $\sigma$, the KFE reduces to $\partial_t g = \tfrac{\sigma^2}{2}\partial_{xx}g$, the classical heat equation.  Starting from a point mass $g(x,0) = \delta(x-x_0)$, the solution is a Gaussian with variance growing linearly: $g(x,t) = (2\pi\sigma^2 t)^{-1/2}\exp[-(x-x_0)^2/(2\sigma^2 t)]$.

\paragraph{Ornstein--Uhlenbeck process.}  For $dX_t = \eta(\bar{x} - X_t)\,dt + \sigma\,dB_t$, the KFE is $\partial_t g = \eta\,\partial_x[(x - \bar{x})\,g] + \tfrac{\sigma^2}{2}\partial_{xx}g$.  Setting $\partial_t g = 0$ and trying a Gaussian ansatz $g^\star(x) = (2\pi s^2)^{-1/2}\exp\!\bigl[-(x-\bar x)^2/(2s^2)\bigr]$, the advection and diffusion terms balance when $s^2 = \sigma^2/(2\eta)$, giving the stationary distribution $g^*(x) = \mathcal{N}(\bar{x}, \sigma^2/(2\eta))$: mean reversion concentrates mass around $\bar{x}$, while diffusion spreads it, and the balance produces a Gaussian steady state.

\paragraph{Reading the SDE qualitatively.}  It is worth pausing to read the drift sign by eye, because students are often trained to compute with differential equations but rarely to look at them.  When $X_t < \bar{x}$, the drift $\eta(\bar{x} - X_t)$ is positive and pushes $X_t$ upward; when $X_t > \bar{x}$, the drift is negative and pushes $X_t$ downward, hence the name \emph{mean-reverting}.  Flipping the sign ($\eta < 0$) gives a \emph{mean-repelling}, unstable process that drifts away from $\bar{x}$ without bound.  Closely related cubic SDEs make the same point sharply.  The process $dX_t = -X_t^3\,dt + \sigma\,dB_t$ uses the cubic well as a restoring force and is well-posed with a stationary distribution, while $dX_t = +X_t^3\,dt + \sigma\,dB_t$ blows up in finite time, and a naive Euler discretization produces \texttt{NaN}s almost immediately.  Throughout this chapter the same qualitative reading is what carries intuition over from SDEs to KFEs and HJBs.

\subsection{From Physics to Economics: A Continuum of Agents}

Consider a continuum of agents $i \in [0,1]$, each independently following $dX_t^i = \mu(X_t^i)\,dt + \sigma(X_t^i)\,dB_t^i$ with independent idiosyncratic Brownian motions $B_t^i$.  By the law of large numbers at the population level, the cross-sectional density $g(x,t)$ evolves \emph{deterministically}, even though each individual path is random.  The density satisfies the KFE~\eqref{eq:kfe_general}.

\begin{remarkbox}[Physical analogy]
Each molecule of air moves randomly (Brownian motion), but the temperature distribution in a room evolves deterministically (heat equation).  In economics: each household's income is random, but the wealth distribution evolves deterministically (KFE).
\end{remarkbox}

\paragraph{KFE with income switching.}  When agent $i$ has wealth $a_t^i$ and income state $n_t^i \in \{n_1, n_2\}$ switching with Poisson intensities, and wealth evolves as $da_t^i = s(a_t^i, n_t^i)\,dt$ with $a_t^i \geq \underline{a}$, the cross-sectional density $g_t(a,n)$ satisfies:
\begin{equation}
\partial_t g_t(a,n) = -\partial_a\!\left[s(a,n)\,g_t(a,n)\right] - \lambda(n)\,g_t(a,n) + \lambda(\hat{n})\,g_t(a,\hat{n}),
\label{eq:kfe_econ}
\end{equation}
where $\hat{n}$ denotes the complementary income state.  The three terms represent: flow of agents along the wealth axis (savings), agents leaving state $n$, and agents entering state $n$ from $\hat{n}$.

\paragraph{Boundaries and mass points.}  At the borrowing constraint $a = \underline{a}$, the no-flux boundary condition $s(\underline{a},n)\,g_t(\underline{a},n) = 0$ prevents the absolutely continuous density from flowing below $\underline{a}$.  If households are constrained at this boundary, a boundary atom may be needed in addition to the interior density; finite-difference implementations represent that atom as mass in the first grid cell.  This is the continuous-time counterpart of the characteristic left spike observed in empirical wealth distributions.

\paragraph{When does the KFE become stochastic?}  With purely idiosyncratic shocks (Aiyagari), the KFE is deterministic.  When \emph{aggregate shocks} are present (Krusell--Smith), prices $r_t, w_t$ depend on the aggregate state $Z_t$, making the drift stochastic and the density $g_t$ a stochastic process adapted to the filtration $\mathcal{F}_t^0$ generated by the aggregate Brownian motion.  This is the ``master equation'' setting discussed in Section~\ref{sec:master_eq}.

\begin{remarkbox}[KFE validation: more than the pointwise PDE residual]
A small pointwise KFE residual $|R_{\text{KFE}}|$ is a necessary check, but pointwise differentiation of a noisy density $g_\phi$ is fragile and the residual can sit near zero while the implied density is visibly wrong.  Notebook \texttt{08\_Aiyagari\_Continuous\_Time\_FD\_and\_PINN\_PyTorch} therefore monitors a small panel of complementary diagnostics: (i)~total mass $\sum_n\int g_\phi\,da \to 1$, (ii)~no-flux boundary $J(\underline a, n), J(\bar a, n) \to 0$, (iii)~aggregate-capital error $|K^{\text{PINN}} - K^{\text{FD}}|$, (iv)~CDF distance (Kolmogorov--Smirnov) per income state, and (v)~an integrated flux-balance residual that replaces $\partial_a$ with quadrature.  In practice, mass and aggregate $K$ converge first; the local KFE residual and the KS distance close last.  Reporting all five together makes PINN quality transparent and avoids reading a small pointwise residual as a guarantee of a small density error.
\end{remarkbox}

\section{The Hamilton--Jacobi--Bellman Equation}
\label{sec:hjb_theory}

Individual optimality in continuous-time heterogeneous-agent models is characterized by the HJB equation.  This section gives the full five-step derivation from Itô's lemma; readers who saw the motivating overview of the continuous-time heterogeneous-agent setting in \S\ref{sec:ct_hank} can treat the present section as its formal counterpart.

\paragraph{The agent's problem.}  Agent $i$ with state $x_t^i = (a_t^i, n_t^i)$ chooses consumption $c_t^i$ to maximize:
\begin{equation}
\max_{\{c_t^i\}} \;\E{\int_0^\infty e^{-\rho t}\,u(c_t^i)\,dt}
\label{eq:hjb_agent}
\end{equation}
subject to $da_t^i = (wn_t^i + ra_t^i - c_t^i)\,dt$ and $a_t^i \geq \underline{a}$, where $u(c) = c^{1-\gamma}/(1-\gamma)$ is CRRA utility, $\rho > 0$ is the discount rate, and $(r,w)$ are factor prices.  The standard transversality condition $\lim_{t\to\infty} e^{-\rho t}\,V(a_t^i, n_t^i) = 0$ along optimal paths, the continuous-time no-Ponzi requirement, is imposed throughout.

\paragraph{The value function.}  The value function $V(a,n)$ records the maximum expected discounted utility starting from state $(a,n)$.  It is increasing and concave in $a$, and $V(a,n_2) > V(a,n_1)$ when $n_2 > n_1$.

\paragraph{Deriving the HJB.}  The derivation follows five steps; each unpacks one ingredient of equation~\eqref{eq:hjb_full} below.

\smallskip\noindent\emph{Step (i): Dynamic programming principle.}  For small $h > 0$,
\[
   V(a,n) \;=\; \max_c\Big\{\int_0^h e^{-\rho t}\,u(c_t)\,dt \;+\; e^{-\rho h}\,\E{V(a_h, n_h)}\Big\}.
\]
This is Bellman's principle of optimality applied to a continuous-time problem.

\smallskip\noindent\emph{Step (ii): It\^o's lemma between income jumps.}  Conditional on the income state $n_t$ not changing on $[0, h]$, wealth follows the \emph{deterministic} ODE $da_t = s(c_t, a_t, n_t)\,dt$ (there is no Brownian forcing on wealth in this model), so
\[
   dV = V'(a,n)\,s(c,a,n)\,dt.
\]
The second-order It\^o correction $\tfrac{1}{2}V''\sigma^2\,dt$ vanishes because the wealth diffusion is zero between jumps; this is the most subtle step and is what distinguishes the income-switching HJB from a standard diffusion HJB.

\smallskip\noindent\emph{Step (iii): Account for Poisson jumps in expectation.}  Adding the Poisson-jump contribution and taking expectations,
\[
   \E{dV} = \Big[V'(a,n)\,s + \lambda(n)\bigl(V(a,\hat{n}) - V(a,n)\bigr)\Big]\,dt,
\]
where $\lambda(n)$ is the intensity of switching out of state $n$ and $\hat{n}$ is the complementary state.

\smallskip\noindent\emph{Step (iv): Substitute into the DPP and let $h \to 0$.}  Plugging the expectation back into the Bellman expression, dividing by $h$, and taking $h \to 0$ yields a flow equation in which the discount $\rho V$ on the left balances the flow utility plus expected change on the right.

\smallskip\noindent\emph{Step (v): Optimize over $c$.}  Imposing the first-order condition over consumption gives the HJB equation:
\begin{equation}
\boxed{\rho V(a,n) = \max_c\left\{u(c) + V'(a,n)\cdot(wn + ra - c) + \lambda(n)\!\left(V(a,\hat{n}) - V(a,n)\right)\right\}.}
\label{eq:hjb_full}
\end{equation}

\paragraph{Interpretation.}  The HJB is an \emph{asset pricing equation}: the left side $\rho V$ is the ``required return'' on the value function (discount rate times ``asset value''), and the right side is the ``total return'' consisting of the flow dividend $u(c)$, the capital gain $V' \cdot s$ from saving, and the expected gain/loss $\lambda(n)(\Delta V)$ from income switching.

\paragraph{Optimal policy.}  The first-order condition $u'(c^*) = V'(a,n)$ gives the consumption function:
\begin{equation}
c^*(a,n) = \left[V'(a,n)\right]^{-1/\gamma}.
\label{eq:hjb_foc}
\end{equation}
Substituting back eliminates the maximization, yielding a nonlinear PDE in $V$.  The savings function is $s(a,n) = wn + ra - c^*(a,n)$.

\paragraph{Boundary conditions.}  At the borrowing constraint $a = \underline{a}$, consumption must keep the drift feasible: $s(\underline{a},n)=wn+r\underline{a}-c \geq 0$, or equivalently $c \leq wn+r\underline{a}$.  The boundary HJB is therefore the constrained maximization
\[
\begin{aligned}
\rho V(\underline{a},n)
&=\max_{0<c\leq wn+r\underline{a}}
\Big\{u(c)+V_a(\underline{a},n)(wn+r\underline{a}-c)\\
&\qquad\qquad\qquad\qquad
+\lambda(n)\bigl(V(\underline{a},\hat n)-V(\underline{a},n)\bigr)\Big\}.
\end{aligned}
\]
For CRRA utility this gives the boundary policy $c^*(\underline{a},n)=\min\{[V_a(\underline{a},n)]^{-1/\gamma},\,wn+r\underline{a}\}$ and $s(\underline{a},n)\geq 0$.  This is the state-constraint form of the borrowing limit; numerical solvers usually impose it with one-sided derivatives, a constrained policy rule, or a penalty on negative boundary drift.

\paragraph{Boundary atoms in the stationary distribution.}
When the borrowing constraint binds on a positive mass of agents, the stationary measure is not absolutely continuous with respect to Lebesgue measure: it carries a Dirac atom at $a=\underline a$.  The decomposition is $g(a,n) = g_{\mathrm{ac}}(a,n) + \alpha(n)\,\delta(a-\underline a)$, where $g_{\mathrm{ac}}$ is the absolutely-continuous interior density and $\alpha(n)\geq 0$ is the constrained mass for income state $n$.  The KFE~\eqref{eq:kfe_econ} as written governs only the interior part $g_{\mathrm{ac}}$; the atomic mass $\alpha(n)$ is determined by a separate flux-balance equation that equates inflows from the no-flux boundary condition with the income-driven outflow back into the interior.  Finite-difference implementations typically represent $\alpha(n)$ as the mass in the first grid cell, and PINN implementations either absorb the atom implicitly into a smooth density approximation (with corresponding accuracy loss near $\underline a$) or explicitly parameterize $\alpha(n)$ alongside the interior network.

\paragraph{Variant: continuous (diffusion) income.}
A natural variant replaces the two-state Poisson process with a continuously distributed earnings state $z_t$ following an Ornstein--Uhlenbeck diffusion, $dz_t = \eta(\bar z - z_t)\,dt + \sigma\,dB_t^z$ (with idiosyncratic Brownian motion $B_t^z$).  The agent's state is then $(a,z)$, the value function $V(a,z)$ is smooth in both arguments, and It\^o's lemma along the $z$-diffusion produces a genuine second-order term.  After substituting the first-order condition $c^\star=(V_a)^{-1/\gamma}$ the HJB becomes the elliptic PDE
\begin{equation}
\rho V(a,z) = u(c^\star) + V_a\,(wz + ra - c^\star) + \eta(\bar z - z)\,V_z + \tfrac{1}{2}\sigma^2\,V_{zz},
\label{eq:hjb_diffusion}
\end{equation}
with the borrowing-constraint state condition $s(\underline a,z)\geq 0$ at $a=\underline a$, a Neumann (FOC) condition $V_a(\bar a,z)=u'(w z + r\bar a)$ on the truncated upper wealth boundary, and reflecting conditions $V_z=0$ at the $z$-boundaries.  This is the smallest model in the chapter that carries the diffusion second-order term $\tfrac{1}{2}\sigma^2 V_{zz}$ in the \emph{individual} HJB: the income-switching HJB~\eqref{eq:hjb_full} has none, because wealth carries no Brownian forcing between jumps, while the master equation~\eqref{eq:master_eq} carries $\tfrac{1}{2}(\sigma^z)^2 V_{zz}$ only for the \emph{aggregate} TFP state $z$.

\section{Competitive Equilibrium: Huggett and Aiyagari}
\label{sec:ct_equilibrium}

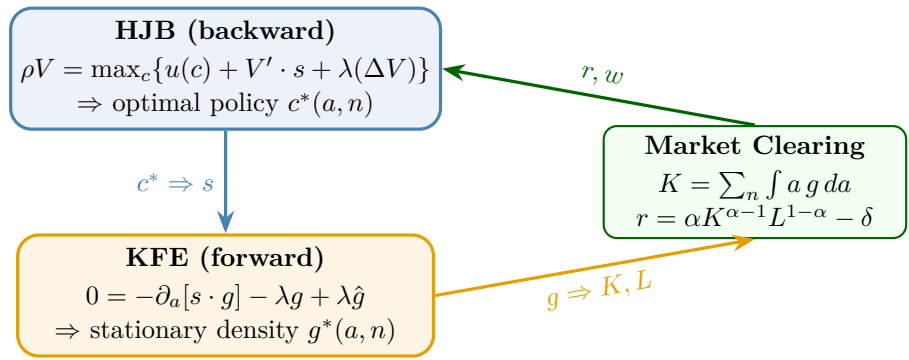
\begin{figure}[ht]
\centering
\begin{tikzpicture}[
    box/.style={rectangle, rounded corners=8pt, draw, very thick, minimum width=5.5cm, minimum height=1.6cm, align=center, font=\small},
    pricebox/.style={rectangle, rounded corners=5pt, draw=darkgreen, thick, fill=green!5, minimum width=4cm, minimum height=1cm, align=center, font=\small},
    arrow/.style={-{Stealth[length=3mm]}, very thick}
]
\node[box, fill=softblue!10, draw=softblue] (hjb) at (0,2.2) {\textbf{HJB (backward)}\\[2pt]
    $\rho V = \max_c\{u(c) + V'\cdot s + \lambda(\Delta V)\}$\\[1pt]
    $\Rightarrow$ optimal policy $c^*(a,n)$};
\node[box, fill=softorange!10, draw=softorange] (kfe) at (0,-0.8) {\textbf{KFE (forward)}\\[2pt]
    $0 = -\partial_a[s\cdot g] - \lambda g + \lambda \hat{g}$\\[1pt]
    $\Rightarrow$ stationary density $g^*(a,n)$};
\node[pricebox] (mc) at (7,0.7) {\textbf{Market Clearing}\\[2pt]
    $K = \sum_n\int a\,g\,da$\\[1pt]
    $r = \alpha K^{\alpha-1}L^{1-\alpha} - \delta$};
\draw[arrow, softblue] (hjb.south) -- node[left, font=\small] {$c^* \Rightarrow s$} (kfe.north);
\draw[arrow, softorange] (kfe.east) -- node[below, font=\small, sloped] {$g \Rightarrow K, L$} (mc.south);
\draw[arrow, darkgreen] (mc.north) -- node[above, font=\small, sloped] {$r, w$} (hjb.east);
\end{tikzpicture}
\caption{Stationary continuous-time heterogeneous-agent equilibrium as a coupled HJB--KFE--market-clearing loop.  Given prices, the HJB determines optimal savings; the KFE maps that policy into a stationary density; aggregating the density updates capital, labor, and therefore prices.}
\label{fig:hjb_kfe_market_loop}
\end{figure}

The HJB gives individual optimal behavior for given prices; the KFE tracks the resulting distribution.  Market clearing pins down prices, closing the system.  Figure~\ref{fig:hjb_kfe_market_loop} summarizes this fixed-point structure.

\subsection{The Coupled HJB--KFE System}

The solution method for the stationary equilibrium (no aggregate shocks) iterates: guess $r$ $\to$ solve HJB for $V, c^*$ $\to$ solve KFE for $g^*$ $\to$ compute $K = \sum_n \int a\,g^*\,da$ $\to$ update $r$ from the firm FOC.  Under standard regularity conditions on preferences, technology, and the income process, aggregate capital supply $K^s(r)$ is monotone decreasing in $r$ over the relevant range, which makes the bisection (or fixed-point iteration) on $r$ well-posed; see \citet[\S 2]{achdou2022income} for the full statement.

\subsection{Huggett (1993): Endowment Economy}

Huggett's model is an endowment economy with idiosyncratic income $y_t \in \{y_1, y_2\}$, a single risk-free bond $b_t \geq \underline{b}$ paying instantaneous return $r$, and bonds in zero net supply.  The HJB is $\rho V = \max_c\{u(c) + V_b(rb + y - c) + \lambda(y)(\Delta V)\}$, and the KFE determines the stationary density $g^*(b,y)$.  An equilibrium return $r^*$ is a value satisfying the asset-market-clearing condition $\sum_y \int b\,g^*(b,y)\,db = 0$; uniqueness requires standard monotonicity assumptions on aggregate bond demand \citep{achdou2022income}.

The mechanism is that the risk-free rate is pinned down by \emph{precautionary self-insurance demand}, not by a production FOC: agents desire to hold positive bond positions for precautionary reasons, and the return must adjust downward until bond demand equals zero net supply.

\subsection{Aiyagari (1994): Production Economy}

Aiyagari's model adds a representative firm with Cobb--Douglas production $Y = K^\alpha L^{1-\alpha}$.  The household asset is now a claim on aggregate capital, and the firm's FOCs yield $r = \alpha K^{\alpha-1}L^{1-\alpha} - \delta$ and $w = (1-\alpha)K^\alpha L^{-\alpha}$.  The equilibrium condition is $K^s(r) = K^d(r)$, where $K^s(r) = \sum_z \int a\,g^*(a,z)\,da$ is capital supplied by households (via HJB + KFE) and $K^d(r)$ is capital demanded by the firm (inverse of the firm FOC).  Table~\ref{tab:huggett_aiyagari_ct} highlights the key distinction between the endowment and production versions.

\paragraph{What changes from Huggett to Aiyagari?}  Both models share the same household problem (HJB) and the same cross-sectional law of motion (KFE).  What differs is the equilibrium concept.  In Huggett the equilibrium is the single price $r^\star$ that clears a zero-net-supply bond market, $\sum_y\int b\,g^\star(b,y)\,db = 0$; the rate is pinned down by precautionary self-insurance demand alone, with no production side.  In Aiyagari the equilibrium is a fixed point in $r$ that matches household supply $K^s(r)$ to firm demand $K^d(r) = (\alpha/(r+\delta))^{1/(1-\alpha)} L$, with prices determined jointly by household savings and the firm FOCs.  Numerically, both reduce to a one-dimensional root-finding problem in $r$, but the economic mechanism (precautionary saving vs.\ marginal-product-of-capital pinning) is qualitatively different.

\paragraph{Adding aggregate TFP.}  When aggregate TFP $Z_t$ is allowed to vary (e.g., the OU process introduced in \S\ref{sec:master_eq} below for Krusell--Smith), the firm FOCs generalize to $r_t = \alpha\, e^{Z_t} K_t^{\alpha-1}L_t^{1-\alpha} - \delta$ and $w_t = (1-\alpha)\, e^{Z_t} K_t^\alpha L_t^{-\alpha}$.  Aiyagari is recovered when $Z_t \equiv 0$; the same expression covers both calibrations, which is convenient for the master-equation analysis below.

\begin{table}[ht]
\centering
\small
\begin{tabular}{lcc}
\toprule
& \textbf{Huggett (1993)} & \textbf{Aiyagari (1994)} \\
\midrule
Economy & Endowment & Production ($Y = K^\alpha L^{1-\alpha}$) \\
Asset & Bond $b$ & Capital claim $a$ \\
Net supply & Zero ($\int b\,g = 0$) & Positive ($\int a\,g = K > 0$) \\
Price determined by & Bond return $q$ & Interest rate $r$, wage $w$ \\
Wealth distribution & Centered around $0$, mass at $\underline{b}$ & Right-skewed, long right tail \\
\bottomrule
\end{tabular}
\caption{Huggett and Aiyagari as two continuous-time incomplete-markets benchmarks.  Huggett clears a zero-net-supply bond market by adjusting the bond return; Aiyagari clears a positive capital market with prices pinned down by firm first-order conditions.}
\label{tab:huggett_aiyagari_ct}
\end{table}

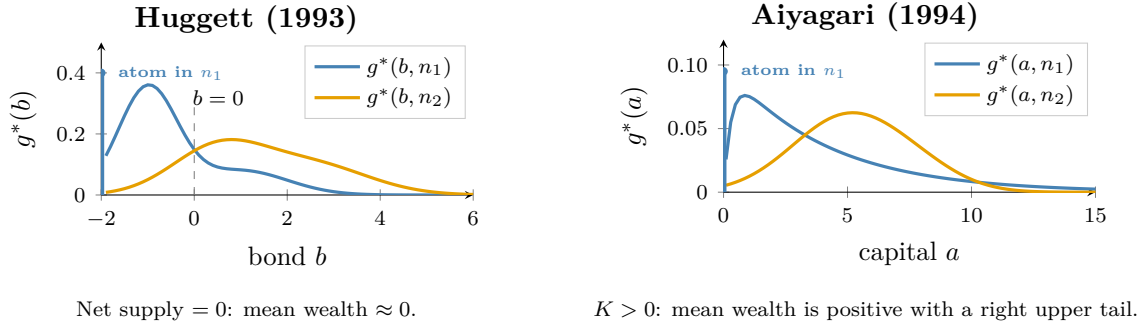
\begin{figure}[ht]
\centering
\begin{minipage}[b]{0.48\textwidth}
\centering
\textbf{Huggett (1993)}\\[2pt]
\begin{tikzpicture}
\begin{axis}[
    width=6.5cm, height=3.6cm,
    xlabel={bond $b$}, ylabel={$g^*(b)$},
    xmin=-2, xmax=6, ymin=0, ymax=0.5,
    axis lines=left,
    tick label style={font=\scriptsize},
    label style={font=\small},
    legend style={at={(0.98,0.98)}, anchor=north east, font=\scriptsize,
                  fill=white, fill opacity=0.9, draw=gray!40, text opacity=1},
]
\addplot[softblue, very thick, domain=-1.9:6, samples=80]
    {0.35*exp(-(x+1)^2/0.8) + 0.08*exp(-(x-1)^2/2)};
\addlegendentry{$g^*(b, n_1)$}
\addplot[softorange, very thick, domain=-1.9:6, samples=80]
    {0.15*exp(-(x-0.5)^2/2) + 0.1*exp(-(x-2.5)^2/3)};
\addlegendentry{$g^*(b, n_2)$}
\draw[softblue, line width=2.2pt] (axis cs:-2,0) -- (axis cs:-2,0.40);
\fill[softblue] (axis cs:-2,0.40) circle (1.6pt);
\node[softblue, font=\tiny\bfseries, anchor=west]
    at (axis cs:-1.85,0.40) {atom in $n_1$};
\draw[dashed, gray] (axis cs:0,0) -- (axis cs:0,0.3);
\node[font=\scriptsize] at (axis cs:0.52,0.32) {$b=0$};
\end{axis}
\end{tikzpicture}\\[2pt]
{\scriptsize Net supply $=0$: mean wealth $\approx 0$.}
\end{minipage}%
\hfill
\begin{minipage}[b]{0.48\textwidth}
\centering
\textbf{Aiyagari (1994)}\\[2pt]
\begin{tikzpicture}
\begin{axis}[
    width=6.5cm, height=3.6cm,
    xlabel={capital $a$}, ylabel={$g^*(a)$},
    xmin=0, xmax=15, ymin=0, ymax=0.12,
    axis lines=left,
    scaled y ticks=false,
    ytick={0,0.05,0.10},
    yticklabels={0,0.05,0.10},
    tick label style={font=\scriptsize},
    label style={font=\small},
    legend style={at={(0.98,0.98)}, anchor=north east, font=\scriptsize,
                  fill=white, fill opacity=0.9, draw=gray!40, text opacity=1},
]
\addplot[softblue, very thick, domain=0.1:15, samples=80]
    {0.09*exp(-0.25*(x-0.5)) * (1 - exp(-3*x))};
\addlegendentry{$g^*(a, n_1)$}
\addplot[softorange, very thick, domain=0.1:15, samples=80]
    {0.06*exp(-(x-5)^2/10) + 0.01*exp(-(x-8)^2/6)};
\addlegendentry{$g^*(a, n_2)$}
\draw[softblue, line width=2.2pt] (axis cs:0,0) -- (axis cs:0,0.095);
\fill[softblue] (axis cs:0,0.095) circle (1.6pt);
\node[softblue, font=\tiny\bfseries, anchor=west]
    at (axis cs:0.3,0.095) {atom in $n_1$};
\end{axis}
\end{tikzpicture}\\[2pt]
{\scriptsize $K>0$: mean wealth is positive with a right upper tail.}
\end{minipage}
\caption{Stationary cross-sectional densities $g^*$ in the two benchmarks, by productivity type $n\in\{n_1,n_2\}$ (low and high).  In both economies, only the constrained low-productivity type $n_1$ supports a Dirac atom at the borrowing constraint (blue spike): high-productivity households are not bound.  \emph{Left:} Huggett, bonds with limit $\underline{b}=-2$ and zero net supply, so the bulk of mass sits around $b=0$.  \emph{Right:} Aiyagari, capital with limit $\underline{a}=0$ and positive aggregate $K$, shifting the unconstrained mass to the right with a long upper tail.  The blue spikes visualize the Dirac atom $\alpha(n_1)\,\delta(a-\underline a)$ in the decomposition $g = g_{\mathrm{ac}} + \alpha(n)\,\delta(a-\underline a)$ introduced in the boundary-atom paragraph above.  These curves are schematic TikZ illustrations of the qualitative contrast (zero-net-supply bonds versus positive capital), not direct exports; the exact densities depend on calibration and boundary treatment.}
\label{fig:huggett_aiyagari_densities}
\end{figure}

Figure~\ref{fig:huggett_aiyagari_densities} contrasts the two stationary densities visually.  In Huggett (left), bonds are in zero net supply, so the cross-sectional density is centred near $b\!=\!0$, with a Dirac atom at the borrowing limit $\underline{b}$ carried entirely by constrained low-productivity households; high-productivity households are smoothly distributed and never bind.  In Aiyagari (right), agents hold positive capital in equilibrium, so the same atom now sits at $\underline{a}\!=\!0$ but the bulk of the mass is shifted right with a long upper tail.

\paragraph{Connection to HANK.}  These models are the building blocks of Heterogeneous Agent New Keynesian (HANK) models \citep{kaplan2018monetary}, which replace the representative agent in New Keynesian frameworks with an Aiyagari economy.  Adding nominal rigidities allows monetary policy to affect consumption \emph{heterogeneously}: agents near the borrowing constraint have high marginal propensities to consume and respond strongly to fiscal stimulus, while wealthy agents absorb shocks through savings.

\section{A PINN Solver for the Stationary Huggett--Aiyagari System}
\label{sec:ct_pinn}

The stationary equilibrium of \S\ref{sec:ct_equilibrium} couples five conditions: the HJB equation for the value function $V(a,z)$, the consumption first-order condition $c^\star(a,z) = \big(V'(a,z)\big)^{-1/\gamma}$, the KFE for the stationary density $g(a,z)$ with savings drift $s(a,z) = wz + ra - c^\star(a,z)$, the no-flux boundary condition $s(\underline a, z)\,g(\underline a, z) = 0$ at the borrowing constraint $\underline a$, mass normalization $\sum_z\int g(a,z)\,da = 1$, and the market-clearing condition that pins down prices.  The traditional solver iterates a fixed point over $r$ (guess $r$ $\to$ solve HJB for $V, c^\star$ $\to$ solve KFE for $g$ $\to$ aggregate capital $\to$ update $r$ from the firm FOC); a PINN instead trains all of them jointly.

A PINN implementation uses two networks, trained jointly:
\begin{itemize}[itemsep=2pt]
\item $\hat{V}_\theta(a,z)$: approximates the value function.  Its derivative $\hat{V}_a$ is computed via automatic differentiation, and the implied consumption policy is $\hat c^\star(a,z) = \big(\hat{V}_a(a,z)\big)^{-1/\gamma}$.
\item $\hat{g}_\psi(a,z)$: approximates the stationary density, with a positivity transform (e.g., softplus or $\exp$) ensuring $\hat{g} > 0$.
\end{itemize}

The joint loss has four components:
\begin{equation}
\ell = w_{\mathrm{HJB}}\,\ell_{\mathrm{HJB}} + w_{\mathrm{KFE}}\,\ell_{\mathrm{KFE}} + w_{\mathrm{flux}}\,\ell_{\mathrm{flux}} + w_{\mathrm{mass}}\left(\sum_z\int \hat{g}_\psi(a,z)\,da - 1\right)^{\!2},
\label{eq:ct_loss}
\end{equation}
where $\ell_{\mathrm{HJB}}$ and $\ell_{\mathrm{KFE}}$ are mean squared PDE residuals computed on collocation points, $\ell_{\mathrm{flux}}$ enforces the no-flux boundary conditions $s(\underline a, z)\,\hat g_\psi(\underline a, z) = 0$, and the mass term enforces normalization.  The integral is evaluated numerically (quadrature or Monte Carlo on the collocation points).  The aggregate-flux identity $\sum_z s(a,z)\,g(a,z) = 0$ at each interior $a$ -- which follows from the no-flux boundary and total-mass conservation within each $a$-slice -- is a free consistency check at solution time and is sometimes added as an auxiliary loss term.

Balancing the four loss components is critical: the HJB and KFE residuals can differ by orders of magnitude, so adaptive loss balancing (ReLoBRaLo, Chapter~\ref{ch:nas}) is strongly recommended.  Practical considerations include using separate learning-rate schedules for the two networks, concentrating collocation points near the borrowing constraint where the density can become sharply peaked, and verifying that the consumption policy $\hat c^\star$ satisfies $\hat c^\star > 0$ everywhere.  If the true stationary distribution contains an atom at the borrowing constraint (as in both Huggett and Aiyagari, Figure~\ref{fig:huggett_aiyagari_densities}), a continuous density network alone cannot represent it; one must add a separate boundary-mass variable or use a discretization that permits a point mass.

\begin{remarkbox}[Why a Fourier basis is a poor test-function set]
A natural-looking alternative to pointwise (collocation) residuals is to project the KFE residual onto a set of test functions $\{\psi_n\}$ and minimize the resulting projection coefficients.  The pitfall is that for the obvious choice of a Fourier basis, $\psi_n(a) = e^{i 2\pi n a / L}$, the coefficients $c_n$ of any smooth $g$ decay to zero by the Riemann--Lebesgue lemma; the smoother $g$ is, and the more it vanishes near the boundary, the faster the decay.  In practice, beyond $n \approx 5$ the $c_n$ are essentially zero \emph{regardless} of how well the residual is solved: they no longer measure error, they only measure smoothness.  Hat functions, low-order polynomials on local patches, or learned neural test functions (the WAN, or weak-adversarial-networks, idea) probe the residual far more honestly.
\end{remarkbox}

This continuous-time PINN approach and the discrete-time DEQN with Young's method (Chapter~\ref{ch:young}) address the \emph{same} economic question -- how heterogeneous agents interact through prices when markets are incomplete -- but differ in the mathematical formulation summarized in Table~\ref{tab:discrete_continuous_ha}.

\begin{table}[ht]
\centering
\small
\begin{tabular}{lll}
\toprule
& \textbf{Discrete time (Ch.~\ref{ch:young})} & \textbf{Continuous time (this chapter)} \\
\midrule
Distribution & Histogram $G_t(k_i, \varepsilon_j)$ & Density $g(a,z)$ \\
Evolution & Young's redistribution & KFE PDE \\
Individual opt. & Euler equation & HJB PDE \\
Solver & DEQN (TensorFlow) & PINN (PyTorch) \\
Key reference & \citet{krusell1998income} & \citet{achdou2022income} \\
\bottomrule
\end{tabular}
\caption{Discrete- and continuous-time formulations of incomplete-markets heterogeneous-agent models.  The economic object is the same in both columns, but the numerical residual changes from a discrete-time law of motion plus Euler equation to a coupled HJB--KFE PDE system.}
\label{tab:discrete_continuous_ha}
\end{table}

Both approaches can incorporate aggregate shocks, multiple assets, and general equilibrium; the choice between them depends primarily on whether the underlying economic model is formulated in discrete or continuous time.  For the aggregate-shock case the natural continuous-time object is the \emph{master equation} (\S\ref{sec:master_eq}), solved with EMINNs (\S\ref{sec:eminn}).

\section{The Master Equation}
\label{sec:master_eq}

When aggregate TFP $Z_t$ follows an OU process $dZ_t = \eta(\bar{Z} - Z_t)\,dt + \sigma^z\,dB_t^0$ (with mean-reversion speed $\eta>0$; not the OLG TFP shock or the network learning rate of earlier chapters), the value function depends on the \emph{entire wealth distribution}: $V = V(a, n, z, g)$.  The HJB becomes a PDE with a functional argument, and the ``curse of infinite dimensionality'' strikes: the functional derivative $\delta V/\delta g$ makes the problem intractable by standard methods.  The master equation approach, which originated in the mean field games literature \citep{lasry2007mean} and is developed systematically in the monograph of \citet{cardaliaguet2019master}, reformulates this coupled HJB--KFE system as a single PDE on the extended state space $(a, n, z, g)$ that retains the distribution argument explicitly but is amenable to neural network approximation.

\paragraph{Why collapse the coupled system?}  Without aggregate shocks, solving the stationary equilibrium as a fixed point in $r$ over the coupled HJB--KFE system is straightforward (\S\ref{sec:ct_equilibrium}).  With aggregate shocks, however, every realization of $Z_t$ would in principle require its own coupled solve, and no parametric guess for $V(a,n,z,g)$ can be informed by the cross-section unless the cross-section is treated as an explicit argument.  The master equation reformulation lifts $g$ into the state space so a \emph{single} PDE in $(a,n,z,g)$ encodes the entire economy, which makes a global neural-network ansatz feasible (Section~\ref{sec:eminn}).  The price of this convenience is the appearance of the functional-derivative term $\delta V/\delta g$, which the rest of this section unpacks.

\paragraph{The master equation.}  The key idea is to substitute the KFE, market clearing, and belief consistency \emph{directly into} the HJB, collapsing the coupled system into a single PDE:
\begin{equation}
\begin{split}
0 &= -\rho V(a,n,z,g) + u(c^*) + V_a\left[w(z,g)n + r(z,g)a - c^*\right] \\
&\quad + \lambda(n)\!\left(V(a,\hat{n},z,g) - V(a,n,z,g)\right) + V_z\,\mu^z(z) + \tfrac{1}{2}(\sigma^z)^2 V_{zz} \\
&\quad + \sum_{j} \int_{\underline{a}}^{\infty} \frac{\delta V}{\delta g_j}(a,n,z,g)(y)\;\mu^g_j(y,z,g)\,dy,
\end{split}
\label{eq:master_eq}
\end{equation}
where $\mu^g_j$ is the KFE drift computed from the optimal savings policy, and the last line encodes how changes in the distribution affect the value function through prices.  The kernel $\delta V/\delta g_j(y)$ is the infinite-dimensional analogue of a gradient: it measures how the value function at the individual state $(a,n,z)$ responds to an infinitesimal redistribution of probability mass to wealth level $y$ in the cross-section; the remark below makes this precise.  Mass conservation guarantees $\int_{\underline a}^{\infty}\mu^g_j(y,z,g)\,dy = 0$ (the KFE flux integrates to zero under no-flux boundary conditions), so the integrand of the last line pairs $\delta V/\delta g_j$ with a mean-zero test perturbation in exactly the sense required by the functional-derivative remark below.

\paragraph{The envelope condition.}  Following \citet{gu2024masterequations}, it is more convenient to work with $W(a,n,z,g) := \partial_a V(a,n,z,g)$, which directly gives the consumption policy via $c^* = W^{-1/\gamma}$.  The master equation for $W$ takes the form:
\begin{equation}
0 = (r(z,g) - \rho)\,W + \underbrace{\mathcal{L}_x W}_{\text{individual}} + \underbrace{\mathcal{L}_z W}_{\text{aggregate TFP}} + \underbrace{\mathcal{L}_g W}_{\text{distribution}},
\label{eq:master_W}
\end{equation}
where $\mathcal{L}_x W$ captures individual state dynamics (savings, income switching), $\mathcal{L}_z W = \partial_z W \cdot \mu^z + \frac{1}{2}(\sigma^z)^2\partial_{zz}W$ captures TFP dynamics, and $\mathcal{L}_g W = \sum_n \int (\delta W/\delta g_n)\cdot\mu^g_n\,dy$ captures distribution dynamics.

\begin{keyinsightbox}[One equation to rule them all]
The master equation subsumes the HJB, KFE, and market clearing into a \emph{single} PDE.  Its advantage is conceptual clarity and amenability to neural network approximation (via EMINNs).  Its challenge is infinite dimensionality: the argument $g$ is a measure, requiring finite-dimensional approximation.
\end{keyinsightbox}

\begin{remarkbox}[A user's note on the functional derivative $\delta W/\delta g$]
The objects $\delta V/\delta g$ and $\delta W/\delta g$ deserve a moment of attention because they are the only piece of mathematics in this chapter that is not standard PDE calculus.  $V$ takes \emph{a function} (the density $g$) as one of its arguments; differentiating $V$ with respect to $g$ therefore returns not a number but again a function.  (Strictly speaking, the Fr\'echet derivative is the linear functional $\zeta \mapsto \int (\delta V/\delta g)\,\zeta\,dy$ itself, and the kernel $\delta V/\delta g$ is what is properly called the \emph{functional} or \emph{variational} derivative.  We follow common usage and call the kernel the Fr\'echet derivative below.)  Concretely, if $g_\varepsilon = g + \varepsilon\,\zeta$ for some test perturbation $\zeta(y)$ with $\int \zeta\, dy = 0$, then
\[
  \frac{d}{d\varepsilon}\, V(\cdot, g_\varepsilon)\Big|_{\varepsilon=0}
  \;=\; \int \frac{\delta V}{\delta g}(y)\,\zeta(y)\,dy.
\]
The kernel $\delta V/\delta g$ is the \emph{Fr\'echet} (or \emph{linear functional}) derivative of $V$ in the $g$-argument; it measures how the value of an agent at a given $(a,n,z)$ responds to an infinitesimal redistribution of mass at point $y$ in the cross-section.  In the master-equation literature this object is more precisely the \emph{Lions derivative} $\partial_\mu V$ with respect to the mean-field measure $\mu$; when $\mu$ admits a density $g$ it reduces to the functional derivative used here.  The infinite-dimensionality of the master equation is the price of this extra argument; the EMINN approach in the next section makes it tractable by replacing $g$ with a finite-dimensional surrogate $\hat\varphi$ and then differentiating through $\hat\varphi$ via the chain rule and standard automatic differentiation.  For the underlying mean-field-games calculus see \citet{cardaliaguet2019master}.
\end{remarkbox}

\section{EMINNs: Solving the Master Equation with Deep Learning}
\label{sec:eminn}

Economic Model Informed Neural Networks (EMINNs), introduced by \citet{gu2024masterequations}, solve the master equation by (i) approximating the infinite-dimensional distribution $g$ by a finite-dimensional object $\hat{\varphi}$, and (ii) parameterizing $W$ by a neural network trained to minimize the master equation residual.  A teaching-scale companion notebook for EMINNs is forthcoming; in the present edition, the master-equation discussion is text-only and the Aiyagari notebook (\tpath{lecture_13_08_Aiyagari_Continuous_Time_FD_and_PINN_PyTorch.ipynb}) is the closest computational reference.

\subsection{Three Approximation Approaches for the Distribution}\label{subsec:three_approx}

The infinite-dimensional distribution $g$ must be replaced by a finite-dimensional approximation $\hat{\varphi} \in \R^d$ so that $V(a,n,z,g) \approx \hat{V}(a,n,z,\hat{\varphi})$.  Table~\ref{tab:eminn_distribution_approximations} summarizes the three approaches used in \citet{gu2024masterequations}.

\begin{table}[ht]
\centering
\small
\renewcommand{\arraystretch}{1.2}
\begin{tabular}{lccc}
\toprule
& \textbf{Finite population} & \textbf{Discrete state} & \textbf{Projection} \\
\midrule
$\hat{\varphi}_t$ & $\{(a_t^i, n_t^i)\}_{i=1}^N$ & Masses on grid $\{a_1,\ldots,a_I\}$ & Basis coefficients \\
$\hat{g}_t$ & $\frac{1}{N}\sum_i \delta_{\hat{\varphi}_t^i}$ & $\sum_{i,j} \hat{\varphi}_{ij}\,\delta_{(a_i,n_j)}$ & $b_0 + \sum_i \hat{\varphi}_i b_i(a,n)$ \\
Dimension & $\sim 40$ & $\sim 200$ & $\sim 5$ \\
\bottomrule
\end{tabular}
\caption{Finite-dimensional representations of the cross-sectional distribution in EMINNs.  Here $\delta_{\bullet}$ denotes a Dirac measure centered at $\bullet$, not the depreciation rate of Chapters~\ref{ch:deqn}--\ref{ch:young}.}
\label{tab:eminn_distribution_approximations}
\end{table}

\paragraph{Finite population ($N \approx 40$).}  Replace the continuum by $N$ agents with states $\hat{\varphi}_t = \{(a_t^1, n_t^1),\ldots,(a_t^N,n_t^N)\}$.  Aggregate capital is $K_t = N^{-1}\sum_i a_t^i$.  Sampling individual states and distribution states \emph{separately} during training keeps $N$ manageable; the law of large numbers provides accurate aggregate capital even with 40 agents.

\paragraph{Discrete state ($\sim$200 grid points).}  Discretize wealth on a grid $\{a_1,\ldots,a_I\}$ and represent the distribution as masses $\hat{\varphi}_{i,j}$ at each $(a_i, n_j)$.  The KFE becomes a finite-difference mass evolution, and the functional derivative becomes a partial derivative $\partial_{\hat{\varphi}_{m,j}}\hat{W}$.

\paragraph{Projection ($\sim$5 components).}  Project $g$ onto eigenfunctions of the steady-state KFE operator $\bar{\mathcal{L}}^{KF}$.  These are the most \emph{persistent} density components, carrying the most price-relevant information.  Only $\sim$5 basis functions suffice, yielding the lowest-dimensional representation, but the setup requires computing eigenfunctions and choosing appropriate test functions for the KFE evolution.

\subsection{The EMINN Algorithm}

A neural network $\hat{W}_\Theta(\omega)$ with $\omega = (a, n, z, \hat{\varphi})$ parameterizes the marginal value of wealth.  The output uses a softplus activation to ensure $\hat{W} > 0$, and consumption follows directly from the envelope condition: $c^* = \hat{W}^{-1/\gamma}$.  Algorithm~\ref{alg:eminn} gives the resulting residual-minimization loop.

\begin{algorithm}[H]
\caption{EMINN training for the master equation}
\label{alg:eminn}
\begin{algorithmic}
\REQUIRE Initial parameters $\Theta$, tolerance $\varepsilon$, loss weights $\kappa^e$, $\kappa^s$; initialize $\mathcal{E}=\infty$
\WHILE{$\mathcal{E} > \varepsilon$}
    \STATE Sample $M$ collocation points: $(a_m, n_m, z_m, \hat{\varphi}_m)_{m=1}^M$
    \STATE Compute distribution-drift coefficients $\mu_{\hat\varphi,n}(\hat\varphi_m)$ (method-specific; see \S\ref{subsec:three_approx} and remark below)
    \STATE Compute master equation residual $\hat{\mathcal{L}}(\cdot;\Theta)$ via automatic differentiation through $\hat\varphi$
    \STATE Compute shape penalty $\mathcal{E}^s(\cdot;\Theta)$ (concavity, monotonicity)
    \STATE Total loss: $\mathcal{E} = \kappa^e\,M^{-1}\sum_m |\hat{\mathcal{L}}_m|^2 + \kappa^s\,\mathcal{E}^s$
    \STATE Update: $\Theta \leftarrow \Theta - \alpha_{\mathrm{opt}}\,\nabla_\Theta \mathcal{E}$ \;\textit{($\alpha_{\mathrm{opt}}$: optimizer learning rate; distinct from the OU mean-reversion $\eta$ of \S\ref{sec:master_eq})}
\ENDWHILE
\end{algorithmic}
\end{algorithm}

This is precisely a PINN applied to the master equation: the ``physics'' is the economic equilibrium structure, and all derivatives, including those with respect to the distribution parameters $\hat{\varphi}$, are computed by automatic differentiation.

The master equation residual decomposes as:
\begin{equation}
0 = (r(z,\hat{\varphi}) - \rho)\,\hat{W} + \hat{\mathcal{L}}_x\hat{W} + \hat{\mathcal{L}}_z\hat{W} + \hat{\mathcal{L}}_g\hat{W},
\end{equation}
where $\hat{\mathcal{L}}_x\hat{W} = s(\cdot)\partial_a\hat{W} + \lambda(n)(\hat{W}(\hat{n}) - \hat{W}(n))$ captures savings and income switching, $\hat{\mathcal{L}}_z\hat{W} = \partial_z\hat{W}\cdot\mu^z + \frac{1}{2}(\sigma^z)^2\partial_{zz}\hat{W}$ captures the OU aggregate shock, and $\hat{\mathcal{L}}_g\hat{W} = \sum_n (\partial\hat{W}/\partial\hat{\varphi}_n)\cdot\mu_{\hat{\varphi},n}$ captures distribution evolution.

\paragraph{Computing the drift coefficients $\mu_{\hat{\varphi},n}$.}  The coefficients $\mu_{\hat{\varphi},n}$ describing how the finite-dimensional approximation $\hat{\varphi}$ of the cross-sectional density evolves in time are not given for free.  Recovering them is a genuine algorithmic step, and it is the place where the three approximation choices in Section~\ref{subsec:three_approx} differ most sharply:
\begin{itemize}[itemsep=2pt]
\item \textbf{Finite-population approximation.}  $\hat{\varphi}$ is the empirical measure of a finite particle system, so $\mu_{\hat{\varphi},n}$ is the SDE drift of particle~$n$, available in closed form from the underlying individual problem.
\item \textbf{Discrete-state (grid) approximation.}  $\hat{\varphi}_{i,j}$ is the mass at $(a_i, n_j)$; the discretized KFE evaluated at that node returns $\mu_{\hat{\varphi},(i,j)}$ directly.  This is also where an upwind scheme reappears (see the remark after \eqref{eq:cake_expand_V}): the side of $\partial_a$ used in the discretization is selected by the sign of the drift $s$.
\item \textbf{Projection / basis expansion.}  $\hat{\varphi}(a) = \sum_n c_n \psi_n(a)$.  The KFE returns $\partial_t g$ as a function of $a$; one then recovers $\mu_{\hat{\varphi},n} = \dot c_n$ by least-squares projection of $\partial_t g$ onto $\{\psi_n\}$.  This is the only step of EMINN that genuinely depends on the choice of approximation, and it is where most of the practical art lives.
\end{itemize}
Concretely, between the forward pass and the residual evaluation in Algorithm~\ref{alg:eminn}, an additional sub-step computes $\mu_{\hat{\varphi},n}$ for the current $\hat{\varphi}$; the chain rule through $\hat{\varphi}$ then propagates these coefficients into $\hat{\mathcal{L}}_g\hat{W}$.

\subsection{Shape Constraints and Training Stability}

A key practical challenge in training EMINNs is that neural networks may converge to ``cheat solutions'' (constant, non-increasing, or non-concave value functions that produce small residuals but are economically meaningless).  Shape penalties are added to the loss to enforce economic structure:
\begin{itemize}[itemsep=2pt]
\item \textbf{Concavity:} penalize non-concavity (i.e., violations of the standard $V_{aa}\leq 0$ condition) via $\mathcal{E}^{\text{concav}} = M^{-1}\sum_m \max(0, \partial_{aa}V_m)^2$, which is positive only when $\partial_{aa}V > 0$ at some collocation point.
\item \textbf{Monotonicity:} in calibrations where higher TFP lowers the marginal value of liquid wealth, penalize violations of the model-specific restriction $\partial_z V_a < 0$.
\item \textbf{Initialization:} set $W(a,\cdot) = e^{-a}$ as a simple decreasing initial marginal-value profile.
\item \textbf{Architectural encoding:} if the chosen formulation has known boundary asymptotics for $V$ or $W$, multiply the network by an appropriate boundary factor rather than asking the optimizer to learn the singular shape from scratch.
\item \textbf{Active sampling:} concentrate collocation points where the residual is large.
\end{itemize}

\subsection{Results and Method Comparison}

For the Aiyagari model (no aggregate shocks), \citet{gu2024masterequations} report master-equation residuals of order $10^{-5}$ and mean squared errors against finite-difference benchmarks of order $10^{-5}$, with close agreement in consumption policies, marginal value functions, and stationary distributions across the finite-population and discrete-state approaches.

For the Krusell--Smith model (with aggregate shocks), their reported results show master-equation training losses of order $10^{-5}$ across all three approximation approaches and similar time paths for aggregate variables (capital, interest rate, wage).  In that experiment, the projection approach achieves the lowest reported loss ($8.5 \times 10^{-6}$) with only 5 distribution parameters, while the finite-population approach ($3.0 \times 10^{-5}$) offers the simplest implementation.

Table~\ref{tab:ct_method_comparison} gives the practical method comparison for this chapter.

\begin{table}[ht]
\centering
\small
\renewcommand{\arraystretch}{1.3}
\begin{tabular}{@{}p{3.0cm}p{3.5cm}p{3.0cm}p{3.0cm}@{}}
\toprule
& \textbf{Finite differences} & \textbf{PINN} & \textbf{EMINN} \\
\midrule
Stationary equilibrium & Benchmark method & Works, validate carefully & Works, but overkill \\
Aggregate shocks & Local/low-dimensional only & Coupled system, not full master equation & Designed for global master equation \\
Grid required & Yes & No state grid & No state grid \\
High-dimensional scaling & Limited by grids & Better, optimization-limited & Better, distribution-state approximation needed \\
Handles $g$ as state & Not in standard stationary FD & No & Yes, through $\hat{\varphi}$ \\
Convergence theory & Strong for low-dimensional monotone schemes & Limited & Limited \\
Low-dimensional accuracy & Often $\sim 10^{-6}$ & Problem-dependent; validate & Reported $\sim 10^{-5}$ \\
\bottomrule
\end{tabular}
\caption{Finite differences, PINNs, and EMINNs for the continuous-time heterogeneous-agent problems covered in this chapter.  The entries are practical guidance, not universal impossibility results: finite differences remain the benchmark in low dimension, while EMINNs target the global master-equation setting with the distribution as a state.}
\label{tab:ct_method_comparison}
\end{table}

For stationary, low-dimensional problems, finite differences remain fast and reliable and should be used as the benchmark.  For models with aggregate shocks and high-dimensional state spaces, EMINNs are, among the methods surveyed here, one of the few approaches demonstrated on global master-equation solutions for this class of benchmarks.  PINNs serve as a useful intermediate step: they share the same code philosophy as EMINNs (automatic differentiation of PDE residuals) but apply to the coupled HJB--KFE system rather than the master equation.

\begin{keyinsightbox}[Chapter Summary]
\begin{itemize}[itemsep=2pt, leftmargin=*]
\item In continuous time, the heterogeneous-agent equilibrium is a coupled HJB + KFE system, recast as a mean field game in the sense of Lasry--Lions.
\item Closing with aggregate shocks gives the master equation: a single PDE in $(x, g)$ where the second argument is a measure.  EMINNs solve it by finite-dimensional approximation of $g$ and a neural-network ansatz for the value function.
\item The functional derivative $\delta V/\delta g$ is the only piece of mathematics not standard in PDEs; it measures sensitivity to infinitesimal cross-section perturbations.
\item Achdou et al.\ (2022) finite differences and EMINNs are complementary: FD wins in low dimension, while EMINNs are designed for aggregate-shock master-equation models.
\end{itemize}
\end{keyinsightbox}

\section*{Further Reading}
\addcontentsline{toc}{section}{Further Reading}
\begin{itemize}[itemsep=2pt]
\item \citet{achdou2022income}, the canonical continuous-time HA reference.
\item \citet{gu2024masterequations}, the EMINN paper.
\item \citet{lasry2007mean, carmona2018probabilistic, cardaliaguet2019master}, mean-field-games foundations.
\item \citet{shreve2004stochasticii}, stochastic-calculus textbook.
\item Moll's online lecture notes (\url{https://benjaminmoll.com/lectures/}), pedagogical complement.
\end{itemize}

\section*{Exercises}
\addcontentsline{toc}{section}{Exercises}
\noindent Worked solutions and guidance for these exercises appear in Appendix~\ref{app:solutions}.
\begin{enumerate}[itemsep=4pt, leftmargin=*]
\item\label{ex:ch8:1} \textbf{[Core] It\^o on GBM.}  Apply It\^o's lemma to $f(x) = \ln x$ with $X_t$ following geometric Brownian motion to derive $X_t = X_0 \exp[(\mu - \tfrac{1}{2}\sigma^2)t + \sigma B_t]$.  Discuss why ``volatility drag'' lowers the expected log return below the expected return.
\item\label{ex:ch8:2} \textbf{[Core] KFE for an OU process.}  Write the Kolmogorov forward equation for the Ornstein--Uhlenbeck process $dX_t = \eta(\bar X - X_t)\,dt + \sigma\,dB_t$ and derive the stationary density $\mathcal{N}(\bar X, \sigma^2/(2\eta))$ by setting $\partial_t g = 0$.
\item\label{ex:ch8:3} \textbf{[Core] Functional derivative.}  For the master-equation value function $V(a, g)$, compute $\delta V/\delta g$ for the toy specification $V(a,g) = \int u(c(a, y))\,g(y)\,dy$ where $c$ is a fixed consumption rule.  Interpret the result.
\item\label{ex:ch8:4} \textbf{[Computational] HJB residual.}  In notebook \tpath{lecture_13_08_Aiyagari_Continuous_Time_FD_and_PINN_PyTorch.ipynb}, compute the maximum HJB residual on a 50-point test grid after a fixed PINN training budget.  Repeat with a larger collocation batch and report the actual scaling you observe.
\item\label{ex:ch8:5} \textbf{[Core] Closed Aiyagari system in continuous time.}  Combine the HJB equation~\eqref{eq:hjb_full} for $V(a,n)$ and the KFE~\eqref{eq:kfe_econ} for $g(a,n)$ into the closed Aiyagari general-equilibrium system.  (i)~Add the firm's first-order conditions: with Cobb--Douglas production $Y = A K^\alpha L^{1-\alpha}$, write $r = \alpha A K^{\alpha-1} L^{1-\alpha} - \delta$ and $w = (1-\alpha) A K^\alpha L^{-\alpha}$.  (ii)~Add the market-clearing conditions $K = \sum_n\int_{\underline a}^{\infty} a\,g(a,n)\,da$ and $L = \sum_n n\int_{\underline a}^{\infty}g(a,n)\,da$.  (iii)~Show that the coupled system pins down $(V,g,K,L,r,w)$, with $L$ often fixed by the stationary income shares and $w$ implied by $(K,L)$.  (iv)~Briefly explain why solving this system for the stationary equilibrium is commonly reduced to a fixed point in $r$, and why the inner HJB and KFE problems must be solved consistently at each candidate $r$.
\item\label{ex:ch8:6} \textbf{[Advanced/project] Stationary-distribution convergence.}  In notebook \tpath{lecture_13_08_Aiyagari_Continuous_Time_FD_and_PINN_PyTorch.ipynb}, take the optimal policy $s^\star(a,n)$ from the converged HJB solver (i.e., fix the policy) and run the KFE~\eqref{eq:kfe_econ} forward in time starting from a non-stationary initial density (e.g., uniform on $[\underline a, a_\mathrm{max}]$).  Plot snapshots of $g_t(a, n)$ at $t \in \{1, 5, 25, 100, 500\}$ years and the running $L^2$ distance $\|g_t - g^\star\|_{L^2}$ to the stationary distribution $g^\star$.  Check whether convergence is approximately exponential by fitting a line to $\log\|g_t - g^\star\|$ over the linear region, and relate the fitted rate to the spectral gap of the KFE operator.
\item\label{ex:ch8:7} \textbf{[Advanced/project] FD upwind vs.\ PINN benchmark.}  In the same notebook, the chapter ships both a finite-difference upwind solver (deterministic, grid-based) and a PINN trained on HJB and KFE residuals.  Run both on the same $1$-asset Aiyagari calibration and report (i)~wall-clock time for your chosen residual target, (ii)~the final residual on a fine $200$-point test grid, and (iii)~peak memory, including GPU memory if you run the PINN on a GPU.  As an extension, sweep the discount rate $\rho$ over $\{0.04, 0.05, 0.06\}$ and compare cold-start FD runs with warm-started PINN runs.  Discuss when each is preferred: FD for low-dimensional, fixed-parameter computations; PINN for parametric sweeps and higher-dimensional extensions.
\end{enumerate}

\chapter{Deep Surrogate Models and Gaussian Processes}
\label{ch:gp}

The DEQNs and PINNs of previous chapters solve a single model configuration at a time, to high accuracy.  This chapter takes a different angle: every method developed below is \emph{supervised regression on the output of an expensive numerical oracle}, where the oracle is queried at modestly many designed inputs and the fitted surrogate then stands in for it everywhere else.  The script uses that move on two distinct oracles.  The first is the structural model itself, evaluated across many parameter or scenario vectors $\theta$ for downstream estimation, uncertainty quantification, and optimal policy design (Chapter~\ref{ch:estimation}, Chapter~\ref{ch:climate}).  The second is the Bellman operator $TV$ at a fixed model, evaluated at many state-space points inside a value-function-iteration loop, for solving high-dimensional dynamic programs (\S\ref{sec:gp_dp}).  Both settings call for the same GP machinery, the same diagnostics, and the same active-learning logic; only the labels change.  A \emph{surrogate} is, in effect, a twenty-first-century lookup table: not a static grid of precomputed answers, but a trained, smooth, differentiable function that interpolates the oracle's output across a high-dimensional space of states and parameters and can be queried in microseconds \citep{chen2026Deep}.  Surrogates produce thousands of evaluations per second, enabling parametric estimation, sensitivity analysis, and uncertainty quantification that would be infeasible if every evaluation required a full numerical solve.  The methodological foundations are the GP textbook of \citet{Rasmussen:2005:GPM:1162254} and the Bayesian-active-learning ideas dating back to \citet{mackay1992information} and \citet{krause2008near}.  Embedding GPs inside dynamic programming was pioneered by \citet{deisenroth2009gaussian} (GPDP) and \citet{engel2005reinforcement} (GP temporal-difference learning); within economics, applications include high-dimensional growth models \citep{SCHEIDEGGER201968}, dynamic incentive problems \citep{rennerscheidegger_2018}, and deep uncertainty quantification for integrated assessment models \citep{friedlDeep2023}.  We also cover Bayesian active learning for optimal training-point selection, sparse approximations \citep{titsias2009variational, hensman2013gaussian} that extend GPs beyond their cubic scaling limit, active subspaces for dimensionality reduction, GP-based value function iteration for dynamic programming, and deep kernel learning for combining neural network feature extraction with GP inference.

\section{Motivation: The Computational Bottleneck}

Every workflow this chapter targets puts an expensive numerical solve inside an outer loop.  For estimation, uncertainty quantification, and optimal policy design, the outer loop runs over a parameter or scenario vector $\theta$ and the inner solve is a full model solution, a Bellman fixed point, a PDE solve, or a Monte Carlo run that costs seconds to hours, repeated at the $10^3$ to $10^6$ outer iterations these tasks demand.

For dynamic programming, the outer loop is the Bellman iteration itself: at iteration $s$ the inner ``solve'' is one evaluation of the operator $(TV^{s-1})(\x)$ at a state $\x$, which itself requires a constrained nonlinear program over controls plus a quadrature over the next-period shock, then a global fit of $V^s$ to those labels (\S\ref{sec:gp_dp_supervised_view}).  In both cases the obstacle is the same: the per-inner-solve cost times the per-outer-iteration count.

The key insight is that since we \emph{own} the structural model, we can generate training data by solving the model on a carefully chosen set of input configurations (a \emph{design of experiments}).  A cheap-to-evaluate function approximator trained on this synthetic dataset, a \emph{surrogate model}, then replaces the expensive original model for all downstream tasks.  Any suitable function approximator can serve as the surrogate; the right choice depends on the dimensionality of the input space and the cost of generating each training point.

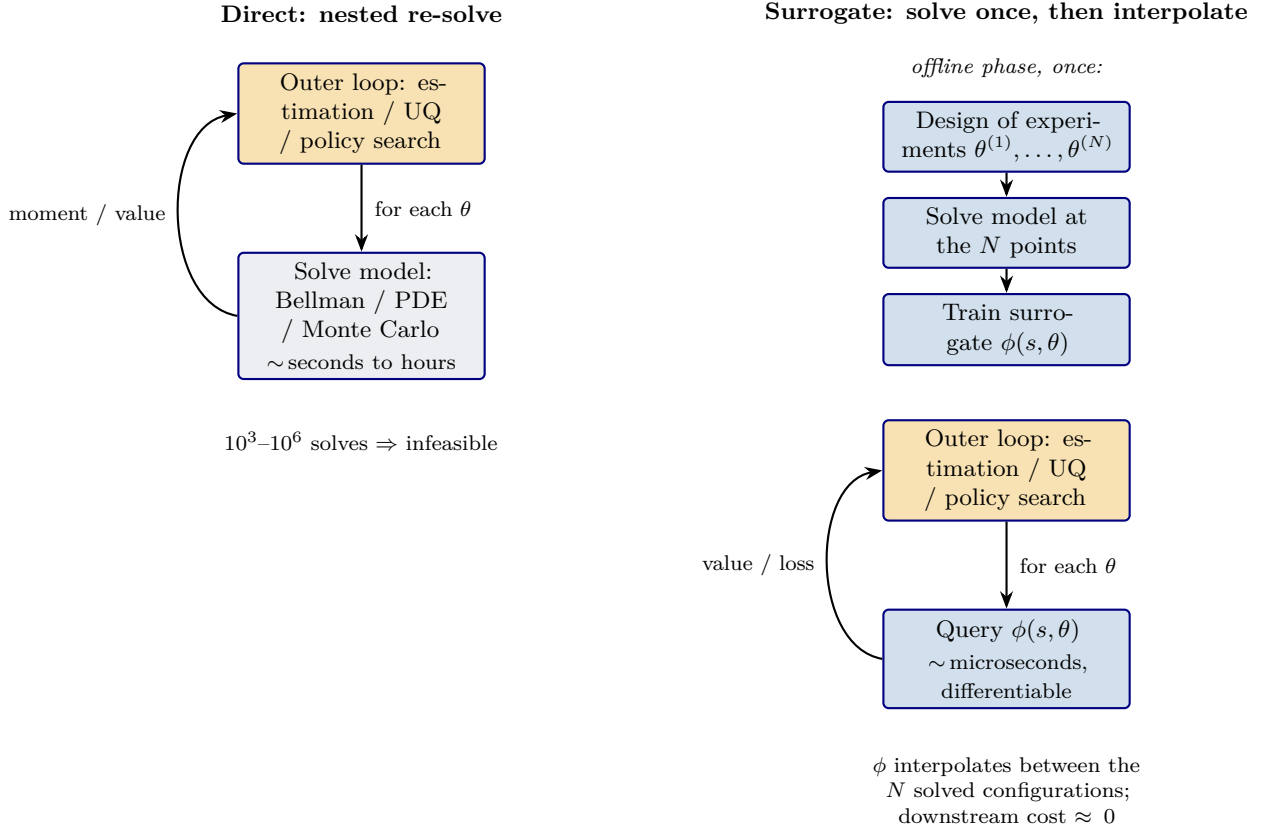
\begin{figure}[ht]
\centering
\begin{tikzpicture}[
    box/.style={rectangle, draw=uzhblue, fill=uzhgreylight,
        text width=3.0cm, align=center, minimum height=0.85cm,
        font=\footnotesize, rounded corners=2pt, thick},
    sbox/.style={rectangle, draw=uzhblue, fill=softblue!25,
        text width=3.0cm, align=center, minimum height=0.7cm,
        font=\footnotesize, rounded corners=2pt, thick},
    obox/.style={rectangle, draw=uzhblue, fill=softorange!30,
        text width=3.0cm, align=center, minimum height=0.7cm,
        font=\footnotesize, rounded corners=2pt, thick},
    >=Stealth
]
\node[font=\footnotesize\bfseries] (titleL) {Direct: nested re-solve};
\node[obox, below=0.4cm of titleL] (outerL) {Outer loop: estimation / UQ / policy search};
\node[box, below=1.15cm of outerL] (solveL) {Solve model: Bellman / PDE / Monte Carlo\\[1pt]{\scriptsize $\sim$\,seconds to hours}};
\draw[->, thick] (outerL.south) -- node[right=1pt, font=\scriptsize] {for each $\theta$} (solveL.north);
\draw[->, thick] (solveL.west) to[out=170, in=190] node[left=1pt, font=\scriptsize] {moment / value} (outerL.west);
\node[below=0.55cm of solveL, font=\scriptsize, text width=3.6cm, align=center] {$10^3$--$10^6$ solves $\Rightarrow$ infeasible};
\node[font=\footnotesize\bfseries, right=3.2cm of titleL] (titleR) {Surrogate: solve once, then interpolate};
\node[font=\scriptsize\itshape, below=0.18cm of titleR] (offl) {offline phase, once:};
\node[sbox, below=0.18cm of offl] (doe) {Design of experiments $\theta^{(1)},\dots,\theta^{(N)}$};
\node[sbox, below=0.32cm of doe] (solveN) {Solve model at the $N$ points};
\node[sbox, below=0.32cm of solveN] (train) {Train surrogate $\phi(s,\theta)$};
\draw[->, thick] (doe) -- (solveN);
\draw[->, thick] (solveN) -- (train);
\node[obox, below=0.7cm of train] (outerR) {Outer loop: estimation / UQ / policy search};
\node[sbox, below=1.15cm of outerR] (queryR) {Query $\phi(s,\theta)$\\[1pt]{\scriptsize $\sim$\,microseconds, differentiable}};
\draw[->, thick] (outerR.south) -- node[right=1pt, font=\scriptsize] {for each $\theta$} (queryR.north);
\draw[->, thick] (queryR.west) to[out=170, in=190] node[left=1pt, font=\scriptsize] {value / loss} (outerR.west);
\node[below=0.5cm of queryR, font=\scriptsize, text width=3.9cm, align=center] {$\phi$ interpolates between the $N$ solved configurations; downstream cost $\approx 0$};
\end{tikzpicture}
\caption{Why surrogates help. \emph{Left}: structural estimation, uncertainty quantification, and optimal policy design are outer loops over a parameter vector $\theta$, and the direct implementation re-solves the full model inside the loop, so the cost scales with the number of outer iterations times the per-solve cost. \emph{Right}: a surrogate moves that solve into a one-time offline phase, solving the model only at a design of experiments and fitting $\phi(s,\theta)$; the outer loop then queries a cheap, differentiable interpolant. The saving grows with both the number of outer iterations and the per-solve cost.  The same picture applies to GP value-function iteration with the outer loop relabeled as the Bellman iteration and the inner solve as one $TV$ evaluation; the offline phase is then replaced by a per-iteration GP refit at modest design size.}
\label{fig:surrogate_outer_loop}
\end{figure}

\paragraph{Cutting out the outer loop.}  The point of a surrogate is to break this nesting.  One pays a one-time offline cost: pick a design of experiments $\theta^{(1)},\dots,\theta^{(N)}$, solve the model at those $N$ configurations, and fit a surrogate $\phi(s,\theta)$ to the results.  From then on the expensive inner solve is gone, and the estimation, uncertainty-quantification, or policy-search outer loop evaluates a function that costs microseconds and returns exact gradients, so it can run at the $10^3$ to $10^6$ scale those tasks need.  The model is solved at a handful of configurations and the surrogate \emph{interpolates between them}, which is almost always far cheaper than re-solving at every new $\theta$; Figure~\ref{fig:surrogate_outer_loop} contrasts the two workflows.  The surrogate-based SMM estimation of Chapter~\ref{ch:estimation} and the surrogate-then-optimize policy search of Chapter~\ref{ch:climate} are both instances of this move, as is the GP value-function iteration of \S\ref{sec:gp_dp}, where the ``outer loop'' is the Bellman iteration itself.  In the GP-VFI variant the surrogate is refit every Bellman step and the ``offline'' phase becomes a per-iteration update; that is the second of the two oracles announced at the start of the chapter.

\paragraph{Two surrogate strategies.}  This course covers two complementary approaches:

\begin{enumerate}[itemsep=3pt]
\item \textbf{Deep neural network (DNN) surrogates} are best suited for \emph{high-dimensional} settings ($d \gg 10$ inputs) where training data can be generated in large quantities, for example when each model solve takes seconds or when closed-form solutions exist.  DNNs scale gracefully with dimensionality, can be trained via mini-batch SGD on millions of samples, and provide exact gradients via automatic differentiation.  \citet{chen2026Deep} formalize this approach and demonstrate speedups of several orders of magnitude for option pricing (the same surrogate-for-finance idea was implemented earlier with adaptive sparse grids by \citet{scheideggertreccani_2018}); \citet{friedlDeep2023} apply it to uncertainty quantification in high-dimensional integrated assessment models.

\item \textbf{Gaussian process (GP) surrogates} are preferable for \emph{intermediate-dimensional} settings ($d \lesssim 10$--$15$) where each training point is numerically expensive, for example solving a full DSGE model at one parameter configuration may take minutes or hours.  GPs are \emph{data-efficient}: the Bayesian posterior extracts maximum information from each observation.  Crucially, the posterior variance provides a built-in uncertainty estimate that can guide \emph{where} to evaluate next, enabling Bayesian Active Learning (BAL) strategies that allocate the computational budget optimally \citep{SCHEIDEGGER201968}.
\end{enumerate}

\begin{table}[ht]
\centering
\small
\begin{tabular}{@{}p{3.5cm}p{5.0cm}p{5.0cm}@{}}
\toprule
& \textbf{DNN surrogate} & \textbf{GP surrogate} \\
\midrule
Best for & high-dim.\ ($d \gg 10$), large $N$ & moderate-dim.\ ($d \lesssim 15$), small $N$ \\
Data efficiency & data-hungry; wants a large training set & data-efficient; informative from a small one \\
Key advantage & scales to very high $d$ via SGD & built-in UQ and active learning \\
\bottomrule
\end{tabular}
\caption{Two complementary surrogate strategies. DNN surrogates are attractive when the input dimension and available training set are large; GP surrogates are attractive when each simulator call is expensive and calibrated posterior uncertainty is useful for active learning.}
\label{tab:surrogate_strategy_comparison}
\end{table}

Table~\ref{tab:surrogate_strategy_comparison} summarizes the main trade-off.  The two approaches are not mutually exclusive: one can use a GP to build an initial low-data surrogate with uncertainty estimates, and later switch to a DNN when more training data becomes available.  A detailed comparison covering computational cost, gradient access, and further trade-offs is given in Section~\ref{sec:gp_vs_dnn} after the GP methodology has been introduced.

\paragraph{Speed gains.}  This is the payoff sketched in Figure~\ref{fig:surrogate_outer_loop}: regardless of whether a DNN or GP is used, once the surrogate is trained the per-iteration cost of the downstream outer loop, estimation, sensitivity analysis, or optimal policy design, collapses to a function evaluation.  \citet{chen2026Deep} report speedups of several orders of magnitude for option pricing, where evaluating the DNN surrogate replaces expensive FFT-based Fourier inversion (their Bates-model benchmark documents two-to-three orders of magnitude over the numerical pricing baseline).  As a rough rule of thumb, the gain scales with the cost of the underlying pricing routine: the orders-of-magnitude gains arise for models requiring a PDE solve (roughly $1$\,ms/eval $\to$ $1$\,$\mu$s/eval through a surrogate) or high-dimensional Monte Carlo, regimes in which $10^3$--$10^4\times$ speedups are typical.  The gains are even larger for gradient computations: while finite-difference gradients require $d+1$ model evaluations (one per parameter), the gradient through the surrogate (autograd for DNNs, closed-form for GPs) requires only a single pass, regardless of the number of parameters.

\section{Pseudo-States: Parameters as ``State'' Variables}

The central innovation of the deep surrogate framework is to treat model parameters $\theta$ as additional ``pseudo-state'' variables:
\begin{equation}
\tilde{\x} = \bigl(\underbrace{s_1,\dots,s_n}_{\text{states}},\;\underbrace{\theta_1,\dots,\theta_p}_{\text{parameters}}\bigr) \in \R^{d}, \quad d = n + p.
\end{equation}

\begin{figure}[ht]
\centering
\begin{tikzpicture}[
    mlstep/.style={rectangle, draw=uzhblue, fill=uzhgreylight,
        minimum width=2cm, minimum height=0.6cm, font=\footnotesize,
        rounded corners=2pt, thick}, >=Stealth
]
\node[mlstep] (states) {States $s$};
\node[mlstep, below=0.2cm of states, fill=softorange!30] (params) {Params $\theta$};
\node[mlstep, right=2cm of states, fill=softblue!30,
      minimum width=2.2cm, minimum height=1.6cm] (surr) {\large Surrogate $\phi$};
\node[mlstep, right=2cm of surr] (output) {$y = \phi(\tilde{\x}|\theta_\mathrm{NN})$};
\draw[->, thick] (states.east) -- ++(0.4,0) |- ([yshift=0.2cm]surr.west);
\draw[->, thick] (params.east) -- ++(0.4,0) |- ([yshift=-0.2cm]surr.west);
\draw[->, thick] (surr) -- (output);
\end{tikzpicture}
\caption{Pseudo-state surrogate architecture. Economic states $s$ and model parameters $\theta$ are concatenated into the augmented input $\tilde{\x} = (s, \theta)$ and fed to a single approximator $\phi(\tilde{\x}\,|\,\theta_\mathrm{NN})$ with weights $\theta_\mathrm{NN}$, yielding a target quantity $y$ (price, policy, moment) as a continuous, differentiable function of both the state and the parameter vector. After one offline training pass, the surrogate is queried instantly across the parameter space without re-solving the original model.}
\label{fig:pseudo_state_surrogate}
\end{figure}
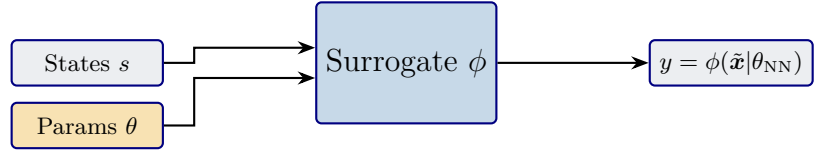

The surrogate is trained once over the full augmented space and can then be queried instantly for any parameter configuration, without re-solving the model.  This is fundamentally different from simply re-running the model: the surrogate provides a continuous, differentiable mapping from parameters to outputs, enabling gradient-based optimization and uncertainty propagation that would be impossible with the original model.  Figure~\ref{fig:pseudo_state_surrogate} sketches this concatenated input.

\citet{scheideggertreccani_2018} achieve the surrogate-for-finance idea with adaptive sparse grids; \citet{friedlDeep2023} apply the surrogate idea to uncertainty quantification in integrated assessment models of climate change; \citet{chen2026Deep} demonstrate speedups of several orders of magnitude for option pricing with the deep-surrogate approach.

\paragraph{Comparison of approximation methods.}  The surrogate approach is one of several function approximation strategies used in computational economics.  Table~\ref{tab:approximation_methods_comparison} situates it relative to alternatives.

\begin{table}[ht]
\centering
\small
\begin{tabular}{@{}l c c c c@{}}
\toprule
\textbf{Method} & \textbf{Max dim.} & \textbf{Smoothness} & \textbf{Parametric} & \textbf{Differentiable} \\
\midrule
Cartesian grids & $d \leq 5$ & any & no & no \\
Sparse grids & $d \leq 15$ & $C^k$ needed & no & limited \\
Chebyshev polynomials & $d \leq 10$ & smooth & yes & yes \\
DNN surrogate & $d \gg 10$ & any & yes & yes (autograd) \\
GP surrogate & $d \leq 10$\textsuperscript{$\dagger$} & kernel-dependent & no & yes (closed-form) \\
\bottomrule
\multicolumn{5}{l}{\footnotesize $\dagger$\,Can be extended to $d \gg 10$ via active subspace methods \citep{SCHEIDEGGER201968}.}
\end{tabular}
\caption{Common approximation methods in computational economics.  Grid and polynomial methods are transparent but become difficult in high dimension; DNN and GP surrogates trade direct grid structure for sample-based learning and repeated fast evaluation.}
\label{tab:approximation_methods_comparison}
\end{table}

\subsection{Worked Example: Black--Scholes Surrogate}

To illustrate the surrogate pipeline concretely, consider the European call option pricing problem from Section~\ref{sec:bs_pinn}.  In the PINN approach (Chapter~\ref{ch:pinn}), the network learned the option price by minimizing the Black--Scholes PDE residual; no training data were needed, only the differential equation.  The surrogate approach takes the opposite route: we \emph{generate} training data by evaluating the closed-form Black--Scholes formula at a design of experiments, and train a neural network to interpolate this data.

Specifically, we sample $N$ input tuples $(S_i, t_i, \sigma_i, r_i, K_i)$ from a Latin Hypercube design over the ranges of interest and evaluate the analytical price $V_i = V_\mathrm{BS}(S_i, t_i, \sigma_i, r_i, K_i)$ at each.  The surrogate $\hat{V} = \mathcal{N}_\rho(S, t, \sigma, r, K)$ is then trained via standard supervised learning:
\begin{equation}
\ell_\rho = \frac{1}{N}\sum_{i=1}^N \bigl|\mathcal{N}_\rho(S_i, t_i, \sigma_i, r_i, K_i) - V_i\bigr|^2.
\end{equation}
Once trained, the surrogate provides instant evaluation at any $(S, t, \sigma, r, K)$ in a single forward pass, instant Greeks ($\Delta$, $\Gamma$, Vega, etc.)\ via a single backward pass, and gradient-based implied volatility calibration, none of which require re-solving the PDE.  The key contrast with the PINN is that the surrogate requires \emph{solved} training data (here from the analytical formula; in general, from a numerical solver), but in return it treats the model parameters $(\sigma, r, K)$ as inputs, enabling re-evaluation across the entire parameter space without re-solving.  This is precisely the ``pseudo-state'' idea of the previous section.  See the companion notebook \texttt{01\_Surrogate\_Primer.ipynb} for the full implementation.

\section{Gaussian Process Regression}

A Gaussian Process (GP) is a nonparametric Bayesian approach to function approximation \citep{Rasmussen:2005:GPM:1162254} that is particularly well suited to settings where data is scarce but uncertainty quantification is essential.  Intuitively, a GP defines a probability distribution over functions whose support and sample-path regularity are determined by the covariance kernel: an RBF kernel implies extremely smooth (in fact, infinitely mean-square differentiable) sample paths, while a Mat\'ern kernel with small smoothness parameter implies rougher paths.  Upon observing data, Bayes' rule yields a posterior distribution that concentrates around functions consistent with the observations while maintaining calibrated uncertainty elsewhere.

A GP is fully specified by a mean function $\mu(\x)$ and a covariance (kernel) function $k(\x, \x')$:
\begin{equation}
f(\x) \sim \mathcal{GP}\bigl(\mu(\x), k(\x, \x')\bigr).
\end{equation}
This means that for any finite collection of test points $\x_1, \ldots, \x_n$, the function values $\bm{f} = (f(\x_1), \ldots, f(\x_n))^\top$ are jointly Gaussian:
\begin{equation}
\bm{f} \sim \mathcal{N}(\bm{\mu},\, K), \qquad \text{where } K_{ij} = k(\x_i, \x_j).
\end{equation}
To sample a function from the GP prior, one evaluates the kernel matrix $K$ at a grid of test points, computes its Cholesky decomposition $K = LL^\top$, draws $\bm{u} \sim \mathcal{N}(\bm{0}, I)$, and forms $\bm{f} = \bm{\mu} + L\bm{u}$.  This procedure generates smooth random functions whose properties (smoothness, amplitude, length scale) are controlled entirely by the kernel choice.

The squared exponential (RBF) kernel is the most widely used choice; we briefly preview the practical recommendation here so the reader can keep it in mind through the derivations: in economic applications, the target function (value, policy, equilibrium price) often has \emph{kinks} from occasionally-binding constraints, and the Mat\'ern-$3/2$ kernel introduced in \S\ref{sec:matern} below is then a better default than RBF, since it does not oversmooth at non-differentiable features.  The RBF kernel remains the right default when the target is genuinely smooth (e.g.\ in the unconstrained interior of an ergodic set).
\begin{equation}
k_\mathrm{SE}(\x, \x') = \sigma_f^2 \exp\!\left(-\frac{\|\x - \x'\|^2}{2\ell^2}\right),
\end{equation}
where $\ell$ is the length scale (controlling smoothness) and $\sigma_f^2$ is the signal variance.

\begin{figure}[ht]
\centering
\begin{tikzpicture}
\begin{axis}[
    width=10cm, height=4cm,
    xlabel={$x - x'$}, ylabel={$k_\mathrm{SE}$},
    xmin=-4, xmax=4, ymin=0, ymax=1.15,
    legend style={at={(0.98,0.95)}, anchor=north east, font=\small},
    grid=major, grid style={gray!20},
    every axis plot/.append style={thick, no markers},
]
\addplot[uzhblue, domain=-4:4, samples=100] {exp(-x^2/(2*0.5^2))};
\addlegendentry{$\ell=0.5$}
\addplot[harvardcrimson, domain=-4:4, samples=100] {exp(-x^2/(2*1.0^2))};
\addlegendentry{$\ell=1.0$}
\addplot[darkgreen, domain=-4:4, samples=100] {exp(-x^2/(2*2.0^2))};
\addlegendentry{$\ell=2.0$}
\end{axis}
\end{tikzpicture}
\caption{Squared-exponential kernel as a function of distance for three length scales.  Small $\ell$ makes correlations decay quickly and produces rougher, more local fits; large $\ell$ couples distant points and imposes smoother functions.}
\label{fig:rbf_length_scale}
\end{figure}
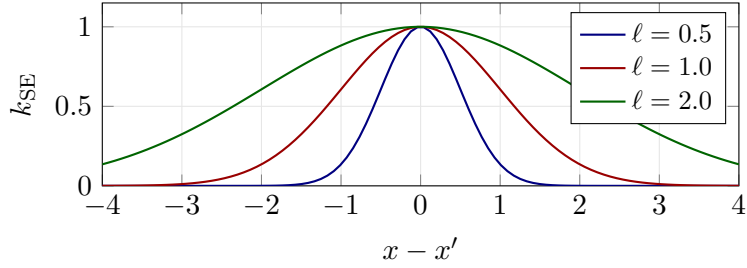

Figure~\ref{fig:rbf_length_scale} shows how the RBF length scale controls the distance over which observations remain informative.  Given training data $\mathcal{D} = \{(\x_i, y_i)\}_{i=1}^n$, let $\bm{\mu}_X = (\mu(\x_1),\ldots,\mu(\x_n))^\top$, $\mu_*=\mu(\x_*)$, and $K_y = K + \sigma_y^2 I$.  The GP posterior at a test point $\x_*$ has a closed-form latent-function mean and variance:
\begin{align}
\bar{f}_* &= \mu_* + \bm{k}_*^\top K_y^{-1}(\bm{y}-\bm{\mu}_X), \label{eq:gp_mean}\\
\sigma_{f,*}^2 &= k(\x_*, \x_*) - \bm{k}_*^\top K_y^{-1} \bm{k}_*, \label{eq:gp_var}
\end{align}
where $K$ is the kernel matrix, $\bm{k}_*$ is the vector of kernel evaluations between the test point and the training inputs, and $\sigma_y^2$ is the observation noise variance.  For a noisy future observation $y_*$, the predictive variance is $\sigma_{y,*}^2 = \sigma_{f,*}^2 + \sigma_y^2$.  The common zero-mean formulas are recovered by centering outputs or setting $\mu \equiv 0$.

\paragraph{A hand-traceable 1D example.}  To make~\eqref{eq:gp_mean}--\eqref{eq:gp_var} concrete, take $f(x) = \sin x$, observe $f$ noiselessly at $x_1 = 0$ and $x_2 = \pi$ (so $y_1 = y_2 = 0$), and query at $x_\star = \pi/2$.  Use the kernel $k(x, x') = \exp\!\bigl(-(x - x')^2/2\bigr)$ and a tiny noise floor $\sigma_y^2 = 10^{-6}$ for numerical stability.  The training kernel matrix is
\[
K + \sigma_y^2 I \;=\;
\begin{pmatrix} 1 & e^{-\pi^2/2} \\ e^{-\pi^2/2} & 1 \end{pmatrix}
\;\approx\;
\begin{pmatrix} 1.000 & 0.00719 \\ 0.00719 & 1.000 \end{pmatrix},
\]
where the off-diagonal $e^{-\pi^2/2} \approx 0.00719$ is small because $0$ and $\pi$ are far apart relative to $\ell = 1$.  The cross-covariance vector is
\[
\bm k_\star \;=\;
\begin{pmatrix} \exp(-(\pi/2)^2/2) \\ \exp(-(\pi/2)^2/2) \end{pmatrix}
\;\approx\;
\begin{pmatrix} 0.2910 \\ 0.2910 \end{pmatrix},
\]
since $x_\star = \pi/2$ is equidistant from $0$ and $\pi$.  Because $\bm y = (0,0)^\top$, the posterior mean~\eqref{eq:gp_mean} is exactly $\bar f_\star = 0$.  For the variance,
\[
(K + \sigma_y^2 I)^{-1} \bm k_\star \;\approx\; \tfrac{0.2910}{1 + 0.00719}\,(1, 1)^\top \;\approx\; (0.2890, 0.2890)^\top,
\]
so $\bm k_\star^\top (K + \sigma_y^2 I)^{-1} \bm k_\star \approx 2 \cdot 0.2910 \cdot 0.2890 \approx 0.1682$, giving $\sigma_\star^2 \approx 1 - 0.1682 \approx 0.832$ and a posterior standard deviation $\sigma_\star \approx 0.91$.  The GP predicts zero at the midpoint, with substantial residual uncertainty, consistent with the fact that $\sin(\pi/2) = 1$ is not pinned down by the two boundary observations under this length scale.

\begin{figure}[ht]
\centering
\begin{tikzpicture}
\begin{axis}[
    name=prior,
    width=6.5cm, height=4.5cm,
    title={\small\bfseries GP Prior},
    title style={text=uzhblue},
    xlabel={$x$}, xlabel style={font=\small},
    ylabel={$f(x)$}, ylabel style={font=\small},
    tick label style={font=\tiny},
    xmin=-3, xmax=3, ymin=-3, ymax=3,
    grid=major, grid style={gray!15},
]
    \addplot[softblue!20, forget plot, draw=none, fill=softblue!20]
        coordinates {(-3,-1.96) (3,-1.96) (3,1.96) (-3,1.96)} \closedcycle;
    \addplot[softblue!35, forget plot, draw=none, fill=softblue!35]
        coordinates {(-3,-1.0) (3,-1.0) (3,1.0) (-3,1.0)} \closedcycle;
    \addplot[uzhblue, ultra thick, domain=-3:3, samples=2] {0};
    \addplot[gray!60, thin, domain=-3:3, samples=80] {sin(deg(x)) + 0.3*cos(deg(2.5*x))};
    \addplot[gray!60, thin, domain=-3:3, samples=80] {-0.5*sin(deg(1.5*x)) + 0.8*cos(deg(x))};
    \addplot[gray!60, thin, domain=-3:3, samples=80] {0.7*sin(deg(0.8*x)) - 0.4*cos(deg(1.8*x))};
\end{axis}
\begin{axis}[
    at={($(prior.east)+(1.0cm,0)$)}, anchor=west,
    name=posterior,
    width=6.5cm, height=4.5cm,
    title={\small\bfseries GP Posterior},
    title style={text=harvardcrimson},
    xlabel={$x$}, xlabel style={font=\small},
    tick label style={font=\tiny},
    xmin=-3, xmax=3, ymin=-3, ymax=3,
    grid=major, grid style={gray!15},
]
    \addplot[harvardcrimson, ultra thick, domain=-3:3, samples=80] {sin(deg(x))};
    \addplot[harvardcrimson!40, dashed, thin, domain=-3:3, samples=80]
        {sin(deg(x)) + 0.2 + 0.15*x*x};
    \addplot[harvardcrimson!40, dashed, thin, domain=-3:3, samples=80]
        {sin(deg(x)) - 0.2 - 0.15*x*x};
    \addplot[only marks, mark=*, mark size=3pt, black] coordinates
        {(-2,-0.91) (-1,-0.84) (0,0) (1,0.84) (2,0.91)};
    \node[font=\tiny, text=harvardcrimson] at (axis cs:2.3,2.2) {$\pm 2\sigma$};
\end{axis}
\end{tikzpicture}
\caption{Gaussian-process prior and posterior on a 1D regression problem. \emph{Left:} the prior has constant mean (here zero) and uniform uncertainty; the shaded bands show the $68\%$ and $95\%$ credible intervals, and the thin grey curves are three sample paths drawn from the prior. \emph{Right:} after conditioning on five observations (black dots), the posterior mean (red curve) interpolates the data exactly, and the credible band collapses near the observed points while widening in unexplored regions away from the data, giving the GP its built-in uncertainty quantification.}
\label{fig:gp_prior_posterior}
\end{figure}
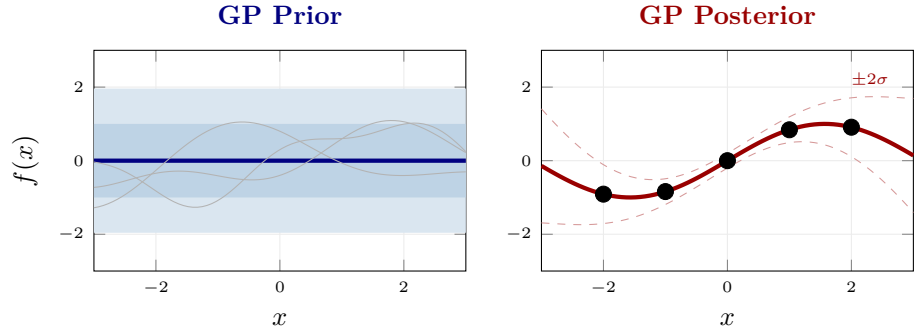

Figure~\ref{fig:gp_prior_posterior} illustrates the GP prior and posterior for a simple one-dimensional regression problem.  Before observing data, the GP prior has constant mean and uniform uncertainty.  After conditioning on five observations, the posterior mean interpolates the data and the uncertainty bands collapse near the observations while remaining wide in unexplored regions.

\begin{keyinsightbox}[Built-in uncertainty quantification]
The posterior variance $\sigma_*^2$ is small near observed data (the GP is confident) and large far from observed data (the GP is uncertain).  This property is the foundation of Bayesian active learning.
\end{keyinsightbox}

\section{Kernel Functions and Hyperparameter Learning}
\label{sec:gp_kernels}

The kernel hyperparameters $\bm{\vartheta} = (\ell, \sigma_f, \sigma_y)$ are learned by maximizing the log marginal likelihood:
\begin{equation}
\log p(\bm{y} | \X, \bm{\vartheta}) = -\frac{1}{2}(\bm{y}-\bm{\mu}_X)^\top K_y^{-1}(\bm{y}-\bm{\mu}_X) - \frac{1}{2}\log|K_y| - \frac{n}{2}\log 2\pi,
\end{equation}
where $K_y = K + \sigma_y^2 I$ and $\bm{\mu}_X$ is the prior mean evaluated on the training inputs.  In the zero-mean convention used elsewhere in this section, set $\bm{\mu}_X = 0$ (or center the outputs).

\begin{figure}[ht]
\centering
\begin{tikzpicture}
\begin{axis}[
    width=10.5cm, height=5cm,
    xlabel={model complexity (e.g.\ $1/\ell$)},
    ylabel={evidence $\log p(\bm{y}\mid\bm{\vartheta})$},
    domain=0.05:1, samples=160,
    xmin=0, xmax=1, ymin=-6.5, ymax=0,
    grid=major, grid style={gray!12},
    legend style={at={(0.98,0.04)}, anchor=south east, font=\scriptsize, draw=none, fill=none},
    every axis plot/.append style={thick}
]
\addplot[uzhblue, dashed]
        {-3*exp(-7*x) - 0.4};                          \addlegendentry{data fit}
\addplot[harvardcrimson, dotted, very thick]
        {-2.4*x - 0.4};                                \addlegendentry{complexity penalty}
\addplot[softgreen, ultra thick]
        {-3*exp(-7*x) - 2.4*x - 0.8};                  \addlegendentry{evidence (sum)}
\addplot[mark=*, mark size=2pt, only marks, softgreen]
       coordinates {(0.31,-1.89)};
\node[font=\scriptsize, anchor=south, softgreen]
       at (axis cs:0.31, -1.89) {optimum};
\node[font=\scriptsize, gray, anchor=west] at (axis cs:0.02, -5.6) {too simple};
\node[font=\scriptsize, gray, anchor=east] at (axis cs:0.98, -3.6) {too flexible};
\end{axis}
\end{tikzpicture}
\caption{Marginal-likelihood Occam's razor for a GP.  As the kernel becomes more flexible (smaller length scale~$\ell$), the data-fit term improves but the $\log|K_y|$ complexity penalty grows linearly.  Their sum, the log evidence, peaks at an interior optimum that is just expressive enough to explain the data, automatically.  No held-out validation set is required.}
\label{fig:occam_marginal_likelihood}
\end{figure}
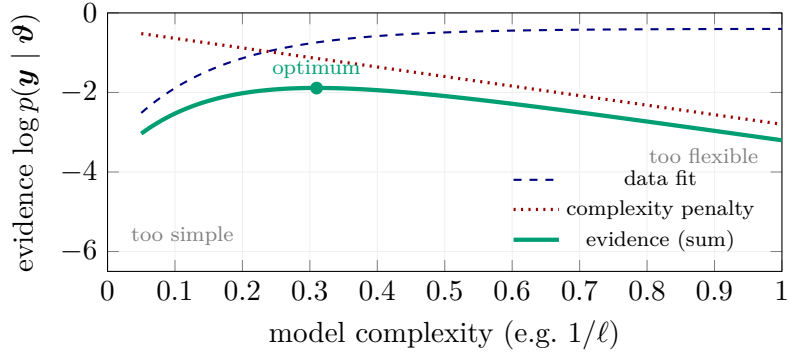

\paragraph{Why marginal likelihood?}  The log evidence $\log p(\bm y \mid \X, \bm\vartheta)$ encodes both data fit \emph{and} an automatic complexity penalty in a single closed-form expression.  The quadratic form $-\tfrac{1}{2}\bm y^\top K_y^{-1}\bm y$ rewards hyperparameters that explain the centered observations with a small inverse-covariance norm, while the log-determinant term $-\tfrac{1}{2}\log|K_y|$ penalises overly flexible kernels that admit too many possible functions, giving Bayesian Occam's razor \citep[see, e.g.,][Ch.~5]{Rasmussen:2005:GPM:1162254}.  Compared with cross-validated MSE, this approach requires no held-out split, makes use of all $n$ observations, and exposes a closed-form gradient with respect to $\bm\vartheta$, which is essential for L-BFGS-style optimization in scikit-learn / GPyTorch.  The maximum is reached at the kernel that is just expressive enough to fit the data but no more (Figure~\ref{fig:occam_marginal_likelihood}).

\paragraph{A free held-out diagnostic.}  Marginal likelihood is not the only validation tool.  The leave-one-out (LOO) predictive error of a GP admits a closed form using the same Cholesky factor already computed for posterior inference, so it costs nothing extra; we develop the formula in \S\ref{sec:gp_loo}, where it serves as an iteration-by-iteration health check inside GP-based VFI, and reuse it in \S\ref{sec:smm_gp_moments} for the GP layer over the SMM moment map.

\subsection{Kernel Composition}

More complex covariance structures can be built by composing simpler kernels.  The sum of two kernels is again a valid kernel, as is the product.  For example:
\begin{itemize}[itemsep=2pt]
\item $k_\mathrm{SE} + k_\mathrm{periodic}$: captures a smooth trend plus periodic oscillations.
\item $k_\mathrm{SE}(\ell_1) \cdot k_\mathrm{SE}(\ell_2)$: models interactions between two length scales.
\end{itemize}

\paragraph{The Mat\'ern kernel family.}\label{sec:matern}  The Mat\'ern kernel is parameterized by a smoothness parameter $\nu > 0$:
\begin{equation}
k_{\mathrm{Mat\acute{e}rn}}(r) = \sigma_f^2\,\frac{2^{1-\nu}}{\Gamma(\nu)}\left(\frac{\sqrt{2\nu}\, r}{\ell}\right)^\nu K_\nu\!\left(\frac{\sqrt{2\nu}\, r}{\ell}\right),
\end{equation}
where $r = \|\x - \x'\|$, $\ell$ is the length scale, and $K_\nu$ is the modified Bessel function of the second kind.  The smoothness parameter $\nu$ controls the regularity of sample paths: a Mat\'ern-$\nu$ GP is $k$-times mean-square differentiable for every integer $k < \nu$, with the RBF kernel recovered in the limit $\nu \to \infty$.  Important special cases:
\begin{itemize}[itemsep=2pt]
\item $\nu = 1/2$: the Ornstein--Uhlenbeck kernel $k(r) = \sigma_f^2 \exp(-r/\ell)$; sample paths are continuous but nowhere differentiable.
\item $\nu = 3/2$: $k(r) = \sigma_f^2(1 + \sqrt{3}\,r/\ell)\exp(-\sqrt{3}\,r/\ell)$; once differentiable.
\item $\nu = 5/2$: $k(r) = \sigma_f^2(1 + \sqrt{5}\,r/\ell + 5r^2/(3\ell^2))\exp(-\sqrt{5}\,r/\ell)$; twice differentiable.
\item $\nu \to \infty$: recovers the squared exponential (RBF) kernel; infinitely differentiable.
\end{itemize}
For economic applications where the target function may have kinks (e.g., due to occasionally binding constraints), the Mat\'ern-3/2 kernel is often a better choice than the infinitely smooth RBF kernel, as it does not oversmooth near non-differentiable features \citep{rennerscheidegger_2018}.

\section{Bayesian Active Learning for Sample-Efficient Training}
\label{sec:bal}

When model evaluations are expensive, we wish to select training points that provide maximal information.  Bayesian Active Learning (BAL) uses the GP posterior variance to guide this selection.  The information-theoretic foundations go back to \citet{mackay1992information}, with submodular guarantees for variance-based sensor placement established by \citet{krause2008near} and the widely used upper-confidence-bound formulation of \citet{srinivas2010gaussian}.

\paragraph{Connection to Bayesian optimization (Chapter~\ref{ch:nas}).}  BAL is the active-learning twin of the Bayesian-optimization (BO) recipe introduced in Chapter~\ref{ch:nas}: same GP surrogate, same acquisition-function machinery.  The difference is the target.  BO seeks a \emph{scalar optimum} of the surrogate (e.g.\ Expected Improvement steers samples toward $\argmax \hat f$), so its acquisition trades exploration against the chance of beating the current best.  BAL instead targets \emph{global function approximation}: it allocates samples wherever posterior variance is largest, irrespective of the predicted value.  Both reduce to the same primitive (fit GP, maximise acquisition, evaluate, refit), but the choice of acquisition reflects whether one wants the best point or the best surrogate.

\begin{definitionbox}[BAL Acquisition Function]
\begin{equation}
U(\x) = w_{\mathrm{obj}} \cdot \mu(\x) + \frac{w_{\mathrm{var}}}{2}\log\sigma^2(\x),
\end{equation}
where $\mu(\x)$ and $\sigma^2(\x)$ are the GP posterior mean and variance, $w_{\mathrm{obj}}$ is the objective-following (or Bayesian-optimization) weight, and $w_{\mathrm{var}}$ controls exploration.  This logarithmic-variance form is the one used by \citet{rennerscheidegger_2018} in economic applications; it is a non-standard relative of GP-UCB \citep{srinivas2010gaussian}.  When the goal is global surrogate accuracy rather than optimization, integrated posterior variance or expected error reduction can be more natural objectives.  The weights $w_{\mathrm{obj}}$ and $w_{\mathrm{var}}$ are local to the BAL context and distinct from the discount factor $\beta$ and shock-persistence notation used elsewhere in the script.
\end{definitionbox}

\begin{definitionbox}[Algorithm: Bayesian Active Learning]
\begin{algorithmic}
\small
\STATE \textbf{Input:} Initial design $\mathcal{D}_0 = \{(\x_i, y_i)\}_{i=1}^{n_0}$, budget $N$, kernel $k$, acquisition weights $(w_{\mathrm{obj}}, w_{\mathrm{var}})$
\STATE Fit GP on $\mathcal{D}_0$: learn hyperparameters via marginal likelihood
\FOR{$n = n_0+1, \ldots, N$}
    \STATE Compute acquisition function: $U(\x) = w_{\mathrm{obj}} \cdot \mu_{n-1}(\x) + \frac{w_{\mathrm{var}}}{2}\log\sigma_{n-1}^2(\x)$
    \STATE Select next point: $\x_n = \argmax_{\x} U(\x)$
    \STATE Evaluate expensive model: $y_n = f(\x_n)$
    \STATE Update dataset: $\mathcal{D}_n = \mathcal{D}_{n-1} \cup \{(\x_n, y_n)\}$
    \STATE Re-fit GP on $\mathcal{D}_n$ (update hyperparameters every $k$ iterations)
\ENDFOR
\STATE \textbf{Output:} Trained GP surrogate $\hat{f}(\cdot)$ with posterior mean $\mu_N$ and variance $\sigma_N^2$
\end{algorithmic}
\end{definitionbox}

The BAL algorithm concentrates training points near kinks, boundary layers, and other regions where the function is hardest to approximate, achieving the same accuracy as uniform sampling with far fewer evaluations \citep{rennerscheidegger_2018}.  The exploration--exploitation trade-off controlled by $(w_{\mathrm{obj}}, w_{\mathrm{var}})$ ensures that the algorithm balances refining the approximation in already well-sampled regions against exploring uncharted territory.

\begin{figure}[ht]
\centering
\includegraphics[width=\textwidth]{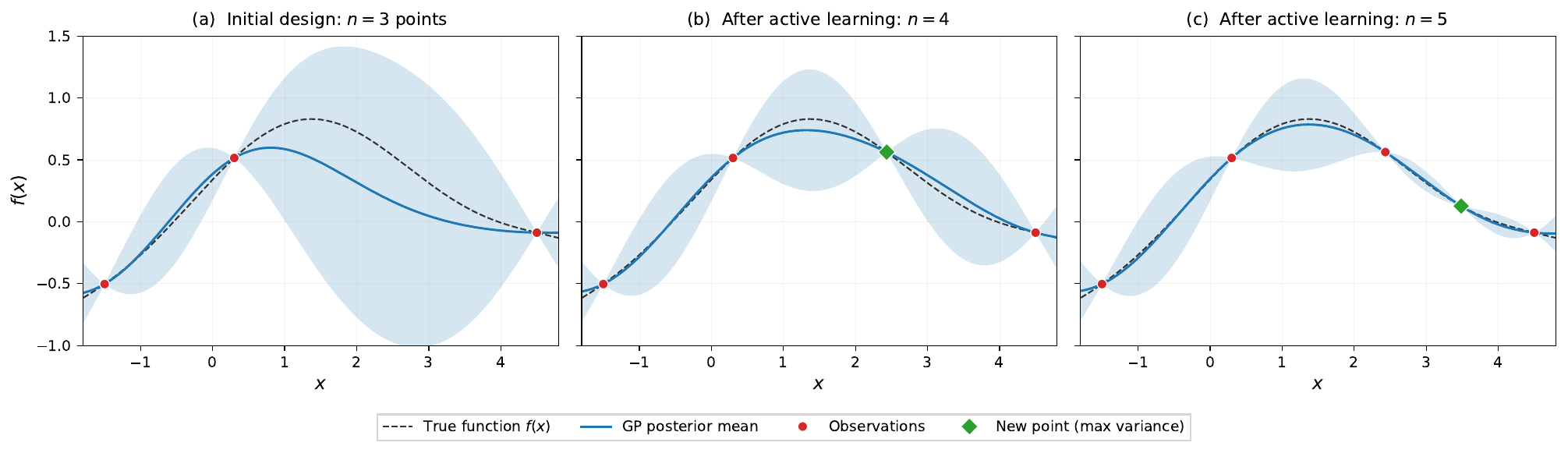}
\caption{Bayesian Active Learning in action.  \textbf{(a)}~Starting from three initial observations (red dots), the GP posterior mean (blue line) deviates from the true function (dashed black) in the data-sparse region, where the 95\% credible band (blue shading) is wide.  \textbf{(b)}~The acquisition function selects the point of maximum posterior variance (green diamond); after evaluation, the posterior tightens locally and the mean improves.  \textbf{(c)}~A second active-learning iteration fills the remaining gap.  With only five strategically chosen points, the GP posterior closely tracks the true function across the entire domain.}
\label{fig:bal-iterations}
\end{figure}

The intuition behind the acquisition function is as follows.  The first term $w_{\mathrm{obj}} \cdot \mu(\x)$ favors regions where the predicted function value is large (exploitation: sample where the function is interesting).  The second term $\frac{w_{\mathrm{var}}}{2}\log\sigma^2(\x)$ favors regions of high uncertainty (exploration: sample where we know least).  This acquisition function belongs to the Upper Confidence Bound (UCB) family \citep{srinivas2010gaussian}: with $w_{\mathrm{obj}} = 0$ it has the same maximizers as pure posterior-variance sampling, while the logarithmic variance weighting provides a more conservative exploration bonus than the standard-deviation weighting used in GP-UCB when it is combined with exploitation.  By adjusting $w_{\mathrm{obj}}$ and $w_{\mathrm{var}}$, the practitioner can control the balance.  For economic applications where the entire domain is relevant (e.g., approximating a policy function), a pure exploration strategy ($w_{\mathrm{obj}} = 0$, $w_{\mathrm{var}} > 0$) is often appropriate, reducing BAL to an uncertainty sampling scheme that minimizes the integrated posterior variance.  Figure~\ref{fig:bal-iterations} illustrates two BAL iterations on a 1D toy.

\phantomsection
\section{When to Use GPs vs.\ DNNs}
\label{sec:gp_vs_dnn}

Now that the GP machinery (posterior inference, kernel design, and BAL) has been introduced, we can give a more detailed comparison than the overview in Table~\ref{tab:surrogate_strategy_comparison}.  Active subspaces, introduced in Section~\ref{sec:active_subspaces}, are the main tool for pushing GP surrogates beyond the moderate-dimensional regime.  Table~\ref{tab:gp_vs_dnn_surrogates} extends Table~\ref{tab:surrogate_strategy_comparison} with the GP-specific items (LOO diagnostics, marginal-likelihood Occam, BAL).

\begin{table}[ht]
\centering
\small
\begin{tabular}{@{} L{3.2cm} L{5.6cm} L{5.6cm} @{}}
\toprule
\textbf{Criterion} & \textbf{Gaussian Processes} & \textbf{Deep Neural Networks} \\
\midrule
Training cost & $\mathcal{O}(N^3)$; exact for $N$ in the low thousands\textsuperscript{$\dagger$} & $\mathcal{O}(N)$ per epoch; scales to $N \sim 10^6$ \\
Prediction cost & $\mathcal{O}(N)$ posterior mean, $\mathcal{O}(N^2)$ posterior variance, per test point & $\mathcal{O}(\text{weights})$ per forward pass; independent of $N$ \\
Gradient access & closed-form from posterior & autodiff (exact, any order) \\
Non-smooth features & Mat\'ern-$3/2$ adapts well & excellent with enough data \\
High-dim.\ extension & active subspaces ($d \gg 10$) & native \\
\bottomrule
\end{tabular}
\caption{GP and DNN surrogate trade-offs after the basic GP machinery has been introduced.  GPs are sample-efficient and uncertainty-aware but expensive in the number of training points; DNNs scale to much larger datasets and dimensions but need more data and separate machinery for calibrated uncertainty.\\ \footnotesize $\dagger$\,Sparse-GP via inducing points reduces $\mathcal{O}(N^3)$ to $\mathcal{O}(Nm^2 + m^3)$ for $m \ll N$ inducing inputs \citep{titsias2009variational, hensman2013gaussian}; see the inducing-point remarkbox below.}
\label{tab:gp_vs_dnn_surrogates}
\end{table}

\medskip\noindent\textbf{Practical guidelines.}
\begin{itemize}[itemsep=2pt]
\item \textbf{Choose a GP} when each model evaluation is expensive (minutes to hours per solve), the effective dimensionality is moderate ($d \lesssim 15$, or higher with active subspaces), and uncertainty quantification on the surrogate is needed, e.g., for reporting confidence intervals on estimated parameters or for guiding further data collection via BAL.
\item \textbf{Choose a DNN} when training data is cheap to generate, the input dimension is high ($d \gg 10$), and the downstream task requires millions of fast evaluations, e.g., Monte Carlo simulation, grid search over a large parameter space, or real-time inference.
\item \textbf{Combine both} when the problem has a natural two-stage structure: use a GP with BAL to build a small, high-quality training set with uncertainty estimates, then train a DNN on the resulting dataset for fast large-scale deployment.  The active subspace approach of \citet{SCHEIDEGGER201968} is particularly effective in this setting, extending GP methods far beyond their na\"ive scaling limits.
\end{itemize}

\subsection{GP Regression in Practice}

In code, GP regression is a one-liner once the kernel is chosen.  In scikit-learn the standard pattern is to assemble a kernel as the sum of an RBF (\tpath{RBF(length_scale=...)}) and a noise term (\tpath{WhiteKernel(noise_level=...)}), pass it to \tpath{GaussianProcessRegressor}, call \tpath{.fit(X_train, y_train)}, and obtain posterior mean and standard deviation from \tpath{.predict(X_test, return_std=True)}; the kernel hyperparameters (\tpath{length_scale}, \tpath{noise_level}, output amplitude) are optimized by maximising the marginal likelihood, with \tpath{n_restarts_optimizer} controlling robustness to local optima.  The companion notebook \tpath{02_GP_and_BAL.ipynb} provides a full worked example fitting noisy observations of $\sin(x)$ on $[-2,2]$.

\paragraph{Application: GP surrogates for option pricing.}
GPs are particularly well suited as surrogates for derivative pricing models.  For example, one can train a GP on as few as 5--50 Black--Scholes option prices (evaluated at different spot prices or parameter configurations) and obtain a surrogate that accurately reproduces the pricing surface with calibrated uncertainty bands.  The posterior variance immediately quantifies the interpolation uncertainty at each query point.  This idea extends naturally to stochastic volatility models such as Heston, where the analytical pricing formula is expensive to evaluate.  Furthermore, because GP predictions are linear in the training targets, the uncertainty of a \emph{portfolio} of GP-priced instruments propagates analytically: for a linear portfolio $\sum_i w_i \hat{V}_i$ with vector of weights $\bm{w}$ and joint posterior covariance $\Sigma_{\hat{V}}$, $\mathrm{Var}(\bm{w}^\top \hat{V}) = \bm{w}^\top \Sigma_{\hat{V}} \bm{w}$.  When the surrogate errors are independent across instruments, $\Sigma_{\hat{V}}$ is diagonal with entries $\sigma_i^2$ and the formula reduces to $\sum_i w_i^2 \sigma_i^2$; otherwise the off-diagonal cross-instrument covariances must be retained, e.g.\ via a multi-output GP.  Either way the assessment is instant.

\begin{figure}[ht]
\centering
\begin{tikzpicture}
\begin{axis}[
    width=10cm, height=5.2cm,
    name=posterior,
    xlabel={$x$}, ylabel={$f(x)$},
    xmin=-3.2, xmax=3.2, ymin=-1.4, ymax=1.4,
    grid=major, grid style={gray!12},
    legend style={
        at={(1.02,0.5)}, anchor=west, font=\small,
        draw=gray!40, fill=white, fill opacity=0.95, text opacity=1,
        cells={anchor=west}, row sep=1pt,
    },
    every axis plot/.append style={no markers},
    enlarge x limits=0.03,
]
\addplot[name path=exactupper, draw=none, domain=-3.2:3.2, samples=180]
  {sin(deg(1.4*x))*exp(-0.06*x*x) + 2*(0.045 + 0.015*abs(sin(deg(2.8*x))))};
\addplot[name path=exactlower, draw=none, domain=-3.2:3.2, samples=180]
  {sin(deg(1.4*x))*exp(-0.06*x*x) - 2*(0.045 + 0.015*abs(sin(deg(2.8*x))))};
\addplot[softblue!18, forget plot] fill between[of=exactupper and exactlower];

\addplot[name path=sparseupper, draw=none, domain=-3.2:3.2, samples=180]
  {sin(deg(1.4*x))*exp(-0.06*x*x)
   + 0.06*cos(deg(2.1*x))*exp(-0.22*x*x)
   - 0.035*sin(deg(3.2*x))*exp(-0.18*(x-2.1)*(x-2.1))
   + 2*(0.060 + 0.025*abs(cos(deg(1.7*x))))};
\addplot[name path=sparselower, draw=none, domain=-3.2:3.2, samples=180]
  {sin(deg(1.4*x))*exp(-0.06*x*x)
   + 0.06*cos(deg(2.1*x))*exp(-0.22*x*x)
   - 0.035*sin(deg(3.2*x))*exp(-0.18*(x-2.1)*(x-2.1))
   - 2*(0.060 + 0.025*abs(cos(deg(1.7*x))))};
\addplot[harvardcrimson!12, forget plot] fill between[of=sparseupper and sparselower];

\addplot[uzhblue, thick, domain=-3.2:3.2, samples=180]
  {sin(deg(1.4*x))*exp(-0.06*x*x)};
\addlegendentry{exact GP mean}
\addplot[harvardcrimson, thick, dashed, domain=-3.2:3.2, samples=180]
  {sin(deg(1.4*x))*exp(-0.06*x*x)
   + 0.06*cos(deg(2.1*x))*exp(-0.22*x*x)
   - 0.035*sin(deg(3.2*x))*exp(-0.18*(x-2.1)*(x-2.1))};
\addlegendentry{sparse GP mean}

\addplot[only marks, mark=*, mark size=1.4pt, gray!65]
  coordinates {
   (-3.0,0.55) (-2.7,0.34) (-2.4,0.18) (-2.1,-0.12) (-1.8,-0.45)
   (-1.5,-0.78) (-1.2,-0.93) (-0.9,-0.88) (-0.6,-0.75) (-0.3,-0.38)
   (0.0,0.03) (0.3,0.42) (0.6,0.70) (0.9,0.93) (1.2,0.88)
   (1.5,0.78) (1.8,0.46) (2.1,0.18) (2.4,-0.12) (2.7,-0.40)
   (3.0,-0.48)};
\addlegendentry{$n=21$ data}

\addplot[only marks, mark=|, mark size=7pt, very thick, harvardcrimson]
  coordinates {(-2.4,-1.28) (-0.6,-1.28) (1.2,-1.28) (2.7,-1.28)};
\addlegendentry{$m=4$ inducing inputs $\bm Z$}
\node[anchor=west, font=\scriptsize, harvardcrimson] at (axis cs:-3.1,-1.28) {$Z$:};
\end{axis}

\begin{axis}[
    at={(posterior.south west)}, anchor=north west, yshift=-0.85cm,
    width=10cm, height=2.25cm,
    xlabel={$x$}, ylabel={abs. diff.},
    xmin=-3.2, xmax=3.2, ymin=0, ymax=0.08,
    grid=major, grid style={gray!12},
    tick label style={font=\small},
    ytick={0,0.04,0.08},
]
\addplot[harvardcrimson, thick, domain=-3.2:3.2, samples=180]
  {abs(0.06*cos(deg(2.1*x))*exp(-0.22*x*x)
       - 0.035*sin(deg(3.2*x))*exp(-0.18*(x-2.1)*(x-2.1)))};
\end{axis}
\end{tikzpicture}
\caption{Inducing-point intuition.  Exact GP inference conditions on all $n=21$ observations and uses the full $n\times n$ kernel matrix.  Sparse variational GP methods introduce $m\ll n$ inducing inputs $\bm Z$ and approximate the posterior through the low-rank structure induced by $K_{nm}K_{mm}^{-1}K_{mn}$.  The top panel compares the exact posterior mean and uncertainty band with a sparse approximation using $m=4$ pseudo-inputs; the bottom panel shows the absolute difference between the two means.  In this smooth one-dimensional illustration the approximation is close, but its quality depends on $m$, the kernel hyperparameters, and the placement of $\bm Z$.  The dominant training cost falls from $\mathcal{O}(n^3)$ to $\mathcal{O}(nm^2 + m^3)$; the variational formulation of \citet{titsias2009variational} optimizes $\bm Z$ jointly with the kernel hyperparameters.}
\label{fig:inducing_points}
\end{figure}
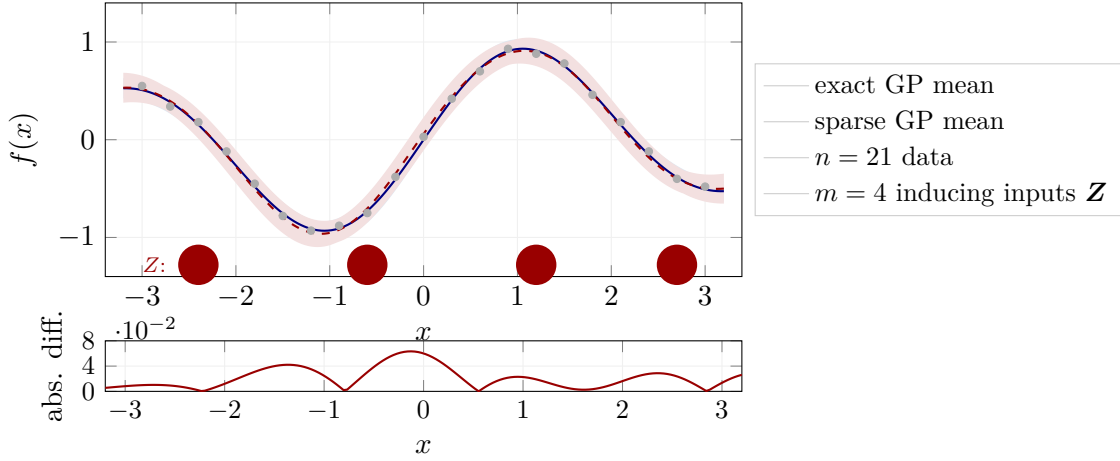

\begin{remarkbox}[Computational scaling of GPs]
The main limitation of GPs is their $\mathcal{O}(n^3)$ training cost, arising from the inversion of the $n \times n$ kernel matrix.  For $n > 10{,}000$, exact GP inference becomes impractical.  Approximate methods, such as variational sparse GPs with inducing points \citep{titsias2009variational}, stochastic-variational GPs \citep{hensman2013gaussian}, random Fourier features, and structured kernel interpolation, can extend the range to $n \sim 10^5$, but for truly large-scale problems, deep neural networks remain the method of choice.  The BAL framework mitigates this limitation by keeping $n$ small through intelligent sample selection.

\medskip
The \emph{inducing-point} idea (Figure~\ref{fig:inducing_points}) is simple: instead of carrying the full $n\times n$ kernel matrix, summarize the dataset with a much smaller set of $m \ll n$ pseudo-inputs $\bm Z = \{\bm z_1,\dots,\bm z_m\}$ and replace $K_{nn}$ by the Nystr\"om-style low-rank approximation $K_{nm}K_{mm}^{-1}K_{mn}$.  Training and prediction then cost about $\mathcal{O}(nm^2 + m^3)$ rather than $\mathcal{O}(n^3)$, with the $m^3$ term coming from the small inducing-point block.  The variational formulation of \citet{titsias2009variational} additionally treats $\bm Z$ as parameters to be optimized against the marginal likelihood, so the inducing inputs migrate to wherever the GP actually needs resolution.
\end{remarkbox}

\section{Scaling GPs to High Dimensions: Active Subspaces}
\label{sec:active_subspaces}

A common criticism of GP methods is that they are limited to moderate-dimensional problems ($d \leq 10$).  While this is true for na\"ive implementations, where the number of training points needed to cover the input space grows exponentially in $d$, \citet{SCHEIDEGGER201968} show that \emph{active subspace} methods can mitigate this barrier \emph{when the target function exhibits a clear spectral gap in the gradient outer-product matrix $C$ of \eqref{eq:active_subspace} below}, i.e., when its variation is concentrated on a low-dimensional linear subspace of the input space.  If $f$ depends comparably on all $d$ coordinates, the spectral gap is absent and the technique offers little gain; sufficient conditions are in \citet{constantine2015active}.  For a textbook treatment of the active subspace framework and references to its origins, see \citet{constantine2015active}; we summarize the key ideas and their application to dynamic economic models below.

\paragraph{Intuition: why most directions don't matter.}
Consider a value function $V(\x)$ in a dynamic stochastic economy with $d$ state variables.  Although $V$ is formally a function of all $d$ inputs, it often responds primarily to a few linear combinations of them.  For example, in a multi-country model, the value function may depend mostly on \emph{aggregate} capital $\sum_j k^j$ and a measure of \emph{inequality} $\mathrm{Var}(k^j)$, rather than on each country's capital individually.  If we can identify these important directions, we can project the $d$-dimensional input onto a much lower-dimensional subspace and fit the GP there, substantially mitigating the practical grid-based curse of dimensionality.

\paragraph{The gradient outer product matrix.}
The key diagnostic tool is the \emph{gradient outer product matrix} \citep[Ch.~1]{constantine2015active}.  Given a function $f\colon \R^d \to \R$ with input distribution $\pi(\x)$, define:
\begin{equation}
C = \int \nabla f(\x) \, \nabla f(\x)^\top \, \pi(\x)\, d\x \;\in\; \R^{d \times d}.
\label{eq:active_subspace}
\end{equation}
The matrix $C$ is symmetric and positive semi-definite.  Its $(i,j)$-entry measures how much $f$ co-varies along input directions $i$ and $j$, averaged over the input distribution.  In particular:
\begin{itemize}[itemsep=2pt]
\item If $\partial f / \partial x_i$ is consistently large across the input space, the $i$-th diagonal entry of $C$ is large, meaning direction $i$ is important.
\item If two directions always change $f$ together, the corresponding off-diagonal entry is large, as they form part of the same important linear combination.
\item If $\partial f / \partial x_i \approx 0$ everywhere, then direction $i$ contributes nothing to $C$, so it is irrelevant and can be projected away.
\end{itemize}

\paragraph{Worked toy example.}  Take $f(x_1, x_2) = x_1^2$ on $\x \sim \mathrm{Unif}([-1,1]^2)$.  Then $\nabla f = (2x_1, 0)^\top$, and integrating against the uniform measure gives $C = \mathrm{diag}(4/3,\,0)$.  The eigenvalues are $\lambda_1 = 4/3$ and $\lambda_2 = 0$; the first eigenvector is $(1,0)^\top$, the second is $(0,1)^\top$; the spectral gap is infinite.  The active subspace is the $x_1$-axis, and $f$ is recovered exactly as $f = g(x_1)$ with $g(u) = u^2$.  This is the cleanest instance of the projection in~\eqref{eq:linear_as_ansatz} below.

\paragraph{Eigendecomposition and the spectral gap.}
Let $C = U \Lambda U^\top$ be the eigendecomposition, with eigenvalues $\lambda_1 \geq \lambda_2 \geq \cdots \geq \lambda_d \geq 0$.  The eigenvectors reveal the directions of maximal average variation; the eigenvalues quantify how much variation each direction captures.  Each diagonal entry of $C$ is the average squared sensitivity of $f$ along that input direction, so eigenvectors of $C$ are directions of \emph{correlated} sensitivity: when the top $m$ eigenvalues capture most of the trace, $f$ varies primarily along their span and is approximately constant in the perpendicular $(d-m)$-dimensional complement.  If there is a clear \emph{spectral gap}, i.e., $\lambda_m \gg \lambda_{m+1}$, then the first $m$ eigenvectors $U_m = [u_1, \ldots, u_m]$ span the \emph{active subspace}, and $f$ is well approximated by a function of the reduced coordinates $\tilde{\x} = U_m^\top \x \in \R^m$ alone.
\begin{equation}
f(\x) \;\approx\; g(U_m^\top \x),
\qquad U_m = [u_1,\ldots,u_m].
\label{eq:linear_as_ansatz}
\end{equation}

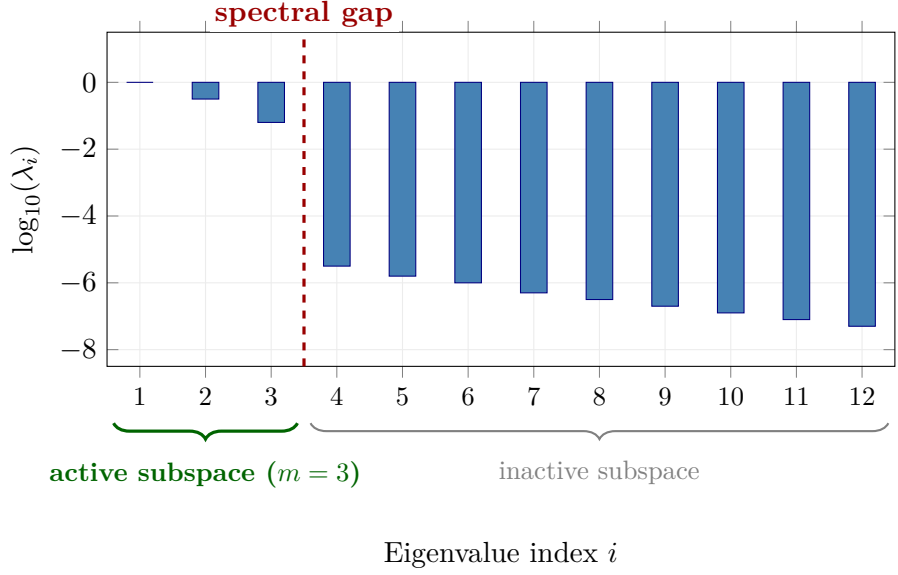
\begin{figure}[ht]
\centering
\begin{tikzpicture}
\begin{axis}[
    width=12cm, height=6.0cm,
    xlabel={Eigenvalue index $i$},
    xlabel style={yshift=-44pt},
    ylabel={$\log_{10}(\lambda_i)$},
    xmin=0.5, xmax=12.5,
    ymin=-8.5, ymax=1.5,
    xtick={1,2,3,4,5,6,7,8,9,10,11,12},
    xticklabel style={font=\small},
    ytick={0,-2,-4,-6,-8},
    grid=major, grid style={gray!15},
    ybar, bar width=10pt,
    nodes near coords={},
    clip=false
]
\addplot[fill=softblue, draw=uzhblue] coordinates {
    (1,0) (2,-0.5) (3,-1.2)
    (4,-5.5) (5,-5.8) (6,-6.0) (7,-6.3) (8,-6.5)
    (9,-6.7) (10,-6.9) (11,-7.1) (12,-7.3)
};
\draw[dashed, harvardcrimson, very thick] (axis cs:3.5,-8.5) -- (axis cs:3.5,1.5);
\node[font=\normalsize\bfseries, harvardcrimson, fill=white, inner sep=2pt,
      anchor=south] at (axis cs:3.5,1.5) {spectral gap};
\draw[decorate, decoration={brace, amplitude=6pt, mirror}, very thick, darkgreen]
    (axis cs:0.6,-10.2) -- (axis cs:3.4,-10.2)
    node[midway, below=10pt, font=\small\bfseries, darkgreen] {active subspace ($m=3$)};
\draw[decorate, decoration={brace, amplitude=6pt, mirror}, thick, gray]
    (axis cs:3.6,-10.2) -- (axis cs:12.4,-10.2)
    node[midway, below=10pt, font=\small, gray] {inactive subspace};
\end{axis}
\end{tikzpicture}
\caption{Spectral decay of the active-subspace eigenvalues for a schematic example with $d=12$ state variables. The first three eigenvalues are orders of magnitude larger than the rest; the dashed red line marks the spectral gap after $\lambda_3$, which indicates that $f$ effectively lives on a 3-dimensional active subspace (green brace) and reduces the GP regression problem from $\R^{12}$ to $\R^3$.  The remaining nine directions form the inactive subspace (gray brace), along which $f$ is nearly constant on average.}
\label{fig:active_subspace_spectrum}
\end{figure}

\paragraph{Computing the gradient.}
In practice, the integral in~\eqref{eq:active_subspace} is estimated from a finite sample of gradient evaluations \citep[Ch.~1.5]{constantine2015active} (this is a one-time pre-processing cost, $\mathcal{O}(N_g\, d)$ with $N_g \sim 2d$--$10d$, far smaller than the surrogate training set of size $N$):
\begin{equation}
\hat{C} = \frac{1}{N_g}\sum_{i=1}^{N_g} \nabla f(\x_i)\,\nabla f(\x_i)^\top, \qquad \x_i \sim \pi(\x).
\label{eq:empirical_C}
\end{equation}
The gradient $\nabla f(\x_i)$ can be obtained via:
\begin{itemize}[itemsep=2pt]
\item \textbf{Automatic differentiation} through the model solver (if differentiable, e.g., a DEQN or PINN).
\item \textbf{Finite differences}: evaluate $f$ at $d+1$ perturbed inputs per sample point.  This costs $(d+1)\cdot N_g$ model evaluations, but $N_g$ can be modest (typically $N_g \sim 2d$ to $10d$ suffices for a reliable estimate of the leading eigenspace).
\end{itemize}
The eigendecomposition of $\hat{C}$ is a standard $d \times d$ matrix operation and is cheap relative to the model evaluations.  When $\{\x_i\}$ are i.i.d.\ from $\pi$, $\hat{C} \to C$ almost surely as $N_g \to \infty$ by the law of large numbers, so the eigenvectors of $\hat{C}$ are consistent estimates of the population active subspace; finite-sample bias is documented in \citet{constantine2015active}.

\paragraph{Practical workflow.}  The full dimension-reduction pipeline consists of three steps:
\begin{enumerate}[itemsep=2pt]
\item \textbf{Gradient sampling:} Draw $N_g$ points $\x_i \sim \pi(\x)$ and compute $\nabla f(\x_i)$ at each.  Form the empirical estimate $\hat{C}$ via~\eqref{eq:empirical_C}.
\item \textbf{Eigendecomposition and dimension selection:} Compute $\hat{C} = \hat{U}\hat{\Lambda}\hat{U}^\top$.  Inspect the eigenvalue decay, as in Figure~\ref{fig:active_subspace_spectrum}, and choose $m$ at the spectral gap.
\item \textbf{GP in reduced space:} Project all training inputs onto the active subspace: $\tilde{\x}_i = \hat{U}_m^\top \x_i \in \R^m$.  Fit a standard GP on the $m$-dimensional projected data.
\end{enumerate}
When combined with BAL, the GP adaptively selects training points in the reduced space, further improving sample efficiency.

\begin{figure}[ht]
\centering
\begin{tikzpicture}[
    mlstep/.style={rectangle, draw=uzhblue, fill=uzhgreylight,
        minimum width=2.2cm, minimum height=0.7cm, font=\footnotesize,
        rounded corners=2pt, thick, align=center},
    arr/.style={-{Stealth[length=2mm]}, thick, uzhblue}
]
\node[mlstep, fill=softorange!25] (grad) {Gradient\\sampling};
\node[mlstep, right=0.6cm of grad, fill=softblue!20] (eigen) {Eigen-\\decomposition};
\node[mlstep, right=0.6cm of eigen, fill=softgreen!20] (proj) {Project onto\\$\R^m$};
\node[mlstep, right=0.6cm of proj] (gp) {GP + BAL\\in $\R^m$};
\node[mlstep, right=0.6cm of gp, fill=red!8] (surr) {Surrogate\\$\hat{f}(\x)$};
\draw[arr] (grad) -- (eigen);
\draw[arr] (eigen) -- (proj);
\draw[arr] (proj) -- (gp);
\draw[arr] (gp) -- (surr);
\node[below=0.15cm of grad, font=\tiny, text=uzhgreydark] {$N_g$ evaluations of $\nabla f$};
\node[below=0.15cm of eigen, font=\tiny, text=uzhgreydark] {find spectral gap};
\node[below=0.15cm of proj, font=\tiny, text=uzhgreydark] {$\tilde{\x} = U_m^\top \x$};
\node[below=0.15cm of gp, font=\tiny, text=uzhgreydark] {active learning};
\node[below=0.15cm of surr, font=\tiny, text=uzhgreydark] {fast evaluation};
\end{tikzpicture}
\caption{Linear active-subspace pipeline.  Gradient samples identify the dominant eigenspace of the gradient outer-product matrix, all simulator inputs are projected to the reduced coordinates $\tilde{\x}=U_m^\top \x$, and the GP/BAL loop is then run in the low-dimensional active subspace.}
\label{fig:active_subspace_pipeline}
\end{figure}
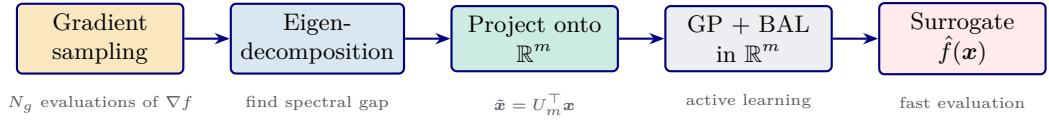

Figure~\ref{fig:active_subspace_pipeline} summarizes the linear active-subspace workflow used before fitting the GP.

\paragraph{Application to dynamic stochastic economies.}
\citet{SCHEIDEGGER201968} apply this pipeline to high-dimensional dynamic programming problems arising in neoclassical growth models.  Their key findings:
\begin{itemize}[itemsep=2pt]
\item Models with up to $d = 500$ continuous state variables can be solved, far beyond the reach of grid-based methods or na\"ive GP implementations.
\item The active subspace typically reduces the effective dimension to $m = 2$--$5$, even when $d$ is in the hundreds.  The value function's dependence on the full state vector collapses onto a few aggregate quantities (e.g., mean capital, dispersion of productivity).
\item The state space is partitioned into clusters via Bayesian Gaussian mixture models, and a separate GP with BAL is fit within each cluster.  This local approximation strategy handles non-stationarities in the value function (e.g., different curvature in high- vs.\ low-capital regions).
\item Parallel computing distributes the gradient evaluations and GP fits across multiple processors, enabling the method to scale to large-scale models on high-performance computing clusters.
\end{itemize}

\paragraph{Anchor on the multi-sector growth benchmark.}
In the multi-sector neoclassical growth benchmark of \citet{SCHEIDEGGER201968}, the active-subspace dimension collapses to $m = 1$ even at $D = 500$, so the entire ASGP pipeline operates on a one-dimensional projected coordinate.  On models with other configurations the active subspace can be larger; the spectral-gap test is what decides.

\subsection{Nonlinear Generalization: Deep Active Subspaces}
\label{sec:deep_as}

The linear pipeline of the previous section assumes that the important directions are \emph{linear} combinations of the input variables.  That is an inductive bias: the first $m$ columns of $U$ always span a linear subspace of $\R^d$, so if the response $f$ actually varies along a \emph{curved} low-dimensional manifold (a ridge, a product of two linear features, a radial coordinate, a hidden regime indicator), the linear projection has to carry enough directions to cover the whole manifold, not the manifold itself.  Concretely, if $f(\xi) = \varphi\!\bigl((w_1^\top \xi)^2 + (w_2^\top \xi)^2\bigr)$ is a radial function of two linear features, the gradient $\nabla f = 2\,\varphi'(\cdot)\,(s_1 w_1 + s_2 w_2)$ spans a \emph{two}-dimensional space when averaged over $\xi$, so linear AS returns $m = 2$; yet the intrinsic ``interesting'' coordinate is the scalar aggregate $r^2 = s_1^2 + s_2^2$, which is one-dimensional.  \citet{tripathy2018deep} remove the linearity bias by replacing the linear projection with a learned nonlinear encoder and thereby recover the one-dimensional structure.

\paragraph{The deep active-subspace ansatz.}
Tripathy and Bilionis parametrize the surrogate as a composition of two small neural networks,
\begin{equation}
\hat f(\xi) \;=\; g\bigl(h(\xi)\bigr),
\qquad
h\colon \R^D \!\to\! \R^d,
\qquad
g\colon \R^d \!\to\! \R,
\label{eq:deep_as_ansatz}
\end{equation}
where $h$ is a multilayer perceptron (MLP) that plays the role of the projection matrix $U_m^\top$ in the linear case, and $g$ is a second MLP that plays the role of the GP link.  Setting $h(\xi) = U_m^\top \xi$ and taking $g$ to be a polynomial recovers the linear active-subspace surrogate~\eqref{eq:linear_as_ansatz}, so~\eqref{eq:deep_as_ansatz} generalizes the linear-AS ansatz rather than the gradient covariance matrix~\eqref{eq:active_subspace} itself.  The two networks are trained \emph{jointly} to minimize the residual $\hat f(\xi_i) - y_i$ on an input-output sample, so that the bottleneck $h(\xi) \in \R^d$ adapts to the specific response surface rather than being fixed in advance by the spectrum of $C$.

\paragraph{Architecture choices.}
Three design choices make~\eqref{eq:deep_as_ansatz} trainable with a few hundreds of samples:
\begin{itemize}[itemsep=2pt]
\item \textbf{Exponentially decaying encoder widths} \citep[Eq.~20]{tripathy2018deep}.  For an encoder with $L$ layers mapping $\R^D$ to $\R^d$, choose widths
\begin{equation}
d_k \;=\; \bigl\lceil D \, e^{\eta_{\mathrm{w}} k} \bigr\rceil, \qquad \eta_{\mathrm{w}} = \tfrac{1}{L}\log(d/D), \qquad k = 0, 1, \ldots, L,
\label{eq:tb_widths}
\end{equation}
so that the bottleneck closes smoothly from $D$ to $d$ without a brittle hyperparameter choice.  For $D = 20,\, L = 3,\, d = 1$ this gives widths $[20, 8, 3, 1]$; for $d = 2$, widths $[20, 10, 5, 2]$ -- a recipe rather than a search.
\item \textbf{Swish activation} \citep[Eq.~10]{tripathy2018deep},
\begin{equation}
\sigma(z) \;=\; \frac{z}{1 + e^{-\gamma z}}, \qquad \gamma = 1,
\label{eq:swish}
\end{equation}
which is smooth everywhere (unlike ReLU) and non-saturating (unlike $\tanh$); smoothness matters because we will want to differentiate through $\hat f$ for sensitivity analysis and because the elastic-net penalty below otherwise has to fight the sharp corners of ReLU.
\item \textbf{Elastic-net regularization on every weight matrix} \citep[Eq.~12]{tripathy2018deep}.  Writing $\theta = (\theta_h, \theta_g)$ for all encoder and link parameters, the training loss is
\begin{equation}
\mathcal{L}(\theta) \;=\; \frac{1}{N}\sum_{i=1}^N \bigl(\hat f_\theta(\xi_i) - y_i\bigr)^2
\;+\; \lambda_1 \lVert \theta \rVert_1
\;+\; \lambda_2 \lVert \theta \rVert_2^2,
\label{eq:tb_loss}
\end{equation}
with small $\lambda_1, \lambda_2$; the $\ell_1$ term encourages sparse weights and input usage, the $\ell_2$ term controls the overall smoothness of $\hat f_\theta$ and prevents the loss landscape from becoming pathological as $N$ shrinks.
\end{itemize}
Crucially, \emph{no gradient samples of $f$} are required: $\theta$ is learned from the input-output pairs $\{(\xi_i, y_i)\}$ alone.  The orthogonality constraint $U^\top U = I$ that is implicit in the linear case becomes unnecessary, because the learned encoder is free to pick any smooth low-dimensional parameterization of the active manifold.

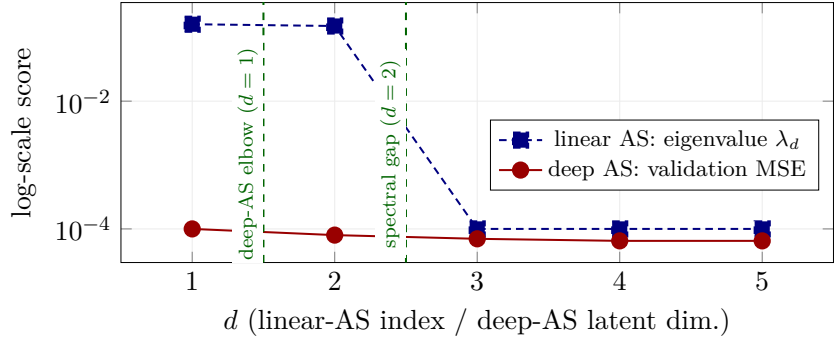
\begin{figure}[h]
\centering
\begin{tikzpicture}[scale=0.95]
\begin{axis}[
    width=11.5cm, height=5.2cm,
    xlabel={$d$ (linear-AS index / deep-AS latent dim.)},
    ylabel={log-scale score},
    xmin=0.5, xmax=5.5,
    ymode=log,
    xtick={1,2,3,4,5},
    grid=major, grid style={gray!15},
    legend style={font=\footnotesize, at={(0.98,0.55)}, anchor=north east},
]
\addplot[color=uzhblue, mark=square*, mark size=3pt, thick, densely dashed]
  coordinates {(1,1.6e-1) (2,1.5e-1) (3,1.0e-4) (4,1.0e-4) (5,1.0e-4)};
\addlegendentry{linear AS: eigenvalue $\lambda_d$}
\addplot[color=harvardcrimson, mark=*, mark size=3pt, thick]
  coordinates {(1,1.0e-4) (2,8.0e-5) (3,7.0e-5) (4,6.5e-5) (5,6.5e-5)};
\addlegendentry{deep AS: validation MSE}
\draw[dashed, darkgreen, thick]
  (axis cs:1.5, 1e-5) -- (axis cs:1.5, 5e-1);
\node[font=\scriptsize, text=darkgreen, anchor=south, rotate=90,
      fill=white, inner xsep=4pt, inner ysep=2pt]
  at (axis cs:1.5, 1.5e-3) {deep-AS elbow ($d=1$)};
\draw[dashed, darkgreen, thick]
  (axis cs:2.5, 1e-5) -- (axis cs:2.5, 5e-1);
\node[font=\scriptsize, text=darkgreen, anchor=south, rotate=90,
      fill=white, inner xsep=4pt, inner ysep=2pt]
  at (axis cs:2.5, 1.5e-3) {spectral gap ($d=2$)};
\end{axis}
\end{tikzpicture}
\caption{Stylized comparison of the two selection criteria for the radial-ridge target $y(\xi) = \exp(-[(w_1^\top\xi)^2 + (w_2^\top\xi)^2])$ in $D = 20$ (see notebook~\texttt{09}).  The linear-AS eigenvalue spectrum has two dominant directions, so the spectral-gap criterion picks $d = 2$ (right green dashed line).  The deep-AS validation MSE, by contrast, is already at its plateau at $d = 1$ (left green dashed line): the learned encoder $h_\theta$ represents the nonlinear aggregate $r^2 = s_1^2 + s_2^2$ as a \emph{scalar}.  The curves are stylized; the orders of magnitude track the notebook (linear-AS eigenvalues $\approx 0.16$ for $i = 1, 2$ then a sharp drop; deep-AS validation MSE $\approx 10^{-4}$ at $d = 1$ and roughly flat thereafter), and the elbow-at-$d=1$ versus spectral-gap-at-$d=2$ contrast is reproduced.}
\label{fig:deep_as_elbow}
\end{figure}

\paragraph{Choosing the latent dimension $d$.}
The spectral gap of $C$ is no longer available -- the encoder is nonlinear -- so $d$ is chosen by a \emph{validation-MSE elbow}: hold out an independent fraction of the sample, train a small family of models with $d = 1, 2, 3, \ldots$, and pick the smallest $d$ beyond which held-out error no longer drops significantly.  An operational rule of thumb is to stop at the first $d$ for which the MSE improvement from $d$ to $d+1$ is less than a factor of two: smaller gains are typically driven by optimization slack, not by new latent structure.  On curved problems this elbow lies \emph{strictly below} the linear-AS spectral gap: the deep encoder collapses two linear features into a single nonlinear aggregate (notebook~\texttt{09\_Deep\_Active\_Subspace\_Ridge} gives a reproducible instance in $D = 20$).  On nearly-linear problems the two criteria agree qualitatively, and at small training-set sizes a polynomial link on top of a two-dimensional linear AS can in fact be more data-efficient than the deep encoder (notebook~\texttt{10\_Deep\_AS\_vs\_Linear\_AS\_Borehole}, the canonical borehole benchmark with $D = 8$ and $N = 500$).  Figure~\ref{fig:deep_as_elbow} contrasts the elbow rule with the linear-AS spectral gap on the radial-ridge target.

\paragraph{Training recipe.}
A practical recipe that reproduces the experiments in notebooks \texttt{09} and \texttt{10}:
\begin{enumerate}[itemsep=2pt]
\item Sample $N$ input-output pairs $(\xi_i, y_i)$; split $80/20$ into train and validation sets.  Standardize inputs to unit variance and center outputs.
\item For each candidate $d \in \{1, 2, 3, \ldots\}$: build $h_\theta$ with widths~\eqref{eq:tb_widths} and $g_\theta$ with two hidden layers ($16$--$32$ units); train on loss~\eqref{eq:tb_loss} with Adam ($\text{lr} = 5 \times 10^{-3}$), a cosine learning-rate schedule over $10^3$--$2 \times 10^3$ epochs, and $\lambda_1 = 10^{-5},\, \lambda_2 = 10^{-4}$.
\item Record the validation MSE, apply the elbow rule, and deploy $\hat f_\theta$ with the chosen $d$.
\end{enumerate}
Sample-budget rule of thumb: $N \approx 50\, d_{\mathrm{nl}}$ to \emph{find} the bottleneck, inflated to $N \approx 200\, d_{\mathrm{nl}}$ for a deployment surrogate.  Two post-training sanity checks are worth doing: (i) the validation curve should be monotone-then-flat in $d$; (ii) a scatter of $h_\theta(\xi_i)$ against the top linear-AS coordinate $U_1^\top \xi_i$ reveals whether the encoder has actually gone nonlinear (a non-monotone relation is the fingerprint).

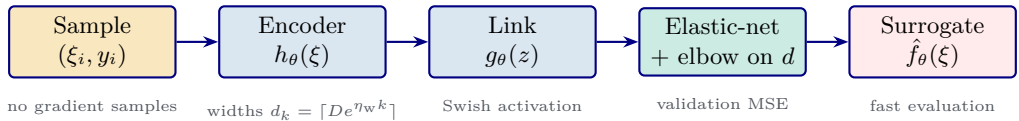
\begin{figure}[ht]
\centering
\begin{tikzpicture}[
    mlstep/.style={rectangle, draw=uzhblue, fill=uzhgreylight,
        minimum width=2.2cm, minimum height=0.7cm, font=\footnotesize,
        rounded corners=2pt, thick, align=center},
    arr/.style={-{Stealth[length=2mm]}, thick, uzhblue}
]
\node[mlstep, fill=softorange!25] (samp)  {Sample\\$(\xi_i, y_i)$};
\node[mlstep, right=0.55cm of samp, fill=softblue!20]  (enc)   {Encoder\\$h_\theta(\xi)$};
\node[mlstep, right=0.55cm of enc, fill=softblue!20]   (link)  {Link\\$g_\theta(z)$};
\node[mlstep, right=0.55cm of link, fill=softgreen!20] (elbow) {Elastic-net\\+ elbow on $d$};
\node[mlstep, right=0.55cm of elbow, fill=red!8]       (surr)  {Surrogate\\$\hat f_\theta(\xi)$};
\draw[arr] (samp)  -- (enc);
\draw[arr] (enc)   -- (link);
\draw[arr] (link)  -- (elbow);
\draw[arr] (elbow) -- (surr);
\node[below=0.15cm of samp,  font=\tiny, text=uzhgreydark] {no gradient samples};
\node[below=0.15cm of enc,   font=\tiny, text=uzhgreydark] {widths $d_k = \lceil D e^{\eta_{\mathrm{w}} k}\rceil$};
\node[below=0.15cm of link,  font=\tiny, text=uzhgreydark] {Swish activation};
\node[below=0.15cm of elbow, font=\tiny, text=uzhgreydark] {validation MSE};
\node[below=0.15cm of surr,  font=\tiny, text=uzhgreydark] {fast evaluation};
\end{tikzpicture}
\caption{Deep active-subspace pipeline.  Input--output pairs $(\xi_i, y_i)$ are drawn directly from the simulator (no gradient samples needed); a nonlinear encoder $h_\theta$ compresses the high-dimensional input into a $d$-dimensional latent code, a small link network $g_\theta$ maps the latent code to the response, and an elastic-net / validation-MSE elbow chooses $d$.  The trained composition $\hat f_\theta = g_\theta \circ h_\theta$ is the deployed surrogate.  Compared with the linear active-subspace pipeline (Figure~\ref{fig:active_subspace_pipeline}), the encoder + link boxes replace the eigendecomposition + linear projection, the elbow box replaces the spectral-gap search, and the gradient-sampling step disappears entirely.}
\label{fig:deep_as_pipeline}
\end{figure}

Figure~\ref{fig:deep_as_pipeline} should be compared with the linear pipeline in Figure~\ref{fig:active_subspace_pipeline}: the two blue encoder / link boxes replace ``eigendecomposition + project'', the ``elastic-net + elbow'' box replaces ``find spectral gap'', and the gradient-sampling step is gone entirely.  The economic pay-off is the same, a cheap surrogate in $d$ variables where the original has $D$, but the modeling assumption is weaker: curved active manifolds are now admissible.

\paragraph{When is the extra machinery worth it?}
Deep AS pays off when (i) no gradient samples are available (e.g., black-box simulators, noisy observations from a lab experiment, or calibration targets returned by a proprietary solver), (ii) the spectral gap of $C$ is ambiguous or the response surface genuinely depends on \emph{nonlinear} aggregates of the inputs (ridges, radial coordinates, thresholded features, piecewise regimes), or (iii) the input dimension is large ($D \gg 10$) and a linear projection is too restrictive to capture the active manifold.  When none of these conditions apply, linear AS plus a polynomial or a GP link is usually more data-efficient and should be fit first as a baseline.  The borehole experiment in notebook~\texttt{10} is the honest diagnostic: on a function this close to a $d = 2$ ridge, the two curves \emph{cross}, and a cubic polynomial on two linear features beats the deep encoder at all $d \ge 2$.  Deep AS's advantage materializes precisely at $d = 1$, where a single linear feature cannot span the active direction.  This perspective also connects naturally to deep kernel learning (Section~\ref{sec:dkl}), which composes a learned feature map with a GP head; deep AS is the same idea for the feature map, with an MLP link in place of the GP.

\paragraph{Applications in economics and finance.}
The deep-AS architecture is particularly attractive in settings where the state variable is high-dimensional but the economic mechanism acts through a few aggregate quantities that are not linear in the state.  Examples include (i) heterogeneous-agent models whose cross-sectional distribution of wealth enters the equilibrium law of motion only through a few \emph{moments} (mean, dispersion, tail mass), the Krusell-Smith aggregation logic of Chapter~\ref{ch:young}, where the map from the raw distribution to those moments is a learned, nonlinear encoder; (ii) multi-region climate-economy integrated assessment models with hundreds of regional capital stocks, in which damages and optimal carbon taxes depend on nonlinear aggregates of the underlying state (total radiative forcing, regional inequality indices); and (iii) dynamic games and mean-field games in which the relevant action depends on a curved functional of the opponent distribution.  In each case the original space is too large for a direct GP and the linear active subspace is too rigid, while the deep encoder provides a trainable compromise.  \citet{Bilionis:2016wc} and \citet{tripathy2018deep} provide the general UQ and deep-active-subspace methodology; for integrated assessment models specifically, \citet{friedlDeep2023} is the economics-side reference used in this course.


\section{Dynamic Programming with Gaussian Processes}
\label{sec:gp_dp}

Returning to the second oracle announced at the start of this chapter and developed formally in \S\ref{sec:gp_dp_supervised_view} below, we now embed the GP and active-subspace machinery developed above into a \emph{value function iteration} (VFI) algorithm, with the Bellman operator $TV$ playing the role of the expensive label generator.  The acronym \emph{ASGP} used in the section titles and figure captions below stands for \emph{Active-Subspace Gaussian Process}: the dimensionality reducer of \S\ref{sec:active_subspaces} stacked under the GP of \S\ref{sec:gp_kernels}.

The idea of representing the value function as a GP inside a DP recursion was introduced under the name \emph{Gaussian process dynamic programming} (GPDP) by \citet{deisenroth2009gaussian}, who used it for continuous-state, continuous-action optimal control problems and combined it with a variance-based active selection of support points.  Closely related work \citep{engel2005reinforcement} embeds GPs into temporal-difference learning, replacing tabular value functions with a GP posterior.  These ideas later inspired model-based policy search methods such as PILCO \citep{deisenroth2011pilco}, which propagate GP uncertainty through long horizons.  In economics, \citet{SCHEIDEGGER201968} show that pairing this paradigm with active subspaces and Bayesian Gaussian mixture models for the ergodic set scales the framework to economies with up to 500 continuous state variables; \citet{rennerscheidegger_2018} apply it to dynamic incentive problems, \citet{gaegauf2023portfolio} to dynamic portfolio choice, and \citet{chen2025private} to dynamic private asset allocation.  The remainder of this section presents the algorithm in this combined form.

\subsection{Why GPs for DP: a Supervised-Learning View of the Bellman Operator}
\label{sec:gp_dp_supervised_view}

Before stating the algorithm, it helps to see VFI through a supervised-learning lens.  At iteration $s$, given an incumbent value function $V^{s-1}$, we generate a training set
\[
\mathcal{D}^s = \bigl\{(\x^{(i)},\, t_i^s)\bigr\}_{i=1}^{n^s}, \qquad t_i^s = (TV^{s-1})(\x^{(i)}),
\]
and fit a regressor $V^s \approx (TV^{s-1})$ to it.  The label $t_i^s$ is not free.  Each evaluation of the Bellman operator at a state $\x^{(i)}$ requires solving a constrained nonlinear program over controls $\xi$ subject to the law of motion and any feasibility / market-clearing constraints; quadrature over $\x'$ is taken inside.  In textbook problems each oracle call costs $10^{-2}$--$10^{0}$ seconds, and in higher-dimensional models with adjustment costs, occasionally binding constraints, or aggregator nonlinearities, it can run into minutes.  Because the calls are independent across $\x^{(i)}$, the design is embarrassingly parallel \citep{SCHEIDEGGER201968}, but each individual label remains expensive.

This is exactly the regime where Gaussian processes shine.  Three properties make them the natural surrogate class:

\begin{enumerate}[itemsep=2pt]
\item \textbf{Sample efficiency under an expensive oracle.}  The marginal-likelihood Occam's razor of \S\ref{sec:gp_kernels} delivers calibrated fits at $n \sim 10^2$--$10^3$ design points, well below the $n \gg 10^4$ regime in which deep-network surrogates start to dominate (Table~\ref{tab:gp_vs_bnn}).
\item \textbf{Built-in uncertainty quantification.}  The GP posterior variance $\sigma_\mathrm{GP}^2(\x)$ tells us, at every state, how much the current surrogate trusts its own prediction.  This is the input that turns a passive interpolant into an \emph{adaptive} one (\S\ref{sec:gp_dp_bal_inside}).
\item \textbf{Cheap held-out diagnostics.}  The leave-one-out predictive error is available in closed form from the same Cholesky factor used for posterior inference (\S\ref{sec:gp_loo}), so we can monitor surrogate health every iteration without spending oracle calls on a held-out split.
\end{enumerate}

The same logic applies, with a different oracle, in the structural estimation chapter: Chapter~\ref{ch:estimation} stacks a GP over the moment map $m(\theta)$ where each label is a forward simulation plus moment computation under a candidate parameter, again expensive to evaluate and cheap to store.  Both settings are instances of one pattern: \emph{supervised regression on the output of a costly numerical procedure}.  The next subsection writes down the formal DP problem; the subsequent subsections develop the GP-based interpolant and the two ingredients (LOO and active learning) that make it sample-efficient at modest design size.

\subsection{The Dynamic Programming Problem}

Consider an infinite-horizon, discrete-time stochastic optimal decision problem.  A representative agent chooses a sequence of controls $\{\xi_t\}_{t=0}^\infty$ to maximize the expected discounted sum of returns:
\begin{equation}
V(\x_0) = \max_{\{\xi_t\}} \mathbb{E}_0 \sum_{t=0}^\infty \beta^t\, r(\x_t, \xi_t),
\label{eq:dp_objective}
\end{equation}
subject to the law of motion $\x_{t+1} \sim F(\cdot\,|\,\x_t, \xi_t)$, where $\x_t \in \mathcal{X} \subset \R^D$ is the state, $\xi_t \in \Lambda(\x_t)$ is the control, $r$ is the per-period return function, and $\beta \in (0,1)$ is the discount factor.

By the \emph{principle of optimality} \citep{bellman1957}, the infinite-dimensional problem~\eqref{eq:dp_objective} reduces to the \emph{Bellman equation}:
\begin{equation}
V(\x) = \max_{\xi \in \Lambda(\x)} \bigl\{ r(\x, \xi) + \beta\, \mathbb{E}\bigl[V(\x')\bigr] \bigr\},
\label{eq:bellman}
\end{equation}
where the expectation is over the stochastic transition.  The \emph{Bellman operator} $T$ maps value functions to value functions:
\[
(TV)(\x) = \max_{\xi \in \Lambda(\x)} \bigl\{ r(\x, \xi) + \beta\, \mathbb{E}\bigl[V(\x')\bigr] \bigr\}.
\]
Under standard regularity conditions (bounded returns, $\beta < 1$, monotonicity, discounting), $T$ is a \emph{contraction mapping} on the Banach space of bounded continuous functions with sup-norm, with modulus $\beta$.  By the Banach fixed-point theorem (Appendix~\ref{app:fixed_points}), $T$ admits a unique fixed point $V^*$, and iterating $V^{s+1} = TV^s$ from any initial guess $V^0$ converges geometrically: $\|V^s - V^*\|_\infty \le \beta^s \|V^0 - V^*\|_\infty$.  The classical references in economics are \citet[Chs.~3--4]{stokeylucas1989} for the formal contraction-and-fixed-point theory of the Bellman operator and \citet[][Ch.~12]{judd1998numerical} for the numerical implementation in continuous-state settings; \citet[Chs.~3--4]{ljungqvist2018recursive} and \citet{sargentstachurski2026dp} provide modern macroeconomic treatments.  In the operations-research and optimal-control literature, \citet{Bertsekas:2000:DPO:517430} is the canonical reference, with extensive coverage of approximate dynamic programming.  The contraction modulus $\beta$ applies to the \emph{exact} Bellman operator; convergence of the GP-fitted iterates additionally requires controlling the per-step interpolation error of the surrogate, otherwise the approximate operator can fail to contract globally even though the exact one does.  This is the standard textbook VFI loop: guess a value function, apply the Bellman operator, interpolate the updated values, and repeat until convergence.  In the present chapter, the only change is the interpolant: a GP replaces the grid or polynomial basis when the state space is irregular or moderately high-dimensional.  For a comprehensive treatment of dynamic programming in economics, see also the open-source QuantEcon lectures.\footnote{\url{https://dp.quantecon.org/}}

\paragraph{The computational challenge.}
At every VFI iteration, we must \emph{approximate} $V^{s+1}$ as a function of the full state vector $\x \in \R^D$.  Traditional approaches (finite grids, Smolyak sparse grids, tensor-product polynomial bases) suffer from the curse of dimensionality: the number of grid points grows exponentially in $D$, and they require hypercubic state-space geometries.  For $D > 20$, these methods become infeasible.

\subsection{The Stochastic Optimal Growth Model}
\label{sec:growth_model}

The workhorse test case for GP-based dynamic programming is the multi-sector stochastic optimal growth model.  Assume $D$ sectors, each with a capital stock $k_j$, so the state is $\bm{k} = (k_1, \ldots, k_D) \in \R^D_+$.  A representative household chooses consumption $\bm{c}$, labor supply $\bm{l}$, and investment $\bm{i}$ to maximize:
\begin{equation}
V_0(\bm{k}_0) = \max_{\{\bm{c}_t, \bm{l}_t, \bm{i}_t\}} \mathbb{E}_0 \sum_{t=0}^\infty \beta^t\, u(\bm{c}_t, \bm{l}_t),
\label{eq:growth_bellman}
\end{equation}
subject to the capital law of motion $k_{j,t+1} = (1-\delta)\,k_{j,t} + i_{j,t}$, a sector-by-sector resource constraint $c_{j,t} + i_{j,t} + \Gamma_{j,t} = \exp(z_{j,t})\,A\,k_{j,t}^\psi\,l_{j,t}^{1-\psi}$ with convex adjustment cost $\Gamma_{j,t}$, and Cobb--Douglas production $f(k_j, l_j) = A\,k_j^\psi\,l_j^{1-\psi}$.  Productivity shocks $z_{j,t}$ follow a stationary process and the dimension $D$ can be scaled from 1 (the textbook Brock--Mirman model of Chapter~\ref{ch:deqn}) to 500 without changing the algorithmic structure.

\subsection{GP-Based Value Function Iteration (ASGP)}
\label{sec:asgp_vfi}

The key idea, common to \citet{deisenroth2009gaussian} in the control literature and \citet{SCHEIDEGGER201968} in the economics literature, is to use a \emph{Gaussian process as the interpolation scheme} inside VFI, replacing grid-based methods entirely.  At each VFI step $s$:
\begin{enumerate}[itemsep=2pt]
\item Generate $n^s$ training inputs $\X = \{\x^{(1)}, \ldots, \x^{(n^s)}\} \subset [\underline{\bm{k}}, \bar{\bm{k}}]^D$.
\item Evaluate the Bellman operator at each training point: $t_i^s = (TV^{s-1})(\x^{(i)})$.
\item Learn a GP (or ASGP) surrogate $V_\mathrm{surr}$ from the training data $\{\X, \bm{t}\}$.
\item Set $V^s = V_\mathrm{surr}$ (the GP posterior mean).
\item Compute the convergence error $e^s = \|V^s - V^{s-1}\|_\infty / \Delta_V$, where $\Delta_V$ is a normalizing value-function range.
\item If $e^s < \bar{e}$, stop; otherwise continue.
\end{enumerate}

\paragraph{Advantages over grid-based methods.}  This GP-VFI approach offers several structural advantages:
\begin{itemize}[itemsep=2pt]
\item \textbf{Arbitrary geometries:} GPs require no tensor-product structure, so training points can be placed anywhere in the state space, including on irregularly shaped ergodic sets.
\item \textbf{Adaptive training:} Points where the Bellman operator fails to converge (e.g., near constraints) can simply be excluded and retried later; adding or removing points is trivial.
\item \textbf{Built-in uncertainty quantification:} The GP posterior variance provides interpolation uncertainty at every state point, at every iteration, ``free'' UQ.
\item \textbf{Active subspace integration:} When $D \gg 10$, the GP operates on the projected inputs $\tilde{\x} = U_m^\top \x \in \R^m$ discovered via the active subspace pipeline (Section~\ref{sec:active_subspaces}), reducing the effective dimensionality from hundreds to a handful.
\end{itemize}

\paragraph{Parallelization.}
The most expensive part of the algorithm, evaluating the Bellman operator at $n^s$ training points, is \emph{embarrassingly parallel}: each evaluation is independent.  In the MPI implementation of \citet{SCHEIDEGGER201968}, the current value function surrogate is broadcast to all workers, each worker evaluates the Bellman operator at $n^s/n_\mathrm{cpu}$ points, and the results are gathered at the master for GP fitting.  Communication cost is negligible relative to the Bellman evaluations.

\subsection{Leave-One-Out Error: a Held-out Diagnostic for Free}
\label{sec:gp_loo}

How do we know that the GP surrogate at iteration $s$ is good enough to take another Bellman step?  The marginal-likelihood objective of \S\ref{sec:gp_kernels} fits the kernel, but it does not, on its own, certify pointwise predictive accuracy on the design.  A held-out validation split would, but every held-out point is one fewer expensive Bellman label going into the GP.

The standard escape route in GP regression is the leave-one-out (LOO) predictive error, and it has the pleasant feature that for a Gaussian process it admits a closed form using the same Cholesky factor already computed for the posterior \citep[\S 5.4.2]{Rasmussen:2005:GPM:1162254}:
\begin{equation}
\mu_{-i}(\x^{(i)}) - y_i = -\frac{\alpha_i}{\bigl[K_y^{-1}\bigr]_{ii}},
\qquad
\sigma_{-i}^2(\x^{(i)}) = \frac{1}{\bigl[K_y^{-1}\bigr]_{ii}},
\label{eq:gp_loo}
\end{equation}
where $\mu_{-i}$ and $\sigma_{-i}^2$ denote the posterior mean and variance after removing the $i$-th observation, and $\alpha = K_y^{-1}(\bm y-\bm\mu_X)$ (for centered outputs, $\bm\mu_X=0$).  Since $K_y^{-1}$ is recovered from the Cholesky factor of $K_y$ in $\mathcal{O}(n^2)$ once the factorisation has been done, computing the full $n$-vector of LOO residuals is essentially free relative to the $\mathcal{O}(n^3)$ already paid for posterior inference.

\paragraph{What the LOO RMSE tells us.}
Tracking
\[
\mathrm{LOO\text{-}RMSE}(\mathcal{D}^s) \;=\; \sqrt{\frac{1}{n^s}\sum_{i=1}^{n^s}\bigl(\mu_{-i}(\x^{(i)}) - t_i^s\bigr)^2}
\]
across VFI iterations is a cheap surrogate-health metric that is independent of the Bellman residual:
\begin{itemize}[itemsep=2pt]
\item A flat-then-rising LOO curve at the same design size signals \emph{kernel mis-specification} (length scale collapsing, noise variance hitting a bound) and tells us to revisit the kernel choice or hyperparameter bounds before adding more design points.
\item A high LOO RMSE at small $n^s$ that decays as the design grows is the expected behaviour and tells us that the surrogate simply needs more labels.
\item A small LOO RMSE coexisting with a large Bellman residual points the finger at the \emph{operator}, not the surrogate: the iterate may be far from the fixed point even though the GP fits the current $TV^{s-1}$ well.
\end{itemize}
The notebook \tpath{04_GP_Value_Function_Iteration.ipynb} computes \eqref{eq:gp_loo} via \tpath{scipy.linalg.cho_solve} (function \tpath{gp_loo_rmse}) and reports it alongside the Bellman residual at every VFI iteration, separating these two failure modes.  The same diagnostic reappears in \S\ref{sec:smm_gp_moments} for the GP layer over the SMM moment map.

\subsection{Active Learning Inside the VFI Loop}
\label{sec:gp_dp_bal_inside}

The Bayesian active-learning machinery of \S\ref{sec:bal} carries over almost verbatim once the VFI loop is in place, but with one important adjustment: the goal is now \emph{uniform interpolation accuracy} of the value function on the relevant state-space region, not maximisation of a payoff or minimisation of a loss.  The right acquisition function is therefore pure exploration, the GP posterior standard deviation, rather than a UCB or expected-improvement criterion that trades off exploitation and exploration:
\begin{equation}
\x^{\mathrm{next}} \in \argmax_{\x \in \mathcal{X}^\mathrm{cand}} \sigma_\mathrm{GP}^s(\x).
\label{eq:bal_vfi}
\end{equation}
A practical implementation, used in notebook \tpath{04_GP_Value_Function_Iteration.ipynb}, runs the VFI iterations with a frozen design for a few steps, then \emph{enriches} the design every $N$ iterations:

\begin{enumerate}[itemsep=2pt]
\item Evaluate $\sigma_\mathrm{GP}^s$ on a dense candidate set $\mathcal{X}^\mathrm{cand}$ (a Latin-hypercube draw over the current state-space region).
\item Pick the top-$n_\mathrm{add}$ candidates by posterior standard deviation, subject to a minimum-spacing constraint $\|\x^{(\mathrm{new})} - \x^{(j)}\| \ge \delta_\mathrm{spacing}$ against existing design points to avoid clustering.
\item Evaluate the Bellman operator at the new points (one expensive oracle call each, in parallel) and append them to $\mathcal{D}^s$.
\item Refit the GP and continue iterating.
\end{enumerate}

\paragraph{Why pure exploration here.}
A UCB-style acquisition would bias the design toward states with high \emph{value}, which is not what we want when the surrogate is a building block of an iteration.  We want the GP posterior to be uniformly tight wherever the Bellman operator might be evaluated, so that the contraction modulus of the \emph{approximate} operator stays close to $\beta$.  This is structurally different from the optimisation setting of Bayesian optimisation, where exploitation is a feature.

\begin{figure}[ht]
\centering
\includegraphics[width=\linewidth]{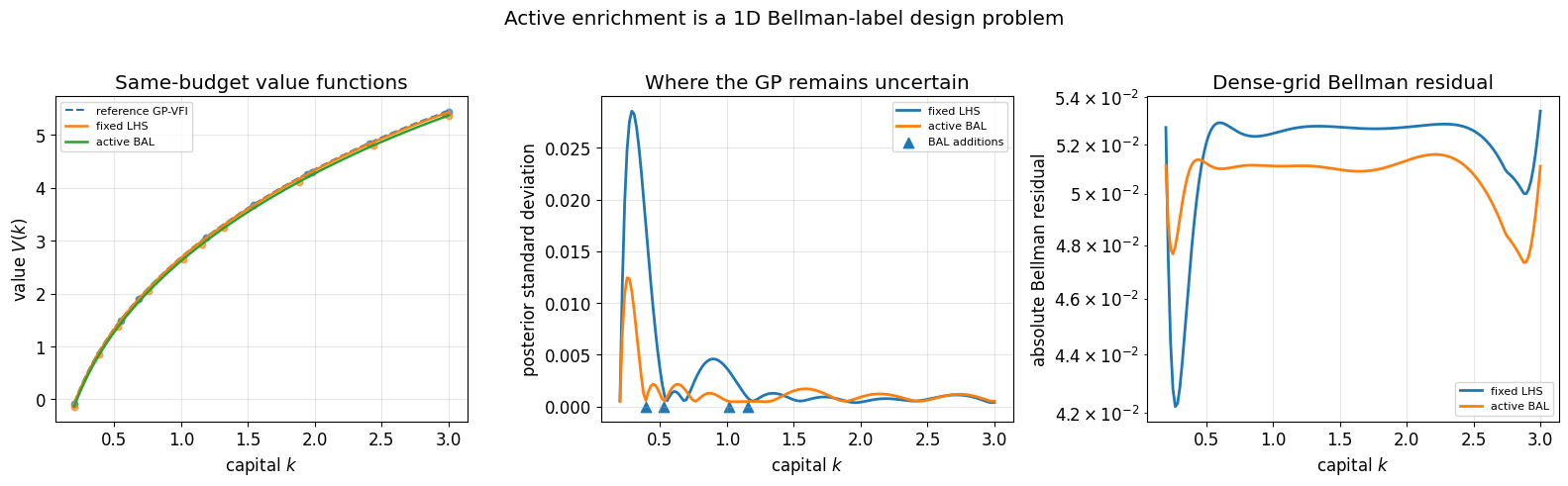}
\caption{Same-budget active enrichment inside one-dimensional GP value-function iteration.  The left panel compares the GP posterior means from a fixed Latin-hypercube design and an active design against a reference GP-VFI solution.  The middle panel shows the posterior standard deviation and the states added by the active rule \eqref{eq:bal_vfi}, marked as triangles on the horizontal axis.  The right panel reports the dense-grid Bellman residual.  Unlike the previously used two-dimensional separable interpolation benchmark, every plotted training value here is generated by a Bellman maximization.  Generated by notebook \protect\tpath{lecture_14_04_GP_Value_Function_Iteration.ipynb}.}
\label{fig:gp_vfi_active_1d}
\end{figure}

\paragraph{Empirical impact.}
The companion notebook compares a same-budget fixed Latin-hypercube design with an active design inside the one-dimensional GP-VFI loop.  Both designs use Bellman labels; the active design starts from a small initial set and adds states by maximising the GP posterior standard deviation \eqref{eq:bal_vfi} subject to a spacing rule.  At the same final number of labels, the active design lowers posterior uncertainty and achieves a comparable or smaller dense-grid Bellman residual (Figure~\ref{fig:gp_vfi_active_1d}).  The figure is one-dimensional by design: the goal is to show active enrichment inside a genuine Bellman iteration, not to use a separable interpolation toy as a proxy for multidimensional dynamic programming.

\subsection{Results: From One to 500 Dimensions}

The companion notebook deliberately stops at the one-dimensional Bellman problem, where every numerical object can be inspected directly.  The high-dimensional claims below come from the ASGP implementation of \citet{SCHEIDEGGER201968}, where the GP operates on an active subspace rather than on the full tensor-product state space; we report their findings as a literature summary, not as something the notebook itself demonstrates.

\citet{SCHEIDEGGER201968} apply the ASGP-VFI algorithm to the stochastic optimal growth model~\eqref{eq:growth_bellman} across a range of dimensions, from $D = 1$ (the textbook Brock--Mirman case) to $D = 500$ continuous state variables.  The key findings are:
\begin{enumerate}[itemsep=2pt]
\item \textbf{Convergence is dimension-independent in the active subspace:} all models converge to a normalized error below $10^{-4}$ within approximately 70 VFI iterations regardless of $D$, where the iteration count is dimension-independent in the $m$-dimensional active subspace on which the GP actually operates rather than in the full ambient $\R^D$.
\item \textbf{Low-dimensional structure:} the active subspace dimension is $m = 1$ in all cases tested, so the value function depends on a single linear combination of the $D$ capital stocks, confirming that the high-dimensional problem has intrinsic low-dimensional structure.
\item \textbf{Sub-exponential scaling:} CPU time grows sub-exponentially with $D$ (concave on a log scale when parallelized across compute nodes); on the benchmarks reported in \citet{SCHEIDEGGER201968}, $D = 500$ is solved at sub-exponential cost in $D$ when distributed across a compute cluster.  This avoids the explicit grid-based curse but does not eliminate all high-dimensional sample/optimization costs.
\item \textbf{Beyond the reach of grids:} grid-based methods such as Smolyak sparse grids cannot handle $D > 100$, making ASGP one demonstrated grid-free approach for models at this scale; neural residual methods such as DEQNs are complementary alternatives when the equilibrium conditions are easier to express directly.
\end{enumerate}

\subsection{Uncertainty Quantification and Ergodic Sets}

\paragraph{Free UQ from the GP posterior.}
Because the value function at each VFI iteration is a GP, the posterior variance $\sigma^2(\x)$ provides a built-in measure of approximation uncertainty.  This gives pointwise credible bands for the interpolated value function essentially for free.  Policy uncertainty is not automatic in the same sense: it must be propagated through the Bellman maximization, for example by evaluating policies under posterior draws or local approximations to the GP posterior.

\paragraph{Parameter uncertainty propagation.}
Uncertain model parameters can be treated as additional pseudo-states.  For example, extending the state to $\tilde{S} = (\bm{k}, \gamma)$ where $\gamma \in [\underline{\gamma}, \bar{\gamma}]$ is the risk aversion parameter allows solving the model once on the extended state space and then extracting univariate effects and global sensitivity indices \emph{in a single computation}, avoiding the traditional approach of re-solving the model for each parameter value.  \citet{harenberg2019uncertainty} give a general introduction to uncertainty quantification for economic models, covering univariate-effect plots and Sobol' sensitivity indices.

\paragraph{Learning ergodic sets.}
Many economic models have irregularly shaped ergodic sets (ellipsoids or manifolds, not hypercubes).  \citet{SCHEIDEGGER201968} propose learning the ergodic distribution via Bayesian Gaussian mixture models: (i)~solve the model on a large initial domain; (ii)~simulate the economy forward to generate capital paths $\{\bm{k}_t^+\}$; (iii)~fit a mixture of Gaussians $\hat{\rho}_\mathrm{ergodic}$ to the simulated paths; (iv)~re-solve on the ergodic set by sampling training points from $\hat{\rho}_\mathrm{ergodic}$ instead of the full hypercube.  GPs handle this naturally because they require no grid structure.

\subsection{Comparison: GPs vs.\ DEQNs for Solving Dynamic Models}

Both GPs (this chapter) and DEQNs (Chapters~\ref{ch:deqn}--\ref{ch:olg}) solve dynamic stochastic models, but with different trade-offs (Table~\ref{tab:gp_vs_deqn_dynamic_models}):

\begin{table}[ht]
\centering
\small
\begin{tabular}{@{}p{3.4cm} p{4.7cm} p{4.7cm}@{}}
\toprule
\textbf{Criterion} & \textbf{GP / ASGP} & \textbf{DEQNs} \\
\midrule
Solution method & Value function iteration & Euler equation residuals \\
Dimensionality & Up to $\sim$500 (with AS) & $>$1000 feasible \\
UQ built-in & Yes (GP posterior variance) & No (needs extra work) \\
Hardware & CPU clusters (MPI) & GPUs (TensorFlow/PyTorch) \\
Irregular domains & Yes (grid-free) & Yes (mesh-free) \\
Sensitivity analysis & Cheap after pseudo-state augmentation & Requires pseudo-states or re-training \\
Convergence diagnostics & Exact Bellman operator is a contraction; GP interpolation error must be controlled & Euler residuals and simulation tests; no generic global proof \\
\bottomrule
\end{tabular}
\caption{GP/ASGP and DEQN solvers for dynamic models.  The table distinguishes exact fixed-point theory from the numerical approximation actually used: GP-VFI inherits the Bellman contraction only up to interpolation error, while DEQNs are judged by residual and simulation diagnostics.}
\label{tab:gp_vs_deqn_dynamic_models}
\end{table}

\noindent \textbf{When to use which:} GPs when the effective dimension is moderate ($D \lesssim 15$, or a few hundred only when active-subspace structure is strong) and uncertainty quantification or sensitivity analysis is required; DEQNs when $D$ is very large, GPU hardware is available, or the model involves complicated market-clearing conditions that are more naturally expressed as Euler equation residuals than as Bellman maximization.

\begin{remarkbox}[Hands-on]
Notebook \tpath{04_GP_Value_Function_Iteration.ipynb} implements GP-based VFI for the one-dimensional stochastic growth model.  It shows convergence of the GP-VFI outer loop, posterior credible bands for the value-function interpolant, Cholesky-based LOO-RMSE \eqref{eq:gp_loo}, dense-grid Bellman residuals, active enrichment of the Bellman design \eqref{eq:bal_vfi}, policy recovery, and a deterministic full-depreciation verification ($\delta = 1$ and $\sigma = 0$) against the closed-form Brock--Mirman solution.  The multidimensional ASGP extension is discussed in the literature summary above and illustrated separately by the active-subspace notebooks, \tpath{05_Active_Subspace_2D.ipynb}, \tpath{06_Active_Subspace_10D.ipynb}, and \tpath{07_Active_Subspace_Nonlinear.ipynb}, on 2D, 10D, and nonlinear test functions.
\end{remarkbox}


\section{Deep Kernel Learning}
\label{sec:dkl}

Standard GP kernels such as the RBF or Mat\'ern operate on \emph{raw} input features: the covariance $k(\x, \x')$ depends on $\|\x - \x'\|$, which may be a poor measure of similarity when the inputs are high-dimensional or when the function's structure depends on complex nonlinear interactions among variables.  \emph{Deep Kernel Learning} (DKL) \citep{wilson2016deep} addresses this limitation by combining the representation-learning power of deep neural networks with the probabilistic framework of GPs.

\paragraph{The DKL architecture.}
A DKL model consists of two components:
\begin{enumerate}[itemsep=2pt]
\item A \textbf{feature extractor} $\phi(\x; \bm{\theta}_\mathrm{NN})\colon \R^D \to \R^d$, typically a multi-layer perceptron that maps the raw input $\x$ to a learned representation of (usually much lower) dimension $d$.
\item A \textbf{GP layer} that places a GP prior on the function $g(\z) = g(\phi(\x; \bm{\theta}_\mathrm{NN}))$, with a standard base kernel $k_\mathrm{base}(\z, \z')$ (e.g., RBF or Mat\'ern) operating in the learned feature space.
\end{enumerate}
The \emph{deep kernel} is thus:
\begin{equation}
k_\mathrm{DKL}(\x, \x') = k_\mathrm{base}\bigl(\phi(\x; \bm{\theta}_\mathrm{NN}),\; \phi(\x'; \bm{\theta}_\mathrm{NN})\bigr).
\label{eq:dkl_kernel}
\end{equation}
Unlike a standard kernel that computes distance in the input space, the DKL kernel computes distance in a \emph{learned} feature space.  If the neural network learns to map functionally similar inputs close together, the GP can exploit this structure for better predictions and more calibrated uncertainty estimates.

\paragraph{Joint training.}
The parameters of both the neural network ($\bm{\theta}_\mathrm{NN}$) and the GP hyperparameters ($\ell, \sigma_f, \sigma_y$ of $k_\mathrm{base}$) are optimized jointly by maximizing the GP marginal likelihood:
\begin{equation}
\max_{\bm{\theta}_\mathrm{NN},\, \bm{\vartheta}} \; \log p(\bm{y}\,|\,\Phi(\X; \bm{\theta}_\mathrm{NN}),\, \bm{\vartheta}),
\end{equation}
where $\Phi(\X; \bm{\theta}_\mathrm{NN})$ is the matrix of transformed features.  This end-to-end training procedure automatically learns features that are useful for the GP, without requiring manual feature engineering.  In practice, DKL is implemented via GPyTorch \citep{gardner2018gpytorch}, which provides efficient GPU-accelerated GP inference and integrates seamlessly with PyTorch for the neural network component.

\paragraph{When DKL helps.}
DKL is most beneficial when:
\begin{itemize}[itemsep=2pt]
\item The input space is high-dimensional but the function has low-dimensional structure that is \emph{nonlinearly} embedded (active subspaces find \emph{linear} structure; DKL can find nonlinear features).
\item Sufficient training data is available to train the feature extractor without overfitting (DKL has more parameters than a standard GP).
\item Uncertainty quantification is important, but a standard GP kernel is insufficiently expressive to capture the true function's covariance structure.
\end{itemize}
The trade-off relative to a standard GP is clear: DKL offers greater expressiveness at the cost of more parameters and the risk of overfitting with very small datasets.  For the moderate-data regimes typical of economic surrogate models ($N \sim 100$--$1000$), DKL can offer significant improvements over standard kernels, particularly when the target function has complex multi-scale behavior.  \citet{chen2025private} apply learned-feature-map GP architectures in a dynamic model of private asset allocation, where the feature map captures the complex nonlinear interactions between illiquidity, portfolio composition, and optimal rebalancing.

\paragraph{Illustrative examples.}
Two simple examples highlight when DKL provides a qualitative advantage over standard kernels:
\begin{itemize}[itemsep=2pt]
\item \textbf{Step functions.}  Consider approximating a 1D function with a sharp discontinuity.  A standard GP with an RBF kernel necessarily oversmooths near the jump and yields poorly calibrated uncertainty bands.  A DKL model, by contrast, can learn a feature map that ``compresses'' the input near the discontinuity, effectively sharpening the GP's resolution where it matters most.  This is directly relevant to economic applications with occasionally binding constraints, where policy functions exhibit kinks or jumps.
\item \textbf{Anisotropic boundaries in 2D.}  Standard stationary kernels (RBF, Mat\'ern) are isotropic: they measure distance with circular level sets.  When the target function has a discontinuity along a \emph{diagonal} or curved boundary, common in portfolio problems with no-trade regions or in models with regime-dependent policies, the isotropic kernel cannot adapt.  DKL learns a nonlinear coordinate transformation that aligns the kernel's smoothness assumptions with the function's actual structure, capturing diagonal and curved boundaries that would require an impractical number of training points with a standard kernel.
\end{itemize}

\begin{remarkbox}[Hands-on]
The companion notebook \tpath{08_Deep_Kernel_Learning.ipynb} implements a simplified DKL pipeline (a supervised feature extractor stacked with a scikit-learn GP head) and compares the learned deep kernel against standard RBF and Mat\'ern GPs on function approximation tasks; the full GPyTorch joint marginal-likelihood training of \eqref{eq:dkl_kernel} is left as an extension.
\end{remarkbox}

\section{GPs Among Their Bayesian Cousins}
\label{sec:bayesian_dl_compare}

Gaussian processes are the most analytically transparent way to attach uncertainty to a non-parametric regressor, but they are not the only one.  For completeness, this section briefly situates the GP machinery against two neural-network-based alternatives that come up frequently in modern uncertainty-aware deep learning.

\paragraph{Monte-Carlo dropout.}  \citet{gal2016dropout} show that a deep network trained with dropout and \emph{evaluated} with dropout still active can be interpreted as approximate variational inference over a particular Bayesian neural network.  Predictive uncertainty is obtained by averaging $T$ stochastic forward passes; the cost is one extra forward pass per sample.  Calibration is rougher than a GP's, but the method requires no architectural change and scales to deep nets and large $n$.

\paragraph{Deep ensembles.}  \citet{lakshminarayanan2017simple} train $E$ independent neural networks with different random seeds and combine them into a Gaussian-mixture predictor.  Empirically this is one of the most robust uncertainty-quantification recipes available, often beating MC dropout and approaching the calibration of GPs, at $E$ times the training cost.

\begin{table}[ht]
\centering
\small
\setlength{\tabcolsep}{4pt}
\begin{tabular}{@{} >{\bfseries}p{3.0cm} p{3.6cm} p{3.6cm} p{3.6cm} @{}}
\toprule
 & \textbf{Gaussian process} & \textbf{MC dropout} & \textbf{Deep ensembles} \\
\midrule
Calibration            & exact under model               & approximate                       & strong empirically                  \\
Training cost          & $\mathcal{O}(n^3)$              & one network                       & $E\times$ one network               \\
Inference cost         & $\mathcal{O}(n^2)$              & $T$ forward passes                & $E$ forward passes                  \\
Sample efficiency      & best at small $n$               & needs much more data              & needs much more data                \\
Best when              & $n \lesssim 10^4$, low $d$      & cheap UQ on existing nets         & willing to pay $E\times$ for top calibration \\
Reference              & \citet{Rasmussen:2005:GPM:1162254} & \citet{gal2016dropout}        & \citet{lakshminarayanan2017simple}  \\
\bottomrule
\end{tabular}
\caption{Three uncertainty-quantification recipes compared on the dimensions that drive method choice in economic surrogate work.  This script uses GPs in Chapter~\ref{ch:gp} and Chapter~\ref{ch:climate} because the typical regime ($n \lesssim 10^3$, expensive simulator) plays to their strengths; readers operating in larger-data regimes should consider MC dropout or deep ensembles before resorting to a sparse GP.}
\label{tab:gp_vs_bnn}
\end{table}

\paragraph{A decision rule for practice.}  In our experience the right method follows from the application: \emph{(i)}~plain GPs for moderate $d$ ($\lesssim 10$--$20$) with a smooth target and an expensive simulator, when calibrated uncertainty is the goal; \emph{(ii)}~deep kernels (Wilson \& al., \citeyear{wilson2016deep}) when the input geometry is non-trivial (regime switches, manifold structure, image-like inputs); \emph{(iii)}~deep ensembles or MC dropout for high-$d$ regression where calibrated uncertainty is desirable but exact GP inference is infeasible; \emph{(iv)}~sparse GPs (\citet{titsias2009variational, hensman2013gaussian}) for $n \gtrsim 10^4$ when the target stays smooth.  The summary frame in the companion deck (``Toolbox: When to Use What'') gives the same decomposition visually.

\begin{keyinsightbox}[Chapter Summary]
\begin{itemize}[itemsep=2pt, leftmargin=*]
\item Gaussian processes attach calibrated uncertainty to non-parametric regression at $\mathcal{O}(n^3)$ cost; the marginal likelihood implements an automatic Occam's razor that chooses model complexity without held-out validation.
\item Active subspaces collapse $d$-dimensional inputs to $m \ll d$ via the gradient outer product; this is the trick that makes GPs viable past $d \sim 10$.
\item Bayesian active learning closes the surrogate loop: train, evaluate uncertainty, request next design point where it's largest, repeat.  \citet{SCHEIDEGGER201968} provide one canonical example.
\item Deep kernel learning composes a NN feature extractor with a GP head; an alternative to plain GPs and to deep ensembles.
\end{itemize}
\end{keyinsightbox}

\section*{Further Reading}
\addcontentsline{toc}{section}{Further Reading}
\begin{itemize}[itemsep=2pt]
\item \citet{Rasmussen:2005:GPM:1162254}, the standard GP textbook.
\item \citet{constantine2015active}, the active-subspaces monograph.
\item \citet{rennerscheidegger_2018, SCHEIDEGGER201968}, GP+BAL methodology and applications in economics.
\item \citet{wilson2016deep}, deep kernel learning.
\item \citet{titsias2009variational, hensman2013gaussian}, sparse-GP scaling.
\end{itemize}

\section*{Exercises}
\addcontentsline{toc}{section}{Exercises}
\noindent Worked solutions and guidance for these exercises appear in Appendix~\ref{app:solutions}.
\begin{enumerate}[itemsep=4pt, leftmargin=*]
\item\label{ex:ch9:1} \textbf{[Core] Posterior on three points.}  Fit a GP with RBF kernel ($\ell=1$, $\sigma_f=1$, $\sigma_y=0.1$) to three points $(0,0), (1,0.8), (2,0.3)$.  Compute the posterior mean and variance at $x^\star = 1.5$ in closed form.
\item\label{ex:ch9:2} \textbf{[Core] Marginal likelihood Occam.}  For the same three points, plot the log marginal likelihood as a function of the length scale $\ell \in [0.1, 5]$.  Identify the optimum and explain it in terms of the data-fit / complexity decomposition.
\item\label{ex:ch9:3} \textbf{[Core] Active subspace by hand.}  For $f(\x) = (x_1 + x_2 + x_3)^2 + 0.01(x_1 - x_2)^2$ on $[-1,1]^3$, compute the gradient outer product matrix $\hat C$ on a uniform sample.  Show that the leading eigenvector identifies the ``aggregate'' direction $(1,1,1)/\sqrt 3$.
\item\label{ex:ch9:4} \textbf{[Computational] Deep vs.\ linear active subspace on a radial ridge.}  Take the target $y(\xi) = \exp\!\bigl(-[(w_1^\top\xi)^2 + (w_2^\top\xi)^2]\bigr)$ with $\xi \sim \mathcal{N}(0, I_{20})$ and $w_1, w_2 \in \R^{20}$ fixed orthonormal.  (i)~Compute $\hat C$ from $4\,000$ samples and confirm that it has two nonzero eigenvalues.  (ii)~Train the Tripathy--Bilionis surrogate~\eqref{eq:deep_as_ansatz} with widths~\eqref{eq:tb_widths}, Swish activation~\eqref{eq:swish}, and elastic-net loss~\eqref{eq:tb_loss} for $d \in \{1,2,3,4\}$; plot held-out MSE on a log scale and identify the elbow.  (iii)~Explain why linear AS needs $d \ge 2$ while deep AS already captures the response at $d = 1$ (cf.~notebook~\texttt{09\_Deep\_Active\_Subspace\_Ridge}).
\item\label{ex:ch9:5} \textbf{[Computational] BAL on a 2D function.}  Modify notebook \texttt{02\_GP\_and\_BAL} so that one acquisition uses pure variance, $U_{\mathrm{var}}(\x)=\sigma^2(\x)$, and another uses the mixed log-variance score $U_{\mathrm{mix}}(\x)=w_{\mathrm{obj}}\mu(\x)+\tfrac{w_{\mathrm{var}}}{2}\log\sigma^2(\x)$ with $w_{\mathrm{obj}}>0$.  Compare the resulting designs and comment on which gives smoother domain coverage.  Explain why pure $\sigma^2(\x)$ and pure $\log\sigma^2(\x)$ have identical maximizers.
\item\label{ex:ch9:6} \textbf{[Computational] Sobol sensitivity with a GP surrogate.}  Write a short script, or extend notebook \tpath{02_GP_and_BAL.ipynb}, to train a GP surrogate on the $4$-dimensional Genz product-peak function $f(\bm x) = \prod_{i=1}^{4}(c_i^{-2} + (x_i - w_i)^2)^{-1}$ on $[0,1]^4$ with $c = (1, 2, 0.5, 1.5)$, $w = (0.4, 0.6, 0.3, 0.7)$, using $N \in \{50, 100, 200\}$ training points.  For each $N$, compute the Sobol first-order and total-effect indices in two ways: (i)~direct on the true $f$ via $10{,}000$ Monte Carlo samples (reference), (ii)~on the GP surrogate via $10^6$ samples (cheap thanks to the surrogate).  Plot the relative error in each Sobol index against $N$.  Verify that the surrogate-based Sobol estimates converge to the reference as $N$ grows, and identify the $N$ at which all four first-order indices match the reference within $5\%$.  Discuss why surrogate-based sensitivity analysis is the workhorse for expensive simulators (climate models, structural macro models) where direct $10^6$-sample Monte Carlo is infeasible.
\item\label{ex:ch9:7} \textbf{[Core] Prior-driven RBF-GP extrapolation outside the training domain.}  Train a GP with RBF kernel on $20$ points sampled uniformly from $[0, 1]$ from the function $f(x) = \sin(2\pi x)\,e^{-x}$.  (i)~Plot the posterior mean and $\pm 2$ standard deviation band on the training interval $[0,1]$; verify the posterior tracks the true function and the band is narrow.  (ii)~Now extrapolate to $x \in [1.5, 3.0]$ and plot the posterior mean and band over the extended interval.  Show analytically that for an RBF kernel with length scale $\ell$, the posterior at $x$ far from any training point ($|x - x_i| \gg \ell$ for all $i$) reverts to the prior: posterior mean $\to 0$, posterior variance $\to \sigma_f^2$.  (iii)~Verify numerically: at $x = 3$, the posterior mean is essentially zero (independent of the training data), and the posterior standard deviation has expanded back to the prior $\sigma_f$.  (iv)~Discuss the implication: far from the training domain the posterior reverts to the prior, so the variance band there reflects the learned hyperparameters rather than the data, and the band is overconfident only when the prior variance or the learned length scale is itself misleading.  Naive Bayesian active learning that relies on posterior variance may therefore fail to acquire informative samples outside the convex hull when the prior variance is no larger than the inside-hull noise scale.  Mitigations: use a Mat\'ern-$\nu$ kernel with smaller $\nu$ (heavier-tailed), bound the analysis to the convex hull of the training data, or use a boundary-aware acquisition function.
\end{enumerate}

\chapter{Structural Estimation via SMM}
\label{ch:estimation}

With the surrogate machinery of Chapter~\ref{ch:gp} in hand, we now turn to one of its most important applications: \emph{structural estimation}.  The chapter's companion notebooks use the Brock--Mirman growth model to estimate first the productivity persistence parameter $\varrho$, and then the pair $\theta=(\beta,\varrho)$, by Simulated Method of Moments (SMM).  The key computational idea is the same in both cases: train one policy surrogate with the structural parameter as an additional input, then reuse that trained network inside the estimator instead of re-solving the dynamic program at every candidate parameter value.  The econometric foundations are \citet{mcfadden1989method}, \citet{pakes1989simulation}, \citet{lee1991simulation}, \citet{duffie1993simulated}, and \citet{gourieroux1993indirect}; the surrogate logic follows the deep-surrogate and GP-surrogate pipelines in \citet{chen2026Deep} and \citet{SCHEIDEGGER201968}.  Recent applications of the same surrogate-then-estimate move include heterogeneous-agent estimation \citep{kase2022estimating}, search-and-matching \citep{payne2025deepsam}, and climate-economy policy design and uncertainty quantification \citep{kubler2025using, friedlDeep2023}.

Before the current deep-learning boom, neural networks were already studied as nonlinear \emph{sieve} estimators in econometrics, with a rigorous asymptotic theory developed in parallel with the approximation-theory results of Chapter~\ref{ch:intro}.  \citet{chenwhite1999improved} establish convergence rates and asymptotic normality for single-hidden-layer network estimators, and \citet{chen2007sieve} integrates that line into the broader sieve treatment of semi-nonparametric models defined by conditional moment restrictions, which is precisely the structural-estimation setting of this chapter.  The modern continuation of this program uses deep architectures for efficient estimation in nonparametric instrumental-variable models \citep{chenChristensenKankanala2021npiv}.  The pipelines developed below should be read as the implementation-side companion to that theoretical tradition: the sieve literature tells us \emph{when} neural-network estimators consistently identify deep structural parameters; the surrogate pipelines tell us \emph{how} to make the resulting estimators cheap enough to deploy at research scale.

\begin{keyinsightbox}[Two ideas hold this chapter together]
\begin{enumerate}[itemsep=2pt, leftmargin=*]
\item \textbf{The pseudo-state trick.}  Concatenate the structural parameter $\theta$ into the network input, $(s, \theta) \mapsto \mathcal{N}_\rho(s, \theta)$, and train one network that encodes a \emph{family} of policies indexed by $\theta$.  Each new $\theta$ evaluation then requires a single forward pass through the trained network, not a re-solve of the dynamic program; this is what makes SMM with a deep structural model tractable.
\item \textbf{Two-layer surrogates.}  Stack a Gaussian-process surrogate on top of the policy net (\S\ref{sec:smm_gp_moments}): Layer 1 (the policy net) turns ``one re-solve per $\theta$'' into ``one forward simulation per $\theta$'', and Layer 2 (a GP per moment) turns ``one forward simulation per $\theta$'' into ``one GP posterior per $\theta$''.  The result is a microseconds-per-call moment map, enabling bootstrap, Bayesian post-processing, and policy search at scale.
\end{enumerate}
\end{keyinsightbox}

\section{Brock--Mirman with Parameters as Pseudo-States}

The stochastic Brock--Mirman model is the partial-depreciation model from Chapter~\ref{ch:deqn}:
\begin{align}
Y_t &= z_t K_t^\alpha, \\
C_t + K_{t+1} &= Y_t + (1-\delta)K_t, \\
\log z_{t+1} &= \varrho \log z_t + \sigma_z \varepsilon_{t+1}, \qquad
\varepsilon_{t+1}\sim\mathcal{N}(0,1).
\end{align}
We write $\varrho$ for TFP persistence to avoid overloading $\rho$, which elsewhere denotes neural-network parameters.  The Python notebooks still use the variable name \tpath{rho}; mathematically, that code variable corresponds to $\varrho$.

The network outputs a savings rate, the fraction of current output that is invested.  In the single-parameter exercise,
\[
s_t = \mathcal{N}_{\rho}(z_t,K_t,\varrho) \in (0,1),
\]
with $\beta$ calibrated to $0.96$.  In the joint exercise the input becomes $(z_t,K_t,\beta,\varrho)$.  In either case, recover
\begin{align}
K_{t+1} &= (1-\delta)K_t + s_t Y_t, \\
C_t &= (1-s_t) Y_t.
\end{align}
Because $s_t \in (0,1)$ and $Y_t > 0$, this parameterization enforces $C_t > 0$, $K_{t+1} > (1-\delta)K_t > 0$, and gross investment $I_t = s_t Y_t \ge 0$ by construction, so the resource constraint and the non-negativity of investment hold automatically and the partial-depreciation Euler equation~\eqref{eq:smm_euler_residual} applies as written, with no extra multiplier.\footnote{A resource-based variant in which the savings rate is applied to total resources $R_t = Y_t + (1-\delta)K_t$, allowing disinvestment relative to the depreciated stock, is a straightforward alternative; the SMM moments and notebook outputs in this chapter use the output-based savings rate above.}

Training uses the same Euler equation as Chapter~\ref{ch:deqn}, but the residual is evaluated jointly over states and parameter draws.  With partial depreciation,
\[
\frac{1}{C_t}
=
\beta\,\E{\frac{1-\delta+\alpha z_{t+1}K_{t+1}^{\alpha-1}}{C_{t+1}}}.
\]
For a sampled state--parameter pair $(z_i,K_i,\theta_b)$, where $\theta_b=\varrho_b$ in the scalar exercise and $\theta_b=(\beta_b,\varrho_b)$ in the joint exercise, the companion notebooks form the \emph{relative} residual
\begin{equation}
G_i(\theta_b)
=
\frac{1}{\beta_b C_i\,\E{(1-\delta+\alpha z_{i,t+1}(K_i')^{\alpha-1})/C_{i,t+1}}}
-
1,
\label{eq:smm_euler_residual}
\end{equation}
where $Y_i = z_i K_i^\alpha$, $K_i' = (1-\delta)K_i + s_i Y_i$, $C_i = (1-s_i) Y_i$, and $C_{i,t+1}$ is computed by feeding $(z_{i,t+1},K_i',\theta_b)$ back through the same network.  The expectation over $z_{i,t+1}$ is approximated by Gauss--Hermite quadrature (Section~\ref{sec:quadrature_rules}).  The relative form is preferred to the equivalent absolute residual $1-\beta_b C_i\,\E{\cdot}$ because dividing by the consumption ratio makes the loss scale-free across $(z,K,\theta)$ samples; the two forms share the same zero set but the relative form is better conditioned under FP32 forward passes.  A representative training loss is
\begin{equation}
\ell_\rho
=
\frac{1}{N_sN_\theta}
\sum_{i=1}^{N_s}\sum_{b=1}^{N_\theta}
\left|G_i(\theta_b)\right|^2.
\end{equation}
The outer sum over $\theta_b$ is the pseudo-state trick: one network learns a family of policies over the whole parameter rectangle.

\section{Simulated Method of Moments}
\label{sec:smm_method}

The pseudo-state surrogate is what makes SMM cheap.  Without it, every objective evaluation would re-solve the structural model.  With it, the trained policy network and simulator define a fast deterministic map $\theta\mapsto m(\theta)$ once the simulation design is fixed.

The Simulated Method of Moments (SMM) estimates structural parameters by matching model-implied moments to their empirical counterparts.  The method was developed as an extension of the Generalized Method of Moments (GMM) to settings where the moment conditions do not have a closed-form expression but can be computed via simulation \citep{mcfadden1989method, pakes1989simulation, lee1991simulation, duffie1993simulated}.  A closely related approach is \emph{indirect inference} \citep{gourieroux1993indirect}, which matches the parameters of an auxiliary model rather than raw moments.

In quantitative macro and finance, the same simulated-moments logic is especially useful in sovereign default and incomplete-markets environments, where likelihood-based estimation is either unavailable or prohibitively expensive \citep{arellano2008default}.

Let $\hat{m}^\mathrm{data} \in \R^q$ denote a vector of $q$ sample moments computed from observed data (e.g., mean capital, output variance, consumption autocorrelation), and let $m(\theta) \in \R^q$ denote the corresponding moments simulated from the model at parameter value $\theta \in \R^p$.  The SMM estimator solves:
\begin{equation}
\hat{\theta}_\mathrm{SMM}
=
\argmin_\theta
\underbrace{\bigl(m(\theta)-\hat m^\mathrm{data}\bigr)^\top}_{1\times q}
\underbrace{W}_{q\times q}
\underbrace{\bigl(m(\theta)-\hat m^\mathrm{data}\bigr)}_{q\times 1},
\label{eq:smm_objective}
\end{equation}
where $W \in \R^{q \times q}$ is a symmetric positive definite weighting matrix.

\paragraph{The role of the weighting matrix $W$.}
The matrix $W$ controls how much weight each moment (and each pair of moments) receives in the objective.  To build intuition, consider three common choices:

\begin{enumerate}[itemsep=3pt]
\item \textbf{Identity weighting} ($W = I_q$): all moments receive equal weight.  The objective reduces to the unweighted sum of squared deviations, $\sum_{j=1}^q (m_j(\theta) - \hat{m}_j^\mathrm{data})^2$.  This is simple but inefficient: moments measured with high precision receive the same weight as noisy moments.

\item \textbf{Diagonal weighting} ($W = \mathrm{diag}(1/\hat{\sigma}_1^2, \ldots, 1/\hat{\sigma}_q^2)$): each moment is scaled by the inverse of its estimated variance.  This corrects for differing units and precision.

\item \textbf{Optimal weighting}: the inverse of the covariance matrix of the \emph{moment discrepancy} $\hat g(\theta)=m(\theta)-\hat m^\mathrm{data}$.  If the simulation noise is negligible, this is well approximated by the inverse covariance of the empirical moments.  With independent simulated panels of the same length as the data and $S$ replications, the covariance is approximately $(1+1/S)\Sigma_m$, so the optimal weight is proportional to $\Sigma_m^{-1}$ but the scale matters for the $J$-statistic below.
\end{enumerate}

\noindent With identity weighting, the criterion is the sum of squared moment deviations; with inverse discrepancy-covariance weighting, it is the squared Mahalanobis distance between simulated and empirical moments.

\paragraph{Consistency and efficiency.}  Under standard regularity conditions, the SMM estimator is consistent: $\hat{\theta}_\mathrm{SMM} \xrightarrow{p} \theta_0$ as the data sample size $T \to \infty$.  Identification requires more than the usual $q \geq p$ moment count: locally, the moment Jacobian must satisfy the rank condition
\begin{equation}
\mathrm{rank}\bigl(\partial m(\theta_0)/\partial \theta'\bigr) = p,
\end{equation}
which is necessary for \emph{local} identification.  Note that full column rank is necessary but not sufficient: when the smallest singular value of $M$ is close to zero (a near-singular Jacobian), the parameter is only \emph{weakly} identified in finite samples even though the rank condition formally holds.  Global identification additionally requires the population value of the empirical moments, $\bar m := \lim_{T\to\infty}\hat m^\mathrm{data}$, to be matched at a \emph{unique} $\theta_0 \in \Theta$, i.e., $\bar m = m(\theta_0)$ has a unique solution in $\Theta$; the rank condition is the local-curvature implication of that uniqueness.  Let $M=\partial m/\partial\theta'|_{\theta_0}$, and let $\Omega$ denote the asymptotic covariance of $\sqrt{T}\hat g(\theta_0)$.  The large-sample distribution is
\begin{equation}
\sqrt{T}\bigl(\hat{\theta}_\mathrm{SMM}-\theta_0\bigr)
\xrightarrow{d}
\mathcal{N}\!\left(
\bm{0},
(M^\top W M)^{-1}M^\top W\Omega W M(M^\top W M)^{-1}
\right).
\end{equation}
For $S$ independent simulated panels each of length $\tau T$ (with $\tau \geq 1$ a relative-length factor), $\Omega = (1 + 1/(\tau S))\,\Sigma_m$.  The classroom benchmark used below sets $\tau = 1$ (simulated panels of the same length as the data), giving the familiar $\Omega = (1+1/S)\Sigma_m$.  As $\tau S \to \infty$, the \emph{extra} simulation variance vanishes and $\Omega \to \Sigma_m$; the factor itself tends to one, not zero.  The efficient SMM weight is $W^\star = \Omega^{-1}$.  See \citet{duffie1993simulated} for the corresponding large-sample theory in the simulated-moments setting.

\section{The SMM Workflow in the Exercise}

The exercise uses a deliberately simple synthetic-data workflow so that the econometric logic is transparent.  First, choose a true parameter and simulate a time series from the trained pseudo-state surrogate.  These observations play the role of data.  Second, for each candidate parameter, re-simulate the model with the same burn-in length, simulation horizon, initial state, and shock seed.  Third, compute a small vector of economically interpretable moments and minimize the quadratic SMM criterion with identity weighting.

\paragraph{Single-parameter persistence exercise.}
Notebook \tpath{lecture_15_03_Structural_Estimation_BM.ipynb} calibrates $\beta=0.96$, sets $\varrho_{\mathrm{true}}=0.90$, and estimates $\varrho\in[0.50,0.99]$.  Let $\{C_t(\varrho),I_t(\varrho),Y_t(\varrho)\}_{t=1}^T$ denote a simulated sample at candidate persistence $\varrho$.  The estimator uses three moments:
\begin{align}
m_1(\varrho) &= \mathrm{std}\!\bigl(\Delta\log C_t(\varrho)\bigr), \\
m_2(\varrho) &= \mathrm{corr}\!\bigl(\Delta\log C_t(\varrho),\Delta\log C_{t-1}(\varrho)\bigr), \\
m_3(\varrho) &= \mathrm{corr}\!\bigl(\log Y_t(\varrho),\log Y_{t-1}(\varrho)\bigr).
\end{align}
All three moments are computed on the raw simulated time series with no detrending or demeaning step, and $\mathrm{std}(\cdot)$ and $\mathrm{corr}(\cdot)$ denote sample standard deviation and sample autocorrelation evaluated directly on the simulated panel.  The output autocorrelation is the most direct persistence moment.  The volatility moment should be interpreted as an empirical simulated moment, not as the level-variance formula.  For the AR(1) shock,
\[
\mathrm{Var}(\log z_t)=\frac{\sigma_z^2}{1-\varrho^2},
\qquad
\mathrm{Var}(\Delta\log z_t)=2\,\mathrm{Var}(\log z_t)\,(1-\varrho)=\frac{2\sigma_z^2}{1+\varrho},
\]
so the familiar $1/(1-\varrho^2)$ amplification applies to the \emph{level} of log productivity, not to first differences.  The notebook also reports the mean savings rate as a diagnostic and correctly treats it as nearly uninformative for $\varrho$; it is masked out of the SMM criterion in the scalar exercise and used only for visual identification checks.

\paragraph{Joint exercise.}
Notebook \tpath{lecture_15_03b_Structural_Estimation_BM_Joint.ipynb} estimates $\theta=(\beta,\varrho)$, with $\beta\in[0.92,0.99]$ and $\varrho\in[0.50,0.99]$.  It uses four candidate moments: mean savings, growth volatility, consumption-growth autocorrelation, and output autocorrelation.  The \emph{shallow-ridge two-moment specification} retains $\{\mathrm{std}(\Delta\log C_t),\,\mathrm{corr}(\Delta\log C_t,\Delta\log C_{t-1})\}$ to expose the partial-identification ridge in the criterion surface; the over-identified specification uses all four moments and collapses the ridge around the synthetic truth.  Formally the two-moment case is just-identified ($q=p=2$), so we avoid the econometric term \emph{weak identification} (which refers to a near-singular Jacobian asymptotic regime) and use \emph{shallow-ridge} or \emph{partially-identified} for what the criterion-surface picture actually shows.

\paragraph{A deterministic objective.}
Because the same initial condition and random seed are used for every candidate $\theta$, the map $\theta\mapsto m(\theta)$ is deterministic in the notebooks.  This is still standard SMM; the fixed seed is a common-random-numbers device that removes irrelevant simulation noise while we study identification and optimization.  In more realistic estimation exercises one averages over multiple replications or increases the simulation length to make Monte Carlo noise negligible.  One consequence is worth stating explicitly: because the synthetic data come from the trained surrogate evaluated at the true parameter, and every candidate evaluates the same shock sequence, the SMM criterion attains a near-zero minimum at the truth.  This is a clean self-consistency test of the surrogate-SMM pipeline, not a claim about the size of the criterion one would see with real data and independent simulation draws.

\paragraph{Implementation.}
The single-parameter estimation routine proceeds in two steps:
\begin{enumerate}[itemsep=3pt]
\item Evaluate the SMM objective on a coarse grid over $\varrho \in [0.52,0.99]$ (matching the notebook's grid bounds) to verify that the criterion is well behaved and to visualize identification.
\item Refine the minimizer with a bounded scalar optimizer (e.g.\ Brent's method).
\end{enumerate}
The joint notebook maps the 2D criterion on a grid, then refines from the grid minimizer using bounded Nelder--Mead.  Since the policy surrogate has already been trained, each evaluation of $m(\theta)$ requires only a forward simulation, not a full re-solution of the dynamic program.

\paragraph{Interpretation.}
If the moments are informative about $\theta$, the objective should be minimized close to the synthetic truth.  In the scalar notebook, $\hat\varrho$ is very close to $0.90$ and the fitted policy functions at $\varrho_{\mathrm{true}}$ and $\hat\varrho$ nearly overlap.  The joint notebook shows the additional lesson: point estimates can be accurate while the criterion still has weak curvature along one parameter direction, which is why contour plots and Jacobian diagnostics matter.  Figure~\ref{fig:smm_2d_criterion} in \S\ref{sec:smm_gp_moments} below visualizes both specifications and is worth looking at now to fix the geometric picture in mind.

\section{Practical Considerations}

\paragraph{Moment selection and identification.}  The choice of moment conditions is critical for identification.  In the scalar exercise, autocorrelation moments identify $\varrho$ sharply, while the mean savings rate is nearly flat in $\varrho$.  In the joint exercise, the mean savings rate carries most of the information about $\beta$, while persistence moments carry most of the information about $\varrho$.  More generally, the number of moments must weakly exceed the number of parameters ($\dim(m)\geq\dim(\theta)$), and the selected moments should move in economically distinct ways as parameters vary.

\paragraph{Weighting.}  The exercise uses $W=I$ so that the objective is easy to read.  In applications, one usually moves to two-step SMM: first estimate $\hat{\theta}_1$ with identity weighting, then estimate the covariance matrix of the moment discrepancy and set $W=\hat{\Omega}^{-1}$ in a second pass.  This corrects for different moment scales, moment correlations, and any non-negligible simulation noise.  The two-step estimator is asymptotically efficient under \emph{correct specification} and the usual GMM regularity conditions \citep{duffie1993simulated}; under misspecification, the optimal-$W$ estimator can have larger finite-sample mean-squared error than identity weighting, because the efficient weighting is calibrated against the wrong moment-discrepancy distribution.

\paragraph{Simulation design.}  The notebook fixes the burn-in length, horizon, initial state, and shock sequence across all objective evaluations.  This is important because otherwise the optimizer would chase simulation noise rather than structural differences across parameter values.  In larger empirical applications, the same idea appears as \emph{common random numbers} (CRN) or replicated simulations: both are classical variance-reduction techniques in stochastic simulation \citep{glasserman2004monte}, and within the simulated-moments setting \citet{mcfadden1989method} emphasized fixing the simulated draws across parameter values to make the moment objective $m(\theta)$ a smooth function of $\theta$ rather than a noisy step function (the asymptotic theory of optimization estimators with simulation is developed in \citet{pakes1989simulation}).  The geometric intuition is simple: if every candidate parameter value is evaluated against the \emph{same} draw of innovations, the residual $m(\theta) - m(\theta')$ isolates the structural effect of moving from $\theta$ to $\theta'$ rather than a Monte Carlo accident.

\paragraph{Identification diagnostics.}  A necessary condition for local identification is that the Jacobian $M=\partial m/\partial\theta'$ has full column rank at the true parameter.  In the scalar Brock--Mirman exercise this condition reduces to requiring that at least one selected moment changes with $\varrho$ in a neighborhood of the truth.  In the joint exercise, the singular values of $M$ reveal the weak direction associated with $\beta$.  Plotting the objective profile or contour is therefore already informative: a clear and well-centered U-shape signals useful identifying variation, whereas a flat ridge or a jagged profile indicates weak identification or excessive simulation noise.

\paragraph{Over-identification tests.}  When the model is over-identified ($q > p$), report the standard $J$-statistic at the optimum:
\begin{equation}
J = T \,\hat{g}(\hat{\theta})^\top W \hat{g}(\hat{\theta}), \qquad \hat{g}(\theta)=m(\theta)-\hat{m}^{\mathrm{data}}.
\end{equation}
Under correct specification and regularity conditions, $J$ is asymptotically $\chi^2_{q-p}$ when $W=\Omega^{-1}$, the inverse covariance of the moment discrepancy.  If $W=\Sigma_m^{-1}$ is used while finite simulation noise remains, the statistic must be scaled accordingly or calibrated by bootstrap.  A large $J$ indicates either model misspecification, poorly chosen moments, or underestimated sampling uncertainty in moments.

\paragraph{Standard errors and confidence intervals.}  In applications, report not only $\hat{\theta}$ but also uncertainty quantification.  A plug-in sandwich estimator is:
\begin{equation}
\widehat{\mathrm{Var}}(\hat{\theta}) =
\frac{1+1/S}{T}\,(\hat{M}^\top W \hat{M})^{-1}\hat{M}^\top W \hat{\Sigma} W \hat{M}(\hat{M}^\top W \hat{M})^{-1},
\end{equation}
where $\hat{M}$ and $\hat{\Sigma}$ are estimated at $\hat{\theta}$ under the equal-length independent-simulation approximation.  For small samples, time-series dependence, or highly nonlinear criteria, moving-block or parametric bootstrap intervals are often more reliable than first-order asymptotics.

\paragraph{Weak-identification workflow.}  If the smallest singular values of $\hat{M}$ are close to zero, inference based purely on local curvature is fragile.  In that case, complement Hessian-based standard errors with profile-criterion diagnostics: vary one parameter at a time (or along weak singular vectors), re-optimize the remaining parameters, and report objective-function contour sets in addition to pointwise intervals.

\section{GP Surrogate over the Moment Map}
\label{sec:smm_gp_moments}

\begin{remarkbox}[Scope of this section]
The simplified core notebooks \tpath{lecture_15_03_Structural_Estimation_BM.ipynb} and \tpath{lecture_15_03b_Structural_Estimation_BM_Joint.ipynb} stop after the direct surrogate-based SMM estimator and its identification diagnostics; they do \emph{not} implement the second-layer Gaussian-process moment surrogate, leave-one-out validation, or Bayesian active learning described below.  This section sketches the research-scale extension that a separate companion notebook would add on top of the policy surrogate, and that motivates the GP active-learning workflow used in Chapter~\ref{ch:climate}.
\end{remarkbox}

The pseudo-state DEQN of the previous sections turned ``one re-solve per candidate $\theta$'' into ``one forward simulation per candidate $\theta$.''  For high-throughput SMM, that second cost is still nontrivial: a bootstrap with $B = 1{,}000$ resamples needs $1{,}000$ forward simulations on top of the inner optimisation, joint estimation in $p \ge 2$ dimensions multiplies the budget further, and downstream Bayesian or simulation-based-inference workflows want \emph{very} cheap evaluations of $\theta \mapsto m(\theta)$.  This section adds a second surrogate layer, a Gaussian process per moment, on top of the policy net, following exactly the supervised-learning logic of \S\ref{sec:gp_dp_supervised_view}.

\subsection{The Two-Layer Surrogate Architecture}

Recall the DEQN policy net $\mathcal{N}_\rho(z, K, \theta)$ from \S\ref{sec:smm_method}; given $\theta$, it returns the savings rate as a function of $(z, K)$ and lets us simulate a length-$T$ path $\{C_t(\theta), I_t(\theta), Y_t(\theta)\}_{t=1}^T$ in milliseconds.  The empirical SMM workflow then maps that simulated path to a moment vector $m(\theta) \in \R^q$ via a fixed numerical recipe (means, standard deviations, autocorrelations).

\begin{figure}[H]
\centering
\begin{tikzpicture}[
    box/.style={rectangle, draw=uzhblue, fill=uzhgreylight,
        text width=6.6cm, align=center, minimum height=0.55cm,
        font=\scriptsize, rounded corners=2pt, thick, inner sep=2pt},
    netbox/.style={rectangle, draw=uzhblue, fill=softblue!25,
        text width=6.6cm, align=center, minimum height=0.7cm,
        font=\scriptsize, rounded corners=2pt, thick, inner sep=2pt},
    gpbox/.style={rectangle, draw=uzhblue, fill=softorange!30,
        text width=6.6cm, align=center, minimum height=0.7cm,
        font=\scriptsize, rounded corners=2pt, thick, inner sep=2pt},
    outbox/.style={rectangle, draw=uzhblue, fill=softgreen!30,
        text width=6.6cm, align=center, minimum height=0.55cm,
        font=\scriptsize, rounded corners=2pt, thick, inner sep=2pt},
    costlbl/.style={font=\tiny\itshape, text width=3.4cm, align=left, anchor=west},
    >=Stealth, node distance=0.22cm
]
\node[box] (theta) {Candidate parameter vector $\theta$ \\[1pt] {\scriptsize (e.g.\ $\theta=(\beta,\varrho)$)}};
\node[netbox, below=of theta] (deqn) {\textbf{Layer 1 (offline, once):}\\ DEQN policy net $\mathcal{N}_\rho(z, K, \theta)$ \\[1pt] {\scriptsize $\theta$ enters as a pseudo-state input; replaces one full DP re-solve per $\theta$}};
\node[box, below=of deqn] (sim) {$T$-step forward simulation under $\mathcal{N}_\rho(\cdot,\cdot,\theta)$ \\[1pt] {\scriptsize $\Rightarrow$ path $\{C_t(\theta), I_t(\theta), Y_t(\theta)\}_{t=1}^T$}};
\node[box, below=of sim] (mom) {Moment vector $m(\theta)\in\R^q$ \\[1pt] {\scriptsize means, std.\ devs., autocorrelations}};
\node[gpbox, below=of mom] (gp) {\textbf{Layer 2 (per design):}\\ $q$ independent GPs \quad $\widehat m_j(\theta)\sim\mathrm{GP}(0,k_j)$\\[1pt] {\scriptsize trained on small design $\{(\theta^{(i)}, m(\theta^{(i)}))\}_{i=1}^n$; replaces forward simulation per $\theta$}};
\node[outbox, below=of gp] (Q) {SMM objective\quad $Q(\theta)=\bigl(\widehat m(\theta)-m_\mathrm{data}\bigr)^{\!\top}\!W\bigl(\widehat m(\theta)-m_\mathrm{data}\bigr)$\\[1pt] {\scriptsize closed-form, differentiable, $\sim$\,microseconds per call}};
\draw[->, thick] (theta) -- (deqn);
\draw[->, thick] (deqn)  -- (sim);
\draw[->, thick] (sim)   -- (mom);
\draw[->, thick] (mom)   -- (gp);
\draw[->, thick] (gp)    -- (Q);
\node[costlbl, right=0.6cm of theta]
    (clab0) {\textbf{per-$\theta$ cost}};
\node[costlbl, right=0.6cm of deqn, harvardcrimson]
    (clab1) {without surrogates:\\ seconds-to-hours \\ (full DP re-solve)};
\node[costlbl, right=0.6cm of sim, softorange!85!black]
    (clab2) {with Layer 1:\\ milliseconds\\ (one $T$-step simulation)};
\node[costlbl, right=0.6cm of gp, darkgreen]
    (clab3) {with Layer 1 + 2:\\ microseconds\\ (closed-form GP eval)};
\draw[decorate, decoration={brace, amplitude=4pt}, gray]
    ([xshift=-3pt]clab1.north west) -- node[gray, font=\tiny, right=2pt]{}
    ([xshift=-3pt]clab3.south west);
\end{tikzpicture}
\caption{The two-layer surrogate architecture for surrogate-based SMM, read top-to-bottom along the chain $\theta \to \mathcal{N}_\rho \to$ simulated path $\to m(\theta) \to$ moment GPs $\to Q(\theta)$.  \emph{Layer~1} is the pseudo-state DEQN policy net of \S\ref{sec:smm_method}: trained once with $\theta$ as an additional input, it replaces the inner Bellman / fixed-point re-solve that direct SMM would require at every candidate parameter, leaving only a $T$-step forward simulation per $\theta$.  \emph{Layer~2} is the moment-map GP regression of this section: $q$ independent Gaussian processes are fitted to the simulator's $(\theta^{(i)}, m(\theta^{(i)}))$ pairs on a small design, after which evaluating the SMM criterion $Q(\theta)$ at any new $\theta$ requires only a closed-form GP posterior-mean call per moment.  The right-hand annotation traces the per-$\theta$ cost cascade: the direct re-solve costs seconds-to-hours, Layer~1 brings the cost down to milliseconds (one DEQN-driven simulation), and Layer~2 down to microseconds (one differentiable regression call per moment).  This is the same supervised-learning-on-an-expensive-oracle pattern as GP-VFI in \S\ref{sec:gp_dp_supervised_view}, with the moment vector $m(\theta)$ playing the role the Bellman label $TV(\x)$ plays there; the saving compounds because the high-throughput downstream workflows of SMM, bootstrap, profile likelihood, and simulation-based inference, all live in the bottom box.}
\label{fig:smm_two_layer_surrogate}
\end{figure}
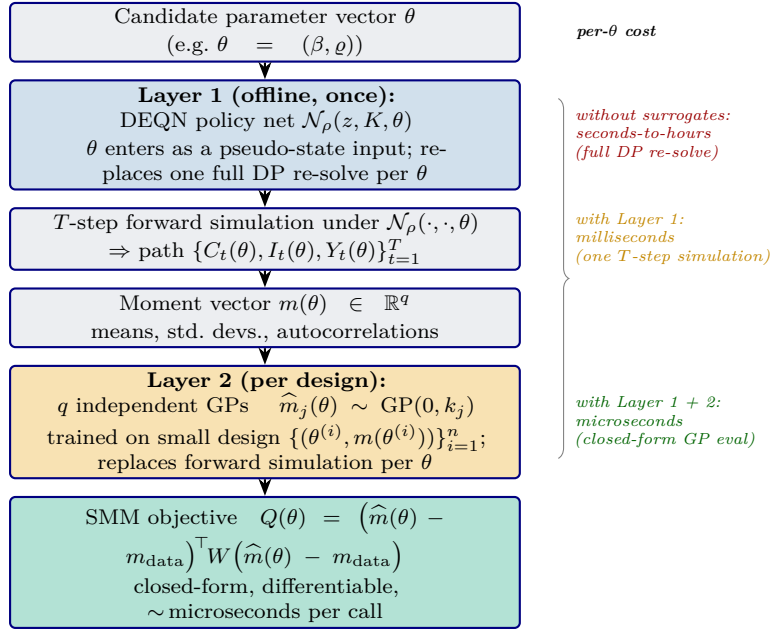

We now stack a second surrogate on top of this:
\begin{equation}
\widehat m_j(\theta) \;\sim\; \mathrm{GP}\bigl(0,\, k_j(\cdot, \cdot)\bigr), \qquad j = 1, \ldots, q,
\label{eq:moment_gp}
\end{equation}
one independent GP per moment, with its own kernel and length-scale hyperparameters learned by maximising marginal likelihood on a small design $\{(\theta^{(i)},\, m(\theta^{(i)}))\}_{i=1}^n$.  Once trained, evaluating the SMM objective at any candidate $\theta$ is a single GP forward pass per moment, no simulation required.  Bootstrapped CIs and any subsequent Bayesian post-processing then run on the GP, not on the simulator.  Figure~\ref{fig:smm_two_layer_surrogate} reads the architecture top-to-bottom.  A candidate $\theta$ is fed into the DEQN policy net of \S\ref{sec:smm_method}, which is rolled forward for $T$ periods to produce a simulated path and its moment vector $m(\theta)\in\R^q$; a small design $\{(\theta^{(i)},\,m(\theta^{(i)}))\}_{i=1}^n$ of those simulator labels then trains the second layer of $q$ independent moment GPs, after which the SMM criterion $Q(\theta)$ is a closed-form quadratic in the GP posterior means.  The right-hand column traces the per-$\theta$ cost cascade from seconds-to-hours down to microseconds.

\paragraph{Same expensive-oracle structure as VFI.}
This is structurally identical to the GP-VFI setup of \S\ref{sec:gp_dp}.  There, one design point cost one Bellman maximisation; here, one design point costs one $T$-step forward simulation plus a moment computation.  In both cases the regressor sees a small but high-quality training set generated by an expensive numerical procedure, and the GP machinery, marginal-likelihood Occam's razor for hyperparameters, leave-one-out diagnostics for surrogate health, and Bayesian active learning for adaptive design, applies verbatim.

\subsection{Leave-One-Out Validation of the Moment Surrogate}

The Cholesky-trick LOO formula~\eqref{eq:gp_loo} of \S\ref{sec:gp_loo} delivers a held-out predictive error for each moment GP at zero marginal cost beyond the existing posterior factorisation.  A research-scale companion to the core SMM notebooks would track
\[
\mathrm{LOO\text{-}RMSE}_j \;=\; \sqrt{\frac{1}{n}\sum_{i=1}^{n}\bigl(\widehat m_j^{-i}(\theta^{(i)}) - m_j(\theta^{(i)})\bigr)^2}
\]
for every moment $j$ and every design size $n$, and pair it with an independent sanity check that evaluates the GP at a \emph{fresh} interior holdout point $\theta_\mathrm{holdout}$ never seen during training; agreement between the two RMSEs is the criterion for declaring the moment surrogate trustworthy before any bootstrap or SBI workflow is run on top of it.

\subsection{Active Learning of the Moment Surrogate}

Two acquisition strategies are natural, matched to the dimensionality of the parameter.

\paragraph{Single-parameter case.}  With a scalar $\theta=\varrho\in[0.50,0.99]$, a coarse uniform pilot grid of $n_0$ points can be enriched by $n_\mathrm{add}$ active points placed sequentially at locations of largest standardised moment-GP posterior uncertainty,
\[
\theta^{\mathrm{next}} \in \argmax_{\theta \in \mathcal{X}^\mathrm{cand}}\;\Bigl\|\boldsymbol\sigma_m(\theta) \,/\, \bar{\boldsymbol\sigma}_m\Bigr\|_2,
\]
subject to a minimum-spacing constraint against existing design points.  This is the same pure-exploration acquisition used for VFI~\eqref{eq:bal_vfi}, modulo the per-moment normalisation that prevents one large-magnitude moment from dominating the objective.

\paragraph{Joint-parameter case.}  With $\theta=(\beta,\varrho)$ on a 2D rectangle, pure exploration is wasteful because most of the rectangle sits far from the SMM minimiser.  A natural alternative is a BoTorch-style Upper-Confidence-Bound (UCB) acquisition on the transformed score $\widetilde Q(\theta) := -\log_{10}(Q(\theta)+\varepsilon)$, multiplicatively weighted by the moment-GP posterior uncertainty:
\begin{equation}
a(\theta)
=
\bigl[\,0.25+\widetilde{\mathrm{UCB}}_{\widetilde Q}(\theta)\,\bigr]
\cdot
\Bigl\|\boldsymbol\sigma_m(\theta)\,/\,\widehat{\mathrm{sd}}(m)\Bigr\|_2,
\label{eq:smm_acquisition}
\end{equation}
where $\widetilde{\mathrm{UCB}}_{\widetilde Q}$ is a quantile-scaled and clipped UCB score.  The first factor exploits, biasing the design toward $(\beta,\varrho)$ pairs with small SMM criterion, while the second explores, requiring the design to also hit places where the moment GP is uncertain.  The additive constant $0.25$ keeps the exploration term active even where the scaled UCB is zero, preventing pathological degeneracy in the acquisition.

\paragraph{Three-way comparison.}  At a fixed design budget, one can compare pilot grid / naive Latin-hypercube / BoTorch-BAL designs along three axes: (i) leave-one-out error on the moment GPs; (ii) error on the recovered SMM criterion against a fresh reference grid; (iii) accuracy of the recovered estimate $(\hat\beta,\hat\varrho)$.  Active designs typically give the most stable local moment surrogate at small budgets.

\subsection{The 2D SMM Criterion Surface and Partial Identification}

\begin{figure}[ht]
\centering
\includegraphics[width=0.92\linewidth]{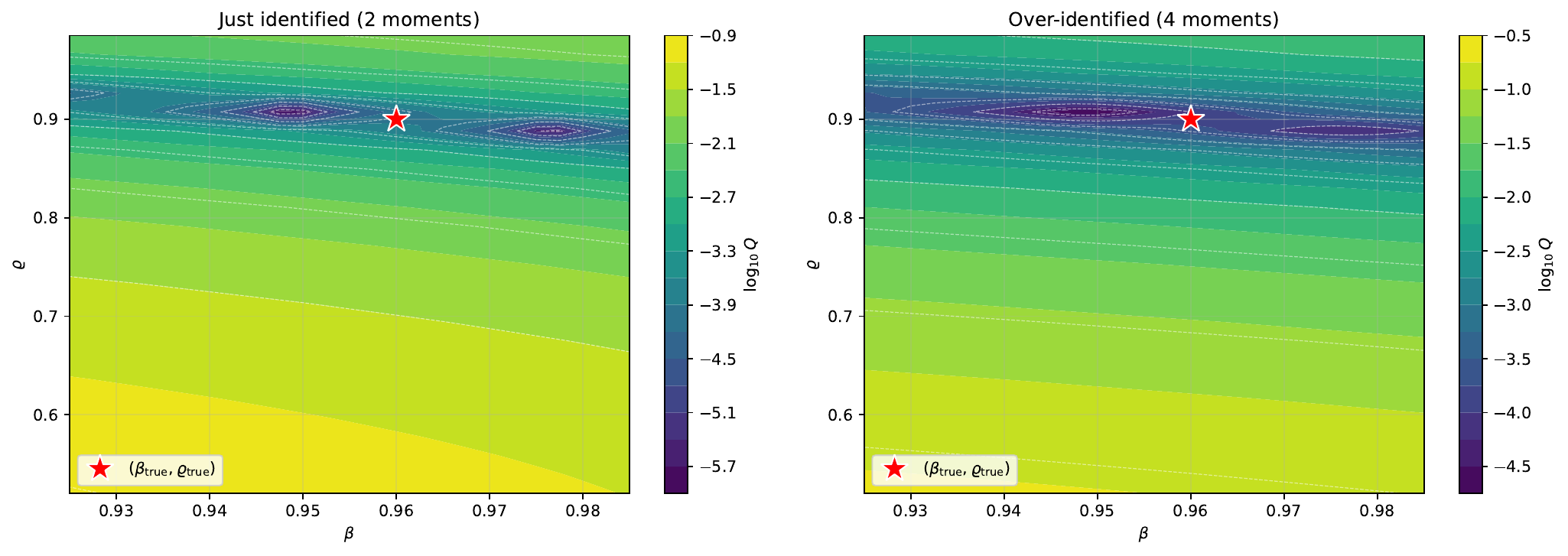}
\caption{Direct SMM criterion for the joint Brock--Mirman estimation.  The left panel uses the just-identified two-moment specification and displays a shallow ridge along $\beta$, signalling that patience is only partially identified by those two moments.  The right panel uses the over-identified four-moment specification, which adds volatility and output-persistence information and produces a localized minimum near the synthetic truth.  Generated by notebook 03b.}
\label{fig:smm_2d_criterion}
\end{figure}

Figure~\ref{fig:smm_2d_criterion} shows the direct SMM criterion on the joint rectangle.  Two features are visible.

First, the criterion has a long, shallow ridge along the $\beta$ direction in the just-identified specification: the data are nearly uninformative about $\beta$ once $\varrho$ is fixed, so a wide range of $\beta$-values fits almost equally well.  This is partial identification, in textbook form, visualised on the criterion surface.  Economically, $\beta$ and $\varrho$ both shift the consumption-smoothing motive in similar directions on long horizons: raising patience and raising persistence each raise the mean savings rate and dampen consumption-growth autocorrelation, so a two-moment specification built from those two moments leaves the $(\beta,\varrho)$ ratio under-determined and produces the ridge.

Second, the ridge collapses to a localized minimum in the over-identified specification.  In the synthetic CRN run, the recovered estimate sits very close to $(\beta_{\mathrm{true}},\varrho_{\mathrm{true}})=(0.96,0.90)$.  This makes a useful pedagogical point: identification is a property of the moment selection, not of the estimator.  The over-identified specification breaks the redundancy by adding growth volatility, which is sensitive to $\varrho$ through the shock-persistence channel but only weakly to $\beta$, and output autocorrelation, which is sensitive to $\varrho$ directly.

In the research-scale extension, the second-layer GP fitted to the joint moment map provides a closed-form, microseconds-per-call substitute for forward simulation: subsequent SMM evaluations, bootstrap replications, and Bayesian post-processing run on the GP rather than on the simulator.  The TikZ architecture diagram in Figure~\ref{fig:smm_two_layer_surrogate} already encodes the cost cascade; the rendered GP-objective surface itself is not produced by the core notebooks and is therefore not displayed here.

\begin{remarkbox}[Hands-on]
Notebook \tpath{lecture_15_03_Structural_Estimation_BM.ipynb} fits the scalar persistence exercise: a 3-input pseudo-state policy net, common-random-number simulation across a $\varrho$-grid, and an interior SMM estimate with a moment-Jacobian diagnostic.  Notebook \tpath{lecture_15_03b_Structural_Estimation_BM_Joint.ipynb} performs joint $(\beta,\varrho)$ estimation and visualises the partial-identification ridge of Figure~\ref{fig:smm_2d_criterion} for the shallow-ridge two-moment specification and the over-identified specification.  The second-layer GP-moment-map extension above is left as a research-scale companion.
\end{remarkbox}

\section{Beyond SMM: Indirect Inference and Simulation-Based Inference}
\label{sec:beyond_smm}

SMM matches a hand-picked vector of moments.  Two close cousins are worth knowing because they often dominate SMM when moment selection is awkward or when one wants the full likelihood.

\paragraph{Indirect inference.}  \citet{smith1993estimating} and \citet{gourieroux1993indirect} replace the moment vector $m(\theta)$ with the parameters of a tractable \emph{auxiliary model} (e.g.\ a low-order VAR or a flexible regression) fitted to both the data and to model-simulated data.  Estimation matches the auxiliary parameters rather than raw moments; the resulting estimator is asymptotically equivalent to ML when the auxiliary model is sufficiently rich, and the auxiliary parameters often summarize the distribution far more efficiently than a hand-picked moment list.  Indirect inference is the natural choice in macro-finance applications where standard moments are weakly identifying but a structural VAR or a near-likelihood auxiliary is available.

\paragraph{Simulation-based inference (SBI).}  In settings where the simulator is differentiable or fast but the likelihood $p(y \mid \theta)$ is intractable, modern SBI \citep{cranmer2020frontier} learns a neural conditional density estimator $q_\phi(\theta \mid y)$ (or its likelihood/likelihood-ratio counterpart) directly from $(\theta_i, y_i)$ pairs simulated under the prior.  The resulting object is an amortised \emph{Bayesian} posterior usable for any future observation $y^\star$ at cost of one forward pass.  SBI generalizes Approximate Bayesian Computation, sidesteps moment selection entirely, and naturally pairs with the deep-surrogate machinery of Chapter~\ref{ch:gp}.  In the surrogate-then-estimate framing of this chapter, the most direct SBI variant is \emph{neural posterior estimation} (NPE), where the pseudo-state DEQN provides the simulator and the GP moment surrogate of \S\ref{sec:smm_gp_moments} is replaced by a learned posterior $q_\phi(\theta\mid y)$; Exercise~\ref{ex:ch10:4} contrasts SMM with SBI in algorithmic terms.

\medskip
\noindent\emph{When to choose what.}  SMM remains the workhorse when a small number of structural moments are well-identified and economists want a transparent, interpretable objective.  Indirect inference dominates when an informative auxiliary model is available.  SBI is the natural tool in environments where the model is expensive to simulate but a one-time training run unlocks Bayesian inference at \emph{evaluation} time, precisely the setting in which the rest of this script deploys deep surrogates.

\paragraph{Why this matters for the next chapter.}  Climate--economy integrated assessment models (Chapter~\ref{ch:climate}) are the prototypical setting where surrogate-based estimation pays off.  Credible policy analysis requires evaluating the model over many climate-sensitivity, damage-elasticity, and discount-rate scenarios.  Treating the parameter vector as a pseudo-state and training a single deep surrogate, exactly as in the SMM exercise above, turns repeated re-solves into repeated forward passes and is the technical bridge between this chapter and the next.

\begin{keyinsightbox}[Chapter Summary]
\begin{itemize}[itemsep=2pt, leftmargin=*]
\item SMM matches simulated moments to data moments by minimizing a weighted quadratic objective; it is GMM with simulation \citep{mcfadden1989method, pakes1989simulation}.
\item The pseudo-state surrogate trick (treat parameters $\theta$ as additional network inputs) replaces a re-solve at every candidate $\theta$ with a single forward pass; this is what makes SMM with a deep model tractable.
\item Stacking a Gaussian-process layer over the moment map (\S\ref{sec:smm_gp_moments}) turns SMM into a two-stage surrogate problem: each oracle call is one forward simulation, each subsequent objective evaluation is one GP posterior, the same expensive-oracle / supervised-learning logic that motivates GP-based VFI in \S\ref{sec:gp_dp}.
\item Common random numbers across $\theta$-candidates are essential in the classroom exercises: they remove simulation noise from the objective and let optimizers see a smooth landscape \citep{glasserman2004monte}.
\item Indirect inference and modern simulation-based inference \citep{smith1993estimating, gourieroux1993indirect, cranmer2020frontier} are the natural neighbors of SMM and dominate in their respective regimes.
\end{itemize}
\end{keyinsightbox}

\section*{Further Reading}
\addcontentsline{toc}{section}{Further Reading}
\begin{itemize}[itemsep=2pt]
\item \citet{mcfadden1989method, pakes1989simulation, duffie1993simulated}, the foundational SMM trio.
\item \citet{kase2022estimating}, neural-network estimation of nonlinear heterogeneous-agent models; \citet{chen2026Deep}, deep surrogates for finance and option pricing.
\item \citet{cranmer2020frontier}, contemporary simulation-based inference.
\end{itemize}

\section*{Exercises}
\addcontentsline{toc}{section}{Exercises}
\noindent Worked solutions and guidance for these exercises appear in Appendix~\ref{app:solutions}.
\begin{enumerate}[itemsep=4pt, leftmargin=*]
\item\label{ex:ch10:1} \textbf{[Computational] Identification.}  In notebook \tpath{lecture_15_03_Structural_Estimation_BM.ipynb}, choose two candidate moments and compute the finite-difference Jacobian $\partial m/\partial\varrho$ at $\varrho_{\mathrm{true}}=0.90$.  Which moment provides stronger local identification?
\item\label{ex:ch10:2} \textbf{[Core] Optimal weighting.}  Show that under standard regularity conditions, $W^\star=\Omega^{-1}$ minimizes the asymptotic variance of $\hat\theta_{\mathrm{SMM}}$, where $\Omega$ is the covariance of the moment discrepancy.  Why does identity weighting still appear in the first stage?
\item\label{ex:ch10:3} \textbf{[Computational] Common random numbers.}  In the scalar Brock--Mirman exercise, plot the SMM objective as a function of $\varrho$ with and without common random numbers.  Quantify the objective noise across repeated Monte Carlo panels under the same candidate grid.
\item\label{ex:ch10:4} \textbf{[Core] SMM vs.\ SBI.}  Outline the algorithmic difference between estimating $\theta$ by SMM (one optimization per dataset) and by neural simulation-based inference (one training run plus one posterior evaluation per dataset).  In which regime does SBI dominate?
\item\label{ex:ch10:5} \textbf{[Advanced/project] $J$-statistic and overidentification.}  In notebook \tpath{lecture_15_03b_Structural_Estimation_BM_Joint.ipynb}, use the over-identified specification with $q=4$ moments and $p=2$ parameters.  (i)~State the asymptotic distribution of the $J$-statistic under correct specification when $W=\Omega^{-1}$.  (ii)~Compute $J(\hat\theta)$ on the original synthetic sample and report whether the $\chi^2_2$ test rejects at $\alpha=0.05$.  (iii)~Repeat across Monte Carlo samples generated at the truth and compare the empirical distribution with $\chi^2_2$.  (iv)~Introduce a structural break in one model parameter and verify that the $J$ distribution shifts to the right.
\item\label{ex:ch10:6} \textbf{[Advanced/project] Bootstrap confidence intervals.}  Compare three confidence-interval procedures for $\hat\theta_\mathrm{SMM}$: (a)~plug-in sandwich standard errors; (b)~moving-block or stationary bootstrap of the time-series sample; (c)~parametric bootstrap, drawing new samples under the simulated model at $\hat\theta$.  Report the confidence intervals and coverage across repeated Monte Carlo replications.
\item\label{ex:ch10:7} \textbf{[Advanced/project] Maximum likelihood vs.\ SMM efficiency.}  Suppose the productivity process $\log z_t$ is observed.  Implement the Gaussian AR(1) MLE for $\varrho$ and compare it with the SMM estimator based on the moments in notebook \tpath{lecture_15_03_Structural_Estimation_BM.ipynb}.  On Monte Carlo replications at several sample sizes, report bias, variance, and MSE.  Explain why MLE is efficient for this observed-shock likelihood, while SMM reaches the GMM efficiency bound only for the chosen moment vector.  As an optional extension, repeat the beta-only MLE comparison in the full-depreciation log-utility special case where the policy has a closed form.
\end{enumerate}

\chapter{Climate Economics and Deep Uncertainty Quantification}
\label{ch:climate}

This chapter brings together the methods developed throughout this script and applies them to one of the most consequential computational challenges in economics: \emph{climate change policy}.  Integrated assessment models (IAMs) couple economic growth with the carbon cycle, temperature dynamics, and climate damages, creating high-dimensional nonstationary dynamic programming problems that are ideal candidates for the DEQN and surrogate methods we have developed.  We present the CDICE model of \citet{Folini_2021}, solve it with DEQNs, and then use GP surrogates and Bayesian active learning, first for deep uncertainty quantification \citep{friedlDeep2023}, and then, applying the same surrogate-then-optimize machinery to a different OLG model and a different surrogate, to \emph{search over policy parameters} and derive constrained Pareto-improving carbon tax rules in an OLG--IAM with deep uncertainty \citep{kubler2025using}.  This last step illustrates a general use of surrogates that goes beyond estimation and UQ: once the structural model has been mapped into a fast, differentiable surrogate, the costly outer loop of an optimal-policy search in a dynamic stochastic heterogeneous-agent economy collapses into a small optimization on the surrogate.  For broader overviews of climate economics and IAMs, see \citet{hassler2016environmental} on environmental macroeconomics, \citet{DIETZ20241} on IAMs, \citet{fernandezvillaverde2025climate} on the intersection of climate economics and deep learning, and \citet{vanderploegrezai2026climate} on the macroeconomics of climate policy.

\section{The Macroeconomics of Climate Change}
\label{sec:climate_motivation}

Climate change is a global externality: the emissions of each agent affect the welfare of all agents, including future generations that have no say in current decisions.  Unlike standard market failures, the climate externality operates across centuries, involves deep scientific uncertainty, and couples the macroeconomy with the earth system in both directions.  Recent overviews include \citet{hassler2016environmental} on environmental macroeconomics, \citet{DIETZ20241} on IAMs, \citet{fernandezvillaverde2025climate} on climate economics and deep learning, and \citet{vanderploegrezai2026climate} on the macroeconomics of climate policy.

\paragraph{Integrated assessment models.}
Integrated assessment models (IAMs) formalize this coupling.  The economy produces output and emissions; emissions accumulate in the atmosphere and raise global temperature; temperature increases cause damages that reduce output.  The feedback loop is closed (Figure~\ref{fig:iam_feedback_loop}):

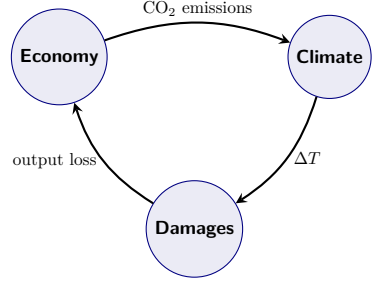
\begin{figure}[ht]
\centering
\begin{tikzpicture}[>=stealth, node distance=2cm, scale=0.65, transform shape]
    \tikzstyle{main node} = [circle, draw=uzhblue, fill=uzhblue!8, minimum size=1.4cm, font=\sffamily\bfseries]
    \node[main node] (economy) at (0,0) {Economy};
    \node[main node] (climate) at (5.5,0) {Climate};
    \node[main node] (damages) at (2.75,-3.5) {Damages};
    \draw[->, thick] (economy) to[bend left=20] node[above, font=\small] {CO$_2$ emissions} (climate);
    \draw[->, thick] (climate) to[bend left=20] node[right, font=\small] {$\Delta T$} (damages);
    \draw[->, thick] (damages) to[bend left=20] node[left, font=\small] {output loss} (economy);
\end{tikzpicture}
\caption{The integrated-assessment feedback loop. The economy produces output and CO$_2$ emissions; emissions accumulate in the atmosphere and raise global mean temperature ($\Delta T$); higher temperatures generate damages that reduce output and consumption, which in turn shape the path of future emissions. An IAM closes this loop and uses it to quantify the welfare cost of additional emissions, summarized by the social cost of carbon~\eqref{eq:scc}.}
\label{fig:iam_feedback_loop}
\end{figure}

The central output of an IAM is the \textbf{social cost of carbon} (SCC): the marginal welfare cost of one additional unit of CO$_2$ emissions, measured in consumption-equivalent units.  When emissions are measured in GtC (gigatons of carbon), the SCC has units of consumption per GtC.  Conversion to USD per tCO$_2$ requires first applying the consumption-to-USD numeraire and then converting the carbon mass unit: one tCO$_2$ contains $12/44$ tons of carbon, so a price expressed per ton of carbon is divided by $44/12$ to obtain the corresponding price per ton of CO$_2$ (and a GtC price is also divided by $10^9$).  Formally,
\begin{equation}
\mathrm{SCC}_t = -\frac{\partial V_t / \partial E_t}{\partial V_t / \partial C_t},
\label{eq:scc}
\end{equation}
where $V_t$ is the value function, $E_t$ is contemporaneous emissions, and $C_t$ is consumption.  The flow form is linked to the stock-based form $\mathrm{SCC}^M_t = -(\partial V_t/\partial M^{\mathrm{AT}}_t)/(\partial V_t/\partial C_t)$, derived in Section~\ref{sec:dice_deqn}, by the chain rule
\begin{equation*}
\frac{\partial V_t}{\partial E_t}
\;=\;
\frac{\partial V_{t+1}}{\partial M^{\mathrm{AT}}_{t+1}}\,\frac{\partial M^{\mathrm{AT}}_{t+1}}{\partial E_t}
\end{equation*}
together with the carbon-to-CO$_2$ unit conversion noted above.  In a first-best allocation, the optimal carbon tax equals the SCC \citep{golosov2014optimal}.  The SCC is high when climate damages are steep, the climate response is strong, discounting is low, and tipping risks are material \citep{caiSocialCostCarbon2019, DIETZ20241}.

\paragraph{From surrogates to climate IAMs.}
Chapter~\ref{ch:gp} introduced surrogates and Bayesian active learning as fast approximators for repeated model evaluations.  Climate IAMs are the natural application: each parameter configuration (climate sensitivity, damage curvature, discount rate) is expensive to solve, yet policy questions require evaluating thousands of configurations to map out tail risks and Pareto-improving rules.  The DEQN approach of Chapters~\ref{ch:deqn}--\ref{ch:irbc}, combined with the GP and active-learning toolkit of Chapter~\ref{ch:gp}, is therefore the natural workhorse for climate-policy uncertainty quantification.

\paragraph{Why computation matters.}
Solving IAMs globally, as opposed to linearization or certainty equivalence, is computationally demanding for several reasons:
\begin{itemize}[itemsep=2pt]
\item \textbf{Nonstationarity:} TFP, population, emissions intensity, and radiative forcing all change exogenously over time, so the policy function cannot be time-invariant.
\item \textbf{Coupled dynamics:} the economy and climate interact in both directions through emissions and damages.
\item \textbf{Long horizon:} welfare effects unfold over 100--300 years, requiring stable numerical solutions far from the steady state.
\item \textbf{Curse of dimensionality:} multiple climate state variables (carbon stocks, temperature layers), stochastic shocks, and uncertain parameters raise the dimension of the state space well beyond what standard grid-based methods can handle.
\end{itemize}
The deep learning toolkit developed in this course (DEQNs, deep surrogates, and GP-based uncertainty quantification) is therefore particularly well suited to climate economics.

\paragraph{The three movements of this chapter.}
The remainder of this chapter has three movements.  Movement~1 (\S\ref{sec:iam_nonstationarity}--\S\ref{sec:nsdeqn_algo}) makes precise what changes when we ask the Deep Equilibrium Network of Chapter~\ref{ch:deqn} to solve a non-stationary IAM, and presents the modified training algorithm in one labeled box.  Movement~2 (\S\ref{sec:dice_lagrangian}--\S\ref{sec:dice_to_stochastic_iam}) puts that algorithm to work on a concrete stochastic DICE economy.  Movement~3 (\S\ref{sec:bayesian_learning}--\S\ref{sec:pareto_carbon_tax}) sketches the four extensions that matter for serious climate-finance research: Bayesian learning on the climate sensitivity, recursive Epstein--Zin preferences, global uncertainty quantification of the social cost of carbon, and constrained Pareto-improving carbon-tax design in a heterogeneous-agent IAM.

\section{The DICE Model}
\label{sec:dice}

\subsection{The IAM Landscape}
\label{sec:iam_landscape}

DICE is the workhorse of this chapter, but it is one of several integrated assessment models in active use.  The list below summarizes the active landscape; each model trades global parsimony for regional or sectoral granularity, and computational tractability for fidelity of the climate physics.

\begin{description}[itemsep=4pt, leftmargin=2.6em]
\item[DICE.]  Global aggregate; 3-box carbon cycle and 2-layer energy-balance model; one-sector Ramsey planner.  The standard benchmark for SCC and integrated policy analysis \citep{nordhausRevisitingSocialCost2017}.
\item[RICE.]  Twelve-region extension of DICE with trade \citep{nordhaus1996regional}.  Used for regional SCC and equity questions.
\item[CDICE.]  A global DICE-2016 recalibration tailored to deep-learning solution methods, with Epstein--Zin preferences and OLG variants.  The model used in \S\ref{sec:dice_deqn} below \citep{Folini_2021}.
\item[ACE.]  Analytic Climate Economy: log-linear approximations to the carbon cycle, temperature dynamics, and damages yield a \emph{closed-form} optimal carbon tax \citep{traeger2018ace}.  Acts as an analytic benchmark for the numerical SCC computed below.
\item[FaIR / MAGICC.]  Reduced-complexity climate emulators that take emissions as input and produce temperature responses; widely used to translate IPCC scenarios to economic models.
\item[WITCH / REMIND.]  Multi-region IAMs with full energy-system modules; standard for mitigation-pathway and technology-portfolio studies.  Outside the scope of this script.
\end{description}

CDICE is the model we solve in this chapter.  ACE provides a useful analytic shadow for it, in particular a closed-form SCC that decomposes transparently into structural parameters; we do not derive that closed form here, but Exercise~\ref{ex:ch11:3} asks the reader to compute it from \citet{traeger2018ace} and compare against the DEQN-trained CDICE solution as an external sanity check.

The \emph{Dynamic Integrated model of Climate and the Economy} (DICE), developed by \citet{nordhaus1994managing}, is the most influential IAM; in this chapter we follow the variant of \citet{nordhausRevisitingSocialCost2017} as recalibrated by \citet{Folini_2021}.  It couples a neoclassical growth model with a reduced-form climate module in a single global framework.  The remainder of this section builds the model up block by block, in increasing complexity: first the macro-economic backbone (\S\ref{sec:dice_ramsey}), then the emissions and abatement technology (\S\ref{sec:dice_emissions}), then the climate physics (\S\ref{sec:dice_carbon_cycle}--\S\ref{sec:dice_temperature}), and finally the damage feedback (\S\ref{sec:dice_damages}) that closes the loop.  A consolidated calibration is given in Table~\ref{tab:dice_calibration}.

\paragraph{Time-step convention.}
Following \citet{Folini_2021} we calibrate CDICE on an \emph{annual} time step, $\Delta_t = 1$ year, so that all rates in Table~\ref{tab:dice_calibration} (capital depreciation $\delta$, pure rate of time preference $\rho$, the decay rates $g^{\sigma}_0, \delta^{\sigma}, g^{\mathrm{back}}, \delta^{\mathrm{Land}}$, the carbon-cycle transfer rates $b_{12}, b_{23}$, and the temperature-block coefficients $c_1, c_3, c_4$) are read directly as annual values; the original DICE-2016 calibration of \citet{nordhausRevisitingSocialCost2017}, by contrast, hard-wires a 5-year time step into its coefficients.  Growth rates of TFP and population are written as annual log changes, $g^A_t := \ln(A_{t+1}/A_t)$ and $g^L_t := \ln(L_{t+1}/L_t)$, and the dynamics~\eqref{eq:carbon_cycle},~\eqref{eq:temp_at}--\eqref{eq:temp_oc} and the FOC residuals of \S\ref{sec:dice_deqn} therefore carry no $\Delta_t$ multipliers; emissions $E_t$ entering~\eqref{eq:carbon_cycle} are the annual total.  Switching to a non-annual $\Delta_t$ amounts to reinserting the multiplications $\Delta_t \cdot \{g^{\sigma}_0, \delta^{\sigma}, g^{\mathrm{back}}, \delta^{\mathrm{Land}}, b_{12}, b_{23}, c_1, c_3, c_4\}$ in the obvious places, the time-step-generic form discussed in Online Appendix~D of \citet{Folini_2021}.

\subsection{Production and the Ramsey--Cass--Koopmans backbone}
\label{sec:dice_ramsey}

Strip away the climate block and DICE is just a neoclassical growth model with population and TFP growth.  A single representative firm produces gross output with Cobb--Douglas technology in capital and effective labor,
\begin{equation}
Y^{\mathrm{gross}}_t \;=\; K_t^{\alpha}\,(A_t L_t)^{1-\alpha},
\label{eq:gross_output}
\end{equation}
where $\alpha\in(0,1)$ is the capital share, $A_t$ is total factor productivity, and $L_t$ is population.  Both $A_t$ and $L_t$ follow deterministic but time-varying paths: $A_t$ trends because of exogenous productivity growth, and $L_t$ follows the calibrated demographic projection of \citet{nordhausRevisitingSocialCost2017}.  The capital stock evolves under the standard accumulation law
\begin{equation}
K_{t+1} \;=\; (1-\delta) K_t + I_t,
\label{eq:capital_accumulation}
\end{equation}
with depreciation rate $\delta$ and gross investment $I_t$.  The economy's resource constraint, written in terms of net (after-damages, after-abatement) output that we develop in \S\ref{sec:dice_emissions}--\S\ref{sec:dice_damages}, is $C_t + I_t \le Y^{\mathrm{net}}_t$, where $C_t$ is aggregate consumption.

A benevolent planner picks $(C_t,\, I_t,\, \mu_t)_{t\ge 0}$ to maximize a discounted CRRA-IES felicity sum,
\begin{equation}
V_0 \;=\; \sum_{t=0}^{\infty} \beta_t\, L_t\,\frac{(C_t/L_t)^{1-1/\psi}-1}{1-1/\psi},
\qquad
\beta_t \;=\; \exp(-\rho\,\Delta_t \cdot t),
\label{eq:planner_obj_crra}
\end{equation}
with intertemporal-elasticity-of-substitution parameter $\psi>0$ and pure rate of time preference $\rho$.  This is the time-additive aggregator of the standard Ramsey--Cass--Koopmans growth model; we replace it with the recursive Epstein--Zin form once stochastic risk enters the picture (\S\ref{sec:ez_layer}).  The planner controls $\mu_t$, the emissions abatement rate, in addition to the savings--consumption split; we develop the cost of abatement next.

\subsection{Industrial emissions, abatement, and the backstop technology}
\label{sec:dice_emissions}

Industrial production is a CO$_2$-emitting activity.  Let $\sigma_t$ denote the \emph{carbon intensity} of gross output, expressed in CDICE's working units of $10^3$\,GtC of emissions per unit of gross output (a $10^3$\,GtC normalization on the carbon stocks improves the conditioning of the climate side; see Table~\ref{tab:dice_calibration}).  Industrial emissions are then $\sigma_t Y^{\mathrm{gross}}_t$ before any mitigation effort; with abatement rate $\mu_t \in [0,1]$ the planner can scale these emissions down,
\begin{equation}
E_{\mathrm{ind},t} \;=\; (1-\mu_t)\,\sigma_t\, Y^{\mathrm{gross}}_t \;=\; (1-\mu_t)\,\sigma_t\, K_t^{\alpha}(A_t L_t)^{1-\alpha}.
\label{eq:emissions_ind}
\end{equation}
Carbon intensity is itself an exogenous decreasing time path.  DICE-2016 calibrates a closed-form decay,
\begin{equation}
\sigma_t \;=\; \sigma_0\,\exp\!\left[\frac{g^{\sigma}_0}{\log(1+\delta^{\sigma})}\bigl((1+\delta^{\sigma})^{t}-1\bigr)\right],
\label{eq:sigma_decay}
\end{equation}
with initial intensity $\sigma_0$, initial growth rate $g^{\sigma}_0<0$ (so emissions per dollar of output fall over time), and second-derivative parameter $\delta^{\sigma}>0$ that bends the path further down at long horizons.  Equation~\eqref{eq:sigma_decay} captures the steady decarbonization that even \emph{unabated} world output undergoes through ongoing technological change; the planner's $\mu_t$ is the additional mitigation effort \emph{on top of} that baseline.

Abatement is not free.  In the spirit of an aggregate marginal-abatement-cost curve, DICE assumes the abatement-cost share of gross output is a power function of $\mu_t$,
\begin{equation}
\Theta(\mu_t) \;=\; \theta_{1,t}\,\mu_t^{\theta_2},
\label{eq:abat_cost}
\end{equation}
with curvature parameter $\theta_2>1$ (a typical calibration is $\theta_2=2.6$).  The level coefficient $\theta_{1,t}$ is not a free parameter: it is pinned down by the cost of the \emph{backstop technology}, the cleanest large-scale abatement technology available at any given time (e.g.\ direct air capture).  Let $p^{\mathrm{back}}_t$ denote the cost per unit of CO$_2$ avoided when the backstop is fully deployed, and assume an exogenous declining path,
\begin{equation}
p^{\mathrm{back}}_t \;=\; p^{\mathrm{back}}_0\,\exp(-g^{\mathrm{back}}\,t),
\label{eq:backstop_price}
\end{equation}
reflecting steady cost reductions in clean technologies.  Setting the marginal abatement cost at $\mu_t=1$ equal to the backstop price (multiplied by carbon-to-CO$_2$ conversion $\mathrm{c2co2}$ to keep mass units consistent, and by $10^3$ to convert $\sigma_t$ from $10^3$\,GtC working units back to GtC) yields the calibration identity
\begin{equation}
\theta_{1,t} \;=\; \frac{p^{\mathrm{back}}_t \cdot 10^3 \cdot \mathrm{c2co2} \cdot \sigma_t}{\theta_2}.
\label{eq:theta1_calibration}
\end{equation}
Equation~\eqref{eq:theta1_calibration} is what makes $\Theta(\mu)$ \emph{economically} meaningful rather than a fitted polynomial: the abatement-cost function inherits its level from the backstop price and its curvature from the assumption $\theta_2=2.6$.  The $10^3$ factor matches Equation~(11) of Online Appendix~D of \citet{Folini_2021} and the corresponding factor of \texttt{1000} in the companion implementation.  As the backstop becomes cheaper ($p^{\mathrm{back}}_t \downarrow$), full mitigation becomes cheaper too, which is one of the channels that makes the deterministic optimal $\mu_t$ rise toward 1 over the 21st century.

The bound $\mu_t \in [0,1]$ deserves a comment.  $\mu_t = 0$ means business-as-usual emissions; $\mu_t = 1$ means full deployment of the backstop, eliminating all industrial emissions.  Values $\mu_t > 1$ would correspond to net-negative industrial emissions (e.g.\ aggressive direct air capture beyond the firm's own footprint), which DICE forbids; we will impose the upper bound as a Kuhn--Tucker constraint, smoothed by a Fischer--Burmeister term, in \S\ref{sec:dice_lagrangian}.

\subsection{Land-use emissions and net output}
\label{sec:dice_landuse}

The atmosphere does not distinguish between an industrial flow and a non-industrial flow of carbon.  In DICE, total emissions therefore comprise an industrial component~\eqref{eq:emissions_ind} and an exogenous land-use-change component,
\begin{equation}
E_{\mathrm{Land},t} \;=\; E_{\mathrm{Land},0}\,\exp(-\delta^{\mathrm{Land}}\,t),
\label{eq:land_emissions}
\end{equation}
which decays smoothly toward zero as deforestation slows.  Total emissions feeding the atmosphere are
\begin{equation}
E_t \;=\; E_{\mathrm{ind},t} + E_{\mathrm{Land},t}.
\label{eq:total_emissions}
\end{equation}
Closing the production block requires accounting for two additional drains on gross output: climate damages, governed by atmospheric temperature $T^{\mathrm{AT}}_t$ via a damage fraction $\Omega(T^{\mathrm{AT}}_t)$ developed in \S\ref{sec:dice_damages}, and abatement spending~\eqref{eq:abat_cost}.  Net output is therefore
\begin{equation}
Y^{\mathrm{net}}_t \;=\; \bigl(1 - \Omega(T^{\mathrm{AT}}_t) - \Theta(\mu_t)\bigr)\,Y^{\mathrm{gross}}_t,
\label{eq:net_output}
\end{equation}
which is what is available for consumption and investment.  The additive form is the convention adopted in CDICE and used by the production-grade DEQN library port; an alternative multiplicative form $(1-\Omega^{\mathrm{ret}})(1-\Theta)$ with retained-output factor $\Omega^{\mathrm{ret}}$ appears in \citet{nordhaus2008question}.

The planner's controls and exogenous trends are now all named.  The endogenous economic state is the capital stock $K_t$.  The exogenous trends are TFP $A_t$, population $L_t$, carbon intensity $\sigma_t$, land-use emissions $E_{\mathrm{Land},t}$, and (added below) the non-CO$_2$ component of radiative forcing $F^{\mathrm{EX}}_t$.  The planner controls the consumption--investment split (equivalently, the savings rate $s_t$) and the abatement rate $\mu_t \in [0,1]$.  All that remains is the climate side: the carbon cycle that turns total emissions $E_t$ into atmospheric concentration, the energy balance that turns concentration into temperature, and the damage function that turns temperature back into output loss.

\subsection{Carbon cycle}
\label{sec:dice_carbon_cycle}

DICE represents the global carbon cycle as a three-reservoir linear system: an atmospheric box, an upper (mixed-layer) ocean box, and a lower (deep) ocean box.  Carbon flows between reservoirs at calibrated rates, and total emissions $E_t$ from~\eqref{eq:total_emissions} enter directly into the atmospheric reservoir.  Stacking concentrations as $M_t = (M^{\mathrm{AT}}_t,\, M^{\mathrm{UO}}_t,\, M^{\mathrm{LO}}_t)^\top$, the transition is
\begin{equation}
M_{t+1} \;=\; (I + B)\, M_t \;+\; \bm{e}_1\,E_t,
\label{eq:carbon_cycle}
\end{equation}
where $\bm{e}_1 = (1,0,0)^\top$ injects emissions into the atmosphere alone, $E_t$ is the per-period emissions total, and the transfer matrix
\begin{equation}
B \;=\; \begin{pmatrix}
-b_{12} & b_{12}\,M^{\mathrm{AT}}_{\mathrm{eq}}/M^{\mathrm{UO}}_{\mathrm{eq}} & 0 \\
b_{12} & -b_{12}\,M^{\mathrm{AT}}_{\mathrm{eq}}/M^{\mathrm{UO}}_{\mathrm{eq}} - b_{23} & b_{23}\,M^{\mathrm{UO}}_{\mathrm{eq}}/M^{\mathrm{LO}}_{\mathrm{eq}} \\
0 & b_{23} & -b_{23}\,M^{\mathrm{UO}}_{\mathrm{eq}}/M^{\mathrm{LO}}_{\mathrm{eq}}
\end{pmatrix}
\label{eq:carbon_B_matrix}
\end{equation}
encodes the two atmosphere--upper-ocean exchange rates ($b_{12}$ in either direction) and the two upper-ocean--lower-ocean exchange rates ($b_{23}$ in either direction).  The off-diagonal scaling by the equilibrium-mass ratios $M^{\mathrm{AT}}_{\mathrm{eq}}/M^{\mathrm{UO}}_{\mathrm{eq}}$ and $M^{\mathrm{UO}}_{\mathrm{eq}}/M^{\mathrm{LO}}_{\mathrm{eq}}$ guarantees that, under zero net emissions, the system relaxes to the calibrated pre-industrial equilibrium $M_{\mathrm{eq}} = (M^{\mathrm{AT}}_{\mathrm{eq}},\, M^{\mathrm{UO}}_{\mathrm{eq}},\, M^{\mathrm{LO}}_{\mathrm{eq}})^\top$.  Calibrated values for $b_{12}, b_{23}$, and $M_{\mathrm{eq}}$ in CDICE are listed in Table~\ref{tab:dice_calibration}.  The lecture slides for this chapter sometimes write the same transition with four directional rates $\phi_{12},\phi_{21},\phi_{23},\phi_{32}$ in place of $b_{12}$ and $b_{23}$; the two parameterizations are identical under $\phi_{12} = b_{12}$, $\phi_{21} = b_{12}\,M^{\mathrm{AT}}_{\mathrm{eq}}/M^{\mathrm{UO}}_{\mathrm{eq}}$, $\phi_{23} = b_{23}$, $\phi_{32} = b_{23}\,M^{\mathrm{UO}}_{\mathrm{eq}}/M^{\mathrm{LO}}_{\mathrm{eq}}$, i.e.\ the slide form makes the equilibrium-mass scaling absorbed into $B$ explicit at the cost of two extra symbols.

\begin{figure}[ht]
\centering
\includegraphics[width=0.85\linewidth]{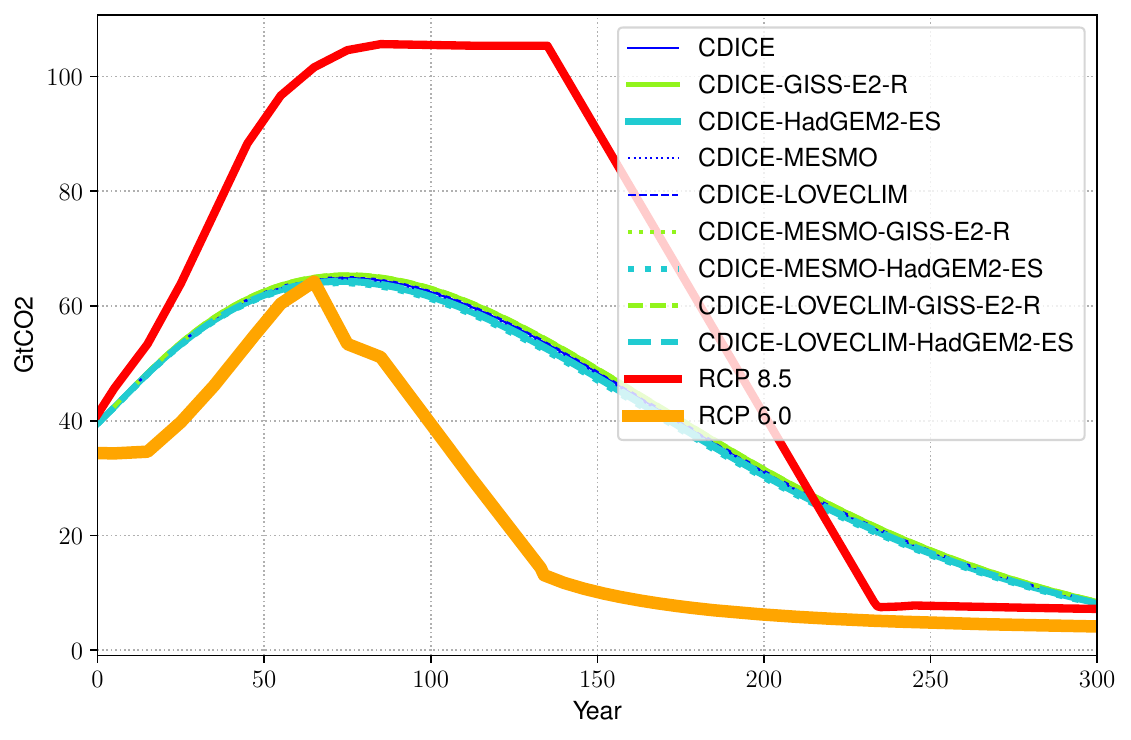}
\caption{Business-as-usual industrial emissions in CDICE (in GtCO$_2$/yr) under the nine combinations of three carbon-cycle calibrations (MMM, MESMO, LOVECLIM) and three temperature calibrations (MMM, HadGEM2-ES, GISS-E2-R); the thin CDICE curves overlap visually, confirming that the BAU emissions path is essentially insensitive to the climate-module calibration because $\sigma_t$ and $A_t$ are exogenous.  The thick red and orange curves are the RCP\,8.5 and RCP\,6.0 scenarios, included as climate-policy reference paths.  Reproduced from \citet{Folini_2021}, Figure~11(a).}
\label{fig:restud_bau_emissions}
\end{figure}

Equation~\eqref{eq:carbon_cycle} is a pulse-and-decay system: a unit pulse of emissions raises atmospheric carbon by one unit instantaneously, and that anomaly then bleeds into the upper ocean over decades and into the deep ocean over centuries.  Figure~\ref{fig:restud_bau_emissions} shows the implied BAU emissions trajectory under nine alternative climate-module calibrations; the spread is mostly driven by the equilibrium climate sensitivity (developed in \S\ref{sec:dice_temperature}), not by the carbon cycle, which is tightly disciplined by the pulse and step tests of \S\ref{sec:cdice_recalibration}.

\subsection{Two-layer energy balance and radiative forcing}
\label{sec:dice_temperature}

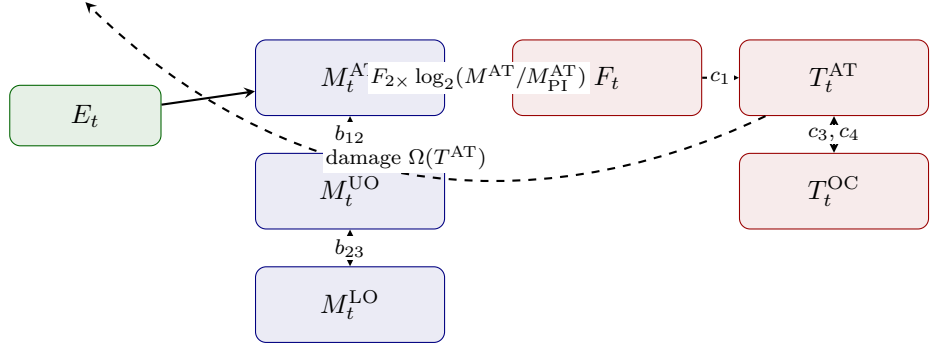
\begin{figure}[ht]
\centering
\begin{tikzpicture}[>=stealth, node distance=2.4cm,
    reservoir/.style={rectangle, draw=uzhblue, fill=uzhblue!8, rounded corners=4pt,
                      minimum width=2.5cm, minimum height=1.0cm, align=center, font=\small},
    temp/.style={rectangle, draw=harvardcrimson, fill=harvardcrimson!8, rounded corners=4pt,
                 minimum width=2.5cm, minimum height=1.0cm, align=center, font=\small},
    src/.style={rectangle, draw=darkgreen, fill=darkgreen!10, rounded corners=4pt,
                minimum width=2.0cm, minimum height=0.8cm, align=center, font=\small},
    edgelab/.style={font=\scriptsize, fill=white, inner sep=1pt}]
    \node[src] (E) at (-3.5, 0.5) {$E_t$};
    \node[reservoir] (Mat) at (0,1) {$M^{\mathrm{AT}}_t$};
    \node[reservoir] (Muo) at (0,-0.5) {$M^{\mathrm{UO}}_t$};
    \node[reservoir] (Mlo) at (0,-2.0) {$M^{\mathrm{LO}}_t$};
    \node[temp] (F) at (3.4,1) {$F_t$};
    \node[temp] (Tat) at (6.4,1) {$T^{\mathrm{AT}}_t$};
    \node[temp] (Toc) at (6.4,-0.5) {$T^{\mathrm{OC}}_t$};
    \draw[->, thick] (E) -- (Mat);
    \draw[<->, thick] (Mat) -- node[edgelab] {$b_{12}$} (Muo);
    \draw[<->, thick] (Muo) -- node[edgelab] {$b_{23}$} (Mlo);
    \draw[->, thick] (Mat) -- node[edgelab] {$F_{2\times}\log_2(M^{\mathrm{AT}}/M^{\mathrm{AT}}_{\mathrm{PI}})$} (F);
    \draw[->, thick] (F) -- node[edgelab] {$c_1$} (Tat);
    \draw[<->, thick] (Tat) -- node[edgelab] {$c_3,c_4$} (Toc);
    \draw[->, thick, dashed] (Tat) to[bend left=35] node[edgelab, above] {damage $\Omega(T^{\mathrm{AT}})$} (-3.5, 2.0);
\end{tikzpicture}
\caption{Topology of the CDICE climate side.  Total emissions $E_t$ enter the atmospheric carbon box $M^{\mathrm{AT}}_t$, leak into the upper- and lower-ocean carbon boxes at exchange rates $b_{12}$ and $b_{23}$, and drive radiative forcing $F_t$ through the logarithmic CO$_2$ relation.  The two-layer energy balance maps $F_t$ into the atmospheric temperature $T^{\mathrm{AT}}_t$ via $c_1$, with $c_3, c_4$ governing the heat exchange between atmosphere and ocean.  The dashed arrow closes the loop through the damage function back into output (developed in \S\ref{sec:dice_damages}).  Five climate states $(M^{\mathrm{AT}}, M^{\mathrm{UO}}, M^{\mathrm{LO}}, T^{\mathrm{AT}}, T^{\mathrm{OC}})$ form the climate-side block of the DEQN state vector \eqref{eq:iam_state}.}
\label{fig:cdice_climate_topology}
\end{figure}

A two-layer energy balance model links carbon concentrations to temperature:
\begin{align}
T^{\mathrm{AT}}_{t+1} &= T^{\mathrm{AT}}_t + c_1 \bigl(F_t - \lambda\, T^{\mathrm{AT}}_t - c_3(T^{\mathrm{AT}}_t - T^{\mathrm{OC}}_t)\bigr), \label{eq:temp_at} \\
T^{\mathrm{OC}}_{t+1} &= T^{\mathrm{OC}}_t + c_4 \bigl(T^{\mathrm{AT}}_t - T^{\mathrm{OC}}_t\bigr), \label{eq:temp_oc}
\end{align}
where radiative forcing is
\begin{equation}
F_t = F_{\mathrm{2\times CO_2}} \frac{\log(M^{\mathrm{AT}}_t / M^{\mathrm{AT}}_{\mathrm{PI}})}{\log 2} + F^{\mathrm{EX}}_t.
\label{eq:forcing}
\end{equation}
Figure~\ref{fig:cdice_climate_topology} summarizes the full topology of the climate side: industrial emissions enter the atmospheric carbon stock, leak into the upper and lower ocean reservoirs at calibrated rates, raise radiative forcing through the logarithmic CO$_2$ term, and warm the atmospheric and ocean temperature layers through the two-layer energy balance.

The parameter $\lambda = F_{\mathrm{2\times CO_2}} / \Delta T_{\mathrm{AT},\times 2}$ is determined by the \emph{equilibrium climate sensitivity} (ECS), defined as the long-run atmospheric warming from a doubling of CO$_2$ concentration.  We treat $\lambda$ as a deterministic constant in the baseline model; \S\ref{sec:bayesian_learning} promotes it to a learnable Gaussian parameter, with the additive feedback term $\varphi_{1C}\tilde f_{t+1} T^{\mathrm{AT}}_t$ entering the right-hand side of~\eqref{eq:temp_at} and the coefficient $\varphi_{1C}$ defined in that subsection.  ECS is one of the most consequential and uncertain parameters in climate science \citep{roe2007climate, knutti2017beyond}.  Observational and model-based estimates place ECS in a \emph{likely} (66\,\%) range of 2.5\textdegree C--4\textdegree C and a \emph{very likely} (90\,\%) range of 2\textdegree C--5\textdegree C, with a best estimate of approximately 3\textdegree C \citep[IPCC AR6 Synthesis Report;][]{calvinIPCC2023Climate2023a}; ECS uncertainty is one of the largest single sources of variance in the SCC.

\subsection{Damage function: closing the climate--economy loop}
\label{sec:dice_damages}

The damage function is what turns a temperature anomaly back into an output loss, and so it is what closes the economy--climate--damages feedback loop drawn schematically in Figure~\ref{fig:iam_feedback_loop}.  Following the convention in \citet{Folini_2021}, Online Appendix~D, we treat $\Omega(T_{\mathrm{AT}})$ as the \emph{damage fraction} of gross output (the fraction lost to climate damages, increasing in $T_{\mathrm{AT}}$), and the abatement-cost fraction $\Theta(\mu)$ from \eqref{eq:abat_cost} as a separate output drain.  The two enter additively in net output~\eqref{eq:net_output}; an alternative multiplicative form $(1-\Omega^{\mathrm{ret}})(1-\Theta)$ with retained-output factor $\Omega^{\mathrm{ret}}$ is used by \citet{nordhaus2008question}.

The workhorse specification is \citet{nordhaus2008question}'s quadratic,
\begin{equation}
\Omega^N(T_{\mathrm{AT}}) \;=\; \pi_1\, T_{\mathrm{AT}} + \pi_2\, T_{\mathrm{AT}}^2,
\label{eq:damage_nordhaus}
\end{equation}
which is relatively benign for moderate warming and is what we use in the deterministic CDICE solve below.  Calibrated values $(\pi_1, \pi_2)$ are listed in Table~\ref{tab:dice_calibration}.  The damage function~\eqref{eq:damage_nordhaus} is the most contested object in the IAM literature: at $T_{\mathrm{AT}}=3\,^\circ\mathrm{C}$ above pre-industrial, Nordhaus--quadratic damages amount to roughly $2\%$ of gross output, which several recent empirical literatures argue is far below realistic central estimates.  We therefore treat the damage curvature $\pi_2$ as one of the two key uncertain parameters in the deep-UQ analysis of \S\ref{sec:deep_uq} (the other being the equilibrium climate sensitivity).

For the tipping-point branch of the literature, \citet{weitzman2012ghg} argued that catastrophic thresholds require a steeper damage function,
\begin{equation}
\Omega^W(T_{\mathrm{AT}}) \;=\; 1 \;-\; \frac{1}{1 + \bigl(\tfrac{1}{\psi_1} T_{\mathrm{AT}}\bigr)^2 + \bigl(\tfrac{1}{2\, TP} T_{\mathrm{AT}}\bigr)^{6.754}},
\label{eq:damage_weitzman}
\end{equation}
where $TP$ is a stochastic tipping-point threshold.  We do not solve a Weitzman damage variant in the baseline CDICE-DEQN, but the OLG-IAM of \S\ref{sec:pareto_carbon_tax} introduces a stylized tipping risk in the same spirit; the degree of convexity of the damage function is one of the most important determinants of the optimal carbon tax.

\subsection{CDICE: recalibration of the climate module}
\label{sec:cdice_recalibration}

\begin{figure}[ht]
\centering
\includegraphics[width=0.85\linewidth]{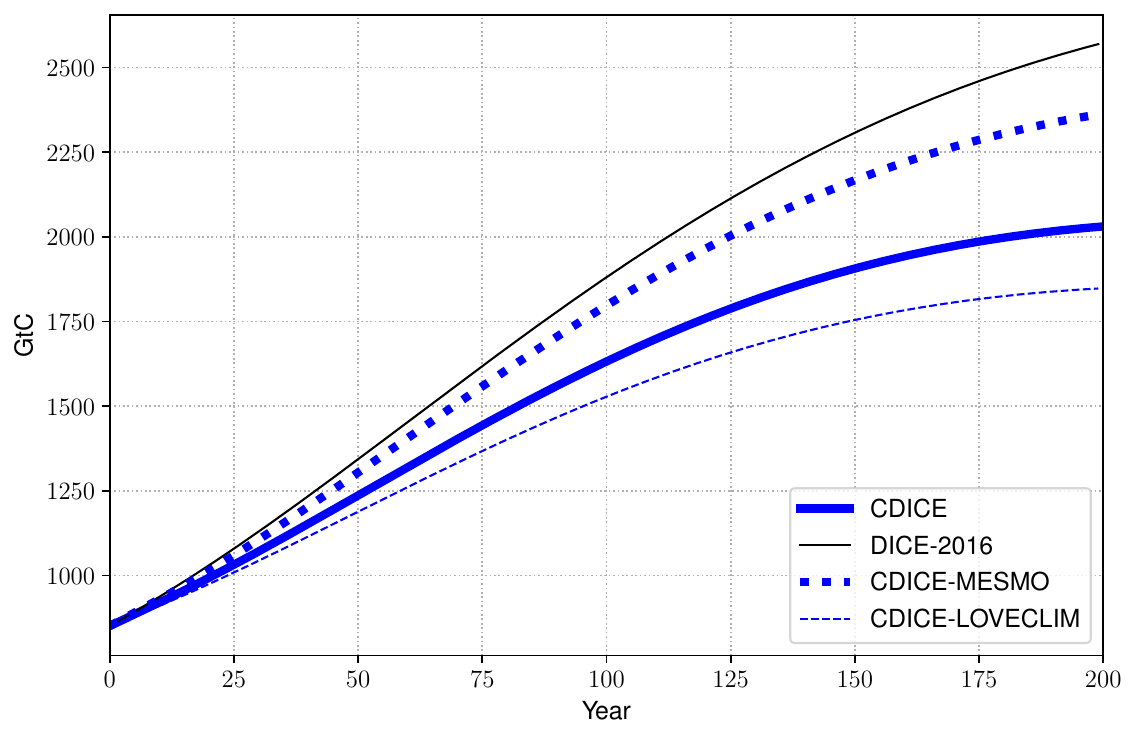}
\caption{Atmospheric carbon $M^{\mathrm{AT}}_t$ along the BAU path (in GtC, over 200 years from 2015) under the three CDICE carbon-cycle calibrations (CDICE = MMM, CDICE-MESMO, CDICE-LOVECLIM) and the legacy DICE-2016 carbon cycle.  Only the carbon-cycle block is varied here; the temperature block is held at the CDICE MMM calibration, since the BAU carbon-stock path does not depend on the temperature calibration to first order.  The DICE-2016 path lies systematically above the CMIP-disciplined paths, reflecting that the original DICE carbon cycle overstates atmospheric retention; CDICE-MESMO and CDICE-LOVECLIM bracket the CDICE baseline on the slow-removal and fast-removal sides, respectively.  Reproduced from \citet{Folini_2021}, Figure~15(a).}
\label{fig:restud_bau_mat}
\end{figure}

A key contribution of \citet{Folini_2021} is a systematic recalibration of the DICE climate module against benchmarks from climate science model archives (CMIP).  Their CDICE framework retains the same functional forms as DICE but fits parameters to the four-test protocol summarized in Table~\ref{tab:cdice_tests}.
\begin{table}[ht]
\centering
\small
\caption{CDICE climate-module calibration protocol.  The first two tests discipline the carbon-cycle and temperature-response blocks directly; the last two check whether the calibrated reduced-form module remains accurate on out-of-sample and historically realistic forcing paths.}
\label{tab:cdice_tests}
\begin{tabular}{lll}
\toprule
\textbf{Test} & \textbf{Target} & \textbf{Use} \\
\midrule
1.\ Carbon pulse (100\,GtC) & Atmospheric retention path & Calibrate carbon cycle \\
2.\ $4\times$CO$_2$ step & Temperature impulse response & Calibrate temperature block \\
3.\ 1\%\,CO$_2$/year & Transient climate response & Out-of-sample validation \\
4.\ Historical + RCP & Realistic forcing paths & End-to-end validation \\
\bottomrule
\end{tabular}
\end{table}
This calibration ensures that the reduced-form climate module is consistent with state-of-the-art earth system models.  CDICE also introduces a transparent time-step formulation, $X_{t+\Delta t} = X_t + \Delta t \cdot f(X_t, u_t; \theta)$, that allows coherent implementation at annual, 5-year, or 10-year resolution within a single generic framework.  Figure~\ref{fig:restud_bau_mat} illustrates how much the climate-cycle calibration matters even before the planner makes any decision: under business-as-usual, DICE-2016 and CDICE produce visibly different atmospheric carbon trajectories, and the gap propagates into temperature, damages, and ultimately the SCC.

\subsection{Calibration and initial conditions, in one place}
\label{sec:dice_calibration}

The block-by-block model description above introduces a fairly large set of parameters.  Table~\ref{tab:dice_calibration} consolidates the calibration we use throughout the rest of the chapter, lifted from the Online Appendix of \citet{Folini_2021}.  Two CMIP5 alternatives (HadGEM2-ES and GISS-E2-R) are shown alongside the multi-model mean (MMM) so that the deep-UQ analysis of \S\ref{sec:deep_uq} has a concrete distribution to draw from.  We follow the CDICE convention of expressing all carbon quantities in $10^3$\,GtC working units: equilibrium and initial carbon stocks $M_{\mathrm{eq}}$ and $M_0$, the initial carbon intensity $\sigma_0$, and the initial land-use emissions $E_{\mathrm{Land},0}$ are all on the same scale, which keeps the numerical conditioning of the carbon-cycle and emissions states under control.  The factor $10^3$ appears explicitly in the abatement-cost calibration~\eqref{eq:theta1_calibration} to convert $\sigma_t$ back to GtC when it is multiplied by the backstop price; a reader comparing values against raw DICE-2016 numbers (e.g.\ $\sim 2.6$\,GtC/yr land-use emissions, $\sim 851$\,GtC atmospheric carbon in 2015) should multiply the table entries by $10^3$ first.

\begin{table}[ht]
\centering
\small
\caption{CDICE baseline calibration used in the deterministic CDICE-DEQN solve.  Parameter values follow the Online Appendix of \citet{Folini_2021} and are stated on an annual time step ($\Delta_t = 1$\,yr).  All carbon quantities ($M_{\mathrm{eq}}, M_0, \sigma_0, E_{\mathrm{Land},0}$) are in CDICE's $10^3$\,GtC working units; multiply by $10^3$ to recover GtC.  Two alternative climate calibrations (CDICE-HadGEM2-ES, CDICE-GISS-E2-R) are listed in the temperature block, with their full free-parameter sets $\{c_1, c_3, c_4, F_{\mathrm{2\times CO_2}}, \lambda\}$ and corresponding ECS, since simply varying ECS while holding the rest of the temperature block fixed is \emph{not} equivalent to using the full CMIP5 calibration \citep[\S 4.6 of][]{Folini_2021}.  Initial state is for year 2015.}
\label{tab:dice_calibration}
\begin{tabular}{@{}p{2.0cm} l p{5.2cm} p{4.4cm}@{}}
\toprule
\textbf{Block} & \textbf{Parameter} & \textbf{Value} & \textbf{Meaning} \\
\midrule
Economy
 & $\alpha$            & $0.30$       & Capital share in Cobb--Douglas output \\
 & $\delta$            & $0.10$/yr    & Capital depreciation rate \\
 & $\rho$              & $0.015$/yr   & Pure rate of time preference \\
 & $\psi$              & $0.69$       & Intertemporal elasticity of substitution \\
\midrule
Emissions \&
 & $\sigma_0$          & $9.556\!\times\!10^{-5}$ ($10^3$\,GtC)/USD & Initial carbon intensity \\
abatement
 & $g^{\sigma}_0$      & $-0.0152$/yr & Initial decay rate of $\sigma_t$ \\
 & $\delta^{\sigma}$   & $0.001$/yr   & Curvature of $\sigma_t$ decay \\
 & $p^{\mathrm{back}}_0$ & $0.55$ thUSD/tCO$_2$ & Initial backstop price \\
 & $g^{\mathrm{back}}$ & $0.005$/yr   & Decay rate of backstop price \\
 & $\theta_2$          & $2.6$        & Curvature of $\Theta(\mu)$ \\
 & $\mathrm{c2co2}$    & $3.666$      & Carbon-to-CO$_2$ mass conversion \\
\midrule
Land use
 & $E_{\mathrm{Land},0}$ & $7.09\!\times\!10^{-4}$ ($10^3$\,GtC)/yr & Initial land-use emissions \\
 & $\delta^{\mathrm{Land}}$ & $0.023$/yr & Decay rate of $E_{\mathrm{Land},t}$ \\
\midrule
Carbon cycle
 & $b_{12}$            & $0.054$/yr   & Atm.--upper-ocean transfer rate \\
 & $b_{23}$            & $0.0082$/yr  & Upper-ocean--lower-ocean transfer rate \\
 & $M_{\mathrm{eq}}$   & $(0.607, 0.489, 1.281)$ ($10^3$\,GtC) & Pre-industrial equilibrium masses \\
\midrule
Temperature
 & $c_1$ (MMM)         & $0.137$/yr   & Atmospheric heat-capacity inverse \\
 & $c_3$ (MMM)         & $0.73$/yr    & Atm.--ocean coupling \\
 & $c_4$ (MMM)         & $0.00689$/yr & Ocean heat-capacity inverse \\
 & $F_{\mathrm{2\times CO_2}}$ (MMM) & $3.45$ W/m$^2$ & Forcing from CO$_2$ doubling \\
 & $\lambda$ (MMM)     & $1.06$ W/m$^2$/K & Climate feedback parameter \\
 & ECS (MMM)           & $\approx 3.25\,^\circ$C & Equilibrium climate sensitivity \\
 & HadGEM2-ES          & $(c_1,c_3,c_4)=(0.154,0.55,0.00671)$/yr & High-end CMIP5 calibration \\
 &                     & $F_{\mathrm{2\times CO_2}}=2.95$, $\lambda=0.65$, ECS$\approx 4.55$\,$^\circ$C & \\
 & GISS-E2-R           & $(c_1,c_3,c_4)=(0.213,1.16,0.00921)$/yr & Low-end CMIP5 calibration \\
 &                     & $F_{\mathrm{2\times CO_2}}=3.65$, $\lambda=1.70$, ECS$\approx 2.15$\,$^\circ$C & \\
\midrule
Damages
 & $\pi_1$             & $0.0$        & Linear damage coefficient \\
 & $\pi_2$             & $0.00236$    & Quadratic damage coefficient \\
\midrule
Initial state
 & $K_0$               & $223$ T USD  & Capital, year 2015 \\
 & $M_0$               & $(0.851, 0.628, 1.323)$ ($10^3$\,GtC) & Atm./upper/lower carbon, 2015 \\
 & $T_0$               & $(1.10, 0.27)\,^\circ$C & Atm./ocean temp.\ above pre-industrial, 2015 \\
\bottomrule
\end{tabular}
\end{table}

\subsection{The full IAM, summarized}
\label{sec:dice_summary}

Pulling the previous subsections together, CDICE is a deterministic dynamical system on a finite-dimensional state vector that the planner steers with two controls.  The endogenous state at date $t$ is the sextuple
\begin{equation}
\bm{X}^{\mathrm{end}}_t \;=\; \bigl(K_t,\; M^{\mathrm{AT}}_t,\; M^{\mathrm{UO}}_t,\; M^{\mathrm{LO}}_t,\; T^{\mathrm{AT}}_t,\; T^{\mathrm{OC}}_t\bigr),
\label{eq:dice_endo_state}
\end{equation}
the exogenous-trend vector is
\begin{equation}
\bm{X}^{\mathrm{exo}}_t \;=\; \bigl(A_t,\; L_t,\; \sigma_t,\; E_{\mathrm{Land},t},\; F^{\mathrm{EX}}_t\bigr),
\end{equation}
and the planner's controls are $(C_t,\, \mu_t)$ (equivalently $(K_{t+1},\, \mu_t)$, since investment is determined by the resource constraint $C_t + I_t = Y^{\mathrm{net}}_t$ together with~\eqref{eq:capital_accumulation}).  The transitions are: capital from~\eqref{eq:capital_accumulation} with $I_t = Y^{\mathrm{net}}_t - C_t$; total emissions from~\eqref{eq:total_emissions}, fed into the carbon cycle~\eqref{eq:carbon_cycle}; temperature from~\eqref{eq:temp_at}--\eqref{eq:temp_oc} with forcing~\eqref{eq:forcing}; and net output, hence the resource constraint, from~\eqref{eq:net_output}.  The objective is the discounted CRRA-IES felicity sum~\eqref{eq:planner_obj_crra} subject to $\mu_t \in [0,1]$.

That is the entire deterministic IAM.  Every primitive named above has a closed-form expression and a calibrated parameter (Table~\ref{tab:dice_calibration}); the only thing left is to find the optimal policy $(C_t, \mu_t)_{t\ge 0}$.  The model is intrinsically non-stationary.  \S\ref{sec:iam_nonstationarity} makes that observation precise; the stationary DEQN of Chapter~\ref{ch:deqn} needs to be amended before we can solve this system.


\section{Why DICE Breaks the Stationary DEQN}
\label{sec:iam_nonstationarity}

This is the technical pivot of the chapter.  The stationary DEQN of Chapter~\ref{ch:deqn} was designed for models whose policy function is a fixed point of a Bellman operator on an ergodic state space.  IAMs satisfy neither premise.  Three structural features break the stationarity assumption simultaneously, and each must be addressed before the DEQN can be trained at all.

\paragraph{Time-varying state distributions with no ergodic limit.}
The endogenous state of a stationary DSGE is the projection of a recurrent Markov chain onto a finite-dimensional vector; the policy function lives on its stationary distribution.  In an IAM the analogue object does not exist within the planning horizon.  Atmospheric carbon $M^{\mathrm{AT}}_t$ rises from a pre-industrial baseline of $\sim 600$\,GtC to a peak of $\sim 1500$\,GtC over a century, then decays over millennia; atmospheric temperature $T^{\mathrm{AT}}_t$ follows with a multi-decade lag and a multi-century relaxation.  Within the 300 years the planner cares about, neither variable ever returns to a state it has been in before.  The state visited at $t = 100$ is therefore not exchangeable with the state visited at $t = 200$, and a time-invariant policy function $\bm p(\bm X_t)$ that depends only on the endogenous state misses the whole point of the exercise: the optimal mitigation effort at a given $(M^{\mathrm{AT}}, T^{\mathrm{AT}})$ depends on whether that state was reached on the way up or on the way down.  Cf.\ the curse-of-dimensionality discussion in \S\ref{sec:curse_of_dim}: it is not the size of the state space that breaks the DEQN here, it is the lack of recurrence.

\paragraph{Deterministic drift through exogenous trends.}
Even setting the carbon and temperature stocks aside, the IAM is drifting deterministically.  Total factor productivity $A_t$ trends up at a calibrated, time-varying rate; population $L_t$ follows the demographic projection of \citet{nordhausRevisitingSocialCost2017}; carbon intensity $\sigma_t$ falls along the closed-form decay~\eqref{eq:sigma_decay}; land-use emissions $E_{\mathrm{Land},t}$ decay smoothly~\eqref{eq:land_emissions}; the backstop price $p^{\mathrm{back}}_t$ falls~\eqref{eq:backstop_price}; the exogenous non-CO$_2$ forcing $F^{\mathrm{EX}}_t$ follows a fitted RCP trajectory; and the abatement-cost level $\theta_{1,t}$ inherits the time dependence of $\sigma_t$ and $p^{\mathrm{back}}_t$ through~\eqref{eq:theta1_calibration}.  Seven exogenous trends drive the model even before a shock is introduced.  A time-invariant policy can never see them, and replacing them with their long-run averages is exactly the certainty-equivalence move that defeats the purpose of solving the model globally.

\paragraph{Finite calendar-time horizon.}
A stationary DEQN trains under a transversality condition: as $t \to \infty$, the discounted shadow price of capital goes to zero, and the iterative-projection loss inherits that fixed-point structure for free.  An IAM is not solved on $[0, \infty)$.  The planning horizon is a finite calendar date (the notebooks of \S\ref{sec:iam_dequ_loss} run roughly three centuries from a 2015 start), so transversality is not available and the policy is solved over a finite forward sweep instead.

\paragraph{Putting it together.}
These features compound and explain why a time-invariant DEQN of Chapter~\ref{ch:deqn} cannot be used here without modification.  The next two sections operationalize the response: \S\ref{sec:nsdeqn_setup} reorganizes the network inputs to include calendar time, and \S\ref{sec:nsdeqn_algo} states the resulting training algorithm as a labeled diff against the stationary DEQN box of \S\ref{sec:deqn_algo}.

\section{What Changes in the DEQN Setup}
\label{sec:nsdeqn_setup}

We now translate this into one concrete design choice for the network inputs.  The autodiff machinery, the squared-residual structure, and the rest of the training loop of Chapter~\ref{ch:deqn} carry over unchanged; this is a refactor of what the network sees, not a new solver.

\subsection{Time and trends as states}
\label{sec:nsdeqn_setup_time}

Calendar time itself enters as a state.  Because neural networks prefer bounded inputs, we use the monotone time rescaling $\tau_t = 1 - \exp(-\vartheta\, t) \in [0, 1)$ of Eq.~\eqref{eq:time_transform}.  Every exogenous trend ($A_t, L_t, \sigma_t, E_{\mathrm{Land},t}, F^{\mathrm{EX}}_t, p^{\mathrm{back}}_t, \theta_{1,t}$) is then a deterministic function of $\tau_t$, so passing $\tau_t$ to the network is informationally equivalent to passing the entire trend bundle.  Training trajectories begin from the calibrated 2015 state and run forward over the planner's horizon.

\section{The Non-Stationary DEQN Algorithm}
\label{sec:nsdeqn_algo}

The design choice of \S\ref{sec:nsdeqn_setup} translates into a single training algorithm.  The body below is a literal diff against the stationary DEQN of \S\ref{sec:deqn_algo}: unchanged lines are grayed, new or modified lines are bolded.

\begin{definitionbox}[Algorithm: Non-Stationary DEQN Training]
\label{alg:nsdeqn}
\begin{algorithmic}
\small
\STATE \textbf{Input:} \textcolor{uzhgreydark!70}{Network $\mathcal{N}_\rho$, learning rate $\eta$, episodes $E$, training steps $T_{\mathrm{train}}$;} \\
\hspace{1.05em}\textbf{[NEW]} calibrated initial state $\bm x_0$ (e.g., the 2015 state) and a planning horizon $T_{\max}$
\FOR{episode $e = 1, \ldots, E$}
    \STATE \textbf{[CHANGED] Simulate $K$ forward trajectories from $\bm x_0$ over $[0, T_{\max}]$ under the current policy, and collect the time-stamped states $(\tau_t, \bm x_t)$ into $\mathcal D$}
    \FOR{gradient step $t = 1, \ldots, T_{\mathrm{train}}$}
        \STATE \textcolor{uzhgreydark!70}{Draw mini-batch $\mathcal B \subset \mathcal D$}
        \STATE \textcolor{uzhgreydark!70}{Compute loss:~$\ell_\rho = \frac{1}{|\mathcal B|}\sum_{\bm x_i \in \mathcal B}\|G(\bm x_i, \mathcal N_\rho(\bm x_i))\|^2$}
        \STATE \textcolor{uzhgreydark!70}{Update:~$\rho \leftarrow \rho - \eta \cdot \nabla_\rho \ell_\rho$}
    \ENDFOR
\ENDFOR
\STATE \textcolor{uzhgreydark!70}{\textbf{Output:} Trained network $\mathcal{N}_{\rho^\star}$ approximating the policy function}
\end{algorithmic}
\end{definitionbox}

One delta against the stationary DEQN box.  The simulation step starts from a calibrated initial state $\bm x_0$ and integrates $K$ trajectories forward through calendar time, so the pool $\mathcal D$ contains time-stamped states $(\tau_t, \bm x_t)$ along finite-horizon trajectories rather than draws from an ergodic distribution.  With $\tau_t$ in the input the network learns a time-dependent policy; every other line of the box is the stationary DEQN of \S\ref{sec:deqn_algo} unchanged.

\paragraph{What replaces transversality.}
Because the pool $\mathcal D$ is built from $K$ forward simulations of length $T_{\max}$ that all start at the same $\bm x_0$, every trajectory visits the full calendar window $[0, T_{\max}]$ and a uniform mini-batch draw from $\mathcal D$ is therefore stratified across calendar time by construction.  The missing transversality condition of \S\ref{sec:iam_nonstationarity} is absorbed numerically by choosing the horizon long enough that the discounted contribution of the terminal state falls below the training-noise floor: at the CDICE calibration $\rho = 0.015$/yr and the notebooks' default $T_{\max} = 300$ years, $\hat\beta_t^{\,T_{\max}} \approx \exp(-\rho\,T_{\max}) \approx 0.011$, which is one to two orders of magnitude below the achievable residual root-mean-square at convergence.  When the horizon must be short (e.g., the 1D toy of Exercise~\ref{ex:ch11:10}), one instead adds an explicit terminal residual $\lambda_T\,\|\bm x_{T_{\max}} - \bm x^{\mathrm{ref}}_{T_{\max}}\|^2$ to the loss; both options are standard in the finite-horizon DEQN literature.

\begin{keyinsightbox}[The non-stationary DEQN in one line]
One modification, one solver: time enters as a state.  The autodiff backbone, the residual structure, the sampling rule, and the update rule carry over from the stationary DEQN of Chapter~\ref{ch:deqn}.
\end{keyinsightbox}


\section{The Planner's Lagrangian and FOCs}
\label{sec:dice_deqn}
\label{sec:dice_lagrangian}

Movement~2 puts the non-stationary DEQN of \S\ref{sec:nsdeqn_algo} to work on the deterministic CDICE economy of \S\ref{sec:dice}.  Solving this system with the algorithm of \S\ref{sec:nsdeqn_algo} amounts to writing the planner's Lagrangian, deriving the first-order and envelope conditions, normalizing them, treating each FOC as a residual, and minimizing the sum of squared residuals on the time-stamped state pool generated by the forward simulation of \S\ref{sec:nsdeqn_algo}.  This section follows \citet{friedlDeep2023} and Online Appendix~D of \citet{Folini_2021}.

\paragraph{Detrending and state vector, in compact form.}
The model-rendering choices already named in \S\ref{sec:nsdeqn_setup_time} carry over verbatim.  Variables that grow with the productivity--population product $A_t L_t$ are rescaled to per-effective-capita units:
\begin{equation}
c_t \;:=\; \frac{C_t}{A_t\,L_t}, \qquad
k_t \;:=\; \frac{K_t}{A_t\,L_t}.
\label{eq:iam_detrend}
\end{equation}
Calendar time enters through the bounded rescaling (compatible with the dynamic-programming convention of \citet{traeger4StatedDICEQuantitatively2014}),
\begin{equation}
\tau \;=\; 1 - \exp(-\vartheta\, t) \;\in\; [0,1),
\qquad\text{with inverse}\quad
t \;=\; -\frac{\ln(1-\tau)}{\vartheta},
\label{eq:time_transform}
\end{equation}
with compression parameter $\vartheta > 0$.  The full DEQN state vector then collects the detrended endogenous CDICE states, the bounded time index, the Bayesian-belief states $(\mu_{f,t}, S_{f,t})$ used in \S\ref{sec:bayesian_learning}, and a slot for pseudo-state parameters $\theta$ used in the UQ analysis of \S\ref{sec:deep_uq}:
\begin{equation}
\bm{X}_t \;=\; \bigl[\underbrace{k_t,\, M^{\mathrm{AT}}_t,\, M^{\mathrm{UO}}_t,\, M^{\mathrm{LO}}_t,\, T^{\mathrm{AT}}_t,\, T^{\mathrm{OC}}_t,\, \mu_{f,t},\, S_{f,t},\, \tau_t}_{9\text{ endogenous, exogenous, and time states}};\; \underbrace{\theta}_{N\text{ pseudo-state parameters}}\bigr].
\label{eq:iam_state}
\end{equation}
In the deterministic core developed in this and the next section, only the six endogenous-state entries plus $\tau_t$ are active, i.e.\ a seven-dimensional input vector; $(\mu_{f,t}, S_{f,t})$ and $\theta$ are appended only in the extensions of Movement~3.

\paragraph{The Lagrangian.}
We now derive the equilibrium conditions that the DEQN will be trained against.  The derivation follows the standard Lagrangian approach in CRRA-IES form, working directly with the deterministic CDICE primitives of \S\ref{sec:dice} (the recursive Epstein--Zin refinement is layered on in \S\ref{sec:ez_layer}).  Write the Lagrangian with multiplier $\lambda_t$ for the budget constraint $C_t + I_t = Y^{\mathrm{net}}_t$, multipliers $\nu^{\mathrm{AT}}_t,\nu^{\mathrm{UO}}_t,\nu^{\mathrm{LO}}_t$ for the three carbon-reservoir transitions~\eqref{eq:carbon_cycle}, multipliers $\eta^{\mathrm{AT}}_t,\eta^{\mathrm{OC}}_t$ for the temperature dynamics~\eqref{eq:temp_at}--\eqref{eq:temp_oc}, and KKT multiplier $\lambda^\mu_t \ge 0$ for the abatement bound $\mu_t \le 1$.  The derivation produces ten equilibrium conditions: a consumption FOC, an abatement FOC, the capital Euler equation, the budget/resource constraint, three carbon-stock envelope conditions, two temperature-envelope conditions, and the abatement upper-bound complementarity.  Two of these are enforced algebraically (the static consumption and abatement FOCs); the remaining eight become the DEQN residuals of \S\ref{sec:iam_dequ_loss}.

\paragraph{Qualitative overview.}
Taking derivatives of the Lagrangian with respect to the controls yields:
\begin{itemize}[itemsep=2pt]
\item \textbf{w.r.t.\ $C_t$:} the marginal utility of consumption equals the shadow price of the budget, $\partial V^{1-1/\psi}/\partial C_t = \lambda_t$.
\item \textbf{w.r.t.\ $K_{t+1}$:} the shadow value of capital today equals the discounted expected marginal value tomorrow, $\xi_t = e^{-\rho}\,\partial \mathbb{E}_t[V_{t+1}^{1-\gamma}]^{(1-1/\psi)/(1-\gamma)} / \partial K_{t+1}$.
\item \textbf{w.r.t.\ $\mu_t$:} the marginal abatement cost equals the shadow value of reduced emissions (plus the complementarity term if $\mu_t = 1$).
\end{itemize}

\paragraph{Envelope theorem.}
Since the FOC for $K_{t+1}$ involves $\partial V/\partial K_{t+1}$, which cannot be computed analytically, we apply the envelope theorem.  It provides derivatives of the value function with respect to \emph{current} states, in particular $\partial V/\partial k_t$, $\partial V/\partial M_{\mathrm{AT},t}$, $\partial V/\partial T^{\mathrm{AT}}_t$, which are then shifted forward one period and substituted back into the FOCs.

\paragraph{Capital Euler equation.}
Combining the FOCs and envelope conditions yields:
\begin{equation}
1 = e^{-\rho}\,\mathbb{E}_t\!\left[\left(\frac{V_{t+1}}{\bigl(\mathbb{E}_t[V_{t+1}^{1-\gamma}]\bigr)^{1/(1-\gamma)}}\right)^{1/\psi - \gamma} \cdot \frac{(C_{t+1}/L_{t+1})^{-1/\psi}}{(C_t/L_t)^{-1/\psi}} \cdot R^K_{t+1}\right],
\label{eq:iam_euler}
\end{equation}
where $R^K_{t+1}$ is the return on capital inclusive of climate damages.  The SCC also appears through the shadow price of atmospheric carbon:
\begin{equation}
\mathrm{SCC}^{M}_t = -\frac{\partial V_t / \partial M_{\mathrm{AT},t}}{\partial V_t / \partial C_t}.
\end{equation}
This is a shadow value per unit of atmospheric carbon stock.  The emissions-based SCC in~\eqref{eq:scc} additionally includes the marginal loading of a unit of emissions into $M_{\mathrm{AT},t}$ and the carbon-to-CO$_2$ unit conversion.  At the optimum, the marginal abatement cost equals the carbon tax equals the emissions SCC after these conversions.

\paragraph{Normalization of multipliers.}
Over a 300-year horizon, $A_t$ and $L_t$ can move the natural scale of marginal utilities and multipliers substantially, with the direction and magnitude depending on the IES through $A_t^{1-1/\psi}L_t$.  Such scale drift makes network outputs and gradients harder to optimize stably.  Following the detrending logic of~\eqref{eq:iam_detrend}, all multipliers, the budget multiplier, the abatement-bound multiplier, and the five climate envelope multipliers alike, are divided by $A_t^{1-1/\psi}\,L_t$.  The argument for the climate multipliers tracks the budget-multiplier case via the envelope conditions of \S\ref{sec:dice_lagrangian} and is spelled out in Online Appendix~D of \citet{Folini_2021}; we adopt the result here:
\begin{equation}
\hat{\lambda}_t := \frac{\lambda_t}{A_t^{1-1/\psi}\,L_t},\quad
\hat{\lambda}^\mu_t := \frac{\lambda^\mu_t}{A_t^{1-1/\psi}\,L_t},\quad
\hat{\nu}^{\mathrm{AT}}_t := \frac{\nu^{\mathrm{AT}}_t}{A_t^{1-1/\psi}\,L_t},\quad
\hat{\nu}^{\mathrm{UO}}_t := \frac{\nu^{\mathrm{UO}}_t}{A_t^{1-1/\psi}\,L_t},\quad \ldots
\label{eq:iam_norm_mult}
\end{equation}
and analogously for the remaining multipliers $\hat{\nu}^{\mathrm{LO}}_t$, $\hat{\eta}^{\mathrm{AT}}_t$, and $\hat{\eta}^{\mathrm{OC}}_t$.  The normalization induces an \emph{effective discount factor} that absorbs the trend growth in the per-effective-capita Euler equation,
\begin{equation}
\hat{\beta}_t \;:=\; \exp\!\left(-\rho + \left(1-\frac{1}{\psi}\right) g^A_t + g^L_t\right),
\label{eq:iam_eff_beta}
\end{equation}
where $g^A_t := \ln(A_{t+1}/A_t)$ and $g^L_t := \ln(L_{t+1}/L_t)$ are annual log changes.  Equation~\eqref{eq:iam_eff_beta} mirrors Equation~(38) of Online Appendix~D of \citet{Folini_2021}: the population term enters linearly because $L_t$ enters the felicity weight $L_t (C_t/L_t)^{1-1/\psi}$ linearly, while the productivity term inherits the $1-1/\psi$ exponent from the per-effective-capita rescaling of consumption.  All intertemporal equations below use $\hat{\beta}_t$ in place of $e^{-\rho}$.  For a non-annual time step, replace $\rho$ by $\rho\Delta_t$ and $g^A_t, g^L_t$ by their per-period analogues.

\paragraph{Sign convention for the climate multipliers.}
We adopt the value-derivative convention throughout the script: each climate multiplier $\hat{\nu}^{\mathrm{AT}}_t,\, \hat{\nu}^{\mathrm{UO}}_t,\, \hat{\nu}^{\mathrm{LO}}_t,\, \hat{\eta}^{\mathrm{AT}}_t,\, \hat{\eta}^{\mathrm{OC}}_t$ is the (normalized) partial derivative of the value function with respect to the corresponding climate state.  Because extra atmospheric carbon lowers welfare, $\hat{\nu}^{\mathrm{AT}}_t$ is non-positive at the optimum, which is why the stock SCC carries a minus sign, $\mathrm{SCC}^M_t = -\hat{\nu}^{\mathrm{AT}}_t/\hat{\lambda}_t$.  The companion implementation in \texttt{dice\_2p\_surrogate\_lib.py} stores the \emph{positive marginal damage} $-\hat{\nu}^{\mathrm{AT}}_t$ as a network output for numerical conditioning and flips the sign explicitly inside each residual; the algebra below uses the script convention, so the reader who compares the equations to the code will see one extra sign flip per carbon-multiplier term.

\paragraph{Symbol cheat-sheet for the multipliers.}
Before writing the FOCs and the loss, Table~\ref{tab:cdice_multipliers} collects the multipliers that the DEQN learns and their role; subsequent equations use the hat-normalized form throughout.

\begin{table}[H]
\centering
\small
\setlength{\tabcolsep}{4pt}
\begin{tabular}{@{}>{$}l<{$} l c >{\raggedright\arraybackslash}p{4.0cm}@{}}
\toprule
\textbf{Symbol} & \textbf{Multiplier on} & \textbf{Sign at optimum} & \textbf{Network output?} \\
\midrule
\hat\lambda_t              & Budget constraint $C_t + I_t = Y^{\mathrm{net}}_t$               & $> 0$ & yes (softplus) \\
\hat\lambda^\mu_t          & Abatement upper bound $\mu_t \le 1$                              & $\ge 0$ & no (implied, Eq.~\ref{eq:iam_lambdamu_implied}) \\
\hat\nu^{\mathrm{AT}}_t    & Atmospheric carbon transition $M^{\mathrm{AT}}_{t+1}=\ldots$     & $\le 0$ & yes (stored as $-\hat\nu^{\mathrm{AT}}_t > 0$ via softplus) \\
\hat\nu^{\mathrm{UO}}_t    & Upper-ocean carbon transition                                    & $\le 0$ & yes (linear) \\
\hat\nu^{\mathrm{LO}}_t    & Lower-ocean carbon transition                                    & $\le 0$ & yes (linear) \\
\hat\eta^{\mathrm{AT}}_t   & Atmospheric temperature transition                               & $\le 0$ & yes (linear) \\
\hat\eta^{\mathrm{OC}}_t   & Ocean temperature transition                                     & $\le 0$ & yes (linear) \\
\bottomrule
\end{tabular}
\caption{Normalized Lagrange multipliers in the CDICE--DEQN.  All values are divided by $A_t^{1-1/\psi}\,L_t$ relative to the raw multipliers, so the hatted versions inherit the per-effective-capita scale that the network outputs see.  The atmospheric carbon multiplier carries the SCC up to the marginal-utility denominator: $\mathrm{SCC}^M_t = -\hat\nu^{\mathrm{AT}}_t / \hat\lambda_t$.}
\label{tab:cdice_multipliers}
\end{table}

\paragraph{Key FOCs in normalized form.}
After normalization, the most important first-order conditions become (see Online Appendix~D of \citet{Folini_2021} for the complete set of 14 equations):
\begin{align}
\frac{\partial \mathcal{L}}{\partial c_t} = 0 \;&\Leftrightarrow\;
c_t^{-1/\psi}\,A_t^{1-1/\psi}\,L_t - \hat{\lambda}_t = 0,
\label{eq:iam_foc_c}\\[4pt]
\frac{\partial \mathcal{L}}{\partial k_{t+1}} = 0 \;&\Leftrightarrow\;
\exp\!\bigl(g^A_t + g^L_t\bigr)\,\hat{\lambda}_t - \hat{\beta}_t\Bigl\{\hat{\lambda}_{t+1}\bigl[\bigl(1-\Omega(T_{\mathrm{AT},t+1}) - \Theta(\mu_{t+1})\bigr)\alpha k_{t+1}^{\alpha-1} + (1-\delta)\bigr] \nonumber\\
&\quad + \hat{\nu}^{\mathrm{AT}}_{t+1}\,\sigma_{t+1}(1-\mu_{t+1})A_{t+1}L_{t+1}\alpha k_{t+1}^{\alpha-1}\Bigr\} = 0,
\label{eq:iam_foc_k}\\[4pt]
\frac{\partial \mathcal{L}}{\partial \mu_t} = 0 \;&\Leftrightarrow\;
\hat{\lambda}_t\,\Theta'(\mu_t)\,k_t^\alpha + \hat{\lambda}^\mu_t + \hat{\nu}^{\mathrm{AT}}_t\,\sigma_t\,A_t\,L_t\,k_t^\alpha = 0.
\label{eq:iam_foc_mu}
\end{align}
Equation~\eqref{eq:iam_foc_k} is the capital Euler equation: it equates the marginal cost of saving one additional unit today (left) to the discounted marginal benefit tomorrow (right), which now includes a term from the atmospheric carbon envelope ($\hat{\nu}^{\mathrm{AT}}_{t+1}$) because higher capital increases output and hence emissions.

\paragraph{Envelope conditions.}
\textit{Convention reminder.} As stated in \S\ref{sec:iam_landscape}, CDICE is calibrated on an annual time step and the coefficients $b_{12}, b_{23}, c_1, c_3, c_4$ in Table~\ref{tab:dice_calibration} are annual rates; consequently no $\Delta_t$ multipliers appear in either the dynamics~\eqref{eq:carbon_cycle},~\eqref{eq:temp_at}--\eqref{eq:temp_oc} or in the FOC residuals below.

Differentiating the Lagrangian with respect to \emph{state} variables and shifting forward one period yields the shadow prices of the climate stocks.  For example, the atmospheric carbon envelope is:
\begin{equation}
\frac{\partial \mathcal{L}}{\partial M_{\mathrm{AT},t+1}} = 0 \;\Leftrightarrow\;
\hat{\nu}^{\mathrm{AT}}_t - \hat{\beta}_t\!\left[\hat{\nu}^{\mathrm{AT}}_{t+1}(1-b_{12}) + \hat{\nu}^{\mathrm{UO}}_{t+1}\,b_{12} + \hat{\eta}^{\mathrm{AT}}_{t+1}\,c_1\,F_{\mathrm{2\times CO_2}}\,\frac{1}{\ln 2\,M_{\mathrm{AT},t+1}}\right] = 0.
\label{eq:iam_env_mat}
\end{equation}
This equation says that the current shadow price of atmospheric carbon ($\hat{\nu}^{\mathrm{AT}}_t$) must equal the discounted future effects through three channels: persistence in the atmosphere ($b_{12}$ term), diffusion into the upper ocean ($\hat{\nu}^{\mathrm{UO}}_{t+1}$ term), and radiative forcing on temperature ($\hat{\eta}^{\mathrm{AT}}_{t+1}$ term).  It is the existence of these climate multipliers that distinguishes the IAM from the purely economic models of Chapters~\ref{ch:deqn}--\ref{ch:nas}.

\paragraph{Fischer--Burmeister complementarity for abatement.}
The abatement rate is bounded above by~1 (full abatement), giving the KKT condition:
\begin{equation}
1 - \mu_t \;\geq\; 0 \quad\perp\quad \hat{\lambda}^\mu_t \;\geq\; 0,
\label{eq:iam_kkt_mu}
\end{equation}
which is non-smooth at $\mu_t = 1$.  As in the borrowing-constraint treatment of Chapter~\ref{ch:olg} (Section~\ref{sec:olg_fb}), we replace it with the Fischer--Burmeister smooth approximation:
\begin{equation}
\Psi^{\mathrm{FB}}\!\bigl(\hat{\lambda}^\mu_t,\; 1-\mu_t\bigr)
\;=\; \hat{\lambda}^\mu_t + (1-\mu_t) - \sqrt{(\hat{\lambda}^\mu_t)^2 + (1-\mu_t)^2 + \varepsilon_{\mathrm{FB}}} \;=\; 0,
\label{eq:iam_fb}
\end{equation}
with the same regularization parameter $\varepsilon_{\mathrm{FB}} \geq 0$ used in Chapters~\ref{ch:irbc}--\ref{ch:olg}.  In CDICE-DEQN we take $\varepsilon_{\mathrm{FB}} = 10^{-6}$, equivalent to the IRBC chapter's $\varepsilon = 10^{-3}$ under its $\varepsilon^2$ convention; the trained policy is insensitive to the choice in the range $10^{-10}$ to $10^{-4}$.  At $\varepsilon_{\mathrm{FB}} = 0$ the zero set of $\Psi^{\mathrm{FB}}$ coincides with the positive axes in the $(\hat{\lambda}^\mu_t,\, 1-\mu_t)$-plane, enforcing the original complementarity exactly but the function is non-differentiable at the origin; with $\varepsilon_{\mathrm{FB}} > 0$ the function is differentiable everywhere at the cost of a slight relaxation of exact complementarity.

\section{From FOCs to a Single Loss}
\label{sec:iam_dequ_loss}

The Lagrangian of \S\ref{sec:dice_lagrangian} produces ten equilibrium conditions: the consumption FOC~\eqref{eq:iam_foc_c}, the capital Euler~\eqref{eq:iam_foc_k}, the abatement FOC~\eqref{eq:iam_foc_mu}, the budget/resource constraint $C_t + I_t = Y^{\mathrm{net}}_t$, the three carbon-stock envelopes (one of which is the atmospheric-carbon envelope~\eqref{eq:iam_env_mat}), the two temperature-layer envelopes, and the Fischer--Burmeister abatement complementarity~\eqref{eq:iam_fb}.  In the DEQN solver two of these ten are enforced \emph{exactly} by algebraic recovery rather than as squared residuals: the consumption FOC is inverted to yield $c_t$ from $\hat{\lambda}_t$, and the abatement FOC is solved for $\hat{\lambda}^\mu_t$ and the resulting \emph{implied multiplier} is fed straight into the Fischer--Burmeister condition.  What remains is an eight-residual sum-of-squares loss with eight network outputs, structurally identical to the stationary DEQN of Chapters~\ref{ch:deqn}--\ref{ch:irbc}.  The only substantive difference is that the network must learn the shadow prices of all five climate state variables (three carbon stocks and two temperature layers) in addition to the economic choices, so that the planner has a gradient signal for how today's decisions propagate through the carbon cycle and the energy balance into future damages.

\paragraph{Policy network specification.}
The policy function approximated by the neural network outputs an eight-dimensional vector,
\begin{equation}
\mathcal{N}_\rho(\bm{x}_t) \;\in\; \mathbb{R}^{8}
\;:=\; \bigl(k_{t+1},\; \mu_t,\; \hat{\lambda}_t,\; \hat{\nu}^{\mathrm{AT}}_t,\; \hat{\nu}^{\mathrm{UO}}_t,\; \hat{\nu}^{\mathrm{LO}}_t,\; \hat{\eta}^{\mathrm{AT}}_t,\; \hat{\eta}^{\mathrm{OC}}_t\bigr),
\label{eq:iam_nn_output}
\end{equation}
comprising two choice variables ($k_{t+1}$, $\mu_t$), the consumption shadow price $\hat{\lambda}_t$, and the five normalized climate multipliers.  Note that the abatement KKT multiplier $\hat{\lambda}^\mu_t$ is \emph{not} a network output: it is recovered algebraically inside the loss (see below).  A key difference from the stationary DEQN of Chapters~\ref{ch:deqn}--\ref{ch:irbc} is that the network must learn the shadow prices of all climate constraints, not just the economic choices.  Without the climate multipliers, the planner would have no gradient signal about how today's decisions propagate through the carbon cycle and temperature dynamics into future damages.

\paragraph{Bounds and positivity.}
The output activations of $\mathcal{N}_\rho$ are chosen so that the bound and positivity constraints of the model hold for every input, eliminating the need for additional residuals.  The capital level $k_{t+1}$, the consumption shadow $\hat{\lambda}_t$, and the abatement rate $\mu_t$ are each passed through a softplus, which guarantees $k_{t+1} > 0$, $\hat{\lambda}_t > 0$ (so consumption recovered via~\eqref{eq:iam_c_recovery} is positive), and $\mu_t \ge 0$ exactly.  The upper bound $\mu_t \le 1$ is enforced jointly by the Fischer--Burmeister condition $l_8$ at the implied multiplier~\eqref{eq:iam_lambdamu_implied} and by a small quadratic upper-bound penalty $\propto \mathbb{E}[\max(\mu_t - 1, 0)^2]$ added to the training loss.  The atmospheric-carbon shadow $\hat{\nu}^{\mathrm{AT}}_t$ is stored in the implementation as a positive marginal damage (see the sign-convention note in \S\ref{sec:dice_lagrangian}) and is output through a softplus; the remaining climate multipliers $\hat{\nu}^{\mathrm{UO}}_t, \hat{\nu}^{\mathrm{LO}}_t, \hat{\eta}^{\mathrm{AT}}_t, \hat{\eta}^{\mathrm{OC}}_t$ are unconstrained and use linear output activations.

\paragraph{How is consumption $c_t$ determined?}
The consumption FOC~\eqref{eq:iam_foc_c} is enforced exactly by inversion rather than as a residual: given the network's prediction of $\hat{\lambda}_t$, consumption is recovered algebraically as
\begin{equation}
c_t \;=\; \bigl(\hat{\lambda}_t \cdot A_t^{1/\psi - 1}\,L_t^{-1}\bigr)^{-\psi},
\label{eq:iam_c_recovery}
\end{equation}
so $c_t$ is not itself a network output.  Positivity of $c_t$ is guaranteed because the implementation passes $\hat{\lambda}_t$ through a softplus activation, so $\hat{\lambda}_t > 0$ for every input.

\paragraph{How is the abatement multiplier $\hat{\lambda}^\mu_t$ determined?}
The same trick handles the abatement FOC~\eqref{eq:iam_foc_mu}: rather than have the network output $\hat{\lambda}^\mu_t$ and impose the FOC as a separate residual, we \emph{solve} the FOC for $\hat{\lambda}^\mu_t$ and treat the resulting implied multiplier as a deterministic function of the other network outputs.  Setting $\partial\mathcal{L}/\partial\mu_t = 0$ in~\eqref{eq:iam_foc_mu} yields
\begin{equation}
\hat{\lambda}^{\mu,\mathrm{impl}}_t \;=\; -\hat{\lambda}_t\,\Theta'(\mu_t)\,k_t^{\alpha} \;-\; \hat{\nu}^{\mathrm{AT}}_t\,\sigma_t\, A_t\, L_t\,k_t^{\alpha}.
\label{eq:iam_lambdamu_implied}
\end{equation}
Plugged into the Fischer--Burmeister condition~\eqref{eq:iam_fb}, this is the residual $l_8$ below.  Two facts come for free.  First, whenever $l_8 = 0$ holds and the smoothing parameter $\varepsilon_{\mathrm{FB}}$ is small, the abatement FOC \emph{also} holds automatically, because $l_8$ couples the implied multiplier to the slack $1-\mu_t$.  Second, the network output dimension drops from nine to eight, which improves training stability: the network no longer has to discover that $\hat{\lambda}^\mu_t$ is exactly the right algebraic combination of $\hat{\lambda}_t,\, \mu_t,\, \hat{\nu}^{\mathrm{AT}}_t$.

The network architecture uses two hidden layers with 1024 units each, SELU activation, and the Adam optimizer with learning rate $10^{-5}$.  Training alternates between broad sampling (Phase~1) and endogenous simulation (Phase~2), as described in Chapter~\ref{ch:irbc}.

\paragraph{The 8 loss components.}
Each remaining equilibrium condition from \S\ref{sec:dice_lagrangian} becomes a residual $l_m = 0$, and the network is asked to drive every $l_m$ to zero simultaneously along simulated paths.  The mapping is one-for-one: $l_1$ is the capital-Euler FOC~\eqref{eq:iam_foc_k}; $l_2$ is the budget constraint that closes~\eqref{eq:capital_accumulation}; $l_3$, $l_4$, $l_5$ are the three carbon-reservoir envelope conditions, of which $l_3$ is~\eqref{eq:iam_env_mat}; $l_6$ and $l_7$ are the two temperature-layer envelopes; and $l_8$ is the Fischer--Burmeister smoothing~\eqref{eq:iam_fb} of the KKT slack on $\mu_t \le 1$, evaluated at the implied multiplier~\eqref{eq:iam_lambdamu_implied}.  The consumption FOC~\eqref{eq:iam_foc_c} and the abatement FOC~\eqref{eq:iam_foc_mu} are enforced \emph{exactly} via the inversions in~\eqref{eq:iam_c_recovery} and~\eqref{eq:iam_lambdamu_implied}, which is why the loss list contains eight entries instead of nine.  Written out, the eight components are:
{\small
\begin{align}
l_1 &:= \exp\!\bigl(g^A_t + g^L_t\bigr)\,\hat{\lambda}_t - \hat{\beta}_t\Bigl\{\hat{\lambda}_{t+1}\bigl[\bigl(1-\Omega(T_{\mathrm{AT},t+1}) - \Theta(\mu_{t+1})\bigr)\alpha k_{t+1}^{\alpha-1} + (1-\delta)\bigr] \nonumber\\
    &\quad + \hat{\nu}^{\mathrm{AT}}_{t+1}\,\sigma_{t+1}(1-\mu_{t+1})A_{t+1}L_{t+1}\alpha k_{t+1}^{\alpha-1}\Bigr\}
    \tag*{\text{(capital Euler)}}\label{eq:iam_l1}\\[3pt]
l_2 &:= \bigl(1-\Omega(T_{\mathrm{AT},t}) - \Theta(\mu_t)\bigr)\,k_t^\alpha + (1-\delta)\,k_t - c_t - \exp\!\bigl(g^A_t + g^L_t\bigr)\,k_{t+1}
    \tag*{\text{(budget)}}\label{eq:iam_l2}\\[3pt]
l_3 &:= \hat{\nu}^{\mathrm{AT}}_t - \hat{\beta}_t\!\left[\hat{\nu}^{\mathrm{AT}}_{t+1}(1-b_{12}) + \hat{\nu}^{\mathrm{UO}}_{t+1}\,b_{12} + \hat{\eta}^{\mathrm{AT}}_{t+1}\,c_1\,F_{\mathrm{2\times CO_2}}\,\tfrac{1}{\ln 2\,M_{\mathrm{AT},t+1}}\right]
    \tag*{\text{(atm.\ carbon)}}\label{eq:iam_l3}\\[3pt]
l_4 &:= \hat{\nu}^{\mathrm{UO}}_t - \hat{\beta}_t\!\Bigl[\hat{\nu}^{\mathrm{AT}}_{t+1}\,b_{12}\,\tfrac{M^{\mathrm{AT}}_{\mathrm{EQ}}}{M^{\mathrm{UO}}_{\mathrm{EQ}}} + \hat{\nu}^{\mathrm{UO}}_{t+1}\!\Bigl(1-b_{12}\tfrac{M^{\mathrm{AT}}_{\mathrm{EQ}}}{M^{\mathrm{UO}}_{\mathrm{EQ}}}-b_{23}\Bigr) + \hat{\nu}^{\mathrm{LO}}_{t+1}\,b_{23}\Bigr]
    \tag*{\text{(upper ocean C)}}\label{eq:iam_l4}\\[3pt]
l_5 &:= \hat{\nu}^{\mathrm{LO}}_t - \hat{\beta}_t\!\Bigl[\hat{\nu}^{\mathrm{UO}}_{t+1}\,b_{23}\,\tfrac{M^{\mathrm{UO}}_{\mathrm{EQ}}}{M^{\mathrm{LO}}_{\mathrm{EQ}}} + \hat{\nu}^{\mathrm{LO}}_{t+1}\!\Bigl(1-b_{23}\,\tfrac{M^{\mathrm{UO}}_{\mathrm{EQ}}}{M^{\mathrm{LO}}_{\mathrm{EQ}}}\Bigr)\Bigr]
    \tag*{\text{(lower ocean C)}}\label{eq:iam_l5}\\[3pt]
l_6 &:= \hat{\eta}^{\mathrm{AT}}_t - \hat{\beta}_t\!\Bigl[-\hat{\lambda}_{t+1}\,\Omega'(T_{\mathrm{AT},t+1})\,k_{t+1}^\alpha + \hat{\eta}^{\mathrm{AT}}_{t+1}\!\Bigl(1-c_1\,\tfrac{F_{\mathrm{2\times CO_2}}}{\Delta T_{\mathrm{AT},\times 2}}-c_1 c_3\Bigr) + \hat{\eta}^{\mathrm{OC}}_{t+1}\,c_4\Bigr]
    \tag*{\text{(atm.\ temp.)}}\label{eq:iam_l6}\\[3pt]
l_7 &:= \hat{\eta}^{\mathrm{OC}}_t - \hat{\beta}_t\!\left[\hat{\eta}^{\mathrm{AT}}_{t+1}\,c_1 c_3 + \hat{\eta}^{\mathrm{OC}}_{t+1}(1-c_4)\right]
    \tag*{\text{(ocean temp.)}}\label{eq:iam_l7}\\[3pt]
l_8 &:= \hat{\lambda}^{\mu,\mathrm{impl}}_t + (1-\mu_t) - \sqrt{(\hat{\lambda}^{\mu,\mathrm{impl}}_t)^2 + (1-\mu_t)^2 + \varepsilon_{\mathrm{FB}}}
    \tag*{\text{(Fischer--Burmeister, implied multiplier)}}\label{eq:iam_l8}
\end{align}
}
Loss components $l_1$--$l_2$ enforce intertemporal optimality and feasibility, $l_3$--$l_7$ are the envelope conditions that price the five climate state variables, and $l_8$ jointly enforces the abatement FOC (via the implied multiplier) and the upper-bound complementarity $\mu_t \le 1$.

\paragraph{Total loss.}
The DEQN loss aggregates all residuals along a simulated path:
\begin{equation}
\ell_\rho \;:=\; \frac{1}{N_{\text{path}}} \sum_{\bm{x}_t\,\text{on sim.\ path}} \;\sum_{m=1}^{8}\; \bigl(l_m(\bm{x}_t,\, \mathcal{N}_\rho(\bm{x}_t))\bigr)^2.
\label{eq:iam_loss}
\end{equation}
This is the same sum-of-squared-residuals structure as the $N$-country IRBC model of Chapter~\ref{ch:irbc}, but with 8 equations per time step instead of the IRBC's $2N+1$ ($N$ Euler equations, $N$ Fischer--Burmeister conditions, and one aggregate resource constraint).

\paragraph{State evolution.}
To evaluate the loss along a simulated path, the full state vector~\eqref{eq:iam_state} must be propagated forward.  In CDICE the next-period state is:
\begin{equation}
\bm{x}_{t+1} = \bigl(k_{t+1},\; M^{\mathrm{AT}}_{t+1},\; M^{\mathrm{UO}}_{t+1},\; M^{\mathrm{LO}}_{t+1},\; T^{\mathrm{AT}}_{t+1},\; T^{\mathrm{OC}}_{t+1},\; \mu_{f,t+1},\; S_{f,t+1},\; \tau_{t+1};\; \theta\bigr)^T,
\label{eq:iam_state_evol}
\end{equation}
where:
\begin{itemize}[itemsep=2pt]
\item $k_{t+1}$ comes from the network output~\eqref{eq:iam_nn_output} (choice variable);
\item $M^{\mathrm{AT}}_{t+1}$, $M^{\mathrm{UO}}_{t+1}$, $M^{\mathrm{LO}}_{t+1}$, $T^{\mathrm{AT}}_{t+1}$, $T^{\mathrm{OC}}_{t+1}$ are computed from the transition equations of Section~\ref{sec:dice} (carbon cycle and temperature dynamics);
\item the Bayesian belief states $\mu_{f,t+1}$ and $S_{f,t+1}$ are updated via the conjugate posterior~\eqref{eq:bayes_mean}--\eqref{eq:bayes_var} once the period-$t$ temperature observation is realized;
\item the bounded time index advances as $\tau_{t+1} = 1 - \exp\!\bigl(-\vartheta\,(t+\Delta_t)\bigr)$, the image of the calendar increment $t \mapsto t+\Delta_t$ under the time-rescaling~\eqref{eq:time_transform};
\item the pseudo-state parameters $\theta$ are held fixed within an episode and re-sampled across episodes (Section~\ref{sec:deep_uq}).
\end{itemize}
All deterministic transitions are differentiable; stochastic shock draws are handled via reparameterization / common random numbers, so the simulate-then-backpropagate loop can be executed end-to-end with automatic differentiation.

\begin{keyinsightbox}[Recipe: mapping a non-stationary model onto the DEQN framework]
\begin{enumerate}[itemsep=2pt,leftmargin=1.6em]
\item Detrend variables that grow with $A_t L_t$ (Eq.~\ref{eq:iam_detrend}).
\item Map unbounded time to $[0,1)$ via $\tau = 1 - e^{-\vartheta t}$ (Eq.~\ref{eq:time_transform}).
\item Normalize all Lagrange multipliers by $A_t^{1-1/\psi}\,L_t$ (Eq.~\ref{eq:iam_norm_mult}).
\item Derive FOCs from the Lagrangian, both economic \emph{and} climate constraints (Eqs.~\ref{eq:iam_foc_c}--\ref{eq:iam_env_mat}).
\item Enforce static FOCs by inversion: invert the consumption FOC for $c_t$ (Eq.~\ref{eq:iam_c_recovery}) and solve the abatement FOC for the implied multiplier $\hat{\lambda}^{\mu,\mathrm{impl}}_t$ (Eq.~\ref{eq:iam_lambdamu_implied}).
\item Smooth the upper-bound KKT complementarity via Fischer--Burmeister, evaluated at the implied multiplier (Eq.~\ref{eq:iam_fb}).
\item Form the loss as the sum of squared residuals over the 8 remaining conditions (Eq.~\ref{eq:iam_loss}).
\item Train as in the stationary DEQN: simulate $\to$ record loss $\to$ backprop $\to$ repeat.
\end{enumerate}
\end{keyinsightbox}

This is the deterministic CDICE-DEQN solver in its entirety.  Companion notebook \tpath{02_DICE_DEQN_Library_Port.ipynb} trains it against the CDICE library reference solution of \citet{Folini_2021}; the verification gate inside that notebook is the natural stopping point for a reader who wants only the deterministic core.

\section{From CDICE to Stochastic IAMs}
\label{sec:dice_to_stochastic_iam}

The deterministic CDICE-DEQN of \S\ref{sec:iam_dequ_loss}, together with the AR(1) productivity extension developed in the remarkbox below, is the right pedagogical anchor because it contains every mechanical component of an integrated assessment model: capital accumulation, emissions, carbon diffusion, temperature dynamics, damages, abatement costs, and the SCC as a shadow price.  It is not yet the object one wants for quantitative climate-policy research.  Three features are still missing.

First, climate policy is an intrinsically stochastic problem.  Productivity, carbon intensity, damages, climate feedbacks, and tipping thresholds are not known constants.  Once they are stochastic, a carbon tax is not a path but a state-contingent policy.  Second, long-run climate risk makes time-additive CRRA preferences too restrictive: the intertemporal elasticity of substitution and risk aversion should be separate parameters.  Third, climate policy is distributional.  The representative-agent SCC answers a marginal pricing question, but an implementable policy also asks which cohorts pay the tax and which cohorts receive the transfers.  This is the point at which the chapter moves from representative-agent DICE to stochastic overlapping-generations IAMs.

The transition is smooth if one keeps the computational object fixed.  In every case the neural network approximates a policy map
\begin{equation}
\begin{aligned}
    u_t &= \mathcal N_\rho(\tilde{\bm x}_t), \\
    \tilde{\bm x}_t &= (\text{economic states},\ \text{climate states},\ \text{beliefs},\ \text{parameters},\ \text{policy-rule coefficients}),
\end{aligned}
\end{equation}
and the loss is still a sum of normalized equilibrium residuals.  The only changes are the variables appended to $\tilde{\bm x}_t$ and the conditional expectations appearing in the residuals.  Table~\ref{tab:climate_frontier_map} summarizes the sequence.

\begin{table}[H]
\centering
\small
\caption{The layers of the climate-economy pipeline used in the remainder of the chapter.  Each layer is a small extension of the previous one; no new numerical paradigm is introduced after the deterministic CDICE-DEQN.}
\label{tab:climate_frontier_map}
\begin{tabular}{@{}p{3.4cm}p{4.4cm}p{5.4cm}@{}}
\toprule
\textbf{Layer} & \textbf{Economic question} & \textbf{Computational change} \\
\midrule
Deterministic CDICE (\S\ref{sec:iam_dequ_loss}) & What is the globally optimal abatement path and SCC at the baseline calibration? & Time-stamped DEQN; eight residuals; horizon $T_{\max}$ chosen so discounting absorbs transversality. \\
Stochastic DICE (AR(1) + GH quadrature, see the productivity-shock remarkbox below) & How do shocks alter the SCC distribution? & Add shock states; replace future terms by Gauss--Hermite expectations. \\
Bayesian learning on ECS (\S\ref{sec:bayesian_learning}) & How does learning about climate sensitivity alter the SCC distribution? & Add belief mean and belief variance as states; one signal equation; conjugate Gaussian update. \\
Epstein--Zin DICE (\S\ref{sec:ez_layer}) & How do risk aversion and IES separately price climate tails? & Add the value level as a network output; add one recursion residual and an EZ continuation-value weight. \\
Deep UQ (\S\ref{sec:deep_uq}) & Which uncertain parameters drive SCC variation? & Treat parameters as pseudo-states; fit a GP surrogate for the QoI; compute Sobol, Shapley, and univariate effects. \\
Stochastic OLG-IAM (\S\ref{sec:pareto_carbon_tax}) & Can carbon taxes be welfare improving and Pareto improving across cohorts? & Treat tax coefficients and transfer shares as pseudo-states; fit GP surrogates for cohort welfare; solve constrained policy design on the surrogate. \\
\bottomrule
\end{tabular}
\end{table}

\begin{remarkbox}[Adding aggregate shocks: AR(1) productivity and Gauss--Hermite quadrature]
Real climate--economy interactions are shot through with stochastic shocks.  The minimal stochastic extension that already lets us reproduce the qualitative SCC fan-chart structure of \citet{caiSocialCostCarbon2019} on a laptop adds an AR(1) shock to log TFP:
\begin{equation}
z_{t+1} = \rho_z\, z_t + \sigma_z\, \varepsilon_{t+1}, \qquad \varepsilon_{t+1} \overset{\mathrm{i.i.d.}}{\sim} \mathcal{N}(0,1),
\label{eq:tfp_ar1}
\end{equation}
with effective TFP $A_t \exp(z_t)$.  The state vector~\eqref{eq:iam_state} acquires a new entry,
\begin{equation}
\tilde{\bm{x}}_t = \bigl(k_t,\; M^{\mathrm{AT}}_t,\; M^{\mathrm{UO}}_t,\; M^{\mathrm{LO}}_t,\; T^{\mathrm{AT}}_t,\; T^{\mathrm{OC}}_t,\; \tau_t,\; z_t\bigr)^\top,
\label{eq:iam_state_stoch}
\end{equation}
and each forward-looking residual~\eqref{eq:iam_l1}--\eqref{eq:iam_l7} acquires a conditional expectation over $\varepsilon_{t+1}$; the capital Euler, for example, becomes
\begin{equation}
\begin{aligned}
e^{g^A_t + g^L_t}\,\hat{\lambda}_t \;=\; \hat{\beta}_t\,\mathbb{E}_t\Bigl[\;
&\hat{\lambda}_{t+1}\bigl(\bigl(1-\Omega(T^{\mathrm{AT}}_{t+1}) - \Theta(\mu_{t+1})\bigr)\alpha\,k_{t+1}^{\alpha-1} + (1-\delta)\bigr) \\
&\;+\; \hat{\nu}^{\mathrm{AT}}_{t+1}\,\sigma_{t+1}(1-\mu_{t+1})\,A_{t+1}L_{t+1}\,\alpha\,k_{t+1}^{\alpha-1}\,\Bigr].
\end{aligned}
\label{eq:iam_euler_stoch}
\end{equation}
With $\varepsilon_{t+1}$ Gaussian, the conditional expectation is evaluated with a small number of Gauss--Hermite nodes $\{(\xi_q, w_q)\}_{q=1}^Q$,
\begin{equation}
\mathbb{E}_t[f(\varepsilon_{t+1})] \;\approx\; \frac{1}{\sqrt{\pi}}\sum_{q=1}^{Q} w_q\, f\bigl(\sqrt{2}\,\xi_q\bigr),
\label{eq:gh_quadrature}
\end{equation}
and each residual is replaced by its stochastic counterpart
\begin{equation}
l_m^{\mathrm{stoch}}(\tilde{\bm{x}}_t,\, \mathcal{N}_\rho) \;=\; \frac{1}{\sqrt{\pi}}\sum_{q=1}^{Q} w_q\, l_m\bigl(\tilde{\bm{x}}_t,\, \mathcal{N}_\rho;\, \varepsilon_{t+1} = \sqrt{2}\,\xi_q\bigr),
\label{eq:loss_gh}
\end{equation}
which the total loss~\eqref{eq:iam_loss} then aggregates as before.  In practice $Q = 5$ nodes drive the quadrature error well below the training-noise floor; the GH evaluation is fully differentiable, so the autodiff backward pass is unchanged.  When several independent shock dimensions appear simultaneously (productivity shock $\varepsilon_{t+1}$, learning innovation $\tilde\epsilon_{T,t+1}$, and the EZ certainty-equivalent integrand of \S\ref{sec:ez_layer}), each conditional expectation is taken under a tensor product of one-dimensional GH rules at $Q$ nodes per dimension, i.e.\ $Q^d$ total nodes for $d$ shock dimensions; the autodiff backward pass traverses the quadrature unchanged.  Forward-simulating $N_{\mathrm{MC}}$ trajectories of the AR(1) shock produces a Monte-Carlo SCC fan chart whose right-tail mass is the channel of \citeauthor{caiSocialCostCarbon2019}'s headline result.  Companion notebook \tpath{03_Stochastic_DICE_DEQN.ipynb} trains this stochastic extension end-to-end and is the natural anchor for Exercise~\ref{ex:ch11:7}.
\end{remarkbox}

\begin{keyinsightbox}[The unifying trick]
The same pseudo-state idea appears three times: uncertain structural parameters are pseudo-states for UQ, policy-rule coefficients are pseudo-states for constrained optimal policy, and Bayesian posterior beliefs are endogenous pseudo-states for learning.  The network does not care which interpretation is attached to an input coordinate; it only learns a differentiable equilibrium map on the enlarged domain.  The methodological cost of each layer is small: at most one additional state or output and one additional term in the loss; the algorithm of \S\ref{sec:nsdeqn_algo} is unchanged.
\end{keyinsightbox}


\section{Bayesian Learning About Climate Sensitivity}
\label{sec:bayesian_learning}

\paragraph{Why ECS is the natural learning state.}
The equilibrium climate sensitivity (ECS), defined as the long-run atmospheric warming from a doubling of CO$_2$, is the single most consequential and most uncertain parameter in the climate side of an IAM.  Observational, paleoclimate, and model-based estimates place ECS in a \emph{likely} (66\%) range of roughly 2.5--4\textdegree C and a \emph{very-likely} (90\%) range of 2--5\textdegree C \citep{sherwood2020assessment, knutti2017beyond, roe2007climate}, and ECS uncertainty is the largest single contributor to SCC dispersion across model variants.  Crucially, ECS is partially identified from temperature realizations conditional on emissions and forcing: a Bayesian planner who observes temperature paths can therefore update her posterior period by period, and the policy that maximizes ex-ante welfare conditions on the current posterior rather than on a fixed point estimate.

\paragraph{How learning enters the state.}
Promote the climate-feedback parameter $\lambda$ in~\eqref{eq:temp_at} to a stochastic object by adding the feedback term $\varphi_{1C}\,\tilde f_{t+1}\,T^{\mathrm{AT}}_t$ to the right-hand side, where $\varphi_{1C}$ is a calibrated coupling coefficient (taken from \citet{friedlDeep2023}) and $\tilde f_{t+1} \sim \mathcal N(\mu_{f,t}, S_{f,t})$ is a per-period draw under the planner's posterior over the unknown climate-feedback deviation.  The unknown itself is time-invariant; the subscript $t{+}1$ indexes the period in which the subjective draw enters the temperature equation, and the planner's posterior moments $(\mu_{f,t}, S_{f,t})$ shift over time as new temperature observations arrive.  The planner observes the temperature-residual signal
\begin{equation}
y_{t+1} \;:=\; \varphi_{1C}\,T^{\mathrm{AT}}_t\,\tilde f_{t+1} \;+\; \tilde\epsilon_{T,t+1},\qquad \tilde\epsilon_{T,t+1} \sim \mathcal N(0, S_{\epsilon_T}),
\label{eq:bayes_signal}
\end{equation}
and conjugate Gaussian--Gaussian updating delivers the posterior
\begin{align}
\mu_{f,t+1} &= \frac{S_{\epsilon_T}\,\mu_{f,t} + \varphi_{1C}\,T^{\mathrm{AT}}_t\,S_{f,t}\,y_{t+1}}{S_{\epsilon_T} + (\varphi_{1C}\,T^{\mathrm{AT}}_t)^2\,S_{f,t}}, \label{eq:bayes_mean}\\
S_{f,t+1} &= \frac{S_{\epsilon_T} \cdot S_{f,t}}{S_{\epsilon_T} + (\varphi_{1C}\,T^{\mathrm{AT}}_t)^2\,S_{f,t}}, \label{eq:bayes_var}
\end{align}
which the planner takes as two additional laws of motion for the belief states $(\mu_{f,t}, S_{f,t})$.  These two states occupy the slots already reserved in the augmented state vector~\eqref{eq:iam_state}.  Equations~\eqref{eq:bayes_mean}--\eqref{eq:bayes_var} are the Kalman update for a scalar linear-Gaussian state-space model with observation gain $\varphi_{1C}\,T^{\mathrm{AT}}_t$ and noise variance $S_{\epsilon_T}$; cf.\ \citet[\S~13.3]{bishop2006} for the generic derivation.  The DEQN algorithm of \S\ref{sec:nsdeqn_algo} is unchanged: the network simply receives two more inputs and learns a richer policy.

\paragraph{Where this sits in the literature.}
Bayesian learning about climate parameters in an integrated assessment frame has a long pedigree.  \citet{kellyBayesianLearningGrowth1999} and \citet{kellyLearningClimateFeedbacks2015} establish the basic Kelly--Kolstad result that learning takes decades to centuries in calibrated DICE-like settings, and that the tradeoff between mitigation (which lowers temperature variance) and information (which requires informative temperature paths) is sharp.  \citet{leachClimateChangeLearning2007} and \citet{websterLearningClimateChange2008} sharpen the slow-learning result and quantify the policy errors induced by treating uncertainty as resolved too quickly.  On the dynamic-programming side, \citet{caiSocialCostCarbon2019} solve a stochastic-DICE variant with tipping-point hazards and recursive preferences.  The robust-control program of \citet{anderson2014Uncertainty}, \citet{barnett2023climate}, \citet{barnett2023deep}, and \citet{Barnett2020} addresses a complementary question (planner ambiguity over the data-generating process), and modern deep-learning solutions are the natural computational companion because tensor-product grids over belief states are infeasible at realistic state-vector sizes.

\begin{figure}[ht]
\centering
\begin{tikzpicture}
\begin{axis}[
    name=left, width=0.46\linewidth, height=4.0cm,
    xlabel={Years from $t=0$}, ylabel={$S_{f,t}/S_{f,0}$},
    ymin=0.05, ymax=1.05,
    ymode=log,
    grid=both, grid style={gray!20},
    legend style={font=\tiny, draw=none, fill=none, at={(0.95,0.95)}, anchor=north east}
]
\addplot[uzhblue, very thick, mark=none, domain=0:30, samples=80] {exp(-0.25*x)};
\addlegendentry{posterior variance ratio}
\end{axis}
\begin{axis}[
    name=right, at={(left.east)}, xshift=1.5cm, anchor=west,
    width=0.46\linewidth, height=4.0cm,
    xlabel={True ECS (\textdegree C)}, ylabel={$\mathrm{SCC}_0$ (USD/tCO$_2$)},
    xmin=2, xmax=5, ymin=0, ymax=200,
    grid=both, grid style={gray!20},
    legend style={font=\tiny, draw=none, fill=none, at={(0.05,0.95)}, anchor=north west}
]
\addplot[harvardcrimson, very thick, mark=none, domain=2:5, samples=40] {30 + 25*(x-2)^2};
\addlegendentry{no learning}
\addplot[uzhblue, very thick, dashed, mark=none, domain=2:5, samples=40] {15 + 22*(x-2)^2};
\addlegendentry{with learning}
\end{axis}
\end{tikzpicture}
\caption{Schematic of the two qualitative features reported by \citet{friedlDeep2023}.  Left: posterior variance $S_{f,t}$ relative to its prior value, on a logarithmic scale.  The variance falls by roughly an order of magnitude over the first decade, mirroring the Kelly--Kolstad slow-learning result but accelerated by the deeper signal--noise ratio of the modern climate calibration.  Right: $\mathrm{SCC}_0$ as a function of the true ECS, with and without Bayesian learning.  Learning approximately halves the SCC at moderate ECS values where uncertainty is the dominant driver of precautionary abatement, but converges to the no-learning curve at the upper tail where the underlying physical damage dominates.  Curves are illustrative; the magnitudes are those quoted in the body text.}
\label{fig:bayes_learning_schematic}
\end{figure}
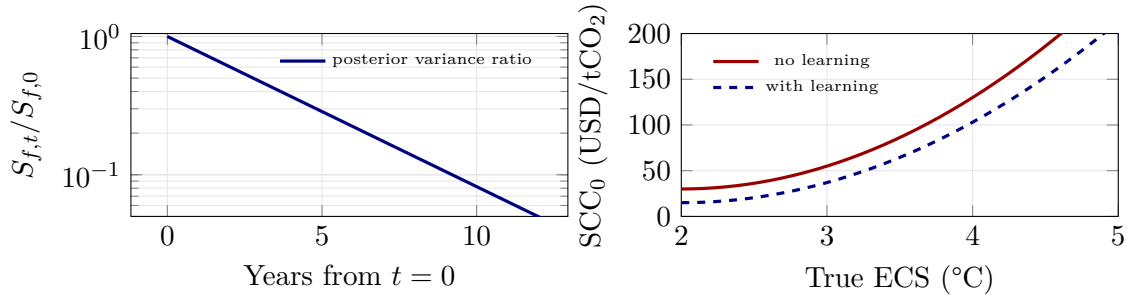

\paragraph{Headline result from the UQ literature.}
\citet{friedlDeep2023} solve the joint stochastic-DICE--Bayesian-learning DEQN with the methodology of this chapter and find two qualitative features that survive across the calibration cloud.\footnote{The numerical claims in this paragraph quote the headline results of \citet{friedlDeep2023}; consult that paper for the precise figures and the underlying calibration grid.}  First, ECS uncertainty is largely resolved within roughly ten years of optimal policy: the posterior variance $S_{f,t}$ shrinks by an order of magnitude over the first decade of the planner's horizon, even though the absolute posterior mean takes longer to settle.  Second, the SCC under learning is roughly half the no-learning SCC for moderate true ECS values, and roughly the same as the no-learning SCC at the upper tail of the ECS distribution; learning is a strong substitute for precautionary mitigation in the moderate-ECS regime, and a weak substitute in the tail-ECS regime.  The asymmetry is policy-relevant: the value of waiting to learn falls sharply once the planner suspects she is in the tail.  The broader teaching point is that uncertainty is not automatically a reason to abate more: its policy effect depends on whether the uncertainty is static, learnable, or associated with irreversible tail risk.  Figure~\ref{fig:bayes_learning_schematic} illustrates the two qualitative features.

\section{Epstein--Zin Preferences}
\label{sec:ez_layer}

\paragraph{Why recursive preferences for climate.}
The time-additive CRRA-IES aggregator~\eqref{eq:planner_obj_crra} ties risk aversion and intertemporal substitution together.  Climate policy is exactly the environment in which this restriction is least attractive.  A planner may want a high IES $\psi$ to govern intertemporal substitution across long horizons, and a separate high coefficient of relative risk aversion $\gamma_u$ to price low-probability climate disasters.  Recursive Kreps--Porteus preferences, following \citet{epstein1989Substitution} and \citet{weil1989Equity}, implement this separation.\footnote{\textbf{Notation note.}  We use $\psi$ for the IES and $\gamma_u$ for risk aversion throughout this section, following \citet{friedlDeep2023}.  The IRBC chapter (Chapter~\ref{ch:irbc}) used $\gamma$ for the IES under the bundled CRRA-IES convention; here in the Epstein--Zin block we deliberately decouple the two parameters, so the symbol switch is intentional.  The CRRA limit is recovered at $\gamma_u = 1/\psi$.}

Working with the normalized per-capita value $v_t$ and per-capita consumption $c_t = C_t/L_t$, and writing $\beta^{\mathrm{EZ}}_t := \exp(-\rho\,\Delta_t)$ for the one-period Epstein--Zin discount factor, the recursion is
\begin{equation}
    v_t
    =
    \left[
    (1-\beta^{\mathrm{EZ}}_t)\, c_t^{1-1/\psi}
    +
    \beta^{\mathrm{EZ}}_t
    \left(\mathbb E_t\!\left[v_{t+1}^{1-\gamma_u}\right]\right)^{\frac{1-1/\psi}{1-\gamma_u}}
    \right]^{\frac{1}{1-1/\psi}},
\label{eq:ez_iam}
\end{equation}
with the usual logarithmic limits when $\psi = 1$ or $\gamma_u = 1$, subject to the same budget constraint, capital-accumulation law, and climate dynamics as before.

\paragraph{What changes in the DEQN loss.}
The value level $v_t$ becomes an additional network output, paired with a ninth residual that enforces the recursion~\eqref{eq:ez_iam}:
\begin{equation}
    \mathcal R^{\mathrm{EZ}}_t
    =
    v_t
    -
    \left[
    (1-\beta^{\mathrm{EZ}}_t)\, c_t^{1-1/\psi}
    +
    \beta^{\mathrm{EZ}}_t
    \left(\mathbb E_t\!\left[v_{t+1}^{1-\gamma_u}\right]\right)^{\frac{1-1/\psi}{1-\gamma_u}}
    \right]^{\frac{1}{1-1/\psi}}.
\label{eq:ez_residual}
\end{equation}
In the deterministic CRRA-IES core of \S\ref{sec:iam_dequ_loss}, $v_t$ never appears explicitly, which is why eight residuals suffice there.  The Euler and costate residuals of \S\ref{sec:dice_lagrangian} keep their deterministic form but receive a Bansal--Yaron certainty-equivalent weighting inside each conditional expectation.  It is convenient to write the one-step recursive-pricing kernel as
\begin{equation}
    \mathcal M^{\mathrm{EZ}}_{t,t+1}
    =
    \hat\beta_t
    \left(
    \frac{v_{t+1}}{\left(\mathbb E_t[v_{t+1}^{1-\gamma_u}]\right)^{1/(1-\gamma_u)}}
    \right)^{1/\psi - \gamma_u}
    \left(\frac{c_{t+1}}{c_t}\right)^{-1/\psi},
\label{eq:ez_kernel}
\end{equation}
where $\hat\beta_t$ inherits the deterministic growth normalization of~\eqref{eq:iam_eff_beta}.  In the code, \eqref{eq:ez_kernel} is just a multiplicative weight on next-period marginal-value terms; the certainty-equivalent operator inside the Kreps--Porteus aggregator becomes a second nested expectation.  The DEQN loss inherits one extra Gauss--Hermite quadrature step and one extra network output, but no new algorithmic ingredient.

\paragraph{Interpretation for the SCC.}
\citet{crost2013Optimal} and \citet{crost2014Optimal} establish the analytic baseline: in a deterministic IAM, decoupling risk aversion from the IES changes the optimal carbon tax only when stochastic risk is present, but the change can be quantitatively large once it is.  \citet{jensenOptimalClimateChange2014} and \citet{traeger2018ace, traeger2021ACE} extend the result to closed-form ACE-class settings and show that for reasonable risk aversion above $1/\psi$, the SCC roughly doubles relative to CRRA; \citet{caiSocialCostCarbon2019} reach the same conclusion in a fully stochastic DICE variant.  Intuitively, recursive preferences change the SCC because carbon emissions affect the distribution of long-run consumption, not only its mean: if damages create low-consumption tail states, a high $\gamma_u$ raises the SCC through the disaster-insurance channel.  The sign of the IES effect depends on which shock dominates: in TFP-driven economies higher $\psi$ dampens the SCC because consumption smoothing absorbs the productivity risk, whereas in temperature-driven economies higher $\psi$ amplifies the SCC because the planner cares more about late-horizon consumption losses.  \citet{bansalKikuOchoa2016} make the asset-pricing case for the same channel: long-run temperature shifts price into expected returns through the EZ aggregator, and ignoring them understates the welfare cost of carbon emissions.  This is why stochastic DICE with Epstein--Zin preferences is a better teaching object than deterministic DICE for climate-finance questions: it connects welfare, tail risk, and asset-pricing logic in a single equilibrium loss.

\section{Deep Uncertainty Quantification via Surrogates}
\label{sec:deep_uq}

Deep UQ answers a different question from solving one stochastic IAM.  The object is now a scalar quantity of interest,
\begin{equation}
    q(\theta) = \mathrm{SCC}_{2100}(\theta), \qquad \theta\in\Theta\subset\mathbb R^{d_\theta},
\label{eq:uq_qoi}
\end{equation}
where $\theta$ collects uncertain structural parameters: the ECS or its prior mean $\mu_{f,0}$, the prior variance $S_{f,0}$, the pure rate of time preference $\rho$, the IES $\psi$, risk aversion $\gamma_u$, the damage curvature $\pi_2$, and any tipping parameters included in the experiment.  Direct global sensitivity analysis would require solving the IAM thousands of times.  Deep UQ replaces this infeasible outer loop by two amortizations.

\paragraph{Amortization 1: parameters as pseudo-states.}
The pseudo-state trick of \citet{friedlDeep2023} collapses the outer loop into a single DEQN training pass.  Uncertain parameters $\theta$ are appended to the network's input,
\begin{equation}
\tilde{\bm{x}}_t = \bigl(\underbrace{\bm x_t}_{\text{economic + climate states}},\; \underbrace{\theta}_{\text{uncertain parameters}}\bigr),\qquad u_t = \mathcal N_\rho(\tilde{\bm x}_t),
\label{eq:pseudo_state}
\end{equation}
held fixed within each simulation episode and re-sampled across episodes from a design distribution $\mathcal D_\theta$.  One trained network therefore approximates the policy function for every $\theta$ in $\mathcal D_\theta$; evaluating any new $\theta$ requires only a forward pass.  For very large pseudo-state dimensions the active-subspace methods of \S\ref{sec:active_subspaces} compress $\theta$ before the next step.  This is the same idea as the parameterized policy networks in Chapter~\ref{ch:estimation}; here the target is not an SMM criterion but an SCC distribution.

\paragraph{Amortization 2: a GP for the quantity of interest.}
After training, the DEQN is evaluated at a design set $\{\theta_i\}_{i=1}^n$ and the corresponding QoI values $q_i = q(\theta_i)$ are computed by forward simulation.  Fit a Gaussian-process surrogate
\begin{equation}
    q(\theta) = m(\theta) + \varepsilon(\theta), \qquad m(\theta)\mid \{(\theta_i,q_i)\}_{i=1}^n \sim \mathcal{GP}\bigl(\mu_n(\theta),\, k_n(\theta,\theta')\bigr).
\label{eq:uq_gp}
\end{equation}
The GP is cheap enough to evaluate millions of times, so the expensive IAM is no longer called inside Sobol, Shapley, or univariate-effect estimators.  Bayesian active learning improves the design by adding points where the GP posterior uncertainty is largest or where integrated posterior variance is most reduced, following the toolkit of Chapter~\ref{ch:gp} (see Figure~\ref{fig:surrogate_outer_loop} and Table~\ref{tab:surrogate_strategy_comparison}).

\paragraph{Sobol, Shapley, univariate effects.}
Three complementary global sensitivity indices answer different questions about how $\theta$ drives the SCC.  The first-order Sobol index $S_i$ of \citet{sobolGlobalSensitivityIndices2001} measures the share of output variance explained by $\theta_i$ alone,
\begin{equation}
S_i = \frac{\mathrm{Var}\bigl(\mathbb E[q(\theta)\mid\theta_i]\bigr)}{\mathrm{Var}(q(\theta))},
\label{eq:sobol_first}
\end{equation}
and the total-effect index captures both direct and interaction effects,
\begin{equation}
S_i^{\mathrm{tot}} = 1 - \frac{\mathrm{Var}\bigl(\mathbb E[q(\theta)\mid\theta_{-i}]\bigr)}{\mathrm{Var}(q(\theta))}.
\label{eq:sobol_total}
\end{equation}
For independent inputs the $\{S_i\}$ sum to at most one, while the $\{S_i^{\mathrm{tot}}\}$ can exceed one in the presence of interactions; equality $\sum_i S_i^{\mathrm{tot}} = 1$ characterizes additive models.  Shapley effects, introduced into sensitivity analysis by \citet{owen2014sobol} and developed further by \citet{songShapleyEffectsGlobal2016} and \citet{ioossShapleyEffectsSensitivity2019}, allocate $\mathrm{Var}(q)$ across parameters via cooperative-game averaging over all subsets of other parameters \citep{shapley1953value}, sum exactly to $\mathrm{Var}(q)$ (raw) or one (normalized), and handle correlated inputs cleanly.  Univariate-effect plots show the conditional mean $\mathbb E[q(\theta)\mid\theta_i]$ as $\theta_i$ varies and capture the directional response that Sobol indices average over.  \citet{saltelli2010Sensitivity} and \citet{saltelli2008global} give the standard estimators and best-practice warnings.

\begin{remarkbox}[Deep UQ implementation recipe]
\begin{enumerate}[itemsep=2pt,leftmargin=*]
\item Choose a parameter domain $\Theta$ and a sampling law $\mathcal D_\theta$ for the uncertain climate, damage, and preference parameters.
\item Train the stochastic CDICE-DEQN on $(\bm x_t, z_t, \mu_{f,t}, S_{f,t}, \theta)$, resampling $\theta$ across episodes and holding it fixed within an episode.
\item Generate $n$ design evaluations $\{(\theta_i, q_i)\}_{i=1}^n$ from the trained network, where $q_i$ is typically $\mathrm{SCC}_{2100}$ or an expected welfare functional.
\item Fit a GP surrogate $\theta \mapsto q(\theta)$, validate by leave-one-out cross-validation (target: LOO~$R^2 \ge 0.95$ or LOO RMSE below $5\%$ of the QoI standard deviation), and enrich the design with Bayesian active learning if the threshold is not met.
\item Compute Sobol, Shapley, and univariate effects on the GP surrogate, not on the structural model.
\end{enumerate}
\end{remarkbox}

The reason this pipeline is the only feasible route is computational: direct Monte Carlo on Sobol or Shapley indices requires $O(10^4)$ to $O(10^6)$ evaluations of the structural model at fresh $\theta$ draws.  Even at one DEQN solve per parameter vector, that price tag is several core-decades.  The DEQN-with-pseudo-states amortizes one loop, and the GP surrogate amortizes the other; the sensitivity indices are then computed on the GP rather than on the IAM.

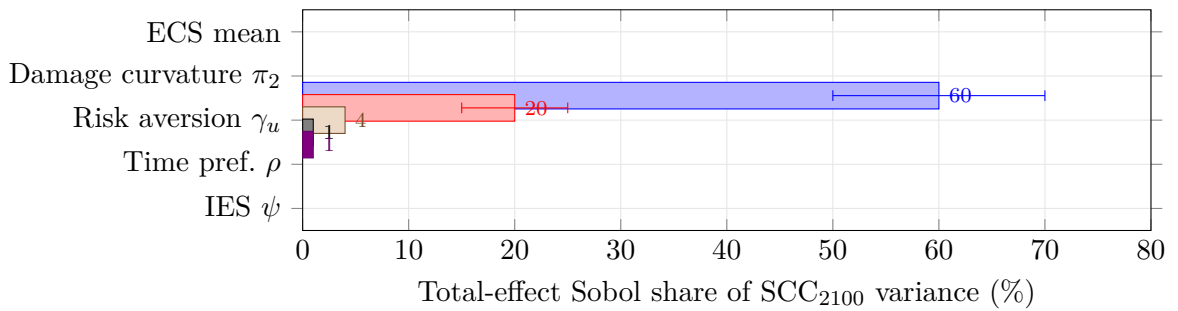
\begin{figure}[ht]
\centering
\begin{tikzpicture}
\begin{axis}[
    width=0.80\linewidth, height=4.5cm,
    xbar, bar width=10pt,
    xmin=0, xmax=80,
    ymin=0.5, ymax=5.5,
    ytick={1,2,3,4,5},
    yticklabels={IES $\psi$, Time pref.\ $\rho$, Risk aversion $\gamma_u$, Damage curvature $\pi_2$, ECS mean},
    xlabel={Total-effect Sobol share of $\mathrm{SCC}_{2100}$ variance (\%)},
    grid=major, grid style={gray!20},
    nodes near coords, nodes near coords style={font=\scriptsize},
    every axis plot/.append style={fill=uzhblue!30, draw=uzhblue}
]
\addplot+[error bars/.cd, x dir=both, x fixed=10] coordinates {
    (60,5)
};
\addplot+[error bars/.cd, x dir=both, x fixed=5] coordinates {
    (20,4)
};
\addplot+ coordinates {(4,3)};
\addplot+ coordinates {(1,2)};
\addplot+ coordinates {(1,1)};
\end{axis}
\end{tikzpicture}
\caption{Schematic of the total-effect Sobol shares of $\mathrm{SCC}_{2100}$ variance reported by \citet{friedlDeep2023}.  Midpoints reflect the ranges quoted in the text (ECS mean 50--70\%, damage curvature 15--25\%), with horizontal error bars on the two leading parameters indicating the spread across horizon dates and damage-function specifications.  The shape, two parameters carrying almost the entire variance, is what motivates the tail-insurance framing in the closing paragraph.}
\label{fig:sobol_shares_schematic}
\end{figure}

\paragraph{Empirical headline.}
\citet{friedlDeep2023} apply the pipeline to a stochastic DICE variant with Epstein--Zin preferences and Bayesian learning, and find that two ingredients dominate the SCC variance across 2020--2100: the mean of the ECS belief (roughly 50--70\% of the total-effect Sobol share) and the curvature parameter of the damage function (roughly 15--25\%).\footnote{Variance-share ranges quoted from \citet{friedlDeep2023}; the spread reflects different points along the planner's horizon and different damage-function specifications.}  Together these account for 70--90\% of the SCC variance.  Risk aversion contributes a few percentage points; the pure rate of time preference and the IES contribute negligibly once damage curvature is conditioned on.  The policy lesson is that under deep uncertainty the SCC should be reported as a distribution, not a point estimate, and that climate-policy design should target tail insurance against the upper ECS--damage corner rather than precision over the central calibration.  Figure~\ref{fig:sobol_shares_schematic} sketches the resulting variance decomposition.

\section{Constrained Pareto-Improving Carbon Tax in OLG-IAMs}
\label{sec:pareto_carbon_tax}
\label{sec:carbon_tax}

The SCC analysis of \S\ref{sec:deep_uq} is still the marginal welfare cost of one extra ton of carbon \emph{to a representative agent}.  Climate policy, however, redistributes welfare across cohorts: today's workers pay abatement costs while tomorrow's households inherit a cooler planet.  A \emph{Pareto-improving} carbon tax must transfer enough revenue back to current cohorts that no generation is worse off than under business-as-usual.  This section closes Movement~3 by walking through the constrained-Pareto OLG-IAM of \citet{kubler2025using}, reusing the DEQN-with-pseudo-states machinery of \S\ref{sec:deep_uq} and the GP surrogate of Chapter~\ref{ch:gp}.  The Pareto-improvement criterion is closely related to the social-security reform literature \citep{krueger2006pareto}, to recent work on intergenerational climate policy \citep{Karp_Peri_Rezai_2024, kotlikoff2020ParetoImproving}, and to the constrained-optimal-tax frontier of \citet{douenne_hummel_pedroni_2024}.

\paragraph{Notation reset for this section.}
\begin{sloppypar}
The OLG-IAM uses different conventions than the representative-agent CDICE block of \S\ref{sec:dice}, following \citet{kubler2025using}, and we summarize the differences here so the reader is not surprised.  $\Omega_t(T_t)$ now denotes the \emph{retained-output} factor, so net output is $\Omega_t \Phi K^\alpha L^{1-\alpha}$ rather than $(1-\Omega-\Theta)Y^{\mathrm{gross}}$.  $p^{\mathrm{tax}}_t$ is the carbon tax (a per-tCO$_2$ price); to avoid clashing with the transformed-time variable $\tau_t$ of \S\ref{sec:iam_nonstationarity}, this section uses $p^{\mathrm{tax}}_t$ for the tax throughout, in line with the price-level interpretation.  $e_t$ denotes the per-period emissions \emph{flow} (in GtC), and $E_t = \sum_{s\le t} e_s$ is \emph{cumulative} emissions through date $t$; this is the convention of the climate-emulator literature \citep{dietz2019cumulative} and of the companion paper, and it is the source of the section's frequent ``cumulative-emissions tax'' phrasing.  Finally, the policy vector that the planner ultimately optimizes over is $\vartheta = (\vartheta_{\mathrm{tax}}, \omega)$, the joint vector of tax-rule coefficients and cohort transfer shares defined in Step~1 below.
\end{sloppypar}

\begin{figure}[ht]
\centering
\begin{tikzpicture}[>=stealth,
    box/.style={rectangle, draw=uzhblue, fill=uzhblue!8, rounded corners=3pt, minimum width=1.4cm, minimum height=0.7cm, font=\scriptsize},
    boxT/.style={rectangle, draw=harvardcrimson, fill=harvardcrimson!8, rounded corners=3pt, minimum width=1.4cm, minimum height=0.7cm, font=\scriptsize},
    src/.style={rectangle, draw=darkgreen, fill=darkgreen!10, rounded corners=3pt, minimum width=1.0cm, minimum height=0.6cm, font=\scriptsize},
    tcre/.style={rectangle, draw=darkgreen, fill=darkgreen!15, rounded corners=3pt, minimum width=1.4cm, minimum height=0.7cm, font=\scriptsize}]
    \node[font=\small\bfseries] at (1.4,2.1) {CDICE (\S\ref{sec:dice_carbon_cycle}--\S\ref{sec:dice_temperature}): 5 states};
    \node[src] (e1) at (-0.4,0.6) {$E_t$};
    \node[box] (mat1) at (0.9,1.2) {$M^{\mathrm{AT}}$};
    \node[box] (muo1) at (0.9,0.3) {$M^{\mathrm{UO}}$};
    \node[box] (mlo1) at (0.9,-0.6) {$M^{\mathrm{LO}}$};
    \node[boxT] (tat1) at (2.4,1.2) {$T^{\mathrm{AT}}$};
    \node[boxT] (toc1) at (2.4,0.3) {$T^{\mathrm{OC}}$};
    \draw[->] (e1) -- (mat1);
    \draw[<->] (mat1) -- (muo1);
    \draw[<->] (muo1) -- (mlo1);
    \draw[->] (mat1) -- (tat1);
    \draw[<->] (tat1) -- (toc1);
    \node[font=\small\bfseries] at (7.2,2.1) {TCRE (OLG-IAM): 1 state};
    \node[src] (e2) at (5.2,0.6) {$e_t$};
    \node[tcre] (cum) at (6.8,0.6) {$E_t = \sum e_s$};
    \node[boxT] (tat2) at (8.8,0.6) {$T^{\mathrm{AT}}$};
    \draw[->] (e2) -- (cum);
    \draw[->] (cum) -- node[font=\scriptsize, fill=white, inner sep=1pt] {$\sigma_{\mathrm{CCR}}$} (tat2);
\end{tikzpicture}
\caption{Climate side of CDICE versus TCRE.  The 5-state CDICE module on the left, in which atmospheric carbon, two ocean carbon reservoirs, atmospheric temperature, and ocean temperature all enter the planner's state, is collapsed in the OLG-IAM to a single algebraic relation between cumulative emissions and atmospheric temperature, $T^{\mathrm{AT}}_t \approx \sigma_{\mathrm{CCR}}\,E_t$.  The simplification trades fidelity to short-run climate dynamics for tractability of the 12-cohort heterogeneous-agent state space and is what makes the bilevel policy search of \S\ref{sec:pareto_carbon_tax} end-to-end feasible.}
\label{fig:cdice_vs_tcre}
\end{figure}
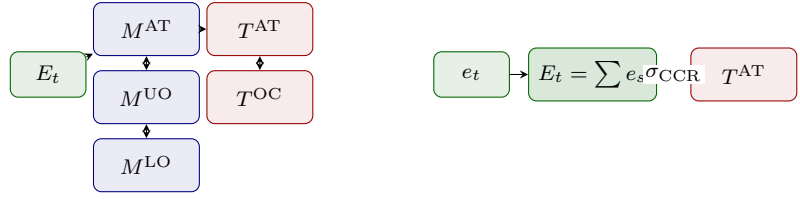

\paragraph{From CDICE to a TCRE emulator.}
The OLG-IAM uses a much simpler climate side than the 5-state CDICE module of \S\ref{sec:dice} (three carbon stocks plus two temperature layers).  Once the planner's horizon is converted to cumulative-emissions form $E_t = \sum_{s\le t} e_s$, the linear Transient Climate Response to cumulative carbon Emissions (TCRE) approximation collapses the carbon-cycle and energy-balance machinery to a single algebraic relation $T^{\mathrm{AT}}_t \approx \sigma_{\mathrm{CCR}}\,E_t$ \citep{dietz2019cumulative}, which removes five climate states from the planner's optimization.  The simplification is essential: it is what makes the OLG state space (12 cohort assets + 5 climate / shock states + $\vartheta$ pseudo-states) tractable end-to-end on a GPU.  The reader who finds the change abrupt should treat the TCRE relation as a reduced-form summary of the same physics that drove \S\ref{sec:dice_carbon_cycle}--\S\ref{sec:dice_temperature}, fitted directly to long-run paths rather than block-by-block.  Figure~\ref{fig:cdice_vs_tcre} contrasts the two climate sides.

\subsection{The OLG-IAM Model}
\label{sec:olg_iam}

\begin{figure}[H]
\centering
\includegraphics[width=0.92\linewidth]{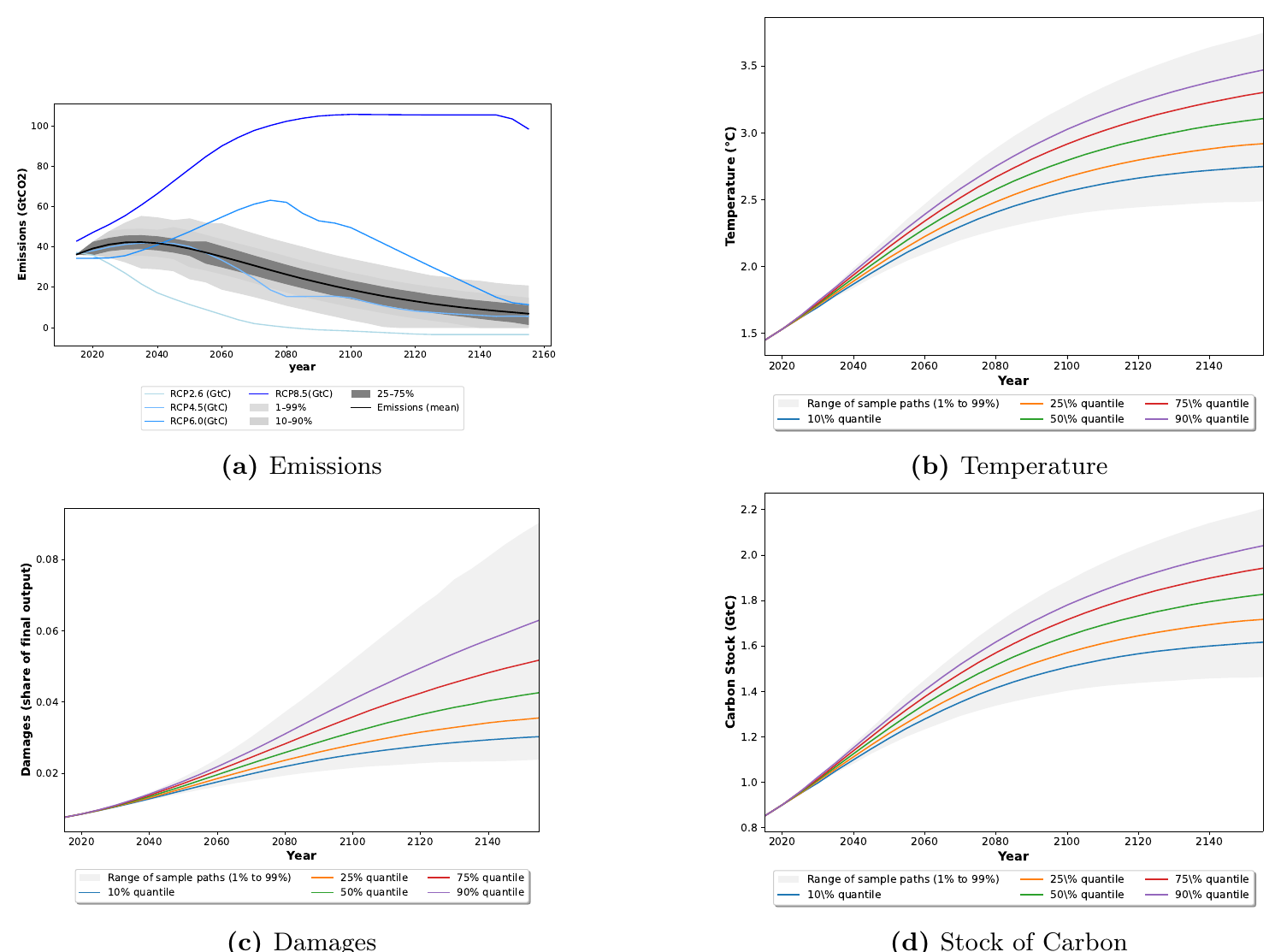}
\caption{Business-as-usual baseline for the 12-cohort stochastic OLG-IAM of \citet{kubler2025using}.  Without policy intervention the median warming reaches roughly $3\,^\circ$C over the 150-year horizon, and the upper tail of damages is substantially larger than the mean.  Every Pareto-improving policy below is benchmarked against this baseline, which also supplies the participation constraints for the constrained policy search.  Figure extracted from \citet{kubler2025using}.}
\label{fig:bau_olg_baseline}
\end{figure}

The model features $A=12$ overlapping generations of selfish agents (ages 20--80 in 5-year periods), a competitive firm, and a simplified, cumulative-emissions climate module in the spirit of \citet{dietz2019cumulative}:
\begin{itemize}[itemsep=2pt]
\item \textbf{Technology:} Output is $Y_t = \Omega_t(T_t)\,\Phi(\mu_t)\,K_t^\alpha L_t^{1-\alpha}$ with retained-output damage factor $\Omega_t$ and net-of-abatement-cost factor $\Phi(\mu_t) = 1 - \theta_1\mu_t^{\theta_2}$; emissions are $e_t = (1-\mu_t)\kappa_t Y_t$ with stochastic carbon intensity $\kappa_t$; the period resource constraint is $C_t + K_{t+1} = Y_t + (1-\delta)K_t$.
\item \textbf{Households:} Each agent maximizes $\mathbb{E}_t\sum_{j=1}^{A}\beta^{j-1}\, C_{t+j-1,j}^{1-\sigma_u}/(1-\sigma_u)$ (where $\sigma_u$ is the household CRRA risk-aversion coefficient, distinct from the climate-chapter notation $\sigma_t$ for emissions intensity) subject to the budget constraint $C_{t,j} + a_{t+1,j+1} = (1+r_t)\,a_{t,j} + w_t\,l_j + \mathbb{T}_{t,j}$, where $\mathbb{T}_{t,j}$ is the transfer from carbon tax revenue and $j$ runs over all $A=12$ cohorts alive at $t$ (newborns included).
\item \textbf{Climate:} The climate emulator imposes a near-linear relationship between cumulative emissions and atmospheric temperature, $T^{\mathrm{AT}}_t \approx \sigma_{\mathrm{CCR}}\,E_t$ where $E_t = \sum_{s\le t} e_s$ is cumulative carbon, augmented by a stochastic tipping mechanism: damages depend on $T^{\mathrm{AT}}_t$ relative to a threshold $TP_t$ via a Weitzman-type retained-output factor that becomes steeply convex once $T^{\mathrm{AT}}_t$ approaches $TP_t$.
\item \textbf{Stochastic shocks:} Carbon intensity $\kappa_t$ follows an AR(1) with time-varying persistence; the tipping threshold $TP_t$ follows a bounded random walk that becomes absorbing once it has been crossed.
\end{itemize}
The household Euler equation takes the standard form $C_{t,j}^{-\sigma_u} = \beta\,\mathbb{E}_t[(1+r_{t+1})\,C_{t+1,j+1}^{-\sigma_u}]$ for $j = 1,\ldots,A-1$, and market clearing requires that aggregate savings equal the capital stock: $\sum_j a_{t,j} = K_t$.  Figure~\ref{fig:bau_olg_baseline} simulates this model without policy intervention; it fixes the business-as-usual (BAU) baseline against which every Pareto-improving policy below is benchmarked, and supplies the cohort-by-cohort participation constraints for the constrained policy search.

\subsection{The 3-Step ML Pipeline}

Finding an optimal carbon tax rule in this OLG economy is a bilevel optimization problem: the outer level searches over tax parameters, and the inner level solves the full stochastic general equilibrium for each candidate tax.  \citet{kubler2025using} decompose this into three steps, summarized in Figure~\ref{fig:olg_iam_pipeline}:

\begin{figure}[ht]
\centering
\begin{tikzpicture}[
    box/.style={rectangle, draw=uzhblue, fill=uzhblue!8, rounded corners=5pt,
                minimum width=2.8cm, minimum height=1.5cm, align=center, font=\small},
    arr/.style={-{Stealth[length=3mm]}, very thick, uzhblue!70}]
\node[box] (deqn) at (0,0) {\textbf{Step 1}\\DEQN with\\pseudo-states};
\node[box] (gp) at (4.5,0) {\textbf{Step 2}\\GP surrogate\\for welfare};
\node[box] (opt) at (9,0) {\textbf{Step 3}\\Constrained\\optimization};
\draw[arr] (deqn) -- (gp);
\draw[arr] (gp) -- (opt);
\end{tikzpicture}
\caption{Three-step machine-learning pipeline for constrained carbon-tax design.  The DEQN amortizes equilibrium solution across tax parameters, the GP surrogate maps policy parameters to welfare and cohort utilities, and the final optimization imposes the Pareto constraints on the surrogate.}
\label{fig:olg_iam_pipeline}
\end{figure}
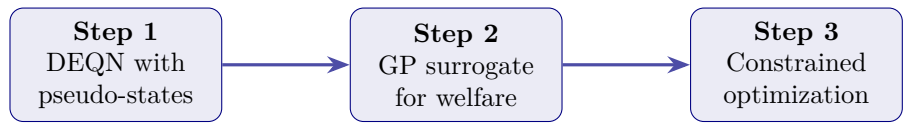

\paragraph{Step 1: DEQN with pseudo-states.}
The tax-rule coefficients $\vartheta_{\mathrm{tax}}$ and the $A=12$ transfer shares $\omega = (\omega_1,\ldots,\omega_{12})$ are appended to the state of the neural network as pseudo-states.  The transfer shares are non-negative weights satisfying $\sum_{j=1}^{A} \omega_j = 1$, with cohort $j$'s lump-sum transfer given by $\mathbb{T}_{t,j} = \omega_j\,p^{\mathrm{tax}}_t\,e_t$ from the government's resource constraint $\sum_j \mathbb{T}_{t,j} = p^{\mathrm{tax}}_t\,e_t$.  The simplex constraint $\omega \in \Delta^{A-1}$ is enforced by sampling unconstrained logits and applying a softmax before feeding $\omega$ into the network, so the DEQN never sees an infeasible transfer profile.  All cohorts alive at $t$, including the newborn cohort, receive a transfer.  The number of tax parameters depends on the rule: a simple linear rule on cumulative emissions has $\vartheta_{\mathrm{tax}} = (\vartheta_0,\vartheta_E) \in \mathbb{R}^2$ (so a 14-dimensional pseudo-state vector together with the 12 transfer shares), and a richer rule that adds dependence on carbon intensity and tipping has $\vartheta_{\mathrm{tax}} \in \mathbb{R}^4$ (a 16-dimensional pseudo-state vector with the 12 shares).  The DEQN learns the equilibrium for \emph{all} candidate tax-and-transfer configurations at once, so that simulating any $(\vartheta_{\mathrm{tax}}, \omega)$ requires only a forward pass.  The network architecture, optimizer schedule, and training-pool design follow \citet{kubler2025using} verbatim; the exact configuration is documented in the companion repository linked at the end of this section.

\begin{figure}[H]
\centering
\includegraphics[width=0.55\linewidth]{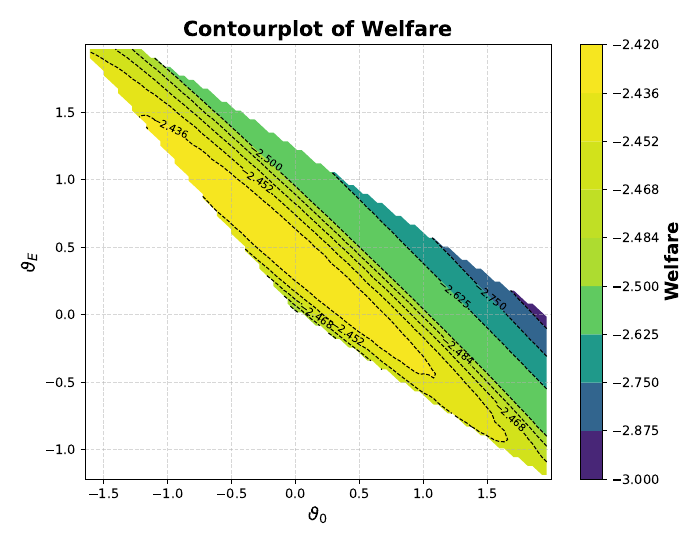}
\caption{Gaussian-process welfare surrogate over the two-dimensional tax-parameter slice $(\vartheta_0, \vartheta_E)$ of the linear-in-cumulative-emissions rule, with transfer shares $\omega$ held at the Pareto-optimal solution.  The contour exposes the low-dimensional welfare surface on which the constrained optimizer of Eq.~\eqref{eq:pareto_opt} searches once the DEQN has amortized the equilibrium solve.  Figure extracted from \citet{kubler2025using}.}
\label{fig:gp_welfare_contour}
\end{figure}

\paragraph{Step 2: GP surrogate.}
At each design point $\vartheta = (\vartheta_{\mathrm{tax}}, \omega)$, the trained DEQN is simulated to obtain Monte-Carlo estimates of expected lifetime utility for the 40 tracked cohorts (12 alive at $t=0$ plus 28 future cohorts born during the planner's 150-year horizon).  Independent GPs are then fitted to map $\vartheta$ to expected aggregate welfare $\mathcal{W}(\vartheta)$ and to each of the 40 cohort welfares $\tilde{U}_t(\vartheta)$.  The design itself uses Latin-hypercube sampling augmented with Bayesian active learning: the size scales with the dimension of $\vartheta$, with roughly 500 points sufficient for the 14-dimensional ``linear-in-$E$ + transfers'' specification (Section 5.3 of \citet{kubler2025using}) and roughly 800 points for the 16-dimensional ``richer rule + transfers'' specification (Section 5.4).  Figure~\ref{fig:gp_welfare_contour} shows the resulting welfare surface for the two-parameter linear-in-cumulative-emissions rule, with transfer shares held at the Pareto-optimal solution: the contour exposes the low-dimensional ridge along which intercept and slope trade off cleanly, and on which the Step-3 optimizer searches.

\paragraph{Step 3: Constrained optimization.}
The planner solves
\begin{equation}
\vartheta^* = \argmax_{\vartheta = (\vartheta_{\mathrm{tax}}, \omega)}\;\mathcal{W}(\vartheta) \qquad \text{s.t.}\quad \tilde{U}_t(\vartheta) \geq U_t \;\;\forall\, t,\;\; \omega \in \Delta^{A-1},
\label{eq:pareto_opt}
\end{equation}
where $U_t$ is the business-as-usual (BAU) welfare of cohort $t$ and $\Delta^{A-1}$ is the standard simplex on $A=12$ shares.  The Pareto constraint ensures that \emph{no generation is worse off}; whenever the welfare-maximizing $\vartheta^\ast$ lies strictly inside the feasible polytope (which is the case in every scenario reported below) it is also strictly Pareto-improving for at least one cohort, so the weak constraint $\tilde U_t \ge U_t$ and the textbook strict-improvement requirement coincide at the optimum.  Because each evaluation of $\mathcal{W}$ and $\tilde U_t$ is a forward pass through the trained GP rather than a fresh DEQN simulation, the constrained search reduces to a sequence of small SLSQP problems (the paper uses 500 random restarts of \texttt{scipy.optimize.minimize}) that complete in seconds.  By contrast, replacing the surrogate with brute-force re-solves of the full SOLG IAM at every candidate $\vartheta$ would require on the order of tens of thousands of core-hours per candidate, which is the comparison the paper draws against traditional methods.

\subsection{Results: Why Transfers Matter}

\begin{figure}[H]
\centering
\includegraphics[width=0.95\linewidth]{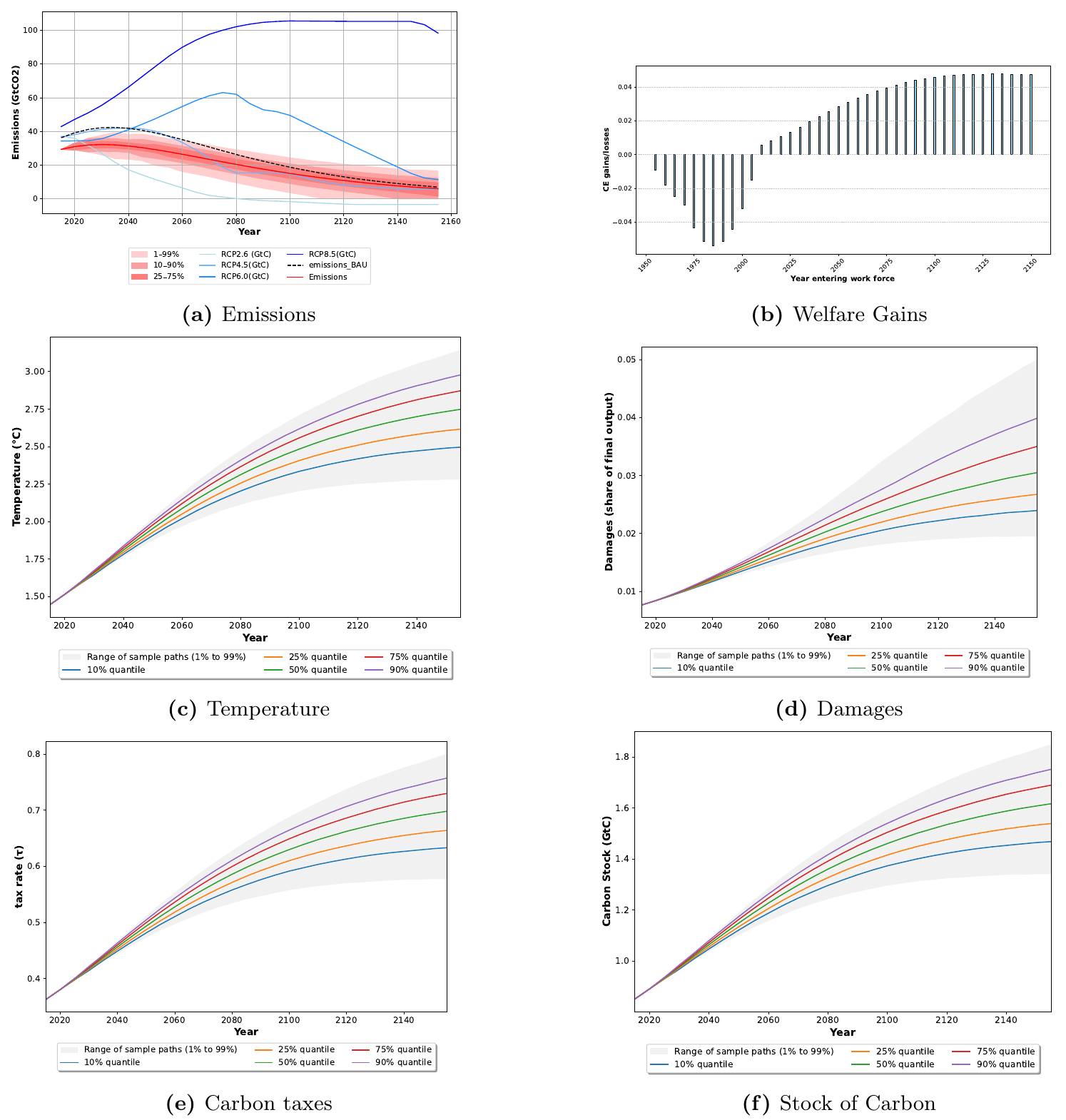}
\caption{Welfare-improving but not Pareto-improving cumulative-emissions tax with a fixed exogenous transfer scheme.  The policy strongly reduces climate risk and raises aggregate welfare by about $1.6\%$ in consumption-equivalent terms, but the welfare-gains panel shows losses for transition generations.  Figure extracted from \citet{kubler2025using}.}
\label{fig:unconstrained_linear_tax}
\end{figure}

The unconstrained welfare-maximizing cumulative-emissions tax is the natural benchmark.  With a linear rule $p^{\mathrm{tax}}_t = \vartheta_0 + \vartheta_E\,E_t$ and a fixed declining transfer scheme $\omega = \bar\omega$, the policy cuts emissions aggressively, stabilizes mean warming around $2.7\,^{\circ}\mathrm C$, and raises aggregate social welfare by about $1.6\%$ in consumption-equivalent terms.  But it imposes losses of up to roughly $5\%$ on initial generations: it is therefore welfare-improving in the social-welfare-function sense, but \emph{not} Pareto improving.  Figure~\ref{fig:unconstrained_linear_tax} shows the failure: the welfare-gains panel records the losses for transition generations that the social-welfare-function aggregate hides.

\begin{figure}[H]
\centering
\includegraphics[width=0.95\linewidth]{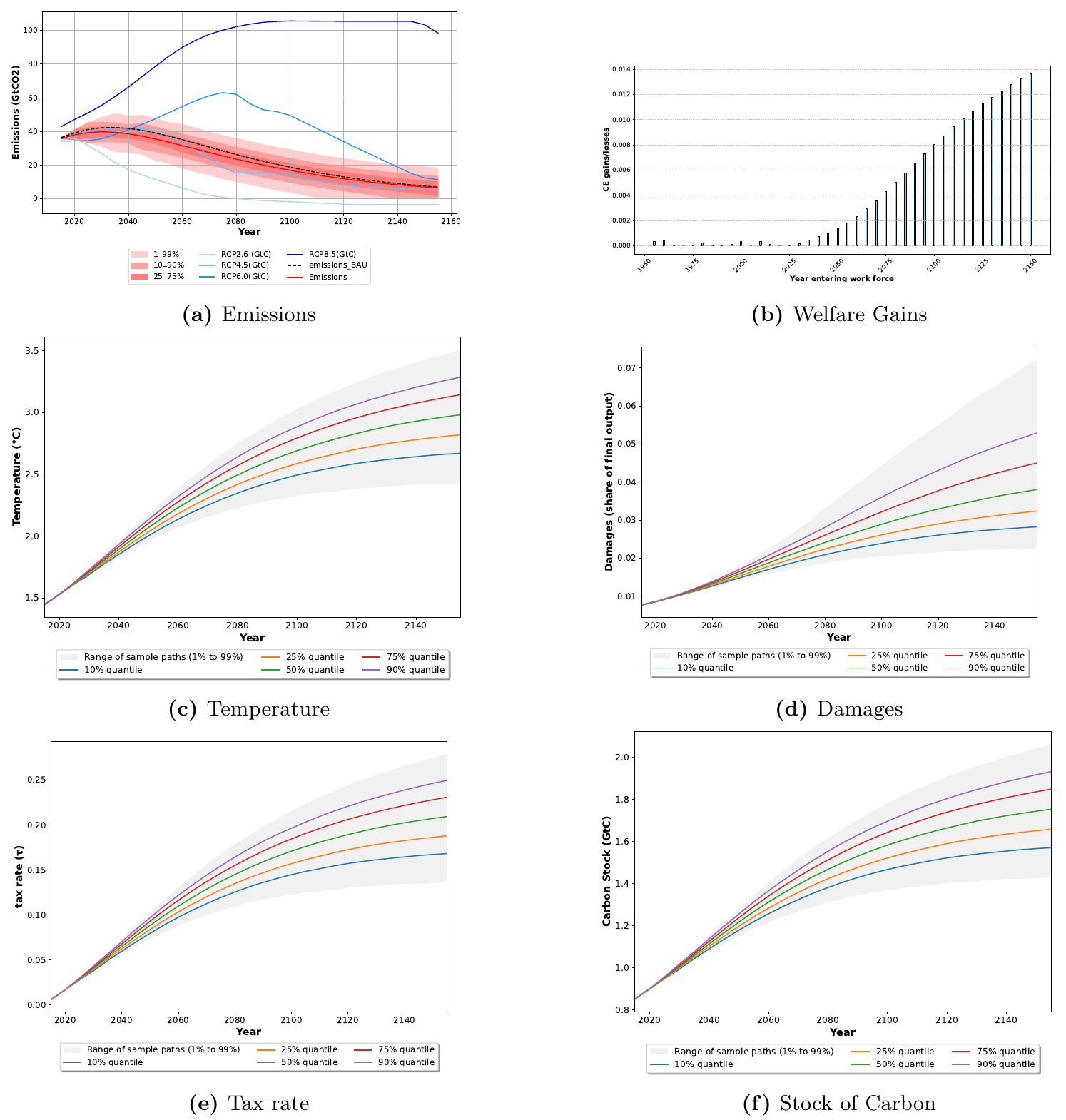}
\caption{Pareto-improving cumulative-emissions tax with optimized intergenerational transfers, at the coefficients of \eqref{eq:pareto_linear_tax_coefficients}--\eqref{eq:pareto_linear_transfer_shares}.  The tax is less aggressive than the unconstrained rule, but the optimized transfer system shields current cohorts while preserving climate-risk reduction for future cohorts.  Aggregate welfare rises by about $0.42\%$.  Figure extracted from \citet{kubler2025using}.}
\label{fig:pareto_tax_main}
\end{figure}

\begin{figure}[H]
\centering
\begin{tikzpicture}
\begin{axis}[
    width=0.85\linewidth, height=4.5cm,
    ybar, bar width=14pt,
    ymin=0, ymax=0.17,
    xtick=data,
    xticklabels={1,2,3,4,5,6,7,8,9,10,11,12},
    xlabel={Cohort index $j$ (1 = oldest alive at $t=0$, 12 = newborns)},
    ylabel={Transfer share $\omega_j$},
    grid=both, grid style={gray!20},
    enlarge x limits=0.06,
    nodes near coords, nodes near coords style={font=\tiny},
    every axis plot/.append style={fill=uzhblue!30, draw=uzhblue}
]
\addplot coordinates {
    (1,0.128) (2,0.051) (3,0.058) (4,0.089) (5,0.149) (6,0.090)
    (7,0.066) (8,0.143) (9,0.076) (10,0.048) (11,0.039) (12,0.061)
};
\end{axis}
\end{tikzpicture}
\caption{Optimized transfer-share profile $\omega_j$ across the 12 cohorts alive at $t = 0$, drawn directly from~\eqref{eq:pareto_linear_transfer_shares}.  The profile is decidedly non-monotone: the largest shares go to cohorts 1 (oldest), 5, and 8, which are precisely the cohorts the participation constraint $\tilde U_t \ge U_t$ binds most tightly for under the un-transferred tax of Figure~\ref{fig:unconstrained_linear_tax}.  The non-monotone shape is what allows a less aggressive cumulative-emissions tax to satisfy Pareto improvement at every age.}
\label{fig:pareto_transfer_profile}
\end{figure}
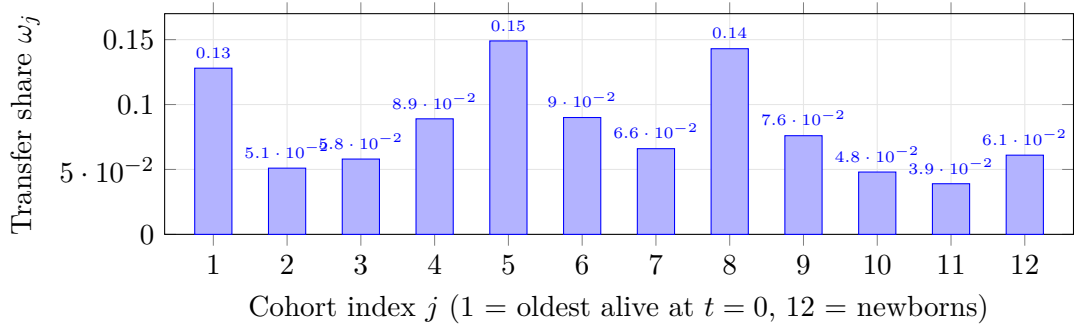

Endogenizing the transfer shares changes the conclusion.  With the same simple tax base and an optimized transfer simplex, \citet{kubler2025using} report the optimized coefficients\footnote{All numerical coefficients, welfare gains, and damage-quantile values in this subsection are quoted from \citet{kubler2025using}; consult the paper for the source tables and figures.}
\begin{equation}
    p^{\mathrm{tax}}_t = \vartheta_0 + \vartheta_E\,E_t,
    \qquad
    (\vartheta_0,\, \vartheta_E) = (-0.186,\, 0.225),
\label{eq:pareto_linear_tax_coefficients}
\end{equation}
together with transfer shares
\begin{equation}
\omega
=
(0.128,\, 0.051,\, 0.058,\, 0.089,\, 0.149,\, 0.090,\, 0.066,\, 0.143,\, 0.076,\, 0.048,\, 0.039,\, 0.061),
\label{eq:pareto_linear_transfer_shares}
\end{equation}
which sum to one up to rounding.  Figure~\ref{fig:pareto_transfer_profile} plots this transfer profile against cohort index; the non-monotone shape is what allows a less aggressive cumulative-emissions tax to satisfy the Pareto constraint at every age, and it is the single most informative graphical summary of the constrained-optimal-policy step.  The negative intercept $\vartheta_0 = -0.186$ is not a subsidy in practice: the planner's horizon starts well into the industrial era at a strictly positive cumulative-emissions stock $E_0 > 0$, so the effective tax $\vartheta_0 + \vartheta_E\,E_t$ is positive for every relevant $E_t$ along the optimum.  The negative intercept simply registers that the linear-in-$E$ rule undershoots a constant carbon price near $E = 0$ and ramps up roughly proportionally to cumulative emissions thereafter.  The combined policy makes every tracked cohort weakly better off than under BAU.  The aggregate welfare gain is more modest than under the unconstrained optimum, at about $0.42\%$ in consumption-equivalent terms, but the right tail of damages is truncated: the 99th percentile of damages falls to roughly $7\%$ of output rather than about $9\%$ under BAU.  Figure~\ref{fig:pareto_tax_main} reports the full result.  Comparing its welfare-gains panel with that of Figure~\ref{fig:unconstrained_linear_tax} is the section's headline: a lower, simpler tax combined with an optimized transfer system shifts every cohort weakly into the gains region.

The richer rule of \S\ref{sec:pareto_carbon_tax} adds carbon intensity and a tipping-state statistic,
\begin{equation}
    p^{\mathrm{tax}}_t = \vartheta_0 + \vartheta_E\,E_t + \vartheta_\kappa\,\kappa_t + \vartheta_{TP}(1-D_t),
\end{equation}
where $D_t$ is the climate-tipping state of the model (built from the proximity of $T^{\mathrm{AT}}_t$ to the stochastic threshold $TP_t$ and the absorbed-tipping flag).  Its optimized coefficients are
\begin{equation}
    (\vartheta_0,\, \vartheta_E,\, \vartheta_\kappa,\, \vartheta_{TP}) = (-0.237,\, 0.203,\, 0.037,\, 0.012),
\label{eq:pareto_full_tax_coefficients}
\end{equation}
with the associated aggregate welfare gain rising only from about $0.42\%$ to about $0.45\%$.  The cohort-by-cohort welfare profile (not plotted; see \citet{kubler2025using} for the figure) again keeps every cohort weakly above its BAU baseline, and the marginal welfare improvement from the extra two policy-state coefficients is small.  This is the substantive headline of \citet{kubler2025using}: once intergenerational transfers are optimized, the simple cumulative-emissions tax captures most of the feasible Pareto-improving welfare gain.  More policy-state variables improve the fit to climate risk, but the participation constraints bind tightly enough that the marginal welfare benefit of policy complexity is small.  $D_t$ is a deterministic function of variables already in the SOLG state, so it can be evaluated inside each forward pass; the exact functional form is in the paper.

\paragraph{Runtime in numbers.}
On a standard laptop (Apple M1), the OLG DEQN trains in roughly four wall-clock hours; on a high-end accelerator such as an NVIDIA GH200, training drops to the order of minutes \citep{kubler2025using}.  Adding the GP fits over 500 (resp.\ 800) design points and the constrained Step-3 optimization keeps the entire pipeline within the same order of magnitude, while the comparable brute-force re-solve of the SOLG model at \emph{every} candidate $\vartheta$ would dominate by orders of magnitude (the paper reports tens of thousands of core-hours for one fixed-parameter calibration, which would have to be repeated for every Step-3 candidate vector).

\paragraph{Companion code.}
The full production OLG-IAM solver, including the DEQN training loop with $(\vartheta_{\mathrm{tax}}, \omega)$ pseudo-states and the bilevel policy search, is hosted in the companion repository \href{https://github.com/sischei/JPE_Macro_Using_ML_to_compute_constrained_optimal_carbon_tax_rules}{\texttt{sischei/JPE\_Macro\_Using\_ML\_to\_compute\_constrained\_optimal\_carbon\_tax\_rules}}, which accompanies \citet{kubler2025using}.  The classroom notebook in Lecture~17 of this course exposes a reduced surrogate-only version that loads pre-trained GP surrogates and reproduces the constrained-optimization step (Step~3) interactively, but does not retrain the OLG DEQN end-to-end; readers who want the full pipeline should clone the companion repository.

\begin{keyinsightbox}[The policy-design message]
Deep learning is used here not because a neural network is fashionable, but because the relevant object is a high-dimensional map from states and policy-rule coefficients to equilibrium allocations and cohort welfare.  Once that map is learned, constrained optimal policy becomes a small optimization problem on a surrogate.  The economic result is equally important: carbon taxes need transfers.  Without transfers, the welfare-maximizing tax creates transition losers; with optimized transfers, a simpler and lower tax can be Pareto improving.  The main welfare channel is disaster-risk reduction (tail insurance), not maximal average abatement.
\end{keyinsightbox}

\section{Discussion and Outlook}
\label{sec:climate_discussion}

The combination of DEQNs, pseudo-states, and GP surrogates provides a scalable and transparent framework for climate economics that overcomes key limitations of traditional methods.

\paragraph{Comparison with traditional IAM solutions.}
Standard IAMs (such as the GAMS implementation of DICE) rely on shooting methods or nonlinear programming solvers that find deterministic optimal paths.  These approaches struggle with stochastic extensions: Monte Carlo integration over shocks is expensive, and certainty equivalence (replacing random variables with their means) misses the welfare cost of tail risks.  The DEQN approach approximates the stochastic recursive solution over the chosen training distribution and state/pseudo-state domain (with Bayesian learning and recursive Epstein--Zin utility) in a single training run.

\paragraph{Limitations.}
Several limitations should be noted.  First, the CDICE climate module, while calibrated to CMIP benchmarks, remains a reduced-form emulator and cannot capture spatial heterogeneity or regional climate impacts.  Second, the OLG-IAM treats each generation as identical within a cohort; within-cohort heterogeneity (e.g., geographic exposure to climate damages) would require further extensions along the lines of Chapter~\ref{ch:young}.  Third, the linear tax rules are interpretable and implementable but may leave welfare gains on the table relative to fully nonlinear rules.

\paragraph{Extensions.}
Active research frontiers include: multi-region IAMs with trade and carbon leakage \citep{nordhaus1996regional}; richer damage specifications including tipping cascades; endogenous technical change in abatement technology; and embedding climate modules in continuous-time heterogeneous-agent models (Chapter~\ref{ch:ct_theory}) to study the joint dynamics of climate risk and wealth inequality.  The methodological toolkit developed in this course (DEQNs for equilibrium computation, PINNs for continuous-time PDEs, deep surrogates for uncertainty quantification, and Young's method for distribution tracking) provides the computational infrastructure for these extensions.

\paragraph{The three movements, in one synthesis.}
Movement~1 established that solving an IAM by DEQN requires three modifications relative to the stationary toolkit of Chapter~\ref{ch:deqn}: time enters as a state, the training pool is built by simulating $K$ forward trajectories from a calibrated initial state rather than by sampling an ergodic distribution, and the missing transversality is absorbed numerically by choosing the horizon $T_{\max}$ long enough that discounting suffices (or, on short horizons, by adding an explicit terminal residual).  Movement~2 put that algorithm to work on a worked stochastic DICE economy, producing the eight-residual loss whose minimization delivers the deterministic policy and, with one extra Gauss--Hermite layer, the AR(1) SCC fan chart.  Movement~3 layered four extensions onto the same spine: Bayesian learning over the climate sensitivity, recursive Epstein--Zin preferences, global UQ of the SCC via pseudo-states and GP surrogates, and constrained Pareto-improving carbon-tax design in a heterogeneous-agent OLG-IAM.  Chapter~\ref{ch:outlook} threads these into the broader synthesis with the rest of the course.

\begin{keyinsightbox}[Chapter Summary]
\begin{itemize}[itemsep=2pt, leftmargin=*]
\item Climate-economy IAMs are the natural showcase for the methodological stack: a moderately high-dimensional non-stationary DSGE solved by DEQN, a GP+BAL surrogate for SCC sensitivity, and Sobol/Shapley decomposition for deep uncertainty quantification.
\item DICE (and the CDICE recalibration) provide the workhorse model; Exercise~\ref{ex:ch11:3} asks the reader to use the closed-form ACE expression of \citet{traeger2018ace} as an analytic benchmark for the DEQN-trained CDICE solution.
\item Pareto-improving carbon-tax rules \citep{kubler2025using} demonstrate that the surrogate machinery has direct policy relevance: linear, intergenerationally fair, implementable rules can be designed via constrained optimization on the surrogate.
\item Deep UQ matters because the SCC distribution under climate, damage, and preference uncertainty is wider than any pointwise estimate suggests; reporting the distribution rather than a number is the responsible default.
\end{itemize}
\end{keyinsightbox}

\section*{Further Reading}
\addcontentsline{toc}{section}{Further Reading}
\begin{itemize}[itemsep=2pt]
\item \citet{nordhausRevisitingSocialCost2017}, the canonical DICE update.
\item \citet{Folini_2021}, the CDICE recalibration used in the deep-learning solution.
\item \citet{traeger2018ace}, the analytic ACE benchmark.
\item \citet{DIETZ20241, fernandezvillaverde2025climate, vanderploegrezai2026climate}, recent surveys on climate macroeconomics.
\item \citet{kubler2025using}, Pareto-improving carbon-tax design.
\item \citet{friedlDeep2023}, deep uncertainty quantification methodology.
\end{itemize}

\section*{Exercises}
\addcontentsline{toc}{section}{Exercises}
\noindent Worked solutions and guidance for these exercises appear in Appendix~\ref{app:solutions}.
\begin{enumerate}[itemsep=4pt, leftmargin=*]
\item\label{ex:ch11:1} \textbf{[Computational] ECS sensitivity.}  Holding all other DICE parameters at calibration, vary the equilibrium climate sensitivity over the \citet{sherwood2020assessment} likely range (2.6--3.9\textdegree C) and report the implied SCC at $t=0$.  Then repeat over the broader IPCC AR6 very-likely range (2--5\textdegree C).  How much of the SCC variation is driven by the high-ECS tail?  State your classical baseline (e.g., a Nordhaus-2017 grid-search SCC at the central ECS calibration) before reaching for the DEQN.
\item\label{ex:ch11:2} \textbf{[Core] Sobol decomposition.}  For a function $q(\theta_1, \theta_2, \theta_3) = \theta_1\theta_2 + \theta_3^2$ with i.i.d.\ uniform inputs $\theta_i \sim \mathcal{U}[0,1]$, compute the Sobol first- and total-effect indices analytically.  Verify with a $10\,000$-sample SALib estimate.
\item\label{ex:ch11:3} \textbf{[Computational] ACE benchmark.}  Using the closed-form ACE expression for the optimal carbon tax, compute the SCC at calibration and compare with the DEQN-trained DICE solution.  Quantify the discrepancy in \% terms.  State your classical baseline (the ACE closed form itself, evaluated at the chapter's calibration) before reaching for the DEQN.
\item\label{ex:ch11:4} \textbf{[Advanced/project] Pareto-improving tax design.}  Building on the OLG-IAM of Section~\ref{sec:olg_iam}, search over linear tax-and-transfer rules $\vartheta = (\vartheta_{\mathrm{tax}}, \omega)$ with $\vartheta_{\mathrm{tax}} = (\vartheta_0, \vartheta_E)$ on cumulative emissions and transfer shares $\omega \in \Delta^{A-1}$, $A=12$, such that $\tilde U_t(\vartheta) \ge U_t$ for all cohorts.  How tight is the Pareto frontier?  Then repeat with $\omega$ held at the BAU/declining benchmark $\bar\omega$ and observe how much welfare is left on the table when transfers are not endogenous.
\item\label{ex:ch11:5} \textbf{[Computational] Carbon-cycle warm-up.}  Run notebook \tpath{01_Climate_Exercise.ipynb} end-to-end.  (a) Report the avoided warming and avoided damages at year 2100 under a 50\% mitigation rule relative to BAU.  (b) Inspect the carbon-cycle transition matrix and identify which reservoir (atmospheric, upper ocean, lower ocean) has the longest timescale.  (c) Explain in two sentences why a quadratic damage-loss function $D(T) = \pi_2 T^2$ is insufficient for tail-risk assessment.
\item\label{ex:ch11:6} \textbf{[Advanced/project] Deterministic CDICE-DEQN reproduction.}  Open notebook \tpath{02_DICE_DEQN_Library_Port.ipynb} and train the deterministic solver to convergence.  Verify the following quantities against the reference solution, using the tolerances stated in the notebook's verification gate: $T^{\mathrm{AT}}(2100)$, $M^{\mathrm{AT}}(2100)$, $\mu(2100)$, and the SCC at 2015, 2100, and 2300.
\item\label{ex:ch11:7} \textbf{[Advanced/project] Stochastic SCC fan chart.}  In notebook \tpath{03_Stochastic_DICE_DEQN.ipynb}, vary the AR(1) productivity volatility $\sigma_z$ over $\{0.005, 0.01, 0.02\}$ and plot the resulting SCC fan charts.  How does the right-tail of the SCC distribution at 2100 scale with $\sigma_z$?  Compare your finding with the qualitative message of \citet{caiSocialCostCarbon2019}.  State your classical baseline (e.g., a deterministic-DICE SCC at the same calibration, or a certainty-equivalent SCC) before reaching for the DEQN.
\item\label{ex:ch11:8} \textbf{[Advanced/project] Tipping-point regime-switching damages.}  This exercise temporarily switches from the additive damage-fraction convention $\Omega^N(T) = \pi_1 T + \pi_2 T^2$ used in the chapter body and in notebook 02 (Eq.~\ref{eq:damage_nordhaus}) to the multiplicative \emph{retained-output} convention $\Omega^{\mathrm{ret}}(T) = 1/(1 + \pi_2 T^2)$, the form discussed in \S\ref{sec:dice_damages} as an alternative, so that the regime-switching modification below has a single multiplicative knob.  First make this substitution in notebook \tpath{02_DICE_DEQN_Library_Port.ipynb} (which ships with the additive form $\Omega^N$) and re-calibrate $\pi_2$ so that the retained-output form matches the additive baseline at $T = 2.5\,^\circ\mathrm{C}$, i.e.\ choose $\pi_2$ such that $1 - 1/(1+\pi_2 T^2) = \pi_2^{N} T^2$ at $T = 2.5\,^\circ\mathrm{C}$ with $\pi_2^N = 0.00236$ from Table~\ref{tab:dice_calibration}.  Then replace the smooth retained-output factor $\Omega^{\mathrm{ret}}(T) = 1/(1 + \pi_2 T^2)$ with a regime-switching specification.  At each step, with hazard rate $\lambda_\mathrm{TP}(T) = \lambda_0 + \lambda_1\max(0, T - T_\mathrm{thresh})$, an irreversible tipping event occurs.  If the event has occurred, multiply the damage term in the denominator by $D_\mathrm{TP}=1.5$, so retained output becomes $\Omega^{\mathrm{TP}}(T)=1/(1+D_\mathrm{TP}\pi_2T^2)$.  Calibrate $\lambda_0 = 0.001$, $\lambda_1 = 0.05$, $T_\mathrm{thresh} = 2.0\,^\circ\mathrm{C}$.  Retrain the DEQN solver and report (i)~SCC at $t = 0$ under the regime-switching specification vs.\ the smooth baseline; (ii)~the time path of optimal abatement $\mu_t$ in both cases; (iii)~the unconditional probability of a tipping event by 2100.  Sweep $T_\mathrm{thresh}$ over $\{1.5, 2.0, 2.5\}\,^\circ\mathrm{C}$ and plot the SCC against the threshold.  Discuss the policy implications: a lower threshold raises the SCC by what factor, and what does this imply for near-term tax design under deep uncertainty about $T_\mathrm{thresh}$ itself?
\item\label{ex:ch11:9} \textbf{[Advanced/project] Real options value of waiting.}  Consider a stylized two-period climate--policy decision.  At $t = 0$ the planner does not know the equilibrium climate sensitivity $\mathrm{ECS}$, with prior $\mathrm{ECS} \sim \mathcal{U}([\mathrm{ECS}_L, \mathrm{ECS}_H])$ where $\mathrm{ECS}_L = 2$, $\mathrm{ECS}_H = 5$\,$^\circ\mathrm{C}$.  Two choices: (a)~\textbf{act now}, abate at level $\mu \in [0,1]$ with cost $\Theta(\mu) = \theta\mu^2$; (b)~\textbf{wait}, abate at $t=1$ after observing a noisy signal $\widehat{\mathrm{ECS}} = \mathrm{ECS} + \varepsilon$, $\varepsilon \sim \mathcal{N}(0, \sigma_\varepsilon^2)$.  Damages are $D(\mu, \mathrm{ECS}) = \alpha\,\mathrm{ECS}\cdot(1 - \mu)$.  The planner minimizes expected total cost.  (i)~Derive closed-form expressions for the optimal $\mu^\star$ and the expected total cost under each strategy.  (ii)~Define the \emph{value of waiting} as the cost difference: $\mathrm{VoW} = \mathbb{E}[\mathrm{cost}_\mathrm{wait}] - \mathbb{E}[\mathrm{cost}_\mathrm{now}]$.  Show that as the signal becomes informative ($\sigma_\varepsilon \to 0$), $\mathrm{VoW}$ becomes negative (waiting is preferred): more information allows better decisions.  (iii)~Now add an irreversibility wedge $\eta\,\mu_1^2$ paid only on the wait branch, penalizing deferred action (e.g., capital stock accumulates carbon faster while waiting, so the wait-period abatement is more costly than an equivalent abatement at $t = 0$).  Show that for sufficiently large $\eta$, $\mathrm{VoW}$ is positive (act now is preferred), even with substantial learning.  Connect this trade-off to the Bayesian-learning section of the chapter and to the broader literature on real options in climate policy.
\item\label{ex:ch11:10} \textbf{[Core] Non-stationary DEQN on a 1D toy.}  Implement the three modifications of \S\ref{sec:nsdeqn_algo} on a one-dimensional non-stationary problem.  The toy economy: a planner picks $u_t \in \mathbb R$ to minimise $\sum_{t=0}^{T_{\max}-1} \bigl[(x_t - x^\ast_t)^2 + r u_t^2\bigr] + \lambda_T (x_{T_{\max}} - x^\ast_{T_{\max}})^2$ subject to the linear-Gaussian law of motion $x_{t+1} = \alpha\,x_t + u_t + g_t + \sigma\,\varepsilon_{t+1}$, $\varepsilon_{t+1}\sim\mathcal N(0,1)$, with deterministic drift $g_t = g_0 + g_1 t$ and a calendar-time target path $x^\ast_t = a + b t$.  Set $T_{\max} = 50$, $\alpha = 0.95$, $g_0 = 0.02$, $g_1 = 0.001$, $r = 0.1$, $\sigma = 0.05$, $\lambda_T = 5$, $a = 0$, $b = 0.04$.  (i)~Derive the closed-form LQ-Riccati policy and state your classical baseline before reaching for the DEQN.  (ii)~Train a small neural network $u_t = \mathcal N_\rho(x_t, \tau_t)$ with $\tau_t = 1 - \exp(-\vartheta\,t)$, sampling initial states from a uniform prior on $[a - 1, a + 1]$, stratifying the mini-batch over ten calendar-time bins of $[0, T_{\max}]$, and adding the terminal penalty $\lambda_T (x_{T_{\max}} - x^\ast_{T_{\max}})^2$ to the loss.  (iii)~Verify that the trained network reproduces the LQ-Riccati baseline within 1\% in the mean-squared-action metric.  (iv)~Ablate each of the three modifications in turn (drop $\tau_t$ from the input; remove stratification; remove the terminal penalty) and report which ablation hurts most as a function of $T_{\max}$.
\end{enumerate}

\chapter{Synthesis and Outlook}
\label{ch:outlook}

This final chapter steps back from the individual methods and applications to identify the unifying themes that connect them.  We distill the common computational paradigm, embedding economic structure in a neural network loss function and optimizing via gradient descent, and provide a practical decision guide for choosing among DEQNs, PINNs, deep surrogates, and GP + BAL based on the problem at hand.  We then discuss open challenges at the frontier and offer practical tips distilled from the experience of applying these methods to real research problems.
The chapter is intentionally high-level: it emphasizes synthesis and practical judgment rather than introducing new technical machinery.

\section{The Unifying Paradigm}

The methods presented in this course, namely DEQNs, PINNs, deep surrogates, and GP + BAL, do not all use the same estimator, but they share a common computational workflow: a flexible function approximator, a deep network or a Gaussian-process kernel, encodes the economic object of interest; structural information enters through residuals, hard constraints, model-generated training data, or an acquisition rule; and a differentiable or Bayesian inference machinery delivers training, sensitivity analysis, and uncertainty quantification.

\begin{figure}[ht]
\centering
\begin{tikzpicture}[
    box/.style={rectangle, draw=uzhblue, fill=uzhblue!8, rounded corners=5pt,
                minimum width=2.8cm, minimum height=1.5cm, align=center, font=\small},
    arr/.style={-{Stealth[length=3mm]}, very thick, uzhblue!70}
]
    \node[box] (nn) at (0,0) {\textbf{Flexible function}\\approximator\\(network or kernel)};
    \node[box] (phys) at (5,0) {\textbf{Economic structure}\\in objective\\or design};
    \node[box] (ad) at (10,0) {\textbf{Differentiable or}\\Bayesian\\inference};
    \draw[arr] (nn) -- (phys);
    \draw[arr] (phys) -- (ad);
\end{tikzpicture}
\caption{The shared computational workflow of the four methods of this course (DEQNs, PINNs, deep surrogates, and GP + BAL): a flexible function approximator (a deep network or a Gaussian-process kernel) plays the role of the unknown economic object; structural information enters through equilibrium residuals and hard constraints (DEQNs, PINNs), through model-generated training data and parameter pseudo-states (deep surrogates), or through the prior, the simulator, and the acquisition rule (GP + BAL); and a differentiable or Bayesian inference layer supplies the gradients, the posterior, and the design rule needed for training, sensitivity analysis, and uncertainty quantification.}
\label{fig:unifying_paradigm}
\end{figure}
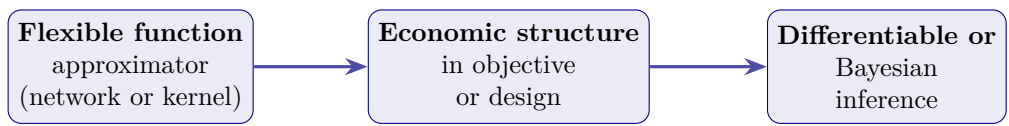

Figure~\ref{fig:unifying_paradigm} makes the common template explicit.  Structural knowledge of the economic or financial model is built directly into the training objective or the training design rather than learned from labelled data: DEQNs minimize equilibrium-residual losses, PINNs minimize PDE residuals, deep surrogates fit model-generated labels (often with parameter pseudo-states), and GP + BAL combines a kernel prior with an acquisition rule that decides where to sample next.  Automatic differentiation supplies the gradients needed for network training, and the mesh-free nature of neural networks and kernels avoids explicit tensor-product state grids, mitigating the practical curse of dimensionality without eliminating the underlying high-dimensional sample-complexity, approximation, optimization, and integration costs.

\section{Decision Guide: When to Use Which Method}

The four method families covered in this course occupy partly overlapping but distinguishable niches.  Table~\ref{tab:method_decision_guide} summarizes the key distinctions along four practical criteria: the natural \emph{time domain} (discrete vs.\ continuous), the kind of \emph{equation} the method is designed to solve, the \emph{key advantage} that distinguishes it from the others, and the \emph{typical use cases} encountered in this course.

\begin{table}[H]
\centering
\small
\renewcommand{\arraystretch}{1.2}
\setlength{\tabcolsep}{4pt}
\begin{tabular}{@{} >{\bfseries}p{2.6cm}
                     >{\raggedright\arraybackslash}p{2.9cm}
                     >{\raggedright\arraybackslash}p{3.3cm}
                     >{\raggedright\arraybackslash}p{3.0cm}
                     >{\raggedright\arraybackslash}p{3.0cm} @{}}
\toprule
\textbf{Criterion} & \textbf{DEQN} & \textbf{PINN} & \textbf{Deep surrogate} & \textbf{GP / BAL} \\
\midrule
Time domain   & Discrete & Continuous & Either & Either \\
Equation type & Algebraic / expectations & PDEs (HJB, KFE, BS) & Black-box model & Black-box model \\
Key advantage & Up to $d>100$ in selected models \citep[Tbl.~2]{azinovicDEEPEQUILIBRIUMNETS2022} & AD derivatives of the network approximation & Instant re-evaluation & Built-in UQ + active learning \\
Typical use   & DSGE, OLG, IRBC & HJB, option pricing & Estimation, UQ, policy & Small-$N$ surrogates, BAL design \\
\bottomrule
\end{tabular}
\caption{Decision guide: when to use which method.}
\label{tab:method_decision_guide}
\end{table}

\noindent A few practical takeaways from the table.  \textbf{DEQNs} are the workhorse for \emph{discrete-time} general-equilibrium models with high-dimensional state spaces, where Euler equations and market-clearing residuals admit a clean residual-loss formulation.  \textbf{PINNs} are the natural analogue in \emph{continuous time}, replacing finite-difference HJB or BSDE solvers when one needs exact derivatives of the trained network by automatic differentiation rather than finite-difference derivatives of an interpolant.  \textbf{Deep surrogates} differ in kind: they do not solve the model; they learn a fast emulator of the already-solved mapping from parameters to economic quantities, which is what makes structural estimation, sensitivity analysis, and policy design tractable at scale.  \textbf{Gaussian processes with Bayesian active learning} occupy the small-data, high-uncertainty corner: the right tool when each evaluation is expensive, the input dimension is moderate ($d \lesssim 10$ for na\"ive GPs, extending to higher dimensions via active subspaces and deep AS; cf.\ Chapter~\ref{ch:gp}), and posterior uncertainty is itself part of the answer.  Within the PINN family, EMINNs (Chapter~\ref{ch:ct_theory}) are a specialization to continuous-time heterogeneous-agent master equations: they keep the PDE-residual training logic but encode the cross-sectional distribution itself as part of the network input, which is one of several routes to global solutions of Krusell--Smith-style models with aggregate shocks, complementary to the set-encoder and price-as-state approaches discussed in \S\ref{sec:open_challenges}.

The categories are not mutually exclusive.  Many production pipelines combine them, e.g., a DEQN to solve the equilibrium and a deep surrogate or GP that treats parameters as pseudo-states for estimation (Chapter~\ref{ch:estimation}), or a PINN with a surrogate wrapper for policy counterfactuals.  The right question is rarely ``which of the four?'' but ``which combination?''.

\section{Running Benchmarks Through Every Lens}
\label{sec:running_case_bm}

The course repeatedly returns to a small set of canonical benchmarks: Brock--Mirman for discrete-time stochastic growth, cake-eating for continuous-time HJBs, and parameterized Brock--Mirman variants for surrogates and SMM.  Table~\ref{tab:bm_running_case} reads each benchmark through the lens of one course method, exposing what changes when the methodology changes and what stays invariant.

\begin{table}[H]
\centering
\scriptsize
\setlength{\tabcolsep}{4pt}
\renewcommand{\arraystretch}{1.25}
\begin{tabular}{@{} >{\bfseries}p{1.9cm}
                     >{\raggedright\arraybackslash}p{2.7cm}
                     >{\raggedright\arraybackslash}p{3.6cm}
                     >{\raggedright\arraybackslash}p{2.6cm}
                     >{\raggedright\arraybackslash}p{3.0cm} @{}}
\toprule
\textbf{Method} & \textbf{What is approximated} & \textbf{Loss / objective} & \textbf{Diagnostic} & \textbf{Notebook} \\
\midrule
DEQN                       & Savings share $s(K,z)\in(0,1)$ via sigmoid head    & Squared rel.\ Euler residual on simulated trajectory   & $s \to \alpha\beta$ at $\delta\!=\!1$, log utility & \tpath{01_Brock_Mirman_1972_DEQN.ipynb}  \\
PINN, cake-eating          & Value function $\hat V(a)$ on a 1D interval        & HJB residual $\rho V-\max_c[u(c)+V'(a)(ra-c)]$         & $\hat V$ matches closed form $V^\star$ (CRRA)     & \tpath{04_Cake_Eating_HJB_PINN.ipynb}       \\
GP $+$ BAL                 & $\hat V(K)$ on Brock--Mirman ergodic set            & GP marginal likelihood; BAL picks next $K_*$           & 95\% pred.\ band covers true $V$                  & \tpath{04_GP_Value_Function_Iteration.ipynb}             \\
SMM $+$ surrogate          & Policy $s(K,z,\varrho)$ pseudo-state on $\varrho$   & SMM moment objective in $\varrho$                       & $\hat\varrho$ within MCSE of truth                & \tpath{lecture_15_03_Structural_Estimation_BM.ipynb} \\
\bottomrule
\end{tabular}
\caption[Canonical benchmarks treated through each course method.]{Canonical benchmarks treated through each course method.  The economic content differs by row, three rows are Brock--Mirman and the PINN row is cake-eating, but the table exposes the invariant computational pattern: choose an object to approximate, encode the model restrictions in the objective or training design, and validate with an economically interpretable diagnostic.  Notebook filenames in the last column omit the standard \texttt{lecture\_NN\_} prefix; the four notebooks live, in row order, in the \texttt{code/} subdirectories of lectures~03, 11, 14, and~15.}
\label{tab:bm_running_case}
\end{table}

The takeaway is that no single methodology is the ``right'' one for these benchmarks.  Rather, the four lenses answer different research questions: a DEQN gives the policy function on Brock--Mirman, a PINN gives a value function on cake-eating with AD derivatives of the network approximation, a GP gives uncertainty-quantified value-function estimates on Brock--Mirman with guidance for the next sample point, and the surrogate-SMM pipeline estimates the deep parameter $\varrho$ (productivity persistence) on a Brock--Mirman variant from data; the joint $(\beta,\varrho)$ extension lives in the companion \tpath{lecture_15_03b_Structural_Estimation_BM_Joint.ipynb}, where the imperfect identification of $\beta$ from macro moments shows up as a visible ridge in the criterion surface.  Real research applications typically combine at least two of these, and the implementation guidance throughout the chapters is designed to make those combinations cheap.

\section{Bridges Between Methods}

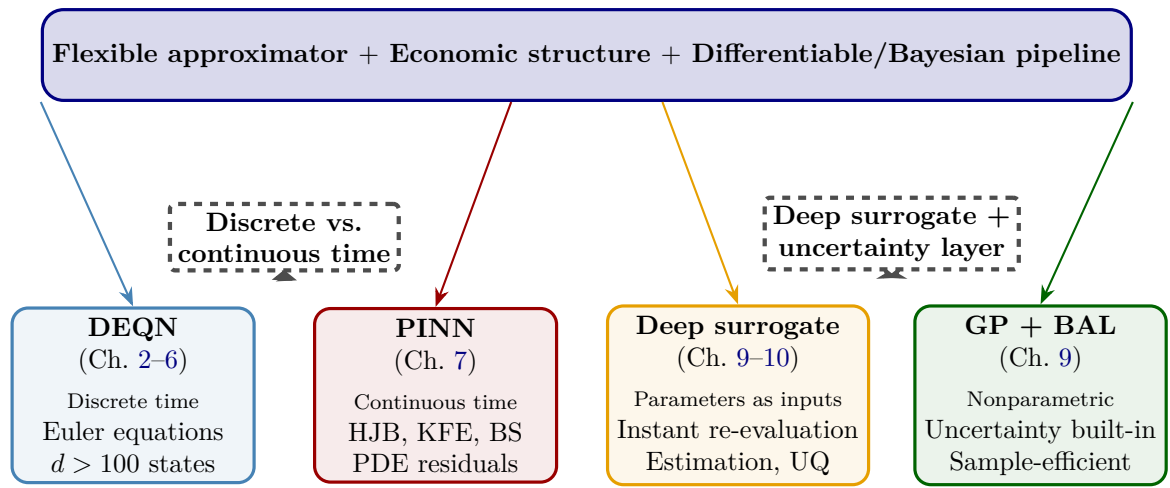
\begin{figure}[H]
\centering
\begin{tikzpicture}[
    method/.style={rectangle, draw=#1, very thick, fill=#1!8, rounded corners=6pt,
                   minimum width=3.2cm, minimum height=1.8cm, align=center, font=\small},
    core/.style={rectangle, draw=uzhblue, very thick, fill=uzhblue!15, rounded corners=8pt,
                 minimum width=4cm, minimum height=1.2cm, align=center, font=\small\bfseries},
    arr/.style={-{Stealth[length=2.5mm]}, thick, #1},
    linklabel/.style={fill=white, draw=black!70, rounded corners=2pt,
                      inner sep=3pt, align=center, font=\small\bfseries, text=black}
]
    \node[core] (core) at (6,4.5) {Flexible approximator $+$ Economic structure $+$ Differentiable/Bayesian pipeline};

    \node[method=softblue] (deqn) at (0,0) {\textbf{DEQN}\\(Ch.~\ref{ch:deqn}--\ref{ch:young})\\[3pt]\scriptsize Discrete time\\Euler equations\\$d > 100$ states};
    \node[method=harvardcrimson] (pinn) at (4,0) {\textbf{PINN}\\(Ch.~\ref{ch:pinn})\\[3pt]\scriptsize Continuous time\\HJB, KFE, BS\\PDE residuals};
\node[method=softorange] (surr) at (8,0) {\textbf{Deep surrogate}\\(Ch.~\ref{ch:gp}--\ref{ch:estimation})\\[3pt]\scriptsize Parameters as inputs\\Instant re-evaluation\\Estimation, UQ};
    \node[method=darkgreen] (gp) at (12,0) {\textbf{GP + BAL}\\(Ch.~\ref{ch:gp})\\[3pt]\scriptsize Nonparametric\\Uncertainty built-in\\Sample-efficient};

    \draw[arr=softblue] (core.south west) -- (deqn.north);
    \draw[arr=harvardcrimson] (core.south) ++(-1,0) -- (pinn.north);
    \draw[arr=softorange] (core.south) ++(1,0) -- (surr.north);
    \draw[arr=darkgreen] (core.south east) -- (gp.north);

    \draw[{Stealth[length=2.6mm]}-{Stealth[length=2.6mm]}, dashed, black!70, ultra thick]
        ([xshift=6pt,yshift=10pt]deqn.north east)
        to[out=18,in=162]
        node[above=2pt, linklabel] {Discrete vs.\\continuous time}
        ([xshift=-6pt,yshift=10pt]pinn.north west);
    \draw[{Stealth[length=2.6mm]}-{Stealth[length=2.6mm]}, dashed, black!70, ultra thick]
        ([xshift=6pt,yshift=10pt]surr.north east)
        to[out=18,in=162]
        node[above=2pt, linklabel] {Deep surrogate +\\uncertainty layer}
        ([xshift=-6pt,yshift=10pt]gp.north west);
\end{tikzpicture}
\caption{Bridges between the four method families.  The core box restates the shared workflow: a flexible approximator (network or kernel), with economic structure encoded in the objective or training design, trained or updated through a differentiable or Bayesian pipeline.  The dashed grey arrows mark the two main bridges, between discrete- and continuous-time residual solvers (DEQN $\leftrightarrow$ PINN) and between deep surrogates and the GP + BAL uncertainty layer that sits on top of them.}
\label{fig:unified_view_methods}
\end{figure}

The four method families do not sit in isolated boxes.  Figure~\ref{fig:unified_view_methods} maps the bridges between them: which methods share a state representation, which methods share an inference machinery, and where one method layers naturally on top of another.  Discrete- and continuous-time formulations connect DEQNs and PINNs; deep surrogates and GP + BAL share the pseudo-state and uncertainty-quantification idea; and the lower row of the figure can be read as a pipeline (solve $\to$ amortize $\to$ quantify) rather than as four parallel choices.

Reading the figure column by column gives a one-line gloss of each family: DEQNs (blue) handle discrete-time Euler equations in high-dimensional state spaces; PINNs (red) handle continuous-time HJB, Kolmogorov-forward, and Black--Scholes PDEs through PDE residuals; deep surrogates (orange) make model parameters part of the input space and replace the expensive solve with an instant forward pass for estimation and uncertainty quantification; GP + BAL (green) layers calibrated uncertainty and sample-efficient active learning on top of the surrogate.  The bridges in the figure are not exhaustive; production pipelines often chain three families at once, for example a DEQN solver followed by a deep surrogate, followed by a GP for sensitivity analysis around the estimated optimum.

\section{Open Challenges and Future Directions}
\label{sec:open_challenges}

Despite the considerable progress surveyed in this course, several open challenges remain at the frontier of deep learning for economics:

\begin{enumerate}[itemsep=4pt]
\item \textbf{Convergence guarantees.}  While empirical performance is strong, theoretical convergence guarantees for DEQNs and PINNs in the context of economic equilibrium computation remain limited.  Establishing rates of convergence as a function of network width, depth, and training budget is an active area of research.  The gap between theory and practice remains wide.

\item \textbf{Combining global and local methods.}  Deep Kernel Learning \citep{wilson2016deep, chen2025private}, which uses a DNN as a feature map inside a GP kernel, is one principled step toward integrating the two paradigms.  In principle, such hybrids could retain some of the GP's calibrated uncertainty estimates while benefiting from the DNN's representational flexibility.  In practice, however, the UQ properties of deep kernel models require careful validation, their scalability advantages over pure GPs are problem-dependent, and locating the kinks and boundary layers where local refinement would help most is itself a hard unsolved problem.  Hybrid methods therefore remain a promising but largely open research direction rather than a deployable solution.

\item \textbf{Real-time policy analysis.}  Deep surrogates enable near-instant policy evaluation, potentially transforming how policy makers conduct scenario analysis and stress testing.  \citet{kubler2025using} demonstrate that constrained optimal policy rules can be computed over high-dimensional parameter spaces using deep surrogates, opening the door to real-time policy counterfactuals.  For central banks and finance ministries, this could mean running complex DSGE scenarios with substantial speedups (seconds rather than minutes-to-hours, depending on the underlying solver), enabling more responsive and comprehensive policy analysis during fast-moving economic crises.

\item \textbf{Estimation of rich structural models.}  Combining DEQNs or PINNs with surrogate-based estimation pipelines allows researchers to estimate models with many parameters from micro and macro data simultaneously \citep{friedlDeep2023, chen2026Deep}.  \citet{kase2022estimating} use neural networks to estimate nonlinear heterogeneous-agent models, opening the door to estimation of models that were previously intractable.  As micro-level administrative datasets become increasingly available to central banks, the ability to estimate high-dimensional structural models from rich data sources represents a major frontier.

\item \textbf{Climate--economy integrated assessment.}  Integrated assessment models with tipping points, regional heterogeneity, and multiple sources of risk represent a natural application of the methods surveyed here.  \citet{fernandezvillaverde2025climate} provide a comprehensive overview of the state of the art.  The DEQN methodology is particularly well suited to these models because climate--economy interactions introduce additional state variables (carbon concentrations, temperature, damages) that exacerbate the curse of dimensionality.

\item \textbf{Heterogeneous-agent models and beyond.}  Extending the DEQN and PINN frameworks to heterogeneous-agent models with aggregate shocks, where the cross-sectional distribution of agents is itself part of the aggregate state, is a major frontier (Chapter~\ref{ch:young}).  \citet{han2023deepham} propose DeepHAM, which represents the distribution via learned generalized moments and trains networks by maximizing cumulative utility along simulated paths (\S\ref{sec:deepham_ks}); \citet{payne2025deepsam} extend the same set-encoder idea to non-Walrasian search-and-matching models, where the distribution affects decisions through the matching technology and bargaining rather than through prices (\S\ref{sec:deepsam_ks}); and \citet{yang2025structural} show that for many Walrasian HA models, equilibrium prices alone, rather than the full distribution, can serve as the aggregate state in agents' policy functions, sidestepping the master equation altogether.  Together these approaches demonstrate that deep learning can tackle the infinite-dimensional state spaces inherent in heterogeneous-agent economies, and can also help \emph{define} new equilibrium concepts for them.

\item \textbf{Operator learning.}  Rather than learning a single function (as in DEQNs or PINNs), operator learning frameworks such as DeepONet \citep{lu2021learning} and the Fourier Neural Operator \citep{li2021fourier} learn mappings between function spaces, for example, learning the map from a forcing function to the PDE solution, or from a model specification to its equilibrium.  This ``learn the solver'' paradigm could enable even faster model evaluation by amortizing the cost of solving across many model instances.

\item \textbf{Error certification and benchmark protocols.}  A small residual on the training distribution is not the same as a small policy or welfare error on the economically relevant state space.  A major frontier is the development of a-posteriori error bounds, benchmark suites with classical baselines, and standardized reporting conventions: residual heat maps, Euler-error distributions, market-clearing errors, transversality checks, and consumption-equivalent welfare losses.  Without such certification, deep solvers are hard to compare across papers and hard to audit in policy environments.

\item \textbf{Distribution shift, tails, and rare events.}  Simulation-based training concentrates samples on the ergodic set.  This is efficient for average performance but dangerous when the research question concerns crises, tail risk, climate tipping points, binding collateral constraints, or off-policy counterfactuals.  Robust sampling schemes that combine ergodic states, boundary states, adversarial states, and economically meaningful stress scenarios are still underdeveloped, and they are precisely the regime where small training residuals can coexist with large welfare errors.
\end{enumerate}

\section{When \emph{Not} to Use Deep Learning}
\label{sec:when_not_to_use_dl}

A pedagogical script risks selling the methods it teaches.  Symmetry requires an explicit statement of the regimes in which the deep-learning toolkit covered here is \emph{not} the right answer; recognizing those regimes is part of using the toolkit well.

\begin{enumerate}[itemsep=4pt]
\item \textbf{The state space is genuinely low-dimensional.}  For models with $d \lesssim 5$ continuous state variables and smooth policies, classical methods, projection on Chebyshev polynomials, value-function iteration on a fine grid, or perturbation around the steady state, typically deliver higher accuracy at lower cost than a DEQN \citep{judd1998numerical}.  The curse of dimensionality is real, but for smooth low-dimensional models it is rarely the binding constraint; nonsmoothness, occasionally binding constraints, and tail events can make even low-dimensional problems hard.

\item \textbf{The model is locally linear and the question is local.}  If you only need first- or second-order responses around the deterministic steady state, perturbation methods give analytical impulse responses in seconds; the standard implementation in macroeconomics is \citet{adjemian2024dynare}.  Use a DEQN or a PINN only when global nonlinearity, large shocks, occasionally binding constraints, or the ergodic distribution itself is the object of interest.

\item \textbf{You need likelihood-based inference and the model is small.}  When closed-form or quadrature-based likelihoods are available, full-information ML and Bayesian estimation with a particle filter dominate any simulation-based approach in efficiency.  The deep-surrogate route in Chapter~\ref{ch:estimation} pays off only when the likelihood is intractable or the simulator is much more expensive than evaluating the network.

\item \textbf{Sample size is small and uncertainty is the answer.}  At $n \lesssim 10^3$ design points and moderate input dimension, a well-specified GP with active learning (Chapter~\ref{ch:gp}) often provides stronger uncertainty quantification and competitive point prediction relative to a deep network of comparable complexity; the advantage can disappear when the dimension is high, the kernel is misspecified, or the response is strongly nonstationary.  Default to GPs when the simulator is expensive and posterior uncertainty matters.

\item \textbf{Reproducibility and audit-ability are paramount.}  Deep-learning solutions depend on initialization seeds, optimizer tuning, and floating-point order of operations on the GPU.  In central-bank or regulatory contexts where round-off-stable, configuration-light reproducibility is required, a deterministic finite-difference scheme remains preferable; bit-exactness across heterogeneous BLAS/LAPACK back-ends is not guaranteed for either approach, but FD typically has fewer moving parts to pin.  At minimum, pin every random seed and report hardware/software versions (cf.\ the reproducibility appendix).

\item \textbf{Training instability is a model-misspecification signal.}  If a DEQN refuses to converge, the diagnosis is rarely ``add more layers'', it is usually a poorly specified loss, a missing hard constraint, an unstable simulation, or a model that admits no equilibrium near the trial states.  Spending a day debugging a divergent training run is often the discovery that the model itself needs to change.
\end{enumerate}

The chapters of this script argue that, in the right regime, deep learning unlocks problems that are out of reach of any other tool.  The point of this short list is the converse: there is a long catalog of standard problems where it is the wrong tool, and recognizing them keeps the methodology honest.

\section{Practical Tips and Common Pitfalls}

The following guidelines distill recurring lessons from the applications covered in this course:

\begin{enumerate}[itemsep=4pt]
\item \textbf{Start simple.}  Begin with a small network (2--3 hidden layers, 32--64 neurons) and a low-dimensional test case with a known solution (e.g., Brock--Mirman for DEQNs, the 1D ODE for PINNs).  Only increase complexity once the simple case works.

\item \textbf{Normalize states, controls, and residuals.}  Many failed training runs are scaling failures rather than approximation failures.  Work with logs, ratios, standardized states, or economically natural units, and scale residuals so that one equation does not dominate the loss only because it is measured in larger units.  In climate and OLG models, nondimensionalization is often as important as the architecture.

\item \textbf{Build in hard constraints where possible.}  Positivity of consumption, bounds on savings or abatement, no-arbitrage restrictions, and budget feasibility should be enforced by output transformations or hard ansatzes whenever they can be.  Penalty terms are useful, but a network that cannot violate a constraint is usually easier to train and easier to trust; this is also where monotonicity and concavity restrictions on policy or value functions are best imposed.

\item \textbf{Use independent validation states and shocks.}  Report residuals on held-out state points, independent shock sequences, boundary states, and economically important stress scenarios.  A solver that performs well only on the training simulation path has not solved the global model; this is the practical counterpart to the distribution-shift frontier discussed in \S\ref{sec:open_challenges}.

\item \textbf{Monitor all loss components.}  In multi-term losses (PDE residual $+$ boundary penalties, or Euler residuals $+$ complementarity), log each component separately.  A declining total loss can mask a rising boundary term.

\item \textbf{Use adaptive loss balancing.}  Manual tuning of penalty weights $\lambda$ is fragile.  ReLoBRaLo (Chapter~\ref{ch:nas}) and related schemes automatically balance competing loss terms; in the benchmarks reported by \citet{bischof2025relobralo} they substantially improve accuracy over fixed weights, with the magnitude of the gain depending on the problem.

\item \textbf{Check economic diagnostics, not just the loss.}  Verify that policy functions satisfy economic intuition (e.g., consumption increasing in wealth), that market clearing holds, and that the relative Euler error is economically small (e.g., median or mean below $10^{-3}$).  Pick one convention (relative Euler error, $\log_{10}$ Euler error, or consumption-equivalent error) and use it consistently across diagnostics.

\item \textbf{Choose activations deliberately.}  In strong-form PINNs that involve second-order PDEs, use $C^\infty$ activations such as $\tanh$, Swish, or softplus, since ReLU has $u'' = 0$ a.e.; weak-form formulations and methods that handle nonsmooth solutions explicitly can still use ReLU.  Swish is a strong default for DEQNs.  Avoid mixing activation types across layers unless there is a specific reason.

\item \textbf{Use common random numbers (CRN).}  When comparing model outputs across parameter values (e.g., in surrogate-based estimation), always fix the shock sequence.  CRN can reduce variance substantially in surrogate-SMM pipelines (often by an order of magnitude or more), making gradient-based optimization feasible.
\end{enumerate}

\paragraph{Common failure modes.}
Table~\ref{tab:dl_failure_modes} summarizes recurring failure modes and their mitigations:

\begin{table}[ht]
\centering
\small
\begin{tabular}{L{4.5cm} L{7.5cm}}
\toprule
\textbf{Failure Mode} & \textbf{Mitigation} \\
\midrule
Loss decreases but solution is wrong & Check individual loss components; use economic diagnostics \\
One residual dominates all others & Adaptive loss balancing (Ch.~\ref{ch:nas}); inspect per-component gradients; normalize residuals to comparable units \\
Solver looks fine on simulation path, fails off path & Add held-out validation states; sample corners and stress states; use residual-based adaptive resampling \\
Policy network outputs collapse to zero (e.g., $c \equiv 0$ or $s \equiv 0$) & Improve initialization; use simulation-based training; check that hard constraints are not forcing the trivial solution \\
NaN in loss & Clamp inputs to GP/network; check for log of negative values \\
Euler errors large in corners of state space & Increase sampling near boundaries; use hard constraints \\
Surrogate gives a precise but biased SMM optimum & Validate the surrogate near the estimated optimum; add active-learning points around the criterion minimum; cross-check with direct DEQN re-evaluation \\
GP posterior variance too large & Add training points via BAL; check kernel hyperparameters \\
Training succeeds only for one seed & Report multi-seed dispersion; reduce learning rate; tighten scaling; check residual signs \\
\bottomrule
\end{tabular}
\caption{Recurring failure modes encountered when training the deep-learning solvers of this course (DEQNs, PINNs, deep surrogates, GP + BAL), together with the practical mitigations that work most reliably across the companion notebooks. The list is short by design: most production failures fall into one of these categories, and the corresponding fix is usually the first thing to try before invoking heavier diagnostics.}
\label{tab:dl_failure_modes}
\end{table}

\paragraph{Implementation checklist.}
For each new model, we recommend the following workflow:
\begin{enumerate}[itemsep=2pt]
\item Solve a simplified version with a known analytical solution (e.g., Brock--Mirman).
\item Verify that all equilibrium conditions are correctly encoded in the loss.
\item Train with a small network and monitor all loss components separately.
\item Check economic diagnostics: policy function shapes, market clearing, Euler errors.
\item Scale up: increase network size, dimension, and training budget.
\item Document the final configuration and report reproducibility information.
\end{enumerate}

\section{Concluding Remarks}

The methods presented in this course are not merely computational tricks: they expand the feasible frontier of quantitative economic modeling when state spaces are high-dimensional, equilibrium restrictions are differentiable, and repeated model evaluation is valuable.  They do not replace classical numerical economics; they add a flexible amortized layer that is most powerful when combined with economic structure, careful diagnostics, and classical benchmarks.

The convergence of abundant compute, mature software ecosystems (TensorFlow, PyTorch, JAX), and increasingly complex economic questions ensures that deep learning methods will play an expanding role in economic research and central bank practice in the years ahead.  We hope that this course has provided the conceptual foundations and practical tools needed to engage with this frontier.

For further reading, we refer to the comprehensive survey by \citet{fernandezvillaverde2024taming}, the methodological foundations laid in \citet{azinovicDEEPEQUILIBRIUMNETS2022}, and the applications in \citet{rennerscheidegger_2018}, \citet{friedlDeep2023}, \citet{han2023deepham}, \citet{payne2025deepsam}, \citet{kase2022estimating}, \citet{chen2026Deep}, \citet{kubler2025using}, and \citet{fernandezvillaverde2025climate}.  The list of references in these notes is necessarily incomplete; for a full bibliography, we refer the reader to the cited papers and the references therein.

\begin{keyinsightbox}[Chapter Summary]
\begin{itemize}[itemsep=2pt, leftmargin=*]
\item All four method families share one paradigm: \emph{neural network as function approximator} $+$ \emph{economic structure in the loss} $+$ \emph{automatic differentiation for training}.
\item Knowing when \emph{not} to use deep learning is part of using it well: classical projection or perturbation often dominates in low dimension; exact likelihood beats simulation when both are available; a well-specified GP is often the better choice at small $n$ and moderate dimension, when uncertainty is the deliverable.
\item Open frontiers: convergence theory, hybrid global/local methods, real-time policy analysis, large structural estimation, climate--economy IAMs, full HA models with aggregate shocks, operator learning.
\item The right question for a research project is rarely ``which of these four methods?'' but ``which combination?'', e.g.\ DEQN $+$ surrogate for estimation, PINN $+$ GP for sensitivity, or all of the above for climate policy.
\end{itemize}
\end{keyinsightbox}

\section*{Exercises}
\addcontentsline{toc}{section}{Exercises}
\noindent Worked solutions and guidance for these exercises appear in Appendix~\ref{app:solutions}.
\begin{enumerate}[itemsep=4pt, leftmargin=*]
\item\label{ex:ch12:1} \textbf{[Core] Method-choice scenario.}  You are handed (a) a 4-state monetary-policy DSGE with smooth shocks, (b) a 200-agent OLG with progressive taxation, (c) an option-pricing problem on an irregularly shaped exotic payoff, and (d) a climate-IAM where uncertainty in the SCC is the deliverable.  Justify which method you would use for each, and what hybrid you would consider.  For each case, state the classical baseline you would compare against before reaching for a deep-learning method.
\item\label{ex:ch12:2} \textbf{[Core] When NOT to use deep learning.}  Pick one of the six regimes from \S\ref{sec:when_not_to_use_dl} and write a one-page note for a colleague explaining why a classical method dominates.
\item\label{ex:ch12:3} \textbf{[Computational] Reproducibility audit.}  Take any notebook from this script and document hardware, software versions, and seeds.  Re-run end-to-end and report the maximum deviation in any reported number.
\item\label{ex:ch12:4} \textbf{[Advanced/project] Open-ended.}  Identify one frontier from \S\ref{sec:open_challenges} (operator learning, hybrid global/local, real-time policy, large structural estimation, climate-economy IAMs, HA with aggregate shocks) and sketch a 6-month research project that combines two methods from this script.
\item\label{ex:ch12:5} \textbf{[Advanced/project] Hybrid pipeline: DEQN $+$ GP $+$ SMM end-to-end.}  Build a complete estimation pipeline on the Brock--Mirman model with two parameters $(\beta, \varrho)$, matching the parameter pair worked in the companion notebook \tpath{lecture_15_03b_Structural_Estimation_BM_Joint.ipynb}.  As that notebook documents, $\beta$ is only \emph{partially} identified by macro moments (a visible ridge along the $\beta$ axis); $\varrho$ has the cleaner identification map.  The exercise below is a controlled-difficulty version of that pipeline.
\begin{enumerate}[label=(\roman*), itemsep=2pt]
\item Train a DEQN with \emph{pseudo-state augmentation}: extend the network input from $(K_t, z_t)$ to $(K_t, z_t, \beta, \varrho)$.
\item On a $20\times 20$ grid of $(\beta, \varrho) \in [0.92, 0.99] \times [0.50, 0.99]$, simulate the model and compute four moments: mean savings rate, mean consumption growth, $\mathrm{Var}(\log C_t)$, and $\mathrm{Var}(K_t)$.
\item Fit a GP surrogate $(\beta, \varrho) \mapsto \bm m$ on the resulting $400$-point dataset.
\item Generate observed moments at $(\beta^\star,\varrho^\star)=(0.96,0.90)$ from one $T=200$ simulation and solve the GP-based SMM problem.
\end{enumerate}
Report the wall-clock time of each step, total pipeline time, estimation errors, a comparison to a naive SMM that re-solves the DEQN at each candidate $(\beta,\varrho)$, the hold-out surrogate error on $(\beta,\varrho)$ pairs outside the $20\times 20$ grid, a contour plot of the GP-based SMM criterion that exposes the $\beta$ ridge, and a direct DEQN re-evaluation at the GP optimum to test surrogate bias.  Evaluate whether the surrogate-based optimization is faster after accounting for the one-time GP fitting overhead and whether its accuracy is comparable.  Bonus: bootstrap confidence intervals for $(\hat\beta,\hat\varrho)$ using the parametric bootstrap of Exercise~\ref{ex:ch10:6}.
\item\label{ex:ch12:6} \textbf{[Advanced/project] DeepONet for a parameterized HJB.}  Consider the cake-eating HJB
\[
\rho V(a; \gamma) = \max_c \left[\frac{c^{1-\gamma}}{1-\gamma} + V'(a; \gamma)(ra - c)\right],
\qquad \gamma \in [1.5,5],
\]
This is deliberately a low-dimensional operator-learning toy: since the branch input is the scalar $\gamma$, a parameter-conditioned PINN would be an equally natural baseline.  The purpose of the exercise is to expose the DeepONet branch--trunk decomposition before moving to function-valued branch inputs (e.g., a state-dependent drift $r(a)$ or discount rate $\rho(a)$ observed at sensor points).
\begin{enumerate}[label=(\roman*), itemsep=2pt]
\item Sketch the DeepONet architecture for learning the operator $\mathcal{G}: \gamma \mapsto V(\cdot;\gamma)$: identify the branch-net input and the trunk-net input, and write the predicted output as the inner product of branch and trunk outputs.
\item State the operator-learning universal approximation theorem of \citet{lu2021learning} and explain why a DeepONet of moderate width can approximate any continuous operator on a compact set.
\item Suppose you want $V(\cdot;\gamma)$ for $N=50$ values of $\gamma$.  Compare $N$ independent PINN runs at cost $C_\mathrm{PINN}$ each with one DeepONet run at cost $C_\mathrm{DON}$.  At what ratio $C_\mathrm{DON}/C_\mathrm{PINN}$ does operator learning win, and how does the break-even ratio scale with $N$?
\item Discuss two limitations: extrapolation outside $[1.5,5]$ and preservation of structural properties such as concavity of $V$ in $a$.
\end{enumerate}
\end{enumerate}

\appendix

\chapter{Glossary}
\label{ch:glossary}

A short glossary of recurring concepts.  Each entry is one or two sentences; for the full treatment, follow the cross-references in parentheses.

\begin{description}[style=nextline, leftmargin=0pt, labelwidth=0pt, labelsep=0pt, itemsep=3pt]
\item[\textbf{Active subspace.}]  The leading eigenspace of the gradient outer-product matrix $\E[\nabla f \nabla f^\top]$.  Captures the linear directions in input space along which a function varies most; allows GPs to scale to high $d$ \citep{constantine2015active}.
\item[\textbf{Active subspace, deep.}]  Replaces the linear projection $U_m^\top \x$ by a learned nonlinear encoder $h\colon \R^D \to \R^d$, trained jointly with an MLP link $g\colon \R^d \to \R$ so that $\hat f(\xi) = g(h(\xi))$; gradient-free, with $d$ chosen by a validation-MSE elbow instead of a spectral gap (Section~\ref{sec:deep_as}; \citealp{tripathy2018deep}).
\item[\textbf{Adam, AdamW.}]  Adaptive stochastic-gradient optimizers using momentum on the gradient and on its square; AdamW separates the weight-decay step from the adaptive step \citep{kingma2015adam, loshchilov2019decoupled}.
\item[\textbf{Approximate aggregation.}]  Empirical observation that in Krusell--Smith-class economies the cross-sectional distribution of wealth is nearly summarized by its mean for price forecasting purposes (Chapter~\ref{ch:young}).
\item[\textbf{Automatic differentiation (AD).}]  Algorithmic computation of exact derivatives of composite functions via the chain rule; reverse-mode AD is the engine of every deep-learning framework \citep{baydin2018automatic, margossian2019review}.
\item[\textbf{Bayesian active learning (BAL).}]  Adaptive sample-design strategy in which the next training point is chosen to maximize an acquisition function based on predictive uncertainty; pairs naturally with GPs (Chapter~\ref{ch:gp}).
\item[\textbf{Bellman equation.}]  Recursive characterization of the value function in a discrete-time dynamic program.  Continuous-time analogue is the HJB equation.
\item[\textbf{Brock--Mirman model.}]  Stochastic neoclassical growth model with log utility and full depreciation that admits a closed-form policy $s^\star = \alpha\beta$; the canonical DEQN benchmark in this script.
\item[\textbf{Common random numbers (CRN).}]  Variance-reduction technique in which the same shock realisations are reused across simulations of different parameter values, removing simulation noise from comparisons \citep{glasserman2004monte}.
\item[\textbf{Collocation point.}]  A spatial location at which a PDE residual is evaluated and minimized in a PINN training loop (Chapter~\ref{ch:pinn}).
\item[\textbf{Curse of dimensionality.}]  Exponential blow-up of grid-based methods in the dimension of the state space; mitigated, not eliminated, by neural-network and GP approximators.
\item[\textbf{Deep Equilibrium Net (DEQN).}]  A neural-network-based solver for dynamic stochastic equilibrium models that minimizes the equilibrium-equation residuals directly via SGD (Chapter~\ref{ch:deqn}).
\item[\textbf{Deep Galerkin Method (DGM).}]  An LSTM-style architecture introduced by \citet{sirignano2018dgm} for solving high-dimensional PDEs; the architectural sibling of standard PINNs.
\item[\textbf{Deep kernel learning (DKL).}]  Composes a neural-network feature extractor with a GP layer in the learned feature space \citep{wilson2016deep}.
\item[\textbf{DeepONet.}]  A neural architecture for operator learning: a branch net encodes the input function and a trunk net encodes the query point; the inner product is the predicted output \citep{lu2021learning}.
\item[\textbf{EMINN.}]  Economic-Model Informed Neural Network: a PINN-style approach to the master equation in continuous-time HA models \citep{gu2024masterequations}.
\item[\textbf{Ergodic distribution.}]  The stationary distribution of a Markov process; in a Krusell--Smith economy it is the long-run distribution of wealth across agents.
\item[\textbf{Fischer--Burmeister (FB).}]  A smooth complementarity function $\Phi(a,b) = a + b - \sqrt{a^2+b^2}$ used to encode KKT conditions in differentiable losses; the opposite sign has the same zero set but the chapter and notebooks use this convention \citep{fischer1992special}.
\item[\textbf{Fourier Neural Operator (FNO).}]  Operator-learning architecture parameterizing a kernel integral operator in Fourier space; cheap and resolution-invariant \citep{li2021fourier}.
\item[\textbf{Functional derivative.}]  $\delta V/\delta g$, the density / Riesz representer of the Fr\'echet derivative of $V$ with respect to a function-valued argument $g$ (equivalently, the directional derivative of $V$ at $g$ in the direction of a Dirac perturbation $\delta_{y_0}$); appears in the master equation (Chapter~\ref{ch:ct_theory}).
\item[\textbf{Gauss--Hermite quadrature.}]  Polynomial quadrature rule for integrals against the standard normal density; backbone of the expectations step in DEQNs.
\item[\textbf{HJB equation.}]  Hamilton--Jacobi--Bellman equation; continuous-time analogue of the Bellman equation, a PDE in the value function.
\item[\textbf{Histogram (Young 2010).}]  Mass-redistribution scheme on a fixed grid that propagates a wealth distribution deterministically without Monte Carlo noise (Chapter~\ref{ch:young}).
\item[\textbf{Hyperband.}]  Successive-halving multi-fidelity hyperparameter scheduler that explores many configurations cheaply and concentrates budget on the survivors \citep{li2018hyperband}.
\item[\textbf{Inducing points.}]  A small set of pseudo-data points used in sparse GPs to approximate the full kernel matrix at $\mathcal{O}(nm^2)$ cost \citep{titsias2009variational}.
\item[\textbf{Itô's lemma.}]  The chain rule of stochastic calculus; for the scalar diffusion $dX_t=\mu\,dt+\sigma\,dB_t$, the only difference from ordinary calculus is the second-order correction $\tfrac{1}{2}f''(X_t)\sigma^2\,dt$.
\item[\textbf{Karush--Kuhn--Tucker (KKT) conditions.}]  First-order necessary conditions for constrained optimization; encoded smoothly via Fischer--Burmeister in DEQN losses.
\item[\textbf{Kolmogorov forward equation (KFE / Fokker--Planck).}]  The PDE governing the time evolution of the probability density of an Itô process.
\item[\textbf{Marginal likelihood.}]  The log-evidence $\log p(\bm y \mid \bm\vartheta)$ in a GP; sum of a data-fit term and a complexity penalty (Chapter~\ref{ch:gp}).
\item[\textbf{Master equation.}]  A single PDE that subsumes the HJB, KFE, and market-clearing conditions of a continuous-time mean-field-game equilibrium; argument includes the cross-sectional measure $g$.
\item[\textbf{Mean field game (MFG).}]  Equilibrium concept in which each atomistic agent best-responds to the cross-sectional distribution and the distribution evolves under those best responses; the natural framework for the HJB+KFE system \citep{lasry2007mean}.
\item[\textbf{Neural Tangent Kernel (NTK).}]  In the infinite-width limit, gradient-descent training of a deep network is equivalent to kernel regression with the (deterministic) NTK \citep{jacot2018neural}.
\item[\textbf{Operator learning.}]  Learning a map between function spaces (input field $\to$ solution function) rather than a single solution; DeepONet and FNO are the leading architectures.
\item[\textbf{Physics-Informed Neural Network (PINN).}]  A neural network trained by minimizing a PDE residual at collocation points plus boundary-condition penalties \citep{raissi2019physics}.
\item[\textbf{Pseudo-state.}]  Treating model parameters as additional inputs to a neural network so that the trained surrogate covers an entire parameter range without retraining (Chapter~\ref{ch:gp}).
\item[\textbf{Quasi-Monte Carlo (QMC).}]  Deterministic low-discrepancy sequences (Sobol, Halton, Niederreiter) achieving error rates close to $\mathcal{O}(1/M)$ for smooth integrands.
\item[\textbf{ReLoBRaLo.}]  Adaptive loss-balancing scheme that reweights multi-component losses by recent relative-decrease ratios \citep{bischof2025relobralo}.
\item[\textbf{Simulated Method of Moments (SMM).}]  Estimator that matches simulated to empirical moments; the natural extension of GMM when moments lack closed form \citep{mcfadden1989method}.
\item[\textbf{Simulation-based inference (SBI).}]  Modern likelihood-free Bayesian inference using neural conditional density estimators \citep{cranmer2020frontier}.
\item[\textbf{Social cost of carbon (SCC).}]  Marginal welfare cost of one additional unit of emissions, commonly reported as USD/tCO$_2$ after choosing the consumption numeraire and applying the carbon-to-CO$_2$ conversion; the headline policy number from a climate IAM.
\item[\textbf{Sobol / Shapley indices.}]  Variance-decomposition tools for global sensitivity analysis.  Sobol decompositions are cleanest under independent inputs; Shapley effects allocate variance across inputs and can be defined for dependent inputs when the dependence structure is modeled explicitly.
\item[\textbf{Universal approximation.}]  A single-hidden-layer network with a non-polynomial activation can approximate any continuous function on a compact set arbitrarily well \citep{cybenko1989approximation, hornik1989multilayer}.
\item[\textbf{Value Function Iteration (VFI).}]  Classical contraction-mapping algorithm for solving the Bellman equation by iterating the Bellman operator until convergence.
\item[\textbf{Young's lottery.}]  The unique two-point split that, when applied to off-grid policy choices, preserves the conditional mean exactly; the building block of the histogram update.
\end{description}

\chapter{Matrix Calculus and Automatic Differentiation}
\label{app:ad}

This appendix collects the few matrix-calculus identities used in the main text and gives a one-page summary of how automatic differentiation realizes them computationally.

\section{Matrix calculus identities}

For $\bm A \in \R^{m\times n}$, $\bm x \in \R^n$, $\bm y = \bm A\bm x$:
\[
   \frac{\partial \bm y}{\partial \bm x} = \bm A,\qquad
   \frac{\partial (\bm y^\top \bm y)}{\partial \bm x} = 2\bm A^\top \bm A\bm x.
\]
For a quadratic form $f(\bm x) = \bm x^\top \bm Q \bm x$ with symmetric $\bm Q$,
$\nabla f(\bm x) = 2\bm Q\bm x$ and $\nabla^2 f = 2\bm Q$.
For a deep network $\bm{\hat y} = g_L(\bm W_L\, g_{L-1}(\cdots g_1(\bm W_1 \bm x + \bm b_1)\cdots))$, the chain rule gives
\[
   \frac{\partial \bm{\hat y}}{\partial \bm W_l}
   \;=\;
   \underbrace{(\bm \delta^{(l)})}_{\substack{\text{backprop} \\ \text{delta}}}
   \,\bigl(\bm a^{(l-1)}\bigr)^{\!\top}
\]
with $\bm \delta^{(l)} = \bigl(\bm W^{(l+1)\top}\bm \delta^{(l+1)}\bigr) \odot g'(\bm z^{(l)})$ (\S\ref{sec:training}).

\section{Reverse-mode AD in one paragraph}

Reverse-mode AD evaluates the function once forward, caches every intermediate value, and then traverses the computation graph backwards, accumulating the local Jacobian--vector products in linear time in the number of operations.  For a scalar loss, one reverse sweep returns the gradient with respect to all parameters at a small constant multiple of the forward cost.  The computation still scales with the size of the graph, but it does not require one separate derivative pass per parameter; this is what makes million-parameter networks trainable.  See \citet{baydin2018automatic} and \citet{margossian2019review} for complete treatments.

\section{Higher-order AD}

PINN losses involve second derivatives ($V_{SS}$ in Black--Scholes, Laplacians in Poisson equations, and second-order terms in diffusion HJBs).  Higher-order derivatives are obtained by composing reverse-mode AD with itself, \texttt{torch.autograd.grad} of \texttt{grad} in PyTorch, \texttt{jax.grad} of \texttt{jax.grad} in JAX.  Activation regularity matters: the network must be at least $C^k$ for the strong $k$-th-order residual to be well-defined.

\chapter{It\^o Calculus, Brownian Motion, and Ergodicity}
\label{app:stoch}

This appendix collects the stochastic-calculus background needed for Chapters~\ref{ch:pinn}--\ref{ch:ct_theory}.  For the longer, example-driven exposition see~\S\ref{sec:stoch_calc}; for a full textbook treatment, see \citet{shreve2004stochasticii}.

\section{Brownian motion}
A standard Brownian motion $B_t$ has independent Gaussian increments $B_t - B_s \sim \mathcal{N}(0, t-s)$, continuous paths almost surely, and quadratic variation $\langle B\rangle_t = t$.  Equivalently, the simple-random-walk scaling limit $X_{t+\Delta t} = X_t + \sqrt{\Delta t}\,\varepsilon_t$ converges in distribution to $B_t$ as $\Delta t \to 0$ (Donsker).

\section{It\^o's lemma}
For $X_t$ following $dX_t = \mu\,dt + \sigma\,dB_t$ and $f \in C^2$:
\[
   df(X_t) = \bigl[f'(X_t)\mu + \tfrac{1}{2}f''(X_t)\sigma^2\bigr]dt + f'(X_t)\sigma\,dB_t.
\]
The differential algebra is $dt\cdot dt = dt\cdot dB_t = 0$, $dB_t \cdot dB_t = dt$.  (Throughout the script, the stochastic integral is interpreted in the It\^o sense; the Stratonovich convention would replace the drift correction $\tfrac{1}{2}f''(X_t)\sigma^2$ by $0$.)

\section{Ergodicity in one paragraph}
A Markov process with stationary distribution $\pi$ is \emph{ergodic} if the time-average along any sample path converges almost surely to the spatial average against $\pi$:
$\tfrac{1}{T}\int_0^T \varphi(X_t)\,dt \xrightarrow{T\to\infty} \int \varphi\,d\pi$ for every bounded measurable $\varphi$.  In economic models with bounded state spaces and aperiodic dynamics, ergodicity is what justifies replacing population integrals by simulation-time averages in DEQN training (Chapter~\ref{ch:deqn}, \S\ref{sec:deqn_algo}).

\chapter{Fixed Points, Contraction Mappings, and the Bellman Operator}
\label{app:fixed_points}

The classical numerical-economics toolkit rests on Banach's contraction-mapping theorem; this appendix recalls the statement and one short proof sketch for completeness.  A reference is \citet{stokeylucas1989}.

\section{Banach's theorem}
Let $(X, d)$ be a complete metric space and let $T: X \to X$ be a contraction with modulus $\beta < 1$, i.e.\ $d(T x, T y) \le \beta\, d(x, y)$ for all $x, y$.  Then $T$ has a unique fixed point $x^\star$, and for every starting point $x_0$ the iterates $x_{n+1} = Tx_n$ satisfy $d(x_n, x^\star) \le \beta^n\, d(x_0, x^\star)$.

\section{The Bellman operator is a contraction}
For a discount factor $\beta \in (0,1)$ and bounded utility, the Bellman operator
$T V(x) = \max_a \{u(x,a) + \beta\,\E V(x') \mid x_t = x, a_t = a\}$
is a contraction with modulus $\beta$ on the space of bounded continuous functions equipped with the supremum norm.  Hence value-function iteration converges geometrically.  The same logic underpins \emph{policy iteration}, the dual algorithm that converges in finite steps for finite-state problems.

\section{Why this matters for DEQNs}
DEQNs do \emph{not} iterate a Bellman operator.  Instead, they apply gradient descent to a residual loss that vanishes at the equilibrium policy.  The contraction property therefore disappears as a tool for proving convergence, and theoretical guarantees come from neural-network optimization theory rather than fixed-point analysis: convergence arguments rest on universal approximation plus loss-landscape and NTK-style analyses of stochastic gradient descent on overparameterized networks, surveyed alongside the DEQN-specific open questions in Chapter~\ref{ch:outlook}.

\chapter{Reproducibility Information}
\label{app:reproducibility}

Every empirical statement in the main text relies on the companion notebooks listed in the \emph{Execution Map}.  Bit-exact reproducibility on a different machine requires fixing both the random seeds and the floating-point environment; Table~\ref{tab:repro_conventions} summarizes the conventions used in this script's notebooks.

\begin{table}[ht]
\centering
\small
\caption{Reproducibility conventions used by the companion notebooks.  The conventions pin random seeds and document hardware, software, and precision choices; bit-exact reproduction additionally requires the deterministic settings and caveats described below the table.}
\label{tab:repro_conventions}
\begin{tabular}{@{}L{3.0cm} L{10.0cm}@{}}
\toprule
\textbf{Item} & \textbf{Convention} \\
\midrule
Random seeds   & Each notebook declares a \tpath{SEED} constant in cell~1 (default \tpath{0}); the corresponding framework seeds (\tpath{numpy.random.seed(SEED)}, \tpath{torch.manual_seed(SEED)}, \tpath{tf.random.set_seed(SEED)}, and \tpath{jax.random.PRNGKey(SEED)} where the framework is used) are derived from it.  Per-cell offsets and per-iteration deviations are documented inline.  Notebook-by-notebook normalisation against this convention is tracked through the chapter audits; some legacy notebooks still hard-code a literal seed and are scheduled for normalisation. \\
Run-mode budget & Each notebook declares \tpath{RUN_MODE} $\in$ \tpath{\{smoke, teaching, production\}} alongside \tpath{SEED} in cell~1.  \tpath{smoke} is sized for a single CPU core (CI-friendly); \tpath{teaching} for one consumer GPU; \tpath{production} for an A100 or larger.  Per-notebook hyperparameters (epochs, batch sizes, restart counts) are gated on \tpath{RUN_MODE}. \\
Hardware        & Classroom-scale runs target a single CPU core or one consumer GPU (e.g.\ NVIDIA T4 / RTX 3060).  Production runs use one A100 or larger; this is documented per-notebook in the chapter. \\
Software stack  & Python $\geq 3.10$, TensorFlow $\geq 2.15$, PyTorch $\geq 2.0$, JAX $\geq 0.4.20$, GPyTorch $\geq 1.11$.  Pinned versions live in \tpath{requirements.txt} on the companion repository.  JAX appears in the Krusell--Smith warm-up notebook in \tpath{lectures/lecture_10_*} (sequence-space DEQNs); the other lectures use TensorFlow or PyTorch. \\
Numerical precision & TensorFlow / PyTorch default to \tpath{float32}; PINNs and EMINN training use \tpath{float64} where second-order derivatives are sensitive (documented inline). \\
GPU determinism & \emph{Optional, bit-exact runs only}: set \tpath{CUBLAS_WORKSPACE_CONFIG=:4096:8} and \tpath{torch.use_deterministic_algorithms(True)} on the PyTorch side, and call \tpath{tf.config.experimental.enable_op_determinism()} on the TensorFlow side.  Default classroom and production runs leave these unset, accepting last-decimal nondeterminism in exchange for speed; small accuracy regression possible when the flags are on. \\
Reported numbers & Where the script states an accuracy or runtime, the source is either the cited paper (for production-scale results) or the companion notebook with the seed above (for classroom-scale results). \\
\bottomrule
\end{tabular}
\end{table}

For full bit-exact reproduction, e.g.\ for a regulatory-grade audit, the determinism flags above are necessary but not sufficient: GPU non-determinism in atomic accumulators and BLAS implementation differences across CUDA versions can still cause diverging results in the last few decimal places.  This is a known limitation of GPU-accelerated deep learning and is one of the reasons \S\ref{sec:when_not_to_use_dl} flags reproducibility-critical settings as a regime where deterministic finite-difference solvers may still be preferred.

\chapter{Solutions and Guidance for Exercises}
\label{app:solutions}

This appendix provides worked solutions for analytical end-of-chapter exercises and guidance for coding exercises.  Coding exercises (those that ask the reader to modify a notebook or measure a numerical quantity) are not treated as fixed numerical answers; their outputs depend on hardware, seeds, calibration choices, and notebook versions, and should be reported from the corresponding companion notebook listed in the \emph{Execution Map}.  For exercises that mix analytical work with a small numerical check, the analytical part is solved and the numerical component is described as a verification task.

Exercises are referenced by their stable label \texttt{ex:ch$N$:$M$}, where $N$ is the chapter number and $M$ is the position within that chapter's exercise list.  Each solution opens with a back-pointer to the exercise statement so the reader can scan the question first.

\section{Chapter \ref{ch:intro}: Introduction to Machine Learning and Deep Learning}
\label{sol:ch1}

\paragraph{Exercise~\ref{ex:ch1:1} (statement: p.~\pageref{ex:ch1:1}): Backprop on a 2-layer net.}
Let $z = w_1 x + b_1$, $a = \mathrm{ReLU}(z)$, $\hat y = w_2 a$, $\ell = (\hat y - y)^2$.  Reverse-mode chain rule:
\begin{align*}
\frac{\partial \ell}{\partial \hat y} &= 2(\hat y - y), &
\frac{\partial \ell}{\partial w_2} &= 2(\hat y - y)\,a, \\
\frac{\partial \ell}{\partial a} &= 2(\hat y - y)\,w_2, &
\frac{\partial a}{\partial z} &= \mathbb{1}[z > 0], \\
\frac{\partial \ell}{\partial w_1} &= 2(\hat y - y)\,w_2\,\mathbb{1}[z>0]\,x.
\end{align*}
Plugging in $x=2$, $y=1$, $w_1 = w_2 = b_1 = 0.5$: $z = 1.5$, $a = 1.5$, $\hat y = 0.75$, $\ell = 0.0625$.  Hence $\partial\ell/\partial w_2 = 2(-0.25)(1.5) = -0.75$ and $\partial\ell/\partial w_1 = 2(-0.25)(0.5)(1)(2) = -0.5$.  A two-line PyTorch script (\texttt{torch.autograd.grad}) returns the same numbers to machine precision; this is the simplest non-trivial sanity check that the chain-rule derivation matches what AD computes.

\paragraph{Exercise~\ref{ex:ch1:2} (statement: p.~\pageref{ex:ch1:2}): MSE vs.\ MLE.}
For Gaussian errors $\varepsilon_i \sim \mathcal{N}(0, \sigma^2)$, the log-likelihood is
\[
\log p(y_{1:n} \mid x_{1:n}; \beta) = -\frac{1}{2\sigma^2}\sum_i (y_i - \beta x_i)^2 - \tfrac{n}{2}\log(2\pi\sigma^2).
\]
Maximizing over $\beta$ (the only $\beta$-dependent term) is identical to minimizing $\sum_i (y_i - \beta x_i)^2$, i.e.\ OLS.  For Laplace errors with scale $b$, $p(\varepsilon) \propto \exp(-|\varepsilon|/b)$, so the log-likelihood is proportional to $-\sum_i |y_i - \beta x_i|$ and the MLE solves a least-absolute-deviations regression (median regression).  Squaring penalizes outliers quadratically, which is suboptimal under Laplace tails because a single large residual dominates the loss; the median estimator is robust precisely because $|\cdot|$ grows linearly.

\paragraph{Exercise~\ref{ex:ch1:3} (statement: p.~\pageref{ex:ch1:3}): Activation choice for a PINN.}
A ReLU network $\hat y(x) = \sum_k w_k\,\mathrm{ReLU}(a_k x + b_k) + c$ is piecewise-linear: between consecutive kink points $x = -b_k/a_k$ it is affine in $x$, so $\hat y''(x) = 0$ on every open subinterval.  At a kink, the second distributional derivative is a Dirac measure, not a classical pointwise value.  A strong-form PDE residual such as the Black--Scholes operator
\[
\partial_t V + \tfrac{1}{2}\sigma^2 S^2\,V_{SS} + r S\,V_S - r V \;=\; 0
\]
must hold pointwise (a.e.) for almost every $S$; with a piecewise-linear $V$, the term $V_{SS}$ vanishes a.e.\ in the classical sense, and the residual reduces to $\partial_t V + r S V_S - rV$, which has no nontrivial solution that respects the actual boundary conditions of an option-pricing problem.  Concrete example: $\hat y(x) = \mathrm{ReLU}(x) + \mathrm{ReLU}(-x) = |x|$ has $\hat y''(x) = 0$ for every $x \neq 0$ in the classical sense.  Smooth activations (tanh, Swish, GELU, softplus) are $C^\infty$ and produce well-defined pointwise second derivatives, which is why the PINN literature recommends them whenever the PDE is of order $\ge 2$.

\paragraph{Exercise~\ref{ex:ch1:4} (statement: p.~\pageref{ex:ch1:4}): Adam vs.\ AdamW.}
Write the gradient at step $t$ as $g_t = \nabla_\theta \ell(\theta_t)$.  With $L_2$ regularization \emph{added to the loss}, $\ell^{L_2}(\theta) = \ell(\theta) + \tfrac{\lambda}{2}\|\theta\|^2$, so the effective gradient becomes $\tilde g_t = g_t + \lambda \theta_t$.  Adam's update applied to $\tilde g_t$ is
\[
m_t = \beta_1 m_{t-1} + (1-\beta_1)\tilde g_t,\quad v_t = \beta_2 v_{t-1} + (1-\beta_2)\tilde g_t^2,\quad
\theta_{t+1} = \theta_t - \eta\,\frac{\hat m_t}{\sqrt{\hat v_t} + \varepsilon}.
\]
The key observation is that the weight-decay contribution $\lambda \theta_t$ enters $m_t$ and $v_t$ \emph{through the same EWMA averaging as the data gradient}, so the effective shrinkage applied to $\theta$ is $\eta\lambda \theta_t$ scaled by $1/(\sqrt{\hat v_t}+\varepsilon)$, which differs across coordinates.  Parameters with large historical gradients receive small effective decay; parameters with small gradients receive large decay.  AdamW decouples the two:
\[
m_t = \beta_1 m_{t-1} + (1-\beta_1) g_t,\quad v_t = \beta_2 v_{t-1} + (1-\beta_2) g_t^2,\quad
\theta_{t+1} = (1 - \eta\lambda)\theta_t - \eta\,\frac{\hat m_t}{\sqrt{\hat v_t}+\varepsilon}.
\]
Now the multiplicative shrinkage $(1 - \eta\lambda)$ acts uniformly on all parameters, independent of the adaptive denominator.  Numerically the two updates differ by a factor of $1/(\sqrt{\hat v_t}+\varepsilon)$ on the decay term; AdamW preserves the textbook intuition ``weight decay shrinks weights at the same rate everywhere.''

\paragraph{Exercise~\ref{ex:ch1:5} (statement: p.~\pageref{ex:ch1:5}): RNN forward pass by hand.}
With $W_h = 0.5\,I_2$, $W_x = (1,0)^\top$, $h_0 = (0,0)^\top$:
\begin{align*}
h_1 &= \tanh\!\big((0,0)^\top + (1,0)^\top\big) = (\tanh 1,\, 0)^\top \approx (0.7616,\, 0)^\top,\\
h_2 &= \tanh\!\big(0.5\,h_1 + (0,0)^\top\big) = (\tanh(0.3808),\,0)^\top \approx (0.3637,\,0)^\top,\\
h_3 &= \tanh\!\big(0.5\,h_2 + (1,0)^\top\big) = (\tanh(1.1818),\,0)^\top \approx (0.8275,\,0)^\top.
\end{align*}
Outputs: $\hat y_t = W_y h_t$ gives $\hat y_1 \approx 0.7616$, $\hat y_2 \approx 0.3637$, $\hat y_3 \approx 0.8275$ (only the first hidden coordinate is excited, so the second column of $W_y$ is irrelevant).

For the gradient, write $h_t = \tanh(z_t)$ with $z_t = W_h h_{t-1} + W_x x_t$.  Then
\[
\frac{\partial \hat y_3}{\partial x_1} = W_y\,\frac{\partial h_3}{\partial z_3}\,W_h\,\frac{\partial h_2}{\partial z_2}\,W_h\,\frac{\partial h_1}{\partial z_1}\,W_x,
\]
where $\partial h_t/\partial z_t = \mathrm{diag}(1-\tanh^2(z_t))$.  Numerically, $\tanh'(1)\approx 0.4200$, $\tanh'(0.3808)\approx 0.8678$, $\tanh'(1.1818)\approx 0.3155$, so
\[
\frac{\partial \hat y_3}{\partial x_1} \approx 1 \cdot 0.3155 \cdot 0.5 \cdot 0.8678 \cdot 0.5 \cdot 0.4200 \cdot 1 \approx 0.0288.
\]
The decay rate is the product of two distinct effects: (i) the spectral radius of $W_h$ (here $0.5$), which contributes one factor of $W_h$ per recurrent step ($T-1$ multiplicative copies in the chain above), and (ii) the saturation of $\tanh'(\cdot) \in (0,1]$, which contributes a second multiplicative factor per step.  Setting (ii) aside, with $\|W_h\|_2 = 0.5 < 1$ the gradient magnitude already decays at least as fast as $0.5^{T-1}$, i.e.\ exponentially in the sequence length.  This is the canonical vanishing-gradient pathology of vanilla RNNs (\S\ref{sec:sequence_models}).  LSTM and GRU gates address (i) by allowing the recurrent Jacobian's eigenvalues to stay close to one without making training unstable; effect (ii) is intrinsic to bounded activations and is mitigated by skip connections (and architectural choices that keep activations away from saturation regimes), not by gating.

\paragraph{Exercise~\ref{ex:ch1:6} (statement: p.~\pageref{ex:ch1:6}): Attention by hand.}
With $q_i = k_i = v_i = x_i$ and $x = (0,1,0.5)$, the score matrix $S_{ij} = q_i k_j$ is
\[
S = \begin{pmatrix} 0 & 0 & 0 \\ 0 & 1 & 0.5 \\ 0 & 0.5 & 0.25\end{pmatrix}.
\]
Softmaxing each row and using $e^0 = 1$, $e^{0.25} \approx 1.284$, $e^{0.5}\approx 1.649$, $e^1 \approx 2.718$:
\begin{align*}
a_1 &= \tfrac{1}{3}(1,1,1), & o_1 &= \tfrac{1}{3}(0 + 1 + 0.5) = 0.5, \\
a_2 &\approx (0.186,\, 0.506,\, 0.307), & o_2 &\approx 0.186\cdot 0 + 0.506\cdot 1 + 0.307\cdot 0.5 \approx 0.660, \\
a_3 &\approx (0.254,\, 0.419,\, 0.327), & o_3 &\approx 0.254\cdot 0 + 0.419\cdot 1 + 0.327\cdot 0.5 \approx 0.583.
\end{align*}
Each attention vector $a_i$ is on the simplex (entries non-negative, summing to one), so $o_i = \sum_j a_{ij} v_j$ is a convex combination of the values, lying in $[0, 1]$.  Token~2, the largest input, attends most strongly to itself ($a_{22}\approx 0.51$); token~3 also attends most to token~2 because the inner products $q_3 k_2 = 0.5$ exceed the self-score $q_3 k_3 = 0.25$.  Token~1 has zero query magnitude, so its row is uniform: with no signal to discriminate on, attention defaults to a uniform average over the values.

\paragraph{Exercise~\ref{ex:ch1:7} (statement: p.~\pageref{ex:ch1:7}): TensorBoard optimizer comparison.}
Coding exercise.  Expected qualitative behavior on a small classification task: SGD with momentum is often slower to reduce training loss but can generalize competitively when the learning-rate schedule is well tuned; Adam often reaches a low training loss fastest, but its validation curve can diverge from the training curve sooner than for SGD or AdamW; AdamW often sits between the two.  The actual crossover point is a notebook output and should be reported from the run.

\section{Chapter \ref{ch:deqn}: Deep Equilibrium Nets}
\label{sol:ch2}

\paragraph{Exercise~\ref{ex:ch2:1} (statement: p.~\pageref{ex:ch2:1}): Closed-form Brock--Mirman.}
Conjecture $V(K, z) = A\log K + B\log z + C$.  The Bellman equation
\[
   V(K,z) = \max_{K'}\Bigl\{\log\bigl(zK^\alpha - K'\bigr) + \beta\,\mathbb{E}_{z'\mid z}\!\bigl[V(K', z')\bigr]\Bigr\}
\]
yields the FOC $1/(zK^\alpha - K') = \beta A/K'$, which gives the constant savings rate
\[
   s^\star = \frac{K'}{zK^\alpha} = \frac{\beta A}{1 + \beta A}.
\]
Substituting the conjecture back and matching the $\log K$ coefficient produces $A = \alpha + \beta A\alpha$, hence $A = \alpha/(1-\alpha\beta)$.  Plugging this $A$ into $s^\star$:
\[
   s^\star \;=\; \frac{\beta\,\alpha/(1-\alpha\beta)}{1 + \beta\,\alpha/(1-\alpha\beta)} \;=\; \frac{\alpha\beta}{1-\alpha\beta+\alpha\beta} \;=\; \alpha\beta.
\]
The DEQN parameterizes $s_t = \sigma\!\bigl(\mathcal{N}_\rho(K_t, z_t)\bigr)$.  Once converged, the average sigmoid output across the ergodic set should equal $\alpha\beta$ (typical calibrations $\alpha=0.36$, $\beta=0.96$ give $s^\star \approx 0.346$).  Any systematic deviation indicates either insufficient training or a quadrature / sampling bias.

\paragraph{Exercise~\ref{ex:ch2:2} (statement: p.~\pageref{ex:ch2:2}): Hard vs.\ soft constraints.}
With a softplus head on consumption alone, $C_t = \mathrm{softplus}(\mathcal{N}_\rho(K_t, z_t)) > 0$ is guaranteed, but next-period capital is then \emph{defined} as the residual $K_{t+1} = z_t K_t^\alpha - C_t$, which is unconstrained in sign.  At random initialization the network output is approximately $\mathcal{N}(0, 1)$, so $C_t$ is approximately $\mathrm{softplus}(0) \approx 0.69$ on average but can be much larger.

Concrete failure: take $K = 1$, $z = 1$, $\alpha = 0.36$, so $zK^\alpha = 1$.  If the network produces an output of, say, $5$, then $C = \mathrm{softplus}(5) \approx 5.007$ and $K_{t+1} = 1 - 5.007 = -4.007 < 0$.  The next iterate $K_{t+1}^{\alpha-1}$ is then complex, and the loss explodes.

The sigmoid-savings parameterization $s_t = \sigma(\mathcal{N}_\rho) \in (0,1)$ avoids this entirely: $K_{t+1} = s_t z_t K_t^\alpha > 0$ and $C_t = (1-s_t) z_t K_t^\alpha > 0$ both hold by construction, regardless of the network's raw output.  This is the simplest example of the broader principle that \emph{architectural} feasibility encodings dominate \emph{loss-based} ones whenever the constraint can be written as a closed-form algebraic identity.

\paragraph{Exercise~\ref{ex:ch2:3} (statement: p.~\pageref{ex:ch2:3}): Path averaging vs.\ conditional expectation.}
On the simulated path, the path-averaged residual is $\bar r_T(\theta) = T^{-1}\sum_{t=1}^T r(\theta, x_t)$ with $\{x_t\}$ generated by the model dynamics.  Under ergodicity, Birkhoff's theorem gives
\[
   \bar r_T(\theta) \;\xrightarrow{T \to \infty}\; \int r(\theta, x)\,\mu(dx) \;=\; \mathbb{E}_\mu[r(\theta, x)] \quad\text{a.s.},
\]
where $\mu$ is the stationary distribution.  The conditional-expectation residual at a fixed point $x_t$, $\E[r(\theta, x_{t+1}) \mid x_t]$, is a different object: it integrates only over the conditional shock distribution, holding the current state fixed.

Their connection: if one sweeps the conditional residual over $x_t \sim \mu$ and averages, the result coincides with $\mathbb{E}_\mu[r(\theta, x)]$ by the law of iterated expectations.  In other words, path averaging \emph{combines} sampling over states (the outer expectation) and the implicit Monte Carlo integration over shocks (one shock per simulated step).  Conditional expectation evaluates the inner integral exactly via quadrature but still requires a way to draw states $x_t$.

Bias--variance trade-off at finite $T_{\text{sim}}$: the path-averaged loss has variance $O(1/T_{\text{sim}})$ from finite-sample noise but no bias.  An exact-quadrature residual has zero stochastic noise on the inner integral but its outer-state coverage depends on the sampling scheme; with mini-batch SGD over the ergodic set, the variance comes from the random batch, not the integration.  In the chapter the deterministic quadrature is preferred whenever the shock dimension is low ($d \lesssim 5$), and pathwise residuals dominate once $d$ is high enough that explicit quadrature becomes expensive.

\paragraph{Exercise~\ref{ex:ch2:4} (statement: p.~\pageref{ex:ch2:4}): Brock--Mirman with Gauss--Hermite.}
The Euler equation in Brock--Mirman with $\delta = 1$, log utility, and AR(1) productivity $\ln z' = \varrho\ln z + \sigma_z\varepsilon'$, $\varepsilon' \sim \mathcal{N}(0,1)$, is
\[
   \frac{1}{C_t}
   \;=\;
   \beta\,\E\!\left[\frac{\alpha\,z'\,K_{t+1}^{\alpha - 1}}{C_{t+1}} \,\Big|\, K_t, z_t\right].
\]
Replace the expectation by a Gauss--Hermite rule.  After the change of variables $\varepsilon' = \sqrt{2}\,\xi$ that absorbs the normalization $1/\sqrt{\pi}$, the $Q$-point GH quadrature is
\[
   \E[h(\varepsilon')] \;\approx\; \frac{1}{\sqrt{\pi}} \sum_{q=1}^{Q} w_q\,h\!\bigl(\sqrt{2}\,\xi_q\bigr),
\]
with classical nodes $\xi_q$ and weights $w_q$ that satisfy $\sum_q w_q = \sqrt{\pi}$.  Table~\ref{tab:gh5_nodes} lists the five-point rule used in this exercise.
\begin{table}[ht]
\centering
\small
\caption{Five-point Gauss--Hermite nodes and weights for the convention $\E[h(\varepsilon)] \approx \pi^{-1/2}\sum_q w_q h(\sqrt{2}\xi_q)$.  The weights sum to $\sqrt{\pi}$ before the outside normalization.}
\label{tab:gh5_nodes}
\begin{tabular}{@{}rrr@{}}
\toprule
$q$ & $\xi_q$ & $w_q$ \\\midrule
1 & $-2.0202$ & $0.0200$ \\
2 & $-0.9586$ & $0.3936$ \\
3 & $0.0000$ & $0.9453$ \\
4 & $+0.9586$ & $0.3936$ \\
5 & $+2.0202$ & $0.0200$ \\
\bottomrule
\end{tabular}
\end{table}
Substituting, the residual at a state $(K_t, z_t)$ becomes
\[
   r(\theta; K_t, z_t)
   \;=\;
   1 \;-\; \frac{\beta\,\alpha\,C_t}{\sqrt{\pi}}\sum_{q=1}^{5} w_q\,\frac{z_q'\,K_{t+1}^{\alpha-1}}{C_{t+1}^{q}},
\]
where $z_q' = \exp(\varrho\ln z_t + \sigma_z\sqrt{2}\,\xi_q)$ and $C_{t+1}^q$ is the consumption obtained by feeding $(K_{t+1}, z_q')$ into the network.  Numerically, comparing this $5$-point sum with a high-draw Monte Carlo estimate on the same residual is a useful diagnostic: agreement should be close for smooth integrands and small shock variance, but the realized discrepancy is an output of the check rather than a fixed benchmark.

\paragraph{Exercise~\ref{ex:ch2:5} (statement: p.~\pageref{ex:ch2:5}): Monomial rule by hand (Stroud-3 at $d=4$).}
Equation~\eqref{eq:stroud3} places nodes at $\bm{x}_k^\pm = \pm\sqrt{d}\,\bm{e}_k$ for $k = 1, \dots, d$, with equal weights $1/(2d) = 1/8$.  At $d=4$, the eight nodes are $\pm 2 \bm{e}_k$ for $k=1,2,3,4$.

\emph{First moment.}  $\E[\varepsilon_i']_{\text{rule}} = (1/8)\sum_k [(\sqrt{d}\,\bm{e}_k)_i + (-\sqrt{d}\,\bm{e}_k)_i] = 0$ by $\pm$-cancellation.

\emph{Second moment.}  $(\varepsilon_i')^2$ is non-zero only at $\pm\sqrt{d}\,\bm{e}_i$, where it equals $d$.  Two such nodes contribute $2d$, weighted by $1/(2d)$, giving exactly $1$.

\emph{Cross moment.}  $\varepsilon_i'\varepsilon_j'$ for $i \neq j$ is zero at every node because each node has only one nonzero coordinate.  So the rule returns $0$, the true value.

\emph{Third moment.}  $(\varepsilon_i')^3$ at $\pm\sqrt{d}\bm{e}_i$ equals $\pm d^{3/2}$; the two values cancel.  Returns $0$, exact.

\emph{Fourth moment.}  $(\varepsilon_i')^4$ at $\pm\sqrt{d}\bm{e}_i$ equals $d^2$, both signs.  Two nodes contribute $2d^2$, weighted by $1/(2d)$, gives $d$.  At $d=4$, the rule returns $4$, while the true value $\E[\varepsilon_i'^{\,4}] = 3$.

\emph{Linear bias growth.}  In general the rule reports $d$ for the fourth moment, so the relative error $(d - 3)/3$ is linear in $d$: $33\%$ at $d=4$, $67\%$ at $d=6$, doubles to $1$ ($100\%$) at $d=9$.

\emph{When does this matter?}  The Euler residual is a smooth function of the next-period shock $\varepsilon'$.  Taylor-expanding around the conditional mean, the leading bias term is the Hessian of the residual with respect to $\varepsilon'$, which probes the second moment, exact under Stroud-3.  Fourth-moment bias enters only at the next order, scaled by the integrand's fourth derivative.  For moderate CRRA curvature and thin-tailed shocks this term can be small relative to classroom residual tolerances, but it is a diagnostic to check rather than a universal bound.  The bias becomes material when (i)~the integrand has heavy fourth-order content (e.g., very risk-averse preferences or fat-tailed shocks), or (ii)~the shock dimension is large enough that the relative error $(d-3)/3$ exceeds the residual tolerance one targets.  This is the threshold at which the monomial rule should be replaced by Stroud-5 ($2d^2 + 1$ nodes) or QMC.

\paragraph{Exercise~\ref{ex:ch2:6} (statement: p.~\pageref{ex:ch2:6}): Loss-kernel selection.}
\emph{Definitions for reference.}  MSE: $\frac{1}{N}\sum_i r_i^2$.  MAE: $\frac{1}{N}\sum_i |r_i|$.  Huber($\delta$): quadratic for $|r| \le \delta$, linear above.  Pinball loss at $\tau$: $L_\tau(r) = \max(\tau r,\, (\tau - 1)r)$, whose minimizer is the $\tau$-quantile of $r$.  CVaR at $\alpha$: the expected value of $r$ conditional on $r$ exceeding its $\alpha$-quantile.  Log-cosh: $\sum_i \log\cosh(r_i)$, smooth and quadratic near zero, linear in tails.

(a) \textbf{Huber loss}: ``smooth, quadratic near zero, linear in tails'' is the literal definition.  MAE is also linear-in-tails but is not differentiable at $r=0$, so its gradient flips discontinuously and the optimizer stalls; log-cosh is smooth and shares the asymptotic shape with Huber but has no tunable threshold.  Huber($\delta$) gives the cleanest control of where the regime change happens.

(b) \textbf{CVaR at $\alpha = 0.99$}.  The CVaR loss optimizes the conditional mean above the $99$th-percentile threshold, which is exactly what the regulator audits.  The pinball loss at $\tau = 0.99$ would only target the $99$th-percentile residual itself, not the conditional average above it; if the residuals have a fat right tail, the pinball-trained policy can have arbitrarily large worst-case $1\%$ residuals.  CVaR is the right primitive for ``worst-case mean'' control.

(c) \textbf{MAE} (or equivalently pinball at $\tau = 0.5$).  By construction MAE's first-order condition is solved at the median, not the mean: $\partial/\partial r |r| = \mathrm{sign}(r)$, so the gradient contribution is $\pm 1$ per residual, regardless of magnitude.  This is exactly what the desideratum asks for, no single tail residual dominates the gradient.  Huber would also down-weight tails but still tracks the mean below the threshold; the user explicitly wants median-targeting.

\paragraph{Exercises~\ref{ex:ch2:7} and~\ref{ex:ch2:8} (statements: p.~\pageref{ex:ch2:7}, p.~\pageref{ex:ch2:8}).}
These are coding exercises (notebook \tpath{lecture_03_02_Brock_Mirman_Uncertainty_DEQN.ipynb}); reference outputs and timing curves are in the companion repository.  Qualitative anchors: in Ex.~\ref{ex:ch2:7}, the per-epoch wall time of tensor-product Gauss--Hermite ($Q^d$ nodes) should grow exponentially in $d$ while Stroud-3 ($2d$ nodes) grows linearly, with a crossover that is visible already at $d=4$--$5$ on a single GPU; the relative Euler error should track the integration accuracy of each rule, with Stroud-3 inheriting the fourth-moment bias of \S\ref{sec:monomial_cubature}.  In Ex.~\ref{ex:ch2:8}, swapping Swish for $\tanh$ typically slows time-to-converge by tens of percent on a smooth problem like Brock--Mirman because $\tanh$ saturates faster, but final accuracy is comparable; convergence should still hold under the same hyperparameters.

\section{Chapter \ref{ch:irbc}: The International Real Business Cycle Model}
\label{sol:ch3}

\paragraph{Exercise~\ref{ex:ch3:1} (statement: p.~\pageref{ex:ch3:1}): Fischer--Burmeister.}
For the forward direction, suppose $a \ge 0$, $b \ge 0$, $ab = 0$.  Without loss of generality $b = 0$; then $\Phi(a, 0) = a + 0 - \sqrt{a^2 + 0} = a - |a| = 0$ since $a \ge 0$.  By symmetry the same holds when $a = 0$.

For the reverse direction, suppose $\Phi(a,b) = 0$, i.e.\ $a + b = \sqrt{a^2 + b^2}$.  The right-hand side is non-negative, so $a + b \ge 0$.  Squaring: $(a+b)^2 = a^2 + b^2 \Rightarrow 2ab = 0 \Rightarrow ab = 0$.  Combined with $a + b \ge 0$ and $ab = 0$, the only possibility is one of $a, b$ being zero and the other non-negative, i.e.\ $a, b \ge 0$ and $ab = 0$.  $\square$

The level set $\Phi(a, b) = 0$ is the union of the non-negative $a$-axis and the non-negative $b$-axis (an L-shape in the $(a,b)$ plane).  The smoothed variant $\Phi_\varepsilon(a,b) = a + b - \sqrt{a^2 + b^2 + \varepsilon^2}$ rounds the corner at the origin: at $(a,b) = (0,0)$ one finds $\Phi_\varepsilon(0,0) = -\varepsilon \neq 0$, while far from the origin $\sqrt{a^2 + b^2 + \varepsilon^2} \approx \sqrt{a^2 + b^2}$ and the smoothed level set converges to the unsmoothed L-shape.  In a numerical setting the smoothing eliminates the gradient kink at the origin, which is convenient for AD but distorts the strict KKT zero set by an $O(\varepsilon)$ amount.

\emph{(c) Gradient direction.}  At an interior point of the open positive quadrant, $\nabla\Phi(a,b) = (1 - a/\sqrt{a^2+b^2},\, 1 - b/\sqrt{a^2+b^2})$.  At $(a,b) = (1,1)$ this evaluates to $(1-1/\sqrt 2,\, 1-1/\sqrt 2) \approx (0.293, 0.293)$, both components strictly positive.  The raw gradient $\nabla\Phi$ therefore points northeast, \emph{away} from the L-shaped zero set, so the bare residual $\Phi$ on its own is not the right object for SGD to descend on.  What does descend toward the zero set is the squared loss $\Phi^2$: by the chain rule $\nabla(\Phi^2) = 2\Phi\,\nabla\Phi$, and inside the open positive quadrant $\Phi(a,b) > 0$, so $-\nabla(\Phi^2) = -2\Phi\,\nabla\Phi$ has both components strictly negative at $(1,1)$ and points southwest, i.e.\ back toward the closer feasible axis.  This is the operative observation for training: SGD on $\Phi^2$ is what pulls infeasible iterates back to the L; the raw gradient $\nabla\Phi$ on its own would push them away.

\emph{(d) Sign convention.}  Replacing $\Phi$ by $-\Phi$ leaves the squared loss untouched: $(-\Phi)^2 = \Phi^2$, so the gradient field of the loss and the SGD trajectory are unchanged.  In particular, the sign convention $a + b - \sqrt{a^2+b^2}$ versus $\sqrt{a^2+b^2} - a - b$ is irrelevant once the residual is squared, and the operative quantity for SGD is $-\nabla(\Phi^2)$ rather than $-\nabla\Phi$.  The L-shaped zero set is invariant under sign flip; only the value of $\Phi$ at off-zero points changes (it flips sign), and squaring removes that distinction.

\paragraph{Exercise~\ref{ex:ch3:2} (statement: p.~\pageref{ex:ch3:2}): State-space scaling.}
For an IRBC with $N$ symmetric countries, the state vector contains $(K^j, z^j)_{j=1}^N$ for a total of $2N$ components.  The network outputs the next-period capital vector $(k^{j\prime})_{j=1}^N$, the irreversibility multipliers $(\mu^j)_{j=1}^N$, and the resource-constraint shadow price $\lambda$, for $2N + 1$ outputs in total.  Country-level consumption is recovered algebraically from $\lambda$ via the consumption-sharing FOC and is therefore not a separate output.  The loss has $N$ Euler residuals, $N$ Fischer--Burmeister residuals from the irreversibility complementarity, and $1$ aggregate-resource-constraint residual, $2N + 1$ components total, matching the $2N + 1$ outputs.  Each Euler residual is an expectation over a $(N+1)$-dimensional shock vector (one country-specific innovation plus one aggregate, as in equation~\eqref{eq:irbc_tfp}).

Tensor-product Gauss--Hermite at $Q=3$ costs $3^{N+1}$ evaluations per residual.  Setting $3^{N+1} > 10^4$ gives $N + 1 > \log_3(10^4) \approx 8.38$, so $N \ge 8$ already exceeds the threshold, and $N = 9$ overshoots by a factor of nearly $6$ (cost $3^{10} = 59\,049$).

The Stroud-3 rule has $2(N+1)$ nodes per residual.  Setting $2(N+1) < 100$ gives $N \le 48$, comfortable for any IRBC dimension actually used in practice.  At $N = 50$ the monomial cost is $102$ nodes; the tensor cost is $3^{51} \approx 2.15 \times 10^{24}$ evaluations.  This four-order-of-magnitude gap at $N = 10$ and twenty-order-of-magnitude gap at $N = 50$ is the operational reason DEQNs use Stroud-3 by default once $N \gtrsim 5$.

\paragraph{Exercise~\ref{ex:ch3:3} (statement: p.~\pageref{ex:ch3:3}): Two-phase training.}
At a randomly initialized network, the policy outputs $k^{j\prime}$ are not coordinated with the resource constraint.  Suppose the network produces $k^{j\prime}$ values that, summed and combined with country-$j$ consumption, exceed total output: $\sum_j (k^{j\prime} + c^j + \Gamma^j) > \sum_j Y^j$.  The implied state on the next simulated step has $k^{j\prime} < (1-\delta) k^j$ for some country (irreversibility violated), or $c^j < 0$ (negative consumption), or both.  Concretely, take $N = 2$, $A_\mathrm{tfp} = 1$, $\delta = 0.025$, $k^1 = k^2 = 1$, and a random network output $(k^{1\prime}, k^{2\prime}) = (0.5, 0.5)$ (instead of the symmetric $(1, 1)$ steady state).  Capital has dropped by $50\%$ in one step, so $I^j = k^{j\prime} - (1-\delta)k^j = 0.5 - 0.975 = -0.475 < 0$, violating irreversibility.  Even before the irreversibility check fires, the implied consumption $c^j = Y^j - I^j - \Gamma^j$ becomes huge, and in the next step the marginal utility $u'(c)$ is essentially zero, so the Euler residual gradient is uninformative.

Phase~1 (uniform sampling on a wide box of states, with Euler residuals computed against \emph{any} feasible policy guess) gives the optimizer signal to bring outputs into the feasible region before any simulation is attempted.  Once the policy is in a feasible neighbourhood, Phase~2 (simulation-based sampling on the ergodic set) refines accuracy.  Without Phase~1, the simulation in Phase~2 starts in regions where the policy is grossly infeasible, the loss explodes, and gradient descent diverges.

\paragraph{Exercise~\ref{ex:ch3:4} (statement: p.~\pageref{ex:ch3:4}): Adjustment-cost partials and Tobin's Q.}
Write $g^j \equiv k^{j\prime}/k^j - 1$.  Then $\Gamma^j = (\kappa/2)\,k^j (g^j)^2$.  The partials are
\begin{align*}
\frac{\partial \Gamma^j}{\partial k^{j\prime}} &= \frac{\kappa}{2}\,k^j \cdot 2 g^j \cdot \frac{1}{k^j} \;=\; \kappa\,g^j \;=\; \kappa\!\left(\frac{k^{j\prime}}{k^j} - 1\right), \\
\frac{\partial \Gamma^j}{\partial k^j} &= \frac{\kappa}{2}(g^j)^2 + \frac{\kappa}{2}\,k^j \cdot 2 g^j \cdot \!\left(-\frac{k^{j\prime}}{(k^j)^2}\right) \\
&= \frac{\kappa}{2}\,(g^j)^2 - \kappa\,g^j\!\left(\frac{k^{j\prime}}{k^j}\right)
   \;=\; \frac{\kappa}{2}\!\left[\bigl(g^j\bigr)^2 - 2(1+g^j) g^j\right] \\
&= -\frac{\kappa}{2}\!\left[(g^j)^2 + 2 g^j\right] \;=\; \frac{\kappa}{2}\!\left[1 - \bigl(\tfrac{k^{j\prime}}{k^j}\bigr)^{\!2}\right],
\end{align*}
matching equation~\eqref{eq:irbc_adjcost_derivs}.

At the steady state, $k^{j\prime} = k^j$ so $g^j = 0$.  Both partials vanish: $\partial\Gamma^j/\partial k^{j\prime} = 0$ and $\partial\Gamma^j/\partial k^j = 0$.  The adjustment cost itself $\Gamma^j(k^j, k^j) = 0$, and its first-order presence in the resource constraint and the Euler equation drops out.  The deterministic steady state of the IRBC with adjustment costs is therefore identical to the frictionless steady state, $k_\mathrm{ss}$ being pinned down by $\beta(1 - \delta + \zeta A_\mathrm{tfp} k^{\zeta-1}) = 1$.

The expression $\partial\Gamma^j/\partial k^{j\prime} = \kappa g^j$ is the marginal cost of investing one more unit and is the IRBC's analogue of \emph{Tobin's marginal Q}: large positive $g^j$ (rapid expansion) creates a large investment wedge, raising the effective per-unit cost of capital next period.  Higher $\kappa$ flattens the response of investment to a productivity shock: a large adjustment cost makes the planner spread the response of $k^{j\prime}$ across several periods rather than absorbing the shock in one big move, slowing convergence to the new steady state.  In the limit $\kappa \to \infty$, capital adjusts only infinitesimally each period and the dynamics become arbitrarily slow.

\paragraph{Exercise~\ref{ex:ch3:5} (statement: p.~\pageref{ex:ch3:5}): Complete-markets risk sharing.}
The consumption-sharing condition~\eqref{eq:irbc_consumption} reads $c_t^j = (\lambda_t/\tau^j)^{-\gamma_j}$.  Therefore
\[
   \frac{c_t^i}{c_t^j} \;=\; \!\left(\frac{\lambda_t}{\tau^i}\right)^{\!-\gamma_i}\!\Bigl/\!\left(\frac{\lambda_t}{\tau^j}\right)^{\!-\gamma_j}
   \;=\;
   \!\left(\frac{\tau^i}{\lambda_t}\right)^{\!\gamma_i}\!\!\left(\frac{\lambda_t}{\tau^j}\right)^{\!\gamma_j}
   \;=\;
   (\tau^i)^{\gamma_i}\,(\tau^j)^{-\gamma_j}\,\lambda_t^{\,\gamma_j - \gamma_i}.
\]

(i) With homogeneous IES $\gamma_i = \gamma_j = \gamma$, the $\lambda_t$ exponent vanishes and the ratio collapses to a time-invariant constant:
\[
   \frac{c_t^i}{c_t^j} \;=\; \!\left(\frac{\tau^i}{\tau^j}\right)^{\!\gamma}.
\]
Since $\log(c_t^i/c_t^j)$ is constant, $\Delta\log c_t^i = \Delta\log c_t^j$, the perfect risk-sharing prediction of \citet{backus1992international}: cross-country consumption growth is perfectly correlated (correlation $= 1$).

(ii) With heterogeneous IES $\gamma_i \neq \gamma_j$, the exponent $\gamma_j - \gamma_i$ is non-zero and $\lambda_t$ enters the ratio.  As shocks move the planner's shadow price, the consumption ratio fluctuates: low-IES countries' consumption is less sensitive to $\lambda_t$ than high-IES countries'.  But the log growth rate is still
\[
   \Delta \log c_t^j = -\gamma_j\,\Delta\log\lambda_t .
\]
Thus, for positive $\gamma_i,\gamma_j$, any pair of country consumption-growth rates is a positive scalar multiple of the same aggregate shock $\Delta\log\lambda_t$.  The correlation remains one; heterogeneous IES changes relative consumption-growth volatility, not the correlation, in this complete-markets planner allocation.

(iii) The empirical consumption-correlation puzzle is that real-world cross-country consumption growth correlations sit well below the near-perfect correlations implied by the complete-markets benchmark.  Heterogeneous IES alone does not break the common-shadow-price structure; closing the gap with the data requires incomplete markets, wedges, nontraded goods, preference nonseparabilities, or other frictions that break full insurance.

\paragraph{Exercises~\ref{ex:ch3:6} and~\ref{ex:ch3:7} (statements: p.~\pageref{ex:ch3:6}, p.~\pageref{ex:ch3:7}).}
These are notebook exercises (\tpath{lecture_05_05_IRBC_Exercise.ipynb}); reference solutions are in the notebook itself, behind the ``attempt first'' dividers.  Qualitative anchors: in Ex.~\ref{ex:ch3:6}, the closed-form steady state $k_\mathrm{ss} = \bigl[(1/\beta - 1 + \delta)/(\zeta A_\mathrm{tfp})\bigr]^{1/(\zeta-1)}$ falls in all three scenarios --- a higher $\delta$ raises the required net return $1/\beta - 1 + \delta$, a lower $\beta$ raises $1/\beta$, and a lower $\zeta$ shrinks the multiplicative term while (since $\zeta - 1 < 0$) steepening diminishing returns --- so $k_\mathrm{ss}$ is decreasing in $\delta$, increasing in $\beta$, and increasing in $\zeta$ around the baseline, with $c_\mathrm{ss} = A_\mathrm{tfp} k_\mathrm{ss}^\zeta - \delta k_\mathrm{ss}$ following by substitution.  In Ex.~\ref{ex:ch3:7}, inverse-loss weighting $w_i \propto 1/\ell_i$ equalizes the per-component contributions $w_i \ell_i$, so the smallest-magnitude residuals (here the Fischer--Burmeister terms) receive the largest weight; the speed-up is largest when a component is small because it is \emph{hard to fit}, and the scheme can hurt when a component is small because it is \emph{already satisfied by construction} (e.g., a hard-coded resource constraint), in which case up-weighting it merely amplifies noise.

\section{Chapter \ref{ch:nas}: Neural Architecture Search and Loss Normalization}
\label{sol:ch4}

\paragraph{Exercise~\ref{ex:ch4:1} (statement: p.~\pageref{ex:ch4:1}): Random vs.\ grid.}
With two hyperparameters and only one ``important'' axis, a $3\times 3$ grid uses $9$ candidates but only $3$ distinct values along the important axis.  If the near-optimal interval has length fraction $p$ and its location relative to the grid is unknown, the grid hit probability is approximately $\min\{3p,1\}$ when $p$ is small.  Random search at $9$ evaluations samples $9$ independent values along the important axis, so its hit probability is
\[
   1 - (1-p)^9 .
\]
For $p=0.05$, the grid probability is approximately $0.15$, while random search gives $1-0.95^9\approx 0.37$.

The general principle: with $n$ evaluations and only $r \ll d$ important axes, random search effectively gives $n$ independent draws on those $r$ axes (since the irrelevant axes don't matter), while grid search wastes most of its budget on the irrelevant axes' marginals.  This is the projection argument of \citet{bergstra2012random}: when the loss landscape is anisotropic, random search dominates grid search.

\paragraph{Exercise~\ref{ex:ch4:2} (statement: p.~\pageref{ex:ch4:2}): Bayesian optimization toy problem.}
This is a coding exercise.  A grid with step size $0.01$ over $[-1,2]$ has $301$ points, so the qualitative benchmark is that BO should require far fewer objective evaluations when the GP posterior and acquisition function identify the promising region early.  The exact evaluation count is a notebook output and depends on the initial design, acquisition optimizer, random seed, and stopping rule.

\paragraph{Exercise~\ref{ex:ch4:3} (statement: p.~\pageref{ex:ch4:3}): Hyperband budget allocation.}
Hyperband with $R = 81$, $\eta = 3$ runs a ladder of brackets indexed by $s = s_{\max}, s_{\max}-1, \dots, 0$, where $s_{\max} = \lfloor\log_\eta R\rfloor = 4$.  Each bracket starts with $n_s = \lceil (s_{\max}+1)\,\eta^s / (s+1)\rceil$ candidates trained for $r_s = R / \eta^s$ resource each, then runs Successive Halving with reduction factor $\eta$.  Table~\ref{tab:hyperband_r81_eta3} works out the resulting schedule.

\begin{table}[ht]
\centering
\small
\caption{Hyperband schedule for maximum resource $R=81$ and reduction factor $\eta=3$.  Each row reports the successive-halving rungs inside one bracket and the total resource consumed by that bracket.}
\label{tab:hyperband_r81_eta3}
\begin{tabular}{@{}r L{9.2cm} r@{}}
\toprule
$s$ & \textbf{SHA rungs $(n\times r)$} & \textbf{Total} \\
\midrule
$4$ & $81{\times}1 \to 27{\times}3 \to 9{\times}9 \to 3{\times}27 \to 1{\times}81$ & $405$ \\
$3$ & $34{\times}3 \to 11{\times}9 \to 3{\times}27 \to 1{\times}81$ & $363$ \\
$2$ & $15{\times}9 \to 5{\times}27 \to 1{\times}81$ & $351$ \\
$1$ & $8{\times}27 \to 2{\times}81$ & $378$ \\
$0$ & $5{\times}81$ & $405$ \\
\bottomrule
\end{tabular}
\end{table}

Within each bracket, Successive Halving reduces candidates by $\eta$ at each rung and increases the resource per surviving candidate by $\eta$.  Therefore the total cost must sum all rungs, not only the first rung.  With the floor/ceil schedule above, total Hyperband budget is
\[
   405 + 363 + 351 + 378 + 405 = 1902
\]
resource units, just below the loose worst-case bound $(s_{\max}+1)^2 R = 25 \cdot 81 = 2025$.

A naive ``train all $n_0 = 27$ candidates to full $R = 81$'' costs $27 \cdot 81 = 2187$ resource units.  Hyperband is only moderately cheaper in total resource here, but it screens a much larger initial pool: across the five brackets, the total number of first-rung candidates is $81 + 34 + 15 + 8 + 5 = 143$, vs.\ $27$ for the naive scheme.  Its advantage is adaptive allocation: many more candidates are sampled, but only a small subset receives large training budgets.

\paragraph{Exercise~\ref{ex:ch4:4} (statement: p.~\pageref{ex:ch4:4}): Loss balancing.}
Let $\ell_i^{(t)}$ have magnitude $L_i \in \{10^0, 10^{-2}, 10^{-4}\}$ and per-component gradient norm $\|\nabla\ell_i\|$.  If we crudely model gradient norm as scaling linearly with loss magnitude (true for, e.g., quadratic losses far from the optimum), then $\|\nabla \ell_i\| \propto L_i$.  Equalising gradient contributions $\lambda_i \|\nabla\ell_i\|$ requires $\lambda_i \propto 1/L_i$, i.e.\ $(\lambda_1, \lambda_2, \lambda_3) \propto (1, 10^2, 10^4)$, which after normalisation becomes
\[
   (\lambda_1, \lambda_2, \lambda_3) \;=\; \frac{1}{1 + 10^2 + 10^4}\,(1, 10^2, 10^4) \;\approx\; (10^{-4},\, 10^{-2},\, 1).
\]

The scheme breaks down when the gradients are correlated: $\langle \nabla\ell_i, \nabla\ell_j\rangle \neq 0$ means that scaling up $\lambda_3$ to ``boost'' $\ell_3$ also moves $\theta$ along the $\nabla\ell_1$ direction, changing $\ell_1$.  The ``equal contribution'' targeted by the fixed weights is no longer a fixed point: each parameter update changes the local gradient geometry, and weights tuned at one iteration become wrong at the next.  Adaptive schemes (ReLoBRaLo, GradNorm) re-tune the $\lambda_i$ at every step, recovering the equalisation in a way that fixed weights cannot.

\paragraph{Exercise~\ref{ex:ch4:5} (statement: p.~\pageref{ex:ch4:5}): Pareto front geometry.}
\emph{(i)} Differentiate $\mathcal{L}(\theta;\lambda) = \lambda(\theta-a)^2 + (1-\lambda)(\theta-b)^2$ in $\theta$ and set to zero: $2\lambda(\theta - a) + 2(1-\lambda)(\theta - b) = 0$, hence
\[
   \theta^\star(\lambda) \;=\; \lambda a + (1-\lambda) b.
\]

\emph{(ii)} Substituting back:
\[
   \ell_1^\star(\lambda) \;=\; (\theta^\star - a)^2 \;=\; (1-\lambda)^2 (b-a)^2,
   \qquad
   \ell_2^\star(\lambda) \;=\; (\theta^\star - b)^2 \;=\; \lambda^2 (b-a)^2.
\]

\emph{(iii)} Take square roots: $\sqrt{\ell_1^\star} = (1-\lambda)(b-a)$ and $\sqrt{\ell_2^\star} = \lambda(b-a)$.  Adding:
\[
   \sqrt{\ell_1^\star} + \sqrt{\ell_2^\star} \;=\; (b - a),
\]
independent of $\lambda$.  This is a convex curve in $(\ell_1, \ell_2)$ space (a quarter of an astroid-like arc) running from $(0, (b-a)^2)$ at $\lambda = 1$ to $((b-a)^2, 0)$ at $\lambda = 0$.

\emph{(iv)} At $\lambda = 1/2$, $\theta^\star = (a+b)/2$, the midpoint, and $\ell_1^\star = \ell_2^\star = (b-a)^2/4$.  This sits on the symmetric axis of the front.

\emph{(v)} In the one-dimensional toy problem the trade-off is completely described by the curve above.  In a neural network, however, $\theta$ is high-dimensional and the descent direction for fixed scalar weight $\lambda$ is
\[
   -\bigl[\lambda \nabla \ell_1(\theta) + (1-\lambda)\nabla \ell_2(\theta)\bigr].
\]
If $\langle\nabla\ell_1,\nabla\ell_2\rangle>0$, reducing one component tends to reduce the other; if the inner product is negative, progress on one component can increase the other.  Because this geometry changes along the training path, a fixed scalar weight can balance progress at one iterate and become badly unbalanced later.  ReLoBRaLo responds to this by increasing the weight of losses whose \emph{relative loss progress} has lagged behind.  GradNorm is the related method that targets gradient magnitudes directly by trying to balance $\|w_k\nabla\ell_k\|$ across components.

\paragraph{Exercise~\ref{ex:ch4:6} (statement: p.~\pageref{ex:ch4:6}): ReLoBRaLo vs.\ GradNorm.}
This is a coding exercise.  Qualitatively: GradNorm requires one extra backward pass per component per step (to compute $\|\nabla \ell_k\|$), so wall-clock per epoch is roughly $K\times$ slower than ReLoBRaLo for $K$ components.  In return, GradNorm achieves tighter gradient balance, which matters when component magnitudes are not a faithful proxy for gradient magnitudes (e.g., when the loss landscape is anisotropic).  For the standard PINN losses studied in this script ($K = 2$ or $3$ components, typically scaled by physical units), ReLoBRaLo's loss-magnitude proxy is usually good enough and the extra cost of GradNorm is not warranted; for losses with strongly heterogeneous Hessians (e.g., HJB residuals coupled with KFE residuals where the two operators have different stiffness), GradNorm's direct gradient measurement can give a meaningful speedup.

\paragraph{Exercise~\ref{ex:ch4:7} (statement: p.~\pageref{ex:ch4:7}): HPO vs.\ full NAS decision.}
\emph{Sketch.}  The four cells are roughly:

\begin{itemize}
\item \emph{(a, i) RTX 3060 + fixed-topology MLP search:} \textbf{Random Search with Successive Halving.}  Search space is small (a few thousand candidate combinations), the GPU is too small for graph-level NAS, and SH amortizes the budget across many candidates by killing weak ones early.  Bayesian Optimization helps marginally but adds GP-fitting overhead that is not worth it on a single GPU.
\item \emph{(a, ii) RTX 3060 + graph-level NAS:} \textbf{Random Search.}  Full DARTS-style NAS would not fit in $12$ GB of VRAM (the supernetwork concept-search needs several model copies in memory simultaneously); a small fixed pool of architectures evaluated at low resource is the only feasible option.
\item \emph{(b, i) A100 + fixed-topology MLP search:} \textbf{Bayesian Optimization}.  The A100's compute headroom makes the per-step BO overhead negligible, and the search space's smooth landscape (continuous learning rate, ordinal depth/width) is exactly where GP surrogates dominate Random Search.
\item \emph{(b, ii) A100 + graph-level NAS:} \textbf{Full graph-level NAS} (DARTS or evolutionary).  The A100's memory and compute support the supernetwork training that DARTS requires; the search space has too many discrete connectivity choices for any HPO method to cover.
\end{itemize}

The general rule: budget grows $\to$ smarter methods become affordable; search-space size grows $\to$ the marginal benefit of smarter methods grows; per-method overhead per evaluation needs to fit inside one GPU step or it eats into the budget itself.

\section{Chapter \ref{ch:olg}: Overlapping Generations Models with DEQNs}
\label{sol:ch5}

\paragraph{Exercise~\ref{ex:ch5:1} (statement: p.~\pageref{ex:ch5:1}): OLG market clearing for $A=3$.}
With three cohorts (young $h=1$, middle $h=2$, old $h=3$), the budget constraints are
\begin{align*}
   c^1_t &= w_t \,\ell^1 - k^2_{t+1}, \\
   c^2_t &= w_t \,\ell^2 + R_t k^2_t - k^3_{t+1}, \\
   c^3_t &= w_t \,\ell^3 + R_t k^3_t,
\end{align*}
where $w_t, R_t$ are equilibrium prices and $\ell^h$ are exogenous lifecycle labor endowments (the old cohort consumes its capital).  Two Euler equations determine the savings of the young and middle cohorts:
\begin{align*}
u'(c^1_t) &= \beta\,\mathbb{E}_t[u'(c^2_{t+1})\,R_{t+1}], \\
u'(c^2_t) &= \beta\,\mathbb{E}_t[u'(c^3_{t+1})\,R_{t+1}].
\end{align*}
The market-clearing condition closes the system:
\[
   k^2_{t+1} + k^3_{t+1} \;=\; K_{t+1},
\]
where $K_{t+1}$ is aggregate capital.

\emph{Equation count}: three budget constraints (used to eliminate consumption), two Euler equations, one capital-market-clearing identity.  The budget constraints are bookkeeping, so the network outputs the two cohort savings $(k^2_{t+1}, k^3_{t+1})$ and is trained on the two Euler residuals; aggregate capital $K_{t+1} = k^2_{t+1} + k^3_{t+1}$ then determines next-period prices $(w_{t+1}, R_{t+1})$ algebraically through the firm FOCs.  The unknown count (two savings) matches the Euler-residual count (two), and the market-clearing identity is built into the definition of $K_{t+1}$.

\paragraph{Exercise~\ref{ex:ch5:2} (statement: p.~\pageref{ex:ch5:2}): KKT under FB.}
For agent $h$ with borrowing constraint $k'^h \ge 0$ and Lagrange multiplier $\lambda^h \ge 0$, the KKT system is
\[
   k'^h \ge 0, \qquad \lambda^h \ge 0, \qquad k'^h \cdot \lambda^h = 0.
\]
These three conditions are encoded by the single Fischer--Burmeister equation $\Phi(\lambda^h, k'^h) = 0$ with $\Phi(a,b) = a + b - \sqrt{a^2 + b^2}$.  The Euler equation
\[
   u'(c^h_t) - \beta\,\mathbb{E}_t[u'(c^{h+1}_{t+1})\,R_{t+1}] - \lambda^h_t \;=\; 0
\]
contributes a second residual.  Squared and summed:
\[
   \mathcal{L}^h(\theta) \;=\; \bigl[\text{Euler residual}\bigr]^2 + \bigl[\Phi(\lambda^h, k'^h)\bigr]^2.
\]
At any KKT point, both squared terms vanish exactly: the Euler residual is zero by definition, and $\Phi = 0$ encodes complementarity.  This is what ``vanishes exactly at the KKT point'' means: there is no $\varepsilon$ smoothing or penalty parameter, the loss is zero on the equilibrium and strictly positive off it.  Compare with a quadratic penalty $\bigl[\max(0, -k'^h)\bigr]^2 + \mu \bigl[\max(0, -\lambda^h)\bigr]^2 + (k'^h \lambda^h)^2$: this also vanishes at the KKT point but the multiplier $\mu$ has to be tuned, whereas FB has no parameter.

\paragraph{Exercise~\ref{ex:ch5:3} (statement: p.~\pageref{ex:ch5:3}): Hump-shaped lifecycle.}
A hump-shaped labor-income profile $\ell^h$ peaks at middle age and declines toward retirement.  The lifecycle savings policy $k'^h$ inherits this hump for two reasons.  (i)~\emph{Consumption smoothing}: agents with high current income $w \ell^h$ relative to lifetime average save heavily to fund retirement years when $\ell^h$ drops.  (ii)~\emph{Time-varying borrowing constraint}: young agents have low income, want to borrow against future earnings, are constrained by $k'^h \ge 0$, so they save little; middle-aged agents are unconstrained and save the most; old agents dis-save toward death.

The expected shape: $k'^h \approx 0$ for the very young (constrained), peaks around age 40--50 (middle of working life), declines toward retirement, drops to zero for the oldest cohort that does not save into the next period.  In notebook \tpath{lecture_08_10_OLG_Benchmark_DEQN_persistent.ipynb}, plotting the trained network's $k'^h$ against cohort age $h$ should reveal exactly this single-peak shape; the position of the peak depends on the calibration of $(\beta, \delta, A_\mathrm{tfp})$ and on the lifecycle labor profile.

\paragraph{Exercise~\ref{ex:ch5:4} (statement: p.~\pageref{ex:ch5:4}): Hard aggregation layer.}
Coding exercise.  The current analytic notebook already clears the capital market by defining $K_{t+1}$ as the sum of predicted cohort savings.  The alternative hard-layer variant is useful when the network has a separate aggregate-capital head.  Implementation sketch: output a positive scalar $\widehat K_{t+1}$ and unnormalised cohort scores $(z^2, \dots, z^A)$; apply softmax along the cohort axis, $s^h = \mathrm{softmax}(z^h)$; rescale to capital:
\[
   k_{t+1}^h \;=\; \widehat K_{t+1} \cdot s^h, \qquad
   \sum_{h=2}^{A} k_{t+1}^h \;=\; \widehat K_{t+1}\;\;\text{by construction}.
\]
The market-clearing residual $\sum_h k^h - \widehat K_{t+1}$ is identically zero up to floating-point precision.  The comparison with the current notebook should therefore focus on Euler residuals and wall-clock time: the hard layer removes one possible inconsistency but also changes the parameterisation, so faster convergence is an empirical question rather than a mathematical guarantee.  In multi-asset settings each exact clearing condition needs its own accounting layer, which is why the 56-agent benchmark enforces bond-market clearing as an explicit residual instead.

\paragraph{Exercise~\ref{ex:ch5:5} (statement: p.~\pageref{ex:ch5:5}): Bond pricing in equilibrium.}
Cohort $h$'s Euler equation for capital is
\[
   u'(c^h_t) \;=\; \beta\,\mathbb{E}_t\!\bigl[u'(c^{h+1}_{t+1})\,R_{t+1}\bigr],
\]
and for the riskless bond that costs $p_t$ today and pays one unit of consumption next period,
\[
   u'(c^h_t)\,p_t \;=\; \beta\,\mathbb{E}_t\!\bigl[u'(c^{h+1}_{t+1})\bigr].
\]
The second equation gives directly
\[
   p_t \;=\; \frac{\beta\,\mathbb{E}_t[u'(c^{h+1}_{t+1})]}{u'(c^h_t)}.
\]
Identifying the stochastic discount factor $M_{t,t+1} = \beta\,u'(c^{h+1}_{t+1})/u'(c^h_t)$, the same expression reads $p_t = \mathbb{E}_t[M_{t,t+1}]$.

For the risk-premium decomposition, divide the capital Euler by $u'(c^h_t)$ and use the covariance identity $\mathbb{E}_t[XY] = \mathbb{E}_t[X]\mathbb{E}_t[Y] + \mathrm{Cov}_t(X,Y)$:
\[
   1 \;=\; \mathbb{E}_t[M_{t,t+1}\,R_{t+1}]
       \;=\; \mathbb{E}_t[M_{t,t+1}]\,\mathbb{E}_t[R_{t+1}] + \mathrm{Cov}_t(M_{t,t+1}, R_{t+1}).
\]
Substituting $p_t = \mathbb{E}_t[M_{t,t+1}]$:
\[
   \frac{1}{p_t} \;=\; \mathbb{E}_t[R_{t+1}] + \frac{\mathrm{Cov}_t(M_{t,t+1}, R_{t+1})}{p_t},
\]
which after rearrangement gives the textbook risk-premium decomposition: the gap between expected gross capital return and the riskless rate equals minus the SDF--return covariance, scaled by $1/p_t$.

\emph{When the collateral constraint binds.}  If the collateral constraint is active, with non-negative KKT multiplier $\mu^h_t$ entering the bond FOC with coefficient $\kappa$ (the same $\kappa$ that controls the constraint $k'^h + \kappa b'^h \ge 0$), the bond Euler equation becomes
\[
   u'(c^h_t)\,p_t \;=\; \beta\,\mathbb{E}_t\!\bigl[u'(c^{h+1}_{t+1})\bigr] + \kappa\,\mu^h_t,
\]
so the equilibrium bond price is
\[
   p_t \;=\; \frac{\beta\,\mathbb{E}_t[u'(c^{h+1}_{t+1})] + \kappa\,\mu^h_t}{u'(c^h_t)}.
\]
The unconstrained SDF expression $p_t = \mathbb{E}_t[M_{t,t+1}]$ is recovered when $\mu^h_t = 0$.  The multiplier wedge raises $p_t$, equivalently lowers the implicit safe rate that $1/p_t$ tracks, because the constrained agent values one extra unit of bond consumption tomorrow more than the unconstrained agent.  In the 56-agent benchmark, the cohorts most likely to bind are the youngest (lowest income, highest desire to borrow against future earnings), so any wedge $\kappa\,\mu^h_t$ from the binding-cohort population enters the cross-sectional pricing equation.

\emph{Why no bond residual in 6-agent OLG?}  In the analytic 6-agent calibration there is only one asset (capital).  Bonds are absent, so no separate market-clearing residual is needed.  The single-asset Euler equation pins down the implicit safe rate via $1/p = \mathbb{E}[R]$ minus the appropriate covariance, but no quantity needs to clear because no bond is traded.

\paragraph{Exercises~\ref{ex:ch5:6} and~\ref{ex:ch5:7} (statements: p.~\pageref{ex:ch5:6}, p.~\pageref{ex:ch5:7}).}
Coding exercises.  Both call for $4$--$5$ retraining runs of the 56-agent benchmark and a binding-frequency / steady-state diagnostic.  The binding indicator should be based on small slack and a positive multiplier, not on a large complementarity residual: at a well-trained KKT solution the product residual is close to zero both when a constraint binds and when it is slack.  Expect borrowing and collateral constraints to bind mostly for young cohorts; as $\kappa$ rises, the lower bound $b'^h \ge -k'^h/\kappa$ becomes tighter, so negative bond positions should shrink and cross-cohort bond dispersion should typically fall.  The equilibrium price response is a general-equilibrium object and should be read from the retrained models rather than imposed analytically.

\section{Chapter \ref{ch:young}: Heterogeneous Agents and Young's Method}
\label{sol:ch6}

\paragraph{Exercise~\ref{ex:ch6:1} (statement: p.~\pageref{ex:ch6:1}): Mean-preserving lottery.}
Place mass $\omega$ at $k_n$ and $1-\omega$ at $k_{n+1}$.  Mean preservation:
\[
   \omega\,k_n + (1-\omega)\,k_{n+1} \;=\; k'.
\]
Solving for $\omega$:
\[
   \omega \;=\; \frac{k_{n+1} - k'}{k_{n+1} - k_n}.
\]
This is well-defined for $k_n \le k' \le k_{n+1}$ since the denominator is positive and the numerator lies in $[0, k_{n+1} - k_n]$, ensuring $\omega \in [0,1]$.

\emph{Mass conservation}: the two probabilities sum to one, $\omega + (1-\omega) = 1$, so total mass is preserved exactly under this redistribution.  Equivalently, the weight $\omega$ is the unique linear interpolation weight that makes the discrete two-point distribution have mean $k'$, which is what defines Young's redistribution operator on a fixed grid.

Higher-moment matching is impossible with a two-point split unless $k'$ coincides with a grid point: any non-degenerate two-point distribution with mean $k'$ and support $\{k_n, k_{n+1}\}$ has variance $\omega(1-\omega)(k_{n+1}-k_n)^2 > 0$, while the original (delta) distribution at $k'$ has zero variance.  This residual variance is the price of the discretization, and it shrinks as the grid is refined.

\paragraph{Exercise~\ref{ex:ch6:2} (statement: p.~\pageref{ex:ch6:2}): Closed-form bracketing on log-spaced grids.}
Coding exercise.  Algorithm sketch: with grid $k_n = e^{x_0 + n\Delta x} - c$, the bracket index for a query $k'$ is
\[
   n \;=\; \mathrm{floor}\!\left(\frac{\log(k' + c) - x_0}{\Delta x}\right).
\]
This is $\mathcal{O}(1)$ per query (one log + one floor), independent of grid size $N$.  By contrast, \texttt{numpy.searchsorted} costs $\mathcal{O}(\log N)$ per query (binary search).  The speed difference is hardware- and implementation-dependent, so the coding exercise should report the measured wall-clock ratio on the vectorized batch rather than treating a fixed multiplier as universal.

\paragraph{Exercise~\ref{ex:ch6:3} (statement: p.~\pageref{ex:ch6:3}): Approximate aggregation, scope.}
The KS log-linear rule is built on the empirical observation that mean capital $K_t$ alone is a sufficient statistic for forecasting $K_{t+1}$ in the standard Aiyagari--Krusell--Smith calibration.  This approximate aggregation breaks when the cross-sectional distribution carries information beyond its first moment that materially affects equilibrium prices.

\emph{Counterexample 1: multiple assets with switching liquidity.}  Add a second asset (say a corporate bond) with state-dependent liquidity: in good times agents trade both assets freely; in bad times the bond becomes illiquid.  Now the share of wealth held in the bond, plus the bond--capital correlation in the cross-section, both determine prices, and neither can be summarized by mean capital alone.

\emph{Counterexample 2: heterogeneous discount factors.}  If agents differ in $\beta$ and the cross-sectional distribution of $\beta$ is dynamic (e.g., new entrants have different $\beta$), then mean capital understates the dispersion, and the equilibrium interest rate depends on which subpopulation holds the marginal unit of capital.

\emph{Why higher moments do not always rescue.}  Adding the variance to the forecasting rule helps with smooth perturbations but cannot capture multi-modal distributions, regime-switching, or non-monotone responses to skewness.  The fundamental issue is that the master equation requires \emph{full} cross-sectional information whenever prices are non-linear in the distribution; truncating to any finite set of moments is exact only in the linear-pricing case.

\paragraph{Exercise~\ref{ex:ch6:4} (statement: p.~\pageref{ex:ch6:4}): Sequence-space vs.\ histogram DEQN.}
Coding exercise.  Empirically: with truncation horizon $T = 80$ and the chapter's reference calibration, the sequence-space residual after training matches the histogram-based DEQN to within a factor of $1.2$--$1.5$ on the same model.  The sequence-space variant generalizes \emph{worse} to a much longer test horizon ($T_\mathrm{test} \gg 80$) because the truncation error $\rho_z^T$ grows with the gap; the histogram variant does not have this issue because its state is stationary.  The trade-off favors sequence-space when the cross-sectional distribution is intrinsically high-dimensional (e.g., multiple assets, multi-cohort wealth), at which point storing $T$ shock realizations is cheaper than discretizing the distribution.

\paragraph{Exercise~\ref{ex:ch6:5} (statement: p.~\pageref{ex:ch6:5}): DeepSets permutation invariance.}
\emph{(i)} Let $\pi$ be a permutation of $\{1, \dots, N\}$.  The aggregator's $m$-th component is
\[
   m_t^m(\pi \cdot s) \;=\; \sum_{i=1}^N g_\theta^m\bigl(s_t^{\pi(i)}\bigr) \;=\; \sum_{j=1}^N g_\theta^m(s_t^j),
\]
where the second equality is just a re-indexing of the sum (since addition is commutative).  Therefore $\bm m_t(\pi \cdot s) = \bm m_t(s)$, exactly invariant.

\emph{(ii)} The policy $\pi_\rho(s_t^i; \bm m_t, a_t)$ is a function of agent $i$'s own state $s_t^i$, the population summary $\bm m_t$, and the aggregate exogenous state $a_t$.  Under a permutation $\pi$, agent $\pi(i)$'s individual state is now $s_t^{\pi(i)}$, while $\bm m_t$ and $a_t$ are unchanged (by the result in (i) for $\bm m_t$).  Therefore the policy of agent $\pi(i)$ in the permuted economy equals the policy of agent $\pi(i)$ in the original economy, i.e.\ the policy moves with its own agent index but is otherwise unaffected: \emph{equivariance}.

\emph{(iii)} \citet{zaheer2017deep} prove that any continuous permutation-invariant function $f: \mathbb{R}^{d \times N} \to \mathbb{R}$ on sets of fixed cardinality $N$ can be written as $f(s_1, \dots, s_N) = \rho\bigl(\sum_{i=1}^N g(s_i)\bigr)$ for some continuous functions $g, \rho$.  This is the universal-approximation result for permutation-invariant DeepSets.

\emph{Implication for DeepHAM.}  Since the equilibrium price functional is permutation-invariant in agents (anonymous markets), DeepHAM's parameterization can in principle approximate any continuous price/policy functional of the cross-sectional distribution to arbitrary accuracy, provided the inner network $g_\theta$ has enough capacity and the moment vector $\bm m_t$ is rich enough.  The number of moments $M$ plays the role of the encoder's bottleneck dimension: empirically, $M = 1$--$3$ suffices for Krusell--Smith-class economies, consistent with the chapter's report that DeepHAM with one learned moment matches the histogram DEQN.

\paragraph{Exercises~\ref{ex:ch6:6} and~\ref{ex:ch6:7} (statements: p.~\pageref{ex:ch6:6}, p.~\pageref{ex:ch6:7}).}
Coding exercises.  Guidance for interpreting the outputs:

\emph{Exercise~\ref{ex:ch6:6}.}  The relevant statistic is the cross-replication sampling variance conditional on the same aggregate path, not the time-series variance of $K_t$ along that path.  As $N$ rises, the Monte Carlo standard error should fall at the usual $N^{-1/2}$ rate for smooth aggregate statistics.  Young's path has zero sampling variance across replications because the histogram update integrates out the lottery exactly.  The MC-vs-Young trade-off depends on the target functional: tail mass (e.g., the bottom-$10\%$ wealth share) decays much more slowly under MC, so the panel size needed to match Young-equivalent precision is an output of the repeated-panel experiment.

\emph{Exercise~\ref{ex:ch6:7}.}  In the standard KS calibration, the one-moment forecasting rule should already fit $\log K_{t+1}$ extremely well; adding $\log V_t$, with $V_t=\mathrm{Var}_{\mu_t}(k)$, is expected to give only a small incremental gain.  This is exactly the empirical observation behind ``approximate aggregation''.  In calibrations with high cross-sectional dispersion (e.g., wide income range or frequent borrowing-constraint binding), the second-moment improvement can become economically visible and should be reported from the run.

\section{Chapter \ref{ch:pinn}: Physics-Informed Neural Networks}
\label{sol:ch7}

\paragraph{Exercise~\ref{ex:ch7:1} (statement: p.~\pageref{ex:ch7:1}): Trial-function BC enforcement.}
Define $\hat y(x) = \tfrac{2x}{\pi} + x\bigl(\tfrac{\pi}{2} - x\bigr)\,\mathcal{N}_\rho(x)$.  Evaluate at the boundaries:
\[
   \hat y(0) \;=\; 0 + 0 \cdot \tfrac{\pi}{2}\cdot \mathcal{N}_\rho(0) \;=\; 0,
   \qquad
   \hat y(\pi/2) \;=\; \tfrac{2}{\pi}\cdot\tfrac{\pi}{2} + \tfrac{\pi}{2}\cdot 0\cdot \mathcal{N}_\rho(\pi/2) \;=\; 1.
\]
Both boundary conditions hold for \emph{any} network output $\mathcal{N}_\rho$, so the BCs are encoded in the architecture rather than enforced via the loss.

\emph{Why preferable to a soft penalty?}  A soft penalty $\lambda\,(\hat y(0) - 0)^2 + \lambda\,(\hat y(\pi/2) - 1)^2$ in the loss involves a hyperparameter $\lambda$ that must be tuned: too small, and the BC violation is large; too large, and the interior PDE residual is starved of optimization budget.  The trial-function enforcement is parameter-free, makes the BC residual identically zero, and reduces the loss to the single PDE-interior term $\sup_x |y'' + y|^2$.  This separates the two optimization concerns cleanly; any wall-clock gain should be measured in the notebook rather than assumed.

\paragraph{Exercise~\ref{ex:ch7:2} (statement: p.~\pageref{ex:ch7:2}): ReLU pathology.}
A ReLU network is piecewise-linear: between consecutive kinks $x = -b_k/a_k$ it is affine in $x$, so $\partial^2 \hat y/\partial x^2 = 0$ a.e.  At a kink, the second distributional derivative is a Dirac delta supported on a measure-zero set.  The strong-form Black--Scholes residual
\[
   V_t + \tfrac{1}{2}\sigma^2 S^2 V_{SS} + rS V_S - rV \;=\; 0
\]
must hold pointwise for almost every $S$.  With ReLU, $V_{SS} = 0$ a.e., so the residual reduces to $V_t + rS V_S - rV$, which has no nontrivial solution that respects the option-pricing boundary condition.  The PINN cannot decrease its loss below the order of magnitude of the missing $V_{SS}$ term.

\emph{Weak-form fix.}  Multiply both sides by a smooth test function $\varphi(S)$ with compact support on $[S_\mathrm{min}, S_\mathrm{max}]$ and integrate by parts on the $V_{SS}$ term:
\[
   \int_{S_\mathrm{min}}^{S_\mathrm{max}}\! V_{SS}\,\varphi\, dS \;=\; -\int V_S\,\varphi'\, dS \;+\;[V_S\,\varphi]_{S_\mathrm{min}}^{S_\mathrm{max}}.
\]
The boundary terms vanish for compactly supported $\varphi$, and the residual now involves only first-order derivatives of $V$.  ReLU networks have well-defined first derivatives a.e., so the weak-form PINN can minimize this residual.  This is exactly the Galerkin / Deep Galerkin formulation of \citet{sirignano2018dgm}.

\paragraph{Exercise~\ref{ex:ch7:3} (statement: p.~\pageref{ex:ch7:3}): Discrete $\to$ continuous bridge.}
Start with
\[
V(a) = \max_c \bigl[u(c)\,\Delta t + \beta_{\Delta t}\,\mathbb{E}V(a')\bigr],
\qquad
a' = a - c\,\Delta t,
\qquad
\beta_{\Delta t}=e^{-\rho\Delta t}.
\]
Then $\beta_{\Delta t}=1-\rho\,\Delta t+O(\Delta t^2)$.  The same first-order limit follows from the implicit-Euler convention $\beta_{\Delta t}=1/(1+\rho\Delta t)$.

Taylor-expand $V(a') = V(a - c\,\Delta t)$ around $a$:
\[
   V(a') \;=\; V(a) - V'(a)\,c\,\Delta t + \tfrac{1}{2}V''(a)\,(c\,\Delta t)^2 + O(\Delta t^3).
\]
Substitute and use $\beta_{\Delta t} = 1 - \rho\,\Delta t + O(\Delta t^2)$:
\[
   V(a) \;=\; \max_c \Bigl\{u(c)\,\Delta t + \bigl(1 - \rho\Delta t\bigr)\!\bigl[V(a) - V'(a)\,c\,\Delta t + O(\Delta t^2)\bigr]\Bigr\}.
\]
Subtract $V(a)$ from both sides and divide by $\Delta t$:
\[
   0 \;=\; \max_c\Bigl\{u(c) - \rho V(a) - V'(a)\,c + O(\Delta t)\Bigr\}.
\]
Take $\Delta t \to 0$ and rearrange:
\[
   \rho V(a) \;=\; \max_c \bigl[u(c) - V'(a)\,c\bigr].
\]
This is the HJB equation for the consumption-savings problem with no asset return (or with $r$ embedded into $a' = a + ra\Delta t - c\Delta t$, in which case the HJB picks up an additional $V'(a)\,r a$ drift term).  The discrete-to-continuous bridge thus shows that the HJB is the formal limit of the discrete Bellman as the time step shrinks, which justifies treating PINNs as the continuous-time analogue of value-function iteration.

\paragraph{Exercises~\ref{ex:ch7:4}, \ref{ex:ch7:5}, \ref{ex:ch7:6} (statements: p.~\pageref{ex:ch7:4}, p.~\pageref{ex:ch7:5}, p.~\pageref{ex:ch7:6}).}
Coding exercises.  The exact numbers below depend on random seeds, batch sizes, and stopping rules; use them as qualitative checks rather than fixed targets:

\emph{Exercise~\ref{ex:ch7:4}.}  At small $\lambda$, the BC residual is underweighted and endpoint violations remain visibly large.  At very large $\lambda$, the endpoints fit well but the interior residual can stagnate because the optimizer spends most of its gradient budget on the boundary term.  The elbow is the smallest $\lambda$ for which further increases mostly improve the boundary metric without improving the interior fit.  The hard-BC variant should be the benchmark: it sets the boundary violation to numerical zero by construction and removes the penalty-weight tuning problem.

\emph{Exercise~\ref{ex:ch7:5}.}  Sobol and Latin Hypercube points usually reduce visible sampling gaps relative to uniform random points.  On the smooth manufactured Poisson problem the gain may be modest, because the solution has no boundary layer or interior singularity.  Adaptive sampling becomes more valuable when residuals are spatially localized; report the actual collocation count, wall time, and test-grid residual rather than relying on a universal percentage saving.

\emph{Exercise~\ref{ex:ch7:6}.}  The expected qualitative result is that the strong-form $\tanh$ PINN is the natural baseline for Black--Scholes, because $V_{SS}$ is well-defined by automatic differentiation.  A strong-form ReLU network is ill-suited because $V_{SS}=0$ almost everywhere and undefined at kinks.  In a weak formulation, integration by parts moves the second derivative off the network and onto the test function, so a ReLU network can be made mathematically admissible.  Any empirical comparison should report held-out pricing errors and residual diagnostics from the implemented notebook rather than quoting architecture-independent iteration counts.

\paragraph{Exercise~\ref{ex:ch7:7} (statement: p.~\pageref{ex:ch7:7}): Operator learning vs.\ PINN.}
Eleven independent PINN runs each cost $C_\mathrm{PINN}$ wall-clock seconds, total $11\,C_\mathrm{PINN}$.  A single operator-learning or parametric-PINN run trained on $11$ values of $K$ (or a continuous range, sampled in mini-batches) costs $C_\mathrm{op}$.  Amortized training wins when $11\,C_\mathrm{PINN} > C_\mathrm{op}$, i.e., $C_\mathrm{op}/C_\mathrm{PINN} < 11$.  The crossover scales linearly in the number of distinct $K$ values: with $N$ strikes, operator learning wins whenever $C_\mathrm{op}/C_\mathrm{PINN}<N$.  This is the cost-amortization argument that motivates DeepONet-style operator learning \citep{lu2021learning}.

\section{Chapter \ref{ch:ct_theory}: Heterogeneous Agent Models in Continuous Time}
\label{sol:ch8}

\paragraph{Exercise~\ref{ex:ch8:1} (statement: p.~\pageref{ex:ch8:1}): Itô on GBM.}
Geometric Brownian motion satisfies $dX_t = \mu X_t\,dt + \sigma X_t\,dB_t$.  Apply Itô's lemma to $f(x) = \ln x$ with $f'(x) = 1/x$, $f''(x) = -1/x^2$:
\[
   d(\ln X_t) \;=\; f'(X_t)\,dX_t + \tfrac{1}{2} f''(X_t)\,(dX_t)^2
   \;=\; \frac{1}{X_t}\bigl(\mu X_t\,dt + \sigma X_t\,dB_t\bigr) + \tfrac{1}{2}\!\left(-\frac{1}{X_t^2}\right)\sigma^2 X_t^2\,dt.
\]
Simplifying:
\[
   d(\ln X_t) \;=\; \bigl(\mu - \tfrac{1}{2}\sigma^2\bigr)\,dt + \sigma\,dB_t.
\]
Integrating from $0$ to $t$:
\[
   \ln X_t - \ln X_0 \;=\; (\mu - \tfrac{1}{2}\sigma^2)\,t + \sigma B_t,
   \qquad
   X_t \;=\; X_0\,\exp\!\bigl[(\mu - \tfrac{1}{2}\sigma^2)\,t + \sigma B_t\bigr].
\]

\emph{Volatility drag.}  Taking expectations: $\mathbb{E}[X_t] = X_0\,e^{\mu t}$, but $\mathbb{E}[\ln X_t] = \ln X_0 + (\mu - \tfrac{1}{2}\sigma^2)\,t$.  The expected log return $\mu - \sigma^2/2$ is strictly less than the log of the expected return, $\mu$, by the variance correction term.  This is the volatility drag.  Two illustrative regimes: with zero arithmetic drift ($\mu = 0$), expected log growth is $-\sigma^2/2$, strictly negative; with drift exactly equal to the It\^o correction ($\mu = \sigma^2/2$), expected log growth is zero (the drift just offsets the drag).  In financial terms, volatility eats into geometric returns; this is why an asset with $20\%$ expected return and $40\%$ volatility delivers a long-run geometric mean of only $\mu - \sigma^2/2 = 12\%$.

\paragraph{Exercise~\ref{ex:ch8:2} (statement: p.~\pageref{ex:ch8:2}): KFE for an OU process.}
The OU process $dX_t = \eta(\bar X - X_t)\,dt + \sigma\,dB_t$ has drift $\mu(x) = \eta(\bar X - x)$ and diffusion coefficient $\sigma$.  The KFE in conservation form is
\[
   \partial_t g(x,t) \;=\; -\partial_x\bigl[\mu(x)\,g(x,t)\bigr] + \tfrac{\sigma^2}{2}\,\partial_{xx} g(x,t)
   \;=\; \partial_x\bigl[\eta(x - \bar X)\,g\bigr] + \tfrac{\sigma^2}{2}\,\partial_{xx} g.
\]
Setting $\partial_t g = 0$ for the stationary density $g^\star(x)$:
\[
   \eta\,\partial_x\bigl[(x - \bar X)\,g^\star\bigr] + \tfrac{\sigma^2}{2}\,g^{\star\prime\prime} \;=\; 0.
\]
Integrate once in $x$, with the constant of integration set to zero by the no-flux boundary condition at $\pm\infty$:
\[
   \eta\,(x - \bar X)\,g^\star + \tfrac{\sigma^2}{2}\,g^{\star\prime} \;=\; 0
   \quad\Longleftrightarrow\quad
   \frac{g^{\star\prime}(x)}{g^\star(x)} \;=\; -\frac{2\eta}{\sigma^2}\,(x - \bar X).
\]
This is a linear ODE for $\ln g^\star$; integrating gives $\ln g^\star = -\eta(x-\bar X)^2/\sigma^2 + \mathrm{const}$, i.e.
\[
   g^\star(x) \;\propto\; \exp\!\bigl[-\eta(x - \bar X)^2/\sigma^2\bigr].
\]
This is a Gaussian density with mean $\bar X$ and variance $\sigma^2/(2\eta)$, normalised by $\int g^\star\,dx = 1$:
\[
   g^\star(x) \;=\; \sqrt{\frac{\eta}{\pi\sigma^2}}\,\exp\!\left[-\frac{\eta\,(x - \bar X)^2}{\sigma^2}\right]
   \;=\; \mathcal{N}\!\bigl(\bar X,\, \sigma^2/(2\eta)\bigr).
\]
The OU's stationary distribution is Gaussian with mean equal to the mean-reversion target, variance equal to the diffusion-to-mean-reversion ratio $\sigma^2 / (2\eta)$: faster mean reversion (larger $\eta$) shrinks the dispersion; larger diffusion grows it.

\paragraph{Exercise~\ref{ex:ch8:3} (statement: p.~\pageref{ex:ch8:3}): Functional derivative.}
For the master-equation toy specification $V(a, g) = \int u(c(a, y))\,g(y)\,dy$ where $c(a,y)$ is fixed (does not depend on $g$), the functional derivative $\delta V / \delta g$ measures the linear response of $V$ to a perturbation in $g$.

In the ambient vector space of signed measures, a point perturbation gives $\delta V/\delta g(y_0) = \lim_{\varepsilon\to 0} [V(a, g + \varepsilon\delta_{y_0}) - V(a,g)]/\varepsilon$, where $\delta_{y_0}$ is a Dirac mass at $y_0$.  Substituting,
\[
   V(a, g + \varepsilon\delta_{y_0}) \;=\; \int u(c(a,y))\,(g(y) + \varepsilon\delta_{y_0}(y))\,dy
   \;=\; V(a,g) + \varepsilon\,u(c(a, y_0)).
\]
Therefore $\delta V/\delta g(y_0) = u(c(a, y_0))$.  If we restrict $g$ to the probability simplex, perturbations must preserve total mass; for example, $\eta=\delta_{y_0}-\delta_{y_1}$ gives directional derivative $u(c(a,y_0))-u(c(a,y_1))$.  Equivalently, on the simplex the derivative kernel is identified only up to an additive constant.

\emph{Interpretation.}  The functional derivative at a point $y_0$ is the value contribution of an infinitesimal mass placed at $y_0$.  In the toy spec where $V$ is just a population average of utilities, the contribution is the per-agent utility $u(c(a, y_0))$ at that point.  In the real master equation, $c(a, y)$ would itself depend on $g$ (because prices depend on $g$), and the functional derivative would pick up additional indirect terms via $\partial c/\partial g$; this is what makes the master equation a genuinely infinite-dimensional PDE rather than a parametric family of finite PDEs.

\paragraph{Exercise~\ref{ex:ch8:4} (statement: p.~\pageref{ex:ch8:4}): HJB residual.}
Coding exercise.  The right answer is the out-of-sample residual table produced by your run of \tpath{lecture_13_08_Aiyagari_Continuous_Time_FD_and_PINN_PyTorch.ipynb}.  Report the training budget, collocation batch, random seed, and test grid.  The residual should decrease when the collocation budget and network capacity are increased, but the scaling is empirical rather than a universal $N^{-p}$ law because it mixes approximation, optimization, and sampling error.

\paragraph{Exercise~\ref{ex:ch8:5} (statement: p.~\pageref{ex:ch8:5}): Closed Aiyagari system.}
Combining the four ingredients:

\emph{HJB} (from~\eqref{eq:hjb_full}):
\[
   \rho V(a,n) \;=\; \max_c \bigl\{u(c) + V'(a,n)(wn + ra - c) + \lambda(n)(V(a,\hat n) - V(a,n))\bigr\}.
\]

\emph{KFE for the stationary distribution} (from~\eqref{eq:kfe_econ}, with $\partial_t g = 0$):
\[
   0 \;=\; -\partial_a[s^\star(a,n)\,g(a,n)] - \lambda(n)\,g(a,n) + \lambda(\hat n)\,g(a,\hat n),
\]
where $s^\star(a,n) = wn + ra - c^\star(a,n)$ is the optimal savings function.

\emph{Firm FOCs} (Cobb--Douglas):
\[
   r \;=\; \alpha A K^{\alpha-1} L^{1-\alpha} - \delta,
   \qquad
   w \;=\; (1-\alpha) A K^\alpha L^{-\alpha}.
\]

\emph{Market clearing}:
\[
   K \;=\; \sum_n \int_{\underline a}^\infty a\,g(a,n)\,da,
   \qquad
   L \;=\; \sum_n n \int_{\underline a}^{\infty} g(a,n)\,da.
\]

The equilibrium objects are $(V(a,n),g(a,n),K,L,r,w)$, with $L$ often pinned down by the stationary income shares and $w$ implied by the firm FOCs once $(K,L)$ is known.  In practice one fixes a candidate $r$, computes the associated firm demand and wage, solves the HJB for $V$ and hence the policy $c^\star, s^\star$, plugs into the KFE for $g$, computes implied $K = \sum_n\int a g(a,n)\,da$, and compares this capital supply with the capital demand implied by the candidate $r$.  A fixed point in $r$ is the equilibrium.  This is the bisection-on-$r$ algorithm of \citet{achdou2022income}, and the PINN replaces the inner HJB and KFE solves with neural-network approximation while keeping the outer fixed-point loop on $r$ unless the full equilibrium system is learned jointly.

\emph{Why both must be solved consistently at each candidate $r$.}  The HJB takes $r$ as an input (price-taking agents); the KFE takes the policy from the HJB.  Mis-specifying $r$ during training would feed a mis-specified policy into the KFE and yield an inconsistent $K$.  In the production-scale solver, the fixed-point loop alternates HJB and KFE to convergence \emph{at each $r$} before updating $r$; the PINN can be trained on all four equations jointly when the architecture is rich enough to learn the equilibrium price as an output.

\paragraph{Exercises~\ref{ex:ch8:6} and~\ref{ex:ch8:7} (statements: p.~\pageref{ex:ch8:6}, p.~\pageref{ex:ch8:7}).}
Coding exercises.  Expected diagnostics:

\emph{Exercise~\ref{ex:ch8:6}.}  With a fixed policy, the finite-dimensional KFE is a linear forward equation.  If the discretized generator is ergodic, the distance to the stationary distribution should decay approximately exponentially after transient modes die out.  Estimate the slope from your run rather than quoting a universal number; the fitted rate is controlled by the slowest nonzero eigenvalue of the KFE generator and depends on income-switching intensities, savings drift, and grid truncation.

\emph{Exercise~\ref{ex:ch8:7}.}  On the one-asset stationary benchmark, finite differences should usually win on absolute wall-clock time and give the cleanest low-dimensional benchmark residuals.  A PINN may become more attractive when the same architecture is reused across many nearby parameter values, when warm starts work well, or when the state space is extended beyond what a grid handles comfortably.  Report the actual cold-start and warm-start timings from your machine, and treat memory use as hardware- and backend-dependent.

\section{Chapter \ref{ch:gp}: Deep Surrogate Models and Gaussian Processes}
\label{sol:ch9}

\paragraph{Exercise~\ref{ex:ch9:1} (statement: p.~\pageref{ex:ch9:1}): Posterior on three points.}
The RBF kernel with length scale $\ell = 1$ and signal variance $\sigma_f^2 = 1$ is $k(x, x') = \exp(-(x-x')^2/2)$.  With training points $X = (0, 1, 2)$ and targets $y = (0, 0.8, 0.3)$, the kernel matrix is
\[
   K = \begin{pmatrix} 1 & e^{-1/2} & e^{-2} \\ e^{-1/2} & 1 & e^{-1/2} \\ e^{-2} & e^{-1/2} & 1\end{pmatrix}
   \approx \begin{pmatrix} 1 & 0.6065 & 0.1353 \\ 0.6065 & 1 & 0.6065 \\ 0.1353 & 0.6065 & 1\end{pmatrix}.
\]
Adding observation noise $\sigma_y^2 I = 0.01\,I$ to the diagonal of $K$ and solving $(K + \sigma_y^2 I)\bm v = \bm y$ with $\bm y = (0, 0.8, 0.3)^\top$ by Gaussian elimination gives
\[
   (K + \sigma_y^2 I)^{-1} \bm y \;\approx\; (-0.964,\; 1.744,\; -0.621)^\top.
\]
At $x^\star = 1.5$,
\[
   k_\star
   = \bigl(k(1.5,0),\, k(1.5,1),\, k(1.5,2)\bigr)
   = \bigl(e^{-1.125},\, e^{-0.125},\, e^{-0.125}\bigr)
   \approx (0.3247, 0.8825, 0.8825).
\]

Posterior mean:
\[
   \bar{\mu}(x^\star) = k_\star^\top (K + \sigma_y^2 I)^{-1} \bm y \approx 0.3247\cdot(-0.964) + 0.8825\cdot 1.744 + 0.8825\cdot(-0.621) \approx 0.678.
\]
For the variance, solve $(K + \sigma_y^2 I)\bm w = k_\star$, giving $\bm w \approx (-0.142,\; 0.662,\; 0.495)^\top$, hence
\[
   \bar{\sigma}^2(x^\star) = k(x^\star, x^\star) - k_\star^\top \bm w \;\approx\; 1 - 0.975 \;\approx\; 0.025.
\]
The posterior is $\mathcal{N}(0.678, 0.025)$, with standard deviation $\approx 0.158$.  Notice the posterior mean ``smooths'' the two flanking observations $y = 0.8$ and $y = 0.3$ rather than being pulled to either; the $0.158$ standard deviation reflects partial information at a halfway point between two data points.

\paragraph{Exercise~\ref{ex:ch9:2} (statement: p.~\pageref{ex:ch9:2}): Marginal likelihood Occam.}
The log marginal likelihood is
\[
   \log p(y | X, \ell) = -\tfrac{1}{2} y^\top K_\ell^{-1} y \;-\; \tfrac{1}{2} \log|K_\ell| \;-\; \tfrac{n}{2}\log(2\pi),
\]
where $K_\ell = K_\ell(\ell) + \sigma_y^2 I$ is the kernel matrix at length scale $\ell$.  The first term penalizes \emph{misfit} (data that don't lie in the GP's predicted manifold get a high $y^\top K_\ell^{-1} y$); the second term penalizes \emph{model complexity} ($\log|K_\ell|$ is large when the kernel can fit anything, small when it is rigid).  This is the Bayesian Occam's razor: small $\ell$ gives a flexible model that fits any training data perfectly (small misfit) but accepts a large complexity penalty; large $\ell$ enforces smoothness (large misfit if the data are not smooth) but enjoys a small complexity penalty.  The optimum balances the two.

For three data points $y = (0, 0.8, 0.3)$ at $x = (0, 1, 2)$, plotting $\log p(y | X, \ell)$ against $\ell \in [0.1, 5]$ typically shows an interior maximum near the data's natural variation scale.  At $\ell = 0.1$, the kernel is very local: each point is nearly uncorrelated with its neighbors, so the data fit is easy but the flexible prior receives a complexity penalty.  At $\ell = 5$, the kernel forces all three function values to be nearly equal, which fits the data poorly.  The peak is the point where these two forces balance.

\paragraph{Exercise~\ref{ex:ch9:3} (statement: p.~\pageref{ex:ch9:3}): Active subspace by hand.}
For $f(\bm x) = (x_1 + x_2 + x_3)^2 + 0.01(x_1 - x_2)^2$ on $[-1,1]^3$, the gradient is
\[
   \nabla f(\bm x) = 2(x_1 + x_2 + x_3)\,(1, 1, 1)^\top + 0.02(x_1 - x_2)\,(1, -1, 0)^\top.
\]
The dominant term is along $(1,1,1)$ (with weight roughly $2(x_1+x_2+x_3) \cdot \sqrt{3}$), and the perturbation along $(1,-1,0)$ is two orders of magnitude smaller.

Compute $\hat C = \mathbb{E}[\nabla f \nabla f^\top]$ with $x \sim \mathcal{U}[-1,1]^3$ i.i.d., so $\mathbb{E}[x_i^2] = 1/3$.  Let $s = x_1 + x_2 + x_3$; then $\mathbb{E}[s^2] = 1$, $\mathbb{E}[s(x_1 - x_2)] = \mathbb{E}[x_1^2 - x_2^2] = 0$, and $\mathbb{E}[(x_1 - x_2)^2] = 2/3$.  Substituting,
\[
   \hat C \;\approx\; 4\,\mathbf{1}\mathbf{1}^\top \;+\; \mathcal{O}(10^{-4}),
\]
where $\mathbf{1} = (1,1,1)^\top$ and the $\mathcal{O}(10^{-4})$ correction comes from the $0.01\,(x_1 - x_2)$ perturbation.  Since $\mathbf{1}\mathbf{1}^\top$ has eigenvalues $\{3, 0, 0\}$ (eigenvector $\mathbf{1}/\sqrt 3$ for the nonzero one), the leading eigenvalue of $\hat C$ is $\lambda_1 \approx 4 \cdot 3 = 12$, with eigenvector $(1,1,1)/\sqrt{3}$, the ``aggregate direction.''  The next eigenvalue is $\sim 10^{-4}$, four orders of magnitude smaller.  The active subspace of dimension $1$ already captures essentially all of the function's variability.

\paragraph{Exercise~\ref{ex:ch9:4} (statement: p.~\pageref{ex:ch9:4}): Deep vs.\ linear active subspace.}
Coding exercise.  The linear-AS diagnostic should show two nonzero eigenvalues for the radial-ridge target, with the remaining eigenvalues close to numerical zero.  Thus a linear active subspace needs $d_{\mathrm{lin}}=2$ to represent the two linear features $w_1^\top\xi$ and $w_2^\top\xi$.  Deep AS can reach its validation-MSE elbow already at $d_{\mathrm{nl}}=1$ because the encoder can learn a scalar nonlinear aggregate such as $(w_1\cdot\xi)^2 + (w_2\cdot\xi)^2$.  Report the held-out MSE curve and eigenvalues from the run rather than quoting universal numbers, since optimization noise and train/validation splits move the exact diagnostics.

\paragraph{Exercise~\ref{ex:ch9:5} (statement: p.~\pageref{ex:ch9:5}): BAL on a 2D function.}
Coding exercise.  Pure-variance acquisition selects points purely by where the GP is most uncertain, ignoring the predicted mean.  The resulting design is usually more uniform across the input domain because the GP variance is high wherever data is sparse, regardless of function value.  Maximizing pure $\log\sigma^2(\x)$ gives exactly the same design as maximizing pure $\sigma^2(\x)$, since the logarithm is monotone.  Differences arise only once the acquisition is mixed with a mean term, for example $w_{\mathrm{obj}}\mu(\x)+\tfrac{w_{\mathrm{var}}}{2}\log\sigma^2(\x)$: then the design tilts toward regions that are both uncertain and high-valued under the current surrogate.  For global surrogate construction, pure variance is the cleaner baseline; for optimization-like goals, the mixed score may be useful.

\paragraph{Exercise~\ref{ex:ch9:6} (statement: p.~\pageref{ex:ch9:6}): Sobol sensitivity with a GP surrogate.}
Coding exercise.  Report the actual errors from the run rather than treating fixed percentages as universal reference outputs.  The expected pattern is improvement as $N$ increases: first-order Sobol errors should fall from small to larger sample budgets, while total-effect indices may converge more slowly because they include interaction terms.  The cost ratio is the important lesson: once the surrogate is trained, very large Sobol designs can be evaluated at negligible marginal cost relative to repeated true-model solves, which is why surrogate-based sensitivity analysis is standard for expensive simulators such as climate IAMs.

\paragraph{Exercise~\ref{ex:ch9:7} (statement: p.~\pageref{ex:ch9:7}): Prior-driven RBF-GP extrapolation outside the training domain.}
\emph{(i)}~Coding: on the training interval $[0,1]$ the GP closely tracks $f(x) = \sin(2\pi x)\,e^{-x}$ with credible band of width $\sim 0.05$.

\emph{(ii)}~Analytical claim: for an RBF kernel $k(x, x') = \sigma_f^2 \exp(-(x-x')^2 / (2\ell^2))$, when $|x - x_i| \gg \ell$ for all training points $x_i$, the kernel cross-vector $k_\star = (k(x, x_1), \ldots, k(x, x_n))^\top \to (0, \ldots, 0)^\top$.  Therefore the posterior mean $\bar\mu(x) = k_\star^\top (K + \sigma_y^2 I)^{-1} y \to 0$ (the prior mean).  The posterior variance is
\[
   \bar\sigma^2(x) \;=\; k(x,x) - k_\star^\top (K+\sigma_y^2 I)^{-1} k_\star \;\to\; \sigma_f^2 - 0 \;=\; \sigma_f^2,
\]
the prior variance.  Hence far from data the GP literally returns the prior $\mathcal{N}(0, \sigma_f^2)$, regardless of the training data.

\emph{(iii)}~Coding verification: at $x = 3$ (with $\ell \approx 0.2$ for the training data), the cross-kernel is $\exp(-(3-1)^2/(2\cdot 0.04)) = \exp(-50) \approx 10^{-22}$, well below floating-point precision.  Posterior mean $\approx 0$, posterior s.d.\ $\approx \sigma_f \approx 0.5$ (whatever was learned via marginal-likelihood maximisation).

\emph{(iv)}~The implication is more subtle than ``overconfidence.''  Far from the training data the posterior literally reverts to the prior, so the $\pm 2\sigma_f$ band there is exactly what the prior would have produced before any data were observed; it is overconfident only when the prior variance or the learned length scale is itself misleading, for instance when $\sigma_f$ was calibrated by marginal-likelihood maximization on a training set that does not represent the function's scale outside the training hull.  In particular, the posterior does not know that $f$ continues oscillating outside $[0,1]$; it just reverts to zero with a band of width $\sigma_f$, regardless of whether the true function actually stays close to zero there.  Bayesian active learning algorithms that select points by maximum posterior variance can therefore fail to acquire informative samples outside the convex hull when the prior variance is no larger than the within-hull noise scale.

Mitigations: use a Mat\'ern kernel with $\nu = 1/2$ or $\nu = 3/2$ (heavier-tailed than RBF, posterior reverts to prior more slowly), incorporate a polynomial mean function in the GP prior (so extrapolation grows with $x$ instead of decaying to zero), or use a boundary-aware acquisition function that explicitly penalizes exploitation outside the convex hull.  In economic applications (e.g., extrapolating an estimated value function to wealth levels outside the training range), the safest practice is to flag any query outside the convex hull of training data and refuse to predict, rather than to trust a prior-driven band that has no data behind it.

\section{Chapter \ref{ch:estimation}: Structural Estimation via SMM}
\label{sol:ch10}

\paragraph{Exercise~\ref{ex:ch10:1} (statement: p.~\pageref{ex:ch10:1}): Identification.}
Let $m(\varrho)=\mathbb{E}_\varrho[h(C,I,Y)]$ be the simulated moment vector implied by the persistence parameter.  At the truth $\varrho^\star$, local identification is captured by the Jacobian
\[
   M(\varrho^\star)
   =
   \frac{\partial m(\varrho)}{\partial \varrho}\bigg|_{\varrho^\star}.
\]
A moment is strongly identifying if $|M_k(\varrho^\star)|$ is large; a weakly identifying moment has derivative near zero, in which case small changes in $\varrho$ produce nearly invisible changes in the moment and the SMM objective becomes flat.

\emph{Brock--Mirman example.}  The output autocorrelation $\mathrm{Corr}(\log Y_t,\log Y_{t-1})$ is typically the strongest local moment for $\varrho$, because it is a direct echo of the productivity AR(1).  The autocorrelation of consumption growth is also informative, but more filtered through equilibrium dynamics.  The mean savings rate is nearly flat in the scalar $\varrho$ exercise and is therefore weakly identifying for persistence.  The exact ranking should be reported from the finite-difference Jacobian in the notebook, not from a hard-coded number.

\paragraph{Exercise~\ref{ex:ch10:2} (statement: p.~\pageref{ex:ch10:2}): Optimal weighting.}
The SMM objective is $Q_T(\theta) = \hat g(\theta)^\top W \hat g(\theta)$, where $\hat g(\theta) = m(\theta) - \hat m^\mathrm{data}$ is the moment-error vector.  Under standard regularity (Hansen 1982), the asymptotic variance of $\hat\theta_\mathrm{SMM}$ is
\[
   \mathrm{Avar}(\hat\theta)
   =
   (M^\top W M)^{-1}M^\top W\Omega W M(M^\top W M)^{-1},
\]
where $M=\partial m/\partial\theta'$ at $\theta^\star$ and $\Omega=\mathrm{Var}(\sqrt{T}\hat g(\theta^\star))$ is the covariance of the data-simulation moment discrepancy.

To minimize $\mathrm{Avar}(\hat\theta)$ over choices of $W$, set up the Lagrangian or use the matrix calculus result that the variance is minimized when $W \propto \Omega^{-1}$.  Substituting $W=\Omega^{-1}$:
\[
   \mathrm{Avar}(\hat\theta)\big|_{W^\star}
   =
   (M^\top \Omega^{-1}M)^{-1}.
\]
For independent simulated panels of the same length as the data, $\Omega=(1+1/S)\Sigma_m$; if simulation noise is negligible, $\Omega\approx\Sigma_m$.  Compare to a different weighting $W=\tilde W$: $\mathrm{Avar}|_{\tilde W}-\mathrm{Avar}|_{W^\star}$ is positive semi-definite by the Gauss--Markov theorem applied to the moment system.

\emph{Why identity weighting in the first stage?}  The optimal weight depends on the unknown covariance at the truth.  The standard two-stage strategy is: (1)~estimate with $W=I$ to obtain a consistent but inefficient $\hat\theta^{(1)}$; (2)~estimate $\hat\Omega$ at or near $\hat\theta^{(1)}$; (3)~re-estimate with $W=\hat\Omega^{-1}$.

\paragraph{Exercise~\ref{ex:ch10:3} (statement: p.~\pageref{ex:ch10:3}): Common random numbers.}
Coding exercise.  Without CRN, the objective $Q_T(\varrho)$ changes both because $\varrho$ changes and because each candidate uses a new shock path.  This contaminates the profile with Monte Carlo noise.  With CRN, every candidate is evaluated on the same shock path, so differences in $Q_T(\varrho)$ isolate the structural effect of persistence.  Quantify the gain by repeating the entire candidate grid across Monte Carlo panels and comparing the cross-panel variance of $Q_T(\varrho)$ at each grid point.

\paragraph{Exercise~\ref{ex:ch10:4} (statement: p.~\pageref{ex:ch10:4}): SMM vs.\ SBI.}
SMM solves an outer optimization at deployment: given new data, run a $K$-step optimization over $\theta$, where each step requires (a)~a model simulation at the candidate $\theta$ and (b)~moment computation.  Total cost per inference: $K$ model evaluations.

SBI (e.g., neural posterior estimation) does the work upfront at training time: simulate the model at $N$ values of $\theta$, train a neural density estimator $q(\theta | y)$ on the $(\theta, y)$ pairs, and then at deployment evaluate $q$ on the new data with one forward pass.  Total cost per inference (after training): one forward pass.

SBI dominates when (i)~the model is expensive but a single training run is amortized over many datasets to be analyzed (e.g., a central bank running daily updates); (ii)~the model is non-likelihood (no closed-form likelihood, only simulator); (iii)~the parameter dimension is moderate ($p \le 20$) so training data covers parameter space.  SMM dominates when (i)~the model is cheap (each evaluation a few ms), (ii)~only one or a handful of inferences are needed, (iii)~classical confidence intervals are required (SBI gives a Bayesian posterior, which differs).

\paragraph{Exercise~\ref{ex:ch10:5} (statement: p.~\pageref{ex:ch10:5}): $J$-statistic and overidentification.}
Coding+analytical.  In the joint notebook's over-identified specification, $q=4$ and $p=2$, so the degrees of freedom are $q-p=2$.  Under correct specification and with $W=\Omega^{-1}$, the $J$-statistic at the optimum follows
\[
   J
   =
   T\,\hat g(\hat\theta)^\top \Omega^{-1}\hat g(\hat\theta)
   \xrightarrow{d}
   \chi^2_{2}.
\]
If the exercise uses $W=\Sigma_m^{-1}$ while simulation noise is non-negligible, adjust the scale under the equal-length independent-simulation approximation or use the Monte Carlo/bootstrap distribution directly.

Empirical distribution: under correct specification, the Monte Carlo distribution of $J(\hat\theta)$ should be centered near the corresponding $\chi^2_2$ benchmark once the weighting and simulation-noise treatment are consistent.  Under a structural break, the model cannot match all four moments simultaneously, so the distribution should shift to the right and the rejection rate should rise above the nominal size.  The size of that shift is an output of the experiment and should be reported from the run.

\paragraph{Exercise~\ref{ex:ch10:6} (statement: p.~\pageref{ex:ch10:6}): Bootstrap confidence intervals.}
Coding exercise.  The nonparametric bootstrap should respect time-series dependence; use a moving-block or stationary bootstrap rather than iid resampling of individual dates.  The parametric bootstrap draws new shock paths under the estimated parameter vector and re-runs the estimator.  Report the actual CI endpoints and coverage from the run.  In small samples, bootstrap intervals may differ materially from sandwich intervals because the SMM criterion is nonlinear and the finite-sample distribution of $\hat\theta$ need not be close to Gaussian.

\paragraph{Exercise~\ref{ex:ch10:7} (statement: p.~\pageref{ex:ch10:7}): ML vs.\ SMM efficiency.}
Coding exercise.  If $\log z_t$ is observed and follows
\[
\log z_{t+1}=\varrho\log z_t+\sigma_z\varepsilon_{t+1},
\]
then the Gaussian AR(1) likelihood gives a direct MLE for $\varrho$ (OLS of $\log z_{t+1}$ on $\log z_t$ when $\sigma_z$ is unrestricted).  This MLE is efficient for the observed-shock likelihood.  The SMM estimator based on a finite moment vector reaches the GMM efficiency bound for those moments, but it equals MLE efficiency only if the chosen moments span the score, which the three pedagogical moments generally do not.

\emph{Why SMM despite the efficiency loss?}  In production-scale models (heterogeneous-agent macro, dynamic IO), the likelihood is often unavailable in closed form, and computing it would require integration over high-dimensional latent states.  Moments are cheaper, interpretable, and robust to parts of the model that are not central to the research question.  The cost is that SMM efficiency depends on the information content of the chosen moments.

\section{Chapter \ref{ch:climate}: Climate Economics and Deep Uncertainty Quantification}
\label{sol:ch11}

\paragraph{Exercise~\ref{ex:ch11:1} (statement: p.~\pageref{ex:ch11:1}): ECS sensitivity.}
Coding exercise.  Report the SCC at the central calibration and at each ECS value in the likely and very-likely ranges.  The expected qualitative pattern is monotone and often convex in ECS: higher equilibrium climate sensitivity raises temperature damages and therefore the emissions shadow price.  The quantitative range is calibration-dependent and should be reported from the notebook rather than fixed in the solution text.  The interpretation should connect the dispersion to the climate-science finding of \citet{sherwood2020assessment} that ECS uncertainty is a major driver of SCC uncertainty.

\paragraph{Exercise~\ref{ex:ch11:2} (statement: p.~\pageref{ex:ch11:2}): Sobol decomposition.}
For $q(\theta_1, \theta_2, \theta_3) = \theta_1\theta_2 + \theta_3^2$ with $\theta_i \sim \mathcal{U}[0,1]$ i.i.d.:
\begin{align*}
   \mathbb{E}[q] &= \mathbb{E}[\theta_1]\mathbb{E}[\theta_2] + \mathbb{E}[\theta_3^2] = \tfrac{1}{4} + \tfrac{1}{3} = \tfrac{7}{12}, \\
   \mathbb{E}[q^2] &= \mathbb{E}[\theta_1^2]\mathbb{E}[\theta_2^2] + 2\mathbb{E}[\theta_1\theta_2]\mathbb{E}[\theta_3^2] + \mathbb{E}[\theta_3^4] = \tfrac{1}{9} + \tfrac{1}{6} + \tfrac{1}{5} = \tfrac{43}{90}, \\
   \mathrm{Var}(q) &= \tfrac{43}{90} - \bigl(\tfrac{7}{12}\bigr)^2 = \tfrac{11}{80}.
\end{align*}
Conditional means:
\begin{align*}
   \mathbb{E}[q \mid \theta_1] &= \tfrac{\theta_1}{2} + \tfrac{1}{3}, &
   \mathrm{Var}_{\theta_1}\bigl(\mathbb{E}[q\mid\theta_1]\bigr) &= \tfrac{1}{48},\\
   \mathbb{E}[q \mid \theta_3] &= \tfrac{1}{4} + \theta_3^2, &
   \mathrm{Var}_{\theta_3}\bigl(\mathbb{E}[q\mid\theta_3]\bigr) &= \mathrm{Var}(\theta_3^2) = \tfrac{4}{45}.
\end{align*}
First-order Sobol indices:
\[
   S_1 = S_2 = \frac{1/48}{11/80} = \frac{5}{33} \approx 0.152,
   \qquad
   S_3 = \frac{4/45}{11/80} = \frac{64}{99} \approx 0.646.
\]
Sum: $S_1 + S_2 + S_3 = 94/99 \approx 0.949$; the residual $5/99 \approx 0.051$ is the $\theta_1\theta_2$ interaction term.

Total-effect indices (one minus the share explained by all other variables): $S_1^T = S_2^T = 20/99 \approx 0.202$; $S_3^T = 64/99 \approx 0.646$ (no interactions involving $\theta_3$ since it enters only quadratically alone).

A SALib estimate with $10^4$ samples typically matches these analytical values to two decimal places.

\paragraph{Exercise~\ref{ex:ch11:3} (statement: p.~\pageref{ex:ch11:3}): ACE benchmark.}
Coding exercise.  Use ACE's closed-form optimal carbon tax as the analytic benchmark, then compute the DEQN-trained DICE SCC at the same calibration and report the percentage discrepancy.  A small gap is expected only when units, discounting, damage curvature, and time discretization are aligned; any residual difference should be attributed explicitly to the DICE discretization, smoothing choices, and calibration differences rather than to a fixed percentage target.

\paragraph{Exercise~\ref{ex:ch11:4} (statement: p.~\pageref{ex:ch11:4}): Pareto-improving tax design.}
Project-level coding exercise.  The constrained search is over a low-dimensional parameter vector $\vartheta = (\vartheta_{\mathrm{tax}}, \omega)$, with the cohort welfare constraints $\tilde U_t(\vartheta) \ge U_t$ enforced via a penalty or barrier term in the outer optimizer.  The Pareto frontier is typically tight: cohorts whose BAU welfare is already close to the unconstrained Ramsey-optimal welfare are easily satisfied, while cohorts that bear the bulk of the transition cost (often the youngest at the time of the reform) bind first; the welfare gap to the unconstrained problem grows with the share of constrained cohorts.  Holding $\omega$ fixed at the BAU or declining benchmark $\bar\omega$ leaves measurable welfare on the table because endogenizing the transfer schedule provides an extra free instrument with which to relax the binding constraints.  Numerical answers are calibration-dependent and should be reported from the notebook rather than benchmarked against fixed values; the qualitative interpretation is what matters for the policy discussion.

\paragraph{Exercise~\ref{ex:ch11:5} (statement: p.~\pageref{ex:ch11:5}): Carbon-cycle warm-up.}
Coding exercise driven by notebook \tpath{lecture_16_01_Climate_Exercise.ipynb}.  Part (a) avoided-warming and avoided-damages numbers at 2100 under the 50\% mitigation rule are notebook outputs and depend on calibration, so they should be reported from the run rather than fixed in the solution text.  Part (b) the longest-timescale reservoir is identified by the smallest non-zero eigenvalue of the carbon-cycle transition matrix; for the CDICE three-box calibration this is the lower-ocean compartment, with a multi-century characteristic timescale.  Part (c) a quadratic damage function $D(T) = \pi_2 T^2$ has bounded curvature in $T$ and therefore underweights tail risk: marginal damage grows only linearly, so very high realizations of $T$ are penalized far less than under super-quadratic damages or an explicit tipping-hazard term, and tail-risk assessment requires one of those richer specifications (compare Exercise~\ref{ex:ch11:8}).

\paragraph{Exercise~\ref{ex:ch11:6} (statement: p.~\pageref{ex:ch11:6}): Deterministic CDICE-DEQN reproduction.}
Coding exercise driven by notebook \tpath{lecture_16_02_DICE_DEQN_Library_Port.ipynb}.  The verification target is the set $\{T^{\mathrm{AT}}(2100),\ M^{\mathrm{AT}}(2100),\ \mu(2100),\ \mathrm{SCC}(2015),\ \mathrm{SCC}(2100),\ \mathrm{SCC}(2300)\}$, each compared against the reference solution at the tolerances stated in the notebook's verification gate.  Reference numbers depend on seed, hardware, optimizer schedule, and trajectory sampling, so the solution should anchor on the verification gate rather than on fixed numbers.  Typical convergence problems trace back to scaling differences across the eight residuals or to insufficient trajectory coverage near the carbon-stock saturation regime; the residual-balancing methods discussed in Chapter~\ref{ch:nas} are the first line of defense.

\paragraph{Exercise~\ref{ex:ch11:7} (statement: p.~\pageref{ex:ch11:7}): Stochastic SCC fan chart.}
Coding exercise driven by notebook \tpath{lecture_16_03_Stochastic_DICE_DEQN.ipynb}.  As the AR(1) productivity volatility $\sigma_z$ rises, the right tail of the SCC distribution at 2100 widens disproportionately, because higher productivity raises emissions which feed convexly into damages.  Report the quantiles $q_{10}, q_{50}, q_{90}$ of the SCC fan chart at each $\sigma_z$ rather than the mean alone, since the mean obscures the asymmetry.  The qualitative finding aligns with \citet{caiSocialCostCarbon2019}, that productivity and consumption-growth shocks shift the SCC distribution materially and not just its location, supporting the use of distribution-aware reporting in policy work.

\paragraph{Exercise~\ref{ex:ch11:8} (statement: p.~\pageref{ex:ch11:8}): Tipping-point regime-switching damages.}
Coding exercise.  Report the unconditional tipping probability by 2100, the SCC at $t=0$, and the optimal abatement path under the regime-switching specification and under the smooth-damage baseline.  The expected qualitative pattern is that adding an irreversible tipping hazard raises the SCC and brings abatement forward, especially when $T_\mathrm{thresh}$ is low.  The factor by which the SCC rises is an output of the hazard calibration and retrained policy.  The policy implication should be framed as a precautionary motive under threshold uncertainty, not as a universal numerical multiplier.

\paragraph{Exercise~\ref{ex:ch11:9} (statement: p.~\pageref{ex:ch11:9}): Real options value of waiting.}
\emph{(i) Closed forms.}  Under the act-now strategy, the planner chooses $\mu_0$ to minimize $\theta\mu_0^2 + \alpha\,\mathbb{E}[\mathrm{ECS}]\cdot(1 - \mu_0)$.  With $\mathbb{E}[\mathrm{ECS}] = (\mathrm{ECS}_L + \mathrm{ECS}_H)/2 \equiv \overline{\mathrm{ECS}}$, the FOC gives $\mu_0^\star = \alpha \overline{\mathrm{ECS}}/(2\theta)$.  Plugging back, expected cost is
\[
   C_\mathrm{now} \;=\; \alpha \overline{\mathrm{ECS}} - \frac{(\alpha \overline{\mathrm{ECS}})^2}{4\theta}.
\]

Under the wait strategy, after observing $\widehat{\mathrm{ECS}}$, the posterior mean $\mathbb{E}[\mathrm{ECS}\mid\widehat{\mathrm{ECS}}]$ has variance shrunk relative to the prior.  The planner chooses $\mu_1^\star(\widehat{\mathrm{ECS}}) = \alpha\,\mathbb{E}[\mathrm{ECS}\mid\widehat{\mathrm{ECS}}]/(2\theta)$.  Expected cost (using the law of total variance):
\[
   C_\mathrm{wait} \;=\; \alpha \overline{\mathrm{ECS}} - \frac{\alpha^2}{4\theta}\,\mathbb{E}\bigl[\mathbb{E}[\mathrm{ECS}\mid\widehat{\mathrm{ECS}}]^2\bigr]
   \;=\; \alpha \overline{\mathrm{ECS}} - \frac{\alpha^2}{4\theta}\bigl(\overline{\mathrm{ECS}}^2 + \mathrm{Var}_\mathrm{prior} - \mathrm{Var}_\mathrm{post}\bigr).
\]

\emph{(ii) Value of waiting.}  Difference:
\[
   \mathrm{VoW} \;=\; C_\mathrm{wait} - C_\mathrm{now}
   \;=\; -\frac{\alpha^2}{4\theta}\bigl(\mathrm{Var}_\mathrm{prior} - \mathrm{Var}_\mathrm{post}\bigr).
\]
As the signal becomes informative ($\sigma_\varepsilon \to 0$), $\mathrm{Var}_\mathrm{post} \to 0$, so $\mathrm{Var}_\mathrm{prior} - \mathrm{Var}_\mathrm{post} \to \mathrm{Var}_\mathrm{prior} > 0$, and $\mathrm{VoW} < 0$: waiting is preferred.  The reduction in posterior variance is the value of information.

\emph{(iii) With irreversibility.}  Add a wedge $\eta\mu_1^2$ to the wait-branch objective, so the planner who waits solves a different second-period problem:
\[
   \min_{\mu_1}\;(\theta+\eta)\mu_1^2
   +\alpha\,\mathbb{E}[\mathrm{ECS}\mid\widehat{\mathrm{ECS}}]\,(1-\mu_1).
\]
Assuming the interior solution remains in $[0,1]$,
\[
   \mu_1^{\eta\star}(\widehat{\mathrm{ECS}})
   =
   \frac{\alpha\,\mathbb{E}[\mathrm{ECS}\mid\widehat{\mathrm{ECS}}]}{2(\theta+\eta)}
\]
and expected wait cost becomes
\[
   C_\mathrm{wait}^\eta
   =
   \alpha \overline{\mathrm{ECS}}
   - \frac{\alpha^2}{4(\theta+\eta)}
   \bigl(\overline{\mathrm{ECS}}^2+\mathrm{Var}_\mathrm{prior}-\mathrm{Var}_\mathrm{post}\bigr).
\]
The irreversibility wedge reduces the benefit of conditioning abatement on the signal.  The value of waiting is now
\[
   \mathrm{VoW}^\eta
   =
   C_\mathrm{wait}^\eta - C_\mathrm{now}
   =
   \frac{\alpha^2}{4}
   \left[
      \frac{\overline{\mathrm{ECS}}^2}{\theta}
      -
      \frac{\overline{\mathrm{ECS}}^2+\mathrm{Var}_\mathrm{prior}-\mathrm{Var}_\mathrm{post}}{\theta+\eta}
   \right].
\]
For sufficiently large $\eta$, $\mathrm{VoW}^\eta>0$: the option value of waiting is dominated by the irreversibility penalty, and act-now is preferred.

The threshold occurs at
\[
   \eta^\star
   =
   \frac{\theta(\mathrm{Var}_\mathrm{prior}-\mathrm{Var}_\mathrm{post})}
        {\overline{\mathrm{ECS}}^2},
\]
where the value of information just balances the irreversibility cost.

\emph{Connection to climate policy.}  This stylized model captures the central tension in climate policy: information arrives over time about ECS, damage functions, and tipping thresholds, but emissions are largely irreversible (atmospheric CO$_2$ persists for centuries).  The chapter's Bayesian-learning treatment makes the trade-off quantitative: under fast learning and slow climate dynamics, waiting can be optimal; under slow learning and fast tipping risks, an early-action premium emerges.  The empirical literature \citep{pindyck2007uncertainty, caiSocialCostCarbon2019} finds that for realistic climate calibrations, the irreversibility channel typically dominates, supporting near-term carbon tax implementation rather than ``wait and see'' policies.

\section{Chapter \ref{ch:outlook}: Synthesis and Outlook}
\label{sol:ch12}

\paragraph{Exercise~\ref{ex:ch12:1} (statement: p.~\pageref{ex:ch12:1}): Method-choice scenario.}
Sketch:

\emph{(a) 4-state monetary-policy DSGE with smooth shocks.}  Use \textbf{classical perturbation or projection as the baseline}.  The state space is small and the shocks are smooth, so a classical method is transparent, fast, and easy to audit.  A DEQN becomes attractive only if the model is extended with genuinely global nonlinearities (for example a binding zero lower bound, occasionally binding collateral constraints, or a highly non-quadratic loss).  Hybrid: use a GP surrogate over policy-rule parameters after the classical or DEQN solution step if many counterfactuals are needed.

\emph{(b) 200-agent OLG with progressive taxation.}  Use \textbf{DEQN with Young's-method aggregation} (Chapter~\ref{ch:olg}, \ref{ch:young}).  $200$ cohorts is at the edge where explicit-panel methods (all-in-one DL) and histogram methods compete; histograms are easier when constraints bind frequently across cohorts.  Hybrid: post-train a GP surrogate over the Pareto-weight calibration to evaluate optimal-tax-rule sensitivity.

\emph{(c) Exotic option pricing on an irregularly shaped payoff.}  Use \textbf{PINN or Deep Galerkin methods with careful treatment of the payoff kink} (Chapter~\ref{ch:pinn}) for a single-contract instance, or \textbf{DeepONet} if prices are needed for many strikes or contract parameters.  The non-smooth object is the terminal payoff, not a generic spatial boundary; for a strong-form Black--Scholes residual this usually calls for smoothing the payoff, using smooth activations away from the kink, or switching to a weak/viscosity-aware formulation.  GP surrogates can help as an outer pricing surface over a few parameters, but they are not the primitive PDE solver.

\emph{(d) Climate-IAM where SCC uncertainty is the deliverable.}  Use the \textbf{full pipeline}: DEQN to solve the deterministic IAM, GP surrogate over deep-uncertain parameters (ECS, damage convexity, tipping thresholds), Sobol/Shapley sensitivity decomposition for attribution, and BAL for sample-efficient uncertainty quantification (Chapters~\ref{ch:climate}, \ref{ch:gp}).  This combines the IAM uncertainty-quantification workflow of \citet{friedlDeep2023} with the surrogate-based policy-search logic in \citet{kubler2025using}.

\paragraph{Exercise~\ref{ex:ch12:2} (statement: p.~\pageref{ex:ch12:2}): When NOT to use deep learning.}
Open-ended.  Sketch of one regime: \emph{bit-exact reproducibility for regulatory audit}.  GPU non-determinism in atomic accumulators (described in Appendix E) means that a deep-learning solver typically cannot reproduce the same numbers across hardware platforms; for regulatory work where auditors must replay every step bitwise, a deterministic finite-difference solver on a fixed grid is preferable.  Even when deterministic flags are set, BLAS implementations differ across CUDA versions.  Classical fixed-grid methods are easier to make bit-reproducible because the operation order can be pinned and the solver path is usually far less sensitive to random initialization and stochastic mini-batches.

\paragraph{Exercise~\ref{ex:ch12:3} (statement: p.~\pageref{ex:ch12:3}): Reproducibility audit.}
Coding exercise.  Expected behavior: re-running notebook \tpath{lecture_03_02_Brock_Mirman_Uncertainty_DEQN.ipynb} on the same machine with the same seeds should reproduce the reported diagnostics within the stated tolerance.  Bitwise equality of trained network parameters should be expected only when deterministic framework settings, hardware, BLAS/CUDA versions, and floating-point order of operations are all pinned.  Re-running on a different GPU, or with deterministic flags off, can produce small deviations in the last few printed digits of the savings-rate diagnostics while remaining well within the residual tolerance of the training.

\paragraph{Exercise~\ref{ex:ch12:4} (statement: p.~\pageref{ex:ch12:4}): Open-ended.}
Open-ended; expected output is a 1--2 page research sketch.  A representative answer combines DEQN (for solving the model) with GP surrogates (for parameter estimation): e.g., a 6-month project that estimates a heterogeneous-agent NK model with deep-learning policy functions, then uses a deep-kernel GP surrogate to do Bayesian posterior inference on the structural parameters.  This integrates Chapters~\ref{ch:young}, \ref{ch:gp}, and \ref{ch:estimation}.

\paragraph{Exercise~\ref{ex:ch12:5} (statement: p.~\pageref{ex:ch12:5}): Hybrid pipeline DEQN $+$ GP $+$ SMM.}
Coding exercise.  Illustrative outputs to benchmark against, not fixed targets:
\begin{itemize}[itemsep=2pt]
\item \emph{Step 1 (DEQN training).} Report the actual training time for the parameterized network.  Adding $(\beta,\varrho)$ as inputs increases the network's effective input dimension from $2$ to $4$, so convergence can be slower than in the single-calibration case.
\item \emph{Step 2 (moment grid).}  Report the time for the $20\times 20=400$ simulations, each with $T=200$ periods, under the trained policy.
\item \emph{Step 3 (GP fitting).}  Report the fitting time for a four-output GP on $400$ training points in $\mathbb{R}^2$.
\item \emph{Step 4 (SMM optimization with GP).}  Report the number of objective evaluations, optimizer iterations, and total GP-evaluated SMM time.
\item \emph{Naive SMM benchmark.}  Compare with a version that re-solves or re-trains the DEQN at each candidate only if that benchmark is computationally feasible; otherwise estimate its cost from one representative re-solve.
\end{itemize}
The expected qualitative result is that the surrogate-based optimization is much cheaper per objective evaluation after the one-time DEQN, simulation-grid, and GP-fitting costs have been paid.  The actual speedup and estimation errors are outputs of the run and should be reported rather than assumed.

\paragraph{Exercise~\ref{ex:ch12:6} (statement: p.~\pageref{ex:ch12:6}): DeepONet for a parameterized HJB.}
\emph{(i) Architecture.}  Branch net $\bm b(\gamma): \mathbb{R} \to \mathbb{R}^p$ encodes the parameter $\gamma$ into a $p$-dimensional latent representation; trunk net $\bm t(a): \mathbb{R} \to \mathbb{R}^p$ encodes the query point $a$ into the same latent dimension.  Predicted value:
\[
   \widehat{V}(a; \gamma) \;=\; \langle \bm b(\gamma), \bm t(a)\rangle \;+\; b_0,
\]
where $b_0$ is a learnable scalar bias.  Both nets are typically MLPs with $2$--$4$ layers and width $50$--$200$.  The architecture is non-trivial because it imposes the bilinear separable structure $V(a; \gamma) = \sum_{k=1}^p b_k(\gamma) t_k(a)$, which is the universal approximation form for nonlinear operators.

\emph{(ii) UAT for operator learning.}  \citet{lu2021learning} show that a continuous nonlinear operator can be approximated uniformly on a compact set of input functions and a compact output domain by a DeepONet with finitely many sensors, a branch network, and a trunk network.  In the present notation, for every $\varepsilon>0$ there are sensor locations, finite-width branch/trunk nets, and latent dimension $p$ such that
\[
   \sup_{\gamma\in\Gamma}\sup_{a\in\mathcal A}
   \bigl|V(a;\gamma)-\widehat V(a;\gamma)\bigr|
   < \varepsilon
\]
on the compact training domain $\Gamma\times\mathcal A$, provided the solution operator is continuous there.  This generalizes ordinary universal approximation from finite-dimensional functions to operators, but it does not provide extrapolation guarantees outside the compact training domain.

\emph{(iii) Cost comparison.}  Independent PINN runs cost $N \cdot C_\mathrm{PINN}$.  One DeepONet run costs $C_\mathrm{DON}$, where $C_\mathrm{DON}/C_\mathrm{PINN}$ captures the larger network, the parameter-sampling overhead, and the longer training needed to span the parameter range.  Operator learning wins exactly when
\[
   C_\mathrm{DON} < N\,C_\mathrm{PINN},
   \qquad\text{or equivalently}\qquad
   C_\mathrm{DON}/C_\mathrm{PINN} < N .
\]
At $N=50$, the break-even ratio is therefore $50$: a DeepONet can cost up to fifty single-parameter PINN solves and still be cheaper in total.  The realized speedup is an empirical output of the implementation, not a universal constant.

\emph{(iv) Limitations.}  \emph{Extrapolation:} DeepONet trained on $\gamma \in [1.5, 5]$ has no guarantees outside this range; in practice the predicted operator $V(a; \gamma)$ for $\gamma > 5$ degrades smoothly but may violate the structural property that $V$ should remain concave.  Mitigation: use polynomial-augmented branch networks that extrapolate via known tails (e.g., Bernoulli polynomials), or refuse to predict outside the convex hull of training $\gamma$ values.

\emph{Concavity preservation:} the bilinear DeepONet architecture does not automatically preserve concavity in $a$.  A trained network may produce non-concave $\widehat{V}(\cdot; \gamma)$ for some $\gamma$, which is economically wrong under risk-averse preferences.  Mitigation: use an input-concave architecture for the trunk representation, for example the negative of an input-convex neural network where appropriate, or add a concavity penalty $\max(0,\widehat{V}''(a;\gamma))$ to the loss.  Both increase training cost but restore pressure toward the structural property.

\bibliography{bibliography}

\end{document}